%% file: review.tex
\pdfoutput=1 
\documentclass[twoside,12pt]{book}

\input{review_macros.tex}

\newcommand{\dmmatm}{(2.40\pm0.15)\,10^{-3}\eV^2}
\newcommand{\dmmsun}{(7.58\pm0.21)\,10^{-5}\eV^2}
\newcommand{\ssatm}{1.02\pm 0.04}
\newcommand{\ttsun}{0.484\pm0.048}
\newcommand{\ssCH}{0.07\pm0.04}

\begin{document}

\thispagestyle{empty}
\centerline{ \hfill Version 3, \today\hfill}
\vspace{5mm}
\color{black}
\vspace{0.5cm}
\centerline{\LARGE\bf\color{rossos} Neutrino masses and mixings and...}
 \medskip\bigskip
   \centerline{\large\bf Alessandro Strumia}\vspace{0.2cm}
   \centerline{\em Dipartimento di Fisica dell'Universit\`a di Pisa and INFN, Italia}\vspace{0.4cm}
   \centerline{\large\bf  Francesco Vissani}\vspace{0.2cm}
   \centerline{\em INFN, Laboratori Nazionali del Gran Sasso,
 Theory Group, I-67010 Assergi (AQ), Italy}
\vspace{1cm}
\color{blu}\centerline{\large\bf Abstract}
\begin{quote}\large\indent
We review experimental and theoretical results related to
neutrino physics with emphasis on neutrino masses and mixings,
and outline possible lines of development.


\color{black}
\end{quote}
\vspace{5mm}

\color{black}

\noindent
We try to present the physics in a simple way, 
avoiding unnecessary verbosity, formalisms and details.
Comments, criticisms, etc are welcome.
We want to upgrade  this `review' or `book'  at the light of future developments: therefore
for the moment  we publish it only on
$$\hbox{\url{arXiv.org/abs/hep-ph/0606054}}
\qquad\hbox{ and }\qquad
\hbox{\url{www.pi.infn.it/~astrumia/review.html}}$$
in electronic form.
Nowadays it has several advantages over printed form,
but prevents an official refereeing process.
We asked some experts to privately referee
the parts less connected to our research activity.
The review is organized as follows:
\begin{itemize}
\item Chapter \ref{Introduction}: a brief overview.
\item Chapters \ref{Neutrino}, \ref{Oscillations}, \ref{detecting}: the basic tools.
\item Chapters \ref{atm}, \ref{sun}: the established discoveries (solar and atmospheric).
\item Chapters \ref{oscexp}, \ref{nonOsc}:  future searches for neutrino  oscillation and neutrino  masses. 
\item Chapter \ref{anomalies}: unconfirmed anomalies.
\item Chapters \ref{cosmology}, \ref{supernova}: neutrinos in cosmology and astronomy.
\item Chapters \ref{flavour}, \ref{extra}, \ref{Applications}: speculative lines of development.

\end{itemize}
Acronyms are listed in appendix~\ref{Acronyms}.
Appendix~\ref{Statistics} summarizes
basic facts concerning statistics.
The last pages contain the detailed index.

\newpage

\chapter{Introduction}\label{Introduction}
Neutrinos physics is interesting because
neutrino experiments recently discovered 
something new,
rather than giving only more precise measurements of SM parameters, or
stronger bounds on unseen new physics.
`Solar' and `atmospheric' data 
directly show that lepton flavour is not conserved.


The next step is identifying the new physics responsible of these anomalies.
Theoretical simplicity suggests oscillations of massive neutrinos,
which can fit data provided that mixing angles among the SM  neutrinos are unexpectedly large.
The observed flavour conversions could be produced by other mechanisms.
Present data strongly disfavor alternative exotic possibilities, such as neutrino decay or 
oscillations into extra  `sterile' neutrinos $\nu_{\rm s}$
(i.e.\ light fermions with no SM gauge interactions,
as opposed to the three `active' SM  neutrinos) and
show some hints for the  characteristic  features of oscillations.


\smallskip

Future experiments should confirm and complete this picture.
Oscillations can be directly seen
by precise reactor and long-baseline beam experiments,
that are nowadays respectively testing the solar and atmospheric anomalies.
Other so far unseen oscillation effects 
(`atmospheric' oscillations into $\nu_e$, and CP-violation)
could be discovered soon or never,
depending how large they are.
Future non-oscillation experiments should
detect neutrino masses, and test if they violate lepton number.
Realistically, these developments could be achieved in the next $10$ or $ 20$ years.
Understanding neutrino propagation will allow to do astrophysics, cosmology, geology using
 neutrinos. 

\smallskip

At this point we will still have to understand 
the origin of the neutrino mass scale, $m_\nu \sim (0.01\div 0.05)\eV$.
Presumably neutrino masses are of Majorana type and
are the first manifestation
of a new scale in nature, $\Lambda_L\sim v^2/m_\nu\sim 10^{14}\GeV$.
This could be the mass of new particles,
maybe 3 right-handed neutrinos
with a $3\times 3$ matrix of Yukawa couplings.
Experiments drove the recent progress but cannot
directly test such high energies.
Leptogenesis, $\mu\to e\gamma$ and related processes
could be other manifestations of the new physics behind neutrino masses.

\smallskip

Alternatively, experiments could discover 
something different, e.g.\ some new light particle,
and require a significant change in the above picture.

\bigskip

Before starting, we present a quick overview.
We employ standard, usually self-explanatory notations, precisely defined in the next sections.

\section{Past}
After controversial results, in  1914 Chadwick established that
the electrons emitted in radioactive $\beta$ decays have a continuous spectrum,
unlike what happens in  $\alpha$ and $\gamma$
decays.
It took some time to ensure that, 
if the $\beta$ decay process were $^A_Z{\rm X}\to {}^{~~~A}_{Z-1}{}{\rm X}\, e$ with only two particles in the final state,
energy conservation would unavoidably imply a monochromatic electron spectrum.
(Today we also know that Lorentz invariance requires an even number of fermions in decays).


\index{Pauli}
On 4 december 1930 Pauli proposed a `desperate way out' to save energy conservation,
postulating the existence a new neutral particle, named ``neutron'',
with mass `of the same order of magnitude as the electron mass' and maybe
`penetrating power equal or ten times bigger than a $\gamma$ ray'~\cite{History}.
The estimate of the cross-section was suggested by the old idea that particles emitted in 
$\beta$ decays
were previously bound in the parent nucleus (as happens in $\alpha$ decays) 
--- rather than created in the decay process.
In a 1934 paper containing `speculations too remote from reality'
(and therefore rejected by the journal Nature) Fermi overcame this misconception and
introduced a new energy scale (the `Fermi' or `electroweak' scale)
in the context of a model able of predicting neutrino couplings in terms of
$\beta$-decay lifetimes.
Following a joke by Amaldi, the new particle was renamed neutrino\footnote{In italian
the suffixes for small and large are -ino and -one.
The root of neutrino is `ne$\,$uter', `not either' in latin.
Presently adopted pronunciations differ from the latin-italian pronunciation.
E.g.\ japanese experimentalists recently gave a main contribution to neutrino physics,
but pronounce it as $\includegraphics[width=13mm]{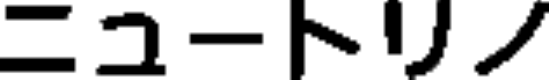}$
(`nyu$\,$to$\,$ri$\,$no'), doubly distorted due to limitations of hiragana (to) and english (nyu) phonetics.}
after that the true neutron had been identified by Chadwick.
Neutrinos were finally directly observed by Cowan and Reines
in 1956 in a nuclear reactor experiment
and found to be left-handed in 1958.

In those years $K^0\leftrightarrow \bar{K}^0$ effects were clarified,
and this lead Pontecorvo to discuss  $\nu\leftrightarrow\bar\nu$ oscillations 
with maximal mixing in a 1957 paper;
maximal $\nu_e \leftrightarrow\nu_\mu$ oscillations of solar neutrinos were considered already in 1967~\cite{Pontecorvo}.
In 1962 $\nu_e \leftrightarrow \nu_\mu$ `virtual transmutations' were mentioned by 
Maki, Nakagawa and Sakata in the context of a wrong 
 model of leptons bound inside hadrons~\cite{MNS}.
For these reasons some authors now name `MNS' (or `MNSP', or `PMNS')
the neutrino mixing matrix, although other authors consider this as improper
as naming `indians' the native habitants of America.
The work that lead to the first
evidence for a neutrino anomaly was done by Davis et al.\
who, using a technique suggested by Pontecorvo~\cite{Cl},
since 1968 measured a $\nu_e$ solar rate smaller than the what predicted by 
Bahcall et al.~\cite{BP}.
Despite significant efforts, up to few years ago, it was not clear if there was a solar neutrino problem
or a neutrino solar problem.
Phenomenologists pointed out a few clean signals possibly produced by oscillations,
but could not tell which ones are large enough to be detected.
Since doing experiments is the really hard job, 
only in 2002 two of these signals have been discovered.
The SNO solar experiment found evidence for $\nu_{\mu,\tau}$ appearance and
the KamLAND experiment confirmed the solar anomaly discovering 
disappearance of $\bar\nu_e$ from  terrestrial (japanese) reactors.

In the meantime, analyzing the atmospheric neutrinos, 
originally regarded as  background for proton decay searches, in 1998 the japanese (Super)Kamiokande experiment~\cite{SKatm}
established a second neutrino anomaly,
confirmed around 2004 by K2K~\cite{K2K},
the first long base-line neutrino beam experiment.

\begin{table}[t]
$$\includegraphics[width=\textwidth]{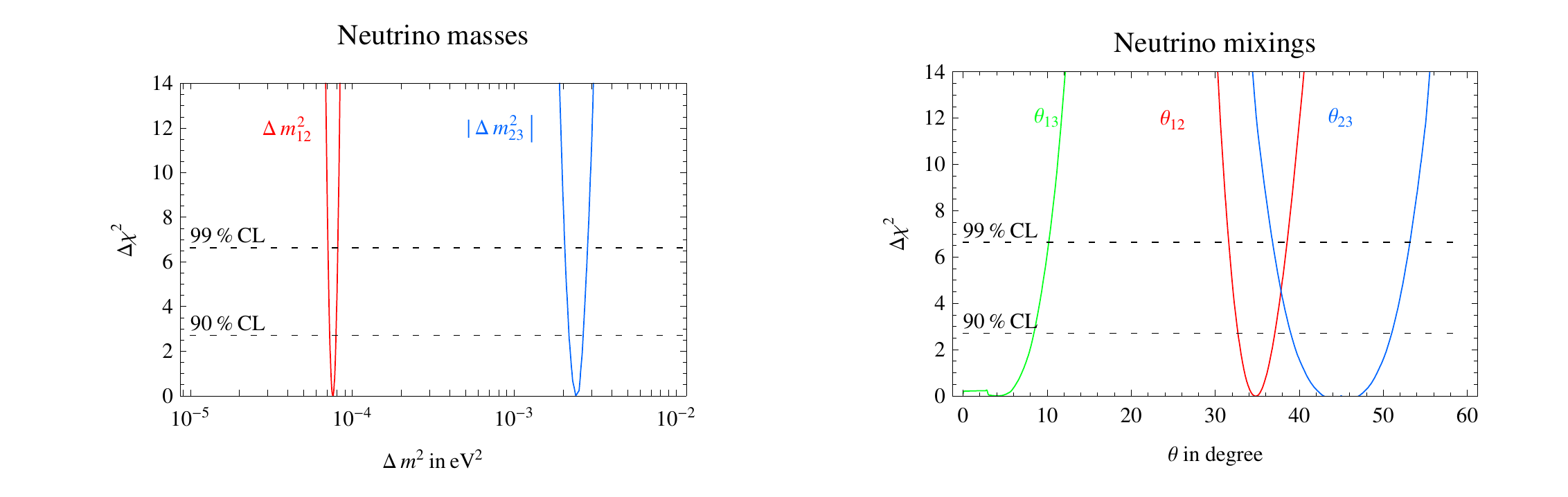}$$
$$\begin{array}{|lrlc|}\hline
\hbox{Oscillation parameter}&\multicolumn{2}{c}{\hbox{central value}} &\hbox{$99\%$ CL range}\\  \hline
\color{rossos}
\hbox{solar mass splitting} & \color{rossos}\Delta m^2_{12} ~=
& \color{rossos}\dmmsun & \color{rossos}(7.1\div 8.1)\,10^{-5}\eV^2 \\
\color{blus}\hbox{atmospheric mass splitting~}  &\color{blus}|\Delta m^2_{23}| ~=
&\color{blus}\dmmatm~ &\color{blus}(2.1\div 2.8) \,10^{-3}\eV^2\\
\color{rossos}
\hbox{solar mixing angle} &\color{rossos} \tan^2 \theta_{12} ~=& \color{rossos}\ttsun
&\color{rossos}31^\circ < \theta_{12}<39^\circ \\
\color{blus}\hbox{atmospheric mixing angle} &\color{blus}\sin^2 2\theta_{23} ~=
&\color{blus}  \ssatm &37^\circ <\theta_{23}< 53^\circ\\
\color{verdes}\hbox{`CHOOZ' mixing angle}  &\color{verdes} \sin^2 2\theta_{13}  ~=&
\color{verdes}\ssCH &\color{verdes} 0^\circ< \theta_{13}<13^\circ\\ \hline
\end{array}$$\vspace{-4mm}
\caption[Experimental values of oscillation parameters]
{\em Summary of present
information on neutrino masses
and mixings from oscillation data.
\label{tab:tab1}}
\end{table}

\section{Present}\label{Present}
Table~\ref{tab:tab1} summarizes the oscillation interpretation of the two established
neutrino anomalies:
\begin{itemize}
\item The {\bf atmospheric} evidence.
{\sc SuperKamiokande}~\cite{SKatm} observed disappearance of $\nu_\mu$ and $\bar\nu_\mu$ atmospheric neutrinos,
with `infinite' statistical significance ($\sim 17 \sigma$).
The anomaly is also seen by  {\sc Macro} and other atmospheric experiments.
If interpreted as oscillations, one needs $\nu_\mu\to \nu_\tau$
with quasi-maximal mixing angle.
The other possibilities,
$\nu_\mu \to \nu_e$  and $\nu_\mu \to \nu_{\rm s}$, cannot explain the anomaly
and can only be present as small sub-dominant effects.
The SK discovery is confirmed by $\nu_\mu$ beam experiments: 
K2K~\cite{K2K} and {\sc NuMi}~\cite{NuMi}.
Table~\ref{tab:tab1} reports global fits for oscillation parameters.


\item The {\bf solar} evidence.
Various experiments~\cite{Cl,Ga,SK,SNO} see a $8\sigma$ evidence for a $\sim 50\%$ deficit of solar $\nu_e$.
The SNO experiment sees a $5\sigma$ evidence for
$\nu_e \to \nu_{\mu,\tau}$ appearance
(solar neutrinos have energy much smaller than $m_{\mu}$ and $m_\tau$, so that
experiments cannot distinguish $\nu_\mu$ from $\nu_\tau$).
The KamLAND experiment~\cite{KamLAND} sees a $6\sigma$ evidence
for disappearance of $\bar\nu_e$ produced by nuclear reactors.
If interpreted as oscillations, one needs a large but not maximal mixing angle,
see table~\ref{tab:tab1}.
Other oscillation interpretations in terms of a small mixing angle enhanced by matter effects,
or in terms of sterile neutrinos, are excluded.
\end{itemize}
There are few unconfirmed anomalies related to neutrino physics.
\begin{itemize}
\item[1.]
{\bf LSND}~\cite{LSND} claimd a $3.8\sigma$ 
$\bar{\nu}_\mu \to \bar{\nu}_e$ anomaly:
{\sc Karmen}~\cite{Karmen} and {\sc MiniBoone}~\cite{MiniBoone} do not confirm the signal,
excluding the na\"{\i}ve interpretations in terms of oscillations with $\Delta m^2 \sim 1\eV^2$ and small mixing.

\item[2.] Preliminary data from {\bf MiniBoone}~\cite{MiniBoone} show a $\sim 3\sigma$
$\nu_\mu\to \nu_e$ anomaly: it cannot be fitted by vacuum oscillations since the anomaly is
concentrated in the lower part of the energy spectrum probed by MiniBoone.
\end{itemize}
If the LSND and {\sc MiniBoone} anomalies are caused by new physics, 
something exotic is needed.
\begin{itemize}

\item[3.] {\bf NuTeV}~\cite{NuTeV} claims a $3\sigma$ anomaly in neutrino {\em couplings}:
the measured ratio between the $\nu_\mu /$iron NC and CC couplings is about $1\%$
lower than some SM prediction.
Specific QCD effects that cannot be computed in a reliable way
could be the origin of the NuTeV anomaly.

\item[4.] A reanalysis~\cite{evid} of the {\bf Heidelberg-Moscow} data~\cite{HM} 
performed by a sub-set of the collaboration claims a hint for
violation of lepton number, the significance varies between $2\sigma$ and $6\sigma$
being inversely proportional to the number of authors who claim it.
The simplest interpretation would be in terms of Majorana neutrino masses,
implying approximatively degenerate neutrinos with mass $m\sim 0.4 \eV$.


\end{itemize}
Furthermore, there are some important constraints
\begin{itemize}
\item LEP data tell that there are {\bf only 3 neutrinos} lighter than $M_Z/2$.
Extra light fermions with no gauge interactions might exist,
and could play the r\^ole of `sterile neutrinos'.

\item 
Together with atmospheric and K2K data
the CHOOZ~\cite{CHOOZ} bound on the $\bar\nu_e$ survival probability 
restricts $\theta_{13}$, the last unseen mixing angle
that induces {\bf $\mb{\nu}_e \leftrightarrow \mb{\nu}_{\mu,\tau}$
oscillations at the atmospheric frequency} $\Delta m^2_{\rm atm}$ to be
$$\sin^2 2 \theta_{13}  = \ssCH. $$

\item 
Under plausible {\em assumptions}, cosmology implies that
{\bf neutrinos are lighter than about 0.2 eV}~\cite{cosmomnu}.
Assuming that neutrinos have Majorana masses,
bounds on $0\nu2\beta$ decay imply $m_\nu\circa{<}1\eV$.
Bounds on $\beta$ decay imply $m_\nu\circa{<}2\eV$.
Astrophysics gives somewhat weaker bounds.

\end{itemize}

\begin{table}[p]
\begin{center}
\begin{tabular}{lcclc}
experiment &  status &name  &start & cost in M\EURtm{} \\
\hline
W\v{C} (3 kton)   &terminated& Kamiokande &1983~~ & 5\\
W\v{C} (50 kton)  &running& SuperKamiokande &  1996~~ & 100\\
W\v{C} (1000 kton)  &proposals&HyperK, UNO? & 2015?~ & 500?\\
Monopole and CR obs.
& terminated & {\sc MaCRO} & 1994 & 40\\
Solar B &running& SNO & 2001~~ & $100+ 500$ (target)\\
Solar Be &construction& Borexino & 2006?~ & 25\\
Solar $pp$ &running& Gallex $\approx$ SAGE/2 & 1991~~ & $1+15$ (target)\\
Solar $pp$ &proposals&many or none&2010?&100??\\
Reactor &terminated&CHOOZ   &  1997~~  &1.5    \\
Reactor &running &KamLAND   & 2002~~ &  20 \\
Reactor &proposal &Double-CHOOZ   & 2009~~ &  10 \\
Long baseline & terminated & K2K & 1999~~ &  (beam)\\
Long baseline & construction &CNGS& 2006~~ &50 (beam) + 80 (detectors)\\
Long baseline & construction &NuMI& 2004~~ & 110 (beam) + 60 (detector) \\
Long baseline & proposal &No$\nu$a& 2011~~ & 160 \\
Long baseline & approved & T2K &2009~& 130\\
Long baseline & proposals & super-beam & 2010? &500?\\
Long baseline & discussions & $\nu$ factory & 2020? &2000?\\[3mm]
CR observatory & construction & Auger & 2006 & 50\\
$\nu$ telescope &  approved &  ANITA & 2007 &  35  \\
$\nu$ telescope & construction & IceCube & 2009? & 150\\
$\nu$ telescope & proposals & KM3NeT & 2012? & 300\\
$\beta$ decay at 0.2 eV &approved& Katrin   & 2012~ & 35   \\
$0\nu2\beta$ at 0.01 eV& proposals &  & 2012? & $70$?   \\
ton-scale DM search & proposals && 2012? & 70?\\
$\nu$ couplings &terminated & NuTeV & 1996 &\\[3mm]
$e\bar{e}$ collider (103 GeV)&terminated & LEP &1989~~ & 1200\\
$e\bar{e}$ collider  (0.5 TeV)&proposals & ILC &2020? & 10000?\\

$pp$ collider (7 TeV)&construction & LHC &2008~ & 3000\\
$pp$ collider (20 TeV)&not approved &SSC & &10000?\\
Satellite & running & WMAP & 2003~~ & 150\\
Satellite & construction & GLAST & 2008 & 200\\
Gravit.\ wave detector & running  & LIGO + VIRGO & 2002 & 250 + 100 \\
Super $B$ factory & discussions & BarbaBelle? & 2015? & 600?\\
Space station &running &  ISS   &  2006? & 100000? \\  

\end{tabular}
\end{center}
  \caption[Main neutrino experiments]{\em Main neutrino experiments.
Costs are estimated approximating 
$\raisebox{-0.1ex}{\hbox{\includegraphics[height=1.6ex]{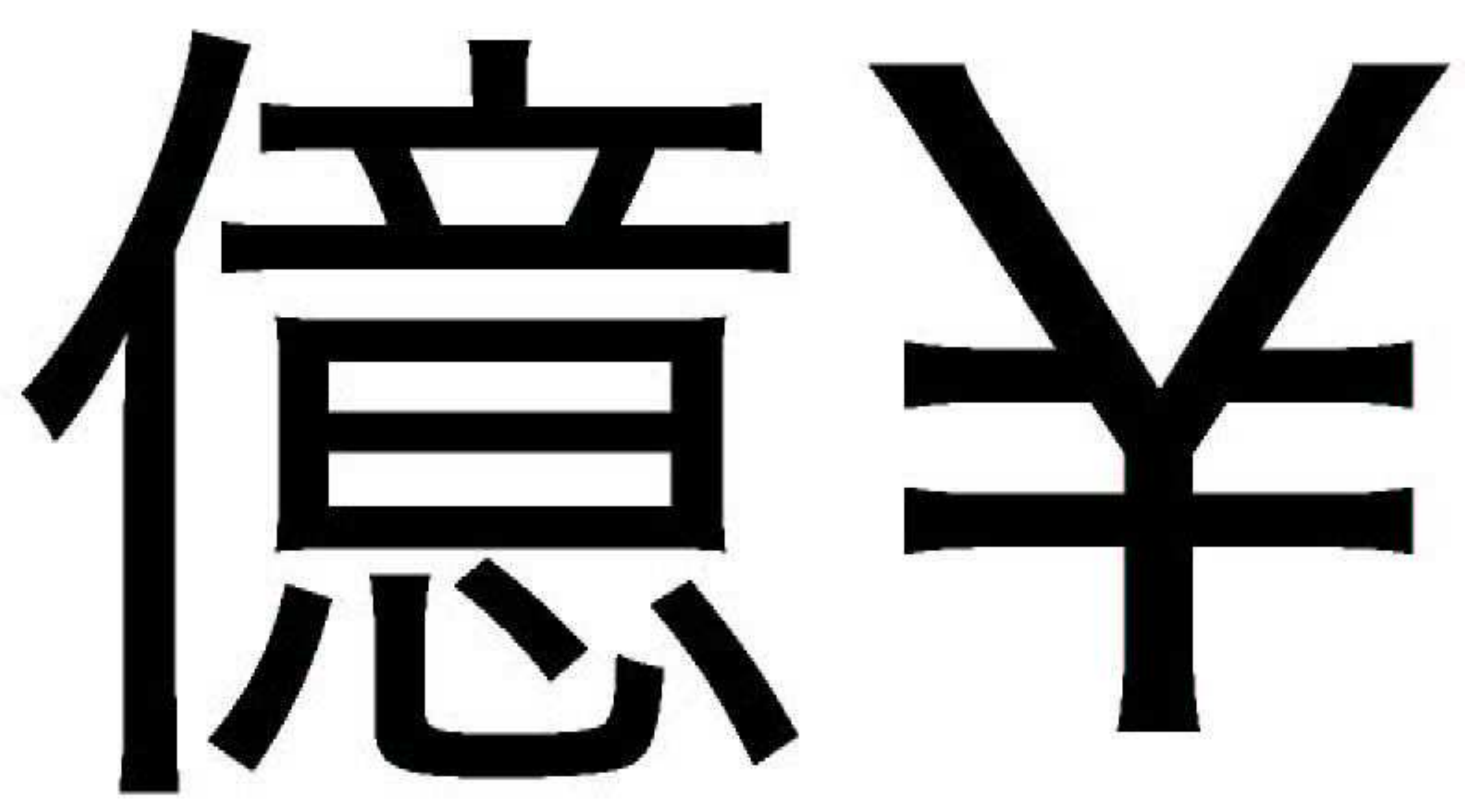}}}
\approx \hbox{\rm M\$}\approx\hbox{\rm M\EURtm}$,
which is about the total life salary of a physicist.
This allows to estimate that manpower costs (not included)
are often comparable to the cost of the experiment.
Some experiments obtained their target material as free rent,
but bigger experiments will have to produce it.
Among the many caveats, we emphasize that
costs of future experiments are rarely underestimated.
  \label{tab:price}}
\end{table}

\begin{figure}[p]
$$\includegraphics[width=16cm]{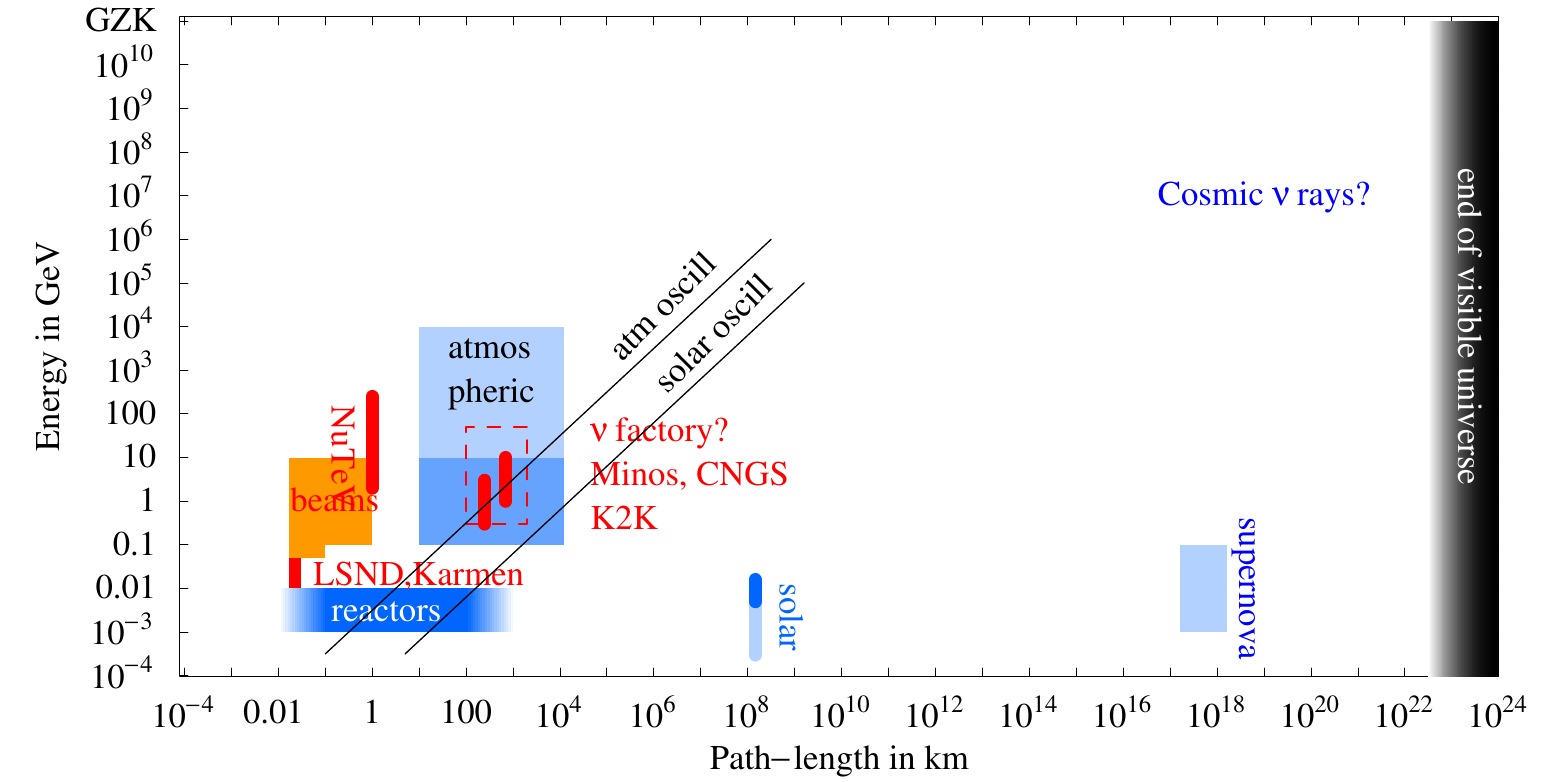}$$
\caption[Explored regions of the $L/E$ plane]{\label{fig:LE}\em Regions 
explored using natural (blue) or artificial (red) neutrino sources.
Solar and atmospheric oscillations occur below the black lines.}
\end{figure}

\begin{figure}[p]
$$\includegraphics[width=16cm]{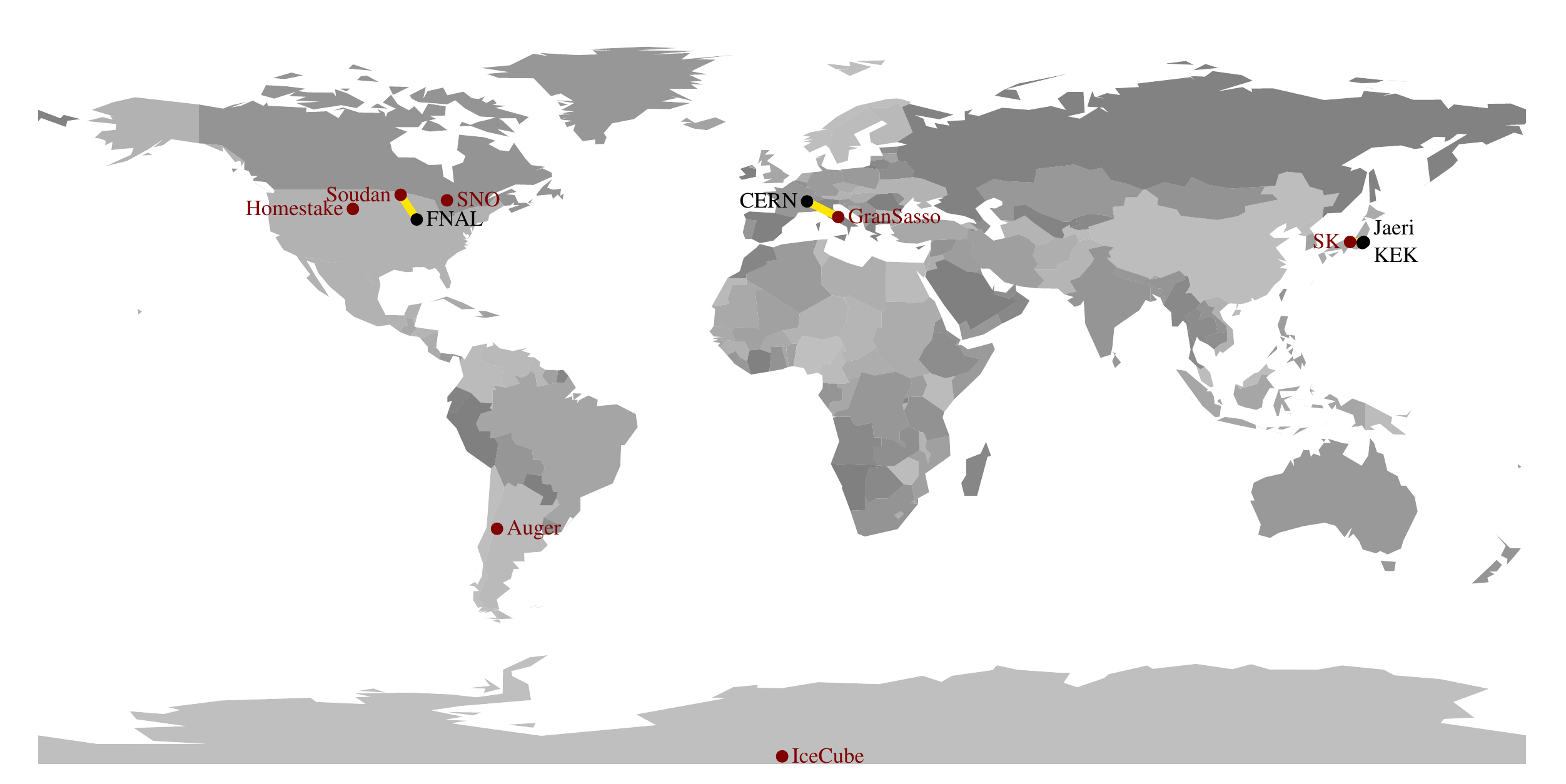}$$
\caption[Location of neutrino experiments]{\label{fig:map}\em Location of main neutrino-related observatories.
}
\end{figure}


\begin{table}
\begin{center}
\begin{tabular}{lcccc}
Source & Mechanism & Flavour & Typical energy & Rate\\ \hline
Sun & nuclear fusion & $\nu_e$ & $(0.1\div 20)\MeV$ & known\\
Atmosphere & $\pi,\mu$ decays & $\nubarnu_{e,\mu}$ & $(0.1\div 1000)\GeV$ & known\\
Big-Bang & thermal & all & $\sim \meV$ & too small\\
Supernov\ae & thermal & all & \circa{<}50\MeV & poorly known\\
Galactic CR & $\pi,\mu$ decays? & $\nubarnu_{e,\mu}$? &up to $10^{6}\GeV$? &poorly known\\
Extragalactic CR & $\pi,\mu$ decays? & $\nubarnu_{e,\mu}$? &up to $10^{11}\GeV$? &poorly known\\
Sun, Earth & DM annihilations & all & $<m_{\rm DM}$ &?\\
Earth & nuclear decays & $\bar\nu_e$ & \circa{<}\hbox{few MeV} & poorly known\\ \hline
Fission reactors & nuclear fission & $\bar\nu_e$ & $\circa{<}10\MeV$ & known\\
Conventional beams & $\pi$ or $\mu$ decay & $\nubarnu_{e,\mu}$ & tunable, $\sim \GeV$ & known\\
$\nu$-factory & $\mu^-$ decay & $\nu_\mu,\bar\nu_e$ & tunable, $\sim10\GeV$ & well known\\
$\nu$-factory & $\mu^+$ decay & $\bar\nu_\mu,\nu_e$ & tunable, $\sim10\GeV$ & well known\\
$\beta$-beams & Nuclear decay & $\nu_e$ or $\bar\nu_e$ & tunable, $\sim\GeV$ & well known\\
\end{tabular}
\end{center}
  \caption[Neutrino sources]{\em Possible natural and artificial neutrino sources.
  \label{tab:sources}}
\end{table}

\section{Future?}
The recent discoveries have been achieved
studying natural phenomena with neutrino detectors.
Solar, atmospheric, supernova, terrestrial and reactor neutrinos are freely available,
and are interesting not only for fundamental physics.
Further progress is expected along these lines, but maybe no new discoveries.
To discover new effects which might affect natural neutrinos in a minor way,
it will be necessary to first produce the physical system to be later studied,
as already customary in collider physics.
Even if significant improvements seem possible,
the necessity of building appropriate neutrino beams
 will make progress more expensive and 
slow, as illustrated in table~\ref{tab:price}.
Fig.\fig{LE} shows the regions of the  ($L,E_\nu$) plane 
which have so far explored, with at least one neutrino flavour.

\begin{itemize}


%




%



\item[2010]
{\bf LHC} begins to collide $pp$ at $\sqrt{s}= 7 \TeV$ aiming at collecting 1/fb
of integrated luminosity, maybe having some indirect relevant impact on neutrino physics.

{\bf IceCube}~\cite{ICECUBE}, located at south pole, tries to start neutrino astronomy at energies larger than about a TeV.
Other projects that use sea water instead of ice are being discussed
(ANTARES, NEMO, NESTOR, to be possibly merged in KM3NET).

Two intense $\nu_\mu$ beam experiments will start:  {\bf T2K}~\cite{T2K} and {\bf NO}\mb{\nu}{\bf A}~\cite{Nova}.
Their main goal is completing the measurement of
neutrino oscillation parameters: $\theta_{13}$,...
The T2K experiment will use SK as detector, that could sooner or later
be upgraded up to a `HyperKamiokande' Mton W\v{C},
relevant also for proton decay and other searches.

The high-intensity beam also allows tests of neutrino couplings and
searches for lepton-flavour violating $\mu$ decays.
The {\bf MEG} $\mu\to e\gamma$ experiment~\cite{muegExp} might give results.

A sensitivity to $\theta_{13} \circa{>}0.05$ could be reached by the 
{\bf DoubleChooz} reactor experiment.
At the same time, the  {\bf Daya Bay}  reactor experiment could start data taking.

\item[2011] Many of the experiments planned for 2010 will realistically start of give data in 2011.

\item[2012]
A sensitivity to neutrino masses $m_\nu \circa{>}0.2\eV$ could be reached by the 
{\bf Katrin}~\cite{Katrin} $\beta$-decay experiment.

Studying CMB weak lensing, the {\bf Planck} satellite experiment could 
test $\sum m_\nu$ with an uncertainty of $0.15\eV$.

Some big {\bf neutrino-less double-beta decay} experiment
will search for violation of lepton number (e.g.\ Majorana $\nu$ masses).
Planned experiments are GERDA and later MAJORANA, CUORE, EXO, MOON, superNEMO.

Some experiment could precisely study {\bf sub-MeV solar neutrinos}.

Somebody believes that on 21 december 2012 the Mayan calendar ends and 
solar neutrinos will start to interact more strongly, melting the Earth and forcing us
to a dramatic revision of current theories.

\item [2020] A {\bf neutrino factory}~\cite{NuFact} or 
{\bf beta-beam}~\cite{BetaBeams} could perform the `ultimate' search
for oscillation effects.

\item [20??] Thousands of neutrino events will be detected
at the next core-collapse galactic {\bf supernova} explosion
allowing to study astrophysics and maybe neutrino oscillations.

\item[2030] december 4: $\nu$ centennial.
This is the only safe expectation.\footnote{Most of the expectations 
for the years 2002---2009 present in our first drafts have been realized.}

\end{itemize}

\newpage
\input{review_basic.tex}\newpage
\input{review_cross.tex}\newpage
\input{review_atmsun.tex}\newpage
\input{review_future.tex}\newpage
\input{review_nonosc.tex}\newpage
\input{review_hints.tex}\newpage
\input{review_cosmo.tex}\newpage
\input{review_astro.tex}\newpage
\input{review_flavour.tex}\newpage
\input{review_behind.tex}\newpage
\appendix
\input{review_appendici.tex}
\input{review_ref.tex}



\newpage 
\footnotesize\listoftables\newpage
\begin{multicols}{2}\listoffigures
 \tableofcontents
\end{multicols}

\end{document}

%% file: review_macros.tex
\def\baselinestretch{1}
\usepackage{graphicx, multicol, color}
\usepackage{marvosym, amsfonts, floatflt,enumerate}
\usepackage{ifpdf}
\ifpdf
\usepackage{hyperref, pdfsync, epstopdf}	
\pdfoutput=1
\else
\usepackage[dvips,bookmarks]{hyperref}	
\fi
\hypersetup{colorlinks,bookmarksopen,bookmarksnumbered,citecolor=verdes,
linkcolor=blus,pdfstartview=FitH,urlcolor=rossos}
\def\hhref#1{\href{http://arxiv.org/abs/#1}{arXiv:#1}} 

\newcommand{\kmwe}{\,{\rm kmwe}}

\newcommand{\mT}{\mb{m}_T}
\newcommand{\mio}[1]{}
\newcommand{\drop}[1]{}

\newcommand{\sW}{s_{\rm W}}

\def\bea{\begin{eqnarray}}
\def\eea{\end{eqnarray}}
\def\be{\begin{equation}}
\def\ee{\end{equation}}
\def\ba{\begin{array}}
\def\ea{\end{array}}

\newcommand{\nubarnu}{\raisebox{1ex}{\hbox{\tiny(}}\overline\nu\raisebox{1ex}{\hbox{\tiny)}}}

\definecolor{rosso}{cmyk}{0,1,1,0.4}
\definecolor{rossos}{cmyk}{0,1,1,0.55}
\definecolor{rossoc}{cmyk}{0,1,1,0.2}
\definecolor{blu}{cmyk}{1,1,0,0.3}
\definecolor{blus}{cmyk}{1,1,0,0.6}
\definecolor{bluc}{cmyk}{1,1,0,0.1}
\definecolor{verde}{cmyk}{0.92,0,0.59,0.25}
\definecolor{verdec}{cmyk}{0.92,0,0.59,0.15}
\definecolor{verdes}{cmyk}{0.92,0,0.59,0.4}
\definecolor{giallo}{cmyk}{0,0,1,0}
\definecolor{gialloverde}{cmyk}{0.44,0,0.74,0}
\newcommand{\ecm}{e\,\mathrm{cm}}
\newcommand{\bAk}[3]{\langle #1|#2|#3\rangle}

\newcommand{\Dsl}{D\!\!\!\raisebox{2pt}[0pt][0pt]{$\scriptstyle/$}}
\newcommand{\Tr}{\hbox{Tr}\,}

\newcommand{\One}{1\hspace{-0.63ex}{\rm I}}
\newcommand{\bk}[2]{\langle#1|#2\rangle}

\newcommand{\km}{\,{\rm km}}
\newcommand{\cm}{\,{\rm cm}}
\newcommand{\m}{\,{\rm m}}
\newcommand{\s}{\,{\rm s}}
\newcommand{\yr}{\,{\rm yr}}
\newcommand{\kg}{\,{\rm kg}}
\newcommand{\K}{{}^\circ {\rm K}}

\newcommand\diag{\hbox{diag}\,}
\newcommand{\BR}{\hbox{BR}\,}
\newcommand{\MGUT}{M_{\rm GUT}}
\newcommand{\Ds}{D\!\!\!\!/\,}
\newcommand{\ds}{\partial\hspace{-1.3ex}/}

\newcommand{\SU}{\hbox{SU}}
\newcommand{\beq}{\begin{equation}}
\newcommand{\eeq}{\end{equation}}

\font\tenrsfs=rsfs10 at 12pt
\font\sevenrsfs=rsfs7
\font\fiversfs=rsfs5
\newfam\rsfsfam
\textfont\rsfsfam=\tenrsfs
\scriptfont\rsfsfam=\sevenrsfs
\scriptscriptfont\rsfsfam=\fiversfs
\def\mathscr#1{{\fam\rsfsfam\relax#1}}
\def\Lag{\mathscr{L}}
\def\mathW{\mathscr{W}}
\newcommand{\mb}[1]{\mbox{\normalsize\boldmath $#1$}}

\newcommand{\fig}[1]{~{\rm \ref{fig:#1}}}
\newcommand{\tab}[1]{~{\ref{tab:#1}}}
\def\circa#1{\,\raise.3ex\hbox{$#1$\kern-.75em\lower1ex\hbox{$\sim$}}\,}
\newcommand{\eV}{\,{\rm eV}}
\newcommand{\meV}{\,{\rm meV}}
\newcommand{\keV}{\,{\rm keV}}
\newcommand{\MeV}{\,{\rm MeV}}
\newcommand{\GeV}{\,{\rm GeV}}
\newcommand{\TeV}{\,{\rm TeV}}
\newcommand{\eq}[1]{~(\ref{eq:#1})} 

\renewcommand{\Re}{\hbox{Re}\,}
\renewcommand{\Im}{\hbox{Im}\,}
\def\be{\begin{equation}}
\def\ee{\end{equation}}
\def\bc{\begin{center}}
\def\ec{\end{center}}
\def\bea{\begin{eqnarray}}
\def\eea{\end{eqnarray}}

\newcommand{\riga}[1]{\noalign{\hbox{\parbox{\textwidth}{#1}}}\nonumber}

\newcommand{\md}[1]{\langle #1 \rangle}
\makeatletter
\newcounter{totcits}
\def\sottonumero{\stepcounter{totcits}}
\def\art{\@ifnextchar[{\eart}{\oart}}
\def\eart[#1]#2#3#4#5#6{{\rm #2}, {\em #3 \rm #4} {\rm (#6) #5 ({\em\hhref{#1}})}\sottonumero}
\def\hepart[#1]#2{{\rm #2, \em\hhref{#1}}\sottonumero}
\newcommand{\oart}[5]{{\rm #1}, {\em #2 \rm #3} {\rm (#5) #4}\sottonumero}

\newcommand{\rep}[2]{#1, #2\sottonumero}

\newcommand{\NP}{Nucl. Phys.}
\newcommand{\Jhep}{{\rm J.HEP}}
\newcommand{\PRL}{Phys. Rev. Lett.}
\newcommand{\PL}{Phys. Lett.}
\newcommand{\PR}{Phys. Rev.}
\newcommand{\book}[5]{{\rm #1}, {\em #2} (#5) {\rm #4, #3}\sottonumero}
\newcounter{alphaequation}[equation]
\def\thealphaequation{\theequation\hbox to
0.6em{\hfil\alph{alphaequation}\hfil}}
\def\eqnsystem#1{
\def\@eqnnum{{\rm (\thealphaequation)}}
\def\@@eqncr{\let\@tempa\relax \ifcase\@eqcnt \def\@tempa{& & &} \or
  \def\@tempa{& &}\or \def\@tempa{&}\fi\@tempa
  \if@eqnsw\@eqnnum\refstepcounter{alphaequation}\fi
\global\@eqnswtrue\global\@eqcnt=0\cr}
\refstepcounter{equation} \let\@currentlabel\theequation \def\@tempb{#1}
\ifx\@tempb\empty\else\label{#1}\fi
\refstepcounter{alphaequation}
\let\@currentlabel\thealphaequation
\global\@eqnswtrue\global\@eqcnt=0 \tabskip\@centering\let\\=\@eqncr
$$\halign to \displaywidth\bgroup \@eqnsel\hskip\@centering
$\displaystyle\tabskip\z@{##}$&\global\@eqcnt\@ne
\hskip2\arraycolsep\hfil${##}$\hfil& \global\@eqcnt\tw@\hskip2\arraycolsep
$\displaystyle\tabskip\z@{##}$\hfil
\tabskip\@centering&\llap{##}\tabskip\z@\cr}

\def\endeqnsystem{\@@eqncr\egroup$$\global\@ignoretrue} \makeatother



%% file: review_basic.tex
\chapter{Neutrino masses}\label{Neutrino}\label{MD}

\section{Massless neutrinos in the SM}
In all observed processes baryon number $B$
and lepton number $L$ are conserved.
Searches for proton decay and for neutrino-less double-beta decay (section~\ref{0nu2beta})
give the dominant constraints.
The SM provides a nice interpretation of these results,
missed by old pre-SM models where $e,\nu,p,n,\pi$ were considered
as point-like fundamental particles.
Yukawa introduced a $pn\pi^-$ coupling in order to account for nuclear forces.
However, the analogous Yukawa-type coupling $p\nu\pi^-$,
obtained by replacing $n$ with 
$\nu$, would give rise to unseen $p$ decay.
These models do not explain why the proton is stable,
but explain why the electron is stable: electric charge is conserved
and the electron is the lightest `charged' particle.
Therefore theorists introduced a new conserved charge, called baryon number $B$,
under which the proton is the lightest charged particle.
This makes the proton stable and forbids all other 
$B$-violating processes, like $pp\to \bar{e}\bar{e}$.

Conservation of lepton number was introduced for analogous reasons.
In particular it forbids unseen $L$-violating neutrino mass terms
without forbidding the neutron mass term.
Exact conservation of $B$ and $L$ was widely considered on the same footing of
conservation of electric charge.

\medskip

The advent of gauge theories and of the SM changed the situation: today
$B$ and $L$ automatically emerge as approximatively conserved charges.

The key difference between $n$ and $\nu$
is that the neutron is a bound state of quarks.
In the SM the Yukawa coupling $pn\pi^-$ arises from renormalizable
strong interactions of quarks,
while the $p\nu\pi^-$ coupling
would correspond to a non renormalizable $qqq\nu$ interaction.
In fact the most general
$\SU(3)_{\rm c} \otimes\SU(2)_L\otimes {\rm U}(1)_Y$
gauge-invariant renormalizable Lagrangian that can be written with the SM fields
(the Higgs  doublet $H$ and the observed fermions:
the three lepton doublets $L = (\nu, \ell_L)$, the lepton singlets $E = \ell_R$,
etc. See table\tab{gAi} at page~\pageref{tab:gAi} for the full list.) beyond `minimal' terms (kinetic and gauge interactions) 
can only contain the following Yukawa and Higgs-potential terms
\begin{equation}
\label{eq:LSM}
\Lag_{\rm SM} = \Lag_{\rm  minimal} +
(\lambda_E^{ij} ~  E^iL^jH^* +
\lambda_D^{ij} D^iQ^jH^*  +  
 \lambda_U^{ij} \,  U^iQ^jH +  
\hbox{h.c.}) + m^2 |H|^2  - \frac{\lambda}{4}|H|^4
\end{equation}
where $i,j = \{1,2,3\}$ are flavour indices.
No term violates baryon number $B$ and lepton flavour $L_e, L_\mu, L_\tau$
(and in particular  lepton number $L = L_e + L_\mu + L_\tau$), that therefore
naturally emerge as {\em accidental} symmetries.\footnote{To be more precise,
quantum
anomalies violate some of these charges in a way which will be relevant only
when discussing baryogenesis in section~\ref{leptogenesi}.
To be less precise, massless neutrinos were already
 suggested, before the SM,  by the $V-A$ structure of weak interactions.}
In the SM there is no need of imposing by hand a stable proton and massless neutrinos.
This line of reasoning leads to more successful predictions:
baryon flavour and CP are violated in a very specific way,
described by the CKM matrix,
giving rise (among other things) to characteristic rates of
$K^0\leftrightarrow \bar{K}^0$, $B^0\leftrightarrow \bar B^0$ transitions.
Since CKM CP violation is accompanied by flavour mixing,
CP-violating effects which do not violate flavour, like electric dipoles,
are strongly suppressed, in agreement with experimental data.\footnote{Most of these theoretical successes would be lost if extensions
 of the SM motivated by the hierarchy `problems', such as the MSSM,
 will be confirmed by future data.}

The Higgs vev breaks $\SU(2)_L\otimes {\rm U}(1)_Y\to {\rm U}(1)_{\rm em}$
\beq \langle H \rangle = (0,v)\qquad\hbox{with}\qquad v\approx 174\GeV,\eeq
and gives Dirac masses to
charged leptons and quarks\footnote{Dirac and Majorana quadri-spinors are
usually presented following the historical development and notation, but this is confusing.
Quadri-spinors are representations of the Lorentz group and of parity,
that was believed to be an exact symmetry.
Since we now know that this is not the case,
it is more convenient to use the basic fermion representations of the Lorentz group:
the 2-dimensional Weyl spinors.
The only Lorentz invariant mass term that can be written with a single Weyl fermion $\psi$ is the Majorana term $\psi^2$.
This mass term breaks a U(1) symmetry $\psi \to e^{i q_\psi \varphi} \psi$ under which
$\psi$ might be charged (it could be electric charge, hypercharge, lepton number, ...).
For example, a Majorana neutrino mass is possible if the electric charge
of neutrinos is exactly zero.
With two Weyl fermions $\psi$ and $\psi'$
one can write three mass terms: $\psi^2$, $\psi^{\prime 2}$ and $\psi\psi'$.
In many interesting cases (all SM fermions, except maybe neutrinos)
the Lagrangian has an unbroken U(1) symmetry (electromagnetism, in the SM) under which
$\psi$ and $\psi'$ have opposite charges, so that
then $\psi\psi'$ is the only allowed mass term.
It is named `Dirac mass term', and one can group $\psi$ and $\psi'$ in one 4-component Dirac spinor
$\Psi = (\psi, \bar\psi')$. 
The electron gets its mass from a Dirac term, that joins two different Weyl fermions
that are therefore named $e_L$ and $e_R$ rather than $\psi$ and $\psi'$.
If one knows what is doing this is the simplest notation.
Since $e_L$ and $e_R$ have opposite electric charges
one usually prefers to use names like  `$\bar{e}_R$' or `$e^c_R$' or `$e^c_L$' in place of `$e_R$'.
For a clean recent presentation of Weyl spinors see~\cite{Haber}.}
mass terms $m_i = \lambda_i v$
$$ m_E~ \ell_R \ell_L + m_D ~ d_R d_L + m_U ~ u_R u_L$$
but neutrinos remain massless.
Within the SM, neutrinos are fully described by the Lagrangian term
$$ \bar{L} i \Ds L$$
i.e.\ a kinetic term plus gauge interactions with the massive vector bosons, $\bar{\nu} Z \nu$
and $\bar{\nu} W \ell_L$.

\begin{figure}
$$\includegraphics[width=16cm]{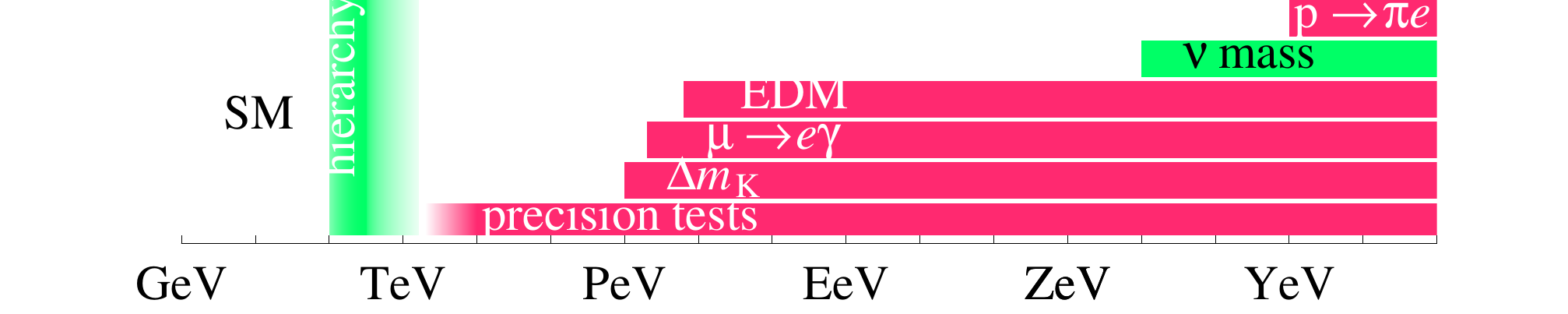}$$
\caption[Bounds on non renormalizable operators]{\em Bounds on the scale $\Lambda$ that suppresses non-renormalizable operators 
that violate $B,L,{\rm CP},L_f, B_f$ and
affect precision data.
Maybe the `hierarchy problem' suggests new-physics around few hundred GeV.\label{fig:NRO}}
\end{figure}

\medskip

\section{Massive neutrinos beyond the SM}
Observations of neutrino masses call for an extension of the SM,
and plausible extensions of the SM suggested neutrino masses.

\medskip

The new physics
can be either lighter or heavier than 100 GeV,
the maximal energy that has been experimentally explored so far.

Since LEP excluded 
new particles coupled to the $Z$ boson and lighter than $M_Z/2$,
the first case can only be realized by adding light {\em right-handed neutrinos\/} $\nu_R$.
Unlike other right-handed leptons and quarks, $\nu_R$ are neutral under all SM gauge interactions:
gauge invariance allows a Majorana mass term for right-handed neutrinos,
$M \nu_R^2/2$, that breaks lepton number.
Neutrinos can acquire Dirac masses like all other fermions if conservation of lepton number
(that in the SM is automatic) is imposed by hand, such that $M=0$.
In such a case, 
the neutrino Yukawa coupling
$\lambda_N \nu_R LH$  
gives the Dirac neutrino mass $m_\nu = \lambda_N v \approx 0.1\eV$ for $\lambda_N\sim 10^{-12}$.

\medskip

Alternatively, generic new physics too heavy for being directly studied
manifests at low energy as non renormalizable operators (NRO),
suppressed by heavy scales $\Lambda$.
NRO give small corrections, suppressed by powers of $E/\Lambda$, to
physics at low energy $E\ll \Lambda$, that is therefore well described
by a renormalizable theory.
The SM is the low energy effective theory of something and we would like to know what 
this something is.
Experimentally, this something can be searched in various ways:
a) going to higher energies;
b) searching for small effects in precision experiments at low energies;
c) searching for small effects enhanced by a large coherence factor;
d) studying rare processes;
e) searching processes that cannot be generated by renormalizable operators.

This last possibility is how the Fermi scale made its first appearance.
The 1896 discovery of radioactivity by Becquerel 
($\beta$-rays were soon identified by Rutherford)
lead Fermi to add to the QED Lagrangian
non renormalizable $pne\nu$ operators 
suppressed by the electroweak scale.

History might repeat now.
Adding NRO to the SM Lagrangian,
$L_e, L_\mu, L_\tau, B$ are no longer accidentally conserved:
\begin{eqnarray}\Lag &=& \Lag_{\rm SM} + \frac{(LH)^2}{2\Lambda_L}  + 
\frac{1}{\Lambda_{B}^2} \bigg[ c_1 (\bar U\bar D)(QL) + c_2 (QQ)(\bar U\bar E) +
+ c_3 (QQ)(QL) + \\ &&  c_4 (Q\tau^a Q)(Q\tau^a L) + c_5 (\bar D\bar U)(\bar U\bar E) +
 c_6 (\bar U\bar U)(\bar D\bar E) + \hbox{h.c.}\bigg]+\cdots.
 \end{eqnarray}
 With only one light higgs doublet there is only one kind of
dimension-5 operator: $(LH)^2 = (\nu h^0 - e_L h^+)^2$.
Inserting the Higgs  vev $v$, this operator gives
a Majorana neutrino mass term, $m_\nu \nu_L^2/2$, with $m_\nu = 
v^2/\Lambda_L\sim 0.1 \eV$ for $\Lambda_L\sim 10^{14\div 15}\GeV$,
close to but below the energy $M_{\rm GUT}\approx 2\cdot 10^{16}\GeV$
where SU(5) gauge unification might happen.
Neutrino masses might be the first manifestation of a new length scale $\Lambda_L$
in nature.

\medskip

The six dimension-6 operators violate $B$ and conserve $B-L$ giving rise
to  proton decay into anti-leptons: $p\to \bar{e}\pi^0, \bar\nu \pi^+,\ldots$
with width $\tau_p^{-1} \sim m_p^5/\Lambda_B^4$.
The first two $B$-violating operators can be mediated by tree-level exchange of heavy vector bosons,
and are predicted by unification models.
The strongest constraint on the proton life-time $\tau_p$ comes from
the
SK experiment, which monitored about $10^{10}$ moles of protons for a few years.
Therefore the present bound is
\beq\tau_p\circa{>} 10^{10} N_A\,\hbox{yr}\approx 10^{34}\,\hbox{yr}\qquad \hbox{i.e.}\qquad
\Lambda_B \circa{>}10^{15}\GeV.\eeq
Observing proton decay would open another window on physics at  high energy scales.

Furthermore, other operators (not shown) give additional sources of CP and hadronic flavour violation,
or affect precision LEP data. Fig.\fig{NRO} summarizes present bounds.
In conclusion, we today have three evidences for non-renormalizable interactions.
Two of them are the solar and neutrino anomalies.
The third one corresponds to case c), and is gravity:
the non renormalizable gravitational couplings, suppressed by $E/M_{\rm Pl}$,
sum coherently over many particles giving the well known Newton force.

\smallskip


\begin{figure}
$$\includegraphics[width=\textwidth]{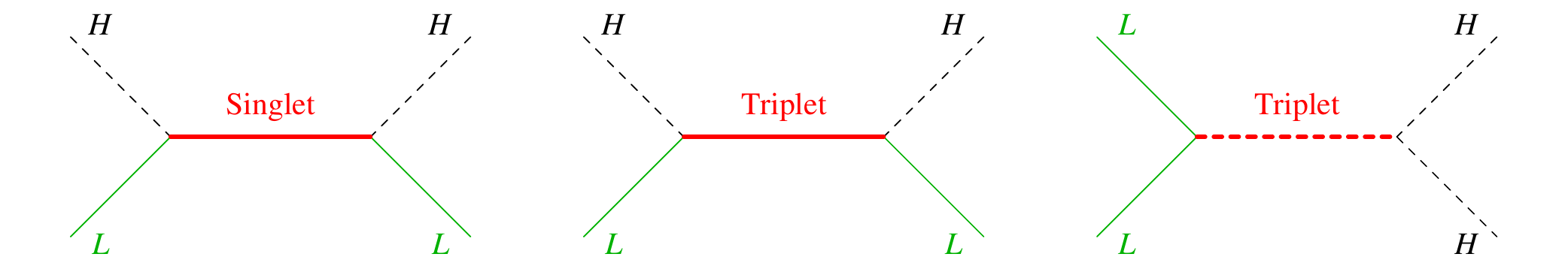}$$
\caption[Mediation of neutrino masses]{\em 
The neutrino Majorana mass operator $(LH)^2$ can be mediated by tree level exchange of: {\rm I)} a fermion singlet (`see-saw'); {\rm II)} a fermion  triplet;
{\rm III)} a scalar triplet.\label{fig:Feyn}}
\end{figure}

\section{See-saw}\index{See-saw}
It is tempting to speculate about which renormalizable extensions of the SM
can generate the Majorana neutrino mass operator $(LH)^2$.
However, the considerations in the previous section indicate that this might be
untestable metaphysics: whatever is the source of the $(LH)^2$ operator,
this operator is all what we can see at low energy; different sources cannot be discriminated.\footnote{We will present our best hopes of making progress on this issue:
 the matter/antimatter asymmetry (that however is only one number) in section~\ref{leptogenesi}
and weak-scale supersymmetry (that however has not yet been discovered)
in section~\ref{LFV}.}

Tree level exchange of 3 different types of new particles can generate neutrino masses:
right-handed neutrinos, and fermion  or scalar $\SU(2)_L$ triplets, as we now discuss.
The first possibility is known as `see-saw', although some authors
apply the same name to all three possibilities.

\subsection{Type I see-saw: extra fermion singlets}\label{ExtraSinglets}
The simplest possibility is adding new fermions with no gauge interactions, that play the r\^ole of `right-handed neutrinos',
$N=\nu_R$. As already anticipated they can have both a Yukawa interaction $\lambda_N$ and a Majorana mass $M_N$:
\beq\label{eq:Lseesaw2}
\Lag = \Lag_{\rm SM} +\bar N_i i\ds N_i + (\lambda_N^{ij} ~  N^i L^jH  + \frac{M_N^{ij}}{2}   N_i N_j +\hbox{h.c.})\eeq
such that neutrinos generically have a $6\times 6$ Majorana/Dirac mass matrix
\beq
\bordermatrix{& \nu_L & \nu_R\cr \nu_L& 0 & \mb{\lambda}^T_N v \cr \nu_R & \mb{\lambda}_N v & \mb{M}_N}\eeq
where bold-face reminds that $\mb{\lambda}_N$ and $\mb{M}_N$ are $3\times 3$ flavour matrices.
The values of $\lambda_N$ and $M_N$ could be related
to the unification scale,
or to supersymmetry-breaking or to the size of extra dimensions
or to some other `fundamental' physics,
but in practice we do not know.
We focus on two interesting extreme limits:

\paragraph{\em Pure Majorana neutrinos.} 
If $M_N\gg \lambda_N v$ 
the full $6\times 6$ mass matrix
 gives rise to 3 (almost) pure right-handed neutrinos with heavy Majorana masses $\mb{M}_N$,
and to 3 (almost) pure left-handed neutrinos with light Majorana masses 
$\mb{m}_\nu = - (v\mb{\lambda}_N )^T  \mb{M}^{-1}_N (v\mb{\lambda}_N)$.

\indent
We now rederive the same result proceeding in a different way.
Integrating out the heavy neutrinos gives
a non-renormalizable effective Lagrangian that only contains the observable low-energy fields.
Fig.\fig{Feyn}a shows that $\nu_R$ exchange generates the Majorana mass operator 
$(L_iH)(L_j H)/2$
with coefficient $-(\mb{\lambda}_N^T  \mb{M}^{-1}_N \mb{\lambda}_N)_{ij}$.
This {\em `see-saw'} mechanism~\cite{seesaw} works naturally and
fits nicely in grand unified extension of the SM.
It generates the 9 measurable neutrino mass parameters (see section~\ref{MD}) from
$\mb{\lambda}_N$ and $\mb{M}_N$, that contain 18 unknown parameters.
Still, it might be not impossible to test it experimentally (sections~\ref{leptogenesi},~\ref{LFV}).

\paragraph{\em Pure Dirac neutrinos.} If $M_N \ll \lambda_N v$ 
the full $6\times 6$ mass matrix gives 3 Dirac neutrinos $\Psi = (\nu_L,\bar{\nu}_R)$
with mass $m_\nu = \lambda_N v$. 
The vanishing of $M_N$ can be justified if conservation of lepton number is {\em imposed} 
(rather than obtained, as in the SM).
In order to get the observed neutrino masses one needs
 $\lambda_N\sim  10^{-12}$ --- much smaller than all other SM Yukawa couplings.
While Majorana masses arise `naturally',  
one needs to `force' the theory in order to get Dirac neutrinos.
Due to these  \ae{}stethical considerations, Majorana neutrinos are considered as more likely.
If Dirac neutrinos will turn out to be the right possibility,
after complementing the SM with a few more fermions one should
understand why they are so surprisingly light.

In general, $M_N$ can be anywhere: e.g.\
$M_N \sim v$ (again giving light Majorana neutrinos, the measured masses are reproduced for
neutrino Yukawa couplings comparable to the electron Yukawa)
or $M_{N}\sim \lambda_N v$ (giving $6$ mixed neutrinos with comparable masses).

\subsection{Type III see-saw: extra fermion triplets}
The extra fermion $N$ added in the previous section could be a $\SU(2)_L$ triplet
with zero hypercharge
rather than a singlet.
The Lagrangian contains analogous $\mb{\lambda}_N$ and $\mb{M}_N$ flavour matrices:
\beq\label{eq:Lseesaw3}
\Lag = \Lag_{\rm SM} +\bar N_i iD\hspace{-1.5ex}/\, N_i +
\bigg[\lambda_N ^{ij} ~ N^a_i  
(L_j \cdot \tau^a\cdot \varepsilon \cdot H)  + \frac{M_N^{ij}}{2}   N_i^a N_j^a+\hbox{h.c.}\bigg].  \eeq
The index $a$ runs over $\{1,2,3\}$,
$\tau^{a}$ are the Pauli matrices
and $\varepsilon$ is the permutation tensor
($\varepsilon_{12}=+1$).
The three components of $N$ are $N^3$ with charge zero and
$(N^1\pm i N^2)/\sqrt{2}$ with charge $\pm 1$.

As long as $M_N \gg v$ (triplets lighter than $M_Z/2$ have been excluded by LEP) 
everything works in the same way:
triplet exchange generates the Majorana mass operator, $(LH)^2$.
This mechanism is sometimes known as `type III see-saw'.

\subsection{Type II see-saw: extra scalar triplet}
We have seen how neutrino masses can be obtained
adding new {\em fermionic} (`matter') fields.
Alternatively, one can add
one {\em scalar} (`Higgs') triplet $T^a$ ($a=\{1,2,3\}$)
with hypercharge $Y_T = 1$
(and so composed by three components with charge $0,+1,+2$), 
such that the most generic renormalizable Lagrangian is
\beq \label{eq:Ltriplet}
\Lag = \Lag_{\rm SM} +|D_\mu T|^2 - M_T^2 |T^a|^2 +\frac{1}{2}
({\lambda}_T^{ij} L^i  \varepsilon \tau^a L^j T^a  +  \lambda_H M_T \, ~H  \varepsilon \tau^a H ~T^{a*} +\hbox{h.c.})\eeq
where $\lambda_T$ is a symmetric flavour matrix,
$\varepsilon$ is the permutation matrix,
and $\tau^a$ are the usual $\SU(2)_L$ Pauli matrices.
Integrating out the heavy triplet generates the Majorana neutrino masses operator $(LH)^2$
(see fig.\fig{Feyn}b) inducing neutrino masses
$m_\nu^{ij}= {\lambda}_T^{ij} \lambda_H v^2/M_T^2$.
This mechanism is sometimes known as `type II see-saw'.
A smaller number of unknown flavour parameters are needed to describe
one extra scalar triplet than the extra fermion scalars or triplets.

\medskip

To conclude, we discuss how these possible sources of neutrino masses are consistent with
plausible extensions of the SM:
gauge unification, supersymmetry and thermal leptogenesis.

SU(5) and SO(10) gauge unification allow to understand the charges of the observed fermions
and suggest a new `unification' scale of about
$10^{16}\GeV$, between $\Lambda_L$ and the Planck scale.
According to this theoretical framework the most attractive way of
generating neutrino masses is adding one right-handed neutrino per family:
it is predicted by SO(10), 
it does not affect running of gauge coupling constants 
and its decays can generate the observed baryon abundancy (section~\ref{leptogenesi}).
Certain unification models naturally accomodate scalar triplets, 
while fermion triplets seem more problematic from this point of view.

Furthermore, all three possibilities are compatible with supersymmetry.
Singlet and triplet fermions can be straightforwardly promoted to superfields.
As well known the SM scalar Higgs must be extended to two
Higgs superfields $H_{\rm d}$ and $H_{\rm u}$ with opposite gauge charges. 
Likewise,  the scalar triplet $T$ must be extended to
two triplet superfields $T$ and $\bar T$ with opposite gauge charges.
In the relevant superpotential 
$$ \mathW = \mathW_{\rm MSSM}  + M_T\, T\bar{T} +\frac{1}{2}(\lambda_T^{ij} L^i L^j T+
 \lambda_{H_{\rm d}} 
H_{\rm d} H_{\rm d}  T+\lambda_{H_{\rm u}}
H_{\rm u} H_{\rm u} \bar T)$$
$\bar T$ does not couple to leptons
and neutrino masses are obtained as
$m_\nu^{ij} = \lambda_T^{ij} \lambda_{H_{\rm u}} v_{\rm u}^2/M_T$.

Finally, all scenarios allow successful thermal leptogenesis (section~\ref{leptogenesi}).

\begin{figure}
$$\includegraphics[width=\textwidth]{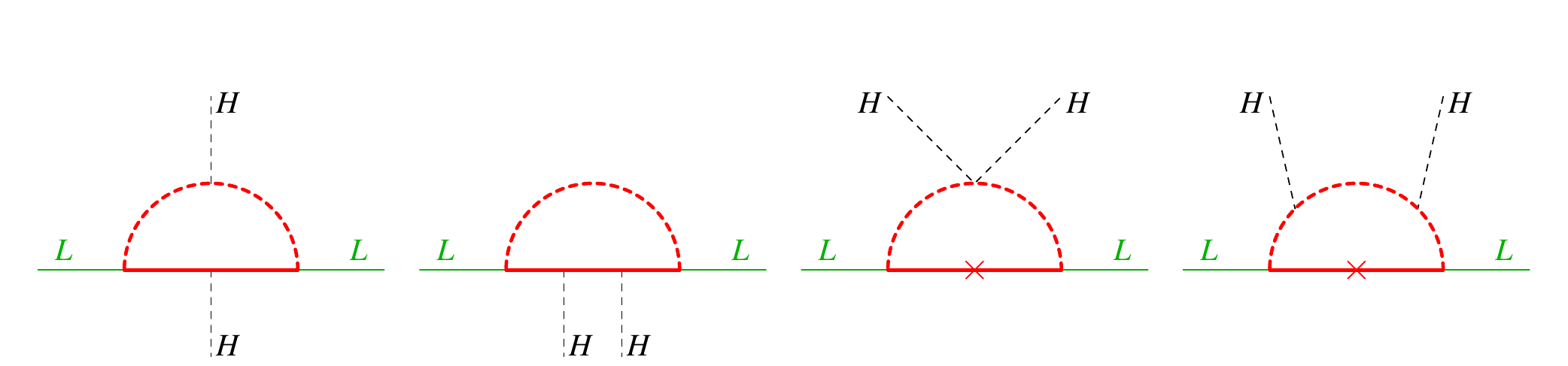}$$
\caption[Loop mediation of neutrino masses]{\em 
The neutrino Majorana mass operator $(LH)^2$ can be mediated by one-loop exchange of various kinds of fermions (red thick continuous line) and scalars (red thick dashed line).\label{fig:FeynLoop}}
\end{figure}


\subsection{Loop mediation of neutrino masses}
Mediation by loop effects can be realized by many ways.
Fig.\fig{FeynLoop} shows the possible one-loop diagrams~\cite{NuSUSY}, 
in each case there are several choices of quantum numbers for the particles in the loop.
For example, one can consider the standard see-saw scenario with a $LNH$ coupling
and replace the Higgs doublet $H$ with another scalar doublet $H'$ with vanishing
vev, coupled to the standard Higgs doublet as $(H^* H')^2+\hbox{h.c.}$: 
neutrino masses arise from the third diagram in fig.\fig{FeynLoop}.
One of the extra particles in the loop ($H'$ or $N$ in the example above)
could be detectably light: neutrino masses remain small
if other extra particles are heavy.

See~\cite{NuSUSY} for alternative speculative possibilities.

\begin{figure}[t]
$$
\includegraphics[width=4cm]{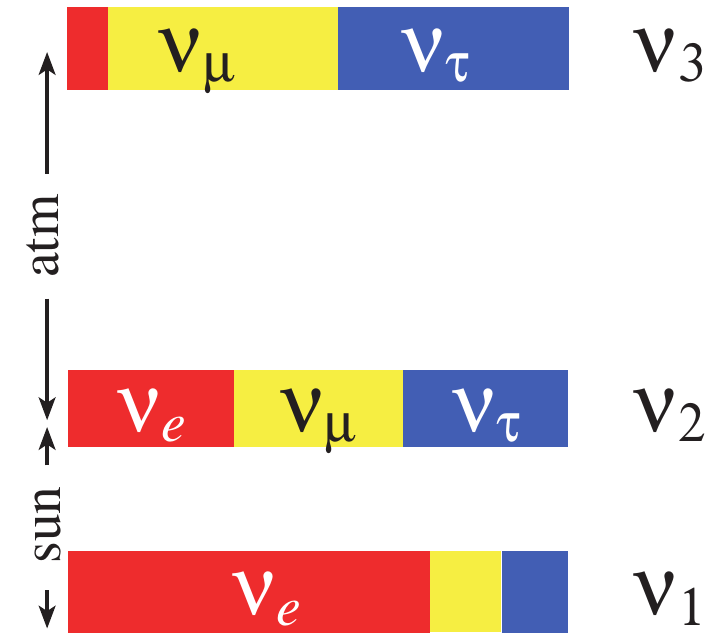}\hspace{4cm}
\includegraphics[width=4cm]{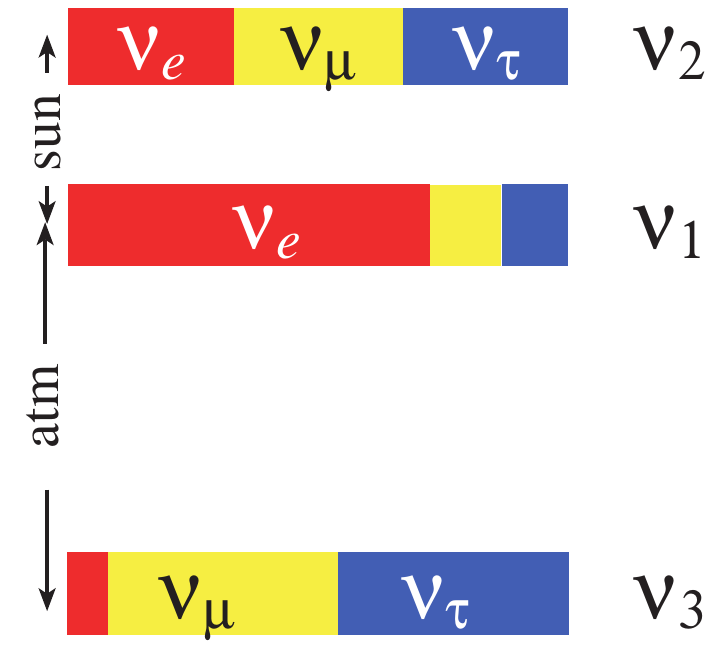}$$
\caption[Normal and inverted neutrino spectra]{\em Possible neutrino spectra: (a) normal (b) inverted.
\label{fig:spectra}}
\end{figure}

\bigskip

We now study in detail the special cases of pure Majorana and Dirac neutrino masses.
We describe
how many and which parameters can be measured in the two cases
by low energy experiments.

\section{Pure Majorana neutrinos}\label{Majorana}\index{Majorana}\index{Mass!Majorana}
We extend the SM by adding to its Lagrangian
the non-renormalizable operator $(LH)^2$ and no new fields.
Below the SU(2)$_L$-breaking scale, $(LH)^2$ just gives rise
to Majorana neutrino masses. 
In this situation, charged lepton masses are described as usual by a complex $3\times 3$ matrix  $\mb{m}_E$,
and neutrino masses by a complex symmetric $3\times 3$ matrix  $\mb{m}_\nu$:
$$-\Lag_{\rm mass} = \ell_R^T \cdot \mb{m}_E \cdot \ell_L + \frac{1}{2}
\nu_L^T \cdot \mb{m}_\nu 
\cdot
\nu_L .$$ How many independent parameters do they contain?
Performing the usual unitary flavour rotations of right-handed $E=\ell_R$ and left-handed
$L=(\nu_L,\ell)$ leptons, that do not affect the rest of the Lagrangian,\footnote{Gauge
interactions are the same in any flavour basis, because
kinetic energy and
gauge interaction originate from the same Lagrangian term, $\bar{L} \Ds L$.
This well known but non-trivial fact rests on solid experimental and theoretical grounds.
}
we reach the standard mass eigenstate basis of charged leptons, 
where 
$m_E = {\rm diag}\,(m_e,m_\mu,m_\tau)$.
It is still possible to redefine the phases of $e_L$ and $e_R$ such that $m_e$ and $m_\nu^{ee}$ are
real and positive; and similarly for $\mu$ and $\tau$.
Therefore charged lepton masses are specified by 9 real parameters and 3 complex phases:
the 3 real parameters $m_e$, $m_\mu$, $m_\tau$;
the 3 real diagonal elements of $m_\nu$;
the 3 complex off-diagonal elements of $m_\nu$.

It is customary to write the mass matrices as
\begin{equation}m_E = \diag (m_e, m_\mu, m_\tau),\qquad
m_\nu = V^* \diag(m_1 e^{-2i\beta},m_2 e^{-2i\alpha},m_3 ) V^\dagger
\end{equation}
where $m_{e,\mu,\tau,1,2,3}\ge 0$.
The neutrino mixing matrix $V$,
that relates the neutrinos with given mass, $\nu_i$,
to those with given flavour, 
\beq\color{rossos} \nu_\ell=V_{\ell i} \nu_i,\eeq
can be written as a sequence of Euler rotations
\begin{equation} V =
R_{23}(\theta_{23}) \cdot
R_{13}(\theta_{13}) \cdot
\hbox{diag}\,(1,  e^{i \phi},1) \cdot
R_{12}(\theta_{12}) 
\label{eq:Vunitary}
\end{equation}
where $R_{ij}(\theta_{ij})$ represents a
rotation by $\theta_{ij}$ in the $ij$ plane
and $i,j=\{1,2,3\}$.
In components
\begin{equation}
\pmatrix{V_{e1}&V_{e2}&V_{e3}\cr
V_{\mu1}&V_{\mu2}&V_{\mu3}\cr
V_{\tau1}&V_{\tau2}&V_{\tau3}} = \pmatrix{ c_{12} c_{13} & c_{13}s_{12}&s_{13} \cr 
-c_{23}s_{12}e^{i\phi} - c_{12}s_{13}s_{23} & c_{12}c_{23}e^{i \phi} - 
s_{12}s_{13}s_{23} & c_{13}s_{23} \cr 
s_{23}s_{12}e^{i\phi} - c_{12}c_{23}s_{13} & -c_{12}s_{23}e^{i \phi} - 
c_{23}s_{12}s_{13} & c_{13}c_{23}} .
\label{eq:Vij}
\end{equation}
Within this standard parameterization\footnote{Other commonly employed parameterizations
have the complex phase in different positions (e.g.\ with complex $V_{e3}$)
or different names for the mixing angles (e.g.\ $\theta_1,\theta_2,\theta_3$ or $\psi,\phi,\omega$
in place of $\theta_{23},\theta_{13},\theta_{12}$).

The ordering of Euler rotations chosen in eq.\eq{Vunitary} frequently naturally occurs in flavor models,
where one starts diagonalizing Yukawa matrices from the 3rd generation,
that has bigger entries. However, neutrinos 
exhibit large mixings and a mild mass hierarchy:
maybe a different parameterization will turn out to have a simpler decomposition,
reflecting some underlying flavor dynamics.

Finally, the number of parameters varies in special points of the parameter space.
As well known, the CP phase $\phi$ becomes unphysical (i.e.\ it can be rotated away by field redefinitions) if any of the three mixing angles $\theta_{13},\theta_{23},\theta_{23}$ vanishes.
Something similar applies also to Majorana phases.

When studying models of quasi-degenerate neutrinos, it is useful to know which entries
of $V$ are already fixed in the limit of degenerate neutrinos.
If neutrinos have the same mass and the same Majorana phase, the whole matrix $V$ becomes unphysical. If neutrinos have the same mass with different Majorana phases,
one can redefine away 3 parameters (remaining with 2 mixing angles and 1 phase)
since, on each couple of generations, the diagonal
neutrino mass matrix is invariant under a
unitary transformation that depends on one free parameter $\theta$:
$$\diag(m,m\, e^{2i\alpha} ) = U \cdot \diag(m,m\, e^{2i\alpha} ) \cdot U^T,\qquad
U(\theta)=\pmatrix{\cos\theta & e^{-i\alpha}\sin\theta\cr e^{i\alpha}\sin\theta & - \cos\theta},\qquad
U U^\dagger = 1.$$
}, 
the 6+3 neutrino parameters are the 3
neutrino mass eigenvalues, $m_1, m_2, m_3$, the 3
mixing angles $\theta_{ij}$ and the 3 CP-violating phases $\phi$, $\alpha$ and $\beta$.
$\phi$ is the analogous of the CKM phase,
and affects the flavour content of the neutrino mass eigenstates.
$\alpha$ and $\beta$ are called `Majorana phases'~\cite{MajoranaPhases} and do not affect oscillations
(see section~\ref{Oscillations}).

We now justify this parameterization.


\begin{enumerate}
\item 
Two parameters, $\theta_{23}$ and $\theta_{13}$,
are necessary to describe the flavour
of the most splitted neutrino mass eigenstate
$$| \nu_3 \rangle = s_{13}  |\nu_e\rangle + 
c_{13}s_{23} |\nu_\mu\rangle + c_{13}c_{23} |\nu_\tau\rangle .  $$
Complex phases can be rotated away by redefining the phases of $L_{e,\mu,\tau}$
and $E_{e,\mu,\tau}$ leaving $m_{e,\mu,\tau}$ real and positive.
Physically, this means that {\em two} mixing angles, $\theta_{23}$ and $\theta_{13}$, give rise to
CP-conserving
oscillations at the larger frequency $\Delta m^2_{23}$.

\item Since the flavours of $|\nu_2\rangle$ and $|\nu_3\rangle$ must be orthogonal,
a single complex mixing angle
(decomposed as one real mixing angle, $\theta_{12}$, plus one relative phase, $\phi$)
are needed to describe the flavour of
$|\nu_2 \rangle =
\sum_\ell V_{\ell 2}^* |\nu_\ell\rangle$.
Since there is no longer any freedom to redefine the phases of $\nu_{e,\mu,\tau}$,
the overall phase of $|\nu_2\rangle$, $\alpha$, is physical.

\item
Finally, no more parameters are needed to describe the flavour of $\nu_1$,
that must be orthogonal to $\nu_2$ and $\nu_3$.
The overall phase of $\nu_1$, $\beta$, cannot be rotated away and is a physical parameter.


\end{enumerate}

Finally, we specify the full allowed range of the parameters.

\begin{figure}[t]
$$\includegraphics[width=6cm]{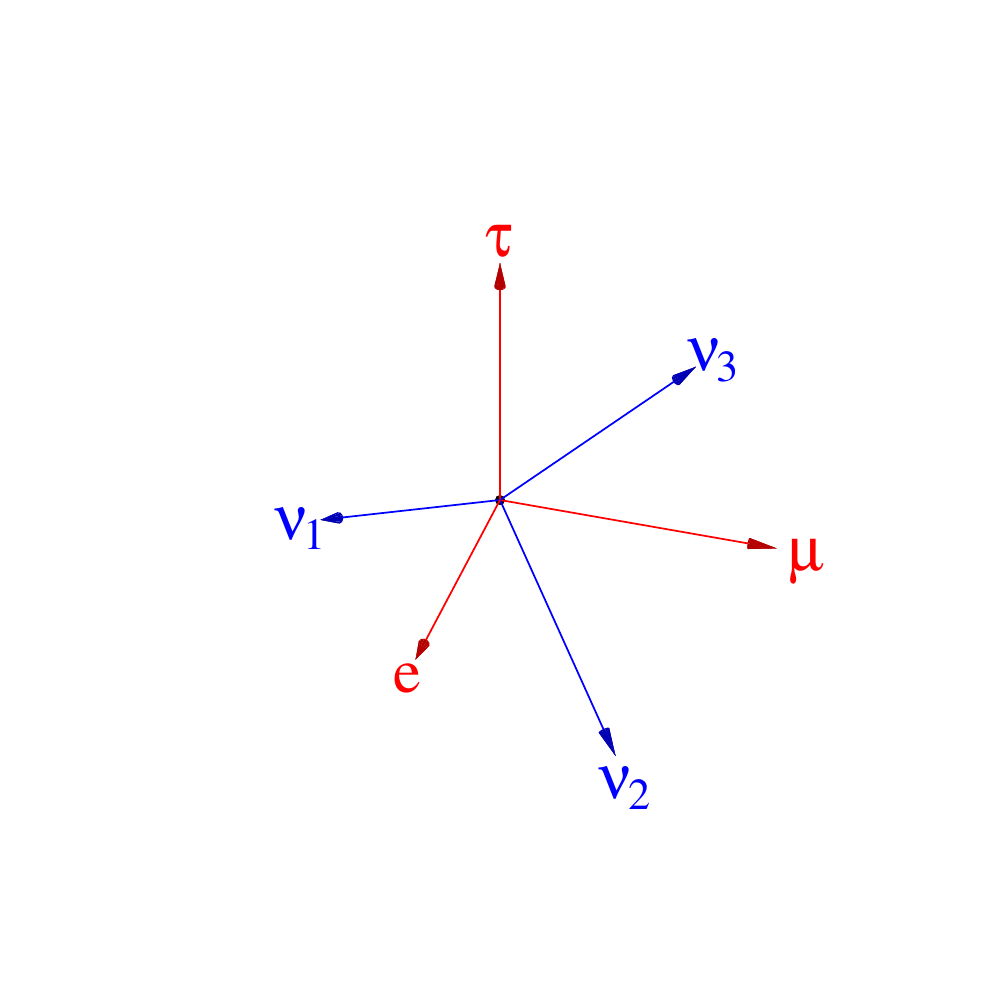}$$\vspace{-1cm}
\caption[The flavour of neutrinos]{\label{fig:NuAxis}\em The neutrino mass eigenstates
$\nu_{1,2,3}$ in 3-dimensional flavour-space.
The neutrino mixing matrix suggested by present data is a rotation with angle $\approx 56^\circ$
along the axis that corresponds to the point of view used in this figure.}
\end{figure}

We order the neutrino masses $m_i$ such that
$m_3$ is the most splitted state and
$m_2 > m_1$, and define
$\Delta m^2_{ij}=m_j^2-m_i^2$.
With this choice, $\Delta m^2_{23}$
and $\theta_{23}$ are the `atmospheric parameters' and
 $\Delta m^2_{12}>0$ and $\theta_{12}$ are the `solar parameters',
whatever the spectrum of neutrinos
(`normal hierarchy' so that
$\Delta m^2_{23}>0$; or
`inverted hierarchy' so that $\Delta
m^2_{23}<0$, see fig.\fig{spectra}).
With this choice the physically inequivalent range of mixing angles is
$$0\le \theta_{12},\theta_{23},\theta_{13}\le \pi/2,\qquad  0\le \phi < 2\pi\qquad
0\le \alpha,\beta \le \pi 
.$$
The flavour composition of the neutrino mass eigenstates $\nu_{1,2,3}$
suggested by present data (table~\ref{tab:tab1})
is indicated in fig.\fig{spectra} in 
a self-explanatory pictorial way.
Fig.\fig{NuAxis} illustrates again the neutrino mixing matrix in
 an alternative, non-standard, way:
 for present best-fit values (we are assuming $\theta_{13}=0$, such that there is no
 CP-violation) the neutrino mixing matrix 
 is a real rotation $V=R_n(\theta) = \exp(i \theta \hat{n}^a T^a)$ 
 with $\theta\approx56^\circ$ along the axis
\beq \hat n =
0.78 |\nu_e\rangle + 0.24 |\nu_\mu\rangle + 0.58 |\nu_\tau\rangle=
0.78|\nu_1\rangle+0.24 |\nu_2\rangle + 0.58|\nu_3\rangle.\eeq

\section{Pure Dirac neutrinos}\label{Dirac}\index{Dirac}\index{Mass!Dirac}
We extend the SM by adding three neutral singlets (one per family),
named ``right-handed neutrinos'', $\nu_R$.
We forbid $\nu_R^2$ mass terms by imposing  conservation of lepton number
(or of its anomaly-free cousin $B-L$).
The most generic renormalizable Lagrangian contains the additional term
\beq \Lag = \Lag_{\rm SM}+\bar\nu_R i\ds \nu_R +
 {\lambda}_N  ~ \nu_R  L H +\hbox{h.c.} \eeq
In this situation, charged lepton masses are described 
as usual by a complex $3\times 3$ matrix  $\mb{m}_E$,
and neutrino masses by a complex $3\times 3$ matrix $\mb{m}_\nu = \mb{\lambda}_N^T v$:
$$-\Lag_{\rm mass} = \ell_R^T \cdot \mb{m}_E \cdot \ell_L + 
\nu_L^T \cdot \mb{m}_\nu 
\cdot
\nu_R$$ 
We have more matrix elements and more fields that can be rotated
than in the pure Majorana case.
One can repeat the steps 1, 2, 3 above,
with the only modification that
the `Majorana phases' can now be rotated away
(reabsorbed in the phases of the $\nu_R$)
leaving only the CKM phase.

In fact, the flavour structure (2 mass matrices for 3 kinds of fields)
is identical
to the well known structure present in quarks
(2 mass matrices for the up and down-type quarks, contained in the 3 fields $u_R$, $d_R$ and $Q=(u_L,d_L)$).
However, a numerical difference makes the physics very different:
neutrino masses are small.
Up and down-type quarks and charged leptons
are produced
in ordinary processes as mass eigenstates,
while neutrinos as flavour eigenstates.
So far, we can produce a $\nu_\mu$,
but we are not able of getting a $\nu_3$.
For this reason, tools analogous to the `unitarity triangle'
(used to visualize CKM mixing among quarks, and useful because
 experiments can measure both its sides and its angles)
have no practical use in lepton flavour.

\bigskip

Before concluding, let us discuss the physical
difference between Majorana and Dirac neutrinos.
In both cases $\nu_\ell = V_{\ell i } \nu_i$ implies $\bar\nu_\ell = V_{\ell i}^* \bar\nu_i$.
While Dirac masses conserve lepton number, that distinguish leptons from anti-leptons,
in the Majorana case there is no Lorentz-invariant distinction between a neutrino and an anti-neutrino.
They are different polarizations of a unique particle that interacts mostly like a neutrino (an anti-neutrino)
when its spin is almost anti-parallel (parallel)  to its direction of motion.
While ideally an anti-neutrino becomes a neutrino,
if seen by an observer that moves faster than it,
in practice these effects are suppressed by $(m_\nu /E_\nu)^2$.
This factor is usually so small that only in appropriately
subtle situations it might be possible to detect it.

\begin{floatingfigure}[r]{4cm}
$$\includegraphics[width=4cm]{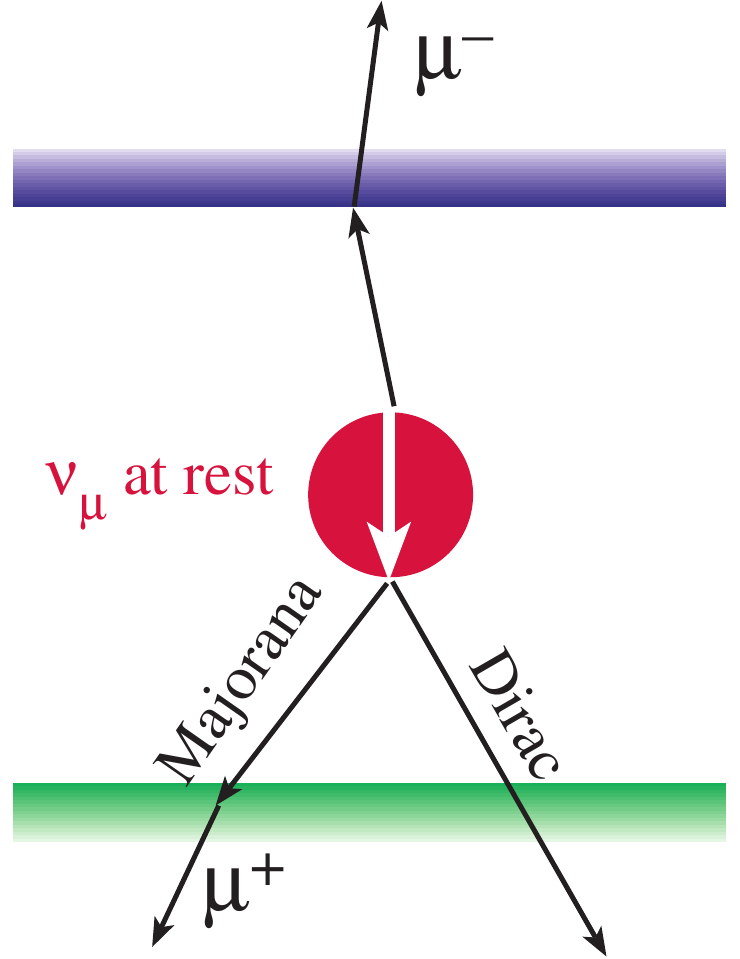}$$
\end{floatingfigure}

A gedanken experiment allows to appreciate the physical 
difference between Majorana and Dirac neutrinos in a simple way.
Suppose that it were practically possible to put at rest a massive
$\nu_\mu$ neutrino with spin-down  in the middle of the room.
If accelerated up to relativistic energies in the up direction,
when it hits the roof can produce a $\mu^-$ trough a CC interaction.
If accelerated up to relativistic energies in the down direction,
when it hits the floor it can produce a $\mu^+$ (if it is a Majorana particle) 
or have no interaction (if it is a Dirac particle).

Coming to realistic experiments, in the next section we show that
oscillation experiments cannot discriminate Majorana from Dirac neutrinos.
No signal induced by neutrino masses other than oscillations
has so far been seen.
It seems that the only realistic hope of experimentally discriminating
Majorana from Dirac neutrino masses
is based on the fact that Majorana masses violate lepton number,
maybe giving a signal in future neutrino-less double $\beta$ decay searches
(section~\ref{0nu2beta}).

\section{Formalism}
One can define the usual neutrino field operators,
neutrino creation and destruction operators, neutrino states,
neutrino wave-functions, etc.
At least in the relevant ultra-relativistic limit
(where $\nu$ and $\bar\nu$ are trivially distinguished), 
there should be no doubt of
how to implement flavour mixing within the standard QFT formalism.
See~\cite{Haber} for a clean presentation of 
Majorana and Dirac fermions in terms of Weyl spinors.

We could skip these details.
However we follow the standard formalism and notation,
and we must warn the reader that it contains one
unfortunate choice that becomes relevant when computing
the sign of CP-violating effects.
The point is that (by convention)
a field operator creates anti-particles
while an anti-field operator creates particles.
As a consequence one must be careful in distinguishing $V$ from $V^*$.
The correct relations between mass eigenstates and flavour eigenstates are:
\beq \begin{array}{lll}
\hbox{Field operators $\nu$:} & \nu_\ell = V_{\ell i}\nu_i, & 
\bar\nu_\ell = V_{\ell i}^* \bar\nu_i\\
\hbox{One-particle states $|\nu\rangle$:} &
|\nu_\ell\rangle = V_{\ell i}^* |\nu_i\rangle &
|\bar\nu_\ell\rangle = V_{\ell i} |\bar\nu_i\rangle\\
\hbox{Wave-functions $\nu(x) \equiv \langle x|\nu\rangle$} & 
\nu_\ell(x) = V_{\ell i}^*\nu_i(x), & 
\bar\nu_\ell(x) = V_{\ell i} \bar\nu_i(x)
\end{array}\eeq
In common-practice all these quantities are often denotes as $\nu$:
to do things properly one should understand physics rather than relying on precise formalisms.

\medskip

\subsection{Inverting the see-saw}
Assuming that three heavy right-handed neutrinos mediate Majorana neutrino masses
according to the see-saw Lagrangian of eq.\eq{Lseesaw2},
the most generic high energy parameters that give rise
to any desired neutrino masses $m_{\nu_i}$ and mixings $V$
can be parameterized as~\cite{CIparam}
\beq \label{eq:seesawParamCI}M_N = \diag(M_1,M_2,M_3),\qquad
\lambda_N = \frac{1}{v} M_N^{1/2} \cdot R \cdot \diag(m_{\nu_1},m_{\nu_2},m_{\nu_3})^{1/2}\cdot V^
\dagger . \eeq
One can always work in the mass eigenstate basis of right-handed neutrinos, where $M_i$
is real and positive.
$R$ is an arbitrary complex orthogonal matrix (i.e.\ $R^T\cdot R=1$),
that can be written in terms of 3 complex mixing angles.
In total the high-energy see-saw theory has 9 real unknown parameters.

\chapter{Oscillations}\label{Oscillations}
We start discussing oscillations in vacuum,
without hiding subtle points and giving practical formul\ae{}.
Next, we discuss oscillations in matter,
and describe how neutrinos oscillate in 
continuously varying density profiles
(e.g.\ in the sun and in supernov\ae).\footnote{
Solar and atmospheric anomalies have been discovered
studying natural sources of neutrinos.
To correctly interpret data one needs to understand how these systems work.
The oscillation formalism was developed around 1980.
The recent experimental progress stimulated
new interest and all these issues have been critically reconsidered.
This generated some healthy confusion, that should not give a wrong impression.
All newly claimed effects turned out to be wrong or already known:
old results have been confirmed~\cite{newOsc}.
Many papers discuss if the standard oscillation phase should be corrected
with extra ${\cal O}(1)$ factors: 
to verify that this is not the case one can simply notice that
the standard oscillation formula reduces to well known physics
in two limiting regimes of small and large oscillation phase
(respectively to first-order perturbation theory
and to multiplication of probabilities, see page~\pageref{ABC}).
These discussions correctly showed that the `standard derivation'
of the vacuum oscillation formula is over-simplified:
following~\cite{stodolsky} we present a simple meaningful derivation.}


\section{Oscillations in vacuum}\index{Oscillation!in vacuum}

One-particle quantum mechanics is the appropriate language for describing
neutrino oscillations.
In all cases of practical interest
neutrino fluxes are sufficiently weak
that multi-particle Fermi-Dirac effects can be neglected.
Concerning this aspect, a neutrino beam is simpler than an electro-magnetic field,
that can be composed by inequivalent configurations of many photons.
Therefore, one should 
\begin{enumerate}
\item {\bf Build a neutrino wave-packet}~\cite{wave},
taking into account the dynamics of the specific process that produces it,
For example, atmospheric and beam neutrinos are mostly produced in $\pi$ and $\mu$ decays.
Solar $\nu_e$ are produced in collisions and decays of light nuclei inside the sun.
Reactor $\bar{\nu}_e$ in decays of fragments of fissioned heavy radioactive nuclei. 
Supernova neutrinos are produced mostly thermally.

\item {\bf Study its evolution}.
Different mass eigenstates acquire different phases,
giving rise to oscillations.
The mass difference also generates other effects.
The lighter mass eigenstate moves faster than the heavier one:
at some point their wavepackets no longer overlap,
destroying oscillations.
While in neutrinos this effect is usually negligible,
the mass differences between quarks are so large
that there are no oscillations between quarks:
e.g.\ the down-type quark $q$ produced  in decays of charmed hadrons,
$c\to q \ell\bar\nu$, is
$|q\rangle = \cos\theta_{\rm C}|d\rangle + \sin\theta_{\rm C} |s\rangle$, 
giving rise to a $\pi$ with probability $\cos^2\theta_{\rm C}$ and
to $K$ with probability $\sin^2\theta_{\rm C}$ --- 
not to $\pi \leftrightarrow K $ oscillations. \label{piK}
(Furthermore the heavier quarks decay fast, while the heavier neutrinos seem to be almost stable).

\item {\bf Compute the observable to be measured},
taking into account what the detector is really doing.
Oscillations are a quantum interference effect.
The necessary coherence is destroyed if the neutrino mass is measured
(for example by measuring the neutrino energy and momentum)
with enough precision to distinguish which one of the
 different neutrino mass eigenvalues has been detected.

\end{enumerate}
We can derive a general and simple result, bypassing the cumbersome
wave-packet analysis,
if we restrict our attention to a {\em stationary} flux of neutrinos
or to experiments that only look at {\em time-averaged} observables~\cite{stodolsky}.
We now show that in these conditions a neutrino wave is {\em fully described by its energy spectrum}
(and of course by its direction, flavour and possibly polarization).
This means that a plane wave is the same thing as a mixture of short wavepackets,
just as the same light can be obtained as a mixture
of circular or linear polarizations.

\begin{figure}
$$\includegraphics[width=18cm]{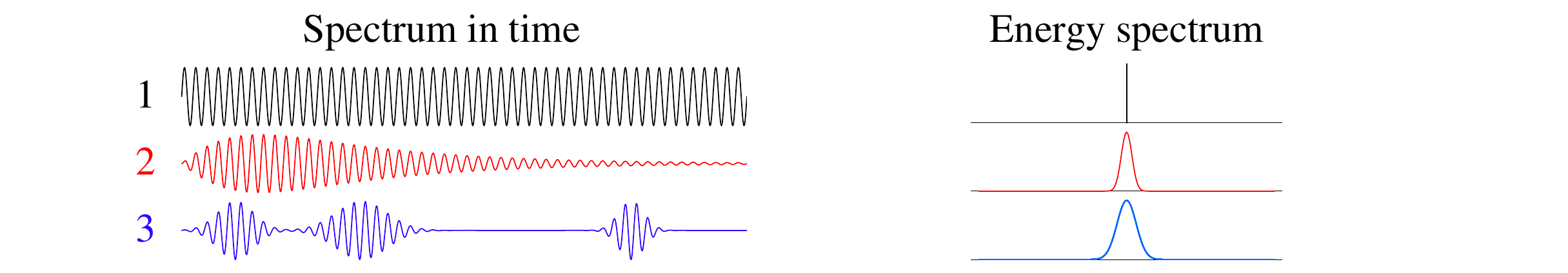}$$
\caption[Neutrino wave]{\label{fig:OscSpettri}\em {\bf 1}: a monochromatic wave;
its energy spectrum is a line.
{\color{rossos}\bf 2}: a pulse of `monochromatic' neutrinos;
its energy spectrum is almost a line.
{\color{blue}\bf 3}: a few wave packets of  `monochromatic' neutrinos;
their energy spectrum is roughly a line.
}
\end{figure}

The basic observation is so simple that it might be difficult to understand it.
It is convenient to work in the basis of eigenstates of the Hamiltonian.
The most generic {\em pure} state is a superposition of them.
In stationary conditions all interferences between states with different 
energy average to zero, $\langle e^{i(E-E')t}\rangle = 0$,
when computing any physical observable.
Therefore the relative phases between neutrinos with different energies are
not observable, and the conclusion follows.
This is illustrated in fig.\fig{OscSpettri}.

We need to generalize this proof to a neutrino flux described
by a density matrix $\rho$.
In fact, let us consider e.g.\ a
neutrino produced in $\pi$ decay, $\pi\to \nu_\mu \bar\mu$.
A wave function describes the  neutrino and the muon.
As usual, when we want we restrict to a subset (the neutrino) of
the full system (neutrino and muon), we are forced to introduce mixed states.
Furthermore, the
particle that produces the neutrino usually interacts
in a non negligible way with the environment
(e.g.\ a stopped $\pi$ at FermiLab, or a $^7{\rm Be}$ in the sun):
using a density matrix for neutrinos is simpler than
studying the wave function of FermiLab, or of the sun.
Again, the result simply follows by the fact that the off-diagonal terms of $\rho$
oscillate in time as $e^{i(E-E')t}$ and therefore average to zero. 
More formally, $i\dot\rho = [H,\rho] = 0$ in stationary conditions, so that
the off-diagonal elements of $\rho$ between states with different energy vanish.
The diagonal elements of $\rho$ tell the neutrino energy spectrum.

\medskip

Our simplifying conditions are valid in all realistic experiments:
an experiment that can measure the time of neutrino detection with
$\Delta t \sim \hbox{ns}$ is not sensible to interference among
neutrinos with $E-E'\gg 1/\Delta t \sim 10^{-6}\eV$, which
is much smaller than any realistic energy resolution.
Deviations from the oscillation probabilities that we now derive are negligible
even when considering a pulsed neutrino beam or a short supernova neutrino burst.

\medskip

Neutrinos with different mass and the same energy oscillate, as we now describe.
We start considering the simplest case, and we do not employ
density matrices, which are the appropriate and convenient formalism for more complex computations.

\subsection{Vacuum oscillations of two neutrinos}
We consider two generation mixing, so that we just have one mixing
angle,
$\theta$, and no CP violation. 
We assume that at the production region, $x\approx 0$, $\nu_e$ are produced
with energy $E$.
To study their propagation it is convenient to utilize the basis of neutrino mass eigenstates
$\nu_{1,2}$, and write  $|\nu(x=0)\rangle =|\nu_e\rangle = \cos\theta |\nu_1\rangle + \sin\theta
|\nu_2\rangle$. Since $\nu_1$ and $\nu_2$ have different masses, the initial
$\nu_e$ becomes some other mixture of $\nu_1$ and $\nu_2$, or equivalently of $\nu_\mu$ and $\nu_e$.
At a generic $x$
$$|\nu(x) \rangle= e^{ip_1 x}  \cos\theta |\nu_1\rangle + e^{i p_2 x} \sin\theta | \nu_2\rangle.$$
The probability of $\nu_\mu$ {\em appearance} at the detection region $x\approx L$ is
\begin{equation}\label{eq:Posc}
 P(\nu_e\to \nu_\mu) = |\langle \nu_\mu | \nu(L)\rangle |^2 = \sin^2 2 \theta
~~\sin^2\frac{(p_2-p_1) L}{2}
\simeq \sin^2 2 \theta ~~\sin^2 \frac{\Delta m^2_{12} L}{4E}. 
\end{equation}
Since in all cases of experimental interest $E\gg m_i$,
in the final passage we have used the ultra-relativistic approximation $p_i = E - m_i^2/2E$,
valid at dominant order in the small neutrino masses and defined $\Delta m^2_{12}\equiv m_2^2 - m_1^2$.
\footnote{We sketch 
the 
standard over-simplified derivation.
It proceeds
writing the
evolution in {\em time} as $|\nu(t)\rangle = e^{-iH t}|\nu(0)\rangle$.
Assuming that neutrinos with different mass have equal momentum,
the hamiltonian is $H \approx p + m m^\dagger/2p$.
This gives the correct final formula, if one does not take into account
that different neutrinos have different velocity.
It is not clear which `time' one should use
(e.g.\ when neutrinos are produced by slow decays), as
no real experiment measures it:
experiments measure the {\em distance} from the production point.

Furthermore,
in many realistic cases neutrinos actually oscillate
in space but not in time,
because their wave-packets have a
much larger spread in momentum
than in energy.
This happens because
the particle that decays into neutrinos
often interacts with a {\em big} environment
and therefore behaves like
a ball that bounces in a box:
it keeps the same energy but changes momentum.

All this discussion applies to oscillations,
not only to neutrino oscillations.}

By swapping the names  of the two mass eigenstates, $\nu_1 \leftrightarrow \nu_2$,
one realizes that the couples ($\theta$, $\Delta m^2_{12}$) and ($\pi/2-\theta$, $ -\Delta m^2_{12}$)
describe the same physics.
On the contrary ($\theta$, $\Delta m^2_{12}$) and ($\pi/2-\theta$, $\Delta m^2_{12}$)
are physically different.
However, eq.\eq{Posc} shows that {\em vacuum oscillations depend only on $\sin^22\theta$} and
do not discriminate these two cases.
Oscillation effects are maximal at $\theta = \pi/4$.

\smallskip

\index{Oscillation!of two neutrinos}
The $\nu_e$ {\em disappearance} probability is
$$ P(\nu_e\to \nu_e) = |\langle \nu_e | \nu(L)\rangle |^2 =1 - P(\nu_e\to \nu_\mu). $$
A convenient numerical relation is found restoring $\hbar$ and $c$ factors:
\begin{equation}\label{eq:Sij}
\color{rossos} S_{ij}\equiv \sin^2 \frac{c^3}{\hbar}\frac{\Delta m^2_{ij} L}{4E}=\sin^2 1.27 \frac{\Delta
m^2_{ij}}{\eV^2}
\frac{L}{\rm Km}\frac{\GeV}{E}.
\end{equation}
The oscillation wave-length is 
\beq \label{eq:lambdaosc}
\lambda = \frac{4\pi E}{\Delta m^2} = 2.48\km \frac{E}{\GeV}\frac{\eV^2}{\Delta m^2_{ij}}.\eeq
Like decays, oscillations are suppressed at large energy by the $m/E$ 
`time-dilatation' Lorentz factor, well known from relativity.
In order to see oscillations one needs neutrinos of low enough energy,
that have small detection cross sections (section~\ref{detecting}).
Furthermore some reactions are kinematically allowed only 
at high enough energies.
For example, a $\nu_\tau$ can be seen
by detecting a scattered $\tau$: 
using the $\nu_\tau e \to \nu_e \tau$ reaction one needs 
$E_{\nu_\tau}> m_\tau^2/2m_e\approx 3\TeV$,
while using
$\nu_\tau n\to \tau p$ one needs
$E_{\nu_\tau}>m_\tau+m_\tau^2/2m_N\approx 3.5\GeV$.

\begin{figure}
$$\hspace{-4mm}\includegraphics[height=7cm]{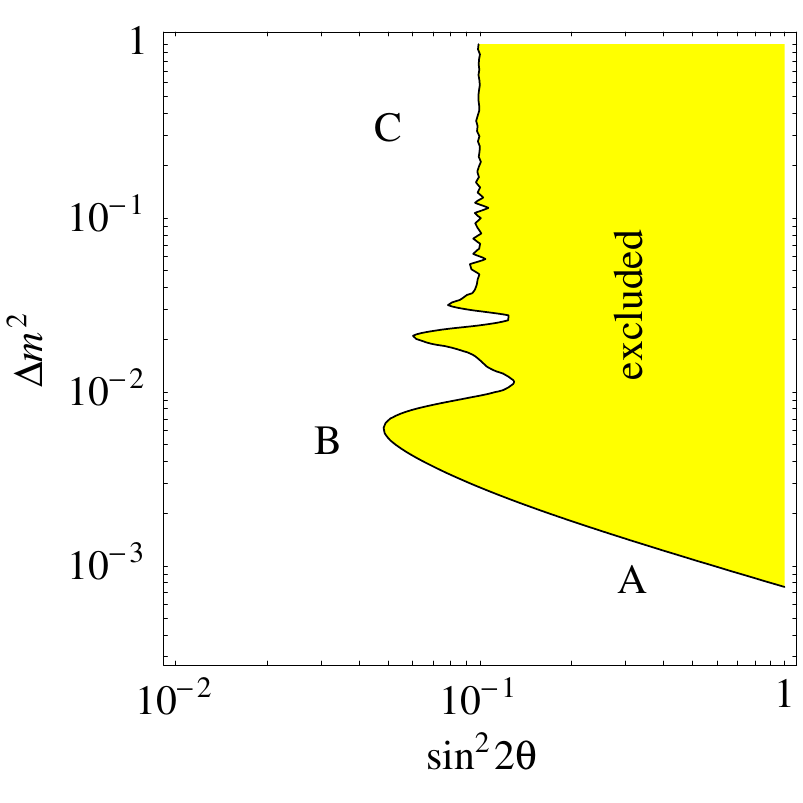}\hspace{5mm}
\raise4mm\hbox{\includegraphics[height=6.4cm,width=10cm]{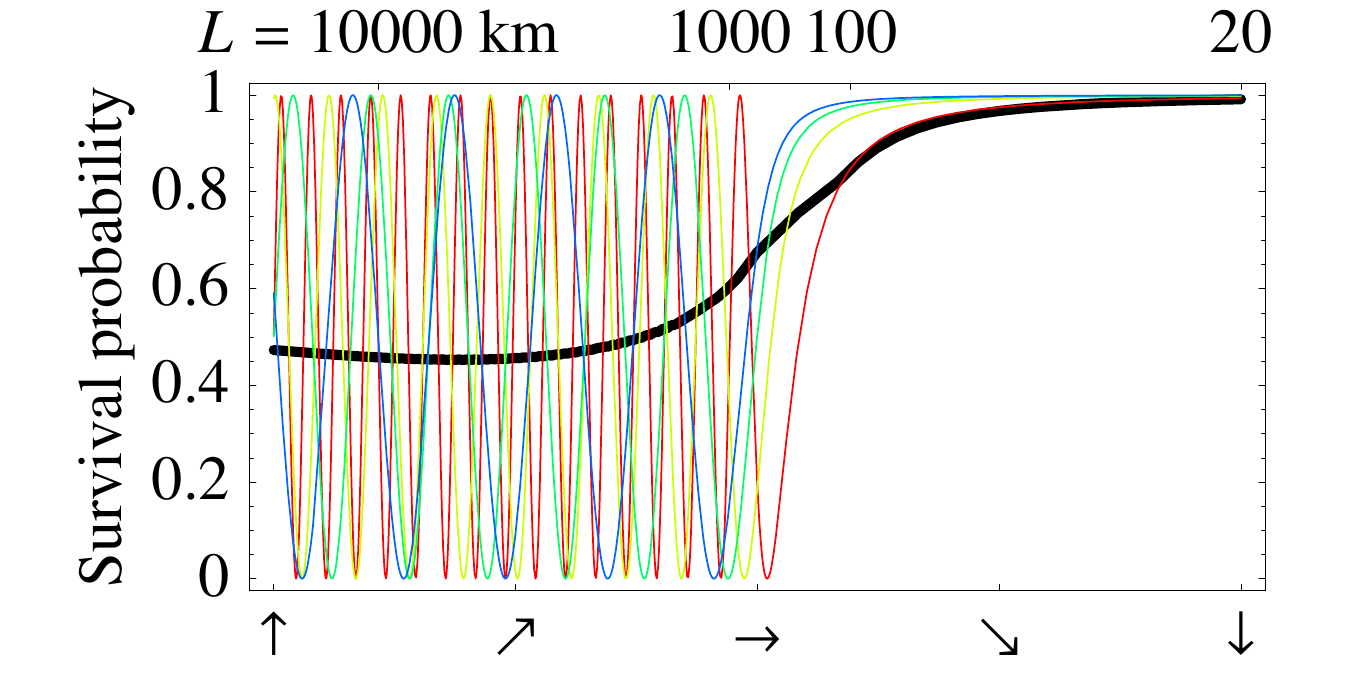}}$$
\caption[Typical ($\Delta m^2,\sin^22\theta)$ plane]{\em (a) Typical bound on oscillations. 
(b) Averaging oscillations over neutrinos with different energies
(here represented with different colors)
gives a smooth survival probability (thick curve).
This plot holds for atmospheric neutrinos,
where the path-length (upper axis) is measured from
the  direction of arrival (lower axis).
\label{fig:tipico}}
\end{figure}

\medskip

\subsection{Limiting regimes}\label{LimitingRegimes}

In a realistic setup, the neutrino beam is not monochromatic,
and the energy resolution of the detector is not perfect:
one needs to average the oscillation probability around some energy range $\Delta E$.
Furthermore, the production and detection regions are not
points: one needs to average around some path-length range $\Delta L$.
Including these effects, in fig.\fig{tipico}a we show a typical 
experimental bound on oscillations. We can distinguish three regions:

\begin{itemize}
\item[A] {\bf Oscillations with short base-line}, where $S_{ij}\ll 1$.
In this limit  oscillations reduce to first-order perturbations:\footnote{Notice
that the standard oscillation factor follows from first-order perturbation theory.
In this simple limit, one could explicitly study
how $e^{i(E-E')t}$ averages to zero, i.e.\
take into account the negligible phenomenon of
oscillations among neutrinos with different energies
$E$ and $E'$.}
$$P(\nu_e\to \nu_\mu)\simeq (H_{e\mu} L)^2\qquad\hbox{with}\qquad
H_{e\mu} \equiv \frac{(m_\nu m_\nu^\dagger)_{e\mu}}{2E_\nu} = \frac{\Delta m^2}{E_\nu} \sin2\theta.$$
This explains the slope of the exclusion region in part A of fig.\fig{tipico}a.
Since $P(\nu_e\to \nu_\mu)\propto L^2$, and
since  going far from an approximatively point-like
neutrino source the neutrino flux decreases as $1/L^2$,
choosing the optimal location for the detector
is usually not straightforward.

\label{ABC}

\item[C] {\bf Averaged oscillations}, where   \index{Oscillation!averaged}
$\langle S_{ij}\rangle = 1/2$
as illustrated in fig.\fig{tipico}b.
In this limit one has
\begin{equation}\label{eq:Pav}
P(\nu_e\to \nu_\mu) = \frac{1}{2}\sin^2 2\theta,\qquad
P(\nu_e\to \nu_e) = 1-\frac{1}{2}\sin^2 2\theta.
\end{equation}
The information on the oscillation phase is lost due to the insufficient experimental
resolution in $E$ or $L$.
Consequently, one can rederive the transition probabilities\eq{Pav} 
by combining {\em probabilities} rather than {\em amplitudes}.
Using the language of quantum mechanics, one refers
   sometimes to this case as the ``classical limit''.
The computation proceeds in full analogy to the our
$\pi/K$ example at page~\pageref{piK},
as illustrated by the following figure:
$$\includegraphics{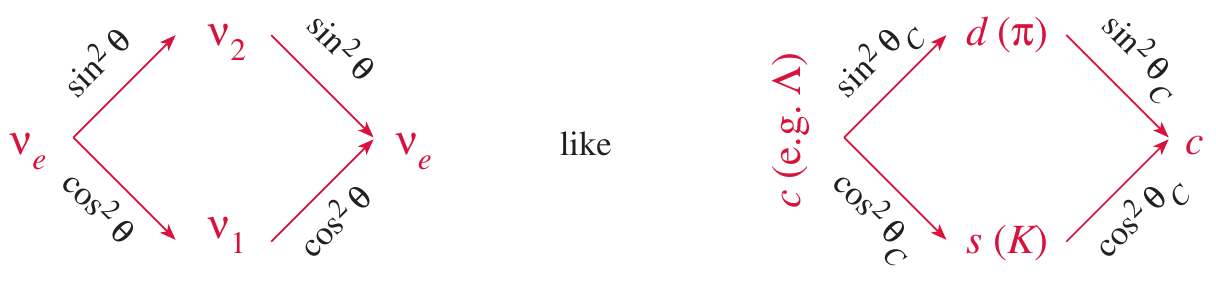}$$
At $x\approx 0$ one produces: 
\begin{itemize}
\item a $\nu_1$ with probability $\cos^2\theta$ (later detected
as a $\nu_\mu$ with probability $\sin^2 \theta$, or as
a $\nu_e$ with probability $\cos^2\theta$),
and 
\item a $\nu_2$ with probability $\sin^2 \theta$ 
(later detected
as a $\nu_\mu$ with probability $\cos^2 \theta$, or as
a $\nu_e$ with probability $\sin^2\theta$).
\end{itemize}
Therefore one obtains the same result as in\eq{Pav}
$$P(\nu_e\to \nu_\mu) = 2\sin^2\theta\cos^2\theta,\qquad
P(\nu_e\to \nu_e) = \sin^4\theta + \cos^4 \theta . $$
We now discuss in more detail how the averaging over the energy spectrum
transforms coherent oscillations into
an incoherent process.
\end{itemize}
This discussion implies that experimental bounds on oscillations
can be approximatively summarized by reporting two numbers:
the upper bound on $\Delta m^2$ assuming maximal mixing,
and the upper bound on $\theta$ assuming large $\Delta m^2$.
This is e.g.\ done in table~\ref{tab:shortbeams}.
If some effect is discovered, 
one can test if it is due to oscillations (and possibly measure oscillation parameters)
by studying how it depends on the neutrino energy and
path-length.  The most characteristic phenomenon appears in the intermediate region.
\begin{itemize}
\item[B] {\bf  The intermediate region}.
Due to the uncertainty $\Delta E$ on the energy $E$
(and possibly on the path-length $L$),
coherence gets lost when neutrinos of different energy
have too different oscillation phases $\phi\sim \Delta m^2 L/E$, i.e.\ when
\begin{equation}\label{eq:d1}
\Delta \phi\approx \frac{\Delta E}{E} \phi \circa{>}1
\end{equation}
Therefore one can see $n \sim E/\Delta E$ oscillations before they average out.
So far, only the KamLAND experiment could clearly observe an oscillation dip.
The energy resolution of the experiment often gives the dominant contribution to the total $\Delta E$.
\end{itemize}
Before concluding, we discuss in greater detail
how the formalism here employed automatically takes into account loss of coherence.
In the alternative wave-packet formalism, coherence is lost when
the wave-packets corresponding to different mass eigenstates
(that move at different velocities $\Delta v\sim \Delta m^2/E^2$)
no longer overlap.
This happens when
\begin{equation}\label{eq:d2}
\Delta v\cdot t \circa{>} \Delta x
\end{equation}
where $\Delta x$ is the size of the wave-packet.

In the stationary case that we are considering,
this phenomenon is accounted by the energy
average over the minimal $\Delta E$ demanded by quantum mechanics,
approximatively equal to $\Delta E\approx 1/\Delta x$,
as dictated by the uncertainty relation $\Delta x\cdot \Delta p\circa{>}\hbar$.
In fact, one can verify that eq.\eq{d1} and\eq{d2} are equivalent.
This point is pictorially illustrated in fig.\fig{OscSpettri}:
computing the Fourier transform of
a sequence of wave-packets (case 3) one finds a broader
energy spectrum than in case 1.

A numerical example shows that this phenomenon is hardly relevant.
A supernova could emit a pulse of neutrinos as short as 
$\Delta t\sim 0.1\s$ (case 2). 
The corresponding quantum uncertainty on their energy is $\Delta E\sim 1/\Delta t\sim 10^{-5}\eV$,
which is much smaller than their typical energy, $E\sim 100\MeV$,
and therefore becomes important only after $\Delta E/E\sim 10^{13}$ oscillations
i.e.\ after oscillating for cosmological distances with wavelength $\lambda =4\pi E/\Delta m^2$.
This same estimate can be reobtained computing the separation between
neutrino wave-packets
$$\Delta v\cdot t \approx L\frac{\Delta m^2}{E^2}\approx 0.1\s
\frac{L}{10^{25}\m}
\frac{\Delta m^2}{3~10^{-3}\eV^2}\bigg(\frac{100\MeV}{E}\bigg)^2$$
and comparing it with the length of the wave-packet,  $\Delta x\sim c \Delta t\sim 0.1\s$.




\subsection{Vacuum oscillations of $n$ neutrinos}
Some results follow from general arguments:
\begin{itemize}
\item {\em Conservation of probability} implies
$$\sum_{\ell'} P(\nu_\ell\to \nu_{\ell'}) = \sum_{\ell'} P(\bar\nu_\ell\to \bar\nu_{\ell'}) =1$$
\item CPT {\em invariance} implies
$$P(\nu_\ell \to \nu_{\ell'}) = P(\bar\nu_{\ell'} \to \bar\nu_\ell)$$

\item In many situations CP {\em invariance} approximately holds and implies
$$P(\nu_\ell \to \nu_{\ell'}) = P(\bar\nu_{\ell} \to \bar{\nu}_{\ell'})$$
Together with CPT-invariance, CP-invariance is equivalent to T {\em invariance}
$$P(\nu_\ell \to \nu_{\ell'}) = P(\nu_{\ell'} \to \nu_\ell)$$
Therefore T-conserving (breaking) contributions are even (odd) in the base-line $L$.
\end{itemize}
Up to an irrelevant overall phase, the transition amplitude is 
\begin{equation}\label{eq:A}
A(\nu_{\ell}\to\nu_{\ell'})=\bk{\nu_{\ell'}}{\nu_{\ell}(L)}=
\langle  \nu_{\ell'}  |U(L)|\nu_{\ell}\rangle=
\sum_i V_{\ell'i} V_{\ell i}^*  e^{2i\varphi_i},\qquad
U(L) = \exp \bigg(-i\frac{m\cdot m^\dagger L}{2E}\bigg)
\end{equation}
where
$\varphi_i\equiv -{m_i^2 L}/{4E}$.
We see that Majorana phases do not affect oscillations.
In the short base-line limit,
approximating $\exp i\mb{\epsilon}t\simeq \One+i\mb{\epsilon}t+{\cal O}(t^2)$
(here $\mb{\epsilon}$ is a flavour matrix),
the oscillation probability
reduces to first-order perturbations
$P(\nu_\ell\to \nu_{\ell'}) = |\epsilon_{\ell\ell'}|^2$ for $\ell\neq \ell'$.

Eq.\eq{A} can be used in numerical computations.
However when $|\varphi_i| \gg 1$
the result rapidly oscillates and it is cumbersome to
compute numerically its mean value,
that usually determines what can be measured.
In the simple case of vacuum oscillations it is possible and
convenient to rewrite eq.\eq{A} in a longer but more useful form.
Using $e^{2i\varphi}=1-2\sin^2 \varphi +i\sin 2\varphi$, from eq.\eq{A} we get
\begin{eqnarray}\nonumber
P(\nu_{\ell}\to\nu_{\ell'})&=&|A(\nu_{\ell}\to\nu_{\ell'})|^2
=\sum_{ij}J_{ij}^{\ell\ell'} (1-2\sin^2 \varphi_{ij}+i\sin2\varphi_{ij})\\ &=&
\underbrace{\delta_{\ell\ell'}-\sum_{i < j}4 \Re  J_{ij}^{\ell\ell'} 
\sin^2 \varphi_{ij}}_{\rm CP-conserving}
-\underbrace{\sum_{i < j}2\Im  J_{ij}^{\ell\ell'}\sin2\varphi_{ij}}_{\rm CP-violating}.\label{eq:PijvacGeneral}
\end{eqnarray}
where
\beq 
\varphi_{ij}\equiv\varphi_i-\varphi_j = \frac{\Delta m^2_{ij} L}{4E},\qquad
\Delta m^2_{ij}\equiv m_j^2 - m_i^2,\qquad
J_{ii'}^{\ell\ell'}=V_{\ell i}^*V_{\ell'i'}^*V_{\ell'i}V_{\ell i'} .\eeq
$\Im  J_{ij}^{\ell\ell'}$ is a rephasing-invariant measure of CP violation, because
it arises from a product of elements of $V$ where
any flavour and any mass-eigenstate  $\nu_\ell$ and $\nu_i$ enters twice,
 once with a complex conjugation, once without.
 The corresponding formul\ae{} for antineutrinos 
are obtained by exchanging
$V\leftrightarrow V^*$, so that in the final formula
only the sign of the CP-violating term changes.
In the limit of averaged oscillations $\langle \sin^2\varphi_{ij}\rangle=1/2$ and
$\langle\sin2\varphi_{ij}\rangle=0$ 
one can reobtain the oscillation probabilities by 
combining probabilities (rather than quantum amplitudes):
\beq \label{eq:Paveraged}
P(\nu_\ell \to \nu_{\ell '}) = P(\bar\nu_\ell \to \bar\nu_{\ell '})=\sum_i |\langle \nu_i|\nu_\ell\rangle|^2
= \sum_i | V_{\ell i} V_{\ell' i}|^2\eeq
which in general depends on the CP-violating phases that affect $|V_{\ell i}|$.

\subsection{Vacuum oscillations of 3 neutrinos}\index{Oscillation!of three neutrinos}
Specializing to the case of 3 neutrinos one gets
\begin{equation}\color{rossos}
P(\nubarnu_{\ell}\to\nubarnu_{\ell'}) =
\delta_{\ell\ell'}+p^{12}_{\ell\ell'} \sin^2\varphi_{12}+p^{13}_{\ell\ell'} \sin^2\varphi_{13}+p^{23}_{\ell\ell'} \sin^2\varphi_{23} \pm
8J
\sin\varphi_{12}\sin\varphi_{13}\sin\varphi_{23}
\sum_{\ell''}\epsilon_{\ell\ell'\ell''} 
\end{equation}
where the $-$ sign holds for neutrinos, the $+$ sign for anti-neutrinos, 
$\epsilon$ is the permutation tensor ($\epsilon_{123}=+1$), and
$$S_{ij} = \sin^2 \varphi_{ij},\qquad
p_{ii'}^{\ell\ell'}=-4\Re V_{\ell i}V_{\ell'i'}V_{\ell'i}^*V_{\ell i'}^*,\qquad\hbox{and in particular}\qquad
p_{ii'}^{\ell\ell}=-4|V_{\ell i}V_{\ell i'}|^2 $$
The CP-violating term is simpler than in eq.\eq{PijvacGeneral} because, with only 3 neutrinos
oscillations depend only on one CP-violating phase and
 \beq\Im J_{ii'}^{\ell\ell'}
=J\sum_{i'',\ell'' }\epsilon_{ii'i''}\epsilon_{\ell\ell'\ell''}\label{eq:J}
\qquad\hbox{where}\qquad
8J\equiv \cos\theta_{13}\sin2\theta_{13}\sin2\theta_{12}\sin2\theta_{23}\cdot\sin\phi.
\eeq
 Up to a sign $J$ equals to twice  the area of the `unitarity triangle' with sides
  $V_{\ell i}V_{\ell i'}^*$ and $V_{\ell'i'}V_{\ell'i}^*$.
  Eq.\eq{J} tells that all such triangles have the same area.
  The maximal value  $|J|=1/6\sqrt{3}$ is obtained in the 
 case known as `trimaximal mixing':
for $\phi = \pi/2$,
  $\theta_{12}=\theta_{23}=\pi/4$ and
$\cos^2\theta_{13}=2/3$
all elements of $V$ have the same modulus $|V_{\ell i}| = 1/\sqrt{3}$,
and oscillation probabilities can reach the borders
of the allowed range $0\le P\le 1$.
(This case is not realized in nature).

We have simplified the CP-violating contribution to $P(\nu_\ell\to\nu_{\ell'})$ using
the trigonometrical identity
$$\sin 2\varphi_{12}+\sin2\varphi_{23}+\sin2\varphi_{31}=4\sin\varphi_{12}
\sin\varphi_{23}\sin\varphi_{13}.$$
As expected the CP-violating contribution
vanishes if $\ell=\ell'$ and  is odd in $L$.
In the small $L$ limit, it is proportional to $L^3$.
It is small when any mixing angle $\theta_{ij}$ or any oscillation phase $\varphi_{ij}$ is small;
it averages to zero when some $\varphi_{ij}\gg 1$.
These properties explain why it is difficult to observe CP-violation.

\medskip
As anticipated in section~\ref{Present},
data indicate that 
$$|\Delta m^2_{13}|\approx| \Delta m^2_{\rm 23}|=\Delta m^2_{\rm atm}\approx 3\cdot 10^{-3}\eV^2,\qquad
\Delta m^2_{\rm 12}=\Delta m^2_{\rm sun}\approx   10^{-4}\eV^2.$$
Therefore it is interesting to consider the limit
$|\Delta m^2_{\rm 23}|\gg \Delta m^2_{12}$,
i.e.\
$S_{13}\approx S_{23}$ so that we can simplify
$$p^{13}_{\ell\ell'}+p^{23}_{\ell\ell'}=-4\Re w^{\ell\ell'}_3(w^{\ell\ell'*}_1+w^{\ell\ell'*}_2)=
-4\Re w_{\ell\ell'}^3(\delta_{\ell
\ell'}-w^{\ell\ell'*}_3).$$
getting
\begin{equation}\label{eq:P3nu}\color{rossos}
P(\nubarnu_\ell\to \nubarnu_{\ell'}) \simeq \left\{\begin{array}{ll}
1-4|V_{\ell1}^2V_{\ell2}^2| S_{12}-4|V_{\ell3}^2|(1-|V_{\ell3}^2|)S_{23} & \hbox{for }\ell=\ell'\\
 - 4 \hbox{Re}[ V_{\ell 1} V_{\ell' 2} V_{\ell' 1}^* V_{\ell 2}^* ] S_{12}
 + 4 | V_{\ell 3}^2 V_{\ell' 3}^2 | S_{23}
 \mp P_{\mbox{\tiny CP}} \sum_{\ell''} \epsilon_{\ell \ell' \ell''}& \hbox{for }\ell\neq\ell'
\end{array}\right.
\end{equation}
where the CP-violating terms becomes
$P_{\rm CP}=8J \sin^2\varphi_{13}\sin\varphi_{12}$.
In this limit vacuum oscillations no longer depend on the sign of $\Delta m^2_{23}$,
which controls if neutrinos have `normal' or `inverted' hierarchy.
Inserting the explicit parametrization of $V$ in eq.\eq{Vij} gives
\begin{eqnsystem}{sys:PatmNonStandard}\label{eq:3nuemu}
 P(\nu_e \to \nu_\mu)   & = & s_{23}^2  \sin^2 2\theta_{13} S_{23} +
c_{23}^2 \sin^2 2 \theta_{12}   \; \underline{S}_{12} -P_{\rm CP}  ,\\
 P(\nu_e\to\nu_\tau)   & = & c_{23}^2  \sin^2 2\theta_{13} S_{23} +
s_{23}^2 \sin^2 2 \theta_{12}   \; \underline{S}_{12}  +P_{\rm CP}  ,\\
P(\nu_\mu \to \nu_\tau) & = & c_{13}^4  \sin^2 2\theta_{23} S_{23} 
\textstyle - s^2_{23} c^2_{23} \sin^2 2 \theta_{12} \; \underline{S}_{12} -P_{\rm CP} ,\label{eq:Pave}\\
 \riga{and}\\[-4mm]
\label{eq:Pee}
 P(\nu_e\to \nu_e)       & = &1- \sin^2 2\theta_{13} \;S_{23}-c^4_{13} \sin^2 2 \theta_{12}   \; {S}_{12} , \\
 P(\nu_\mu \to\nu_\mu) & = &1-4c_{13}^2 s_{23}^2 (1-c_{13}^2 s_{23}^2){S}_{23}
 - c_{23}^4 \sin^2 2 \theta_{12} \; \underline{S}_{12} , \\
 P(\nu_\tau \to\nu_\tau)   & = & 1-4c_{13}^2 c_{23}^2 (1-c_{13}^2 c_{23}^2) {S}_{23}
  - s_{23}^4 \sin^2 2 \theta_{12} \; \underline{S}_{12}.
 \label{eq:Pdiagave}
 \end{eqnsystem}
For simplicity we set $\theta_{13}=0$ in the coefficients of the underlined $\underline{S}_{12}$ terms.
Special interesting limiting cases are:
\begin{itemize}
\item $S_{12}\approx 0$ (the baseline is so short that solar oscillations cannot be seen);
\item $\langle S_{23}\rangle\approx 1/2$ (the baseline is so long that atmospheric oscillations are averaged);

\item  $\langle S_{12}\rangle = \langle S_{23}\rangle = 1/2$
(the baseline is so long that both solar and atmospheric oscillations are averaged).
The survival probabilities are given by eq.\eq{Paveraged} and depend on the CP-phase $\phi$.
\end{itemize}

\begin{figure}[t]
$$\includegraphics[width=11.5cm,height=4cm]{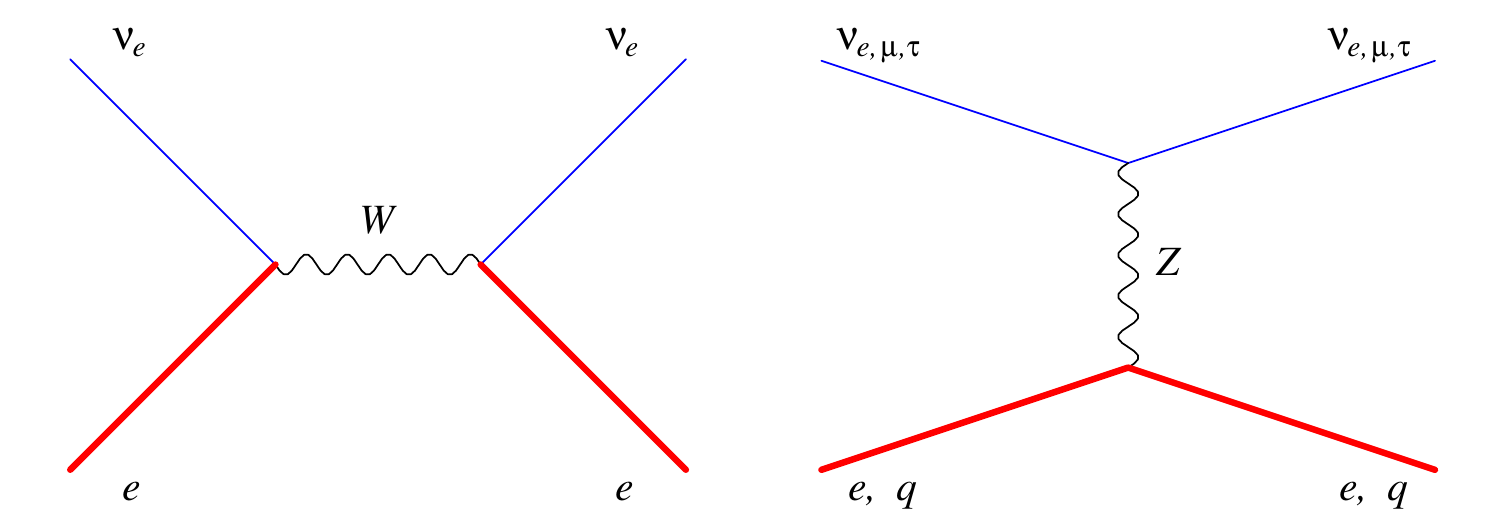}$$
\caption[Matter effects]{\em Interactions of neutrinos with electrons and quarks.\label{fig:materia}}
\end{figure}

\section{Oscillations in normal matter}\label{matterosc}\index{MSW}\index{Oscillations!in matter}
The probability that
a neutrino of energy $E\sim\MeV$ 
gets scattered while crossing the earth is $\sim 10^{-12}$ (section~\ref{detecting}).
Neutrinos of ordinary energies cross the earth or the sun
without being significantly absorbed.
Still, the presence of matter can significantly affect neutrino propagation~\cite{MSW}.
This apparently unusual phenomenon has a well known optical analogue.
A transparent medium like air or water negligibly absorbs light,
but still significantly reduces its speed: $v_{\rm phase} = c/n$,
where $n$ is the `refraction index'.
In some materials or in presence of an external magnetic field
$n$ is different for different polarizations of light,
giving rise to characteristic effects, such as birefringence.
The same thing happens for neutrinos.
Since matter is composed by electrons (rather than by $\mu$ and $\tau$),
$\nu_e$ interact differently than $\nu_{\mu,\tau}$,
giving rise to a flavour-dependent refraction index.
We now compute it and study how oscillations are affected.

{\em Forward} scattering of neutrinos
interferes with free neutrino propagation,
giving rise to refraction.
Scattering of $\nu_\ell$ on electrons and quarks mediated by the $Z$ boson (fig.\fig{materia}b)
is the same for all flavours $\ell=\{e,\mu,\tau\}$, and
therefore does not affect flavour transitions between active neutrinos.
The interesting effect is due to $\nu_e e$ scattering mediated by the $W$ boson (fig.\fig{materia}a),
that is described at low energy by the effective Hamiltonian
(its sign is predicted by the SM)
$$
\mathscr{H}_{\rm eff} = \frac{4 G_{\rm F}}{\sqrt{2}} (\bar\nu_e \gamma_\mu P_L \nu_e)(\bar{e}\gamma^\mu P_L e).$$
In a background composed by non-relativistic and non-polarized electrons and no positrons 
(e.g.\ the earth, and to excellent approximation the sun)
one has
$$\langle \bar{e}\gamma_\mu \frac{1-\gamma_5}{2} e\rangle = \frac{N_e}{2}  (1,0,0,0)_\mu\qquad\hbox{and therefore}\qquad
\langle \mathscr{H}_{\rm eff}\rangle =  \sqrt{2} G_{\rm F} N_e (\bar\nu_e \gamma_0 P_L \nu_e)  $$
where $N_e$ is the electron number density.
Including also the $Z$-contribution\footnote{One needs to evaluate 
quark currents $\bar{q}\gamma_\mu q$
over a background of normal matter.
The result is shown in table~\ref{tab:matter}.
Non obvious but well known properties of the quark currents guarantee that
the proton electric charge is $2q_u+q_d$.
Similarly, one correctly evaluates the average value of the quark currents in terms of proton and neutron number densities $N_n$ and
$N_p$ by simply using $p = uud$ and $n=udd$.
`Ordinary matter' is composed by electrons, protons, neutrons with
 $N_p = N_e$ (no net electric charge)
and $N_n \approx N_p$.
The mass density is $\rho \approx m_p N_p + m_n N_n$.
At tree level matter effects do not distinguish $\nu_\mu$ from $\nu_\tau$;
loop effects generate a small difference of order $(m_\tau/M_W)^2 \sim 10^{-5}$~\cite{MSW}.}, the effective matter Hamiltonian density
in ordinary matter is
\begin{equation}\label{eq:V}
\langle \mathscr{H}_{\rm eff}\rangle =  \bar\nu_\ell A \gamma_0 P_L \nu_\ell\qquad
\hbox{where}\qquad
\qquad
A =\sqrt{2} G_{\rm F}\bigg[N_e \diag(1,0,0) -\frac{N_n}{2} \diag(1,1,1)\bigg]
\end{equation}
is named `matter potential' and is a $3\times 3$ flavour matrix.
If extra sterile neutrinos exist, $A$ becomes a bigger diagonal matrix and all its `sterile' elements vanish.

In table~\ref{tab:matter} we show
the separate contributions to $A$ generated
by matter possibly containing anti-particles.
In order to study neutrino oscillations in the early universe
one more ingredient is needed, as discussed in section~\ref{OscUniverse}.

\medskip

Adding the matter correction to the Hamiltonian density describing free propagation of an ultra-relativistic neutrino,
one obtains a modified relation between energy and momentum, as we will now discuss.
In ordinary circumstances 
the neutrino index of refraction $n$ is so close to one,
$n-1\simeq A/E_\nu \ll 1$,
that optical effects like neutrino lensing are negligible.
On the contrary matter effects significantly affect oscillations,
since $A/(\Delta m^2/E_\nu)$ can be comparable or larger than one.

\begin{table}[t]
$$\begin{array}{c|cc}
\hbox{medium} & A_{\rm CC} \hbox{ for $\nu_e,\bar{\nu}_e$ only}&  A_{\rm NC} \hbox{ for
$\nu_{e,\mu,\tau},\bar{\nu}_{e,\mu,\tau}$}\\  \hline
e,\bar{e} &  \pm \sqrt{2}G_{\rm F} (N_e - N_{\bar e})  & \mp \sqrt{2}G_{\rm F} 
(N_e-N_{\bar{e}}) (1-4\sW^2)/2\\ 
p,\bar{p} &0&  \pm\sqrt{2} G_{\rm F} (N_p-N_{\bar{p}}) (1-4\sW^2)/2\\
n,\bar{n} &0& \mp\sqrt{2} G_{\rm F} (N_n-N_{\bar{n}})/2\\  
\hline
\hbox{ordinary matter} & \pm \sqrt{2}G_{\rm F} N_e  & \mp\sqrt{2} G_{\rm F} N_n/2
\end{array}$$
\caption[Matter potentials]{\em Matter potentials for $\nu$ (upper sign) and $\bar\nu$ (lower sign).
We assumed a non-relativistic and non-polarized background medium,
so that these formul\ae{} do not apply to a background of neutrinos.
\label{tab:matter}}
\end{table}

\subsection{Majorana vs Dirac neutrinos}\label{MDosc}
It is easy to see that pure Majorana neutrinos
oscillate in vacuum in the same way as pure Dirac neutrinos.
Looking only at vacuum oscillations it is not possible to 
experimentally discriminate the two cases.
The additional CP-violating phases present in the Majorana case
do not affect oscillations.

It is less easy to realize that,
in the realistic case of ultrarelativistic neutrinos,
this unpleasant result
continues to hold also for oscillations in matter.
The equations for the neutrino wave-functions in the two cases are


\noindent
\parbox[t]{8cm}{\paragraph{Majorana.}
Neutrinos have only a  left-handed component,
and are described by a single Weyl field $\nu_L$.
Adding the matter term the equation of motion  for the neutrino wave-function is
$$(i \ds  - A \gamma_0) \nu_L = m \bar\nu_L$$
where $m$ is the symmetric Majorana mass matrix.
Squaring, in the ultrarelativistic limit
one obtains  the dispersion relation
$$ (E -A)^2 - p^2 \simeq  m m^\dagger$$
i.e.
$$ p \simeq E - (\frac{m m^\dagger}{2E} +A) .$$
\vspace{1.5cm}}\hfill
\parbox[t]{8cm}{\paragraph{Dirac.}
Neutrinos have both a left and a right-handed component.
Their equation of motion is
$$\left\{\begin{array}{rl}
i\ds \nu_L &= m \bar\nu_R + A \gamma_0 \nu_L\cr
i\ds \bar\nu_R &= m^\dagger \nu_L
\end{array}\right.$$
where $m$ is the Dirac mass matrix.
Eliminating $\nu_R$ and assuming that $A$ is constant one gets
$$[\partial^2 + m m^\dagger + A ~  i\ds\gamma_0] \nu_L = 0.$$
In the ultrarelativistic limit $i\ds\gamma_0 \nu_L\simeq 2i\partial_0 \nu_L$, 
giving the dispersion relation
$$ p \simeq E - (\frac{m m^\dagger}{2E} +A) .$$}

The density of ordinary matter
negligibly changes on a length scale $\sim 1/E$
(which is even typically smaller than an atom)
so that the gradient of $A$ can indeed be neglected.

\medskip

\begin{figure}[t]
$$ \includegraphics[width=7cm]{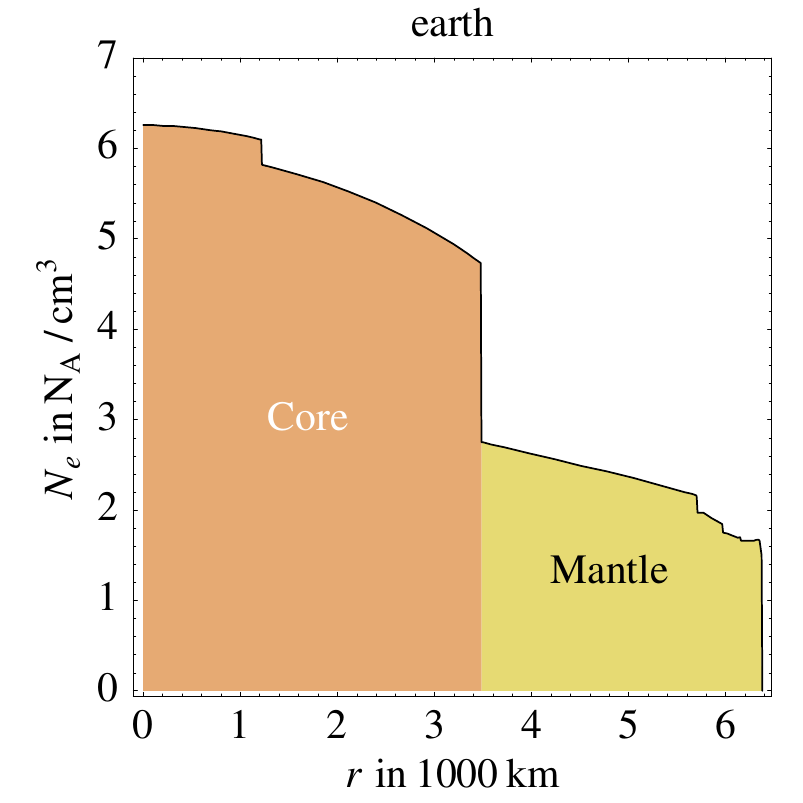}\hspace{1cm} \raise-3mm\hbox{\includegraphics[width=7.3cm]{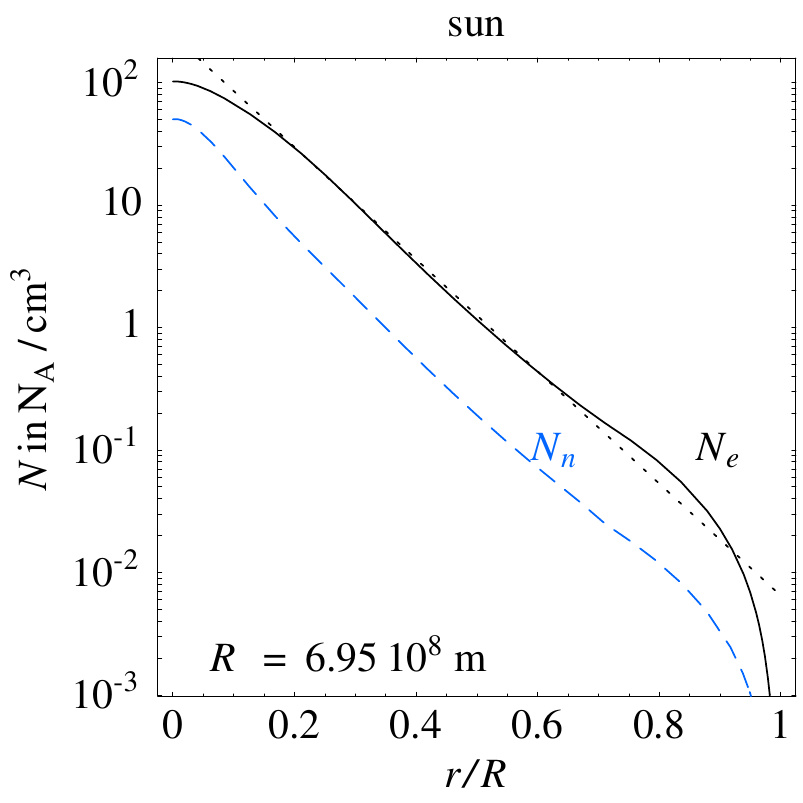}}$$
\caption[Density of the earth and of the sun]{\em Electron number density profile of (a) the earth (b) the sun.
The mass density can be obtained multiplying the number density $N_e$ times the mass
$m_N/Y_e$ present for each electron,
and remembering that
$m_N N_{\rm A} = {\rm gram}$.
 \label{fig:rho}}
\end{figure}

 \noindent
{\em\color{rossos} Summarizing, 
oscillations in matter of ultrarelativistic neutrinos are described by the equation
\begin{equation}\label{eq:m+V}
 i\frac{d}{dx}\nu = H \nu ,\qquad\hbox{where}\qquad
H = \frac{m \cdot m^\dagger}{2E}+A,\qquad \nu = \pmatrix{\nu_e\cr\nu_\mu\cr \nu_\tau},
\end{equation}
that can be solved starting from the production point knowing which flavour is there produced.}
$A$ is given in eq.\eq{V} and 
$mm^\dagger=V^* \cdot \diag(m_1^2,m_2^2,m_3^2) \cdot V^T$
where $V$ is the neutrino mixing matrix and $m_{1,2,3}\ge 0$ are the neutrino mass eigenvalues.
For anti-neutrinos one needs to change $m\to m^* $
(such that $m\cdot m^\dagger$ gets replaced by  $m^\dagger \cdot m = 
 V \cdot \diag(m_1^2,m_2^2,m_3^2) \cdot V^\dagger$;
this induces genuine CP-violating effects) and
$A\to - A$ (the background of ordinary matter breaks CP).
In the case of Majorana neutrinos the mass matrix $m$ is symmetric, so that $m^*=m^\dagger$.

\subsection{Matter oscillations of two neutrinos}
\index{Density!earth}\index{Density!sun}
The matter density can depend on both time and position, but
usually it depends only on the position (e.g.\ in the sun).  Sometimes
it is roughly constant (e.g.\ in the earth mantle\footnote{The
 density profile of the earth, shown in fig.\fig{rho}a, has a few sub-structures~\cite{PREM}.
 The radius is $r=6371\,{\rm km}$.
The continental crust is rigid and has a thickness that varies between 20 to 70 km,
and is made primarily of light elements like potassium, sodium, silicon,
calcium, aluminium silicates.
The mantle is liquid and a has depth of about 2900 km, and is made primarily 
of iron and magnesium silicates.
Density discontinuities of about $6\%$ and $10\%$ are expected to
occur in few-km thick layers at depths of about 400 km and 670 km.
The core is generally believed to be made primarily of iron,
solid in the inner core and liquid in the outer core.
The deepest hole which has ever been dug is only $\sim 10$ km deep.
The above expectations are mostly based on 
seismological data (earthquakes or man-made explosions),
interpreted at the light of our knowledge of physical properties of materials.

To a good approximation one can 
compute neutrino oscillations by considering only the
two main structures (the mantle and the core)
provided that one employs their average density 
over the path followed by the neutrino.
Precise values of the local densities have been computed and are needed for
long-baseline neutrino experiments that hope to discover subtle effects, 
such as CP-violation in neutrino oscillations.

})
and it is convenient to define effective energy-dependent 
neutrino mass eigenvalues $m_{m}^2$, eigenvectors $\nu_m$ and mixing angles $\theta_m$ in
$m$atter by diagonalizing $H$.
These effective oscillation parameters depend on the neutrino energy, 
and of course on the matter density.
In the simple case with only the $\nu_e$ and $\nu_\mu$ flavours
and a mixing angle $\theta$
$$ m m^\dagger = \frac{m_1^2+m_2^2}{2}\pmatrix{1&0\cr0&1}+
\frac{\Delta m^2}{2}\pmatrix{-\cos2\theta & \sin2\theta\cr \sin2\theta & \cos 2\theta}$$
where $\Delta m^2 = m_2^2 - m_1^2$
so that the oscillation parameters in matter are
\begin{equation}\label{eq:paramm}
 \tan2\theta_m = \frac{S}{C},\qquad
\Delta m^2_m = \sqrt{S^2 + C^2},\qquad\hbox{where}\qquad
\begin{array}{l}
S\equiv \Delta m^2 ~\sin2 \theta,\cr
C\equiv \Delta m^2 \cos 2\theta \mp 2\sqrt{2} G_{\rm F} N_e E 
\end{array}
\end{equation}
and $\theta$ and $\Delta m^2$ are the oscillation parameters in vacuum.
The $-$ ($+$) sign holds for $\nu$ ($\bar\nu$) and
$$ V  = \pmatrix{\cos\theta & \sin\theta\cr - \sin\theta & \cos\theta},\qquad
V_m = \pmatrix{\cos\theta_m & \sin\theta_m\cr - \sin\theta_m & \cos\theta_m}$$

\begin{figure}[t]
$$\includegraphics{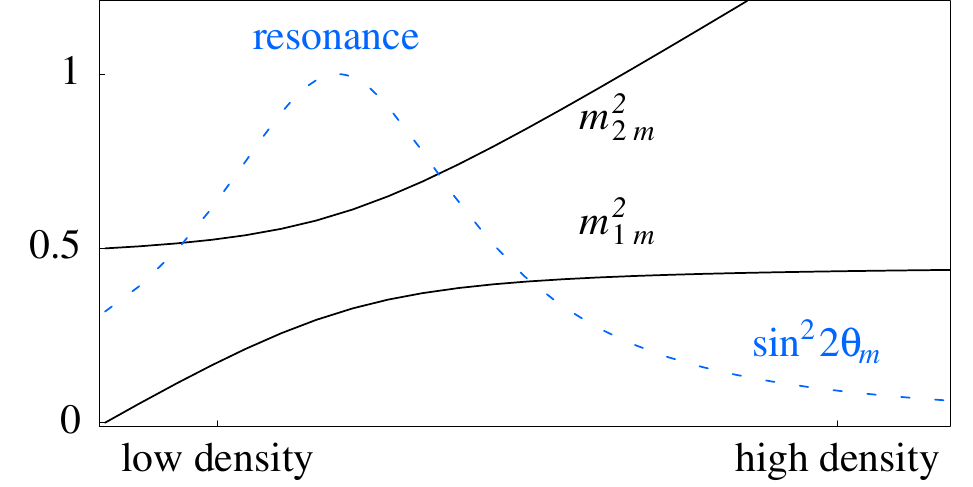}$$
\caption[Mass and mixing in matter]{\em Effective masses and mixing angle in matter for two neutrino flavours as a function of the density.
We take $\theta=0.3$, $\Delta m^2=1/2$ (arbitrary units). \label{fig:risonanza}}
\end{figure}

Fig.\fig{risonanza} shows a  numerical example.
The most noticeable features are:
\begin{itemize}

\item Unlike vacuum oscillations, {\bf matter oscillations distinguish $\theta$ from $\pi/2 - \theta$}.
Consequently $\sin^2 2\theta$ (used in fig.\fig{tipico}a) is no longer a good variable;
it is customary to use $\tan^2 \theta$ in log-scale plots
(because $\tan^2\theta\to 1/\tan^2\theta$ under a reflection $\theta\to \pi/2-\theta$)
and $\sin^2\theta$ in linear-scale plots (because, under the same reflection, $\sin^2\theta \to 1-\sin^2\theta$).
Not caring of the sign of $\theta_m-\pi/4$, eq.\eq{paramm} can be rewritten as
$$\sin^2 2\theta_m = \frac{\sin^2 2\theta}{\lambda^2},\qquad
\Delta m^2_m = \lambda\cdot \Delta m^2 ,
\qquad \lambda = \sqrt{\sin^2 2\theta + \bigg(\cos^2 2\theta \mp \frac{2\sqrt{2}G_{\rm F} N_e E}{\Delta m^2}\bigg)^2}.$$

\item {\bf Resonance}.
If $\Delta m^2 \cos 2\theta > 0$ $(<0)$ the matter contribution can render equal the diagonal elements of the
effective neutrino (anti-neutrino) mass matrix, so that $\theta_m$ can be maximal, $\theta_m = \pi/4$,
even if $\theta\ll 1$.
At the resonance $\Delta m^2_m = \Delta m^2\,\sin2\theta$.
Matter effects resonate at
\begin{equation}\label{eq:Eres}
 E_\nu\sim \frac{\Delta m^2}{2\sqrt{2} G_{\rm F} N_e} = 
3\GeV \frac{\Delta m^2}{10^{-3}\eV^2} \frac{1.5\, {\rm g}/\cm^3}{\rho Y_e}  .
\end{equation}

\end{itemize}
Numerically the matter potential equals
\beq\label{eq:NV}
\sqrt{2} G_{\rm F} N_e = 
0.76 ~10^{-7}\frac{\eV^2}{\MeV}\frac{N_e}{N_{\rm A}/\cm^3}= 
0.76 ~10^{-7}\frac{\eV^2}{\MeV}\frac{Y_e \rho}{{\rm g}/\cm^3}.
\eeq
The typical electron number density of ordinary matter is $N_e \sim 1/{\rm \AA}^3$.
For example, the density of the mantle of the earth is $\rho \approx 3{\rm g}/\cm^3$ and therefore
$N_e = \rho Y_e/m_N \approx 1.5 N_{\rm A}/\cm^3$, where $N_{\rm A}=6.022\,10^{23} $ is the Avogadro number, $m_N$ is the nucleon mass and
$Y_e \equiv N_e/(N_n + N_p) \approx 0.5$ the electron fraction.
Other characteristic densities are $\rho\sim 12 {\rm g}/\cm^3$ in the earth core,
$\rho\sim 100 {\rm g}/\cm^3$ in the solar core,
and $\rho\sim m_n^4\sim 10^{14}{\rm g}/\cm^3$ in the core of a type II supernova.
The density profiles of the earth and of the sun are plotted in fig.\fig{rho}.
\begin{itemize}
\item {\bf Matter-dominated oscillations}.
When neutrinos have high enough energy the matter term dominates: being flavour-diagonal
it suppresses oscillations.
In this situation, neutrinos oscillate in matter
with an energy-independent wave-length $\lambda = \pi/\sqrt{2}G_{\rm F} N_e$.
In the earth mantle $\lambda\sim 3000\km$, comparable to the size of the earth.
\end{itemize}

\begin{figure}[t]
$$\includegraphics[width=\textwidth]{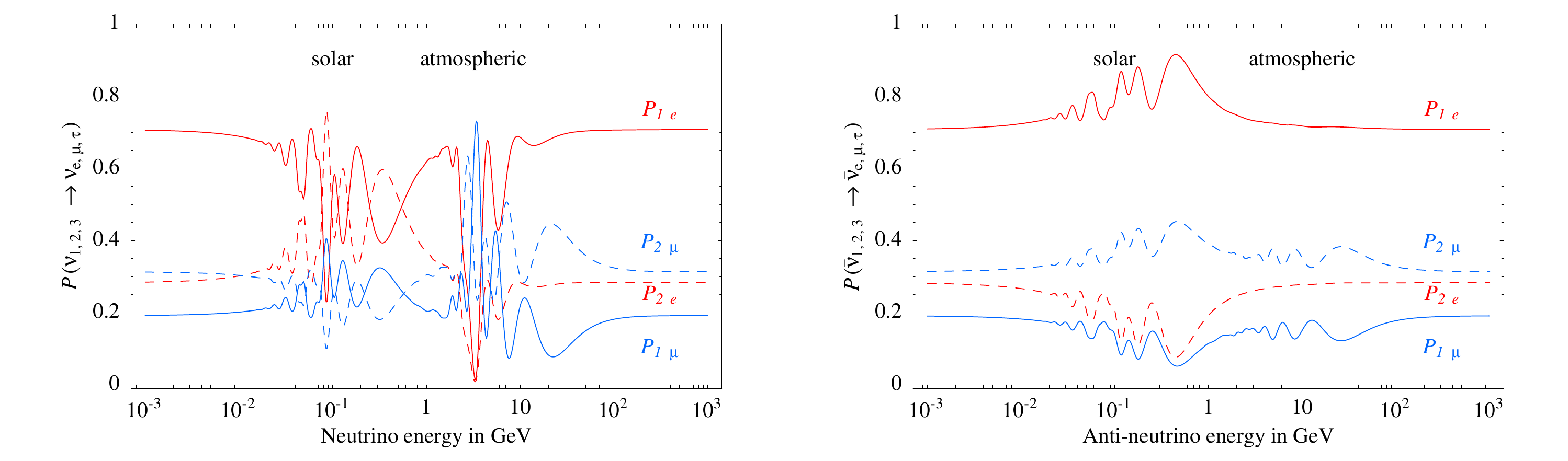}$$
\caption[Earth matter effects]{\em How a neutrino mass eigenstate that
crosses all earth is affected by earth matter effects.
We assumed $\theta_{13}=0.1$, no CP-violation, normal hierarchy and
best-fit solar and atmospheric parameters.
Choosing a much smaller $\theta_{13}$  the `atmospheric' effects around $E_\nu=10$ disappear.
\label{fig:Pif}}
\end{figure}

Let us show the relevance of earth matter effects in a case of practical interest.
Neutrinos produced by distant sources (e.g.\ the sun, supernov\ae,\ldots)
reach the earth as an incoherent mixture of
mass eigenstates, because oscillations
average to zero the coherency among different flavors.
These neutrinos can be detected after having crossed the earth
(indeed in various circumstances only upward-going neutrinos can be detected,
because the background of cosmic rays prevents detection of downward-going neutrinos).
Fig.\fig{Pif} shows how earth matter affects a neutrino mass eigenstate 
that crosses the center of the earth.
The two structures peaked  around $E_\nu \sim 100\MeV$ and around
$E_\nu\sim 10\GeV$ are respectively due to solar and atmospheric oscillations.
Atmospheric effects are present only if $\theta_{13}>0$.

\section{Oscillations in a varying density}\label{MSWresonance}\index{Oscillation!matter enhanced}
In the special cases where the phase in the evolution operator $U={\rm T}\,e^{-i\int_{x_i}^{x_f} H(x) dx}$ is small
(T is the usual time-ordering operator), one can perform a Magnus expansion in $H$~\cite{Petcov}:
\beq \label{eq:Magnus}
U =\exp\left [-\int_{x_i}^{x_f} H(x)dx  - \frac{i}{2} \int_{x_i}^{x_{f}}dx \int_{x_i}^xdy \,[H(x),H(y)]+\cdots\right].
\eeq
The second term vanishes if $H(x)$ is symmetric around the middle point,
as in the case of neutrinos that cross the Earth.


In order to study solar 
and supernova neutrinos
it useful to 
develop an approximation for the oscillation probabilities
for neutrinos produced in the core of the star
(where matter effects are important),
that escape into the vacuum (where matter effects are negligible).
At some intermediate point matter effects can be resonant.

Here, we discuss the case of two neutrino generations in the sun;
later it will be  easy to generalize the discussion.
Briefly, solar neutrinos behave as follows.

\begin{enumerate}\index{Crossing probability}
\item $\nu_e$ are produced in the core of the sun, $r\approx 0$.
The probability of $\nu_e$ being $\nu_{1m}(r\approx 0)$ or $\nu_{2m}(r\approx 0)$ are
$\cos^2\theta_m$ and
$\sin^2\theta_m$ respectively.
When matter effects are dominant $\nu_e\simeq \nu_{2m}$ (i.e.\ $\sin^2\theta_m = 1$).

\item The oscillation wave-length $\lambda$ is much smaller than the solar radius $R$.
Therefore neutrinos propagate for many oscillation wave-lengths:
the phase averages out so that
we have to combine probabilities instead of amplitudes.
If the density changes very slowly (`adiabatic approximation', see below)
each neutrino mass eigenstate will remain the same.
Otherwise neutrinos will flip to the other mass eigenstate
with some level-crossing probability $P_C$ that we will later compute:
$$\nu_{2m}(r\approx 0)\hbox{ evolves to }  \left\{\begin{array}{l}
\nu_{2m}(r\approx R) =\nu_2\hbox{ with probability $1-P_C$}\cr
 \nu_{1m}(r\approx R)=\nu_1 \hbox{ with probability $P_C$}\end{array}\right.$$
(and similarly for $1\leftrightarrow 2$).

\item Neutrinos propagate from the sun to the earth, and possibly inside the earth
before reaching the detector.
For simplicity we start ignoring earth matter effects.
The complete 3-neutrino analysis including earth matter effects
is presented in section~\ref{rho}.

\item Finally, the $\nu_1$ ($\nu_2$) is detected as $\nu_e$ with probability $\cos^2\theta$
($\sin^2\theta$).
\end{enumerate}
\begin{figure}[t]
$$\includegraphics{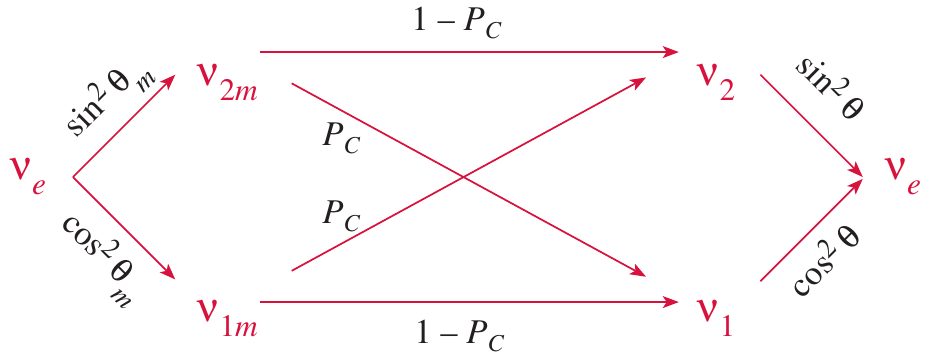}$$
\caption[Propagation steps]{\em Propagation of a neutrino from the sun to the earth,
that leads to eq.\eq{Psun}.
See the text.
\label{fig:propagazione}}
\end{figure}
Combining all these probabilities, as summarized in fig.\fig{propagazione}, one gets~\cite{Petcov}
\begin{equation}\label{eq:Psun}
P(\nu_e \to \nu_e) = \frac{1}{2}+ (\frac{1}{2} - P_C) \cos2\theta \cos2\theta_m
\end{equation}
where $\theta_m$ is the effective mixing angle at the production point.
It is instructive to specialize eq.\eq{Psun} to a few limiting cases:
\begin{itemize}
\item[a)] $P(\nu_e\to \nu_e) = 1 -\frac{1}{2} \sin^22\theta$ (averaged vacuum oscillations)
when matter effects are negligible: $\theta_m = \theta$ and $P_C = 0$.
This case is realized for solar neutrinos at lower energies.
\label{ExampleLMAtext}

\item[b)] $P(\nu_e\to \nu_e) = P_C$ when $\cos 2\theta_{2 m} = -1$ (i.e.\ matter effects dominate so that the heavier effective
neutrino mass eigenstate is
$\nu_{2m}(0)\simeq \nu_e$) and $\theta\ll1$.
\item[c)] $P(\nu_e\to \nu_e) = \sin^2\theta$ when $\cos 2\theta_{2 m} = -1$ and neutrinos propagate adiabatically ($P_C=0$).
This case is realized for solar neutrinos at higher energies.

\item[d)] $P(\nu_e\to \nu_e) = 1 -\frac{1}{2} \sin^22\theta$ 
when $\cos 2\theta_{m} = -1$
and in the extreme non-adiabatic limit ($P_C = \cos^2 \theta$).
This value of $P_C$ can be computed
by considering very dense matter that abruptly terminates in vacuum.
The produced neutrino $\nu_e\simeq \nu_{2m}$
does not change flavour at the transition region,
since it is negligibly short.
Therefore
$P_C = |\langle\nu_e| \nu_1\rangle|^2 = \cos^2\theta$.

To see why $P(\nu_e\to \nu_e) $ is equal to averaged vacuum oscillations
let us follow the neutrino path:
matter effects are very large and block oscillations
around and after the production point,
until they become suddenly negligible.


\begin{figure}[t]
$$\includegraphics[width=0.7\textwidth]{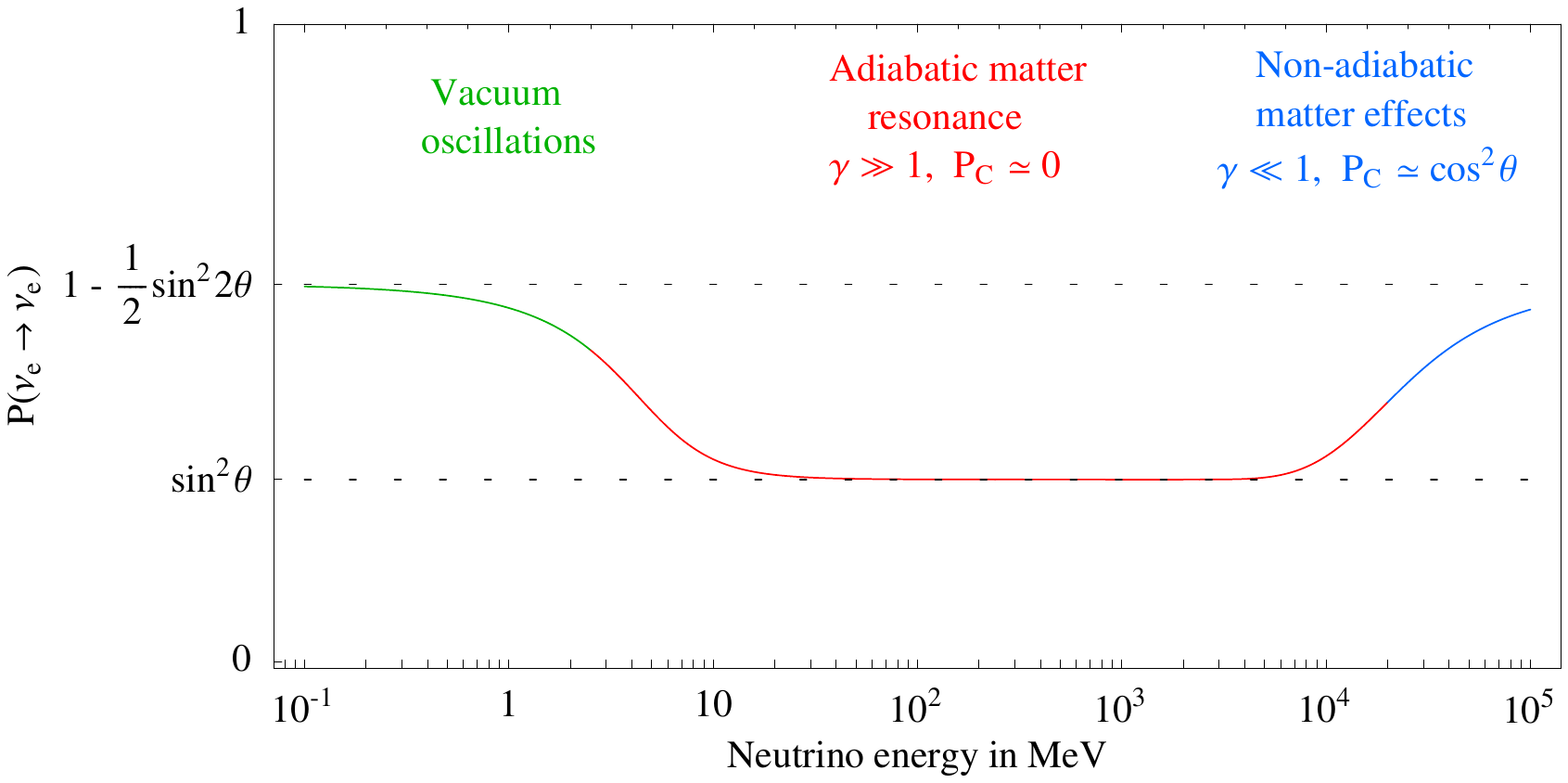}$$
\caption[Example of oscillations]{\em Behavior of $P(\nu_e\to \nu_e)$
that illustrates the limiting regimes {\rm a, c, d} discussed at page~\pageref{ExampleLMAtext}.
{\rm a)} At lower energies matter effects are negligible.
{\rm c)} At intermediate energies matter effects are dominant and adiabatic.
{\rm d)} At higher energies the MSW resonance is no longer adiabatic.
The numerical example corresponds to solar oscillations.
Absorption is neglected.
\label{fig:ExampleLMA}}
\end{figure}

\end{itemize}

In the next section we compute $P_C$.
The main result should be intuitively clear:
$P_C =0$ when the variation of the matter density is smooth enough.
This case was studied in~\cite{MSW} and in the first paper in~\cite{NuSN}.
We will see that this is the case for solar neutrinos, and that
the general discussion in the next section might be necessary in the case of supernova neutrinos,
if $\theta_{13}\sim $ few degrees.

\subsection{Adiabaticity}
We rewrite the evolution equation $i\, d\nu/dx = H(x) \nu$
in the new basis of instantaneous matter mass eigenstates $\nu_m$,
related to the flavour basis by
\beq  \nu = V_m(x) \nu_m,\qquad
V_m = \pmatrix{\cos\theta_m & \sin\theta_m \cr  -\sin\theta_m & \cos\theta_m} .
\eeq
In this new basis the effective Hamiltonian is diagonal,
$H_m = V_m^{-1} H V_m=\diag(m_{1m}^2,m_{2m}^2)/2E$.
However the wave equation contains an additional term 
due to the fact that the new basis is position-dependent:
\begin{equation}\label{eq:inst}
i\frac{d\nu_m}{dx} = \bigg(H_m - i V_m^{-1} \frac{dV_m}{dx}\bigg) \nu_m = 
\pmatrix{m_{1m}^2/2E_\nu & -i\, d\theta_m/dx \cr
i\,d\theta_m/dx & m_{2m}^2/2E_\nu} \nu_m . 
\end{equation}
This basis is convenient for analytic computations because
the off-diagonal term can be relevant only
in the (possibly narrow) interval of $r$ where level crossing occurs.
In the extreme adiabatic limit where the density gradient is small enough that we can always neglect $d\theta_m/dx$,
the level-crossing probability is $P_C=0$.
A resonance is adiabatic if the ratio  between 
the difference in the diagonal elements and the off-diagonal element of\eq{inst} is 
always much larger than one.
Therefore one usually defines
\beq\gamma \equiv \min_x \bigg|\frac{\Delta m_{m}^2 /4E_\nu}{d\theta_m/dx}\bigg|. \eeq
In general the position of the minimum depends on the detailed form of the matter potential $A$.
Whenever the resonance is sharp enough that, around it,  
$A$ can be approximated with a linear function of $x$,
eq.\eq{paramm} shows that the minimum occurs  at the resonance point
(e.g.\ in the case of a $\nu_e/\nu_{\mu,\tau}$ resonance in normal matter,
the resonance condition is $2E_\nu  \sqrt{2} G_{\rm F} N_e(r) =
\Delta m^2  \cos 2\theta$) and $\gamma$ equals 
\beq\gamma =
 \left|\frac{2H_{12}}{\displaystyle \frac{1}{2H_{12}}\frac{d(H_{22}-H_{11})}{dx}}\right|_{\rm res}=
\tilde\gamma \cdot \frac{\sin^2 2\theta}{2\pi \cos2\theta},\qquad
\color{blus}\tilde\gamma =  \frac{\pi \Delta m^2/E_\nu}{
|d\ln A/dx|_{\rm res}}.\eeq
where the second analytic expression holds for a two neutrino system,
and the first expression, written in terms of the `crossing elements' of the
effective Hamiltonian $H$ in the flavour basis,
is useful in numerical computations in generic cases with many neutrinos.

\begin{figure}[t]
$$\includegraphics[width=17cm]{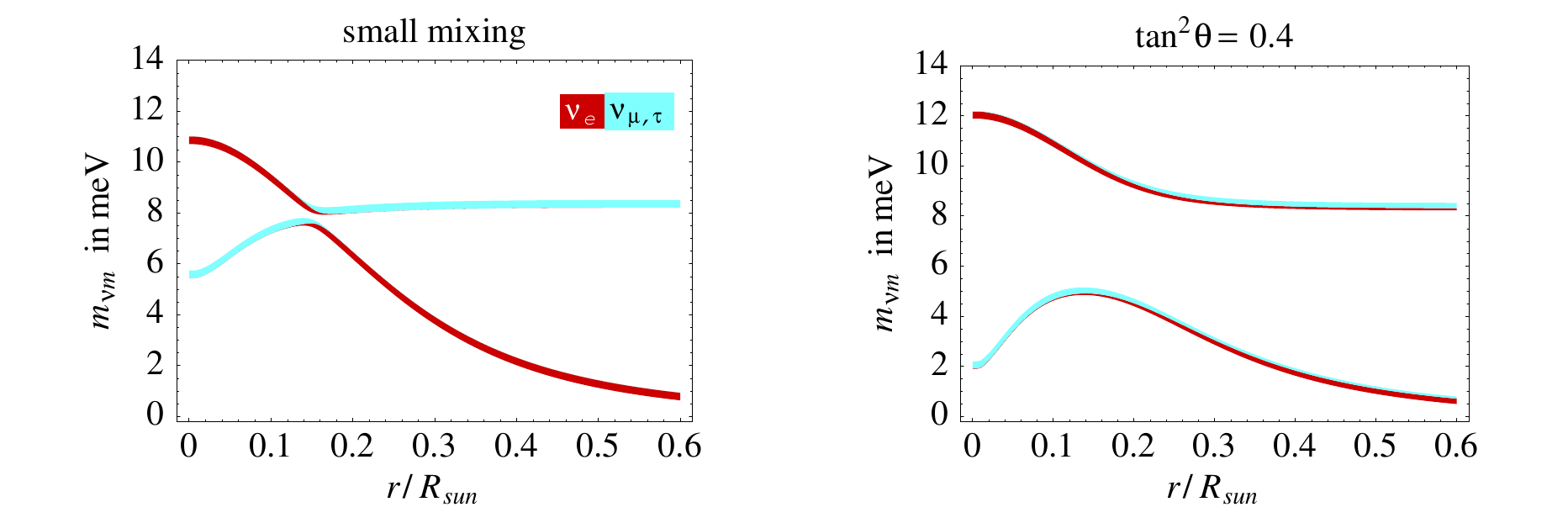}$$
\caption[Example of matter effects]{\em Variation of the matter neutrino eigenstates inside
the sun for $E_\nu = 10\MeV$,
$\Delta m^2 = 7~10^{-5}\eV^2$ and $m_1 = 0$
(i.e.\ $m_2 \approx 8 \meV$).
Dark red (light blue) denotes the $\nu_e$ ($\nu_{\mu,\tau}$) flavour component.
\label{fig:NuSun}}
\end{figure}

A resonance is adiabatic if $\gamma\sim\tilde\gamma\theta^2\gg 1$.
Physically $\tilde\gamma$ is the number of vacuum oscillations
(wave length $\lambda_0 = 4\pi E_\nu/\Delta m^2$) present
in the typical length-scale where the matter potential changes
($r_0 = |d\ln A/dr|^{-1}_{\rm res}$).
In the sun 
$$\tilde\gamma\approx \frac{\Delta m^2/E_\nu}{10^{-9}\eV^2/\MeV}$$
 having used the approximate density
$N_e(r) = 245 N_{\rm A}/\cm^3 \times  \exp (-10.54\,r/R)$ (dashed line in fig.\fig{rho}b).\footnote{We are using the average density.
The transition becomes less adiabatic if $N_e$ fluctuates
on scales comparable to the neutrino oscillation wavelength.
This effect seems negligible in the sun,
where the amplitude of helioseismic fluctuations should be too small.}
The solar neutrino anomaly is due to
oscillations with large mixing angle and $\Delta m^2 \approx 7~10^{-5}\eV^2$.
The level crossing scheme is shown in fig.\fig{NuSun}b
and corresponds to a broad adiabatic resonance:
at $E_\nu\sim 10\MeV$ one has $\gamma\gg1$ and consequently $P_C = 0$.
The resonance ceases to be adiabatic at $E_\nu \circa{>}10\GeV$,
much higher than the maximal solar neutrino energy.
Fig.\fig{ExampleLMA} illustrates the behavior of $P(\nu_e\to \nu_e)$.

\subsection{Analytic approximation for the crossing probability}

$P_C$ can be accurately approximated as follows~\cite{Petcov}.
When the  mixing angle is small there is a narrow resonance
(such as the one shown in fig.\fig{NuSun}a) and one
can approximate any smooth density profile (e.g.\ fig.\fig{rho} show the solar density)
with a simple function, such that the wave equation\eq{inst} can be solved analytically.
By choosing an appropriate simple function that resembles the true one,
this procedure can be applied even to the broader resonances generated by
large mixing angles $\theta\sim \pi/4$ 
(when $\theta > \pi/4$ there is no resonance point,
but adiabaticity is still dominantly violated only in the restricted region
where $A(r)\simeq \Delta m^2/2E_\nu$).
Approximating $A$ with an exponential, $A\propto e^{-r/r_0}$,
and solving\eq{inst} by brute force, one obtains
\begin{equation}\label{eq:PC}\color{rossos}
P_C = \frac{e^{\tilde\gamma \cos^2\theta}-1}{e^{\tilde\gamma} -1},\qquad
\tilde\gamma =  \frac{\pi  r_0 \Delta m^2}{E_\nu}
\end{equation}
This expression was used in analyses of solar neutrinos.
When $\gamma\gg1$ we go back to the adiabatic approximation, $P_C=0$.
In the non-adiabatic limit, $\gamma\ll 1$, one gets $P_C= \cos^2 \theta$ as expected.
If the density profile $A(r)$ is not exactly exponential,
a good approximation is often obtained  by inserting in\eq{PC}  
the gradient $\color{rossos} 1/r_0 = |d\ln A/dr|$ evaluated at the point $r=r_*$
where $\color{rossos}A(r_*) = \Delta m^2/2E_\nu$.

\medskip

Exact solutions are available  also for a few other density distributions.
For $A\propto r^n$ and any $n$ $P_C$  can be written in `double exponential' form 
\begin{equation}\label{eq:PCpower}
P_C =
\frac{\exp\bigg[-\pi\gamma F/2\bigg]-\exp\bigg[-\pi\gamma F/2{\sin^2\theta}\bigg]}
{1-\exp\bigg[-\pi\gamma F/2\sin^2\theta\bigg]}\end{equation}
where the function $F(\theta)$ has a simple expression for
$n=1$ ($F = 1$), for $n=-1$ ($F= (1-\tan^2\theta)^2/(1+\tan^2\theta)$)
and for
$n=-\infty$, which is again the exponential case
($F = 1- \tan^2\theta$ and eq.\eq{PCpower} simplifies to\eq{PC}).
The profile of a supernova is maybe well approximated by $n\approx 3$.

When the resonance is narrow, $\theta\ll1$ as in fig.\fig{NuSun}a,
all above solutions reduce to the
LSZ approximation:
$\color{rossos} P_C\simeq  e^{-\pi\gamma/2}$,
derived using a technique essentially similar to the WKB approximation,
developed in quantum mechanics to compute reflection probabilities.
Basically, it employs the adiabatic approximation and avoids the resonance
region (where it is not valid) by
making a mathematical excursion in the complex $x$ plane.
When  the resonance is not narrow
the LSZ technique gives
$P_C\simeq  e^{-\pi\gamma F/2}$.
The function $F(\theta)$ is expressed as an integral,
that can be evaluated numerically for arbitrary density profiles,
and analytically for various simple profiles.
In this way the functions $F$ in eq.\eq{PCpower} can be
computed systematically.
However, the LSZ
approximation fails in the extreme non-adiabatic limit ($\gamma\ll 1$) giving
$P_C= 1$ rather than $\cos^2\theta$.
Usually one improves the LSZ approximation
by replacing it with the `double exponential' form of eq.\eq{PCpower},
because this turned out to be the exact solution in all cases where it was computed.

\medskip

For anti-neutrinos in normal matter, the matter potential has the opposite sign.
If one also changes the sign of $\Delta m^2$ the oscillation probabilities remain the same.
Therefore the $\bar\nu$ oscillation probabilities are obtained from the above formul\ae{}
valid for neutrinos
by replacing $\theta\leftrightarrow \pi/2-\theta$.

\bigskip

\section{The density matrix and solar oscillations}\label{rho}
In order to make physics as transparent as possible,
so far we studied the simplest situations using the simplest formalism.
Such simple computations can also be done using the  neutrino density matrix,
which becomes necessary in more complicated situations
 and is often convenient even in the simplest situations,
 when physics is more conveniently described by mixed
 rather than by pure quantum states.

In fact, in various circumstances of experimental interest
neutrinos are detected after many oscillations.
The neutrino mass eigenstates $\nu_i$ develop large
phase differences $\varphi_i - \varphi_j$ which depend on  neutrino energy.
As a result the flavour composition of $\nu(E_\nu) \equiv \sum_{i=1}^3 \rho_{ii}(E_\nu) e^{i\varphi_i(E_\nu)} \nu_i$
depends so strongly on neutrino energy 
that only its average value is measured.
However at each $E_\nu$ the pure state $\nu(E_\nu)$ 
has a well defined flavour, that is not measured.
The average flavour can be computed performing
the appropriate average over $E_\nu$: however doing
numerically averages of rapidly oscillating functions
is a  time-consuming task.
It is more convenient to employ the density matrix formalism
that directly describes the `average' neutrino which is observed.

Of course, in simple situations the average is easily performed analytically.
In the case of vacuum oscillations the
regime of `averaged oscillations' was discussed at page~\pageref{ABC},
and generalized in eq.~(\ref{sys:PatmNonStandard}).
In the case of matter oscillations eq.\eq{Psun} contains such average.
The density matrix becomes really useful in more difficult situations 
with multiple $\Delta m^2$ (e.g.\ when `atmospheric'  oscillations are averaged, while
`solar' ones are  not).

Furthermore, in sections~\ref{OscAbs} and~\ref{OscUniverse}
we will consider cases where neutrino scatterings 
are not negligible and break the coherence among neutrino mass eigenstates,
so that propagation of neutrinos must be described
in terms of a neutrino density matrix.

\bigskip

The evolution equation for $\rho_E$,
 the $3\times3$ density matrix of neutrinos with energy $E$, is
\beq\label{eq:dotrhofull}
\frac{d\rho_E}{dx} = -i [H,\rho_E] 
\eeq
where $H$ is given in eq.\eq{m+V}.
The explicit solution is $\rho(L) = e^{-i H L}\rho(0)e^{iHL}$.
In vacuum the solution becomes
\beq\label{eq:rhoevo}
\rho_{ij}^{\rm m}(L) = \rho_{ij}^{\rm m}(0) \exp\bigg[{-i(E_i-E_j)L}\bigg]\qquad
\hbox{and in particular}\qquad
 \rho_{ii}^{\rm m}(L)=\rho_{ii}^{\rm m}(0).
\eeq
The superscript `$^{\rm m}$' denotes that we used the basis of neutrino
mass eigenstates.
Eq.\eq{rhoevo} also holds in uniform matter, in the basis of matter eigenstates.
The mass matrix in the flavour basis is obtained as
$\rho = V^{\rm m}\rho^{\rm m} V^{\rm m\dagger}$.

\medskip

\subsection{Fast average over fast oscillations}\label{AverageOsc}
Since too rapid oscillations cannot be observed, and since computing numerically
the relevant energy-averaged oscillation probabilities is time-consuming,
one often likes to replace the exact oscillation formula
adding something that damps oscillations in the long-baseline limit.
In the simplest case of vacuum oscillations of two neutrinos one can use
$$ P(\nu_e\to \nu_\mu) =  \sin^2 2\theta \frac{ 1- e^{-\epsilon
L \Delta m^4/4E_\nu^2}\cos (\Delta m^2 L/2E_\nu)}{2}$$
with small $\epsilon$,  e.g. $\epsilon =0.01 L$.
We here show how to  correctly extend this approximation to a 
generic three-neutrino case, possibly with matter effects.
It is convenient to consider the evolution of the density matrix in the
basis of matter eigenstates.
Its diagonal elements remain constant, while the phases of off-diagonal elements oscillate.
The correct approximation is 
\beq\label{eq:rhoaverage}
\rho_{ij}^{\rm m}(E_\nu,L) = \rho_{ij}^{\rm m}(E_\nu,0) \exp\bigg[-i(H_i-H_j) L - \epsilon(H_i-H_j)^2L\bigg]\qquad
 \hbox{for $i\neq j$}\eeq
 where $H_i$ are the Hamiltonian eigenvalues.
Off-diagonal entries get suppressed after many oscillations, as desired.
The density profile of the earth is approximatively
piecewise constant,  such that the above formula can be applied in each piece.
For a generic density profile the same average is performed by adding an
appropriate coherency-breaking term to the evolution equation:
\beq\label{eq:dotrhoaverage}
\frac{d\rho}{dx} = -i [H,\rho] - \epsilon [H,[H,\rho]].
\eeq
One can verify that the extra term suppresses
 $\Tr(\rho^2)$, thereby transforming a pure state into a mixed state,
 and that when $H$ is constant acts like in\eq{rhoaverage}.
 Eq.\eq{dotrhoaverage} can be obtained by considering 
 two energies $E_1$ and $E_2$ and suitably approximating
 the equation for the average $\rho = (\rho_{E_1} + \rho_{E_2})/2$.   

\medskip

Some theorists speculate that, at some fundamental level (quantum gravity?),
quantum mechanics might have to be modified adding de-coherence terms.
Possible decoherence terms
are restricted by consistency considerations~\cite{Decoherence}, which suggest the general form
$$\dot \rho =-i[H,\rho] - [D,[D,\rho]]\qquad \hbox{with}\qquad [D,H]=0$$
where $D$ is an arbitrary operator.
Our approximation in eq.\eq{dotrhoaverage} corresponds to  $D\propto H$.
Regarding such term not as a  computational trick but as a real physical effect,
we notice that the simplest choice $D\propto H$ is poorly probed by neutrino oscillations,
because they are a low-energy phenomenon.
Claims that neutrino oscillations allow sensitive tests of de-coherence are based
on appropriate non-minimal choices of $D$.

\subsection{Solar oscillations of three neutrinos}
We here extend the solar oscillation formula, eq.\eq{Psun},
to 3 neutrinos.  Analytical expressions that describe the full evolution
are conveniently obtained using the density matrix formalism.
Such expressions clarify the physics, but are not needed in numerical
computations.  Indeed a simple and systematic way of
performing any computation consists in  evolving the density matrix
step-by-step, starting from its initial value.

$\nu_e$ are produced around the center of sun
where matter eigenstates are given by $\nu_i^{\rm m} = V^{\rm m*}_{\ell i}\nu_\ell$.
The initial density matrix in the matter eigenstate basis is
$\rho^{\rm m} = V^{\rm m\dagger} \Pi_e V^{\rm m}$ where $\Pi_e = \diag(1,0,0)$ is the projector over
$\nu_e$.
In adiabatic approximation the diagonal elements of $\rho^{\rm m}$ remain constant,
while the flavor of matter eigenstates evolves.
The off-diagonal elements acquire oscillation phases.
In many relevant cases phases differences are large and 
therefore amount to average to zero the off-diagonal elements,
resulting in
$\rho^{\rm m} = \diag(|V_{e1}^{\rm m}|^2, |V_{e2}^{\rm m}|^2,|V_{e3}^{\rm m}|^2)$.
In this limit quantum amplitudes reduce to classical probabilities,
fully described by the diagonal elements of $\rho^{\rm m}$.

We must take into account that  adiabaticity can be violated at level crossings.
Solar neutrinos undergo one $\nu_1^{\rm m}/\nu_2^{\rm m}$ level crossing
and exit from the sun as
\begin{eqnarray}
\rho^{\rm m}_S &=&{\rm diag} \pmatrix{1-P_C & P_C &0\cr P_C &1-P_C & 0\cr 0&0&1} \cdot\label{eq:rhoS}
\pmatrix{|V_{e1}^{\rm m}|^2\cr |V_{e2}^{\rm m}|^2\cr |V_{e3}^{\rm m}|^2} \\
	&=&\nonumber
	{\rm diag}\bigg(
	\cos^2 \theta_{13}\bigg[\frac{1}{2}+(\frac{1}{2}-P_C)\cos 2\theta^{\rm m}_{12}\bigg],
	\cos^2\theta_{13}\bigg[\frac{1}{2}-(\frac{1}{2}-P_C)\cos 2\theta^{\rm m}_{12}\bigg],
	\sin^2\theta_{13}\bigg).
\end{eqnarray}
This holds in vacuum where matter eigenstates coincide with vacuum eigenstates.
In the last equation we approximated $\theta_{13}^{\rm m}\simeq\theta_{13}$ because
matter effects negligibly affect the `atmospheric' $\theta_{13}$ mixing. 
Solar neutrinos detected during the day do not cross the earth:
therefore the oscillation probability is
\beq P(\nu_e\to\nu_e,~{\rm day}) = \Tr[\Pi_e \rho_S]= (\rho_S)_{ee} = \sin^4\theta_{13} + 
\cos^4\theta_{13} \bigg[
 \frac{1}{2}+ (\frac{1}{2} - P_C) \cos2\theta_{12} \cos2\theta^{\rm m}_{12} \bigg]
 \eeq
that generalizes eq.\eq{Psun} to $\theta_{13}\neq 0$.

\medskip

\subsection{Earth regeneration}
Neutrinos that reach the earth with density matrix $\rho_E$
can be detected after crossing the earth:
the density matrix at detector  becomes
$\rho_{E}=U \rho_S U^\dagger$ where $U$ is the unitary matrix describing
evolution inside the earth, computed solving eq.\eq{m+V}.
It is often convenient to compute the evolution numerically performing the average 
described in section~\ref{AverageOsc}.
The computation is easily done in mantle/core approximation.
Eq.\eq{rhoaverage} tells how $\rho^{\rm m}$ evolves in a medium with constant density.
At the sharp air/mantle and mantle/core transitions
one simply has to rotate $\rho^{\rm m}$ from the old to the new basis of
matter eigenstates:
when passing from medium $A$ to medium $B$ one has
$\rho^{m}_B = P \rho_A^{\rm m} P^\dagger$.
Notice that the factor $P=V^{\rm m\dagger}_B V^{\rm m}_A$ can be interpreted as
a non-adiabaticity factor, fully analogous to the one introduced in eq.\eq{rhoS}
to account for the possibly non-adiabatic MSW resonance.
Indeed, when levels $i$ and $j$ cross, $P$ is a rotation in the $(ij)$ plane with angle 
$\alpha$ given by $\tan^2\alpha  =P_C/(1-P_C)$,
where $P_C$ is the level-crossing probability.
When level crossing is adiabatic $P_C=0$, $\alpha=0$ and $P = \One$.
When level crossing is fully non-adiabatic $P_C = 1$, $\alpha = \pi/2$
and the matrix $P$ is given by $P=V^{\rm m\dagger}_B V^{\rm m}_A$.
When the off-diagonal elements of $\rho^{\rm m}$ can be neglected
this procedure reduces to combining classical probabilities as in eq.\eq{rhoS}.

\medskip

In the case of solar neutrinos,  earth matter affects neutrinos detected during the night
and one is interested in the oscillation probability
$$P(\nu_e\to\nu_e,~{\rm night}) = (\rho_E)_{ee} = 
\sum_{i=1}^3 P_{ie} \cdot (\rho_S)_{ii},\qquad P_{ie} = |\bAk{\nu_e}{U}{\nu_i}|^2.$$
At solar neutrino energies $E_\nu\circa{<}10\MeV$
earth matter effects negligibly affect the most splitted neutrino $\nu_3$,
so that  $P_{3e}\approx |\bk{\nu_e}{\nu_3}|^2=\sin^2\theta_{13}$.
Eliminating $P_{1e}=1-P_{3e}-P_{2e}$ gives
\begin{equation}\label{eq:Pee3nu}
P(\nu_e\to\nu_e,~{\rm night})
=P(\nu_e\to\nu_e,~{\rm day})+\frac{P_{2e}-c_{13}^2 s_{12}^2}{c^2_{13}\cos 2\theta_{12}}
(1-2P(\nu_e\to\nu_e,~{\rm day}) - 2 s_{13}^2 + 3 s_{13}^4).\end{equation}
This formula describes the effect known as `earth regeneration' of solar neutrinos.
One needs to compute the function $P_{2e}(E_\nu)$ taking into account
the path followed by the neutrino inside the earth.
Neglecting earth matter effects one has $P_{2e}=c_{13}^2s_{12}^2$ and
the survival probability reduces to its `day' value.

\section{Oscillations and absorption}\label{OscAbs}
We study the combined effect of oscillations and absorption,
starting from the simplest case: neutrinos propagate in normal
matter that gets negligibly affected by neutrinos.
So far this case is not relevant for any experimental result:
indeed absorption becomes relevant in the sun at $E_\nu\circa{>}100\GeV$
and in the earth at $E_\nu \circa{>}10\TeV$.
In the future in might become relevant for
a) detection of up-going ultra-high energy neutrinos of cosmic origin.
b) detection of neutrinos generated by annihilations of dark matter particles
clustered around the center of the sun and/or of the earth.

As usual, we can neglect $\nu\leftrightarrow\bar\nu$ spin-flip reactions, suppressed by a factor $(m_\nu/E_\nu)^2$.
Therefore the appropriate formalism  consists in studying the spatial evolution of the $n\times n$ flavour
 {\em density matrices} of neutrinos, $\rho(E_\nu)$, and of anti-neutrinos, $\bar\rho(E_\nu)$
 where $n=3$ or greater if one considers extra sterile neutrinos.
Matrix densities are necessary because
scatterings damp coherencies, so that neutrinos are not in a pure state.
The evolution equation is
\beq \label{eq:drho}\frac{d\rho}{dr} =
- i [{H},\ \rho] +
\left.\frac{d\rho}{dr}\right|_{\rm CC}+
\left.\frac{d\rho}{dr}\right|_{\rm NC}+
\left.\frac{d\rho}{dr}\right|_{\rm in}
\eeq
with an analogous equation for $\bar\rho$.
The first term describes oscillations in vacuum or in matter, and the Hamiltonian
$H$ is given in eq.\eq{m+V}.
The last term represent the neutrino injection due e.g.\ to annihilations of DM particles.
The second and the third term describe the absorption and re-emission 
due to CC and NC scatterings.
Deep inelastic scatterings of $\nubarnu$ on nucleons are the dominant processes at $E_\nu \gg \GeV$
(see section~\ref{nuNucleon>>GeV}).

\medskip

NC scatterings $\nubarnu N \leftrightarrow \nubarnu N$  remove a neutrino from the flux 
and re-inject it with a lower energy. 
So they contribute to the evolution equation as:
\beq\label{eq:NC}\left.\frac{d\rho}{dr}\right|_{\rm NC} = - \int_0^{E_\nu} dE'_\nu 
\frac{d\Gamma_{\rm NC}}{dE'_\nu} (E_\nu,E'_\nu) \rho(E_\nu)+
\int_{E_\nu}^\infty dE'_\nu 
\frac{d\Gamma_{\rm NC}}{dE_\nu} (E'_\nu,E_\nu) \rho(E'_\nu)\eeq
where
\beq\label{eq:Gamma}
\Gamma_{\rm NC}(E_\nu,E'_\nu) = N_p(r)\ \diag\sigma(\nu_\ell p\to \nu_\ell' X)
+N_n(r)\ \diag\sigma(\nu_\ell n\to \nu_\ell' X)\eeq
and $N_{p,n}(r)$ are the proton and neutron profile densities\footnote{
In the sun $N_p/N_n$ varies from the BBN value, $N_p/N_n\sim 7$ present in the outer region 
$r/R_\odot \circa{>}0.3$,
down to $N_p/N_n\sim2$ in the central region composed of burnt $^4$He.
The earth is mostly composed by heavy nuclei, so that $N_p\approx N_n$.}
 The first term describes the absorption:
 the integral over $E'_\nu$ just gives the total NC cross section.
  The second term describes the reinjection of lower energy neutrinos.

\medskip

\begin{figure}[t]
$$\includegraphics[width=\textwidth]{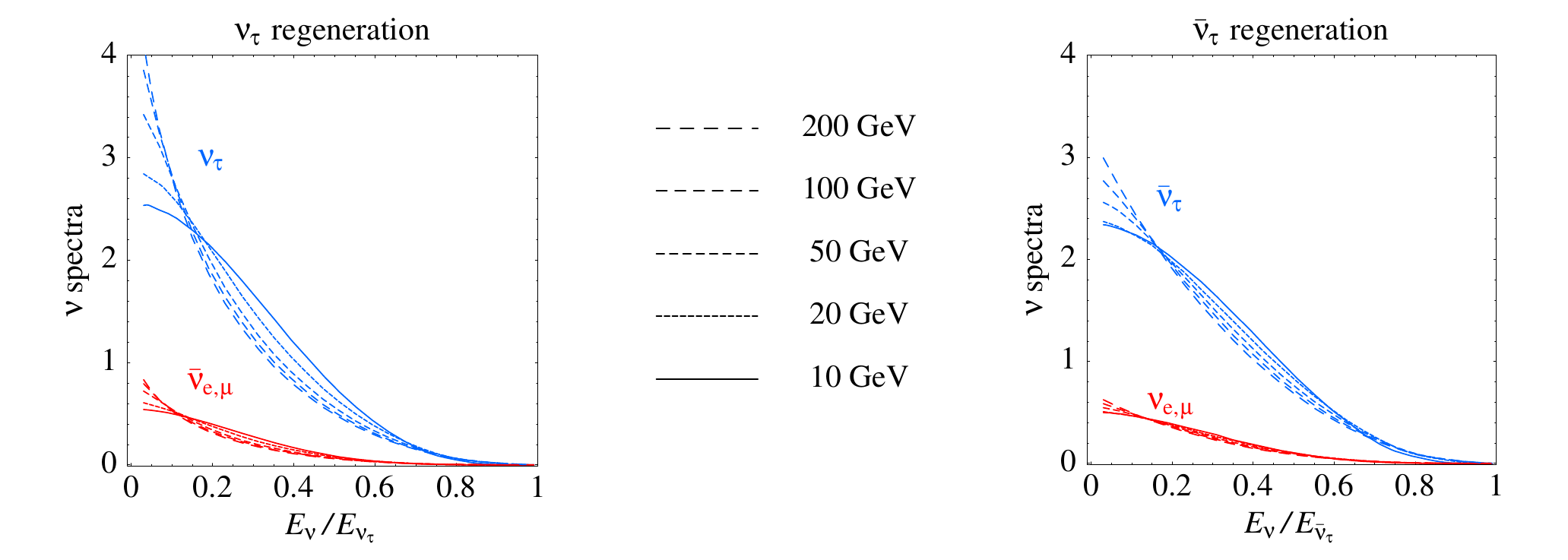}$$
\caption[$\nu_\tau$ regeneration]{\label{fig:TauReg}\em Energy distributions of neutrinos
regenerated by CC scatterings of a $\nubarnu_\tau$ with given energy $E_{\nu_\tau}$,
produced by one $\nubarnu_\tau$/nucleon scattering.
The blue upper curves are $f_{\tau\to\tau}(E_{\nu_\tau},E'_\nu)$, 
and the red lower curves are $f_{\tau\to e,\mu}(E_{\nu_\tau},E'_\nu)$,
plotted for several values of the incident $\nu_\tau$ energy $E_{\nu_\tau}$.
}
\end{figure}

CC scatterings  remove a neutrino and produce a charged lepton together with hadrons:
these particles 
can decay back to neutrinos. While $\mu^\pm$  are long-lived and hadrons have strong interactions,
such they get stopped by ambient matter before decaying,
$\tau^\pm$ can decay promptly re-injecting secondary fluxes of energetic neutrinos: this effect is known
as {\em $\nubarnu_\tau$ regeneration}~\cite{tauRegeneration}.
A $\tau^-$ decays always into $\nu_\tau$  and decays into
$\bar\nu_\mu,\bar\nu_e$ with BR $=0.175$, 
such that $\nubarnu_\tau$ regeneration couples neutrinos with anti-neutrinos.
It is relevant at energies $E_\nu\GeV$: we can assume that all particles are collinear.
The CC contribution to the evolution equation of the density matrices is therefore
\begin{eqnarray}
\left.\frac{d\rho}{dr}\right|_{\rm CC} &=& - \frac{\{\Gamma_{\rm CC},\rho\}}{2}+
\int \frac{dE^{\rm in}_\nu}{E^{\rm in}_\nu}  
 \bigg[ {\Pi}_\tau \rho_{\tau\tau}(E^{\rm in}_\nu) \Gamma_{\rm CC}^\tau(E^{\rm in}_\nu) 
 f_{\tau\to\tau}({E_\nu^{\rm in}},{E_\nu})\\
 & &\qquad\nonumber
 + {\Pi}_{e,\mu} \bar\rho_{\tau\tau} (E^{\rm in}_\nu) \bar\Gamma_{\rm CC}^{\tau} (E^{\rm in}_\nu)
 f_{\bar\tau\to e,\mu}( E^{\rm in}_\nu, E_\nu)
\bigg].  \nonumber
\end{eqnarray}
The first term describes the absorption; the anticommutator arises because loss terms 
correspond to an anti-hermitian effective Hamiltonian.
The second term describes the `$\nu_\tau$ regeneration'. 
In the formul\ae\ above, ${\Pi}_\ell$ are $3\times3$ matrices projecting on the flavor $\nu_\ell$.
The ${\Gamma}_{\rm CC}$, $\bar{{\Gamma}}_{\rm CC}$ matrices express the rates of absorption due to the CC scatterings and are given by
\beq
{\Gamma}_{\rm CC}(E_\nu) = \diag ( \Gamma_{\rm CC}^e,
\Gamma_{\rm CC}^\mu,\Gamma_{\rm CC}^\tau),\quad
\Gamma_{\rm CC}^\ell=
N_p(r)\  \sigma(\nu_\ell p\to \ell X)
+N_n(r)\  \sigma(\nu_\ell n\to \ell X),\\
\eeq
Notice that $\Gamma_{\rm CC}^\tau < \Gamma_{\rm CC}^\mu = \Gamma_{\rm CC}^e$ due to the
kinematical effect of the $\tau$ mass.
The functions $f(E_\nu, E'_\nu)$, plotted in fig.\fig{TauReg},
 are the energy distributions of secondary neutrinos
produced by a CC scattering of an initial neutrino with energy $E_{\nu_\tau}$.

\begin{figure}
$$\includegraphics[width=10cm]{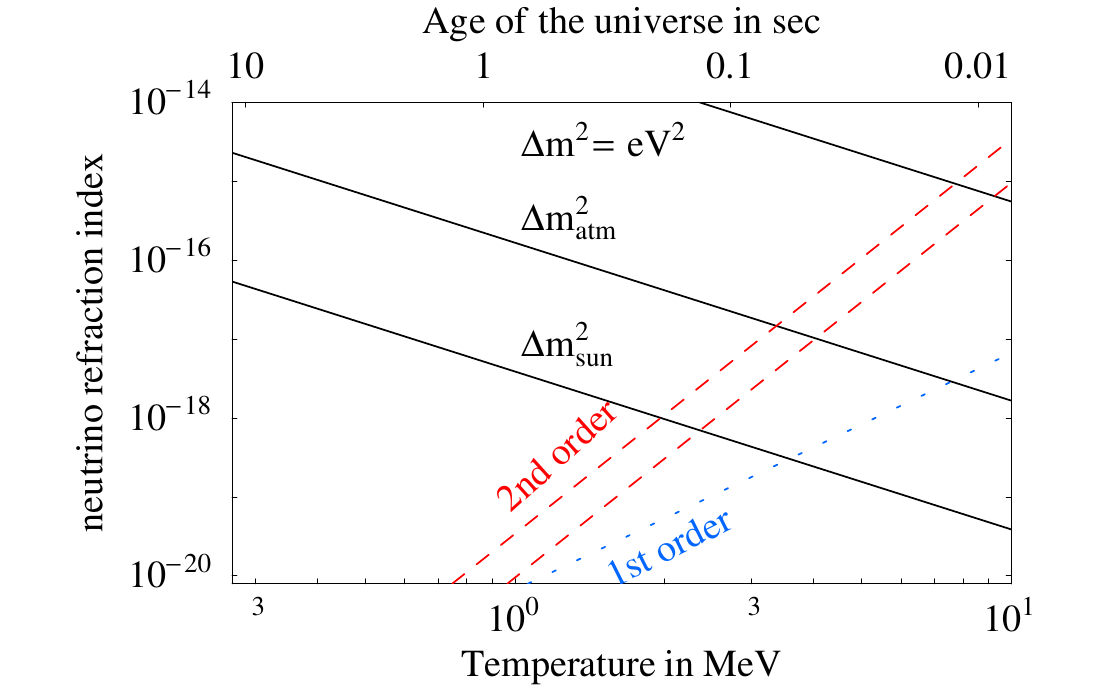}$$
\caption[Matter effects in the early universe]{\label{fig:Rifrazione}\em Contributions to the refraction index $n-1$ from
neutrino masses (continuous lines), matter effects at 2nd order in $G_{\rm F}$
(dashed lines, for $\nu_e$ and
$\nu_{\mu,\tau}$).
Matter effects at 1st order are suppressed by $n_B/n_\gamma$ and therefore negligible.}
\end{figure}

\section{Oscillations in the early universe}\label{OscUniverse}\index{Oscillation!in the universe}
When the universe was cooling down after the big-bang,
weak interactions kept neutrinos in thermal
equilibrium with other particles for about one second,
down to temperatures of few MeV (see section~\ref{cosmology}).
At these energies the observed solar and atmospheric oscillations
are faster than the expansion rate of the universe,
$H \sim T^2/M_{\rm Pl}\sim 1/\hbox{s}$
and give minor corrections to big-bang nucleosynthesis.
If extra sterile neutrinos exist, oscillations could thermalize them,
giving significant observable effects.

The precise study of neutrino oscillations in the early universe
involves new ingredients~\cite{OscUniverse}:
i) the background medium is expanding and contains a significant amount of neutrinos;
ii) neutrino oscillations can modify the background;
iii) new matter effects.

We assume that the medium is in thermal equilibrium and can be
described by macroscopic parameters (such as temperature,
densities of different particles,\ldots ). We neglect inhomogeneities.
We want to study the time evolution of the the neutrino and anti-neutrino
 matrix densities $\rho_p$ and $\bar\rho_p$, where
$p$ is the neutrino momentum.

The relevant interactions at $T\sim \MeV$ are CC neutrino/matter
scatterings,
NC neutrino/matter scatterings and scatterings among 4 neutrinos.
The amplitude of these weak interactions is proportional to the Fermi constant $G_{\rm F}$.
Up to second order in $G_{\rm F}$ the evolution equation of $\rho_p$ can be
found by computing and adding their separate contributions:
\beq\label{eq:dotrhofull2}
\dot\rho_p = -i [E^0_p,\rho_p] + (\dot \rho_p)_{\rm CC}+ (\dot \rho_p)_{\rm NC}+ (\dot \rho_p)_{\rm S}\eeq
where $E^0_p\simeq p + mm^\dagger/2p$ is the energy of a non-interacting ultrarelativistic neutrino with momentum $p$.
We do not write the explicit but lengthy collision integrals.
In practical applications the evolution equations are often approximately mimicked by
\beq\dot\rho_p = -i [E^0_p + A_p,\ \rho_p] - \{\Gamma,(\rho_p-\rho_p^{\rm eq})\}.\eeq
\begin{itemize}

\item The $\Gamma = \hbox{diag}\,(\Gamma_{\nu_e},\Gamma_{\nu_\mu},\Gamma_{\nu_\tau})$ terms,  describe neutrino production and absorption.
The anti-commutator (that approximates a more complicated structure)
describes how collisions tend to drive neutrino to their thermal equilibrium values,
$\rho_p^{\rm eq} = {\rm diag}(n_{\nu_e},n_{\nu_\mu},n_{\nu_\tau})_p$.
A detailed comparison with the full equations\eq{dotrhofull2}
reveals that they can be best approximated by inserting the following
values of the damping coefficients:
\begin{itemize}
\item In the equations for the off-diagonal components of $\rho$, insert the
{\em total scattering rate} $\Gamma_{\rm tot}\sim G_{\rm F}^2 T^5$
because all scatterings damp the coherent interference
between different flavours.

\item In the equations for the diagonal components of  $\rho$,
insert the {\em annihilation} rate 
$\Gamma_{\rm ann}\sim \Gamma_{\rm tot}/10$,
since annihilations are needed to change the number of neutrinos.
\end{itemize}
(Unlike the exact equations, 
the simple approximated equations depend on the choice of flavour basis;
when coherencies among $\nu_\mu$ and $\nu_\tau$ are important, 
one should use their mass eigenstate basis).

\item The $A_p$ terms describe the usual matter effects that arise at first order in $G_{\rm F}$
(to be computed taking into account that the background contains neutrinos)
plus second order terms, which turn out to be the dominant ones.
In fact, when the universe had temperatures $T\gg m_e$
the thermal plasma contained large densities of electrons and positrons,
but they were almost equal:
$N_e \simeq N_{\bar{e}}\sim T^3$.
Therefore the first-order matter effects almost cancel out, being proportional to $N_e - N_{\bar{e}}$.

The second order term in the low-energy expansion of the $W,Z$ propagators,
$$1/(k^2 - M^2) \approx -1/M^2 - k^2/M^4 + \cdots\hbox{ for }k\ll M,$$ 
induces a new contribution to the matter potential,
$A\sim G_{\rm F} (E^2/M^2) ( n_e +  n_{\bar{e}})\sim G_{\rm F}^2 T^5$,
with the same sign for neutrinos and anti-neutrinos.
In a plasma made of photons, electrons and neutrinos the second-order
matter potential is (at $m_e\ll T\ll m_\mu$)
\beq A_p \approx \frac{16\sqrt{2}G_{\rm F} p}{3}\bigg[\frac{\langle E_\nu\rangle}{M_Z^2}
 \diag(n_{\nu_e},n_{\nu_\mu},n_{\nu_\tau}) + 
 \frac{\langle E_e\rangle}{M_W^2}
 \diag(n_{e},0,0)\bigg] \eeq
\end{itemize}
A further approximation allows to replace the equations for $\rho_p$ with
a single equation for the total matrix density. 
Oscillations tend to distort the energy distribution of neutrinos,
while scatterings tend to thermalize it: we can
neglecting spectral distortions and assume that
neutrinos follow a Boltzmann distribution at temperature $T_\nu$,
possibly different from the $\gamma,e,\bar{e}$ temperature $T$.
Inserting  $n_e=2T^3/\pi^2$, $n_\nu = T_\nu^3/\pi^2$ and the average momentum $\langle p\rangle=3T_\nu$
and energy $\langle E_e\rangle = 3T$
one gets, at $T\gg m_e$
\beq A = -\frac{48\alpha_2 T_\nu}{\pi M_W^4}\bigg[T_\nu^4\cos^2\theta_{\rm W}
 \diag(1,1,1) + 2T^4\diag(1,0,0)\bigg] ,\eeq
 $$
  \Gamma_{\rm tot}  \approx  3.6 \; G_{\rm F}^2 \; T^5 \quad \textrm{for $\nu_e,\bar\nu_e$,}\qquad
  \Gamma_{\rm tot}  \approx  2.5 \; G_{\rm F}^2 \; T^5 \quad\textrm{for $\nu_{\mu,\tau},\bar\nu_{\mu,\tau}$},
$$
$$
  \Gamma_{\rm ann} \approx 0.5 \; G_{\rm F}^2 \; T^5 \quad \textrm{for $\nu_e,\bar\nu_e$,}\qquad
  \Gamma_{\rm ann}\approx  0.3 \; G_{\rm F}^2 \; T^5 \quad\textrm{for $\nu_{\mu,\tau},\bar\nu_{\mu,\tau}$}.
$$
We neglected possible neutrino asymmetries, which could be important
only if orders of magnitude larger than the baryon asymmetry.
Otherwise, as illustrated in fig.\fig{Rifrazione},
the normal matter effect becomes dominant only 
at lower temperatures, $T\circa{<}m_e$,
where scatterings become negligible.
If extra sterile neutrinos exist, $A$ and $\Gamma$ become bigger matrices with 
vanishing  `sterile' entries.
The tools here presented are applied to compute
 BBN bounds on extra sterile neutrinos (see section~\ref{nucleos}).
 
 We remark one qualitative feature of the solutions of the Boltzmann equations in\eq{dotrhofull2}:
 let us suppose that one has flavor oscillations together with flavor-conserving scatterings.
 One would qualitatively guess that everything gets thermalized, 
 since the oscillation frequency depends on the energy
 and since scatterings change the energies.
 On the contrary, if scatterings are very fast, the system undergoes collective flavour oscillations
at the thermally-averaged oscillation frequency: 
scatterings force a thermal distribution of energies within each flavor, 
and oscillations change the flavour at fixed energy.

%% file: review_cross.tex


\chapter{Detecting neutrinos}\label{detecting}
Neutrinos only have weak interactions: at ordinary energies they cross the earth
without being absorbed.
Neutrinos can be detected if one has a intense enough flux of neutrinos
and a big enough detector with low enough background
(often achieved by going underground
in order to suppress the cosmic-ray background).
We now discuss what `enough' means in practice.

We recall some numerical constants: $M_W =80.4\GeV$, $M_Z=91.18\GeV$,
$v=174\GeV$, and the Fermi constant
$G_{\rm F} = g_2^2/4\sqrt{2}M_W^2=1/2\sqrt{2} v^2 = 11.66/\TeV^2$.

\index{scattering!neutrino/electron}\label{neutrino/electron}
\section[$\nu$/electron scattering]{Neutrino/electron scattering}
According to the SM, the amplitude for scattering of neutrinos on electrons at rest is $\mathscr{M}\sim G_{\rm F} m_e E_\nu$.
The total cross section is $\sigma \sim |\mathscr{M}|^2/s$, where $s = (P_e + P_\nu)^2$
in terms of the quadri-momenta $P$.
Electrons in atoms have a small velocity $v \sim \alpha_{\rm e.m.}$ and can be considered at rest.
If $E_\nu \ll m_e$ one has $s\sim m_e^2$ and so $\sigma \sim G_{\rm F}^2 E_\nu^2$.
If $E_\nu \gg m_e$ one has $s\sim m_e E_\nu$ and so $\sigma \sim G_{\rm F}^2 m_e E_\nu$.
In the energy range $m_e \ll E_\nu \ll M_Z^2/m_e$, 
the SM prediction at tree level is~\cite{nue}
\begin{equation}
\sigma(\nu_\ell e\to \nu_\ell e) =      \frac{2 m_e E_\nu G_{\rm F}^2}{\pi}  (G_{L\ell}^2 + \frac{1}{3}G_{R\ell}^2),\qquad
\sigma(\bar\nu_\ell e\to \bar\nu_\ell e) =  \frac{2 m_e E_\nu G_{\rm F}^2}{\pi}  (G_{R\ell}^2 + \frac{1}{3}G_{L\ell}^2).
\end{equation}
Only $Z$-exchange contributes to $\nu_{\mu,\tau}$ and $\bar\nu_{\mu,\tau}$ scattering on electrons (see fig.\fig{materia}b).
Therefore when $\ell =\{\mu,\tau\}$ the effective $G_{L,R \ell}$ are equal to the
$\bar{\ell}Z\ell$ couplings, named $g_{L,R \ell}$ and listed in table~\ref{tab:gAi}
in terms of the weak mixing angle $s_{\rm W}^2 \approx 0.223$.
On the contrary $W$ boson exchange contributes to 
$\nu_e e \to \nu_e e$ and $\bar{\nu}_e e \to \bar\nu_e e$ 
scatterings (see fig.\fig{materia}a): therefore 
$G_{Le} = +\frac{1}{2}+\sW^2 \neq g_{Le}$ and $G_{Re} = g_{Re}$,
giving rise to a larger cross section.\footnote{The $\bar{\nu}_e e \to \bar\nu_e e$ process 
is resonantly enhanced at
$s = M_W^2$ i.e., for electrons at rest, at $E_\nu = M_W^2/2m_e = 6.3~10^{6}\GeV$.
$W$ scattering also gives $\nu_\ell e \to \nu_e \ell$ with $\ell=\{\mu,\tau\}$ at 
$E_\nu \circa{>} m_\ell$, while 
$\bar\nu_\ell e \to \bar\nu_e \ell$ violates lepton flavour and 
does not occur.
}
Putting numbers 
\begin{equation}\sigma(\nu_e e) = 0.93 \sigma_0,\qquad
\sigma(\nu_{\mu,\tau} e) = 0.16 \sigma_0,\qquad
\sigma(\bar\nu_e e) = 0.39 \sigma_0,\qquad
\sigma(\bar\nu_{\mu,\tau} e) =0.13 \sigma_0
\end{equation}
where $\sigma_0 =10^{-44}\hbox{cm}^2 \, E_\nu/\hbox{MeV}$ and $E_\nu \gg m_e$.

\smallskip

SK detects solar neutrinos through $\nu e$ scattering.
With $10^{10}$ moles of electrons
($20.000$ ton of water), a flux of $\hbox{few}\times 10^6\,\nu_e/$cm$^2\,$s
with $E_\nu\sim 10\MeV$
(the solar Boron neutrinos), and a $50\%$ efficiency 
SK can detect about $10000\,\nu_e/\hbox{yr}$
(finding that about half of them oscillate away).
SK and SNO can measure $T_e = E_e - m_e$, the kinetic energy of the recoiling electron.
Its kinematically allowed range is $0 \le T_e \le E_\nu/(1+m_e/2 E_\nu)$.
SK and SNO can only detect electrons with $T_e \circa{>} 5\MeV$.
The SM at tree level predicts the energy spectrum of recoil electrons as
\begin{equation}
\frac{d \sigma}{d T_e}(\nu_\ell e\to \nu_\ell e) = \frac{2 G_{\rm F}^2 m_e}{\pi} 
\bigg[G_{L\ell}^2 + G_{R\ell}^2 (1-y)^2 - G_{L\ell} G_{R\ell} \frac{m_e}{E_\nu} y\bigg]\qquad\hbox{where}\qquad
y\equiv \frac{T_e}{E_\nu}.
\end{equation}
The measurement of $T_e$ alone does not allow to reconstruct $E_\nu$, nor allows
to discriminate $\nu_e$ from $\nu_{\mu,\tau}$.
In principle, $E_\nu$ can be reconstructed by measuring $T_e$ and the opening angle $\vartheta_{\nu e}$
between the incident neutrino and the scattered lepton.
However, this angle is small,  $\vartheta_{\nu e}\sim (m_e/E_\nu)^{1/2}$.
When the position of the neutrino source is known (e.g.\ the sun) measuring $\vartheta_{\nu e}$ helps in discriminating
the signal from the background; when it not known (e.g.\ a supernova) measuring the direction of the
scattered $e$ helps in locating the source.

\renewcommand{\arraystretch}{1.2}
\begin{table}
$$\begin{array}{|lccc|}\hline
\hbox{SM fermion}&{\rm U}(1)_Y&\SU(2)_L&\SU(3)_{\rm c}\cr \hline
U = u_R \phantom{*Ã{3\over 5}} & -{2 \over 3} & 1 & \bar{3} \cr
D = d_R \phantom{*Ã{3\over 5}} & \phantom{-}{1 \over 3}& 1 &\bar{3} \cr
E = e_R \phantom{*Ã{3\over 5}} &\phantom{-}1 & 1 &1  \cr
L=(\nu_L, e_L) & -{1 \over 2} & 2 &1\cr
Q=(u_L, d_L) &\phantom{-} {1\over 6}  & 2 & 3\cr \hline
\end{array}\qquad
\renewcommand{\arraystretch}{1.45}
\begin{array}{|c|cc|}\hline
Z\hbox{ couplings}
& g_L & g_R \\ \hline
\phantom{I}^{\phantom{I}^{\phantom{I}}}
\nu_e,\nu_\mu,\nu_\tau \phantom{I}^{\phantom{I}^{\phantom{I}}}
& \frac{1}{2} & 0 \\
\phantom{I}^{\phantom{I}^{\phantom{I}}}
e,\mu,\tau \phantom{I}^{\phantom{I}^{\phantom{I}}}
&-\frac{1}{2}+\sW^2 & \sW^2 \\
\phantom{I}^{\phantom{I}^{\phantom{I}}}
u,c,t \phantom{I}^{\phantom{I}^{\phantom{I}}}
& \phantom{-}\frac{1}{2}-\frac{2}{3}\sW^2 & -\frac{2}{3}\sW^2  \\
\phantom{I}^{\phantom{I}^{\phantom{I}}}
d,s,b \phantom{I}^{\phantom{I}^{\phantom{I}}}
& -\frac{1}{2}+\frac{1}{3}\sW^2 & \frac{1}{3}\sW^2 \\
\hline
\end{array}$$
\caption[$Z$ couplings]{\em The SM fermions and their $Z$ couplings.\label{tab:gAi}}
\end{table}
\renewcommand{\arraystretch}{1}

\section[$\nu$/nucleon scattering]{Neutrino/nucleon scattering}\label{sigmahad}

\subsection{Neutrino/nucleon scattering at $E_\nu \ll \GeV$}
Similarly, the SM amplitude for scattering of neutrinos on nucleons (i.e.\ protons or neutrons) at rest
is $\mathscr{M}\sim G_{\rm F} m_p E_\nu$.
Therefore the total cross section is $\sigma  \sim G_{\rm F}^2  E_\nu^2$ for $E_\nu \ll m_p$
and $\sigma\sim G_{\rm F}^2  m_p E_\nu$ for  $E_\nu \gg m_p$.
In this case neutrino scattering breaks the nucleon (giving pions and nucleons in the final state)
 and the cross section is obtained
by summing the contributions of the individual neutrino/quark sub-processes.
Since $m_p\gg m_e$ neutrino/nucleon has a larger cross-section than neutrino/electron scattering.

\smallskip

At $E_\nu \ll m_p$ (e.g.\ solar and reactor neutrinos),
if one is interested only in CC processes (so that the neutrino is converted into a charged lepton,
that can be detected)
only the reactions $\bar{\nu}_e p \to e^+ n$ and ${\nu}_e n \to e p$ are possible
($\nu_e p\to e^\pm n$ violates either charge or lepton-number), and only the first one
is of experimental interest, because 
it is not possible to build a target containing enough free neutrons,
that would anyway decay.
Enough free protons are obtained using targets made of 
water (${\rm H}_2 {\rm O}$), hydrocarbonic scintillators, etc.

The precise SM prediction is~\cite{IBD}
\begin{equation}
\label{eq:sigmanup}
\sigma(\bar{\nu}_e p \to \bar{e} n) = \frac{G_{\rm F}^2 \cos^2 \theta_{\rm C}}{\pi}(1+3 a^2) E_{\bar{e}} p_{\bar{e}} 
=\frac{2\pi^2E_{\bar{e}} p_{\bar{e}} }{f m_e^5\tau_n}
\approx
0.952 ~10^{-43}\,{\rm cm}^2 \frac{E_e p_e}{\MeV^2}
\end{equation}
where $a = 1.26$ is the axial coupling of nucleon and $f=1.715$ is a phase space factor.
The second expression (obtained by relating $\sigma$ to the neutron lifetime $\tau_n$) is more accurate.
This reaction has a relatively large cross section and allows to reconstruct the neutrino energy.
When $E_\nu \ll m_p$ conservation of energy approximately means $E_{\nu} = E_{\bar{e}} + m_n - m_p = E_e + 1.293\MeV$.
Therefore {\em the neutrino energy can be deduced} by measuring $E_{\bar{e}}$ alone.
Since $E_{\bar{e}} \ge m_e$ this reaction is only possible if $E_\nu \ge m_e + m_n - m_p = 1.804\MeV$
(taking into account the recoil of the neutron, the energy threshold becomes $1.806\MeV$).
An analogous expression holds for $\bar{\nu}_\mu p \to \mu^+ n$ scattering, that 
needs $E_\nu > m_\mu$ --- a too high threshold for reactor and supernova antineutrinos.

\medskip

\begin{figure}
$$\includegraphics[width=0.7\textwidth]{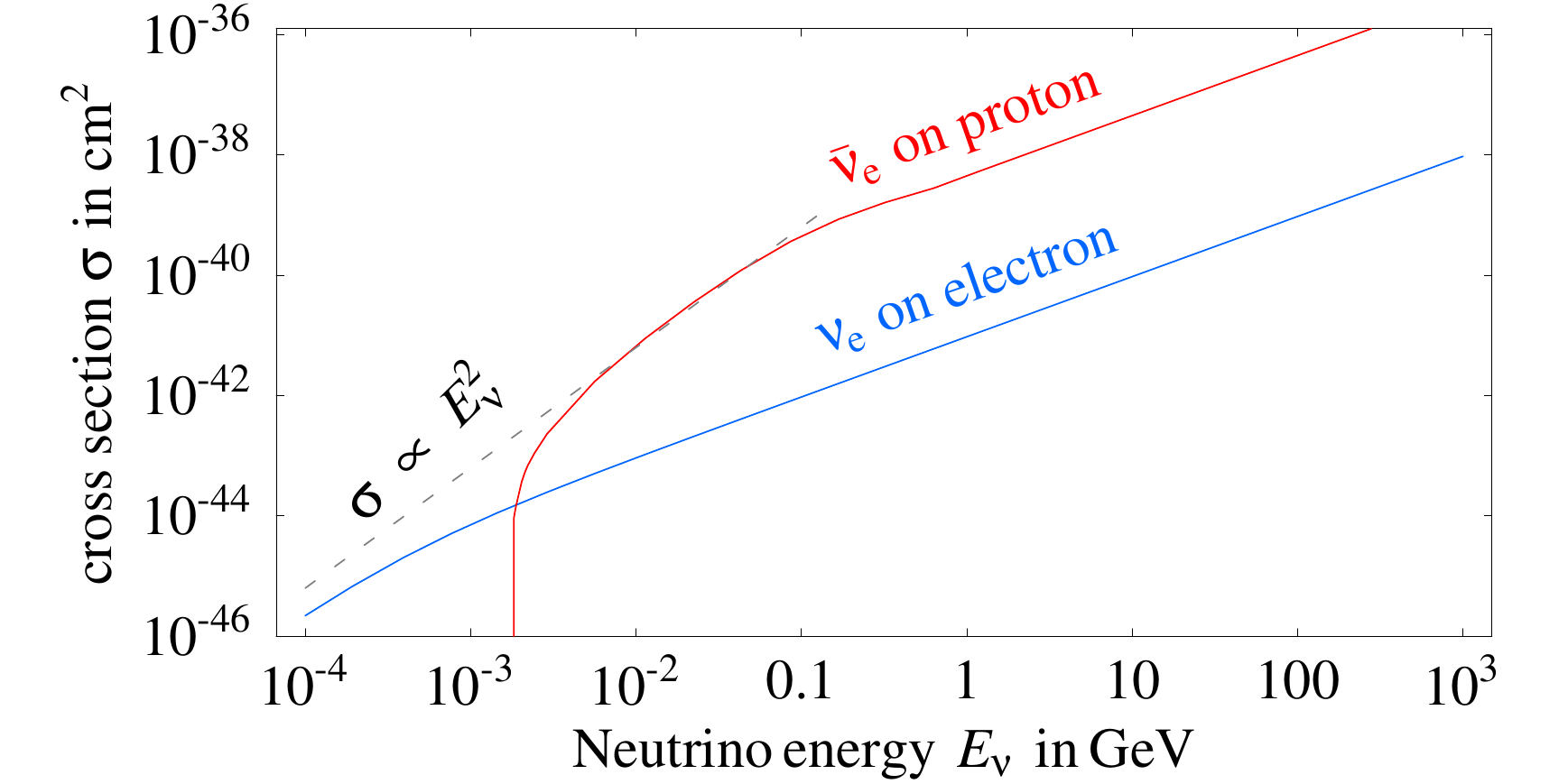}$$
\caption[Neutrino cross sections on $p$ and $e$]{\em Characteristic neutrino cross sections
on target particles at rest with mass $m$:
$\sigma\propto E_\nu^2$ at $E_\nu \circa{<} m$ and
$\sigma\propto E_\nu$ at $E\circa{>} m$.\label{fig:sigma}}
\end{figure}

\subsection{Neutrino/nucleon scattering at $E_\nu \sim \GeV$}
Neutrino/nucleon scatterings at energies comparable
to the nucleon mass depends on the imprecisely known nucleon structure:
as a consequence expressions for
the cross sections are complicated, not very accurate and not reported here.
One sums the  elastic scatterings with a few dominant non-elastic channels,
like one-pion production, computed according to models of form factors.
Fig.\fig{sigma} shows that the low- and the high-energy limits match reasonably well.

This energy range is relevant for experimental studies of neutrino oscillations,
because recoiled particles are emitted at large angles:
as described in section~\ref{K2K} this allows to reconstruct the neutrino energy if 
the direction of the incoming neutrino is known,
e.g.\ in neutrino beam experiments.
These experiments deal with unprecisely known cross sections
by building also a detector near to the neutrino source,
as similar as possible to the far detector used for oscillation studies.

\index{scattering!neutrino/nucleon}
\subsection{Neutrino/nucleon scattering at $E_\nu \gg \GeV$}\label{nuNucleon>>GeV}
At $E_\nu \gg \GeV$ (e.g.\ atmospheric and accelerator neutrinos of higher energy)
the dominant effect is neutrino/quark scattering.
The fact that quarks are bound into a nucleon has no effect
on the total (`inclusive') cross-section:
the quark $q^*$ that collides with neutrinos 
gets an energy much larger than the binding energy
and unavoidably finds some way of escaping from the nucleon.
Typically $q^*$ hadronizes picking a $q\bar{q}$ pair from the vacuum, breaking
the nucleon as $N = qqq^* \to [qqq][\bar{q} q^*] = N\pi$,
giving rise to processes like $\nu p\to \ell^+ n\pi^0$ o $\nu p\to \ell^+ p\pi^-$.

One has to sum over all quark sub-processes,
taking into account the distribution of quarks in the nucleon
(structure functions).
Neutrino/quark scattering is similar to electron/quark, and gives a cross section $\propto G_{\rm F} E_\nu m_N$.
The  effective $\nu_\ell$-quark Lagrangian predicted by the SM at tree level is
\begin{equation}
\label{eq:nuq}
\Lag_{\rm eff} = -
2\sqrt{2}G_F ([\bar{\nu}_\ell \gamma_\alpha \ell_L ][\bar{d}_L \gamma^\alpha u_L ] +\hbox{h.c.}) -
2\sqrt{2}G_F \sum_{A,q} g_{Aq}
[\bar{\nu}_\ell \gamma_\alpha \nu_\ell ][\bar{q}_A \gamma^\alpha q_A]
\end{equation}
where 
$A=\{L,R\}$, $\ell=\{e,\mu,\tau\}$, $q=\{u,d,s,\ldots\}$ and the $Z$ couplings
$g_{Aq}$ are given in table~\ref{tab:gAi} in terms of the 
weak mixing angle $\sW\equiv\sin \theta_{\rm W}$.
Neutrino and CKM mixings appear in the first CC term, 
if one needs to rewrite it in terms of mass eigenstates.

The cross sections for the CC quark sub-processes are
\begin{eqnsystem}{sys:nuqCC}
\frac{d\hat{\sigma}}{dy}(\nu_\ell d\to \ell u) &=&
\frac{d\hat{\sigma}}{dy}(\bar\nu_\ell \bar d\to \bar \ell \bar u) =
\frac{2G_{\rm F}^2(p_\nu\cdot p_d)(p_\ell\cdot p_u)}{\pi (p_\nu\cdot p_d)} R_W\simeq
\frac{G_{\rm F}^2\hat{s}}{\pi} R_W,\\
\frac{d\hat{\sigma}}{dy}(\nu_\ell \bar{u}\to \ell \bar{d}) &=&
\frac{d\hat{\sigma}}{dy}(\bar\nu_\ell u\to \bar\ell d)  =
\frac{2G_{\rm F}^2(p_\nu\cdot p_d)(p_\ell\cdot p_u)}{\pi (p_\nu\cdot p_u)} R_W\simeq
\frac{G_{\rm F}^2\hat{s}}{\pi} (1-y)^2 R_W
\end{eqnsystem}
In the last terms scalar products among quadri-momenta have been
evaluated assuming $E_\nu \gg m_N$ and thereby neglecting quark and lepton masses.
In this limit the factor $y\equiv -\hat{t}/\hat{s}$ has kinematical range $0\le y \le 1$.
In the system where the nucleon is at rest
$y=1-E'/E_\nu$ where $E'$ is the energy of the scattered lepton.\footnote{If $E_\nu$ is not
much greater than $m_N$ the above kinematical formul\ae{} get generalized as follows.
We define the ``parton mass'' as $m\equiv m_N x$ and assume that the final state lepton
is not a $\tau^\pm$, so that we neglect its mass.
The factor $y \equiv -\hat{t}/\hat{s}=(1-E'/E_\nu)/(1+m/2E_\nu)$ ranges between
$0\le y \le (1+m/2E_\nu)^{-2}$. 
The scattering angle $\vartheta$ of the final state lepton
in the rest frame of the nucleon
is given by $1-\cos\vartheta =m(E_\nu-E')/E_\nu E'$
and ranges between $-1\le \cos\vartheta\le 1$.
Its typical value is $\vartheta\sim (m/E_\nu)^{1/2}$.

When the scattered lepton is a $\tau^\pm$ it is also necessary to take into account
its polarization in order to compute its decay products~\cite{sigmatau}.
}

For NC scatterings at $E_\nu\gg m_N$ one has
\begin{eqnsystem}{sys:nuqNC}
\frac{d\hat{\sigma}}{dy}(\nu q\to \nu  q') =\frac{d\hat{\sigma}}{dy}(\bar\nu \bar q\to\bar \nu  \bar q')&=& \frac{G_F^2\hat{s}}{\pi} [g_{Lq}^2+g_{Rq}^2(1-y)^2]R_Z,\\
\frac{d\hat{\sigma}}{dy}(\bar\nu q\to \bar \nu  q') =\frac{d\hat{\sigma}}{dy}(\nu \bar q\to  \nu  \bar q') &=& \frac{G_F^2\hat{s}}{\pi} [g_{Rq}^2+g_{Lq}^2(1-y)^2]R_Z.
\end{eqnsystem}
We neglected quark and lepton masses.
The factors
$R_Z\equiv (1+Q^2/M_Z^2)^{-2}$, $R_W= (1+Q^2/M_W^2)^{-2}$
equal to one when the transferred quadri-momentum $Q^2=-\hat{t}$ 
is much smaller 
than the $W,Z$ masses squared.


In the range $m_N \ll E_\nu \ll M_{W,Z}^2/m_N$ i.e. $\GeV\ll E_\nu\ll 10\TeV$
the total quark CC cross sections are
\beq\hat{\sigma}(\nu_\ell d\to \ell u) =
\hat{\sigma}(\bar\nu_\ell \bar d\to \bar \ell \bar u)=
3\hat{\sigma}(\bar\nu_\ell u\to \bar\ell d) =
3 \hat{\sigma}(\nu_\ell \bar{u}\to \ell \bar{d}) = 
\frac{G_{\rm F}^2\hat{s}}{\pi}. \eeq
$\sqrt{\hat{s}}$ is the center-of-mass energy of the quark sub-processes.
It is given by
$\hat{s}=sx$ , where
$x$ is the fraction of the total nucleon momentum $P$ carried by a quark, $\hat{p} = x P$.

\begin{figure}[t] \begin{center}
\includegraphics{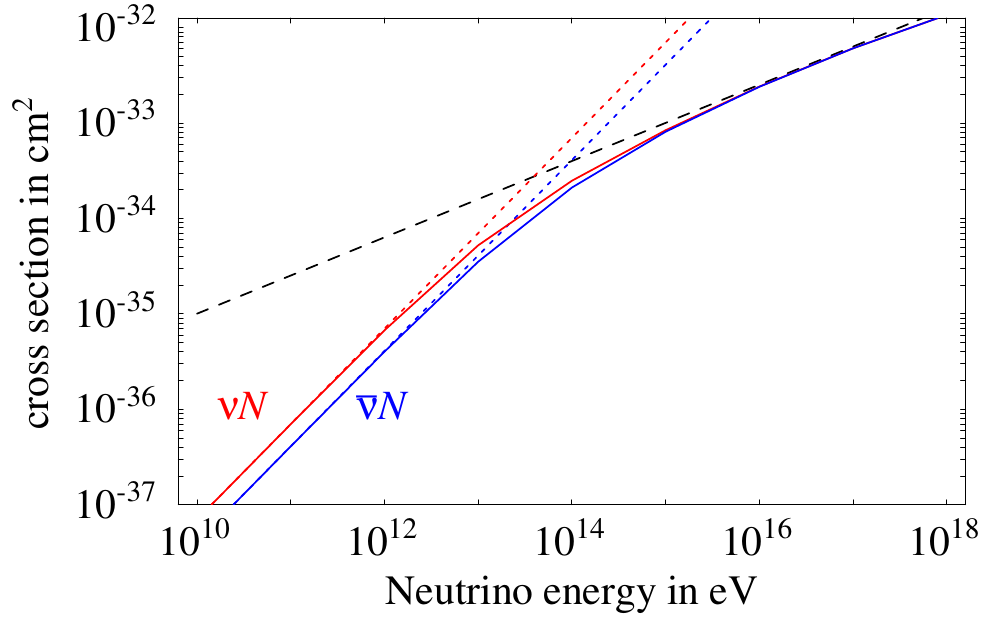}
 \caption[Neutrino/nucleon cross section]{\em Standard Model prediction for
 the total CC cross section of neutrinos (red) and anti-neutrinos (blue) of any flavour 
 on an average nucleon $N$.
 The dotted and dashed lines show the limiting values at $s\ll M_W^2$
 and $s\gg M_W^2$.
 \label{fig:sigmanuN}}
\end{center} \end{figure}

The neutrino/nucleon cross section are found integrating
over the momentum distribution of the quarks, $q(x)$.
The quantities  $p_q=\int_0^1 dx\, x\, q(x)$ are the fraction of the
 the total nucleon momentum carried by the quark $q$.
Their values, renormalized at $Q^2 \sim 10^3\GeV^2$ are
$$
\begin{array}{ll}
p_{\rm u}\hbox{(in the proton)}=
p_{\rm d}\hbox{(in the neutron)}\approx 25\% \quad &
p_{\bar{\rm u}}\hbox{(in the proton)}=
p_{\bar{\rm d}}\hbox{(in the neutron)}\approx 4\%,\\
p_{\rm d}\hbox{(in the proton)}=
p_{\rm u}\hbox{(in the neutron)} \approx 15\%,&
p_{\bar{\rm d}}\hbox{(in the proton)}=
p_{\bar{\rm u}}\hbox{(in the neutron)}\approx 6\%\end{array}
$$
where u (d) indicate that we summed over all types of up-type (down-type) quarks.
We recall that nucleons contain valence quarks together with virtual $q\bar{q}$ pairs and gluons.
Gluons carry about $1/2$ of the total proton momentum.
Anti-quarks carry a fraction $\epsilon\approx 1/5$ of
the momentum carried by quarks; this  fraction increases at higher energy.
The momentum distribution of the neutron is approximatively equal to the one of the proton,
with up and down-type quarks exchanged.
Consequently the total neutrino/nucleon CC cross sections at $s\simeq 2 E_\nu m_N \gg\GeV^2$ are
\beq\begin{array}{ll}
\displaystyle\sigma(\nu_\ell p\to \ell X) \approx \frac{G_{\rm F}^2 s}{\pi}(0.15+\frac{1}{3}0.04),\qquad &
\displaystyle\sigma(\nu_\ell n\to \ell X) \approx \frac{G_{\rm F}^2 s}{\pi}(0.25+\frac{1}{3}0.06)\\[2mm]
\displaystyle\sigma(\bar\nu_\ell n\to \bar\ell X) \approx \frac{G_{\rm F}^2 s}{\pi}(\frac{1}{3}0.15+0.04),&
\displaystyle\sigma(\bar\nu_\ell p\to \bar\ell X) \approx \frac{G_{\rm F}^2 s}{\pi}(\frac{1}{3}0.25+ 0.06).
\end{array}\eeq
Numerically $G_{\rm F}^2 m_N^2/\pi = 1.48~10^{-38}\cm^2$.
These cross sections are plotted in fig.~\fig{sigmanuN} (dotted lines).

\subsection{Neutrino/nucleon scattering at ultra-high energies}\label{nuNUHE}
At energies $E_\nu \circa{>} M_{W,Z}^2/m_N$ the transferred energy becomes
comparable with $M_{W,Z}$ and one must take into account deviations of $R_{W,Z}$ from 1.
The results simplifies again for $E_\nu\gg M_{W,Z}^2/m_N$ (at these high energy the
SM predictions have not been tested: new physics could change the result)
 the total CC and NC partonic cross-sections approach constant values~\cite{IBD}
\beq \hat\sigma_{\rm CC} \simeq \frac{G_{\rm F}^2 M_W^2}{\pi}=\frac{g_2^4}{32\pi M_W^2}
\approx 1.07~10^{-34}\cm^2,\qquad
\hat\sigma_{\rm NC}\simeq \frac{G_{\rm F}^2 M_Z^2}{\pi}[g_{Lq}^2 + g_{Rq}^2].\eeq
These cross sections are not suppressed by $E_\nu$ because they are soft i.e.\
dominated by small $Q\sim M_{W,Z}$.
Indeed, the cross section is $\hat{\sigma}\sim 1/M_{W,Z}^2$ because 
$t$-channel exchange of $W,Z$ bosons gives a Coulomb-like force, exponentially damped
at distances larger than $M^{-1}_{W,Z}$.

To compute the neutrino/nucleon cross section $\sigma$ one has to integrate the partonic cross sections
$\hat\sigma$ times the partonic distribution functions;
in practice $\sigma$ is given by $\hat\sigma$ times the number of 
quarks and anti-quarks that carry a fraction $x$ of the nucleon momentum greater than
$x\circa{>} M_{W,Z}^2/m_N E_\nu$.
So far experiments only probed $x\circa{>}10^{-4}$; HERA data and BFKL theoretical
techniques~\cite{HERABFKL} suggest that
parton distribution functions diverge at $x\to 0$ as $x q(x)\propto x^{-\beta}$ with $\beta\approx 0.5$.
This results into a growth $\sigma \propto E_\nu^\beta$, which seems to violate perturbative unitarity
at $E_\nu \circa{>} 10^{10}\GeV$.
Ignoring this possible problem, the SM prediction is~\cite{IBD}
\begin{eqnsystem}{sys:sigmaUHE}
 \sigma_{\rm CC}(\nu N)\simeq  \sigma_{\rm CC}(\bar\nu N)&\approx& 4\cdot   10^{-32}\cm^2\,
 \bigg(\frac{E_\nu}{10^{10}\GeV}\bigg)^{0.40},\\
 \sigma_{\rm NC}(\nu N)\simeq  \sigma_{\rm NC}(\bar\nu N)&\approx &2 \cdot 10^{-32}\cm^2\,
 \bigg(\frac{E_\nu}{10^{10}\GeV}\bigg)^{0.40}.
 \end{eqnsystem}
 up to a  theoretical uncertainty of about a factor 2.
 These cross sections are plotted in fig.~\fig{sigmanuN} (dashed lines)
 and correspond to an interaction length of about $1000\,{\rm kmwe}$. 
 
 \bigskip

 \index{sphalerons}
Furthermore, 
non-perturbative sphaleronic effects are expected to produce $B+L$-violating scatterings 
 among couples of SM fermions (including neutrinos) with
 cross section suppressed by an exponential tunneling factor~\cite{sigmasphaleron}:
\begin{eqnarray}
\hat\sigma  &=& \frac{1}{M_W^2} \bigg(\frac{2\pi}{\alpha_2}\bigg)^{7/2}\exp\bigg[-\frac{4\pi}{\alpha_2} F(\epsilon)\bigg],\qquad
\epsilon = \frac{\sqrt{\hat {s}}}{4\pi M_W/\alpha_2} \\
&\approx& 5.3\,{\rm barn}\cdot  \exp\bigg[-371. F(\epsilon)\bigg],\qquad\hspace{3ex}
\epsilon \approx \frac{\sqrt{\hat {s}}}{30\TeV}\nonumber\end{eqnarray}
The ``holy-grail function'' $F$ is known for small $\epsilon$, 
$F \simeq 1 - (3\epsilon)^{4/3}/2+\cdots$, and it is not known if at $\epsilon\sim 1$.
If $F$ becomes small enough to avoid the exponential suppression, 
$F\circa{<}0.03$, sphalerons would give a detectable cross-section. 

%
%
%

\section[$\nu$/nucleus scattering]{Neutrino/nucleus scattering}
\index{scattering!neutrino/nucleus}

\subsection{Neutrino/nucleus scattering at $E_\nu \gg \GeV$}
At neutrino energies larger than the nuclear binding energy
neutrino/nucleus scattering reduces to a sum of neutrino/nucleon scatterings.
We give cross-sections for scattering of neutrinos on a typical nucleus ${\cal N}$
at rest which contains $Z$ protons and approximatively $Z$ neutrons.
The CC cross sections are
\beq \sigma(\nu_\ell {\cal N}\to \ell X)\approx \frac{2ZE_\nu}{{\rm GeV}} \times 0.6~10^{-38}~{\rm cm}^2.
\eeq
and
\beq \label{eq:sigmaratio}
r\equiv \frac{\sigma(\bar{\nu}_\ell {\cal N}\to \bar\ell X)}{ \sigma(\nu_\ell {\cal N}\to \ell X)}
=\frac{\epsilon + 1/3}{1+\epsilon/3}\approx 0.5.\eeq
The average angle $\delta\theta$ between the direction of the lepton $\ell$ with respect to the direction of the neutrino is $\delta \theta \approx 0.3\sqrt{m_N/E_\nu}$.
At $E_\nu\circa{>}M_{W,Z}^2/m_p\sim 10\TeV$ the factors $R_{W,Z}$ start to deviate
from unity and must be taken into account.

The NC cross sections are
\begin{eqnarray}
R_\nu&\equiv & \frac{\sigma(\nu_\ell{\cal N}\to \nu_\ell X)}{\sigma(\nu_\ell {\cal N}\to \ell X)} = 
\frac{(g_L^2 + g_R^2/3) + \epsilon (g_R^2 + g_L^2/3)}{1+\epsilon/3} = g_L^2 + r g_R^2\\
R_{\bar{\nu}}&\equiv & \frac{\sigma(\bar{\nu}_\ell{\cal N}\to\bar\nu_\ell X)}{\sigma(\bar{\nu}_\ell {\cal N}\to \bar\ell X)} =
 \frac{(g_R^2 + g_L^2/3) + \epsilon (g_L^2 + g_R^2/3)}{1/3+\epsilon} = g_L^2 + \frac{1}{r} g_R^2
\end{eqnarray}
where
$g_L^2 \equiv g_{Lu}^2 + g_{Ld}^2 = \frac{1}{2}-s_W^2+\frac{5}{9}s_W^4\approx 0.30$,
$g_R^2\equiv  g_{Ru}^2 + g_{Rd}^2 = \frac{5}{9}s_W^4\approx 0.03$.

\begin{figure}[t] \begin{center}
\includegraphics[width=7cm,height=8.2cm,angle=270]{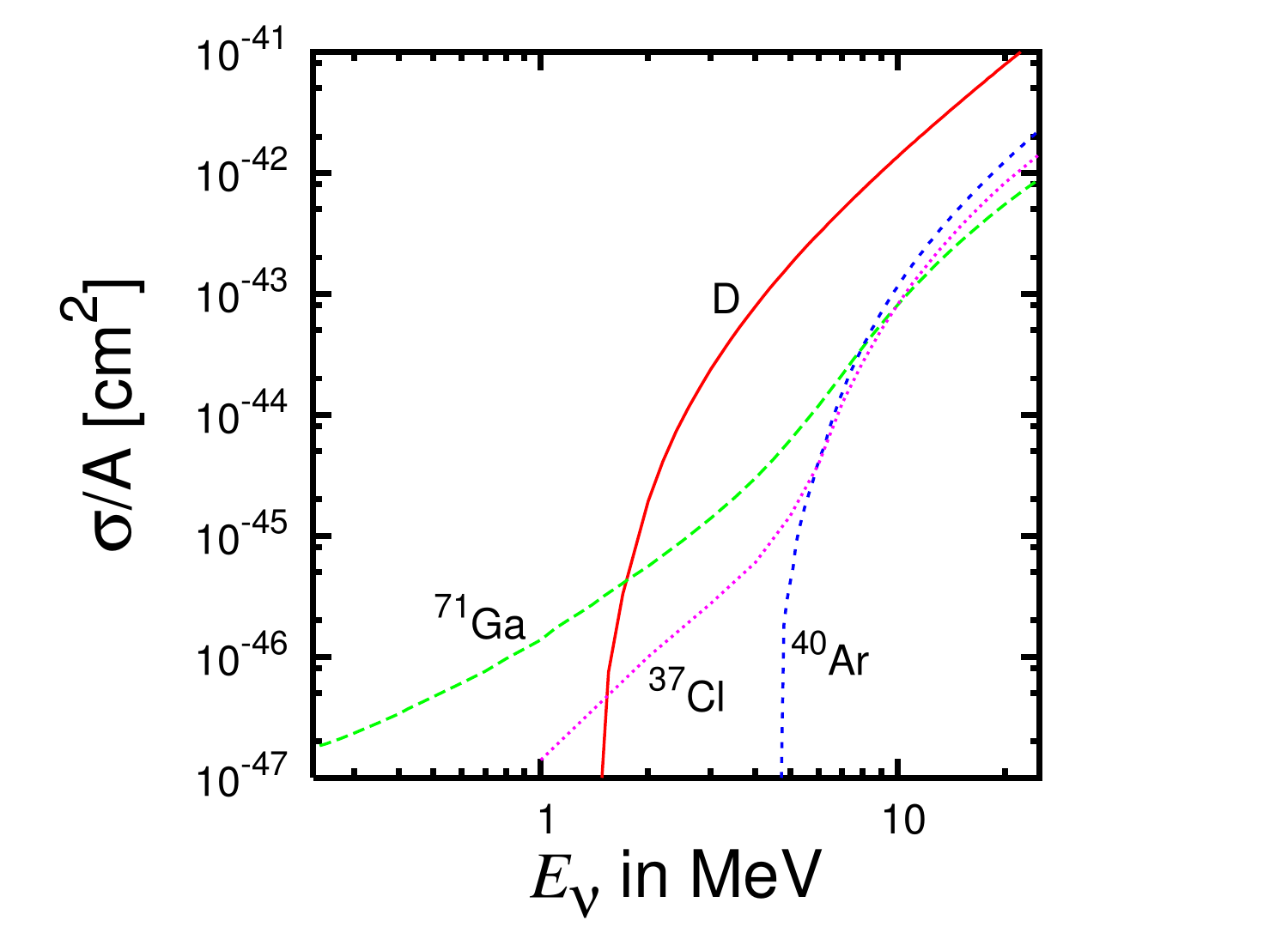}\hskip-13mm\qquad
\includegraphics[width=7cm,height=8.2cm,angle=270]{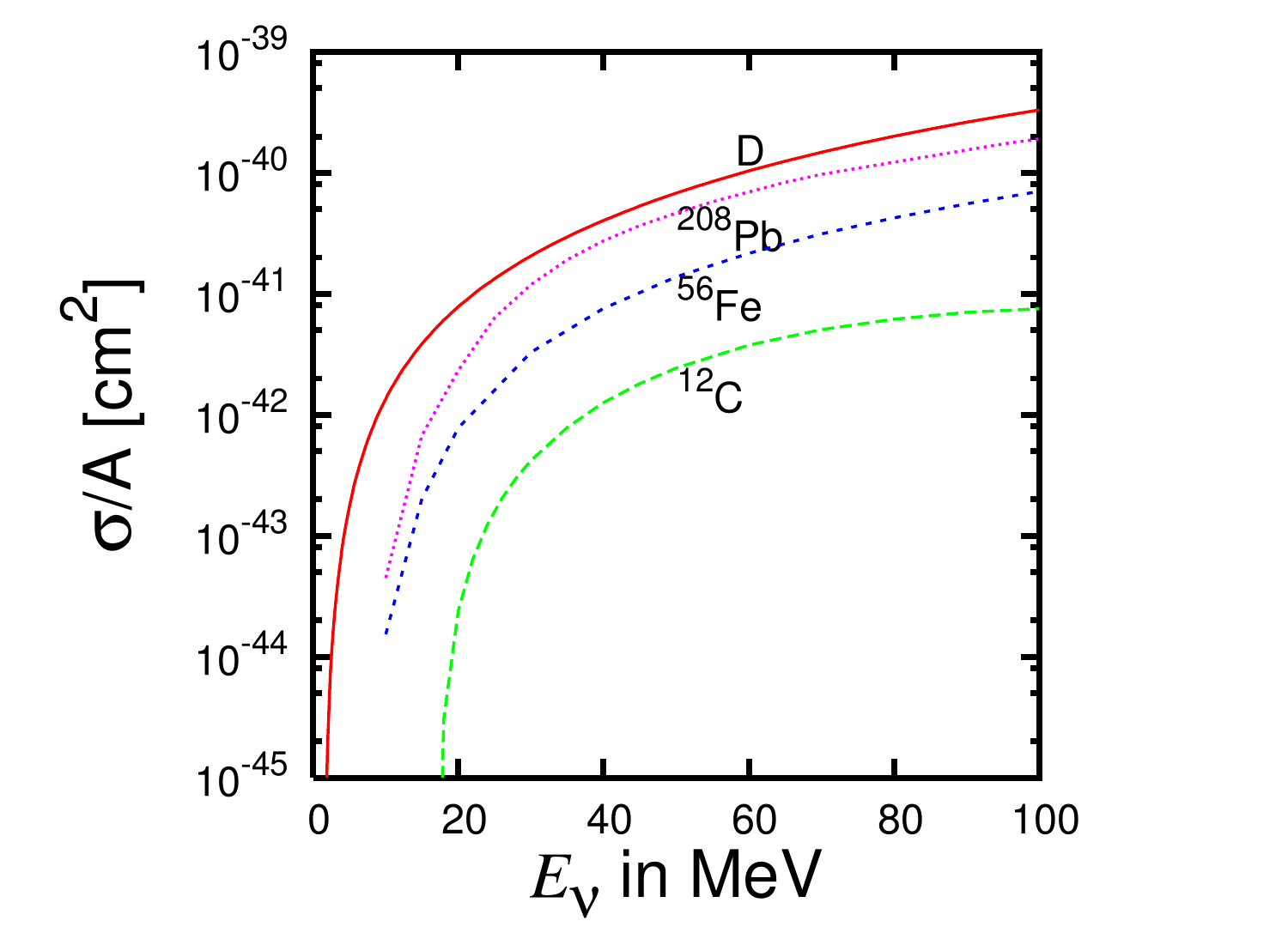}
 \caption[Neutrino/nucleus cross section]{\em Some cross sections of $\nu_e$
with nuclei, normalized to their atomic number $A$.\label{xsecnuce}}
\end{center} \end{figure}

\subsection{Neutrino/nucleus scattering at low energy}
The relevant reactions  of a low energy neutrino with a nucleus $(A,Z)$ are~\cite{bookxsec}
\begin{equation} 
\bar{\nu}_e (A,Z) \to e^+ (A,Z\!-\!1),\qquad \ {\nu}_e
(A,Z) \to e^- (A,Z\!+\!1),\qquad \ {\nu} (A,Z)\to \nu (A,Z) \ .
\end{equation} 
These processes are understood in the following way.
Neutrino interactions excite various energy levels $E_i$ with certain
values of spin, isospin and parity, probing also continuum levels when
above $E_\nu\circa{>} 50\MeV$.  The energy thresholds can be easily
computed from kinematical considerations, knowing the masses of the nuclei
involved. While at high energies $E_\nu \gg m_p$ one has the
forward-peaked incoherent scattering described in the previous section
(cross section $\sigma \sim G_{\rm F}^2 A m_p E_\nu$), at low energy the
cross sections are only slightly directional and increase as $p_e E_e$ for
any level. Beside the vectorial interactions, which are related by CVC to
the electromagnetic currents and are accurately understood theoretically
(Fermi transitions) there are the axial interactions that are less easily
modeled (Gamow-Teller transitions) and have much larger theoretical
uncertainties of the order of 50\% at least. Empirical information,
isospin and crossing invariance can be occasionally used in order to
constrain the transitions better, e.g., connecting the CC interaction
$\nu_e {\cal N}\to {\cal N}' e$ with the $\beta^+$ decay ${\cal N}' \to
{\cal N} \nu_e e^+$ (traditionally the latter information is presented as
a $\log(ft)$ value, where $f$ is the QED final state correction and $t=\ln
2/\Gamma$ is the half-life time).  Deuterium and carbon interactions have
been studied with artificial neutrino beams. Deuterium, chlorine and
gallium nuclei permitted observations and measurements of solar neutrinos
at relatively low energies. Other nuclei have been studied only
theoretically.  In the figures, we provide as an example a few CC $\nu_e$
interactions with nuclei in the energy regions relevant for solar neutrino
detection (left) and for supernov\ae{} (right). Although so far $\nu_e$
supernova events have not yet been observed, and inverse beta decay will
allow to collect large samples of $\bar{\nu}_e$ events, weak interactions
(CC and NC) of neutrinos with $E_\nu\circa{<}200\MeV$ are important for
the dynamics of core collapse supernov\ae. We recall incidentally that the
trapping of muon and tau neutrinos in supernovae is due to neutral current
reactions, that are large due to coherence (namely, they are proportional
to $A^2$).  Some reactions allow to detect NC events: e.g.\ $\nu {\rm
D}\to \nu p n$ (already used by SNO, see section~\ref{SNO}) and $\nu
{}^{12}{\rm C}\to \nu {}^{12}{\rm C}^*$ followed by ${}^{12}{\rm C}^*\to
{}^{12}{\rm C}\gamma(15.1 \MeV)$ where the neutron and the photon allow to
tag the event. At high energy, $E_\nu\circa{>} 50\MeV$, it is common
practice to describe the nucleons in a nucleus adopting a Fermi gas model
or simplified descriptions of the nuclear process.

\section{Neutrino detectors}\label{detectors}
Ideally, one would like to detect neutrinos in real time measuring their energy, direction and flavour.
We have seen how physics ultimately limits what can be done.
We here discuss in a generic way the detection techniques that allow to approach these ultimate limits.
We will give more precise descriptions of the main experiments that have been performed so far
when we will discuss their results.

In practice, one has to find a compromise between 
the various and contrasting needs of an experiment, 
e.g.\ between the wish to have a very `granular'  detector to see all  
the details of the reaction, and the need to monitor a big amount of matter.

Muons are one of the particles possibly produced by neutrino interactions,
and deserve a dedicated discussion, because $\mu^\pm$ can travel for distances
bigger than the detector: their range roughly is
$2.5\,{\rm kmwe}\ln(1+2E_\mu/\TeV)$.
This long range is annoying when $\mu^\pm$ escape from the detector,
and is advantageous because allows to see neutrino interactions 
originated in the material outside the detector:  the event rate is
proportional to the area of the detector, rather than to its volume.
The loss of information on the event is compensated by a gain in the event rate.
Similarly, at ultra-high neutrinos energies neutrinos could be detected using the Earth atmosphere as target
and looking for the macroscopic quasi-horizontal air showers of particles produced by neutrino scatterings
(no positive detection has been claimed so far).


\subsection{Water \v{C}erenkov}
W\v{C} detectors consist of pure water surrounded by photomultipliers,
that see in real time the \v{C}erenkov light emitted by relativistic charged particles scattered by neutrinos.
At neutrino energies less than about a GeV, neutrinos can only scatter $e^\pm $ and $\mu^\pm$.
W\v{C} detectors can distinguish electrons from muons
(because a scattered $\mu$ or $\bar\mu$ produces a clean \v{C}erenkov ring, while
an $e^\pm$ produces a fuzzier ring) but cannot distinguish particles from anti-particles.
$\tau$ leptons fastly decay into hadrons, that cannot be easily studied.
Furthermore, W\v{C} detectors can see the $\gamma$ emitted by nuclear de-excitations or particle decays,
such as $\pi^0\to 2\gamma$. The $\gamma$ are seen as fuzzy rings, like electrons.

Measuring the  \v{C}erenkov light, W\v{C} detectors allow to reconstruct the energy
and the direction of the scattered charged particle(s).

At $E_\nu \sim \GeV$ and if  the position of the neutrino source is known (e.g.\ in neutrino beam experiments)
one can combine these two pieces of information to reconstruct the neutrino energy.
This becomes impossible at larger $E_\nu$, where neutrino make inelastic scatterings
producing unseen neutral particles.
At $E_\nu \gg \GeV$ scatterings are forward peaked, such that the direction of the scattered particle(s) are 
strongly correlated with the direction of incoming neutrino.
Furthermore, $\mu^\pm$ have a long enough range in matter that the material below the detector (e.g.\ rock) acts as a target.
If $E_\nu \ll \GeV$, W\v{C} can see electrons scattered by neutrinos via NC and CC:
(anti)neutrinos with different flavors contribute to this single measurable rate as discussed in section~\ref{neutrino/electron}.
The W\v{C} allowed to detect solar neutrinos down to $E_\nu \sim 5\MeV$.
In section~\ref{SNO} we describe how, replacing water with heavy water, allowed to 
discriminate $\nu_e$ from $\nu_{\mu,\tau}$.

Since water is cheap, the W\v{C} technique allows to build the large detectors needed by neutrino physics.
However, some signals (e.g.\ solar neutrinos, proton decay) can only be seen by placing the detector in underground caverns,
in order to suppress the  background due to cosmic rays: 
this is today the dominant limit on the target mass, that could reach a Mton in some future project.

A variant of the W\v{C} technique (water replaced with heavy water)
allows to observe NC events, as done by 
 SNO and discussed in section~\ref{SNO}.
A variant of the W\v{C} technique allows to reach ${\rm km}^3$ volumes: 
strings of photo-multipliers are inserted in ice or sea-water.
The transparency of the water or ice limits the distance among photo-multipliers,
and a poorly granular detector anyway has a poor energy resolution and high energy threshold
(because particles loose energy before reaching the first photo-multiplier).
Experiments of this kind are being planned as telescopes for neutrinos with $E_\nu\circa{>}\TeV$:
they are mostly sensitive to $\nubarnu_\mu$ and allow to precisely reconstruct their direction.

\subsection{Liquid scintillators}
It is possible to build useful and relatively cheap neutrino 
detectors using liquid organic scintillators. The largest ones in operation are LVD
(total mass: 1 kton, formed by 840 separate tanks) 
and KamLAND (a unique sphere), 
and there is a proposal to build
a detector  30 times larger, LENA. 
The working principle is the 
collection of the light released by charged particles 
propagating in the scintillator, that is  
subsequently amplified by photomultipliers; these detectors
can measure energy well but are not directly useful for 
particle identification. 

Scintillators are adopted for two types of measurements 
involving neutrinos with $E_\nu\sim$ 10 MeV:
the search of neutrinos from supernov\ae, 
performed with relatively simple and stable detectors 
like BUST at Baksan 
in Ukraine and LVD in Gran Sasso; 
the study of low-energy neutrinos from pion decays, done by
Karmen, LSND and MiniBOONE.
Scintillators perform well at low energy, 
and in fact, they are the only possible conventional 
detectors for the low energy neutrinos ($E_\nu \sim$ MeV), 
such as 
(a)~reactor neutrinos (the main experiments are KamLAND , CHOOZ, Bugey); 
(b)~geo-neutrinos (first observation in KamLAND);  
(c)~low energy solar neutrinos, which are the main goal of Borexino.

We now list the main low energy neutrino reactions~\cite{scireac}.
(1)~inverse beta decay $\bar{\nu}_e p\to n e^+$: a scintillator
sees the $e^+$ and can also tag the neutron, because it produces a gamma-ray
thanks to the reaction $ np\to D \gamma$ 
(it is possible to increase the energy of the nuclear gamma-ray by
doping the scintillator with a suitable nucleus, like gadolinium).
(2)~Reactions with carbon 
e.g., the CC reaction $\nu_e {}^{12}{\rm C}\to e^- ~{}^{12}{\rm N}$, 
(3)~Elastic scattering on electrons (that will be used to see beryllium solar neutrinos)
and possibly 
(4)~also on protons (that give sub-MeV events  with supernova neutrinos).
(5)~A last important reaction is with the iron of the support structure.

\subsection{Calorimeters}
Calorimeters measure the total energy of charged and of strongly  
interacting particles produced by neutrino scatterings.
By alternating layers of calorimeters to e.g.\ scintillators one  
gains information on the tracks of the charged particles.
This allows to accurately discriminate CC events due to $\nu_\mu$:
the muons produced by CC scatterings make longer tracks than 
electrons (that give e.m.\ showers) or the hadrons produced in CC or
NC interactions. By magnetizing the calorimeter (that can be made e.g.\ by 
iron) one  can also discriminate charged particles from their anti-particles.
The {\sc NuTeV} (section~\ref{NuTeV}) and {\sc Minos} (section~\ref 
{NuMi}) detectors use this technology.

\subsection{Radio chemical}
This technique was used by the first solar neutrino experiments, and allows to reach
the lowest neutrino energies so far.
A sufficiently large mass of an appropriate target nucleus is put underground
(in order to avoid backgrounds), such that neutrinos induce a nuclear reaction.
The target nucleus is chosen such that the cross section is precisely computable
and such that the produced nuclei can be later counted,
e.g.\ because they decay back in the original nucleus.
These experiments measured solar neutrino rates with $\approx 5\%$ accuracy:
future improvements would need detector  calibrations.

%% file: review_atmsun.tex

\chapter{The atmospheric evidence}\label{atm}
\begin{figure}[h]
$$
\includegraphics[height=7cm,width=7cm]{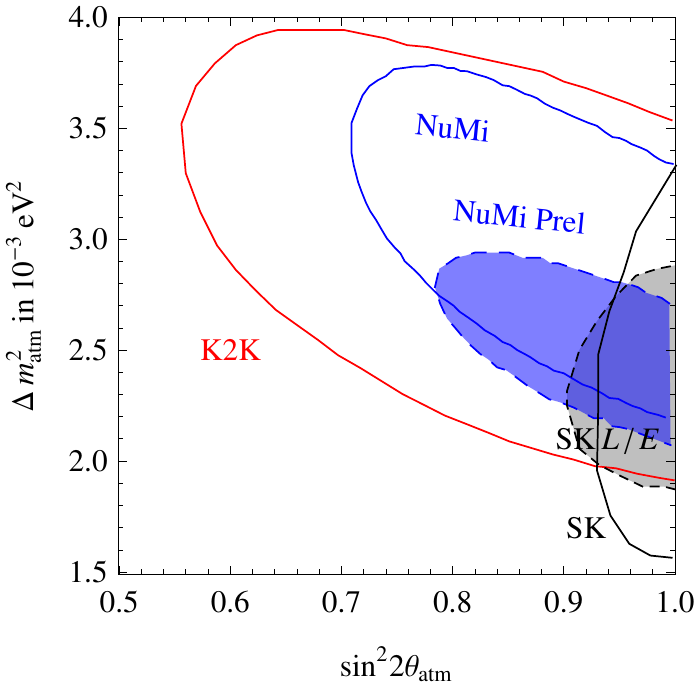}$$
\caption[Fit of atmospheric data]{\label{fig:atmfit} {\bf Atmospheric oscillation parameters}.
\em 
Best-fit regions at 90\% CL from SK (continuous black line), SK $L/E$ analysis (filled gray region), K2K (red line) and NuMi (continuous blue line: published analysis;
filled blue region:  preliminary analysis with more data).}
\end{figure}

The evidence for $\nu_\mu\to\nu_\tau$ oscillations is named `atmospheric' 
because it was established by the SuperKamiokande experiment,
studying atmospheric neutrinos.
{\sc Macro}, IMB, {\sc Soudan2} and {\sc Minos} performed similar measurements.
The K2K and NuMi beam experiments confirm the effect.
The {\sc Chooz}  experiment gives important bounds
on the alternative $\nu_\mu\to \nu_e$ channel.
Fig.\fig{atmfit} shows the result of $\nu_\mu\to\nu_\tau$ oscillation fits.

\section{Atmospheric neutrinos}\label{AtmNu}
\begin{figure}[t]
$$\includegraphics[height=6cm]{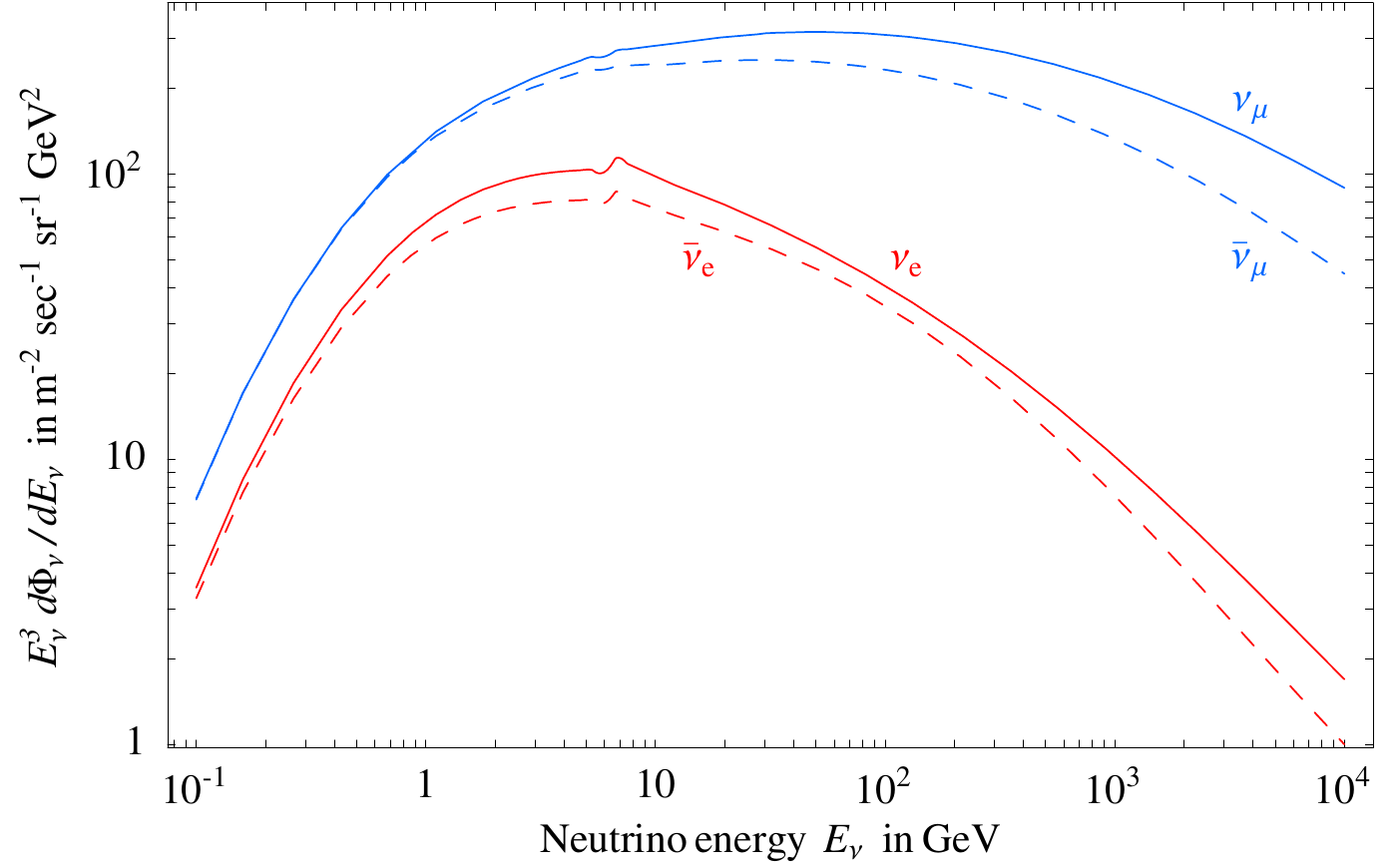}\quad
\includegraphics[height=6cm]{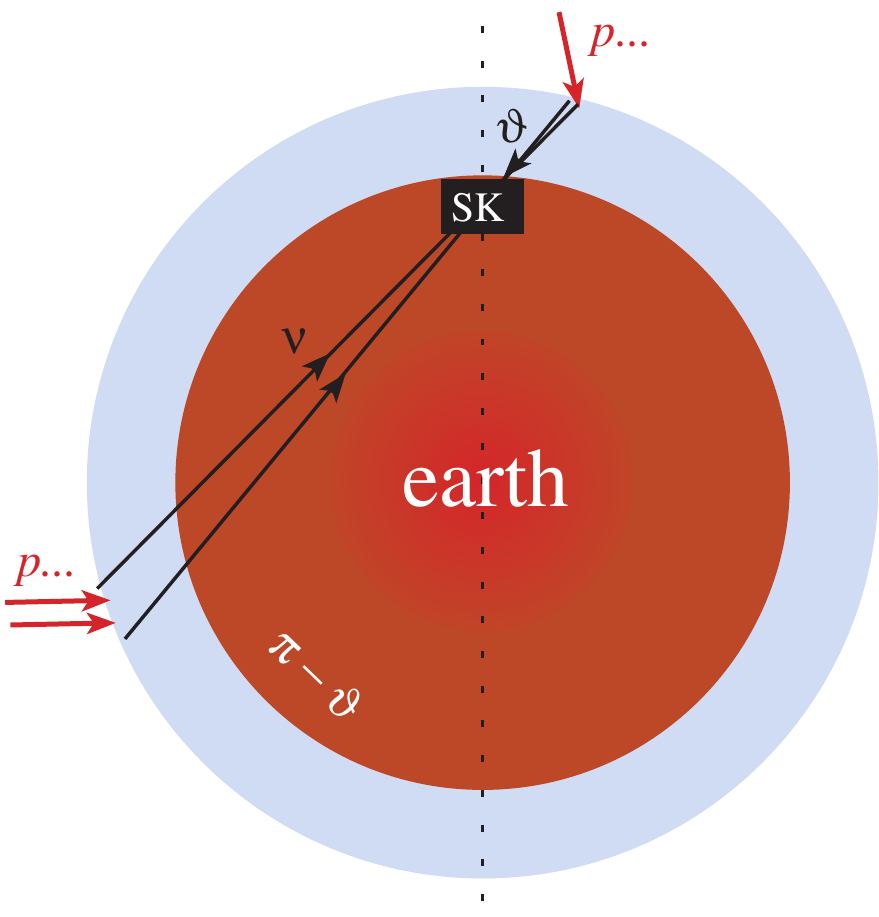}$$
\caption[Predicted flux of atmospheric neutrinos]{\em Flux of atmospheric neutrinos in absence of oscillations,
as predicted by FLUKA~\cite{nuAtm}.
The cartoon at the right shows that,
since the earth is spheric, 
without oscillations the flux of atmospheric neutrinos would
be up/down symmetric.
\label{fig:FLUKA}}
\end{figure}

Atmospheric neutrinos are generated by collisions of primary cosmic rays
(mainly composed by H and He nuclei,  yielding respectively $\sim 82$ \% and $\sim 12$ \% of the nucleons.
Heavier nuclei constitute the remaining fraction).
The process can be schematized in 3 steps~\cite{nuAtm}:
\begin{enumerate}
\item Primary cosmic rays
 hit the nuclei of air in the upper part of the earth atmosphere,
producing mostly pions (and some kaon).

\item
Charged pions decay promptly generating muons and muonic neutrinos:
$$\pi^+ \to \mu^+ \nu_\mu,\qquad \pi^- \to \mu^- \bar{\nu}_\mu$$
(the decay rate into electrons is suppressed by $m_e^2/m_\mu^2$).
The total flux of $\nu_\mu,\bar{\nu}_\mu$ neutrinos is about $0.1/\cm^2\s$ at $E_\nu\sim \GeV$
with a $\sim 20\%$ error (mostly due to the uncertainty in the flux of cosmic rays and in their
hadronic interactions).
At higher energy the flux $d\Phi/d\ln E_\nu$ approximately decreases as $E_\nu^{-2\pm 0.05}$.
The few kaons decay like pions, except that $K\to \pi e^+ \nu_e$ decays are not entirely negligible.

\item The muons produced by $\pi$ decays travel a distance
$$d\approx c \tau_\mu \gamma_\mu \approx 1 \,\hbox{km}  \frac{E_\mu}{0.3\GeV}   $$
where $\tau_\mu$ is the muon life-time and $\gamma_\mu = E_\mu/m_\mu$ is the relativistic dilatation factor.
If all muons could decay
$$\mu^-\to e^-    \bar{\nu}_e \nu_\mu \qquad
\mu^+\to e^+ \nu_e \bar\nu_\mu$$
one would obtain a flux of $\nu_\mu$ and $\nu_e$ in proportion $2:1$,
with comparable energy, larger than $\sim 100\MeV$.
However, muons with energy above few GeV typically collide with the earth
before decaying, so that at higher energy 
the $\nu_\mu : \nu_e$ ratio is larger than 2.
\end{enumerate}
The fluxes predicted by detailed computations is shown in fig.\fig{FLUKA},
at SK location, averaged over zenith angle and ignoring oscillations.


\section{SuperKamiokande}\index{SuperKamiokande}
SK~\cite{SKatm} detects atmospheric neutrinos  through CC scattering on nucleons,
$\nu_\ell N \to \ell N'$.
SK is a cylindrical tank containing 50000 ton of light water surrounded by photomultipliers,
located underground in the Kamioka mine in Japan.
A relativistic charged lepton $\ell$ traveling in water gives rise to a detectable \v{C}erenkov ring
(the scattered proton is not visible because its energy is typically below the W\v{C} threshold).
As discussed in section~\ref{detectors},
SK can distinguish $\nu_\mu$ from $\nu_e$ but cannot distinguish
particles from anti-particles.
The atmospheric fluxes contain a roughly equal number of $\nu$ and $\bar\nu$,
and $\nu$ have roughly a two times larger cross section on nucleons than $\bar\nu$.

Measuring the  \v{C}erenkov light SK reconstructs the energy $E_\ell$
and the direction $\vartheta_\ell$ of the scattered charged lepton.
At high energy $E_\ell \gg m_N$ the scattered lepton roughly keeps the direction of the neutrino,
whose zenith-angle $\vartheta_\nu$ 
is related to the pathlength $L$ by
\beq\label{eq:Latm}
L =\underbrace{ \sqrt{h^2 + 2h r_E + r_E^2 \cos^2\vartheta_\nu}- r_E|\cos\vartheta_\nu|}_{\hbox{\footnotesize in the atmosphere}}
+\underbrace{2r_E|\cos\vartheta_\nu|}_{\hbox{\footnotesize in the earth, if $\cos\vartheta_\nu < 0$}}
\eeq
where $r_E = 6371\km$ is the radius of the earth and $h\sim (15\div 20)\km$ is the height of the atmosphere.
Down-ward going neutrinos ($\cos\vartheta_\nu = 1$) travel $L\sim h$.
Horizontal neutrinos ($\cos\vartheta_\nu = 0$) travel $L\sim \sqrt{2 r_E h}\sim 500\km$.
Up-ward neutrinos ($\cos\vartheta_\nu = -1$) travel $L= 2r_E$.

Measuring $E_\ell$ and $\vartheta_\ell$ is not sufficient to reconstruct the neutrino energy, $E_\nu\circa{>} E_\ell$, since it is not know from which direction one
atmospheric neutrino arrives.
In practice SK can group their data into few big energy `bins' 
defined according to the topology of the events:
\begin{enumerate}
\item{\bf Fully-contained} electron or muon events: the scattered lepton starts and ends inside the detector,
so that its energy can be measured.
These events are conveniently sub-divided into
\begin{itemize}
\item {\bf sub-GeV} events\footnote{The cut is precisely defined in
terms of the amount of \v{C}erenkov light. It is chosen such that sub-GeV $\mu$-like events are mostly
fully contained: the scattered lepton is produced and remains inside the detector.}, with $E_\ell \circa{<} 1.4\GeV$,
and are produced by neutrinos with a typical energy of about a GeV.
The average opening angle between the incoming neutrino and the
detected charged lepton is $\vartheta_{\ell\nu}\sim 60^\circ$: 
the sub-GeV sample has a poor angular resolution:
$E_\nu$ and $L$ can only be estimated.

\item {\bf multi-GeV} events with $E_\ell \circa{>} 1.4\GeV$
are produced by neutrinos with a typical energy of few GeV.
The average opening angle between the incoming neutrino and the
detected charged lepton is $\vartheta_{\ell\nu}\sim15^\circ$ (decreasing at higher energy),
allowing a reasonably precise measurement of $L$.
\end{itemize}
\item {\bf Partially contained muons}: the muon is scattered inside the detector,
but escapes from the detector,
so that its energy cannot be measured.
These events originate from neutrinos with a typical energy only slightly higher than
those giving rise to multi-GeV muons.  Therefore these two classes of events
can be conveniently grouped together.

\item {\bf Up-going stopping muons}: the scattered $\mu$ is produced in the rock below
the detector (so that its energy cannot be measured)
and stops inside the detector.\footnote{This is the first 
method proposed for detecting atmospheric $\nu_\mu$~\cite{markov}.}
The typical energy of parent neutrinos is $E_\nu\sim 10\GeV$.
This technique cannot be used for studying $\nu_e$
(because the scattered electrons shower before reaching the detector) nor for
down-going $\nu_\mu$ (due to the background of cosmic ray muons).

The {\sc Macro} experiments provides additional samples of
partially contained and  up-going stopping muons 
(with average energy of 2.4 and 2.2 GeV, respectively).

\item {\bf Through-going up muons}: the $\mu$ is scattered
in the rock below
the detector and crosses the detector without stopping.
Such events have been observed also by
{\sc Macro} and SNO, which are
competitive with SK because the important parameter
is the surface of the detector, rather than its mass.

These events are produced by neutrinos with a typical energy
between $10\GeV$ and $10\TeV$.
Furthermore, SK could select a sub-sample of higher energy `showring' muons,
as this kind of energy loss in matter becomes dominant above around 1 TeV.
Predictions and direct measurements of cosmic ray primaries at
high energies are difficult~\cite{nuAtm}.

\end{enumerate}
In conclusion, atmospheric neutrinos cover a {\em wide energy range}, from less than a GeV to more than a TeV.
Atmospheric neutrinos also allow to probe a {\em wide range of baselines}, between 10 and 10000 km.
Therefore they are a good probe of oscillations (see fig.\fig{LE}).

\begin{figure}
$$\includegraphics[width=0.95\textwidth]{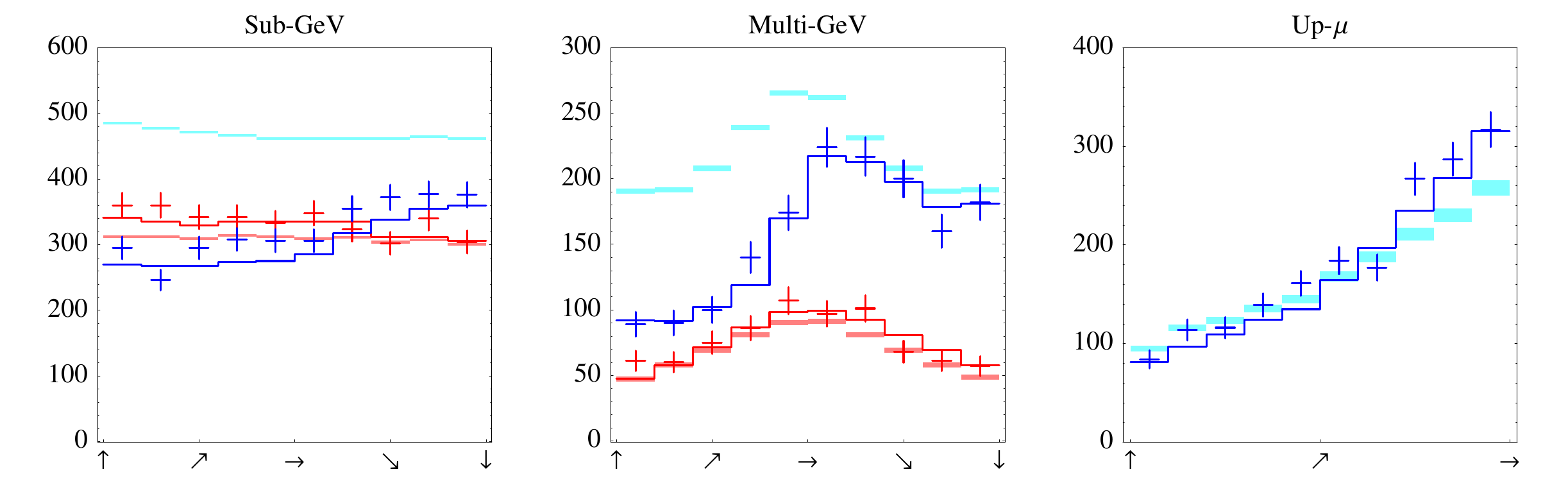}$$
\caption[SK atmospheric data]{\em The main SK data: number of 
{\color{rosso}$e^\pm$ (red)}
and of {\color{blu}$\mu^\pm$ (blue)} events
as function of direction of scattered lepton.
The horizontal axis is $\cos\vartheta$, the cosine of the zenith angle ranging between
$-1$ (vertically up-going events) and $+1$ (vertically down-going events).
Fig.\fig{SKatmdata}c shows high-energy
through-going muons, only measured in the up direction.
The crosses are the data and their errors, 
the thin lines are the best-fit oscillation expectation, 
and thick lines are the no-oscillation expectation:
these are roughly up/down symmetric.
Data in the multi-GeV muon samples are very clearly asymmetric,
while data in the electron samples (in red) are compatible with no oscillations.
\label{fig:SKatmdata}}
\end{figure}

The SuperKamiokandeI data are shown in fig.\fig{SKatmdata}
(1489 days of data taking, terminated by an accident. SK collected
few thousands of events).
The two histograms show the prediction assuming no oscillations,
and the best $\nu_\mu\to\nu_\tau$
oscillation fit, for $\Delta m^2_{\rm atm} = 2.5~10^{-3}\eV^2$ and $\sin^2 2\theta_{\rm atm} = 1$.
The atmospheric neutrino fluxes are varied, within their uncertainties,
to their best-fit values independently in the two cases.
This is why the predicted $e$-like rates differ even if $\nu_e$ do not oscillate:
the no-oscillation fit tries to reproduce the deficit in $\mu$-like events by reducing the
uncertain total flux of $\nubarnu_e$ plus $\nubarnu_\mu$, and by using
a few others of these uncertainties.

\medskip

{\em The ``multi-GeV $\mu$ + PC'' data sample 
shows that a neutrino anomaly is present even without relying
our knowledge of atmospheric neutrino fluxes}.
The crucial point is that since the Earth is a good sphere, in absence of oscillations
the neutrino rate would be up/down symmetric, i.e.\  it depends only on $|\cos\vartheta|$.\footnote{The mountain 
around the detector and especially the
magnetic field of the earth break this symmetry.
But they are minor effects at multi-GeV energy.}
The $dN/d\cos\vartheta_\nu$ spectrum would be flat, if one could ignore that
horizontal muons have more time for freely decaying before hitting the earth,
while vertical muons cross the atmosphere along the shortest path.
This effect produces the peak at $\cos\vartheta_\nu \sim 0$ visible in fig.\fig{SKatmdata}b.

While the zenith-angle distribution of $\mu$ events is clearly asymmetric,
$e$-like events show no asymmetry.
The flux of up-ward going muons is about two times lower than the flux of down-ward muons.
Therefore the data can be interpreted assuming that nothing happens to $\nu_e$
and that $\nu_\mu$ oscillate
into $\nu_\tau$ (or into sterile $\nu_{\rm s}$).
Neglecting earth matter corrections, we assume\index{Oscillation!atmospheric}
\beq\label{eq:Patm}
\color{rosso}P(\nu_e\to\nu_e) = 1\qquad P(\nu_e\leftrightarrow\nu_\mu) = 0\qquad
\color{blu}
P(\nu_\mu\to\nu_\mu) = 1 - \sin^2 2 \theta_{\rm atm} ~\sin^2 \frac{\Delta m^2_{\rm atm} L}{4E_\nu}\eeq
The main result can be approximately extracted from very simple considerations.
Looking at the zenith-angle dependence we notice that
down-ward going neutrinos ($\downarrow$) are almost unaffected by oscillations, while
up-ward going neutrinos ($\uparrow$)  feel almost averaged oscillations,
and therefore their flux is reduced by a factor $1 - \frac{1}{2}\sin^2 2\theta_{\rm atm}$.
This must be equal to the up/down ratio
${N_\uparrow}/{N_\downarrow} = 0.5\pm 0.05$,
so that $\sin^2 2\theta_{\rm atm} =1\pm 0.1$.
Furthermore, multi-GeV neutrinos have energy $E_\nu \sim 3\GeV$, and
according to fig.\fig{SKatmdata}b they begin to oscillate
around the horizontal direction ($\cos\vartheta \sim 0$) i.e.\ at a pathlength of about
$L \sim 1000\,\hbox{km}$.
Therefore $\Delta m^2_{\rm atm} \sim E_\nu/L \sim 3~10^{-3}\eV^2$.

A global fit (performed including also the less safe input from MonteCarlo
predictions of neutrino fluxes) gives the best-fit values shown in fig.\fig{K2K}b.
SK cannot
precisely measure $\Delta m^2_{\rm atm}$ because around the horizontal $L$ depends strongly on $\vartheta_\nu$.

\smallskip

Historically, the measurement of a deficit in the ratio between
$\mu$-like and $e$-like events
gave the first argument for an atmospheric neutrino anomaly.
As explained in section~\ref{AtmNu}, while the overall number of atmospheric
$\nu_\mu$ and of $\nu_e$ cannot be precisely predicted, 
their ratio is predicted to be close to 2 in absence of oscillations.
The measurement of the double ratio
$(N_\mu/N_e)_{\rm exp}/(N_\mu/N_e)_{\rm MC}\sim 0.65$
was considered as the main evidence;
however $(N_\mu/N_e)_{\rm MC}$ significantly deviates from the ideal value 2,
due to the different $\nu_\mu$ and $\nu_e$ energy spectra and experimental cuts.
The main evidence for the atmospheric anomaly is provided by
the zenith-angle dependence of $N_\mu$ discussed above.

\section{Oscillations?}
Here we discuss at which level SK probes if the atmospheric anomaly follows the
 specific energy and pathlength dependence of
$P(\nu_\mu\to\nu_\mu)$ predicted by oscillations.
\paragraph{Energy dependence.}
Fig.\fig{SKatmdata} shows that the anomaly is visible not only in multi-GeV data, but also at lower energy
in sub-GeV data and at higher energy in stopping and through-going muons.
One can see that the anomaly
decreases at higher energy and increases at lower energy, as predicted by $\nu_\mu\to\nu_\tau$ oscillations.
One can be more quantitative:  fitting the SK data assuming
$$P(\nu_\mu \to \nu_\mu ) =
 1- \sin^2 2 \theta  \sin^2 \alpha L E^n
$$
($n=-1$ is predicted by oscillations;
$n=0$ can be obtained from CPT-violating effects; 
$n=1$ from violation of Lorentz invariance)
gives $ n = -1.03 \pm 0.13$~\cite{SKatm}.

\paragraph{Path-length dependence.}  SK can see that
$P(\nu_\mu\to \nu_\mu)$ decreases by $50\%$ when going from short to long baselines.
However it cannot observe the most characteristic feature of oscillations: the first oscillation dip.
As illustrated in fig.\fig{tipico}b at page~\pageref{fig:tipico}, this happens because SK cannot measure the neutrino energy:
the oscillation pattern gets washed when averaging over too different neutrino energies.
While neutrino decay $P(\nu_\mu \to \nu_\mu) = e^{-L m_\nu /E_\nu\tau}$ cannot explain SK data,
one might think that any function that interpolates 
$$P(\nu_\mu \to \nu_\mu \hbox{ at short $L$})=1\qquad\hbox{with}\qquad
P(\nu_\mu \to \nu_\mu \hbox{ at long $L$})\approx1/2$$
provides an acceptable description of SK data.
This is not the case: the dip-less SK data in fig.\fig{SKatmdata} 
are in excellent quantitative agreement with the oscillation prediction.
Global fits of SK data disfavor at about $4\sigma$
alternative $\nu_\mu$ disappearance mechanisms
which do not give rise to a dip in $P(\nu_\mu\to\nu_\mu)$~\cite{SKatm}.
Some concrete examples are
$$P(\nu_\mu \to \nu_\mu ) \simeq \left\{
\begin{array}{ll}
 1- (\sin^2 \theta +\cos^2 \theta \,  e^{-c L/E})^2      &  \hbox{decay of mixed neutrinos~\cite{NuDecay}}  \\
1- \frac{1}{2}\sin^2 2\theta ( 1 - e^{-c L/E^n}) & \hbox{decoherence~\cite{Decoherence}}\\
 |\cos^2 \theta + \sin^2 \theta (1-\hbox{erf}\sqrt{{i c L}/{E}})|^2      &   \hbox{oscillations into 5-dimensional $\nu$~\cite{nuextrad}}
\end{array}\right.$$
where $c$ and $n$ are arbitrary constants.
The case $n=1$ is the least disfavored~\cite{SKatm};
the theoretical considerations of section~\ref{rho} suggest $n=2$.

\medskip

SK reanalyzed their data selecting a sub-sample of `cleanest' events, 
where the resolution in $L/E_\nu$ 
is good enough to see a hint of the first oscillation dip.
This is roughly done by excluding events with energy $E_\mu\circa{<}\GeV$ 
(because at low energy the direction of the scattered muon
is poorly correlated with the direction of the neutrino)
and with reconstructed zenith-angle $|\cos\vartheta_{\nu}|\circa{<}0.2$
(around the horizontal it is difficult to measure the path-length,
because it strongly depends on  $\vartheta_{\nu}$,
see eq.\eq{Latm} and fig.\fig{tipico}b).
In this way SK achieved a $\sim 50\%$ resolution in $L/E_\nu$.
By choosing $L/E_\nu$ bins smaller than the resolution,
SK could produce plots where the first oscillation dip is visible to the naked eye~\cite{SKatm}.
This kind of analysis provides a better resolution in $\Delta m^2_{\rm atm}$ at the expenses
of a poorer resolution in $\theta_{\rm atm}$: the result is shown in fig.\fig{K2K}b.

\medskip

In section~\ref{oscexp} we will discuss which future experiments can more directly test these important issues.

\section{K2K}\label{K2K}\index{K2K}
An artificial long-baseline $\nu_\mu$ pulsed beam is sent from KEK to 
the SK detector, located $L=250 \km$ away in the Kamioka mine.
Since the beam is pulsed, SK can discriminate atmospheric $\nu$ from KEK $\nu_\mu$,
both detected using charged-current scattering on nucleons, as previously discussed.
The neutrino beam was produced by colliding a total of
$9\cdot 10^{19}$ accelerated protons on a target.
A magnetic field is used to collect and focus the resulting $\pi^+$,
obtaining  from their decays a $98\%$ 
pure $\nu_\mu$ beam with an average energy of $E_\nu\sim 1.3\GeV$.
The base-line $L$ and the energy $E_\nu$ have been chosen such that
\begin{itemize}
\item[1.] $\Delta m^2_{\rm atm} L/E_\nu \sim 1$ in order to sit
around the first oscillation dip;
\item[2.] $E_\nu\sim m_p$ in order to have large opening angles between the incoming neutrino and the scattered $\mu$: $\vartheta_{\mu \nu}\sim 1$.
\end{itemize}
Since  the direction of the incoming neutrino is known
(unlike in the case of atmospheric neutrinos),
measuring $E_\mu$ and $\vartheta_{\mu \nu}$ 
SK can reconstruct the neutrino energy 
\beq E_\nu = \frac{m_N E_\ell - m_\mu^2/2}{m_N-E_\ell+ p_\ell \cos \vartheta_{\mu \nu}} \eeq
having assumed that $\nu_\mu  n \to \mu p$ is the dominant reaction.
Around $E_\nu \sim \GeV$ pion production and deep-inelastic scattering
give subleading ($\sim 10\%$) contributions.
Since the neutrino flux and the $\nu_\mu N$ cross section are not precisely computable,
small detectors (mainly a 1 kton W\v{C} and  fine-graned systems)
have been built close to the neutrino source in KEK,
so that oscillations can be seen by comparing SK  data with near detectors.
$\nu_\mu\to\nu_\tau$ oscillations at the atmospheric frequency
should give an energy-dependent deficit of events in the far detector,
according to eq.\eq{Patm}.

\medskip

\begin{figure}
$$\includegraphics[height=7cm,width=7cm]{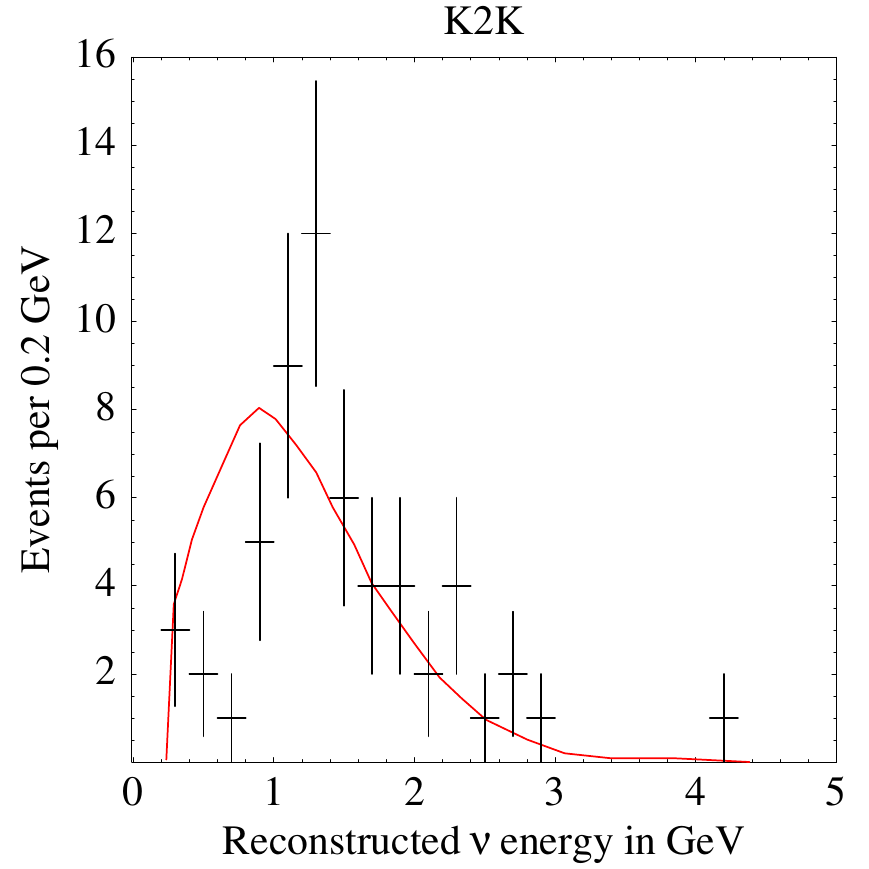}\hspace{1cm}
\includegraphics[height=7cm,width=7cm]{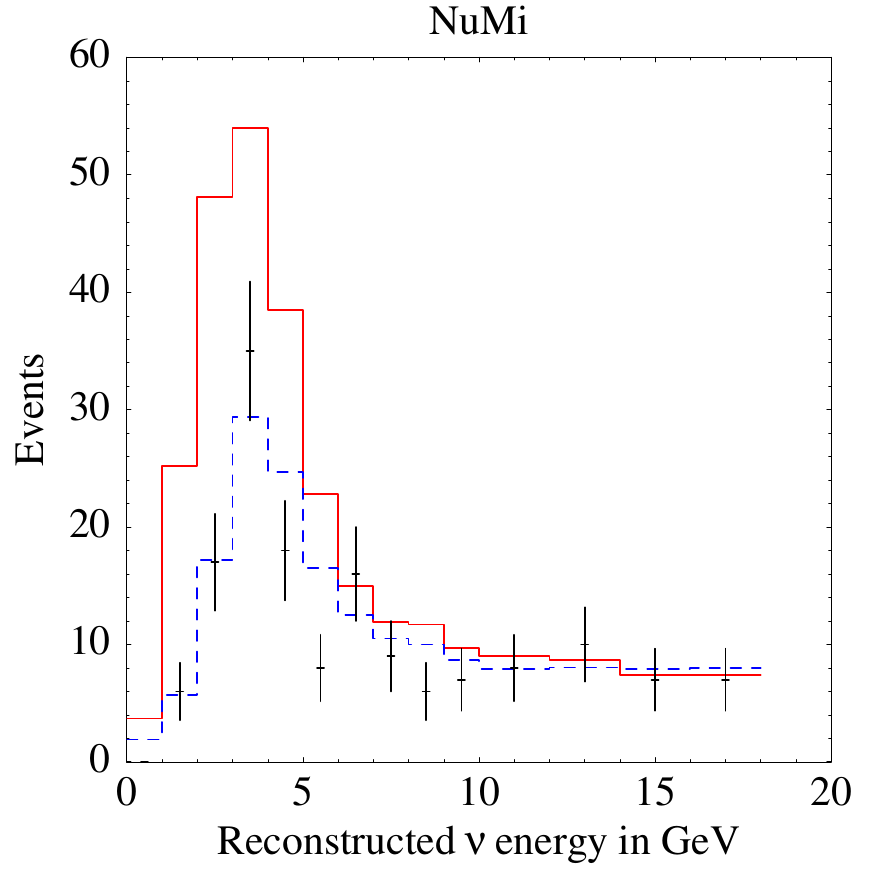}$$
\caption[K2K and NUMI data]{\label{fig:K2K}\em 
Fig.\fig{K2K}a shows the K2K data, and the expectation in absence
of oscillations~\cite{K2K}.
Fig.\fig{K2K}b shows the same thing for {\sc NuMi};
the dashed blue line is the oscillation best fit.
}
\end{figure}

The latest K2K results presented in 2006 and shown in fig.\fig{K2K}
are consistent with the expectations based on SK atmospheric data
and contain a $4.3\sigma$ indication for oscillations.
Concerning the total rate,
one expects in absence of oscillations $158\pm 9$
fully contained events  in the SK fiducial volume
(the uncertainty is mainly due to the far/near extrapolation and to the
error on the fiducial volume).
SK detected 112 events of this kind.
In view of the poorer statistics 
the atmospheric mixing angle is determined much more precisely by SK than by K2K.


The most important K2K result is the energy spectrum:
K2K is competitive on the determination of $\Delta m^2_{\rm atm}$
because, unlike SK, K2K can reconstruct the neutrino energy
and data show a hint of the spectral distortion characteristic of oscillations.
As a consequence K2K suggests a few different local best-fit values of $\Delta m^2_{\rm atm}$,
and the global best fit lies in the region suggested by SK (fig.\fig{atmfit}).

\section{\sc NuMi}\label{NuMi}\index{NuMi}\index{Minos}
The running {\sc NuMi} experiment~\cite{NuMi} is similar to K2K: we emphasize their differences.
A dominantly $\nu_\mu$ pulsed beam is sent from FermiLab to the {\sc Minos} detector,
located 735 km away (fig.\fig{map}). Thanks to the longer base-line, the {\sc NuMi} neutrino beam
has a larger mean energy (around a few GeV) than K2K.
A near detector, functionally identical to the far detector, allows to predict the non-oscillation rate.
Both detectors consist of magnetized steel plates alternated to scintillator strips.
The far detector has a 5.4 kton mass and a magnetic field $B\sim 1.2\,{\rm Tesla}$:
this allows to discriminate particles from anti-particles, and to discriminate NC from CC scatterings.
CC events are selected with $\approx 87\%$ efficiency and $\approx 97\%$ purity.
Backgrounds are suppressed down to a negligible level.

Latest data have been published
studying a total of about $1.3\cdot 10^{20}$ protons-on-target, like the K2K beam,
and finding 215 $\nu_\mu$ events.
Preliminary data  using $2.5~10^{20}$ protons-on-target (563 $\nu_\mu$ events)
have been presented at the
2007 Lepton-Photon conference.
Both analyses show a $\sim 6\sigma$ evidence for a deficit with respect to the number of events expected in absence of oscillations. The deficit appeares at lower energies, $E_\nu \circa{<}5\GeV$ as predicted by oscillations;
however the first oscillation dip is around the minimal energy observable by {\sc NuMi}, so that
it is not clearly visible in data,
see fig.\fig{K2K}b (published data).
These two {\sc NuMi} analyses respectively give
\beq |\Delta m^2_{\rm atm}| = (2.74^{+0.44}_{-0.26})~10^{-3}\eV^2\qquad\hbox{and}\qquad
|\Delta m^2_{\rm atm}| = (2.38^{+0.20}_{-0.16})~10^{-3}\eV^2\hbox{(preliminary})\eeq
compatible with the results from SK and K2K.
The best-fit value of the atmospheric mixing angle is maximal,
compatibly with the more accurate SK determination, see fig.\fig{atmfit}.

Final {\sc NuMi} data should allow to reduce the uncertainty on $\Delta m^2_{\rm atm}$ down to about $\pm 0.1$,
achieve a sensitivity to $\theta_{23}$ slightly worse than SK and
a sensitivity to $\nu_\mu\to\nu_e$ slightly better than CHOOZ.

\section{$\nu_\mu\to \nu_\tau$ or $\nu_\mu\to\nu_{\rm s}$?}
Up to this point,  the atmospheric anomaly could be explained by
the $\nu_\mu\to \nu_\tau$ oscillations or by 
$\nu_\mu\to\nu_{\rm s}$ oscillations, where $\nu_{\rm s}$ is a speculative extra
sterile neutrino, i.e.\ a fermion with no weak interactions.
We assume that $\nu_\mu$ makes mixed active/sterile oscillations at the atmospheric frequency:
$\nu_\mu \to \cos\xi ~\nu_\tau + \sin\xi ~\nu_{\rm s}$, such that the fraction of $\nu_\mu$
that oscillates into $\nu_{\rm s}$ is $f_{\rm s} =  \sin^2 \xi$.
As we now discuss, the SK and {\sc NuMi} data~\cite{SKatm,NuMi} 
favor $\nu_\mu\to \nu_\tau$ and indicate that  $\nu_\mu\to\nu_{\rm s}$ 
can at most give a minor correction: 
\beq f_{\rm s} = 0\pm 0.10.\eeq
(In section~\ref{oscexp} we will discuss which future experiments can further test this important issue).

\begin{enumerate}
\item
{\bf Earth matter corrections at SK} do not affect $\nu_\mu\to \nu_\tau$ oscillations but would
affect $\nu_\mu\to\nu_{\rm s}$ oscillations.
Given the value of $\Delta m^2_{\rm atm}$ this would be a significant effect,
suppressing oscillations at high energy, after possibly having crossed a resonance 
(section~\ref{matterosc}).\footnote{This is not generically true if one considers oscillations into many sterile $\nu_{\rm s}$ --- 
a possibility anyway disfavored by arguments 2.\ and 3.}
According to global fits, stopping and through-going up muons  independently favor $\nu_\mu\to \nu_\tau$ at $3\sigma$.
Another $3\sigma$ hint comes from the analogous effect in  through-going up muons 
detected by the {\sc Macro} experiment at Gran Sasso.

\item {\bf Neutral-current rates at SK}.
`Multi-ring' events (i.e.\ events with two or more separated \v{C}eren\-kov rings
which also satisfy certain other selection cuts) contain a significant fraction of
NC-induced events, about $30\%$ according to the SK MonteCarlo.
The main NC processes which produce multi-ring events are
are $\nu N \to \nu N\pi\pi$ and $\nu N \to \nu N\pi^0\to\nu N \gamma\gamma$.
Multi-ring events are also produced by CC processes, such as $\nu_e N\to e N'\pi$.

While $\nu_\mu\to\nu_\tau$ conversion would not affect NC rates, 
$\nu_\mu\to\nu_{\rm s}$ oscillations
would decrease the number of NC events in a zenith-angle dependent way.
The total number of events does not allow to discriminate the two possibilities because
the relevant cross-sections have not yet been precisely measured and cannot be precisely computed.
The measured up/down asymmetry of multi-ring events
is consistent with up/down symmetric NC events, 
and favors
$\nu_\mu\to\nu_\tau$ with respect to $\nu_\mu\to\nu_{\rm s}$ at about $3.5\sigma$~\cite{SKatm}.

\item {\bf Neutral-current rates at NuMi}.
Similarly, the {\sc Minos} collaboration could measure the total number of active neutrinos from NC rates~\cite{NuMi}.
Assuming that $\nu_\mu$ oscillate at the atmospheric frequency into $\nu_{\rm s}$ and $\nu_\tau$ with
relative fraction  $f_{\rm s}$ and $1-f_{\rm s}$ respectively, they get
$f_{\rm s}=0.28\pm0.28$.

\item {\bf $\tau$ appearance at SK}.
The expected signal rate is about $1\tau/{\rm kton}\cdot{\rm yr}$.
In SK it is difficult to experimentally discriminate $\tau$ events from hadronic events,
which are expected to have a rate about two orders of magnitude larger.
By imposing appropriate selection cuts, SK defines a class of $\tau$-like events.
The resulting sample is not clean: about $8\%$ of them should be composed by
$\tau$ produced by
$\nu_\mu\to\nu_\tau$ oscillations.
SK data show a zenith-angle dependent enhancement of $\tau$-like events,
giving a $2.4\sigma$ hint for $\nu_\tau$ appearance~\cite{SKatm}.
\end{enumerate}
Combining all these hints give a combined $7\sigma$ evidence for $\nu_\mu\to \nu_\tau$ versus $\nu_\mu\to\nu_{\rm s}$.
Fitting SK data in terms of mixed active/sterile oscillations
$\nu_\mu \to \cos\xi ~\nu_\tau + \sin\xi ~\nu_{\rm s}$ gives {\color{blus}$f=\sin^2 \xi = 0\pm 0.10$}~\cite{SKatm}.

\begin{table}[t]
$$
\begin{array}{c|cccc}
\hbox{Isotope} &
^{235}\hbox{U} & ^{239}\hbox{Pu} & ^{238}\hbox{U} & ^{241}\hbox{Pu} \cr \hline
\hbox{Typical relative abundancy $f_i$} &54\% & 33\% & 8\% & 5\% \cr 
\hbox{Energy per fission $E_i$ in $\MeV$ } &201.7 & 205.0 & 210.0 & 212.4 \cr 
a_0 &0.870 & 0.896 & 0.976 & 0.793 \cr 
a_1 &-0.160 &    -0.239 & -0.162 & -0.080 \cr 
a_2 & -0.091 & -0.0981 & -0.079 & -0.1085 
\end{array}$$
  \caption[Reactor neutrinos]{\em Parameters that describe reactor neutrinos.}\label{tab:isotopi}
\end{table}

\section{CHOOZ: $\nu_\mu\to\nu_e$?}\label{CHOOZ}
The atmospheric anomaly cannot be due to $\nu_\mu\to \nu_e$ oscillations,
since no $\nu_e$ appearance effect is observed by SK.
However, 
once that the dominant $\nu_\mu\to\nu_\tau$ oscillations
have transformed the initial flux $N_{\nu_e}:N_{\nu_\mu}:N_{\nu_\tau}\sim1:2:0$ into 
$\sim1:1:1$, this flavour-blind proportion is unaffected by possible further flavour conversions.
Therefore SK cannot set a strong limit on subdominant $\nu_\mu\to \nu_e$ oscillations, 
such as those produced by solar oscillations (at the smaller solar frequency and with a large mixing angle,
see section~\ref{sun}) or by a small mixing angle $\theta_{13}$ at the atmospheric frequency $\Delta m^2_{\rm atm}$.
Detailed analyses confirm this qualitative argument.

While in principle both atmospheric and solar data are sensitive to $\theta_{13}$,
the dominant bound on $\theta_{13}$ is given by the CHOOZ experiment,
which looked for disappearance of $\bar\nu_e$ emitted by nuclear reactors.
The data, shown in fig.\fig{reattore}a, are consistent with no effect.

Reactor $\bar\nu_e$ have an energy of few MeV and CHOOZ was located at
a distance $L\approx 1\km$ from two french reactors: therefore it is sensitive to $\Delta m^2$ down to $10^{-3}\eV^2$,
(see eq.\eq{Sij}), probing
all values of $\Delta m^2$
consistent with the atmospheric anomaly.
Eq.\eq{Pee} shows how oscillations at the atmospheric frequency affect $P(\bar\nu_e\to\bar\nu_e)$;
solar oscillations and earth matter corrections can be neglected.
Taking into account statistical and systematic errors as described in~\cite{CHOOZ},
 fig.\fig{tipico}a at page~\pageref{fig:tipico} shows the
 region of the ($\Delta m^2_{\rm atm},\sin^2 2\theta_{13}$) plane  excluded at $90\%$ CL (2 dof).
 CHOOZ could not see atmospheric oscillations because $\theta_{13}$ is too small.
CHOOZ implies
$\Delta m^2_{\rm atm} < 0.7~10^{-3}\eV^2$ for $\theta_{13}=\pi/4$, and 
$\sin^22\theta_{13}<0.10$ for large $\Delta m^2_{\rm atm}$.
The CHOOZ bound on $\theta_{13}$ strongly depends on $\Delta m^2_{\rm atm}$.
Combining its determination from SK and K2K with CHOOZ,
using the statistical techniques described in appendix~\ref{Statistics}, gives
$\sin ^2 2\theta_{13} =\ssCH$.

\medskip

 Assuming $\theta_{13}=0$, the CHOOZ bound of  fig.\fig{tipico}a also applies
 to the ($\Delta m^2_{\rm sun},\sin^2 2\theta_{\rm sun}$) plane:
CHOOZ could not  see solar oscillations because  $\Delta m^2_{\rm sun}$
 is somewhat too small.
 We conclude this section with a detailed description of reactor experiments,
as the next section starts discussing how KamLAND , another
reactor experiment with longer base-line,
confirms the solar anomaly.

\section{Reactor experiments}\label{reactor}\index{Reactor neutrinos}
Nuclear reactors use neutrons to break heavy nuclei.
Each fission produces other neutrons that sustain the chain reaction,
a few nuclear fragments that decay producing about 6 $\bar\nu_e$,
and of course kinetic energy.
The energy spectrum of $\bar{\nu}_e$ emitted by a nuclear reactor
can be accurately approximated as
$$\frac{dn}{dE} =  \frac{W}{\sum_j f_j E_j} \sum_i   f_i  \exp (a_{0i} + a_{1i} E + a_{2i} E^2)$$
where $E = E_\nu/\MeV$.
The sums run over the isotopes, $i,j=\{^{235}\hbox{U} , ~^{239}\hbox{Pu} , ~^{238}\hbox{U} , ~^{241}\hbox{Pu} \}$,
which fissions produce virtually all the total thermal power $W$.
Their typical relative abundances $f_i$, and
all numerical coefficients are listed in table~\ref{tab:isotopi}, from~\cite{reactors}.
A typical reactor produces a thermal power $W\sim$ few GigaWatt
($\hbox{GW} = 6.24~10^{21} \MeV/\s$)
and neutrinos have a typical energy of about a MeV.
Assuming no oscillations, the $\bar{\nu}_e$ flux at distance $d$ from the reactor is  $dn/dE/(4\pi d^2)$,
and can be predicted
with a few $\%$ error.

\index{CHOOZ}
The CHOOZ experiment in France had $d\approx \km$, optimal for studying
$\bar{\nu}_e$ disappearance induced by oscillations with $\Delta m^2 \approx 6\cdot 10^{-3}\eV^2$.
The KamLAND experiment in Japan is located at distances ranging between
$80$ to few hundred km from several reactors.
Most $\bar{\nu}_e$ come from reactors at $d\sim 200\km$,
optimal for studying
$\bar{\nu}_e$ disappearance induced by oscillations with $\Delta m^2 \approx 3~10^{-5}\eV^2$.

\begin{figure}[t]
$$\raise-1mm\hbox{\includegraphics[width=8cm,height=7.9cm]{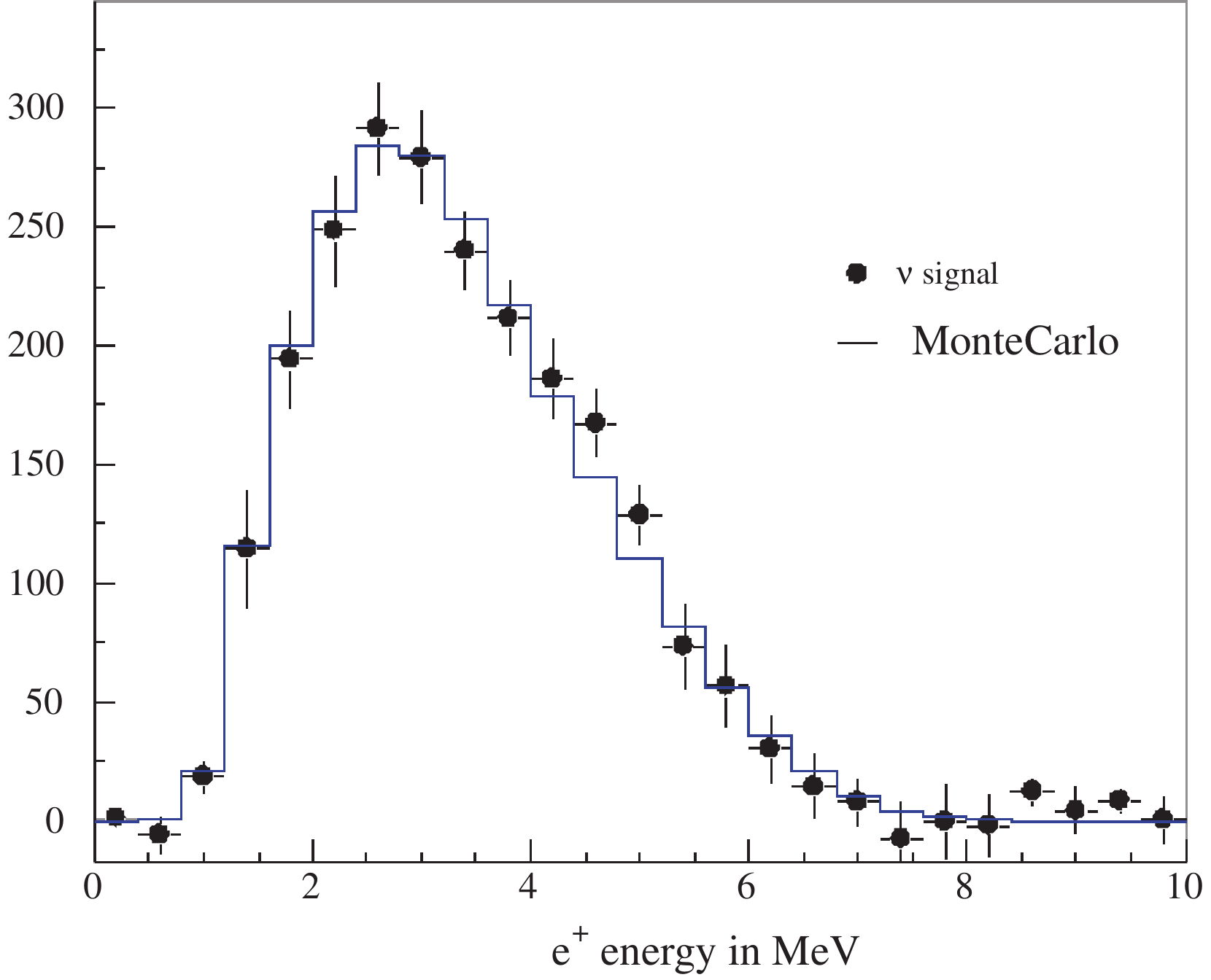}}
\hspace{8mm}\includegraphics[width=8cm,height=8cm]{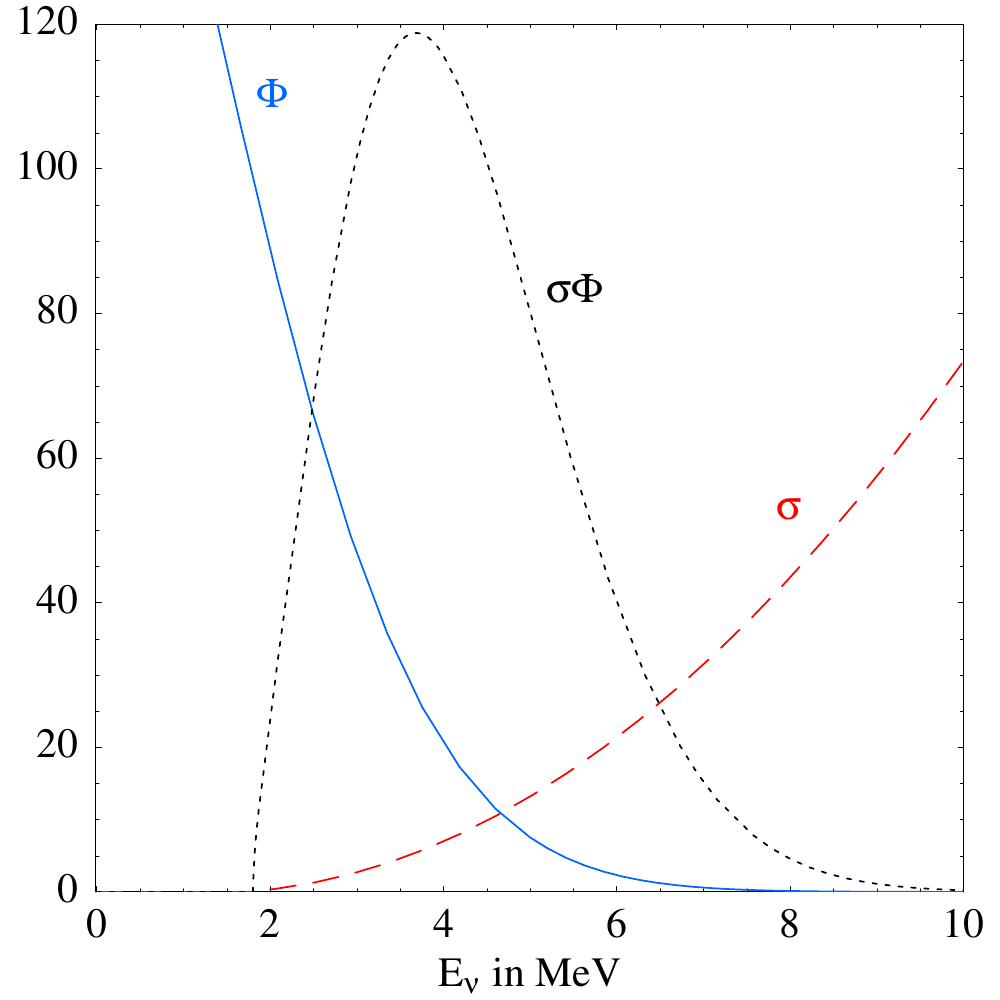}
$$
\caption[CHOOZ data]{\label{fig:reattore}\em 
Fig.\fig{reattore}a shows the final data from CHOOZ~\cite{CHOOZ}.
Fig.\fig{reattore}b shows $\sigma(\bar\nu_e p \to ne^+)$ in $10^{-43}\cm^2$ ({\color{rosso} dashed line}),
the flux of $\bar\nu_e$ in $10^{14}/\cm^2\yr\MeV$ at $1\km$ from a reactor with thermal power $W=1\,{\rm GW}$
({\color{blus} continuous line})
and the resulting interaction rate in $1/{\rm ton}\yr\MeV$ (dotted line).}
\end{figure}


Neutrinos are detected using the gold-plated reaction
$\bar{\nu}_e p \to e^+ n$, discussed in section~\ref{sigmahad}, eq.\eq{sigmanup}.
Using a scintillator it is possible to see both the $\gamma$ ray emitted when the neutron $n$ is captured
and the two $\gamma$ rays with energy $E_\gamma = m_e$ emitted
by the positron $e^+$ as it moves and finally annihilates with a $e$.
In KamLAND the $n$ random walks for $\sim 200\,\mu\s$
before being captured by a proton giving a $\gamma$ ray with $E_\gamma = 2.2\MeV$.
In order to maximize the $n$ capture efficiency and to get more energetic $\gamma$,
Gd was added to the scintillator used in CHOOZ.
The delayed coincidence between the $\gamma$ can be
measurable by photomultipliers with good time and spatial resolution
and allows to select $\bar\nu_e$ events from background.
The total measured energy is $E_{\rm vis} = E_{e^+}+ m_e$, and 
$E_{e^+}$ is related to the neutrino energy by the kinematical relation
$E_\nu = E_{e^+} +m_n -m_p$ (neglecting the recoil of the neutron).

Summing over all reactors $r$ (that emit a power $W_r$ from a distance $d_r$),
 the number of neutrino events with visible energy in any given range
  $E_{\rm vis}^{\rm min} < E_{\rm vis}<E_{\rm vis}^{\rm max}$ is
$$ N =\sum_r
\int dE_\nu ~P_{ee}(d_r/E_\nu) 
\sigma(E_\nu) N_p \frac{dn_r/dE_\nu}{4\pi d_r^2} 
\int_{E_{\rm vis}^{\rm min}}^{E_{\rm vis}^{\rm max}} dE_{\rm vis} 
\frac{e^{-(E_{\rm vis} - E_\nu+0.782\MeV)^2/2\sigma_E^2}}{\sqrt{2\pi}\sigma_E} \varepsilon(E_{\rm vis})
$$
where $\varepsilon$ is the efficiency.
Neglecting small differences between the energy spectra of different reactors
(i.e.\ assuming that all reactors have the same relative isotope abundances $f_i$),
the spectrum of detected neutrinos $dN/dE_\nu$ can be conveniently rewritten as
$$
\frac{dN}{dE_\nu}=n_0  \langle P_{ee}(E_\nu)\rangle  R(E_\nu)$$
where $n_0$ is the rate expected in absence of oscillations,
$\langle P_{ee}(E_\nu)\rangle$ is the averaged survival probability
$$\langle P_{ee}(E_\nu)\rangle =\sum_r p_r P_{ee}(E_\nu/d_r)\qquad
p_r = \frac{W_r/d_r^2}{\sum_{r'}  W_{r'}/d_{r'}^2}$$
and $R$ is the {\em response function} (normalized such that $\int dE_\nu\, R = 1$).
Assuming no oscillations, fig.\fig{reattore}b shows $dN/dE_\nu$.



\chapter{The solar evidence}\label{sun}

\begin{figure}[h]
$$\includegraphics[width=7cm]{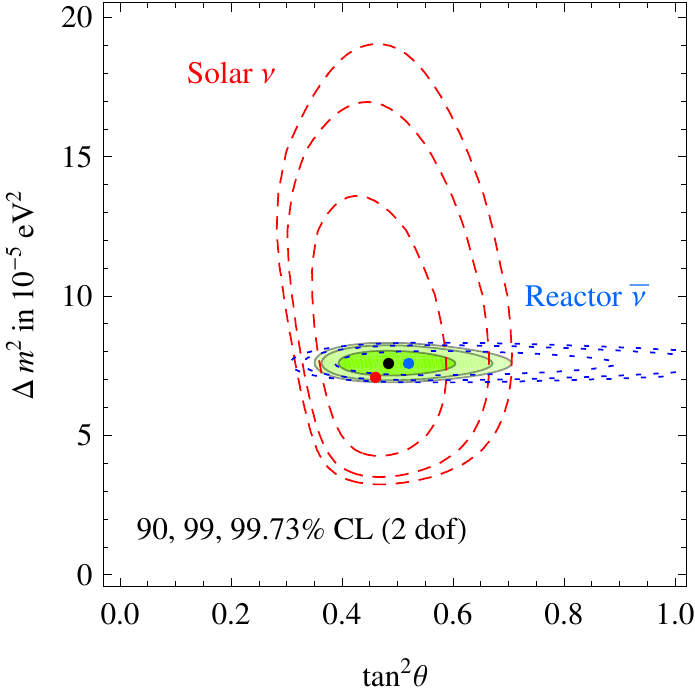}$$
\vspace{-5mm}
\caption[Fit of solar oscillation parameters]{\label{fig:sunfit} {\bf Solar oscillation parameters}.
\em Best-fit regions at $90,~99$ and $99.73\%$ {\rm CL} obtained fitting
solar $\nu$ data (red dashed contours);
reactor $\bar{\nu}$ data that do not distinguish $\theta$ from $\pi/2-\theta$ (blue dotted contours);
all data (shaded region).
Dots indicate the best fit points.
}
\end{figure}

Disappearance of solar $\nu_e$ (and later appearance of $\nu_{\mu,\tau}$) 
gave the first signal of a neutrino anomaly,
 that was therefore named `solar anomaly'.
Oscillations predicted that few different clean experimental signals
could have been detectable, depending on the actual value of the oscillation parameters.
In order to find the true one, all signals have been searched:
therefore we have a large amount of data.
Fits of solar data correctly indicate
that the best-fit solution of the solar neutrino anomaly has
a large mixing angle $\theta_{\rm sun}$
and $\Delta m^2_{\rm sun} \sim 10^{-4}\eV^2$ (`LMA' solution)
but,  as shown in fig.\fig{sunfit}, do not fully esclude other disfavored solutions
with much smaller $\Delta m^2$.
Finally, in 2002 the KamLAND experiment confirmed LMA
oscillations by 
discovering disappearance of $\bar{\nu}_e$ generated by nuclear reactors.
Fig.\fig{sunfit} shows a fit of reactor data, 
a global fit of all solar $\nu$ and reactor $\bar{\nu}$ data.
Reactor data can be understood in a simpler way and in the future should give
the best measurements of $\Delta m^2_{\rm sun}$,
and maybe  of  $\theta_{\rm sun}$.
Therefore we start (contrarily to historical development)
 describing reactor data, and we later discuss solar data.

\section{KamLAND}\label{KamLAND}
\begin{figure}
$$\includegraphics[width=7cm,height=7cm]{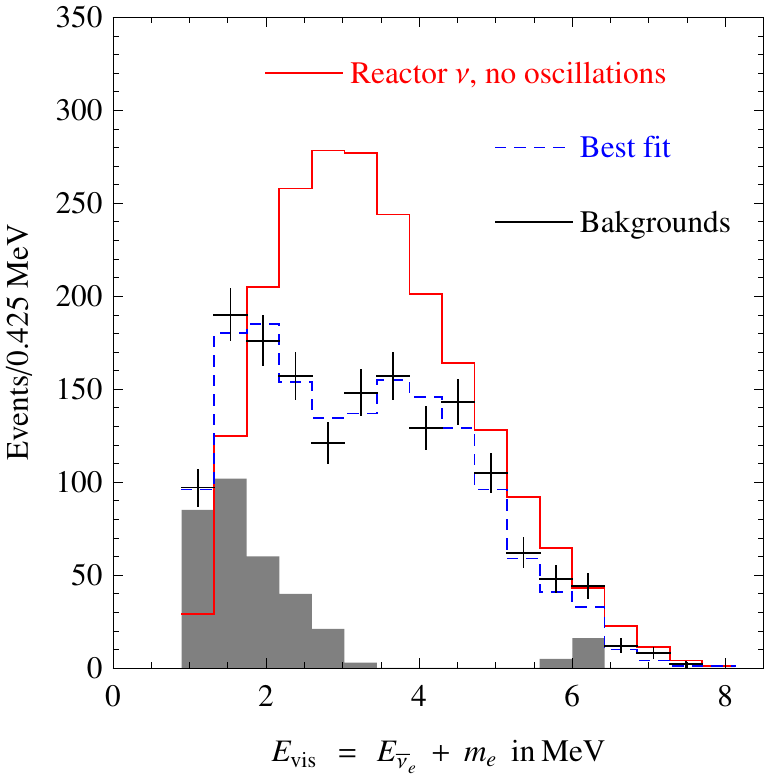}\qquad
\includegraphics[width=7cm,height=7cm]{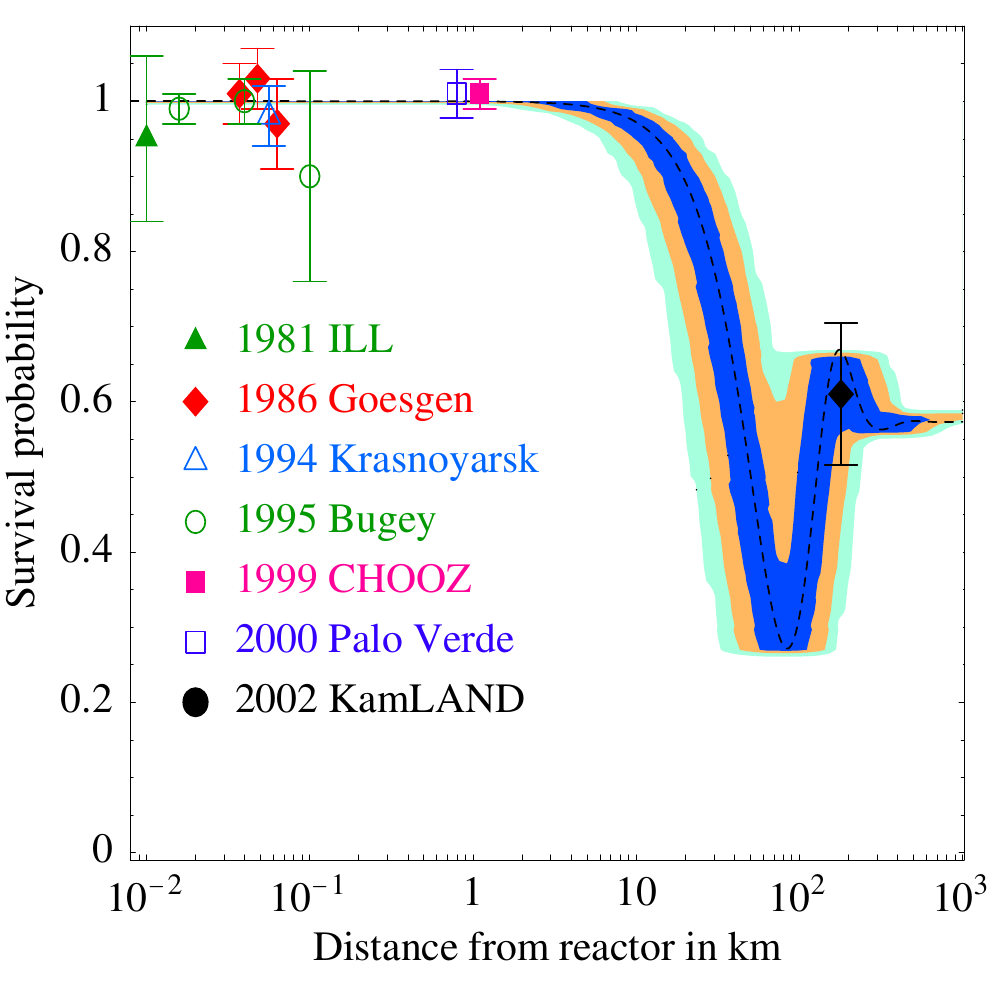}$$
\caption[KamLAND]{\label{fig:KamLAND} {\bf KamLAND}. \em Fig.\fig{KamLAND}a:
the $E_{\rm vis}=E_{\bar{\nu}_e}+m_e$
energy spectrum measured by KamLAND.
Fig.\fig{KamLAND}b: history of reactor experiments and
reduction in the reactor $\bar\nu_e$ flux
as predicted at $1,2,3\sigma$ by a global oscillation fit of solar data.}
\end{figure}

\index{KamLAND}
We presented reactor experiments in section~\ref{reactor}. 
KamLAND~\cite{KamLAND} is a  \v{C}erenkov  scintillator composed by
1 kton of a liquid scintillator (the number of protons,
$8.6~10^{31}$, is about 200 times larger than in CHOOZ)
contained in a spherical balloon surrounded 
by inert oil that shields external radiation.
KamLAND detects $\bar\nu_e$ emitted by terrestrial (mainly japanese) reactors 
using the $\bar\nu_e p \to e^+ n$ reaction.
The detector can see both the positron and the $2.2\MeV$ $\gamma$ ray
from neutron capture on proton.
By requiring their delayed coincidence and being located underground
and having achieved sufficient radio-purity, 
KamLAND reactor data are almost background-free.
As illustrated in fig.\fig{KamLAND}a,
KamLAND only analyzes  $\bar\nu_e$ events with $E_{\rm vis}=E_{e^+} + m_e>2.6\MeV$
(i.e. $E_\nu > 3.4\MeV$) 
in order to avoid a poorly predictable background of  $\bar\nu_e$
generated by radioactive elements inside the earth~\cite{Geonu}
(geoneutrinos are here discussed at page~\pageref{geonu}).
Above this energy threshold 
KamLAND should detect, in absence of oscillations, 
about $500$ events per $\hbox{kton}\cdot\hbox{yr}$,
depending on operating conditions of reactors.
Thanks to previous reactor experiments
the unoscillated $\bar\nu_e$ flux is known with $\sim 3\%$ uncertainty.
This expectation can be checked only partially: 
it is not possible to turn off reactors or
built detectors close to each reactor.

The effect of oscillations is given by 
\beq P(\bar\nu_e\to\bar\nu_e)  = 1 - \sin^2 2\theta_{\rm sun} \sin^2\frac{\Delta m^2_{\rm sun} L}{2E_\nu}
\eeq
up to minor corrections due to earth matter effects and to the small
$\theta_{13}$ atmospheric mixing angle.

The KamLAND  efficiency is about $90\%$.
The latest 2004 data (2881 ton yr) showed,
in the signal region where only $276\pm 23$ background events are expected,
1609 events instead of the $2179\pm89$ signal events
expected in absence of oscillations.
An anomaly is observed at more than $5\sigma$.
As illustrated in fig.\fig{KamLAND}b, this is consistent
with previous reactor data and with expectations from solar data.
KamLAND receives $\bar\nu_e$ from 
many nuclear reactors
located at  different distances.
Most reactors are at $L\approx $180 km,
because KamLAND is located in the Kamioka mine
in a central region of Japan,
while, in absence of big rivers, japanese reactors are build around the coast.

\medskip

More importantly, KamLAND data allow to test if the $\bar\nu_e$ survival
probability depends on the neutrino energy as predicted
by oscillations.
In fact, KamLAND can measure the positron energy with a
$\sigma_E/E = 6.5\%/\sqrt{E/\MeV}$ error.
The neutrino energy is directly linked
to $E_{e^+}$ as $E_{\bar \nu}  \approx E_{e^+} + m_n - m_p$.
KamLAND 2008 spectral data (fig.\fig{KamLAND}a) give a $5\sigma$ indication for oscillation dips:
the first one at $E_{\rm vis}\approx 7\MeV$ (where statistics is poor)
and the second one at $E_{\rm vis}\approx 4\MeV$.
Taking into account the average baseline $L\approx 180\km$,
this second dip occurs at $L/E_{{\bar\nu}_e}\approx 45\km/\MeV$.
This fixes $\Delta m^2_{\rm sun} = 6\pi  E_{{\bar\nu}_e}/L|_{\rm 2nd~dip}\approx 8~10^{-5}\eV^2$. 
The global fit of fig.\fig{sunfit} shows that $\Delta m^2_{\rm sun}$ is presently dominantly
fixed by KamLAND data, which precisely fixes
\beq \Delta m^2_{\rm sun} =(7.58\pm 0.21)~10^{-5}\eV^2,\qquad
\tan^2\theta_{\rm sun}=0.56^{+0.14}_{-0.09}\eeq
with other local minima disfavored at more than $4\sigma$.
Statistical and systematic errors are comparable.
The {\sc Borexino} experiment also observed a deficit in reactor $\bar\nu_e$~\cite{Borexino},
compatible with the more precise {\sc KamLand} result.

%



\section{Solar neutrinos}\index{Solar neutrinos}
How old is the earth?
Around the end of 19th century,
biologists and geologists (like Darwin) suggested that more than 300 Myr were necessary for
natural selection and erosion.
Physicists (like Kelvin)
showed that 
the sun can shine for $(GM^2/R)/(d^2 K_{\rm sun})\sim$ 30 Myr at most, 
using gravitational energy to emit the flux of energy that, at distance $d$ from the sun
 we receive at earth,
$K_{\rm sun} = 8.53\cdot 10^{11}$ MeV cm$^{-2}$ s$^{-1}$.
Theologians (like Lightfoot) believed that the earth was created on 
october 23, $-4004$, at nine o'clock in the morning.

\begin{figure}[t]
$$\includegraphics[width=15cm,height=6.5cm]{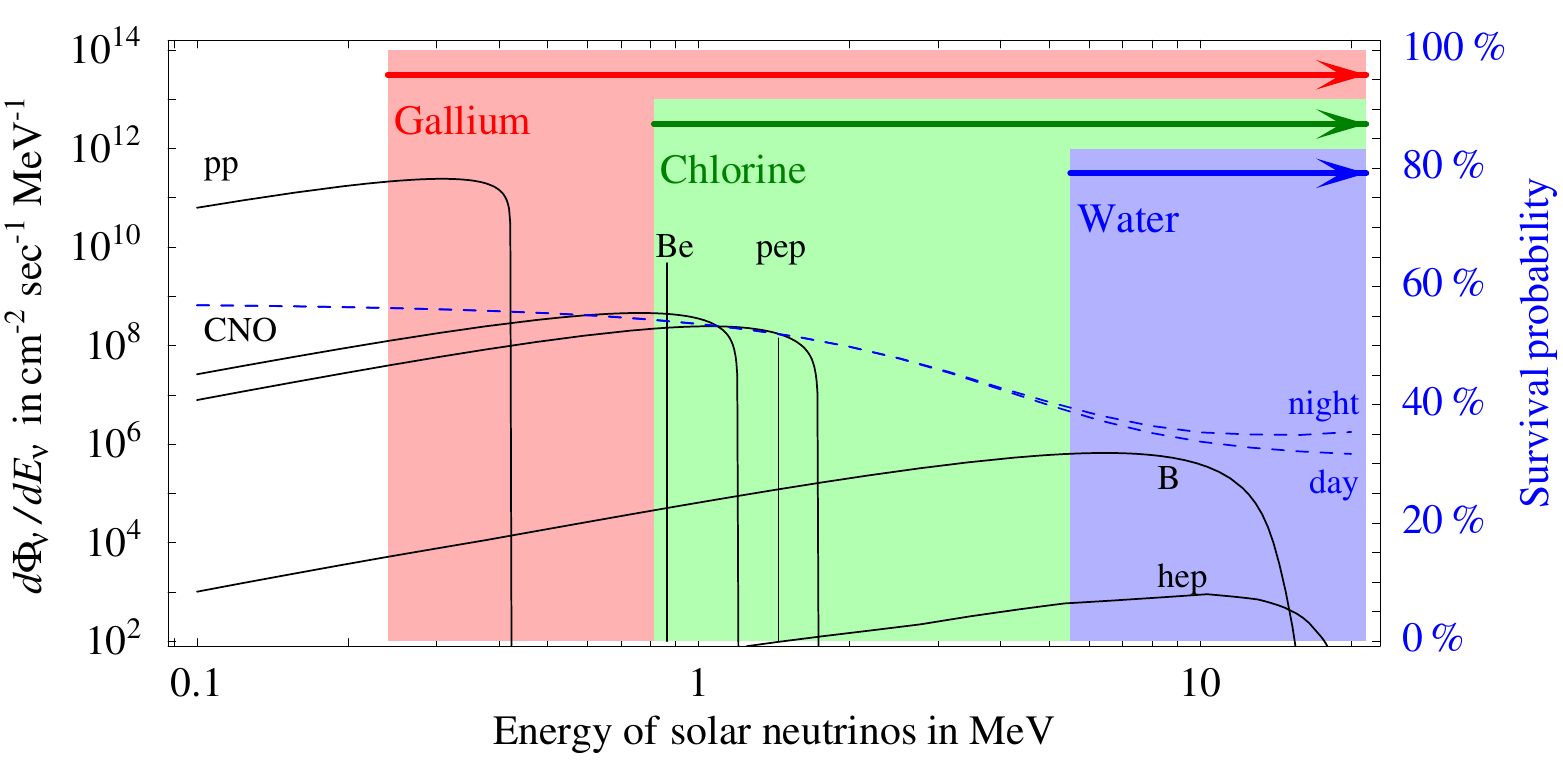}$$
\vspace{-9mm}
\caption[Predicted solar neutrino spectra]{\em The predicted unoscillated spectrum $d\Phi/dE_\nu$
of solar neutrinos, together with
the energy thresholds of the experiments performed so far
and with the best-fit oscillation survival probability $P_{ee}(E_\nu)$ (dashed line).
\label{fig:SpettroSolare}}
$$\includegraphics[width=14cm]{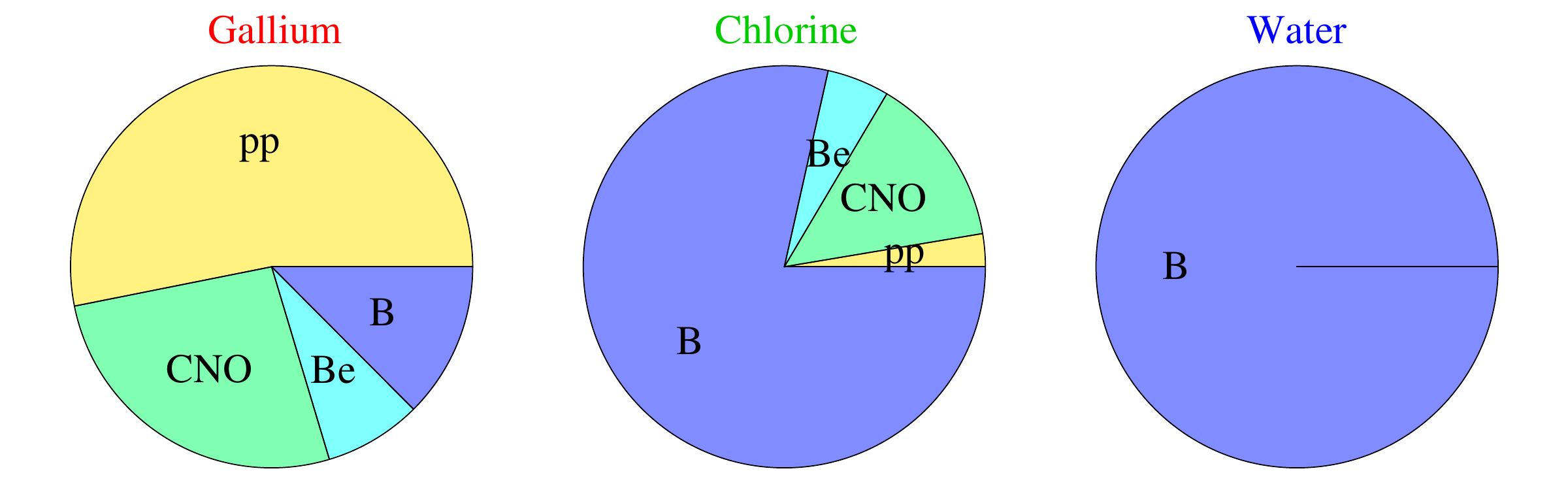}$$
\caption[Predicted solar neutrino rates]{\em The predicted fractional contributions to the neutrino rates 
of present experiments, assuming energy-independent oscillations.
\label{fig:Percentuali}} \end{figure}

\begin{figure}[t]
$$\includegraphics[width=10cm,height=6.5cm]{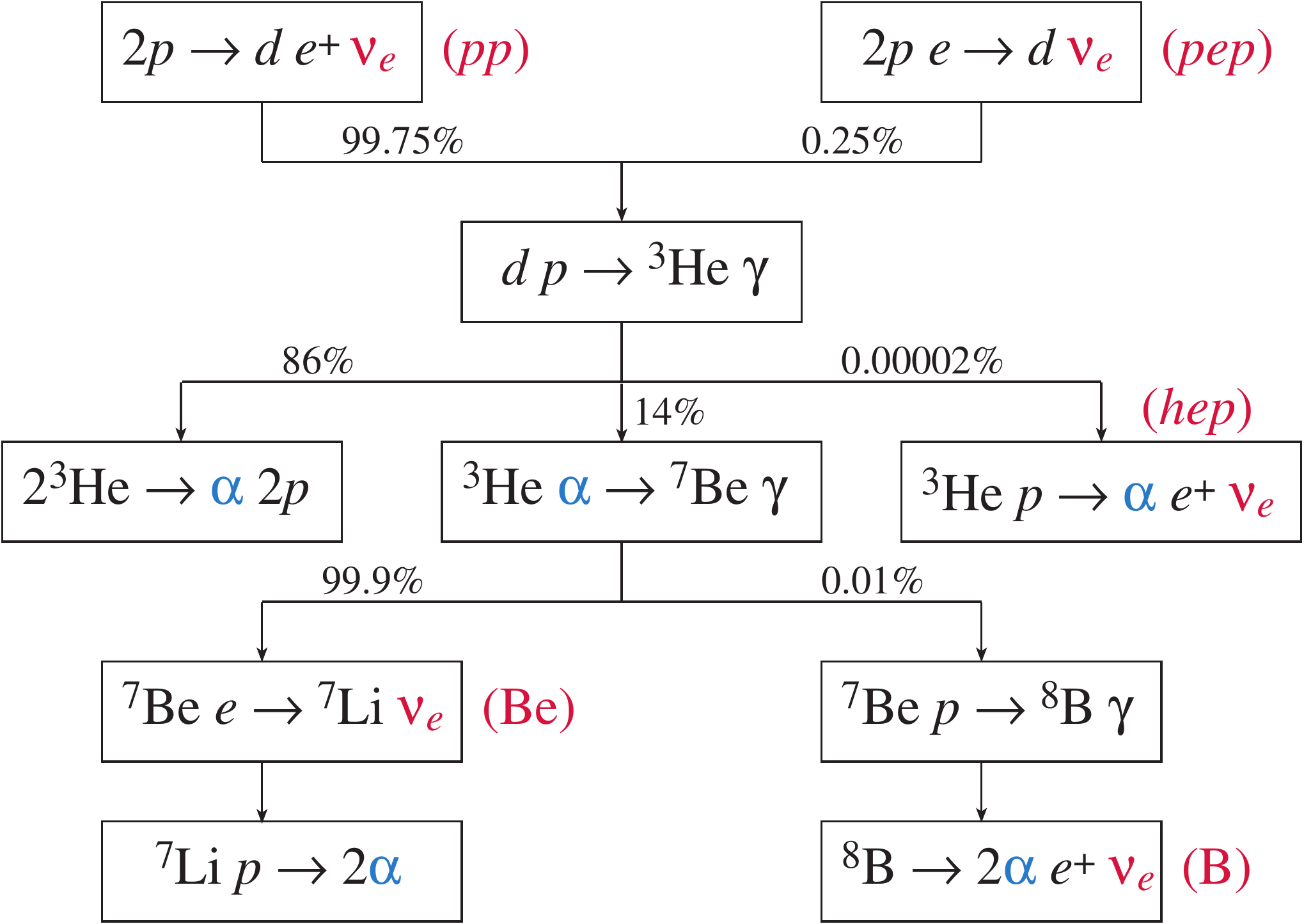}$$
\caption[Nuclear reactions inside the sun]{\em The $4p + 2e \to {}^4{\rm He} +2  \nu_e$ chain.\label{fig:CicloSolare}}
\end{figure}

Biologists and geologists were right. Physicists (like
Aston, Eddington, Gamow, Bethe) later found the missing pieces of the puzzle:
the sun shines thanks to nuclear fusion.
Around the center of the sun, energy and neutrinos are produced
essentially through the ${}^4{\rm He}$ reaction
\begin{equation}
 4p + 2e \to {}^4{\rm He} +2  \nu_e  \qquad (Q=26.7\MeV).
\label{eq:stella}
\end{equation}
The predicted $\nu_e$ spectrum~\cite{BP}, in absence of oscillations,  
is shown in fig.\fig{SpettroSolare}.
The reason of such a complex spectrum is that
the overall reaction\eq{stella}, having 6 particles in the initial state, proceeds in a sequence of steps,
following different routes.
The main routes are summarized in fig.\fig{CicloSolare},
and give rise to five main types of neutrinos.

\begin{table}[!t]
\begin{center}\begin{tabular}{crlcccccc} 
{Flux}&\multicolumn{2}{c}{Reaction}& $ \Phi_\nu$& $\langle E_\nu\rangle$&$E_\nu^{\rm max}$&$Q$ & Cl & Ga\\
&&& $10^{10}/\cm^2\s$& MeV& MeV&  MeV & SNU & SNU\\
\hline
$ pp$&$p  p\!\!\!$ & $\to {\rm ^2H}~\bar{e} ~ \nu_e$& $ 6.0\pm 1\%$ &0.267&0.423&13.10 & 0 & 70.3\\
$ pep$&$p  e p \!\!\!$ & $\to {\rm ^2H}~ \nu_e$&$ 0.014\pm 1\%$ &
\multicolumn{2}{c}{1.445}&11.92 & 0.22 & 2.8\\
$ hep$&${\rm ^3He} ~p \!\!\!$ & $\to {\rm ^4He} ~ \bar{e} ~\nu_e$&$\sim 10^{-6}$ &9.628&18.8&3.737&0.04 & 0.1\\
Be&${\rm ^7Be}  ~e \!\!\!$ & $\to {\rm ^7Li} ~\nu_e$&$0.477 \pm 9\%$ &0.814&0.863&12.60&1.15 &34.2\\
B&${\rm ^8B} \!\!\!$ & $\to {\rm ^8Be} ~\bar{e}~ \nu_e$&$ 0.00050\pm 20\%$ &6.735&16.3&6.630&5.76 & 12.1\\
N&${\rm ^{13}N} \!\!\!$ & $\to {\rm ^{13}C} ~ \bar{e} ~\nu_e$&$ 0.033\pm 20\%$ &0.706&1.20&3.456&0.05 & 2.0\\
O&${\rm ^{15}O} \!\!\!$ & $\to {\rm ^{15}N}~ \bar{e} ~ \nu_e$&$0.026 \pm 20\%$ &0.996&1.73&21.57&0.18 & 2.9 \\
\hline
all & $4p~2e\!\!\!$ & $\to {}^4{\rm He}~2\nu_e$ & --- & --- &---& 26.7 & $7.4\pm 1.3$ & $124\pm 9$
\end{tabular}
\caption[Predicted total solar neutrino fluxes]{\label{tab:flussinu} {\bf Predicted
solar neutrino fluxes} \em
in absence of oscillations.
$\Phi$ is the total $\nu_e$ flux at earth.
$\md{E_\nu}$ ($E_\nu^{\rm max}$) is the mean (maximal) neutrino energy.
$Q$ is the energy released in the reaction.
The last two columns show the contributions to the total rate
measured in Cl and Ga experiments.
$89.7\%$ ($10.3\%$) of the $^7{\rm Be}$ neutrinos are produced in
ground state (excited state) capture and
have energy $0.8631\MeV$ ($0.386\MeV$).}
\end{center}
\end{table}


\begin{enumerate}
\item 
$pp$ neutrinos, generated in the first step $pp\to d e^+ \nu_e$,
 have a large and precisely predicted flux.
However their maximal energy is only $0.42\MeV \sim 2m_p -m_d - m_e$, 
so that it is difficult to detect $pp$ neutrinos.
Since the average solar neutrino energy
is much smaller than $Q$,
most of the energy is carried out from the sun by photons,
that after random-walking in the solar interior for about $10^4$ years,
carry to the earth the well known flux of energy,  $K_{\rm sun}$.
Therefore, the present total neutrino luminosity of the sun is
$\Phi\sim 2 K_{\rm sun}/Q \sim6.5\cdot 10^{10}/\hbox{cm}^2\hbox{s}$.
The precise solar model prediction is 
$\Phi_{pp} = 0.91 \cdot 2 K_{\rm sun}/Q$.

\item $pep$ neutrinos, generated in $pep\to d\nu_e$ collisions,
have a relatively small flux and low energy, 
$E\approx  2m_p +m_e-m_d=1.445\MeV$, 
but are not totally negligible.

The $pep$ and Be neutrinos are almost monochromatic,
because generated by electrons colliding on heavy particles
at temperatures $T\ll m_e$.

\item Be neutrinos have a relatively well predicted and large flux and relatively high
energy and are important for present experiments,
as shown in fig.\fig{Percentuali}.
They are of great interest for future experiments,
mainly because they are almost monochromatic,
$E_{\rm Be} \approx m_{^7\rm Be} - m_{^7\rm Li} - m_e \approx 0.863\MeV$,
allowing interesting measurements (see below).
Thermal motions produce a small asymmetric broadening of the line
(FWHM = $1.6\keV$).

\item B neutrinos are a small fraction of all solar neutrinos,
but can have a relatively large energy, up to $E \circa{<} m_{^8{\rm B}} - 2 m_{\alpha}$.
SK and SNO detect neutrinos with energy larger than about $6\MeV$,
and are consequently sensitive only to B and $hep$ neutrinos.

\item  $hep$ neutrinos have the highest energy,
but are too rare for having significant effects,
given the accuracy of present experiments.

\end{enumerate}
Solar models predict that in stars heavier than the sun 
most of the energy is produced by
a different chain of reactions, named CNO cycle.
In the sun, CNO neutrinos give a minor additional component of solar neutrinos
(fig.\fig{SpettroSolare}).

The energy spectra of the single components 
are essentially determined by kinematics and
do negligibly depend on details of the solar interior.
{\em Solar models play a crucial r\^{o}le in predicting their total fluxes}.
The flux of B neutrinos, produced by the terminal reaction of the chain,
strongly depends on the the solar temperature and cannot be accurately predicted.
Today the dominant uncertainties no longer come from solar parameters,
but from nuclear physics inputs~\cite{sigmasun}:   $S_{17}$
(that parameterizes the $^7{\rm Be}\,p\to {}^8{\rm B}\,\gamma$
 cross-section in fig.\fig{CicloSolare}, thereby fixing the total flux of B neutrinos.
The $^7{\rm Be}\,e$ cross section giving rise to Be neutrinos is larger
and precisely known because
not suppressed by the Coulomb barrier~\cite{S17})
and $S_{34}$
(that parameterizes the $^3{\rm He}\,\alpha\to ^7{\rm Be}\,\gamma$
 cross-section in fig.\fig{CicloSolare}, thereby fixing the total flux of B plus Be neutrinos).
This explains the uncertainties on the fluxes predicted in table~\ref{tab:flussinu}
(fits of solar data take into account correlations among them, as discussed in subsection~\ref{GlobalSunFits}).
Recent measurements of the metallicity of the solar surface found values
different from what predicted by solar models on the basis of helio-seismological data
; this has minor impact on the solar neutrino issue.

\section{Solar neutrino experiments}
We now discuss the main characteristics and results of all
solar neutrino experiments, listed in table\tab{sunexps}.

\subsection{Chlorine}
R.\ Davis leaded the first experiment which detected solar neutrinos and
saw the first hint of a solar neutrino anomaly~\cite{Cl},
using a radiochemical technique suggested by Pontecorvo.
Solar $\nu_e$ induce the reaction
$\nu_e  {}^{37}\hbox{Cl}\to{}^{37}\hbox{Ar} ~ e$, producing
the isotope $^{37}\mathrm{Ar} $.
Such isotopes were separated using their different chemical behavior;
by observing their later decays back to ${}^{37}\hbox{Cl}$ (the half-life is 35 days)
it was possible to count a few atoms in a tank of few hundred tons.

The Cl reaction was employed because its cross section is precisely computable
and its energy threshold, $E_{\nu_e}>0.814\MeV$, is low enough
that the Cl experiment is sensitive to Boron $\nu_e$
with some minor contribution from lower energy solar $\nu_e$ (see fig.\fig{Percentuali} or table\tab{flussinu}).

The measured Cl rate was found to be about 3 times lower than the predicted value,
suggesting an intriguing discrepancy between a pioneering experiment and 
supposedly accurate enough solar models.
In 1972 Pontecorvo commented
``{\em It starts to be really interesting! It would be nice if all this will end with something unexpected from the point of view of particle physics.  Unfortunately, it will not be easy to demonstrate this, even if nature works that way}''.
About 15 years were necessary for a second experiment, and 30 for finally
establishing solar oscillations.

\subsection{Gallium experiments}
The next radiochemical  experiments, SAGE and {\sc Gallex}/GNO~\cite{Ga}
(respectively located in the Baksan and Gran Sasso underground laboratories in Soviet Union and Italy)
employed the reaction
$\nu_e  {}^{71}\mathrm{Ga} \to {}^{71}\mathrm{Ge}  ~ e$
which has the lowest threshold reached so far, $E_{\nu_e}>0.233\MeV$.
As a consequence more than half of the $\nu_e$-induced events is generated
by $pp$ neutrinos, see fig.\fig{Percentuali}. 
Their total flux can be reliably approximated from the solar luminosity
and can be predicted by solar models with $1\%$ error.
After a half-live of $16.5$ days the inverse $\beta$-decay of
${}^{71}\mathrm{Ge} $ produces observable Auger electrons and $X$-rays
with the typical $L$-peak and $K$-peak energy distributions,
giving two different signals used to infer the flux of solar $\nu_e$.
The rate  measured in Gallium experiments is about 2 times lower than the predicted value.
This result made harder to believe that the solar neutrino anomaly was due to
wrong solar model predictions.

The reliability of the Gallium technique was tested using an artificial neutrino source.
Gallium  experiments improved with time,
and both their central values decreased
by about two standard deviations
with respect to the first data.
The final result, averaged over all experiments is
\beq R^{\rm Ga} = (67.6\pm 3.7)\,{\rm SNU}.\eeq


\begin{table}[t]\small
  \centering 
\begin{tabular}{|c|c|c|c|c|c|c|}
\hline Experiment&Reaction&$E_{\rm th}$ (MeV)&\hbox{$\nu$ fluxes}&Running time&
$\displaystyle{ R^{\mathrm{exp}} }$ &${ R^{\mathrm{BP00}} }$\\
\hline\hline\color{verdes}
Homestake&\color{verdes}$\nu_e  {}^{37}\hbox{Cl}\to {}^{37}\mathrm{Ar} ~ e$&$ \color{verdes}0.814 $
&\color{verdes}mainly $^8$B&$\color{verdes}1970-1994$&\color{verdes}$ 2.56 \pm 0.23 $
& \color{verdes} $7.6\pm1.3$\\
\hline\color{red}
SAGE&&&&$\color{red}1990-2003$&$ \color{red}69.1\pm5.7 $&\\
\cline{1-1}
\cline{5-6}
\color{red}GALLEX&$
\color{red}\nu_e  {}^{71}\mathrm{Ga} \to {}^{71}\mathrm{Ge}  ~ e
$&$ \color{red}0.233 $& \color{red}all &\color{red}$1991-1997$&
$ \color{red}77.5\pm7.7 $&  \color{red}$128\pm9$\\
\cline{1-1}
\cline{5-6}
\color{red}GNO&&
&
&\color{red}$1998-2003$&\color{red}$ 62.9\pm 5.9 $ &
\\ \hline 
Borexino && 0.862 & $^7$Be & $2007 - 2008$ &$49\pm 5$ &$74\pm 4$\\ 
\cline{3-4}
\cline{6-7}
&& 3.0 & & $2007 - 2009$ &$2.4\pm0.4$ &\\ 
\cline{1-1}
\cline{3-3}
\cline{5-6}
\color{blus}
Kamiokande
&{\color{blus}$ \nu  e \to \nu  e $}&$\color{blus} 6.75 $&&
$\color{blus}1987-1995$
&
\color{blus}$2.80 \pm 0.36$ &
\\
\cline{1-1}
\cline{3-3}
\cline{5-6}
\color{blus}SK&
&\color{blus}$ 4.75 $&&\color{blus}$1996-2001$&
\color{blus}$ 2.35\pm0.06 $&
\\
\cline{1-1}\cline{3-3}\cline{5-6}
&&$ 5.2 $&${^8\mathrm{B}}$, $hep$&&$ 2.31 \pm 0.21 $&$5.05\pm0.9$\\
\cline{2-3}\cline{6-6}
SNO&$ \nu_e  d \to p  p  e $&$ 6.9 $&&$1999-2003$&$ 1.67\pm0.08 $ & 
\\
\cline{2-3}\cline{6-6}
&$ \nu  d \to p  n  \nu $&$ 2.2 $&&&$ 5.17 \pm 0.38 $&
\\
\hline
\end{tabular}
  \caption[Results of solar neutrino experiments]{\em Solar neutrino experiments.
Rates measured by radiochemical experiments are expressed in  
$\hbox{\rm SNU}\equiv 10^{-36}\,\hbox{interactions per target atom and per second}$,
while SK and SNO report neutrino fluxes in $10^6 \,{\rm cm}^{-2} {\rm s}^{-1}$,
and Borexino gives the $^7{\rm Be}$ rate in interactions per day per 100 ton.
\label{tab:sunexps}}
\end{table}

\subsection{Kamiokande and SuperKamiokande}\index{SuperKamiokande}
SK~\cite{SK} is a 50 kton W\v{C} detector
with a 22.5 kton fiducial volume.
The experiment stopped when an accident damaged its $11146$ photomultipliers.
During its 1496 live days, between 1996 and 2001, 
SK-I collected about 20000 solar neutrinos.
Solar neutrinos are detected via scattering on electrons
 $\nu_{e,\mu,\tau} e\to \nu_{e,\mu,\tau} e$.
SK can measure the kinetic energy $T_e$ and the direction of the scattered electron.
The dominant backgrounds to the solar neutrino signal are $^{222}$Rn in the
water, external $\gamma$ rays and muon-induced spallation products.
As a consequence only data above the cut $T_e > 5.5\MeV$ are used.
Since $T_e\gg m_e$ the electrons are scattered roughly along the direction of the solar neutrino:
before Kamiokande radiochemical experiments could 
count neutrinos but could not verify that they come from the sun.
The direction of the scattered electron allows SK
to discriminate solar neutrino events from a comparable
number of background events.

Although SK cannot distinguish $\nu_e$ from $\nu_{\mu,\tau}$,
all active neutrinos contribute to the total SK solar neutrino rate.
Oscillations suppress the Boron $\nu_e$ flux $\Phi_{\rm B}$
generating fluxes of $\nu_{\mu,\tau}$ and possibly of sterile neutrinos.
Their effect  can be parameterized as:
\beq\label{eq:PeeEta}
\Phi_{\nu_e} = \Phi_{\rm B} P_{ee},\qquad
\Phi_{\nu_{\mu,\tau}}=\Phi_{\rm B}(1-P_{ee})(1-\eta_{\rm s}),\qquad
\Phi_{\nu_{\rm s}} = \Phi_{\rm B}(1-P_{ee})\eta_{\rm s},\eeq
Neglecting the possible dependence of 
$P_{ee}$ and $\eta_{\rm s}$ 
on the neutrino energy,
the SK rate gets suppressed by
$ P_{ee} + 0.155 (1-P_{ee})(1-\eta_{\rm s})$,
where $0.155$ is the value of $\sigma(\nu_{\mu,\tau}e)/\sigma(\nu_e e)$ at
$E_\nu\sim 10\MeV$ (see section~\ref{detecting}).
The measured rate reported in table\tab{sunexps} is not the main SK result.
SK could search for various signals possibly generated by oscillations.
Since none of them was found significant ranges of 
neutrino masses and mixings were excluded,
and global analyses of solar data started favoring the active-only
large mixing angle solution as the true one.
\begin{itemize}
\item The {\em electron energy spectrum} is consistent with an
energy-independent reduction in the flux of B neutrinos.
This excluded SMA and part of VO solutions, which predicted an energy-dependent
survival probability $P_{ee}$. LMA oscillations predict a small slope
comparable to the present experimental accuracy (see fig.\fig{SpettroSolare}).
\end{itemize}
Furthermore, SK is a real-time experiment: this allowed to search for
seasonal and day/night variations in the neutrino rate.
\begin{itemize}
\item Vacuum oscillations with $\Delta m^2\sim 10^{-10}\eV^2$ have a wave-length comparable
to the distance $d$ between the sun and the earth, and therefore
should produce an anomalous seasonal variation of the solar neutrino flux
(beyond the observed trivial variation due to the geometrical $1/d^2$ flux factor).
The SK rate is consistent with no anomalous seasonal variation.
Together with the absence of spectral distortions, this disfavored VO.

\item During the night neutrinos cross the earth before being detected by SK.
Earth-matter corrections significantly affect neutrinos with energy
around the resonant energy of eq.\eq{Eres}:
$$E_\nu^{\rm res} = \frac{\Delta m^2}{2\sqrt{2}G_{\rm F}
N_e^\oplus} \approx 200\MeV \frac{\Delta m^2}{10^{-4}\eV^2}$$
where $N_e^\oplus$ is the electron density of the earth mantle.
The absence of a large day/night asymmetry at $E_\nu\sim 10\MeV$
\beq\label{eq:ADNSK}
A_{\rm DN}^{\rm SK} \equiv 2\frac{\hbox{night rate}-\hbox{day rate}}{\hbox{night rate}+\hbox{day rate}}= 1.8\pm1.6_{\rm stat}\pm 1.3_{\rm syst}\,\%\eeq
excluded oscillations with $\Delta m^2_{\rm sun} \sim 10^{-5\div 7}\eV^2$ and $\theta_{\rm sun}\sim 1$.
LMA predicts a $\sim 2\%$ day/night asymmetry, with a
specific energy and zenith-angle dependence. 
\end{itemize}
Finally, SK puts interesting bounds on exotic processes,
such as time variations of the solar neutrino rate~\cite{Pee(t)} and
conversion into $\bar{\nu}_e$.
The KamLAND 
bound $P(\nu_e\to\bar{\nu}_e) < 2.8~10^{-4}$ at $90\%$ CL~\cite{KamLAND} is stronger than the
SK bound, $P(\nu_e\to\bar{\nu}_e) < 0.8~10^{-2}$ at $90\%$ CL~\cite{SK}.

\begin{figure}[t]
$$\includegraphics[width=8cm,height=5cm]{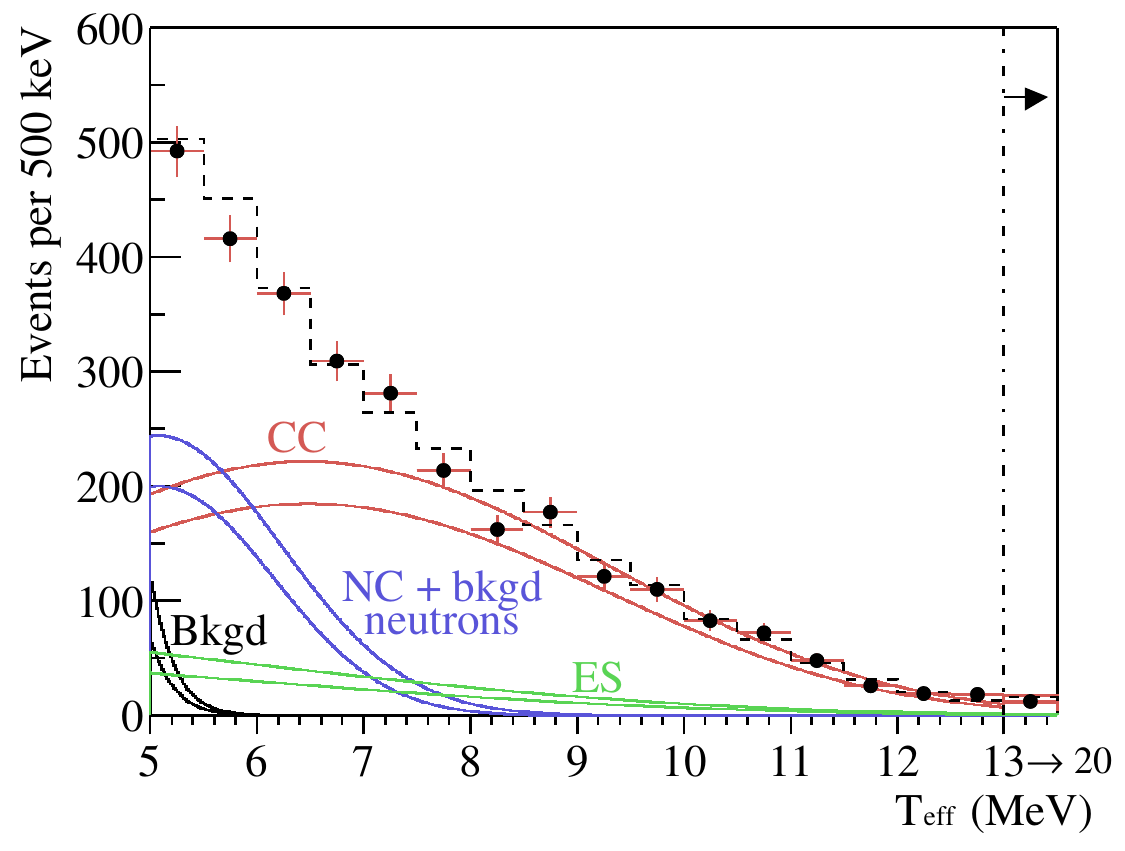}
\raisebox{3mm}{\includegraphics[width=8cm,height=5cm]{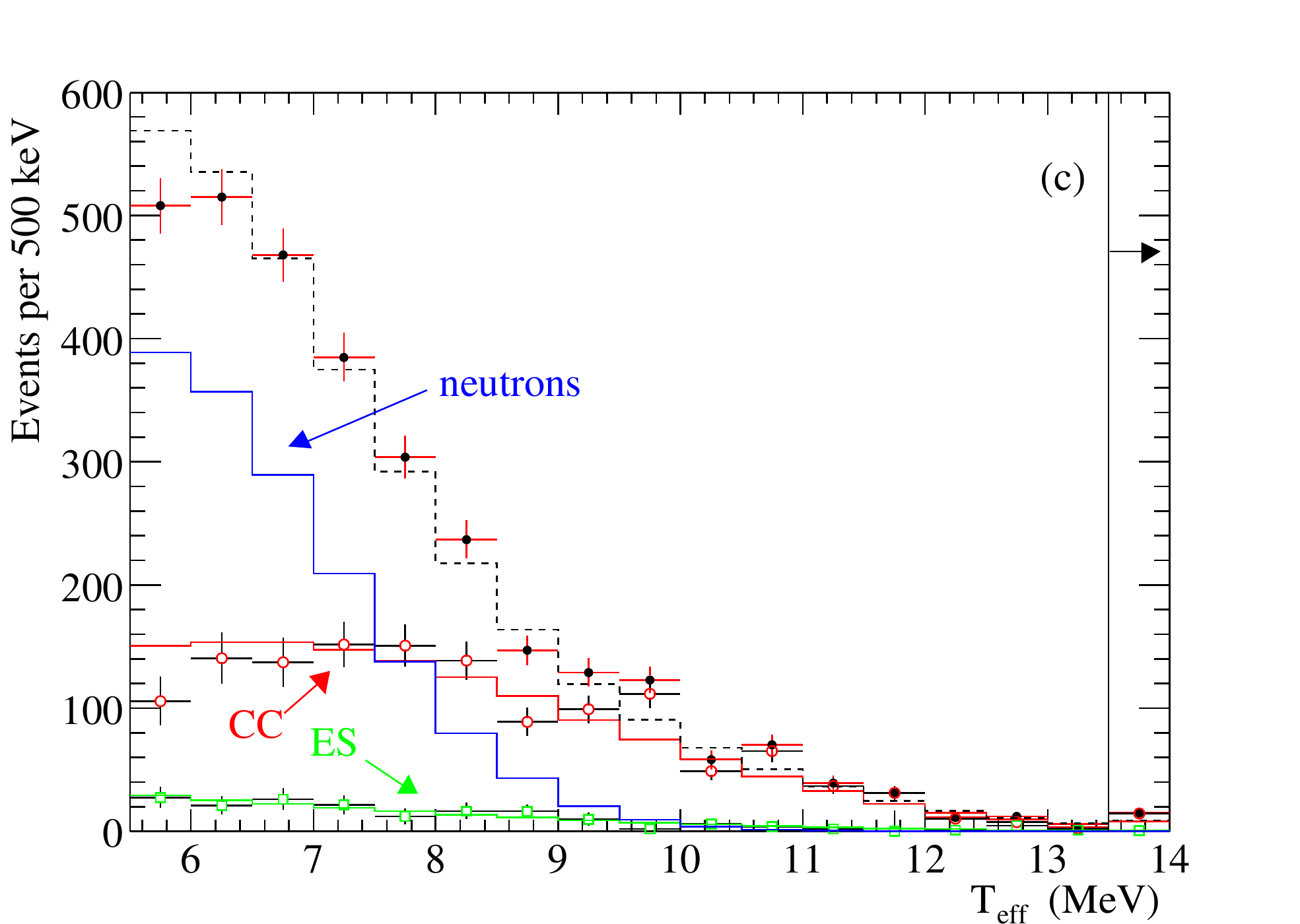}}$$
\caption[SNO]{\label{fig:SNOdata}{\bf SNO}. \em The left (right) figure shows the energy spectra
in the second (third) phase of SNO, decomposed according to their components.
From~\cite{SNO}.}
\end{figure}

\subsection{SNO}\index{SNO}\label{SNO}
SNO~\cite{SNO} is a real-time W\v{C} experiment similar to SK and smaller than it.
The crucial improvement is that SNO employs 1 kton of {\em  salt heavy water}
rather than water, so that neutrinos can interact in different ways,
allowing to measure separately the $\nu_e$ and $\nu_{\mu,\tau}$ fluxes:
SNO is the first solar neutrino {\em appearance} experiment.
\begin{itemize}
\item[{\color{verdes}ES}] Like in SK, $\nu_{e,\mu,\tau}$ can be detected (but not distinguished)
thanks to CC and NC scattering on electrons: $\nu_{e,\mu,\tau} e\to \nu_{e,\mu,\tau} e$.
This allows to measure $\Phi_{\nu_e} + 0.155 \Phi_{\nu_{\mu,\tau}}$
and contrast it with the SSM prediction:
\end{itemize}
SNO events are mostly produced by two interactions not present in SK:
\begin{itemize}
\item[{\color{rossos}CC}] Only $\nu_e$ can produce  $\nu_e d\to pp e$.
SNO sees the scattered electron and measures its direction  and energy.

\item[{\color{blu}NC}] All active neutrinos can break deuterons:
$\nu_{e,\mu,\tau} d\to \nu_{e,\mu,\tau} pn$.
The cross section is equal for all flavours
and has a $E_\nu > 2.2\MeV$ threshold.
About one third of the neutrons are captured by deuterons and give
a $6.25\MeV$ $\gamma$ ray: observing the photo-peak
SNO can detect $n$ with $\sim 15\%$ efficiency.
Adding salt allowed to tag the $n$ with enhanced $\sim 45\%$ efficiency, because
neutron capture by $^{35}$Cl produces multiple $\gamma$ rays.
\end{itemize}
Several handles allow to discriminate ES from CC from NC events.
ES events are not much interesting and can be subtracted since,
unlike CC and NC events,
ES events  are forward peaked.
CC/NC discrimination was performed in different ways before (phase 2)
and after (phase 3) adding salt to heavy water.

As illustrated in fig.\fig{SNOdata}a, in phase 2
SNO mostly discriminated
CC from NC events from their energy spectra:
NC events produce a $\gamma$ ray of known average energy
(almost always smaller than $9\MeV$).
The spectrum of CC events can be computed knowing 
the spectrum of $^8$B neutrinos (oscillations only give a minor distortion).
Phase 2 SNO data imply
\beq
\Phi_{\nu_e}=
1.76\pm 0.06\hbox{(stat)}\pm0.09\hbox{(syst}) ~\frac{10^6}{\cm^2\s} \qquad
\Phi_{\nu_{e,\mu,\tau}}=
5.09\pm 0.44\hbox{(stat)}\pm0.46\hbox{(syst})~\frac{10^6}{\cm^2\s}\eeq
(uncertainties are mildly anti-correlated).
The total flux agrees with the value predicted by solar models,
and the reduced $\nu_e$ flux gives a $5\sigma$ evidence for $\nu_e\to\nu_{\mu,\tau}$
transitions.

After adding salt,
SNO could statistically discriminate events 
from the pattern of photomultiplier-tube hits:
NC events produce  multiple $\gamma$ rays  
and consequently a more isotropic \v{C}herenkov
light than the single $e$ produced in CC and ES scatterings.
Phase 3 SNO data imply
\beq
\Phi_{\nu_e}=
1.59\pm 0.08\hbox{(stat)}\pm0.08\hbox{(syst}) ~\frac{10^6}{\cm^2\s}\qquad
\Phi_{\nu_{e,\mu,\tau}}=
5.21\pm 0.27\hbox{(stat)}\pm0.38\hbox{(syst})~\frac{10^6}{\cm^2\s}\eeq
giving a more accurate and independent measurement of total $\nu_{e,\mu,\tau}$ flux.
SNO finds  $\Phi_{\nu_e}/\Phi_{\nu_{e,\mu,\tau}}<1/2$,
that can be explained by oscillations enhanced by matter effects.

In a successive phase, by adding $^3{\rm He}$ SNO will be able of tagging NC events 
on an event-by-event basis by detecting neutrons via the scattering $n~^3{\rm He}\to p~^3{\rm H}$:
proportional counters allow to see both $p$ and $^3{\rm H}$.

Like SK, SNO can also search for energy-dependent or time-dependent effects.
The day/night asymmetry of the $\nu_e$ flux is found to be
\beq\label{eq:ADNCC} A_{\rm DN}^{\rm CC} = 7.0\pm5.1\%\eeq
assuming zero day/night asymmetry in the $\nu_{e,\mu,\tau}$ flux
(the direct measurement of this asymmetry is consistent with zero up to a $\sim15\%$ uncertainty).

Since oscillations among active neutrinos do not affect the total $\nu_{e,\mu,\tau}$ flux,
enhancing the NC rate makes  the overall SNO
energy spectrum of fig.\fig{SNOdata}b
less sensitive to spectral distortions. 
At the moment SK is more sensitive than SNO to spectral distortions:
SNO has much less statistics than SK, only partially compensated by
the fact that the energy of electrons scattered in CC reactions is
more strongly correlated with the parent neutrino energy than in ES
(because $m_e \ll E_\nu\ll m_d$).

\section{Borexino}
The Borexino~\cite{Borexino} experiment detects
$\nu_e  e\to \nu_e e$ scattering events in real time,
looking at the scintillation light produced in the detector.
The energy spectrum of the recoil electron can be measured with a lower background than previous bigger experiments,
and consequently down to lower energies.
Borexino released data, confirming previous results for the $^8$B rate, and
finding that the the $^7{\rm Be}$ line at $0.862\MeV$ gives
 \beq ( 49\pm 3_{\rm stat} \pm 4_{\rm sys})
 \,  \frac{\hbox{events}}{\hbox{day}\cdot\hbox{100 ton}}.\eeq
This is below the no-oscillation
expectation of $75\pm 4$ events, and consistent with the LMA oscillation
expectation of $49\pm 4$ events (the LOW solution, not yet definitively excluded by solar data alone,
predicts a central value $10\%$ lower than LMA oscillations).
Borexino aims at separately measuring the $pep$ and CNO neutrinos (if it will be possible to control
backgrounds), which have astrophysical interest.
Data about the day/night asymmetry and about seasonal variations have not yet been released.

\section{Implications of solar data}\label{GlobalSunFits}
\index{Oscillation!in the sun}
Oscillations can successfully explain all these data.
No other proposed interpretation can simultaneously explain
the solar $\nu$ data and the reactor $\bar\nu$ data.
Although in figures we show results of  global fits
(performed as described at the end of this section),
thanks to the recent clean SNO and KamLAND data,
main results can now be extracted from simple arguments,
that we discuss in the text.

\subsection{Which oscillation parameters?}
Atmospheric oscillations have a minor or negligible impact on solar neutrino and KamLAND data.
Indeed, $\nu_\mu\leftrightarrow \nu_\tau$ oscillations have no effect
(because  $\nu_\mu$ cannot be distinguished from $\nu_\tau$ when their energy is so
low that they only have NC interactions),
and $\nu_e$ are at most marginally involved in atmospheric oscillations
(the mixing angle $\theta_{13}$ is small, and it is not enhanced by matter effects 
because $G_{\rm F} N_e^{\rm sun}\ll \Delta m^2_{\rm atm}/E_\nu$).
In conclusion, solar  $\nu_e \leftrightarrow \nu_{\mu,\tau}$ oscillations depend
on two oscillation parameters: 
$$\Delta m^2_{\rm sun}\equiv \Delta m^2_{12}\qquad\hbox{ and }\qquad
\theta_{\rm sun}\equiv\theta_{12},$$
and effectively realize oscillations of two neutrinos.
Things become more complicated if one adds extra sterile neutrinos.

\medskip

\subsection{The solar mass splitting} $\Delta m^2_{\rm sun}$ is directly determined by the position of the oscillation dips at
KamLAND, with negligible
contribution from solar experiments.
More precisely, this will be rigorously true in the future.
For the moment solar data are needed to eliminate spurious solutions
mildly disfavored by KamLAND data, as illustrated in fig.~\ref{fig:sunfit}.
KamLAND data also imply a large mixing angle, but its value is more precisely measured by
SNO.

\subsection{Prediction for ${P}({\nu}_e\to{\nu}_e)$}
This implies that oscillations in the sun are adiabatic
to an excellent level of approximation\footnote{Some authors worry that
this conclusion might be invalidated by unexpected inhomogeneities in the sun~\cite{solarinomo}.}, 
that  seasonal variations are
negligible and earth matter corrections are small.
The dashed line in fig.\fig{SpettroSolare} shows the survival probability 
$P(\nu_e\to\nu_e, E_\nu)$ for best-fit values of $\Delta m^2_{\rm sun}$ and $\theta_{\rm sun}$.
Its main features can be understood as follows.
Solar matter effects are negligible  at neutrino energies much lower than
 \begin{equation}\label{eq:noMSW}
E_* \equiv \frac{\Delta m^2_{\rm sun}}{2\sqrt{2}G_{\rm F} N_e^{\rm sun}} 
\approx 4\MeV \frac{\Delta m^2_{\rm sun}}{0.8~10^{-4}\eV^2}
\end{equation}
where $N_e^{\rm sun}$ is the electron density around the solar center. Therefore:
\begin{itemize}
\item[--] Solar neutrinos with  $E_\nu \ll E_*$ (so far probed only by Gallium experiments)
experience averaged vacuum oscillations:
\beq P(\nu_e\to\nu_e,\hbox{small $E_\nu$})\simeq 1-\frac12 \sin^22\theta_{\rm sun}\ge 1/2.\eeq
\item[--]  Solar neutrinos with  $E_\nu \gg E_*$ 
 experience dominant and adiabatic solar matter effects that,
 as discussed in point b.\ of section~\ref{MSWresonance}, 
 convert $\nu_e$ into $\nu_2$:
\beq\label{eq:PeehighEnu}
P(\nu_e\to\nu_e,\hbox{large $E_\nu$})\simeq \sin^2\theta_{\rm sun}.\eeq
\end{itemize} 
These limiting values of $P(\nu_e\to \nu_e)$
do not depend on $\Delta m^2_{\rm sun}$ nor 
on the precise density profile of the sun, that instead determine the value
of the transition energy $E_*$.

In view of this situation, present solar data do not precisely determine $\Delta m^2_{\rm sun}$.
Nevertheless, it is interesting to point out that global fits of solar data alone, 
as shown in fig.\fig{sunfit},
point to a $\Delta m^2_{\rm sun}$ range compatible with KamLAND.
This happens for the following reasons:
\begin{itemize}
\item[--] Smaller values $\Delta m^2_{\rm sun}\sim 10^{-5}\eV^2$ are safely  excluded
because lead to large earth matter effects
 (such as day/night asymmetries) not seen by SK nor SNO.
 \item[--]  Larger value of $\Delta m^2_{\rm sun}$
 are excluded because $E_*$ increases and it becomes impossible to get the SNO
 value of $P(\nu_e\to\nu_e)$, smaller than $1/2$ at $5\sigma$.
\end{itemize}

\begin{figure}[t]
$$
\includegraphics[width=8cm]{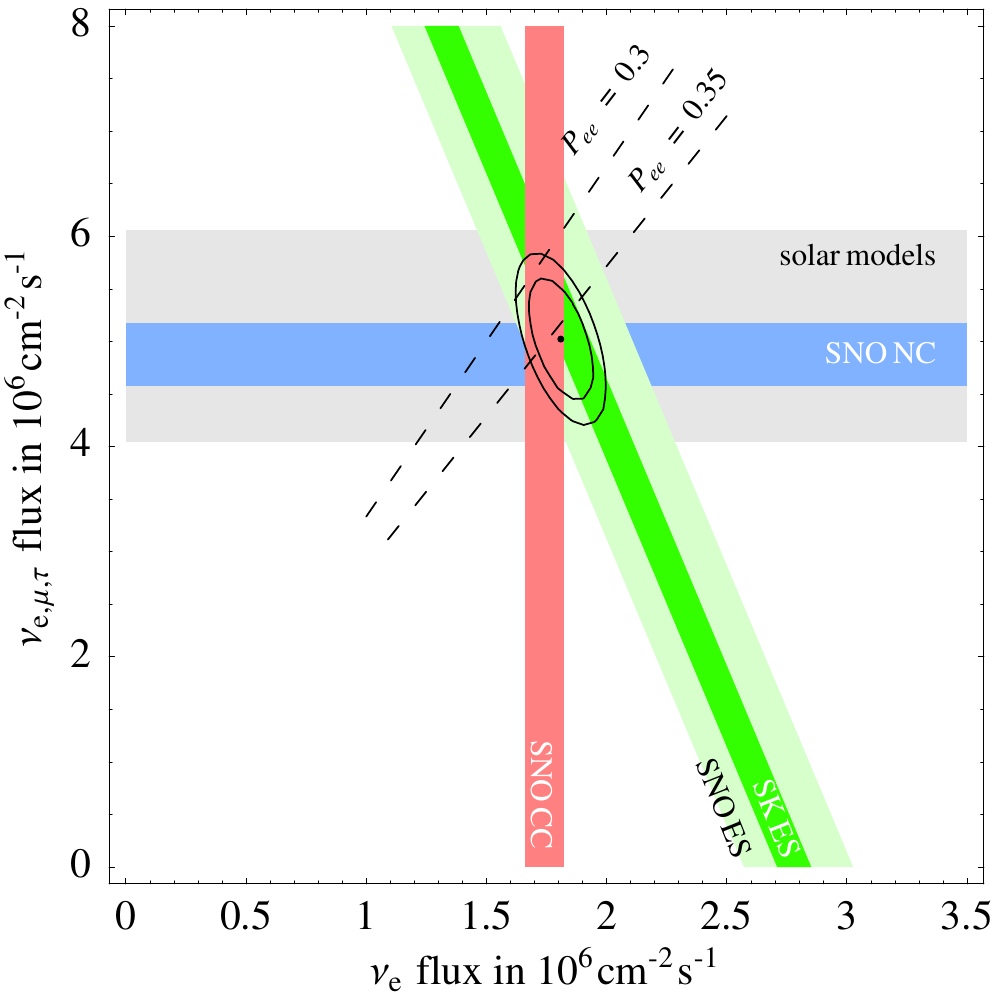}\hspace{1cm}\includegraphics[width=8cm]{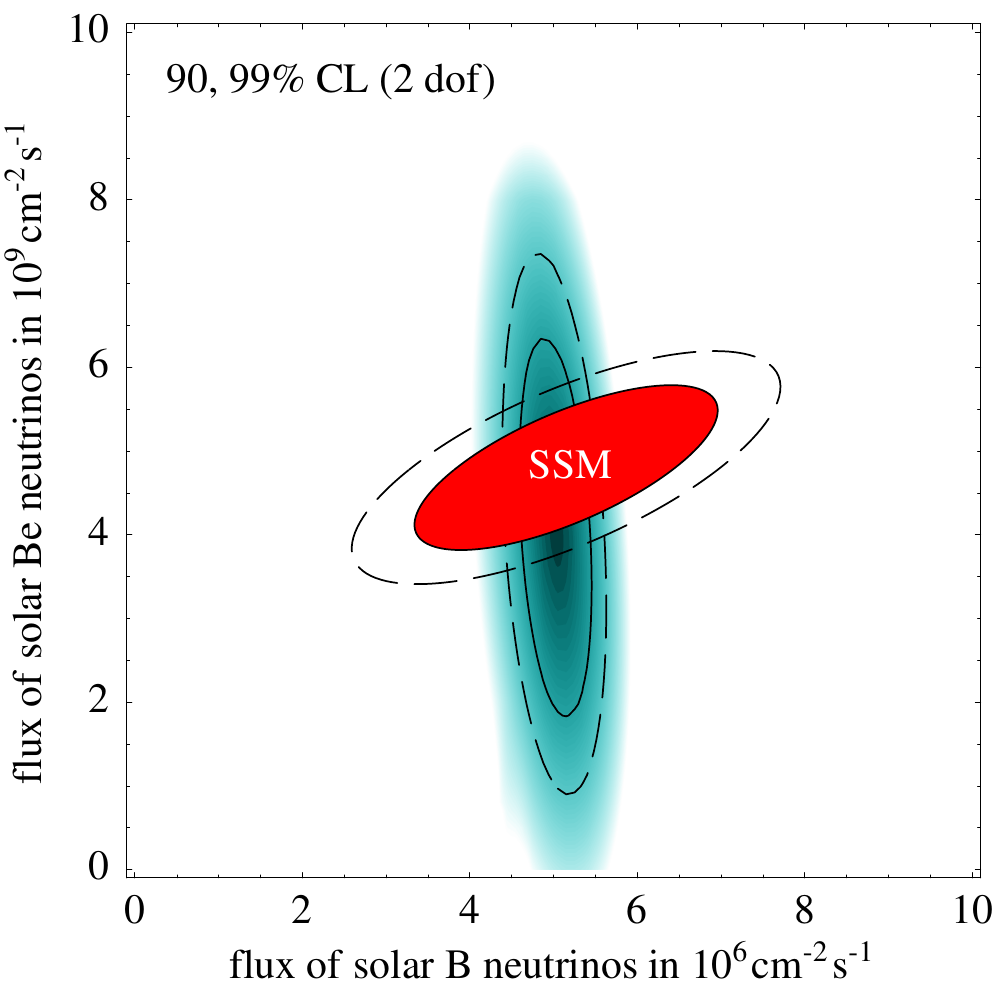}$$
\caption[Fit of solar neutrino fluxes]{\label{fig:fluxes}{\bf Solar fluxes from solar neutrino experiments}. \em 
Fig.\fig{fluxes}a: determinations of the oscillated fluxes of Boron neutrinos.
Fig.\fig{fluxes}b: unoscillated Boron and Beryllium fluxes, as determined by
solar neutrino experiments without using standard solar model (SSM) predictions.
Fig.\fig{fluxes}b also applies to Boron and CNO fluxes.
Contours are drawn at 90 and 99\% CL (2 dof).}
\end{figure}

\subsection{The solar mixing angle} $\theta_{\rm sun}$
is directly determined by SNO measurements of NC and CC solar Boron rates.
Assuming  flavour conversions among active neutrinos, SNO implies
\beq\label{eq:Peesunexp}
\langle P(\nu_e\to\nu_e) \rangle \equiv  {\Phi(\nu_{e}) }/{\Phi(\nu_{e,\mu,\tau}) }=0.357\pm0.030.\eeq
where the average is performed around energies $E_\nu \sim 10\MeV$.
Solar models and SK data provide extra less precise information on
$\langle P(\nu_e\to\nu_e) \rangle $:
fig.\fig{fluxes}a illustrates the consistency of all these determinations.

Eq.\eq{Peesunexp} can be compared with the oscillation prediction for $\langle P(\nu_e\to\nu_e)\rangle$: at $E_\nu\sim 10\MeV$ matter effects are not yet fully dominant, and there is 
a little deviation from the large-energy limit of eq.\eq{PeehighEnu}: one has
\beq\label{eq:Peesunth}
\langle P(\nu_e\to\nu_e) \rangle \approx 1.15\sin^2\theta_{12}.\eeq
Therefore, by combining eq.s\eq{Peesunexp} and\eq{Peesunth}, one infers
\begin{equation}\label{eq:1.15}
 \tan^2\theta_{12}=0.45\pm0.05
 \end{equation}
 which agrees with the results of the global analysis in table~\ref{tab:tab1},
 both in the central value and in its uncertainty.
Notice that the only solar model input that enters
our approximate determination of solar oscillation parameters
is the solar density around the center of the sun,
that controls the $15\%$ correction to $\langle P(\nu_e\to\nu_e)\rangle$ in eq.\eq{1.15}.
This correction factor is comparable 
to the $1\sigma$ uncertainty in $\langle P(\nu_e\to\nu_e)\rangle$:
indeed the associated increase of $P(\nu_e\to\nu_e)$ at smaller $E_\nu$ 
(see fig.\fig{SpettroSolare})
is not visible in the energy spectra measured by SNO and SK.

\subsection{Sterile neutrinos?}
By adding to the data-set
the extra input of the solar model prediction for
the total Boron flux $\Phi_B$ one can extract from SNO measurements
both the solar mixing angle and the parameter $\eta_{\rm s}$
(assumed to be energy-independent, and precisely defined by eq.\eq{PeeEta}) that tells the fraction of
solar neutrinos that possibly oscillate into sterile neutrinos.
One finds
$$
\eta_{\rm s} \approx \frac{\Phi_{\rm B} - \Phi_{\nu_{e,\mu,\tau}}}{\Phi_{\rm B} - \Phi_{\nu_e}}\approx 0\pm 0.2$$
i.e.\ data do not suggest extra sterile neutrinos.
As discussed in section~\ref{Sterile}, $\eta_{\rm s}$ can be
dramatically  energy-dependent,
and the above constraint applies to its value at $E_\nu\sim 10\MeV$
Global fits find a somewhat stringent constraint $\eta_{\rm s}=0 \pm 0.10$,
because to accommodate sterile effects one increases $\Phi_B$ and reduces $\theta_{\rm sun}$
and this tends to give unseen spectral distortions.

\subsection{Effects of a small $\theta_{13}$.}
The main effect of a non vanishing $\theta_{13}$ consists in 
changing the limiting values achieved by $P(\nu_e\to\nu_e)$ in 
the lower-energy regime
and in the higher-energy regime as\footnote{The transition between the two regimes
happens at $E_\nu\sim \hbox{few}\,\MeV$, and the high-energy regime is approached
for $E_\nu \circa{>} 20\MeV$.}
\begin{eqnsystem}{sys:PeeLowHi}\label{eq:PeeHi}
P(\nu_e\to\nu_e,\hbox{large $E_\nu$}) &=&
 \sin^4\theta_{13}+\cos^4\theta_{13}\sin^2\theta_{12},\\
P(\nu_e\to\nu_e,\hbox{small $E_\nu$}) &=& \sum_i |V_{ei}|^4=
 \sin^4\theta_{13}+\cos^4\theta_{13}\bigg[1-\frac{1}{2}\sin^2 2\theta_{12}\bigg].
 \label{eq:PeeLow}
\end{eqnsystem}
This means 
\begin{equation}\label{eq:pred}
P(\nu_e\to\nu_e,\hbox{small $E_\nu$}) \le  1- 2P(\nu_e\to\nu_e,\hbox{large $E_\nu$}) +
2P(\nu_e\to\nu_e,\hbox{large $E_\nu$})^2.
\end{equation}
where the equality applies for $\theta_{13}=0$:
low-energy solar data agree with this oscillation prediction for 
$P(\nu_e\to\nu_e,\hbox{small $E_\nu$})$.
This quantity is presently dominantly determined by
Gallium data and its value can be extracted with a simple approximate argument.
  Subtracting from the total Gallium rate 
\begin{equation}\label{eq:Ga}
(68.1\pm 3.7) \,\hbox{SNU}=R_{\rm Ga} =
R_{pp,pep}^{\rm Ga} +  R_{\rm CNO}^{\rm Ga} +  R_{^7{\rm Be}}^{\rm Ga} + R_{^8{\rm B}}^{\rm Ga} 
\end{equation}
 its  $^8$B contribution 
 (as directly measured by SNO via CC, $R_{^8{\rm B}}^{\rm Ga} = 4.3\pm 1\,\hbox{SNU}$)
 and regarding all remaining fluxes as low energy ones, suppressed by
$P(\nu_e\to\nu_e,\hbox{small $E_\nu$})$, determines it to be $0.57\pm 0.03$.
Alternatively,
by subtracting also the intermediate-energy CNO and Beryllium fluxes, one gets
$P(\nu_e\to\nu_e,\hbox{small $E_\nu$})=0.58\pm0.05$.
We here included only the error on the Gallium rate, which is the dominant error.
This rough analysis shows that the result only mildly depends on how one deals with intermediate energy neutrinos, and on
model-dependent  details of the intermediate region,
thereby suggesting the following general result:
\begin{equation}\label{eq:PeeVO}
P(\nu_e\to\nu_e,\hbox{large $E_\nu$}) =
0.31\pm0.03,\qquad
P(\nu_e\to\nu_e,\hbox{small $E_\nu$}) = 0.58\pm0.04.
\end{equation}
The resulting solar constraint on $\theta_{13}$ is weaker than the CHOOZ constraint.

\subsection{Implications for solar fluxes}
The initial goal of solar neutrino experiments that discovered neutrino oscillations 
was testing models of the sun.
This can be done today that oscillations seem sufficiently well understood, 
and might lead to more unexpected discoveries.
We here  study how well 
present solar neutrino data determine solar neutrino fluxes~\cite{SolarFluxes, solarfits}. 
We recall that the energy spectrum of each single component is safely determined
by simple physics, and we want to measure and test  their total fluxes.
Apparently a detailed global  fit of solar neutrino data
seems needed to address this issue, but --- once again --- 
main results follow from simple considerations:
\begin{itemize}
\item[1.] Assuming the absence of short term (million year) 
variations of the solar energy release, the 
measured solar luminosity allows us to precisely predict the $pp$ flux
(directly related to the smaller $pep$ flux).
Indeed neutrino experiments tell that other 
fluxes are small enough that they negligibly contribute 
to the luminosity constraint, that 
therefore simply fixes the $pp$ fluxes to their solar 
model value.

\item[2.] SNO measured the Boron flux
(and sets an upper bound $\Phi_{hep}<2.3~10^4/{\rm cm}^2\cdot{\rm s}$ on the small $hep$ flux, that we can here neglect)~\cite{SNO}.

\item[3.] Only two kind of experiments,
Gallium and Chlorine, have measured low-energy neutrino fluxes.
Therefore, the  data constrain only two linear 
combinations of low-energy fluxes.

\item[4.]
However, the Chlorine experiment has a poor sensitivity 
to low energy neutrinos.
After subtracting the $\sim80\%$ Boron contribution to the Chlorine rate, 
as directly measured by SNO via CC, the residual low-energy contributions 
to the Chlorine rate is just about $2\sigma$ above zero.

\end{itemize}
Therefore the Chlorine rate carries so little information on low energy fluxes,
that our present knowledge on low-energy fluxes 
is well summarized by a single number: 
their contribution to the Gallium rate.
Starting again from eq.\eq{Ga}, we now subtract from the total Gallium rate $R_{\rm Ga}$ its  $^8$B contribution 
and its $pp,pep$ contributions
(as predicted by solar models and LMA oscillations, see eq.\eq{PeeVO}: $R_{pp,pep}^{\rm Ga} =41.3\pm1.5$), obtaining
 \begin{equation}\label{eq:CNOBe}
 (22.5 \pm 4)\, \hbox{SNU} =R_{\rm CNO}^{\rm Ga} +  R_{^7{\rm Be}}^{\rm Ga}  =
\frac{4.0 \Phi_{^7{\rm Be}} + 4.6 \Phi_{\rm CNO}}{10^{9}/\cm^2{\rm s}} \,\hbox{SNU}.
 \end{equation}
 We have taken into account that, in the standard scenario,
 oscillations suppress both rates by about
 $0.55\pm0.02$ --- a value negligibly different from the
 low-energy limit of $P(\nu_e\to\nu_e)$ of eq.~(\ref{eq:PeeVO}).
 Eq.\eq{CNOBe} means that present data cannot discriminate between Be and CNO,
 and (supplemented by $\Phi_{^7{\rm Be}} >0$) implies
\begin{equation}\label{eq:CNO}
\Phi_{\rm CNO} < 6\cdot 10^9/\cm^2{\rm s}\qquad\hbox{ at $3\sigma$ (1 dof),}\end{equation}
which is one order of magnitude above solar model predictions.
This bound means that the CNO cycle does not give the dominant contribution 
to the total solar luminosity $L_\odot$.
Indeed by converting  the neutrino flux $\Phi_{\rm CNO}$ into
the corresponding energy flux $L_{\rm CNO}$, eq.\eq{CNO} reads
$L_{\rm CNO} \circa{<} 0.1~L_\odot$.
After taking into account the Borexino measurement of the $^7$Be flux, this bound improves to
$L_{\rm CNO} \circa{<} 0.033~L_\odot$.

The main message is that solar neutrino experiments confirm that the sun shines via the $pp$
cycle, while the CNO cycle has at most a subdominant role.
It would be interesting to see its existence; but
the oscillated CNO contribution to the Cl, Ga and (presently) Borexino rates
is slightly smaller than the experimental error
on these rates.   Solar neutrino experiments
cannot confirm that the CNO cycle exist.

\subsection{Global fits}
As previously discussed, thanks to SNO and KamLAND, global fits are no longer essential.
(Various authors might disagree with this view).
 Fig.\fig{sunfit} shows the result of a global $\nu_e\leftrightarrow\nu_{\mu,\tau}$ oscillation fit of 
 all available solar and reactor data.
 For completeness, we conclude by  summarizing how a generic global fit are performed.

Assuming that the solar anomaly is due to oscillations,
one needs to compute neutrino  propagation
in the sun, in the space and in the earth (for neutrinos arrived during the night),
taking into account matter effects in the sun and in the earth,
and seasonal effects due to the eccentricity of the earth orbit.
One needs to average over the neutrino production point in the sun,
as predicted by solar models, using the solar density profile predicted by solar models
(and confirmed by helioseismology).
Neutrinos produced around the center
(on the opposite side) of the sun experience one (two) MSW resonances.

At this point, one can compute the rates measured by the various experiments,
taking into account their cross sections, cuts and energy thresholds.
To include earth matter effects one must know how much time the sun is seen at each zenith-angle
from the sites of the various experiments (there are only minor differences between
Homestake, Gran Sasso, Kamioka,  Sudbury).

Finally, in order to extract the oscillation parameters from data,
one forms a global $\chi^2$,
taking into account the
correlated uncertainties on the solar model predictions\footnote{Solar models predict 
the unoscillated $\nu_e$ fluxes $\Phi_k$ as function of 11 uncertain parameters
$$\lambda_i =\big\{S_{11}, S_{33},S_{34},S_{1,14},S_{17},\hbox{luminosity},Z/X,\hbox{age},\hbox{opacity},
\hbox{diffusion},C_{\rm Be}\big\}$$
measured to be $\lambda_i = \lambda_{i0}$ with uncorrelated one-sigma errors $\sigma_{\lambda_i}$.
We write any quantity $x$ as $x = x_0 + \delta x$, where $\delta x$ represents deviations from
the central value $x_0$.
In Gaussian approximation solar model predictions can be written as
$\Phi_k(\lambda)\simeq \Phi_{0k} (1 + M_{\Phi \,ki} \delta \lambda_i/\lambda_i)$ i.e.\ $\delta \ln\Phi =M_\Phi \cdot\delta \ln \lambda$.
The numerical values of the central values and logarithmic
shift coefficients $M_\Phi$ can be found in the literature~\cite{SolarChiq}.
As discussed in appendix~\ref{Statistics}, eq.\eq{quadrature},
the resulting matrix of correlated uncertainties on solar fluxes is
$$\sigma^2_\Phi = \frac{1}{\Phi_0^T}\cdot 
M_\Phi \cdot \diag (\frac{\sigma_{\lambda_i}^2}{\lambda_i^2})\cdot M_\Phi^T \cdot  \frac{1}{\Phi_0}
\qquad\hbox{so that}\qquad
\chi^2_\Phi(\Phi) =(\Phi-\Phi_0)^T \cdot (\sigma^2_\Phi)^{-1}\cdot (\Phi-\Phi_0).$$
For the purpose of fitting present data, 
this procedure mainly corrects the uncorrelated uncertainties
of table\tab{flussinu} adding a strong correlation between N and O fluxes and some
correlation with Be and B fluxes.},
on the cross sections, and the statistical and systematic experimental errors~\cite{SolarChiq}.
While all Gallium experiments can be condensed into a single rate,
SK and SNO presented rates binned as function of energy and zenith-angle,
forming a total data-set of almost 100 rates.
Each predicted rate $R$ can be computed as a function $T$ of oscillation parameters,
of solar fluxes $\Phi$, and of various systematic parameters $\lambda$
(that take into account uncertainties on the Boron energy spectrum,
on energy scale, resolution in the solar experiments,\ldots).
Marginalizing the global 
\beq\label{eq:chiqsun}
\chi^2_{\rm sun} (\theta,\Delta m^2,\Phi,\lambda) = (R-T)\cdot (\sigma_R^2)^{-1}\cdot (R-T)
+ \chi^2_\Phi(\Phi) + \chi^2_\lambda(\lambda)\eeq
with respect to the nuisance parameters $\Phi$ and $\lambda$ in Gaussian approximation
(appendix~\ref{Statistics}, eq.\eq{quadrature})
gives the $\chi^2_{\rm sun}(\theta,\Delta m^2)$ plotted in fig.\fig{sunfit}.

%% file: review_future.tex
\chapter{Future oscillation experiments}\label{oscexp}

\section{The global oscillation picture and $\theta_{13}$}\label{3nu}
We have discussed the established solar and atmospheric neutrino anomalies.
Few other anomalies, discussed in section~\ref{anomalies},
could be confirmed or refuted by future experiments.
For the moment we ignore them and discuss
how the solar and atmospheric data can be jointly explained in terms
of oscillations between the three SM neutrinos.
The $\Delta m^2$ responsible of the atmospheric anomaly
is larger than the one responsible of the solar anomaly.
Therefore we identify (see fig.\fig{spectra})
$$|\Delta m^2_{13}|\approx |\Delta m^2_{23}|=\Delta m^2_{\rm atm}\approx \dmmatm,\qquad
\Delta m^2_{12}=\Delta m^2_{\rm sun} \approx \dmmsun.$$
A positive $\Delta m^2_{23}$ means that the neutrinos separated
by the atmospheric mass splitting are heavier than those separated
by the solar mass splitting (fig.\fig{spectra}a): this is usually named `normal hierarchy'.
At the moment  this cannot be distinguished from the opposite case (fig.\fig{spectra}b)
usually named `inverted hierarchy'.
As their names indicate, the `normal' case is considered more plausible
than the `inverted' one.

As explained in section~\ref{MD}, the neutrino mixing
matrix contains 3 mixing angles:
two of them ($\theta_{23}$ and $\theta_{13}$) produce oscillations
at the larger atmospheric frequency,
one of them ($\theta_{12}$) gives rise to oscillations at the smaller solar
frequency.
Solar data want a large mixing angle.
The CHOOZ constraint tells that
$\nu_e$ are can only be slightly involved in atmospheric oscillations,
and SK finds that  atmospheric data can be explained by
$\nu_\mu \to \nu_\tau$ oscillations with large mixing angle.
These considerations single out the global solution
$$\theta_{23} =\theta_{\rm atm} \sim 45^\circ \qquad  \theta_{12} = \theta_{\rm sun} \sim 30^\circ,\qquad
\theta_{13}\circa{<}10^\circ ,\qquad \phi = \hbox{unknown}$$
more quantitatively summarized in table~\ref{tab:tab1}.
Nothing is known on the CP-violating phase $\phi$.
If $\theta_{13}= 0$ the solar and atmospheric anomalies depend on different set of
parameters; there is no interplay between them.
A $\theta_{13}\neq 0$ would affect both solar and atmospheric data.
SK alone finds $\sin^2 2\theta_{13} = 0 \pm0.2$~\cite{SKatm}.
Present experimental results show a $2\sigma$ hint for a positive $\sin^22\theta_{13}=\ssCH$, that comes out
by combining the following $1\sigma$ hints:
\beq \sin^2 2\theta_{13} = \left\{ \begin{array}{ll}
0.05\pm 0.05 & \hbox{CHOOZ and atmospheric}\\
0.08\pm0.07 & \hbox{solar and KamLAND}\\
0.15\pm 0.10&\hbox{MINOS}
\end{array}\right.\eeq

The first and most precise determination of $\theta_{13}$ comes from
the CHOOZ reactor experiment, together with atmospheric SK data that constrain $\Delta m^2_{\rm atm}$;
according to some global fits~\cite{theta13fits} SK data suggest a non-zero best-fit value of $\theta_{13}$.
Second, global fits of solar data give a $1\sigma$ hint for a positive $\theta_{13}$.
Third, the recent MINOS results~\cite{NuMi} show a $1.5\sigma$ excess of 35 $\nu_e$ CC events, with respect to the
expected  $27\pm5_{\rm stat}\pm2_{\rm sys}$ background events, that can be due to $\nu_\mu\to \nu_e$ atmospheric oscillations.

\bigskip

In conclusion, all pieces of oscillation data point in the same direction, 
and, up to small corrections, can be analyzed without  performing a 3 neutrino analysis.
Present data can be interpreted in terms of oscillations among the
three SM neutrinos with the oscillation parameters in table~\ref{tab:tab1} (page~\pageref{tab:tab1}).
The corresponding neutrino mixing matrix is
\beq |V| = \bordermatrix{ & \nu_1 & \nu_2 & \nu_3\cr
\nu_e & 0.84 \pm 0.01&  0.54 \pm 0.02 & 0.05\pm0.05 \cr
\nu_\mu &0.38  \pm 0.06 & 0.60\pm 0.06& 0.70\pm 0.06 \cr
\nu_\tau & 0.38 \pm 0.06& 0.60\pm0.06 &0.70 \pm 0.06}\eeq
as pictorially represented in fig.\fig{spectra}.

\section{Known unknowns}
We here {\em assume} that oscillations between the 3 SM neutrinos are the true global picture.
Furthermore we assume that neutrino masses are of Majorana type.
While plausible, this is only an assumption to be tested by future experiments, that could
discover something more, or something different.
If our assumption is true, the goal of future experiments is the reconstruction of the neutrino Majorana mass matrix: as discussed in subsection~\ref{Majorana} we have to measure 9 real parameters:
3 mass eigenvalues, 3 mixing angles and 3 CP-phases.
Two 2 squared mass differences and 2 mixing angles are already partially known.
The main steps of this program are:
\begin{enumerate}

\item Establishing that both the solar and atmospheric anomalies are really due to oscillations.

\item Measuring better and better the solar and atmospheric parameters.
Discovering possible deviations of $\theta_{23}$ from maximal mixing
is an  important but difficult issue.
It is important because maximal mixing could be the result of a new symmetry.
It is difficult because dominant $\nu_\mu \leftrightarrow\nu_\tau$ oscillations are controlled by
$\sin^2 2\theta_{23} =  1  + 0 \epsilon - 4 \epsilon^2 +\cdots$
where $\epsilon = \theta_{23} - \pi/4$.
For example $\sin^2 (2\cdot 40^\circ) \approx 0.97$.
Order $\epsilon$ corrections do not vanish in
subdominant oscillations induced by $\Delta m^2_{\rm 12}$ or $\theta_{13}$,
see eq.~(\ref{sys:PatmNonStandard}).

\item {\bf Discovering the last mixing angle}, $\theta_{13}$ that induces
$\nu_\mu \leftrightarrow \nu_e$ oscillations at the atmospheric frequency.
\end{enumerate}
If a non zero $\theta_{13}$ will be discovered...
\begin{itemize}

\item[4.]
Oscillations in matter allow to discriminate
{\bf the sign of $\Delta m^2_{23}$} (i.e.\
if the atmospheric
anomaly is due to the lightest or heaviest neutrinos, see fig.\fig{spectra}).
If $\Delta m^2_{23}>0$ (normal hierarchy) matter effects
enhance $\nu_\mu\leftrightarrow\nu_e$ oscillations and suppress
$\bar{\nu}_\mu\leftrightarrow\bar{\nu}_e$, while the opposite happens if
 $\Delta m^2_{23}<0$ (inverted hierarchy).

\item[5.] 
The {\bf sign of $\theta_{23}-45^\circ$} (which tells
whether the neutrino state with mass $m_3$
contains more $\nu_\tau$ or more $\nu_\mu$)
can be measured from atmospheric oscillations
by measuring e.g.
$$P(\nu_e\to \nu_e) = 1- \sin^2 2\theta_{13} \sin^2 \frac{\Delta m^2_{23}
L}{4E_\nu}
\qquad\hbox{and}\qquad
P(\nu_\mu\to \nu_e) = \sin^2 \theta_{23}
\cdot [1-P(\nu_e\to \nu_e)]$$
Note that $\nu_\mu$ disappearance experiments alone cannot
distinguish $\theta_{23}$ from $90^\circ-\theta_{23}$,
and that the present bound $\sin^2 2\theta_{23}\circa{>} 0.95$
allows the relatively loose range
$1/3\circa{<}\sin^2\theta_{23}\circa{<}2/3$.

\end{itemize}
Notice that, in line of principle, issues 4.\ and 5.\ are meaningful even if $\theta_{13}=0$,
in view of the presence of the small solar splitting:
both $\theta_{13}$ and $\Delta m^2_{12}$ produce $\nu_\mu\leftrightarrow\nu_e$ oscillations.
However in practice $\Delta m^2_{12}$ is so small that 
planned experiments can answer these issues only if $\theta_{13}$ is large enough.
The following oscillation issue needs $\theta_{13}\neq 0$ both in practice  and in principle:
\begin{itemize}
\item[6.] The {\bf CP-violating phase $\phi$} can be measured in realistic long-baseline oscillation experiments.
\end{itemize}
As discussed at page~\pageref{MDosc},
oscillation experiments cannot access the whole neutrino mass matrix
and
cannot tell if neutrinos have Majorana or Dirac masses.
Oscillations are sensitive to squared neutrino mass differences, 
but not to the overall scale of neutrino masses.
We have measured the charged lepton Dirac masses $m_e$, $m_\mu$, $m_\tau$.
It would be unsatisfactory if instead we knew only the values of
$ m_\tau^2 - m_\mu ^2$ and $m_\mu^2 - m_e^2$.
In the Dirac case oscillation experiments miss only the overall neutrino mass scale.
In the Majorana case they also miss
two CP-violating phases, $\alpha$ and $\beta$.
 \begin{itemize}

\item[7.] In order to access to non oscillation parameters
we need non oscillation experiments (section~\ref{nonOsc}).
Furthermore, neutrino-less-double-beta decay and cosmology are
other realistic ways of clarifying the issue 4 (normal or inverted hierarchy?).

\end{itemize}

\section{Atmospheric experiments}\label{atm future}
There are two possible directions for future improvements: 
bigger detectors, better detectors.

Muons are contained in a bigger detector  up to higher energies.
The {\sc IceCUBE} detector~\cite{ICECUBE} achieves a large km$^3$ volume
at the price of a poor photo-multiplier granularity, such that it can only study
atmospheric $\nubarnu_\mu$ at high energy, $E_\nu\circ{>}\TeV$,
where atmospheric oscillations give small effects.
It can collect about 80000 atmospheric events per year.
A Mton-scale detector~\cite{Mton} (motivated by many other considerations)
could repeat the atmospheric measurements performed by SK with much more statistics.

One possible goal is the search for sub-leading $\mu\leftrightarrow e$ transitions.
In the standard oscillation scenario they can be generated by the $\theta_{13}$ mixing angle
and by solar oscillations.
In view of the different $\Delta m^2$, solar oscillations dominantly affect the sub-GeV $e$-like
angular rate
and atmospheric $\theta_{13}$ oscillations would be most clearly seen in the multi-GeV $e$-like
angular rate.
However the impact of $\mu\leftrightarrow e$ transitions gets suppressed,
for maximal mixing angle,
by a `screening factor' $(r/2-1)$, where
$r$ is the ratio between the unoscillated $\nubarnu_\mu$ and $\nubarnu_e$ atmospheric fluxes.
At sub-GeV energies $r\approx 2$: 
the initial flavour ratio $\nubarnu_e:\nubarnu_\mu:\nubarnu_\tau = 1:2:0$ gets converted by atmospheric oscillations
into the flavour-blind ratio $\nubarnu_e:\nubarnu_\mu:\nubarnu_\tau = 1:1:1$, 
that is not affected by further $\mu\leftrightarrow e$ transitions.
This explains the origin of the `screening factor':
unless $\theta_{\rm atm}$ is significantly non-maximal, atmospheric
experiments are a poor probe of $\mu\leftrightarrow e$ transitions.
An accurate approximate formula that describes the excess of electrons due to
$\mu\leftrightarrow e$ 
solar oscillations and due to $\theta_{13}$ atmospheric oscillations is:
\beq \frac{N_e}{N_e^0}\approx 1 +
(r\cos^2\theta_{\rm atm}-1)\sin^2 2\theta_{12}^m\sin^2 \frac{\Delta m^{2m}_{\rm sun}L}{4E}+
(r\sin^2\theta_{\rm atm}-1)\sin^2 2\theta_{13}^m\sin^2 \frac{\Delta m^{2m}_{\rm atm}L}{4E}+
\cdots \eeq
where $\cdots$ denotes their interference term, 
that depends on the CP-violating oscillation phase $\phi$.
The subscript $^m$ on an oscillation parameter means that it must be computed in matter,
assumed to have constant density.
This effect would be a good probe of $\theta_{\rm atm}-\pi/4$.

Alternatively, one can build a smaller (30 kton?) magnetized iron calorimeter capable of distinguishing
neutrinos from anti-neutrinos~\cite{Monolith}.
This technique allows to measure the direction of scattered $\mu^\pm$ and their energy
(with expected errors of about $\pm 15^\circ$ and $\pm15\%$ respectively)
better than W\v{C}: the $L/E$ of atmospheric $\nubarnu_\mu$ can be reconstructed
a few times better than in SK, such that it would be possible 
to see the first $\approx 2$ atmospheric oscillation dips.
Furthermore, if $\theta_{13}$ is as large as possible, cleanly observing the oscillation pattern 
would help in determining the neutrino mass hierarchy.
On the other hand, W\v{C} detectors have a lower energy threshold and can see scattered $e^\pm$.

\begin{figure}[t]
$$\includegraphics[width=0.8\textwidth]{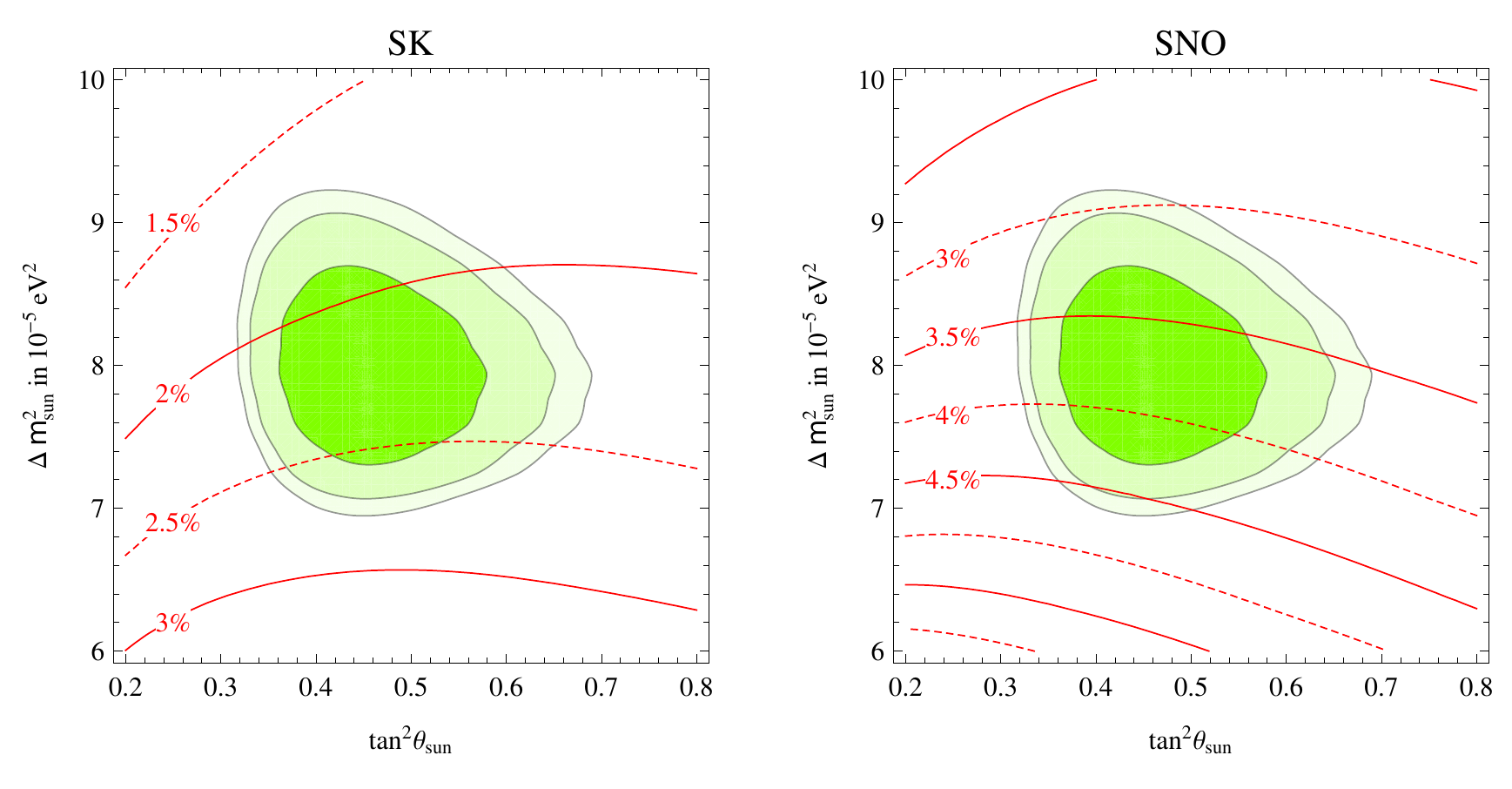}$$
\caption[Day/night asymmetry of solar neutrinos]{\label{fig:DNplot}\em Iso-contours of the expected day/night asymmetry in the total rate of solar neutrinos as measured via ES scatterings at SK and by CC scatterings at SNO.
The green area is the best-fit region at $90,99,99.73\%$ CL.}
\end{figure}

\section{Solar experiments}
The main goals of future soar neutrino experiments seem~\cite{FutureSun}: 
$1)$  detecting some small effect characteristic of oscillations;
$2)$ measuring better and better the oscillation parameters;
$3)$ testing solar model predictions;
$4)$ constraining and possibly discovering unexpected effects.
Taking into account LMA predictions and experimental capabilities,
concrete progress seem possible on the following points.
\begin{itemize}
\item LMA predicts that earth matter effects give a 
{\bf day/night asymmetry},   $$A_{\rm D/N} \sim G_{\rm F} N_e/(\Delta m^2_{\rm sun}/E_\nu)
\sim 3\%\cdot E_\nu/10\MeV,$$
with a characteristic
energy and zenith-angle dependence.
For example, fig.\fig{DNplot} shows the expected day/night asymmetry in the total ES rate at SK and 
in the total CC rate at SNO.
SK and SNO data show a $1\sigma$ hint for this effect, see eq.\eq{ADNSK} and\eq{ADNCC}.
Large neutrino energy and large statistics is needed to see it:
these conditions could be realized by a future SK-like W\v{C} experiment
with a Mton of water~\cite{Mton}.

\item No {\bf other time variation} of solar neutrino fluxes is expected.
Beryllium neutrinos, being almost monochromatic, are a much more sensitive probe
than Boron neutrinos of effects, like seasonal variations, that depend on neutrino energy.
The Borexino experiment (and maybe KamLAND)
should measure the Beryllium flux with significant statistics.
If an effect will be seen, it could be explained by invoking new physics,
such as sterile neutrinos.

\item 
LMA predicts that at lower energy the $\nu_e$ survival probability $P_{ee}$ increases reaching
\begin{equation}\label{eq:PeeLowE}
P_{ee} \simeq 1- \frac{1}{2} \sin^2 2\theta_{\rm sun} = 0.59\pm0.02\qquad\hbox{at $E_\nu\ll E_*$}
\end{equation}
(where $E_*\sim 4\MeV$, see eq.\eq{noMSW}).
Gallium experiments have seen this effect:
global fits disfavor a fully energy-independent  $P_{ee}$.

It is difficult to measure $P_{ee}$ around $E_\nu\sim E_*$ because
there are few solar neutrinos with these energies.
Borexino should give another test of this effect at relatively low neutrino energy.
In the future it might be possible to measure the spectrum of sub-MeV $pp$ neutrinos.
Since solar models can predict their flux with $1\%$ uncertainty
could be used for a {\bf precise measurement of $\theta_{\rm sun}$}.

\end{itemize}

\begin{figure}[t]
$$\includegraphics[width=0.7\textwidth]{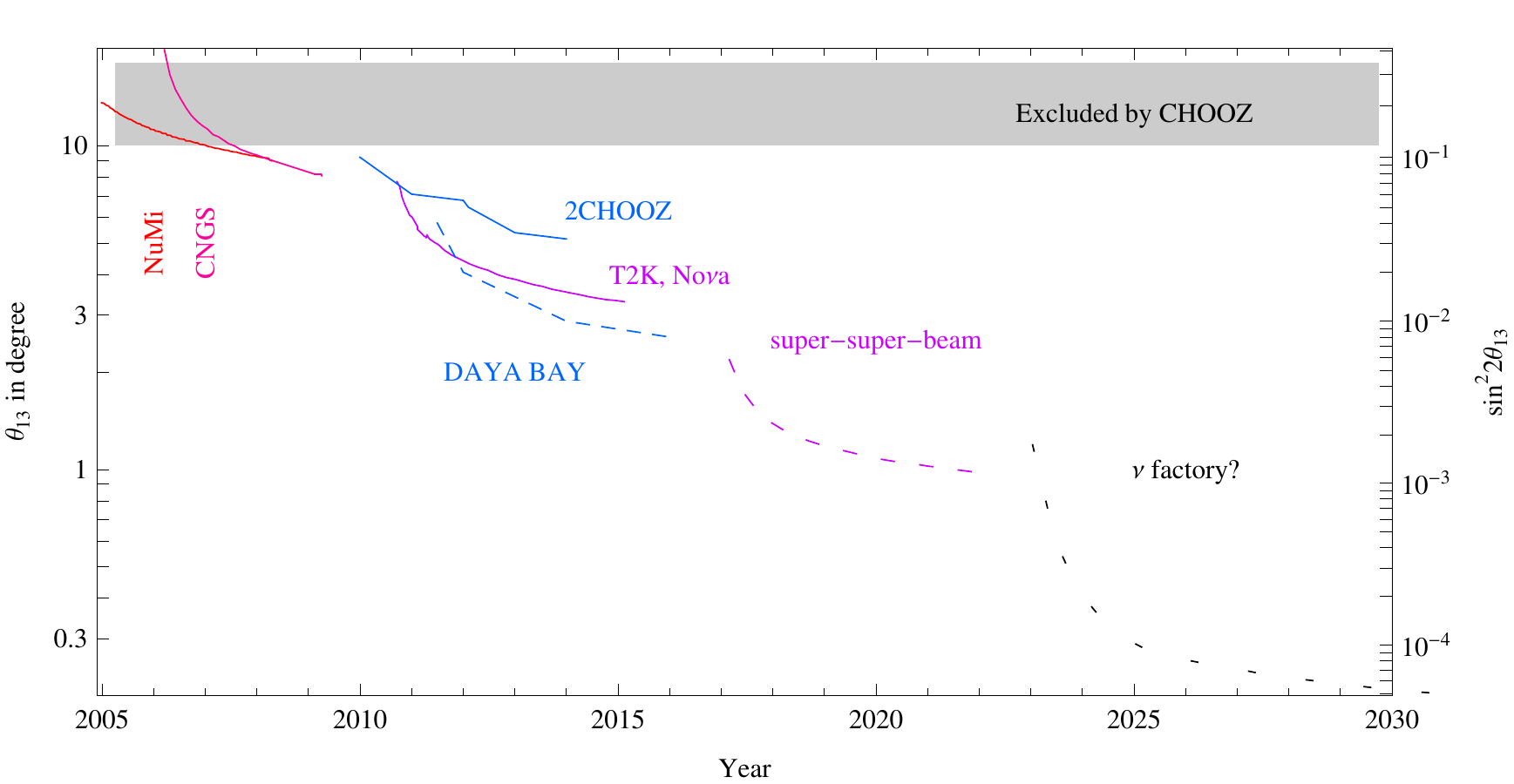}$$
\caption[Future sensitivity to $\theta_{13}$]{\label{fig:q13future}\em Sensitivity to $\theta_{13}$ of planned (continuous lines)
and possible (dashed lines) future experiments.
}
\end{figure}

\section{Reactor experiments}
The oscillation signal is any deviation from the unoscillated $\bar\nu_e$ energy spectrum.
CHOOZ achieved a statistical and a systematic uncertainty of $\pm2.7\%$.
Improved reactor neutrino experiments~\cite{reactors,FutureReactors} are being planned with the main goal
of searching for $\theta_{13}$, that gives $\bar\nu_e$ disappearance effects
well described by
\beq P(\bar\nu_e\to\bar\nu_e)  \simeq  1 - \sin^2 2\theta_{13} \sin^2\frac{\Delta m^2_{\rm 13} L}{4E_\nu}.
\eeq
We can here neglect solar oscillations and earth matter effects.
Sterile neutrinos could produce extra disappearance  effects.

Several projects have been proposed~\cite{FutureReactors}, all based on the same concept:
one near detector (needed in order to reduce systematic errors) and one far detector located
at distance $L\approx (1\div 2)\km$, in order to sit around the atmospheric oscillation dip.
The statistical error then depends on the size of the far detector times  the power of the nuclear reactor,
and a $\pm0.5\%$ level can be reached.
Reducing systematic errors down to the $\pm 0.5\%$ level is considered feasible by the proponents of these projects.
If this goal will be achieved, the {\sc DoubleCHOOZ} proposal (in France)
will provide an increase in the sensitivity to $\theta_{13}$, illustrated in fig.\fig{q13future}, 
{\sc DoubleCHOOZ} is `cheaper and faster' than the long-baseline experiments discussed in the next sections and than other reactor projects, because it will use the neutrino laboratory already built for CHOOZ.
This experiment will start in 2010, and a near detector will be added in 2011, allowing to reduce systematic uncertainties,
hopefully down to $\theta_{13}\approx (4\div 5)^\circ$.
Another more ambitious experiment is under construction at the {\sc Daya Bay} site in China: it aims to reach
a sensitivity down to $\theta_{13}\approx 3^\circ$.
A similar sensitivity could be reached by  {\sc TripleCHOOZ}: adding a bigger third far detector to the CHOOZ site.



Future reactor experiments can also study other issues.
If a $\theta_{13}$ signal is seen, a reactor experiment can measure $|\Delta m^2_{13}|$,
while long-baseline $\nu_\mu$-beam experiments are dominantly sensitive to $|\Delta m^2_{23}|$:
measuring both $\Delta m^2$ with uncertainty smaller than $\Delta m^2_{\rm sun}$ 
would allow to discriminate
direct from inverted neutrino mass hierarchy.
Other techniques can achieve this goal in more realistic ways.
Concerning the solar mixing angle $\theta_{12}$, an improved KamLAND-like reactor experiment
with base-line of about $50\km$
could measure it with $1\sigma$ accuracy comparable to future solar experiments.

\begin{table}[t]
$$\begin{array}{rccccccc}
\hbox{Experiment} & \hbox{from} & \hbox{to} & \hbox{baseline} & \hbox{beam} & \hbox{$\nu$ energy in GeV}  & \hbox{off-axis}& \hbox{start}\\ \hline
\hbox{K2K~\cite{K2K}}      & \hbox{KEK} & \hbox{Kamioka} & 250\,{\rm km} & \nu_\mu &\sim(0.5\div 2) &0 & 1999\\
\hbox{{\sc NuMi}~\cite{NuMi}}      & \hbox{FermiLab} & \hbox{Soudan} &  735\,{\rm km} & \nu_\mu &
\sim (2\div 10) &0 & 2005\\
\hbox{CNGS~\cite{CNGS}} & \hbox{CERN} & \hbox{Gran Sasso} &730\,{\rm km} & \nu_\mu &\sim (5\div 30) & 0 & 2006\\ \hline
\hbox{T2K~\cite{T2K}}      & \hbox{JPARC} & \hbox{Kamioka} & 295\,{\rm km} & \nubarnu_\mu &\approx(0.3\div 1.3) &2^\circ  &2008 \\
\hbox{NO$\nu$A~\cite{Nova}} & \hbox{FermiLab} & \hbox{Ash river} &810\,{\rm km} & \nubarnu_\mu& \approx (1\div 3) &0.8^\circ
& 2010 \\
\hbox{SPL~\cite{SPL}} & \hbox{CERN} & \hbox{Frejus} &130\,{\rm km} & \nubarnu_\mu& \approx (0.1\div 0.5) &?
& ? 
\end{array}
$$
\caption[Long-base-line experiments]{\label{tab:beams}\em
Main characteristics of long-base-line conventional  neutrino beams.}
\end{table}


\section{Conventional neutrino beams}\label{cbeams}\index{NuMi}\index{CNGS}\index{Minos}
Neutrino beam experiments are considered the main next step of
oscillation studies.
K2K  (in Japan), {\sc NuMi} (in USA) and CNGS (in Europe)
are the first long-baseline experiments.
Table~\ref{tab:beams} summarizes their main properties.
They all employ a $\nu_\mu$ beam
(preferred, at least initially, to a $\bar\nu_\mu$ beam because
$\nu_\mu$ detection
cross sections are about two times higher)
produced using the `conventional' technique
(see~\cite{reviews} for a review).
A proton beam with energy $E_p\sim 20 E_\nu$
is sent on a target; 
mesons with positive or negative charge are focused by magnetic horns;
the decays of the resulting charged pions and Kaons produce
the neutrino beam.  Using positively charged  mesons
one gets a $\nu_\mu$ beam,  typically polluted by a few \% of $\bar\nu_\mu$,
$\sim 1\%$ of $\nu_e$ and a few per mille of $\bar\nu_e$.

It is difficult to discuss the r\^ole of these experiments, that
had been planned when SK started,
with the goals of confirming the atmospheric anomaly and
exploring its main properties.
SK already provided first answers to these issues,
and future experiments are now being planned with new goals.

The K2K experiment~\cite{K2K} (already performed and discussed in section~\ref{K2K})
indeed confirmed the atmospheric anomaly, 
finding a result consistent with its oscillation interpretation, and shown how accurately
neutrino beam experiments could measure $\Delta m^2_{\rm atm}$,
if sufficient statistics is accumulated.
A measurement of $\Delta m^2_{\rm atm}$ with $10\%$ accuracy should be performed
by the {\sc NuMi} project~\cite{NuMi} (already stated and discussed in section~\ref{NuMi}).
K2K and {\sc NuMi}  have chosen the neutrino energy which allows to maximize the oscillation effect.

On the contrary, the CNGS project~\cite{CNGS} employs a higher $E_\nu$
(at the price of a lower oscillation probability),
somewhat above the $\nu_\tau\to \tau$ production threshold, 
with the goal of directly confirming the $\nu_\mu\to\nu_\tau$ character of atmospheric oscillations 
by detecting a few $\tau$ appearance events.
The experimental signal of a $\nu_\tau$ is a `kink' i.e.\  two vertices separated
by a distance comparable to $\tau_\tau = 0.086\,{\rm mm}$
(long-lived particles like $K$ and $\pi$ produce a background), that
could be directly seen with a fine-graned emulsion detector.
In practice the detector OPERA is built by alternating\label{OPERA}
emulsion  with some cheaper material (lead), that constitutes most of the detector mass, such that in most events one infers the presence of two separated vertices from the observed tracks.
In other detectors, one can select a class of $\tau$-like events by
appropriate cuts (`statistical analysis') analogously to what SK did for
atmospheric neutrinos.

A long-baseline neutrino beam also allows to test if $\nu_\mu$ travel at the speed of the light 
within $2~10^{-6}$ accuracy,
improving by a factor 20 on the previous constraint. However SN1987A data 
already constrain  the $\bar\nu_e$ velocity with  $2\cdot 10^{-9}$ accuracy
(at the light of oscillations this constraint should now be reinterpreted in terms
of $\bar\nu_{1,2,3}$ velocities).

\medskip

Discovering $\theta_{13}$ is today considered the main goal of future experiments.
These first long-baseline beam projects can moderately improve on the present bound,
as shown in fig.\fig{q13future}.
However, these experiments cannot discriminate the neutrino mass hierarchy, 
and are not sensitive to CP-violation in neutrino oscillations.
Therefore, an approximation for the $\nu_\mu\to\nu_e$
oscillation probability enough accurate for these experiments
can be obtained by neglecting the solar mass splitting in
the vacuum oscillation formula of\eq{3nuemu}, 
and by improving it taking into account that earth matter effects
dominantly affect the frequency of $\nu_\mu\to\nu_e$ oscillations:
\beq\label{eq:PCNGS}
P(\nu_\mu\to\nu_e) \simeq 
\sin^2 2\theta_{13} \sin^2\theta_{23} \frac{\sin^2((1-r)\delta)}{(1-r)^2},
\eeq
where
$\delta = \Delta m^2_{13} L/4E_\nu$ is the atmospheric oscillation phase in vacuum and
\beq\label{eq:rNeDm}
r\equiv \frac{2\sqrt{2}G_{\rm F} N_e E_\nu}{\Delta m^2_{13}}=
\frac{N_e}{1.3\,N_A/\cm^3}\frac{E_\nu}{10\GeV}
\frac{2\cdot 10^{-3}\eV^2}{\Delta m^2_{13}}\eeq
is an adimensional ratio that controls the
relative importance of matter effects.
Notice that $\Delta m^2_{13}\equiv  m_3^2 - m_1^2$, so that
$\delta,r>0$ ($\delta,r<0$) if neutrinos have (normal) inverted mass hierarchy.
$P(\bar\nu_\mu\to\bar\nu_e)$ is obtained from\eq{PCNGS} by replacing $r\to - r$.


The goal of oscillation studies is  not only $\theta_{13}$ but also determining the type of neutrino mass hierarchy,
and possibly discovering CP-violation.  
To reach these extra goals, studying both $\nu_\mu$ and $\bar\nu_\mu$ beams is essential.
We now give an explicit approximate expression for the oscillation probability,
and later discuss the experiments planned to address all these issues.

\subsection{Approximating the oscillation probability}
In all planned beam experiments neutrinos travel
through a matter density which is constant to a good approximation:
the maximal depth $z$ reached by a beam with path-length $L$ much smaller than the earth radius
$r_E$ is
$z \simeq L^2/8r_E^2\sim 20\,{\rm km} (L/1000\,{\rm km})^2$.
The oscillation probabilities can be easily found by numerically computing the 
exponential of the $3\times 3$ matrix $H$ given in eq.\eq{m+V}:
$P(\nu_i \to \nu_f) = \exp (-i L H)_{fi}$.
One can derive a more explicit analytical approximation by splitting $H = H_0 + H_1$
where $H_0$ contains all `large' effects:
($\theta_{\rm atm},\Delta m^2_{\rm atm}$ in the $\mu/\tau$ sector and earth matter effects)
and $H_1$ contains the remaining `small' effects
($\theta_{13}$ and solar oscillations).
The formula
$$e^{-iL(H_0+H_1)}=e^{-iLH_0}+\int_0^1 dx\, e^{-iL(1-x)H_0}\cdot (-iLH_1) \cdot e^{-iLxH_0} + {\cal O}(H_1^2)$$
then allows to expand the survival probabilities  in the small parameters 
$$\varepsilon \equiv \Delta m^2_{\rm 12}/\Delta m^2_{13}\approx \pm 0.04\qquad\hbox{and}\qquad
\varepsilon'\equiv \sin2\theta_{13}\circa{<}0.2.$$
The matrix $H_0$ is immediately exponentiated, because matter effects are diagonal in
the $\mu/\tau$ sector, and does not generate $\nubarnu_e\leftrightarrow\nubarnu_\mu$ oscillations.
These oscillations are generated at first order by $H_1$ as
\beq \label{eq:P=abs^2}
P( \nu_\mu\to \nu_e) \simeq\left|
\varepsilon  e^{i\phi}  \cos\theta_{23}   \sin 2\theta_{12}\frac{e^{-2ir\delta }-1}{2r}+
\varepsilon'  \sin\theta_{23}\frac{e^{-2ir\delta}-e^{-2i\delta}}{2(1-r)}\right|^2.\eeq
where $\delta$ (the atmospheric oscillation phase in vacuum)
and $r$ (that controls matter effects) are defined in \eq{rNeDm}.
In the limit $L\to 0$ eq.\eq{P=abs^2} reduces to the expected 
$P(\nu_e\to \nu_\mu) \simeq |H_{e\mu}L|^2$.
Converting the exponentials to trigonometrical functions gives
\begin{eqnarray}\label{eq:Pmueapprox}
P(\nu_\mu\to\nu_e) &\simeq &
\varepsilon^2 \sin^22\theta_{12}\cos^2\theta_{23} \frac{\sin^2 (r\delta)}{r^2}+
\varepsilon^{\prime2} \sin^2\theta_{23} \frac{\sin^2((1-r)\delta)}{(1-r)^2}+\\
&&+ \varepsilon\varepsilon' \sin (2\theta_{12})\sin(2\theta_{23})
\frac{\sin(r\delta)\sin((1-r)\delta)}{r(1-r)}\cos(\delta+ \phi)\nonumber
\end{eqnarray}
$P(\bar\nu_\mu\to\bar\nu_e) $ is obtained by substituting $r\to -r$ in eq.\eq{Pmueapprox}.
$P(\nu_e\to \nu_\mu)$ is obtained by substituting $\delta \to - \delta$.
$P(\bar\nu_e\to \bar\nu_\mu)$ is obtained by substituting $\delta \to - \delta$ and $r\to -r$.

\medskip

\begin{figure}[t]
$$\includegraphics[width=0.97\textwidth]{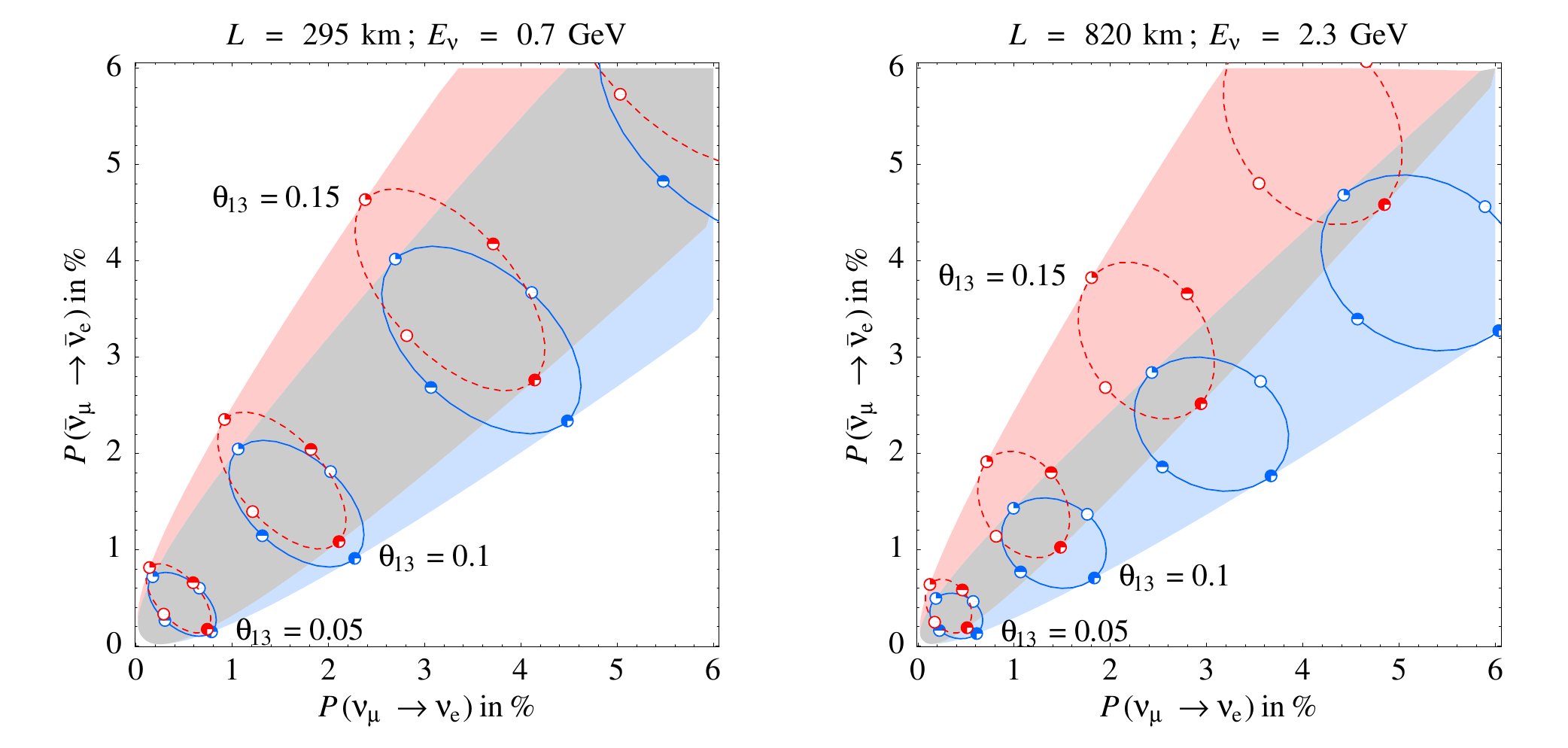}$$
\caption[Sensitivity to CP violation]{\label{fig:Ellissi}\em $\nubarnu_\mu\to\nubarnu_e$ oscillation probabilities
for the present best-fit values of $|\Delta m^2_{13}|$, $\Delta m^2_{12}$, $\theta_{12}$, $\theta_{23}$,
for $\theta_{13}=\{0.05,0.1,0.15,0.2\}$,
$N_e = 1.5\,{\rm moles}/{\rm cm}^3$, and varying the CP-phase $\phi$.
The points correspond to $\phi = 0$ (empty circles),
$\pi/2$ (1/4 filled circles), $\pi$ (half filled circles), $3\pi/4$ (3/4 filled circles).
The continuous blue (red dashed) lines correspond to normal (inverted) mass hierarchy.
The left (right) plot shows a T2K-like (NO$\nu$A-like) configuration.}
\end{figure}

The above expressions confirm that,
since the earth is not CP symmetric, matter effects create a fake CP asymmetry:
($P(\nu_e\to \nu_\mu)\neq P(\bar\nu_e\to\bar\nu_\mu)$ even if $\phi=0$)
and show that they do not create a fake T asymmetry
($P(\nu_e\to \nu_\mu)= P(\nu_\mu\to \nu_e)$ if $\phi=0$).
This happens because we here assumed a constant matter density:
in general the matter density profile gets reversed creating also a fake T asymmetry.
Experiments performed only with $\nubarnu_\mu$ beams must subtract the fake CP asymmetry,
but this is not a problem as earth matter effects are well known.
Detailed experiment-dependent matter profiles can be obtained with the collaboration of  geologists.

At least two measurements are necessary to reconstruct $\theta_{13}$, $\phi$
and the neutrino mass hierarchy~\cite{LBL}.
In fig.\fig{Ellissi} we show how 
 the two measurable $\nu_\mu \to \nu_e$ and $\bar\nu_\mu\to\bar\nu_e$
oscillation probabilities 
(computed for T2K-like and NO$\nu$A-like choices of the neutrino path-length and energy)
depend on  $\theta_{13}$ and $\phi$.
In agreement with the approximate formula of eq.\eq{Pmueapprox},
varying $\phi$ at fixed $\theta_{13}$ approximatively give ellipses.
Ellipses corresponding to normal mass hierarchy are partially shifted
from ellipses corresponding to inverted mass hierarchy, because of earth matter effects.
We see how the amount of overlap depends on $\theta_{13}$ and 
is smaller in the NO$\nu$A-like configuration than in the T2K-like configuration.
This means that
a measurement of these two oscillation probabilities could be or could be not able
of univocally determining $\theta_{13}$ and $\phi$ and the neutrino mass hierarchy,
depending on which value is found.
(A central value outside the theoretically allowed region would falsify the theoretical framework).
Having two different experimental configurations helps in 
measuring oscillation parameters without discrete ambiguities.
For example, the T2K experiment could be modified 
building one of the two planned 0.27 Mton detectors in Kamioka, Japan 
(baseline $L\approx 250\,{\rm km}$)
and the  second one in Korea 
(`T2KK' option, $L\sim 1000\,{\rm km}$): this would improve the sensitivity to the mass hierarchy.
Furthermore, a measurement of $\theta_{13}$ from reactor experiments would
restrict along some ellipse.
These examples show how fig.\fig{Ellissi} allows to visualize the interplay between
different measurements.

Many papers tried to discuss which  configuration is the optimal one,
but the result depends on the unknown parameters we would like to discover.
Therefore experiments will likely be performed at baselines
reasonably fixed by geopolitical considerations and with
a beam energy chosen such that the atmospheric phase is large.

\begin{figure}[t]
$$\includegraphics[width=0.97\textwidth]{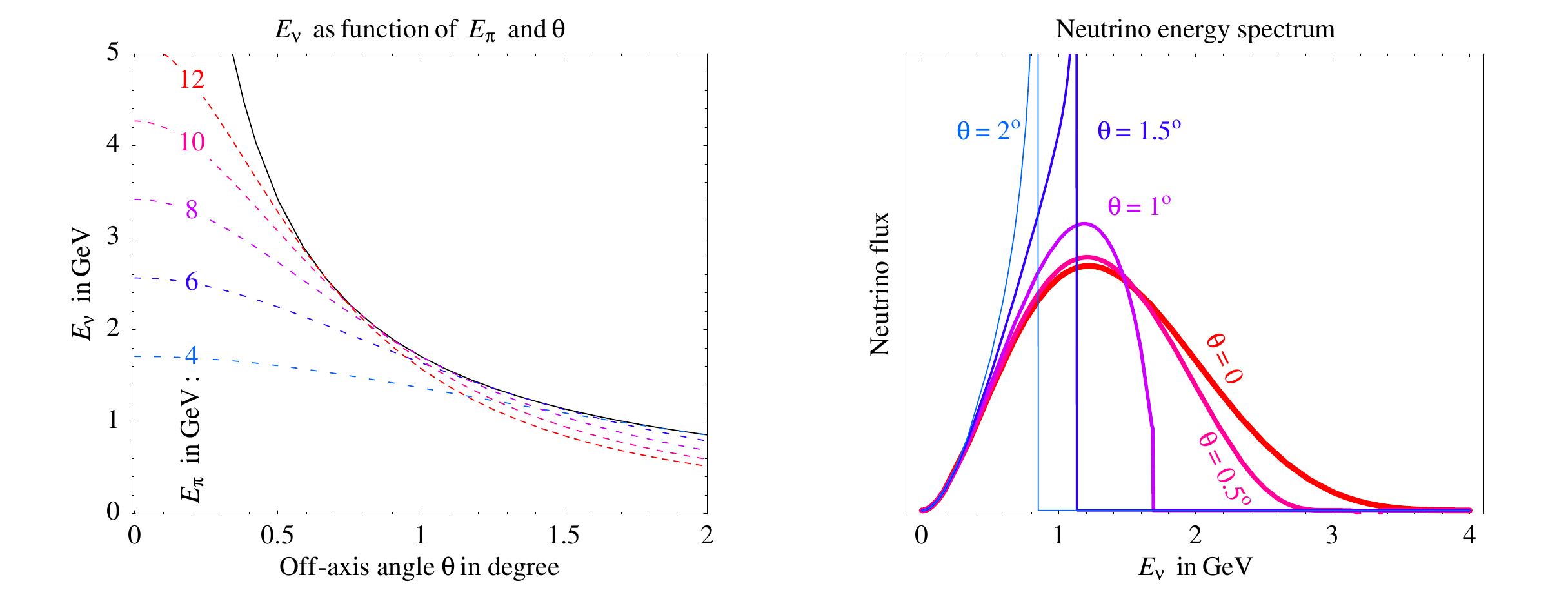}$$
\caption[Off-axis]{\label{fig:OffAxis}\em Fig.\fig{OffAxis}a:
$E_\nu$ produced at  angle $\theta$ with respect to the $\pi$ momentum
by the decay of a $\pi$ with energy $E_\pi$.
The continuous line shows the maximal $E_\nu$.
Fig.\fig{OffAxis}b: neutrino energy spectrum 
produced by $\pi$ decays with a typical energy distribution as
seen off-axis by an angle $\theta$.}
\end{figure}

\subsection{Off-axis super-beams}\index{Super beams}\index{T2K}\index{NO$\nu$A}\index{Off-axis}
Improved experiments, still based on `conventional' $\nubarnu_\mu$ beams,
can make significant progress towards clarifying the known unknowns
of neutrino oscillations.
There are two projects in this direction: T2K~\cite{T2K} and NO$\nu$A~\cite{Nova}.

First of all, present experiments can be improved by building i.) more intense beams;
ii.) bigger detectors. Although the figure of merit of a neutrino beam experiment is
proportional to
$$\hbox{(intensity of the source)} \times \hbox{(size of the detector)}$$
each one of the two factors is separately relevant for other experiments,
such that one has to find the best distribution of resources for a whole set of experiments:
\begin{itemize}
\item[i.)] The factor that presently limits the beam intensity is the target used to convert protons into mesons,
since it gets destroyed by a too intense proton beam.
It seems possible to build proton drivers that can work up to a few MWatt power.
An intense proton driver also produces muons and other particles, allowing to perform
several `high intensity' experiments other than neutrino oscillation studies:  
searches for rare or forbidden processes 
(e.g.\ $\mu\to e$ conversion, $\mu$-EDM, etc.) and
maybe for dark matter~\cite{DMaccumulator}.

\item[ii.)] It seems possible to build a Mton-scale neutrino detector~\cite{Mton}.
A neutrino beam can be pulsed, allowing to relax the background requirements.
However, suppressing the backgrounds would greatly increase the scientific interest of the project,
allowing high-statistics studies of solar neutrinos, atmospheric neutrinos
supernova neutrinos and to search for proton decay.

 \end{itemize}
  The T2K project could hopefully culminate in having a
 Mton `HyperKamiokande' W\v{C} and a beam generated by a 4MW proton driver.
Around this point systematic errors start to dominate over statistical errors,
and further improvements would need new techniques, as discussed in the next sections.
 The NO$\nu$A project can be competitive with T2K on the $\theta_{13}$ issue, but
 the detector will not be underground and can therefore be used almost only 
 for the neutrino beam experiment.
 
 \medskip
 
 Another improvement employed by both T2K and NO$\nu$A consists
 in putting the detector off-axis (with respect to the center of the neutrino beam): in this way
the neutrino beam acquires interesting features that make it more apt for oscillation studies
(more restricted energy) and in particular for $\nu_\mu\to \nu_e$ searches
(reduced $\nu_e$ contamination).
To see this, we recall that the decaying pions have spin 0: therefore in their Center of Mass  (CM) frame 
$\pi\to \mu\nu_\mu$ decays produce an isotropic distribution of $\nu_\mu$
with fixed energy $E_\nu^{\rm CM} = (m_\pi^2- m_\mu^2)/2m_\pi \approx 30\MeV$.
Then kinematics allows to compute the energy of neutrinos emitted
with angle  $\theta$ with respect to the momentum of $\pi$  with energy $E_\pi \gg m_\pi$:
\beq E_\nu = \frac{2 E_\nu^{\rm CM} E_\pi m_\pi}{m_\pi^2 + E_\pi^2 \tan^2\theta}.\eeq
This function is plotted in fig.\fig{OffAxis}a: 
the upper bound $E_\nu < E_\nu^{\rm CM}/\tan\theta$ (continuous line)
is the most notable feature, that can be understood as
\beq  E_\nu \simeq p_\nu^\parallel   = \frac{p_\nu^\perp}{\tan\theta} = \frac{E_\nu^{\rm CM}\sin\theta^{\rm CM}}{\tan\theta}\le \frac{E_\nu^{\rm CM}}{\tan\theta}.\eeq
Therefore neutrinos emitted off-axis have a maximal energy $E_{\rm max}$,
and actually the off-axis neutrino beam has a more narrow energy distribution peaked
just below this maximal energy.
Fig.\fig{OffAxis} shows sample  energy spectra for a few off-axis angles,
computed assuming a $\pi$ beam with a typical energy distribution, 
$dN/dE_\pi \propto (1-E_\pi/E_p)^5$
where $E_p=10\GeV$ is the energy of the proton beam used to produce $\pi$.
(Full MonteCarlo computations take into account other mesons, and the spread
in the meson directions).
We see that going off-axis the flux decreases at higher energy and increases at lower energy.
These features help oscillation experiments,
because neutrinos of higher energy have higher cross-sections and lower oscillation probabilities.
The narrower energy spectrum helps in avoiding that detailed
oscillation features (such as CP-violation) get suppressed by the average over
the energy spread.

Furthermore, going off-axis and selecting events with energy around the peak
allows to reduce the $\nu_e$ background contamination of the beam from about
$1\%$ to about $0.2\%$, see e.g.\ section 3 of the JHF project~\cite{T2K}.
Indeed this background is generated by two sources:
by $K\to \pi e \nu$ and by $\pi\to \mu\to \nu_e$ decays.
and the first source becomes less relevant at the energy where $\pi$ decays accumulate.

A different possible set-up sensitive to CP violation
 consists in sending  a on-axis broad-band $\nubarnu_\mu$ beam
with $E_\nu \sim (1\div 5)\GeV$ to a big ($\sim$Mton) W\v{C} detector located at distance
$L\sim 2000\,{\rm km}$, able of measuring the energy of quasi-elastic events
with $10\%$ accuracy~\cite{BroadBandBeam}.

\index{Neutrino!factory}
\section{Neutrino factory}
Technologies for producing a muon beam had been initially explored with the purpose of
building a muon collider with TeV-scale energy.
This seems impossible because muon decays would produce  TeV-scale neutrinos with 
cross sections low enough that they cannot be shielded, but high enough
to give unacceptable radiation hazards.
Reducing the muon energy down to $E_\mu \sim 10\GeV$
simplifies the technology, and muon decays can be used to obtain a neutrino beam.
This machine, called `neutrino factory'~\cite{NuFact}, is considered as feasible,
on long time-scales (after 2020?) and probably with G\EURtm-scale costs.

The neutrino-factory beam is produced by circulating $\mu^-$ or $\mu^+$ beams
with energy $E_\mu$ in accumulators with km-scale straight sections.
Conventional neutrino beams from $\pi^-$ (or $\pi^+$) decays
are dominantly composed by $\bar\nu_\mu$ (or $\nu_\mu$).
On the contrary, a neutrino beam produced by decays of $\mu^-$ (or $\mu^+$)
consists of  $\nu_\mu + \bar\nu_e$ (or $\bar\nu_\mu + \nu_e$).
This makes easier to build big detectors
dedicated to studies of $\nubarnu_\mu\leftrightarrow\nubarnu_e$ oscillations:
with a conventional neutrino beam their signature is $e^\pm$ appearance;
with a neutrino factory beam their best signature becomes `wrong-sign' $\mu^\pm$ appearance:
e.g.\ $\mu^+$ decays produce $\bar\nu_\mu$ (detected as $\mu^+$)
and $\nu_e$ (detected as $e^-$), that can oscillate
into $\nu_\mu$ (detected as $\mu^-$).
Big detectors with poor granularity are more sensitive to
muons than to electrons, because muons have a much longer range in matter.

The energy  and  flavour spectra of a $\nu$-factory  beam is easily and accurately computed
using the known formul\ae{} that describe
$\mu\to e\bar\nu_e \bar\nu_\mu$ decays.
The resulting flux would be known at per-mille level.
E.g.\ $N_\mu$ decays of unpolarized beamed $\mu^+$ with energy $E_\mu\gg m_\mu$, 
produce the following neutrino fluxes along the beam axis and at distance $L$ from the decay region:
\beq \frac{dN_{\bar\nu_\mu}}{dx~dS}= \frac{N_{\mu^+}}{\pi L^2} \bigg(\frac{E_\mu}{m_\mu}\bigg)^2\cdot
2x^2 (3-2x),\qquad
\frac{dN_{\nu_e}}{dx~dS}= \frac{N_\mu}{\pi L^2} \bigg(\frac{E_\mu}{m_\mu}\bigg)^2\cdot
12 x^2 (1-x)\eeq
where $x = E_\nu/E_\mu$.
These are wide-band spectra with average energies
 $\langle E_{\nu_\mu}\rangle = 7E_\mu/10$ and $\langle E_{\bar\nu_e}\rangle = 3E_\mu/5$.
 $N_\mu$ could be as high as $10^{21}$ per year.
 The flux increases with $E_\mu^2$ because more energetic muons produce
 a more narrow neutrino beam, with opening angle  $\theta\sim m_\mu/E_\mu$.

\medskip

The following channels seem more promising:
\begin{enumerate}
\item {\em `Golden channel'}. $\nu_e\to \nu_\mu$ oscillations are signaled by
 $\mu^-$ appearance,
best seen by  W\v{C} detectors or tracking calorimeters.
Parameterizing the oscillation probability as
$P(\nu_e\to\nu_\mu) \approx |\Delta m^{2}_{\rm eff}(L,E)\cdot L/2E_\nu|^2$
the number of $\mu^-$ events is 
\begin{equation}\label{eq:shortLnuf}
N_{\mu^-} \approx \frac{N_{\mu^+} N_{\rm
kt}\epsilon}{10^{21}}\frac{E_\mu}{70\GeV}\left|\frac{\Delta_{e\mu}^{\rm eff}}{10^{-5}\eV^2
}\right|^2
\end{equation}
where $N_{\rm kt}$ is the size of the detector in kilo$\cdot$tons and
$\epsilon$ is its efficiency.
At short $L\circa{<}500\,{\rm km}$ and large enough energy one simply has
$$\Delta m^2_{\rm eff} \simeq (m\cdot m^\dagger)_{e\mu}\simeq
 \Delta m^2_{\rm atm} \theta_{13}\sin\theta_{23}+
e^{i\phi}\Delta m^2_{\rm sun}\cos\theta_{23}\sin \theta_{12}\cos\theta_{12}$$
and the generic approximation for the oscillation probability is given in eq.\eq{P=abs^2}.

In anti-neutrinos, the related process $\bar\nu_e\to \bar\nu_\mu$ gives a $\mu^+$
signal with a rate about 2 times lower than in neutrinos.

\item  {\em `Silver channel'}.
$\nu_e\to\nu_\tau$ oscillations are signaled by $\nu_\tau$ appearance, that can be
detected as described at page~\pageref{OPERA}.
Detecting both the `golden' and the `silver' channel would allow to test
if CP violation is produced by a unique phase, as predicted in the standard
neutrino scenario (where the two channels should have CP asymmetries with opposite sign).

\item {\em `Bronze channel'}.  While $\nu_\mu \to \nu_e$ searches are considered
as hopeless (because the signal would be $e^-$ appearance),
searches for the $e^+$ produced by
$\bar\nu_\mu\to\bar\nu_e$ oscillations 
could be performed with a magnetized liquid argon TPC,
which is maybe not too unrealistic.

\end{enumerate}
Eq.\eq{shortLnuf} allows to estimate the sensitivity of a neutrino factory.
The sensitivity to $\theta_{13}$ is illustrated in fig.\fig{q13future},
and is accompanied by an interesting sensitivity to CP-violation in neutrino oscillations
and to their mass hierarchy.
The relative merit of a neutrino factory versus a conventional neutrino beam is being actively studied:
a neutrino factory seems to have a better sensitivity if $\theta_{13}$ is small,
while, if $\theta_{13}\circa{>}0.05$, systematic errors dominate and  the situation is not yet fully clarified.
Indeed, neutrino factory plans have been optimized for small values of $\theta_{13}$;
while if $\theta_{13}$ is large  a short baseline would be preferable
to reduce earth matter uncertainties.
The fact that no technology seems clearly preferable to the others possibly means that 
the choice is not much important.

Furthermore, it has been discussed which energy and baseline $L$ allows the `best' experiment.
For example, by choosing   $L=\sqrt{2}\pi/G_{\rm F}N_e \approx 7400\km$ (known as `magic baseline'~\cite{NuFact}; the CERN-INO~\cite{Monolith} distance is 7152 km)
one has
$\sin(r\delta)=0$ such that, for all $E_\nu$, the oscillation probability in eq.\eq{Pmueapprox} simplifies
to its second term only, which no longer depends on $\phi$ nor on solar parameters,
allowing a direct measurement of
$\theta_{13}$. Such a long baseline would require 
building a long inclined and expensive decay tunnel for $\mu^+$;
furthermore uncertainties on the earth density profile might be a serious limitation.


\section{Beta beams}\index{Beta beams}
The neutrino factory concept can be technologically simplified 
(although a km-scale decay ring is still needed)
by replacing muons
with a $\beta$-decaying ionized nucleus~\cite{BetaBeams}, allowing to produce intense, 
perfectly pure and (if desired) pulsed
$\bar\nu_e$ or $\nu_e$ beams.
Therefore one can perform $\nubarnu_e\to\nubarnu_\mu$ searches 
with a W\v{C} detector, or with any other detector unable of identifying the $\mu^\pm$ charge, 

One can choose a nucleus which has a conveniently long life-time (e.g.\ $\tau\sim\sec$),
which produces neutrinos with known energy spectrum
(reconstructed fro measurements of the $e^\pm$ spectrum), and
which can be easily produced (spallation neutrons allow to  get up to $\sim 10^{13}$ nuclei/sec).
By accelerating the ionized nucleus up to energy $E = \gamma m$
one can focus the neutrino beam into a cone with opening angle $\sim 1/\gamma$.
In practice the neutrino rate might be too much limited because:
(i) the number of ions that it is possible to collect limits the neutrino flux;
(ii)  ions have a lower $q/m$ than protons or muons, such that,
the resulting neutrino beta-beam reaches a lower energy than a conventional beam
or a neutrino factory beam.
Let us discuss two concrete proposals
\begin{itemize}
\item A $\bar \nu_e$ beam can be obtained collecting 
$\approx 5~10^{13}$ $^6{\rm He}^{++}$ nuclei/sec
(this number might be too optimistic):
their decay
$^6{\rm He}^{++}\to \hbox{$^6{\rm Li}^{+++}$} ~e^- ~ \bar\nu_e$,
boosted at $\gamma\approx 150$ generates
a $\bar\nu_e$ beam with energy around $0.5\GeV$, that produces
$70$ events per year in a kton detector located  $L\approx 130\km$ away.

\item A $\nu_e$ beam can be obtained collecting
$\approx 10^{12}$ $^{18}$Ne nuclei/sec:
their decay
$^{18}{\rm Ne}\to \hbox{$^{18}{\rm F}$}~e^+~\nu_e$,
boosted at $\gamma\approx 250$ generates
a $\nu_e$ beam with energy around $\GeV$, that produces
$2$ events per year in a kton detector located  $L\approx 130\km$ away.
\end{itemize}
Several variations are possible.
The energies $E = \gamma/m$ quoted above can be reached with a small
accelerator, such as the CERN SPS.
Beta beams with higher energy, $E_\nu\sim\hbox{few GeV}$
need a larger accelerator, like TeVatron or even LHC,
and allow detection after a  longer baseline $L\sim 1000\km$
(possibly using rock as target): earth matter effects become
more important increasing the sensitivity to the neutrino mass hierarchy
and reducing the sensitivity to CP violation.

Beta beams with lower energy, $E_\nu\sim 10\MeV$, allow
nuclear experiments, that study e.g.\ $\nu N$ interactions relevant for supernova physics.
Furthermore, $\beta$-beam techniques also allow to obtain a {\em monochromatic}
neutrino beam, by using a weak decay with an electron in the final state 
(`electron capture', e.g.\ ${\rm C}^{11} e^- \to {\rm B}^{11}\nu_e$),
such that the final state involves only two particles
(while $\beta$-decays have at least three particles in the final state,
and therefore a continuous energy spectrum).



Finally, we mention the somewhat related possibility of using the reaction
$A_{Z-1}\leftrightarrow A_Z\, \bar\nu_e\,e^-$
(where the two atoms $A$ could be $^3{\rm H}$ and $^3{\rm He}$)
for emitting and resonantly detecting monochromatic $\bar\nu_e$ with energy
$E_\nu = M(Z) - M({Z-1})\sim 20\keV$ and $\Delta E_\nu/E_\nu\sim 10^{-17}$~\cite{LowNuResonant}.
If future experiments will be able of observing this process at baselines $\sim 10\cm$,
one can use it for performing searches for $\theta_{13}$ at baselines
$\sim10\,{\rm m}$, discriminating the mass hierarchy, and for testing gravitational red-shift of neutrinos.


%% file: review_nonosc.tex
\chapter{Non-oscillation experiments}\label{nonOsc}
Oscillation experiments are insensitive to 
the absolute neutrino mass scale (parameterized by the mass of the 
lightest neutrino) and to the 2 Majorana phases $\alpha$ and $\beta$.
Other types of experiments can study some of
these quantities and the nature of neutrino masses. They are:
\begin{itemize}
\item $\beta$-decay experiments, that to good approximation probe
$m_{\nu_e}^2\equiv (m\cdot m^\dagger)_{ee}=\sum_i |V_{ei}^2| m_i^2$; 

\item neutrino-less double-beta decay ($0\nu2\beta$) experiments, that probe 
the absolute value of the $ee$ entry of the neutrino
Majorana mass matrix $m$, $|m_{ee}| = |\sum_i V_{ei}^2 m_i|$;

\item cosmological observations 
(Large Scale Structures and anisotropies in the Cosmic Microwave Background), 
that to good approximation probe the sum of neutrino masses,
$ m_{\rm cosmo}\equiv m_1 + m_2 + m_3$.

\end{itemize}
Only $0\nu 2\beta$ probes the Majorana nature of the mass.
The values $|m_{ee}|, m_{\nu_e}, m_{\rm cosmo}$
 are unknown, and can be partially inferred from oscillation data.

Ordering these probes according to their present sensitivities, 
the list is cosmology, $0\nu2\beta$ and finally $\beta$ decay.
Ordering them according to reliability would presumably result into 
the reverse list: cosmological results are based on plausible theoretical assumptions, 
and $0\nu2\beta$ suffers from severe 
uncertainties in the nuclear matrix elements.

Present data contain a few anomalies.
There is a claim that the $0\nu2\beta$
transition has been detected~\cite{evid} (see section~\ref{HM}), 
there is a persisting anomaly in {\sc Troitsk} $\beta$ decay,
and even in cosmology, there is one (weak) claim for  a positive effect. 
None of these hints can be considered as a  discovery of neutrino masses,
but experiments seem not far from reaching the necessary sensitivity.
Existing or planned experiments  will lead to progress in a few years.

\begin{table}[t]
$$\begin{array}{|ccc|cc|}\hline
\hbox{non-oscillation} &\hbox{probed} &\hbox{experimental} &  \hbox{$99\%$ CL range} &  \hbox{$99\%$ CL range}\\
\hbox{parameter}&\hbox{by}&\hbox{limit  at $99\%$ CL}& \hbox{normal hierarchy} &   \hbox{inverted hierarchy}\\ \hline
\hbox{$ee$-entry of $m$}&\hbox{$0\nu2\beta$} &m_{ee}< 0.39\,h\,\eV&(1.1\div 4.5)\meV  & (12\div 57)\meV\\
\hbox{$(m^\dagger m)^{1/2}_{ee}$}&\hbox{$\beta$-decay} & m_{\nu_e}<2.1\,\eV& (4.6\div10)\meV &(42\div 57)\meV \\
\hbox{$m_1+m_2+m_3$} &\hbox{cosmology}& m_{\rm cosmo}\circa{<}0.5\,\eV &(51\div 66)\meV& (83\div 114)\meV\\ \hline
\end{array}$$
\caption[Non-oscillation data]{\em Summary of present constraints on  non-oscillation neutrino mass parameters.
Some $0\nu2\beta$ data are controversial, and  $h\sim 1$ parameterizes uncertain nuclear matrix elements.
The last two columns show the oscillation predictions assuming that the lightest neutrino 
is massless in the two cases of normal (i.e. $m_1\ll m_2\ll m_3$)
and inverted (i.e. $m_3\ll m_1<m_2$) mass hierarchy.
In the opposite limit neutrinos are quasi-degenerate and $|m_{ee}|, m_{\nu_e}, m_{\rm cosmo}$ can be arbitrarily large.
\label{tab:nonosc}}
\end{table}

\section{Cosmology}
There is a non obvious link between cosmological data and neutrino masses,
explained in section~\ref{cosmology} (mainly in its subsection~\ref{CMB}).
Here we give a short summary of results.

Cosmological data roughly probe mostly the
sum of neutrino masses:
$ m_{\rm cosmo}=m_1+m_2+m_3$,
that within standard cosmology
controls the present energy fraction $\Omega_\nu$ in non relativistic neutrinos
as $\Omega_\nu h^2 = m_{\rm cosmo}/ 93.5\eV$, where as usual
$h\approx 0.7$ parameterizes the present value of the Hubble constant as
$h\equiv H_{\rm today}/(100 \hbox{km/s\,Mpc})$.
Cosmology does not distinguish Majorana from Dirac neutrino masses.

In order to convert CMB and LSS data into a constraint on neutrino masses
one needs to assume a cosmological model.
The cosmological constraint~\cite{WMAP} assumes
that the observed structures are generated by Gaussian adiabatic primordial scalar fluctuations
with a constant spectral index $n$
evolved in presence of the known SM particles, of cold dark matter and of a cosmological constant.
This standard model of cosmology seems consistent with all observations.
CMB data alone constrain $m_1+m_2+m_3<2.6\eV$ at 99\% C.L.
LSS data are more strongly affected by neutrino masses,
and give stronger constraints, after assuming that observed luminous matter 
tracks the dark matter density up to a bias factor.
Neutrino masses have the largest impact at scales so small that nowadays
inhomogeneities no longer are a minor correction
to a uniform background, such that computations become difficult and
theoretical uncertainties can become problematic.
Lyman-$\alpha$ data probe inhomogeneities at such small scales and at earlier times,
with imperfect agreement between different groups.
The resulting cosmological constraint
depends on how one deals with these difficulties: more risky approaches
give stronger constraints.
The value reported in table~\ref{tab:nonosc} is a reasonably conservative choice.

In the future the sensitivity to neutrino oscillations will improve thanks to better CMB data
and to new LSS measurements less plagued by potential systematic effects.
If cosmology were simple
(e.g.\ a spectral index $n=1$, no tensor fluctuations,\ldots)
then it seems possible to detect even neutrino masses 
as small as allowed by oscillation data~\cite{WMAP}.
The expected ranges of $m_{\rm cosmo}$ are reported in table~\ref{tab:nonosc}
in the limiting case where the lightest neutrino is massless,
and in fig.\fig{OscNonOsc}a in the general case.
Within standard cosmology and standard neutrinos, a
positive signal is guaranteed if a sensitivity down to $m_{\rm cosmo}\sim 50\meV$ is reached;
furthermore a precise measurement could identify the kind of neutrino mass hierarchy.

\section{Astrophysics}
Time delays between supernova neutrinos allows to constrain neutrino masses~\cite{timedelay}.
We discuss this technique very briefly because it presently gives sub-dominant bounds,
and is seems impossible to reach an interesting enough future sensitivity.

At the next gravitational
collapse of a supernova,
the general strategy will consist in identifying
structures in the time and/or energy distributions of
neutrinos  sensitive to neutrino masses, as the neutronization peak,
the rising (or falling) ramp of the
cooling phase, a hypothetical sharp cutoff due to
black hole formation.
The sensitivity of these approaches has been quantified
in several works, assuming the capabilities of
present detectors (SuperKamiokande, SNO, LVD,\ldots).
The difference in time of flight
between neutrinos and gravitons
will only be sensitive to neutrinos heavier than
about $1\eV$, comparable to present $\beta$-decay bounds.
The difference in time of flight between
different neutrinos will only be sensitive to neutrino
mass {\em differences}
larger than few $10\eV$.
If neutrino emission were suddenly
terminated by black hole formation,
a  measurement of
the difference in time of flight
between neutrinos of different energy
will only be sensitive to neutrino masses larger than few eV.

\section{$\beta$-decay}
\begin{figure}[t]
$$\includegraphics[width=5cm]{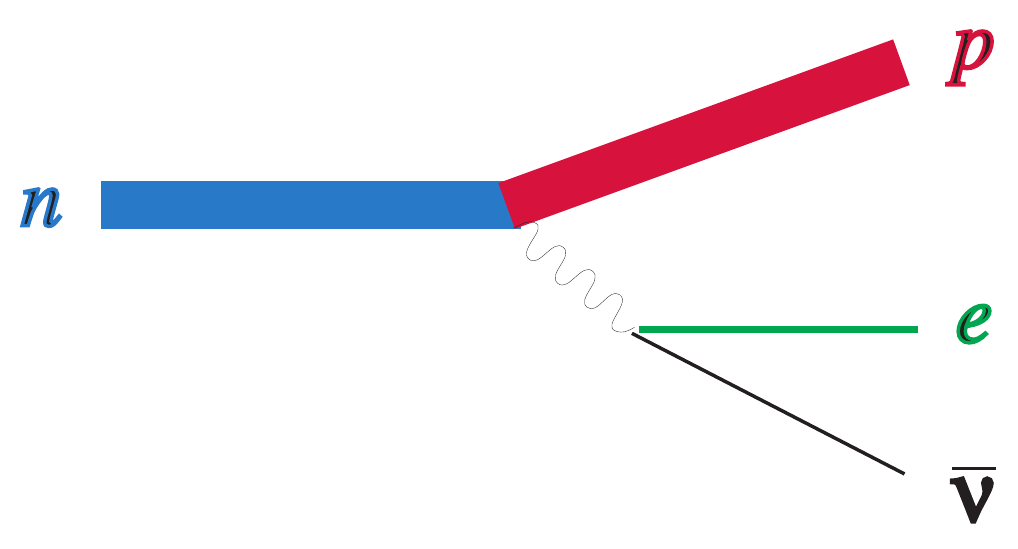}\hspace{0.5cm}\includegraphics[width=8cm]{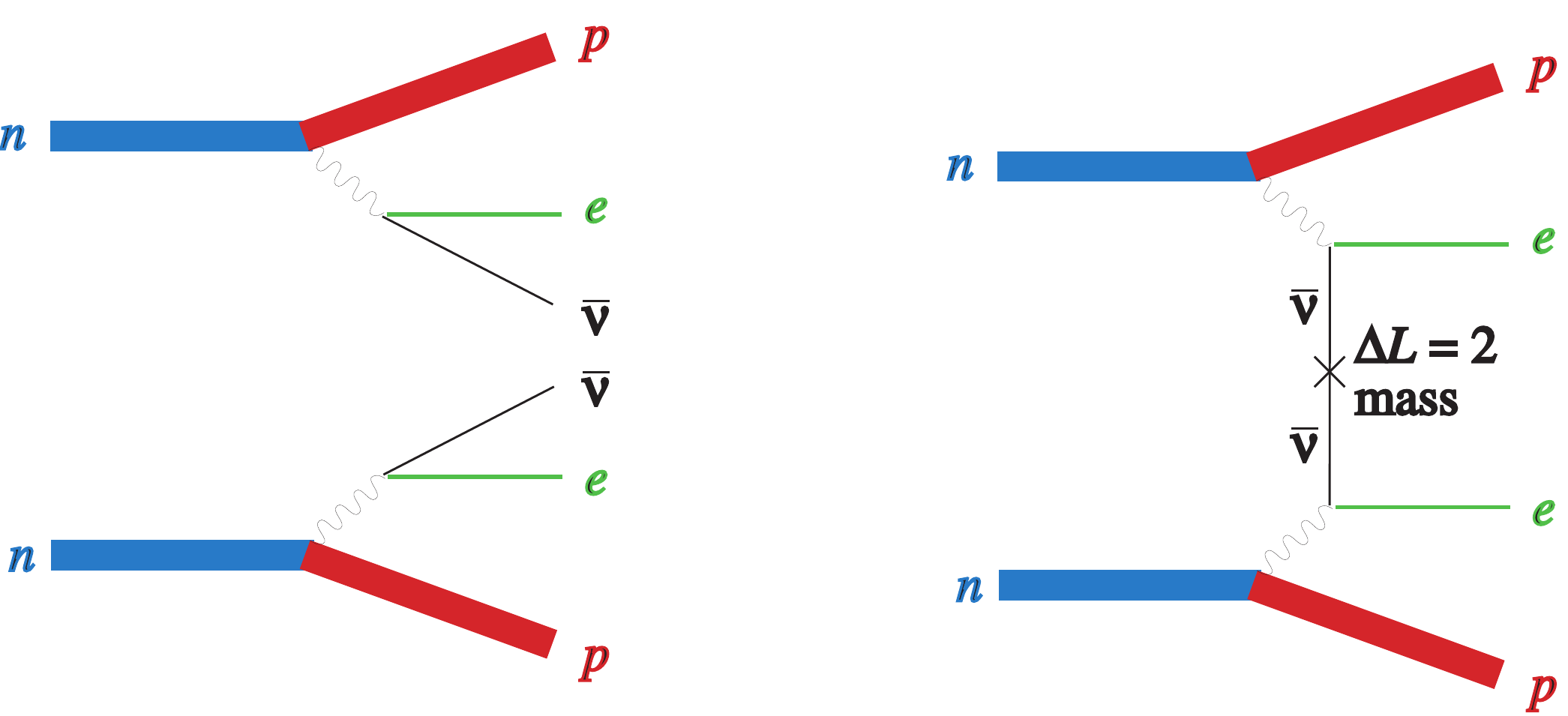}$$
\caption[$\beta$, $2\beta$ and $0\nu2\beta$ decay]{\label{fig:FeynBeta}\em Feynman diagrams for $\beta$ decay, double-$\beta$ decay, and neutrino-less double-$\beta$ decay.}
\end{figure}

\begin{figure}[t]
$$\includegraphics[width=8cm]{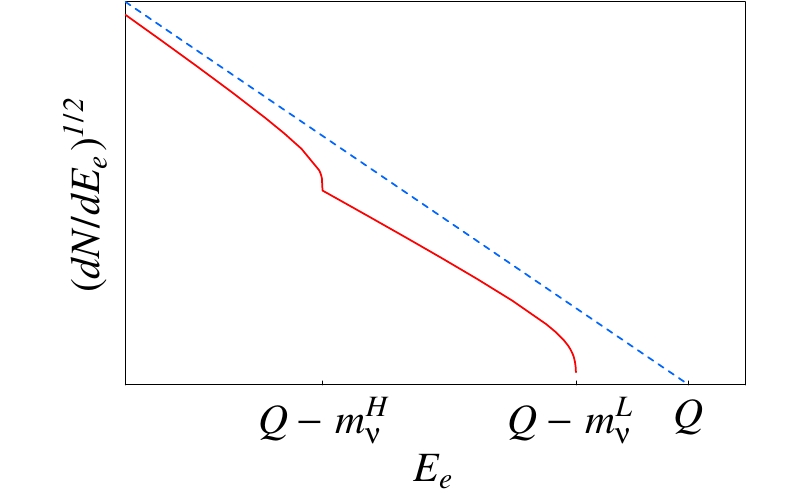}\hspace{1cm}
\includegraphics[width=8cm]{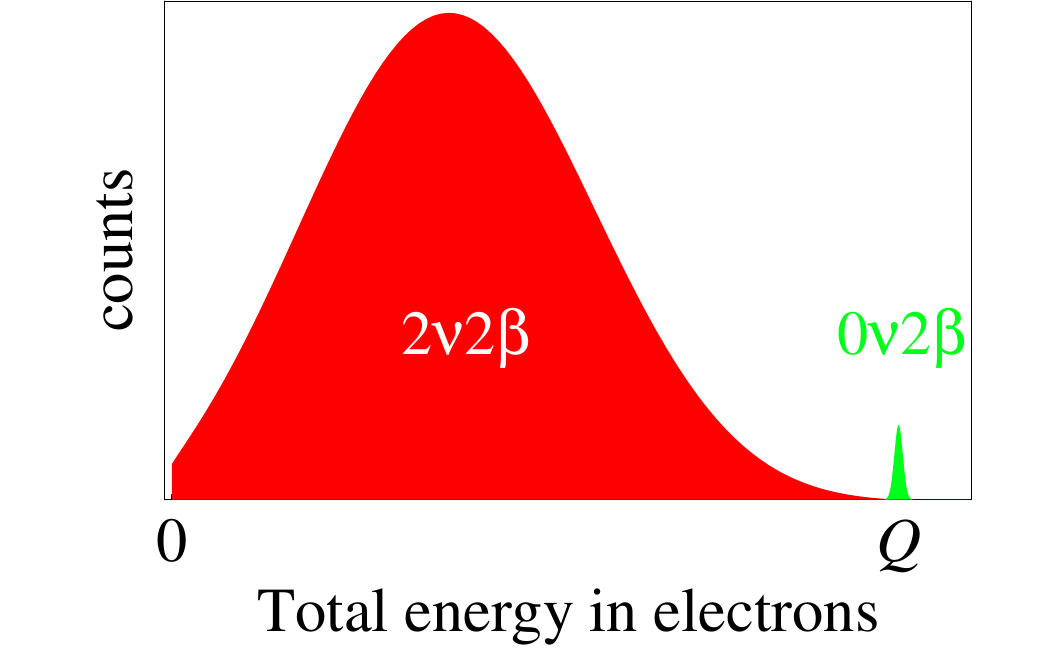}$$
\caption[Spectra of $\beta$ and $0\nu2\beta$ decay]{\em Fig.\fig{beta}a:
$\beta$-decay spectrum close to end-point for a massless (dotted) and massive (continuous line) neutrino.
Fig.\fig{beta}b: $2\nu2\beta$ and $0\nu2\beta$ spectra.
\label{fig:beta}}
\end{figure}

\index{$\beta$ decay}\label{betadecay}
Neutrino masses distort the electron spectrum in the $\beta$-decay  of a nucleus
(i.e.\   $d\to u e \bar{\nu}_e$ at the quark level, and 
$n\to p e \bar{\nu}_e$ at the nucleon level, see fig.\fig{FeynBeta}a).
The most sensitive choice is tritium decay
$$^3{\rm H} \to {}^3 {\rm He}\,  e\, \bar{\nu}_e \qquad (Q= m_{^3\rm H} - m_{^3\rm He} = 18.6\,\rm keV).$$
Energy conservation tells that $E_e \simeq  Q - E_\nu$.
The maximal electron energy is $Q-m_\nu$
(assuming that all neutrinos have a common mass $m_\nu$).
Around its end-point, the electron energy spectrum
is essentially determined by
the neutrino phase space factor $\propto E_\nu p_\nu $.
So
\begin{equation}\label{eq:beta1}
\frac{dN_e}{dE_e} = F(E_e) (Q-E_e)\sqrt{(Q-E_e)^2 - m_{\nu_e}^2}
\end{equation}
where $F(E_e)$ can be considered as a constant.
The signal produced by $m_\nu$ is illustrated in fig.\fig{beta}a.

The fraction of events in the end-point tail is $\propto (m_\nu/Q)^3$
and the relative electron energy resolution needed to be sensitive to neutrino masses
is $\sim m_\nu/Q$, so that  nuclear decays with low $Q$ (and a reasonable life-time) are experimentally preferred.
Older experiments found a fake $4.6\sigma$ evidence for a {\em negative} 
$m_{\nu_e}^2 = -96\pm 21 \eV^2$,
probably because the energy resolution was overestimated.
This was not confirmed by the
most recent experiments {\sc Troitsk} and {\sc Mainz}, that find
\beq
m_{\nu_e}^2=-0.6\pm 2.2 \pm 2.1~\eV^2~\cite{MAINZ}\qquad\hbox{ and }\qquad
m_{\nu_e}^2=-2.3\pm 2.5 \pm 2.0~\eV^2~\cite{TROITSK}.\eeq
Their combined constraint is reported in table~\ref{tab:nonosc}.

\begin{figure}[t]
$$\includegraphics[width=14cm]{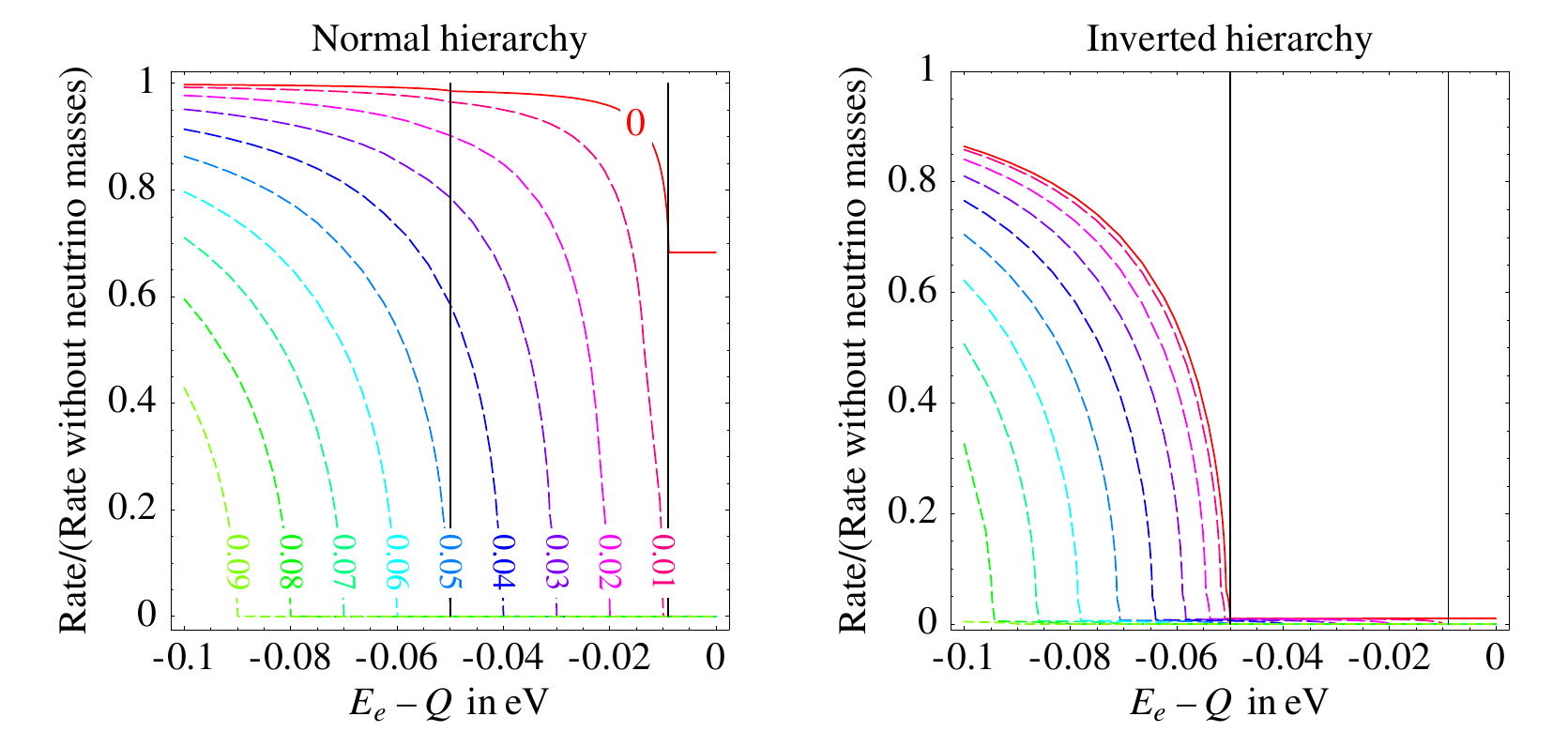} $$
\caption[End-point spectrum in $\beta$-decay]{\em $\beta$-decay spectrum close to end-point 
predicted for best-fit values of oscillation parameters
(we assumed $\theta_{13}=0.1$; the difference with respect to $\theta_{13}=0$ is hardly visible)
and for different values of the lightest neutrino mass, indicated on each curve in the left plot. 
The vertical lines indicate the positions of $(\Delta m^2_{\rm sun,atm})^{1/2}$.
Even without neutrino masses the phase space strongly
suppresses the rate around the end-point.
\label{fig:BetaDecayOsc}}
\end{figure}

\smallskip

The approved experiment
{\sc Katrin} should improve the sensitivity to $m_{\nu_e}$ by one order of magnitude
down to about $0.3\eV$~\cite{Katrin}, thanks to an energy resolution of 1 eV.
New ideas are needed to plan a $\beta$-decay experiment able of reaching 
the neutrino mass scale suggested by oscillation data.
In line of principle, a $\beta$-decay experiment is sensitive
to neutrino masses $m_i$ and mixings $|V_{ei}|$:
\begin{equation}\label{eq:beta2}
\frac{dN_e}{dE_e} = \sum_i  |V_{e i}|^2 F(E_e) (Q-E_e)\sqrt{(Q-E_e)^2 - m_i^2}.
\end{equation}
This is illustrated in fig.\fig{beta}a, where we show the combined effect
of a $H$eavier neutrino with little $e$ component
and of  a $L$ighter neutrino with sizable $e$ component.
Following Kurie we plotted the square root of $dN_e/dE_e$,
that in absence of neutrino masses is a linear function
close to the end-point,
assumed to be known with negligible error.
In fig.\fig{BetaDecayOsc} we show the predicted reduction of the $\beta$-decay rate
around its end-point.
The various curves are for different values of the lightest neutrino mass.

In practice the energy resolution is limited, and only broad features can be seen.
If it is not possible to resolve the difference between neutrino masses,
it is useful to approximate eq.\eq{beta2} with\eq{beta1}
and present the experimental bound in terms of the single effective parameter
\beq\label{eq:mnue}
m_{\nu_e}^2\equiv \sum_i  |V_{ei}^2| m_i^2=
\cos^2 \theta_{13}(m_1^2 \cos^2\theta_{12} +
m_2^2  \sin^2\theta_{12}) + m_3^2 \sin^2\theta_{13}.\eeq
The last equality holds in the standard three-neutrino case.
The expected ranges of $m_{\nu_e}$ are reported in table~\ref{tab:nonosc}
in the limiting case where the lightest neutrino is massless.
From this it is immediate to obtain the ranges corresponding to the generic case of
a non vanishing lightest neutrino mass $m_{\rm lightest}$:
as clear from the definition $m_{\nu_e}^2\equiv (m\cdot m^\dagger)_{ee}$ or from
the more explicit expression in eq.\eq{mnue} one just needs to add $m_{\rm lightest}^2$ 
to $m_{\nu_e}^2$.  The resulting bands at $99\%$ CL\  are plotted in fig.\fig{OscNonOsc}b.

\medskip

Searches for $\nu_\mu$ and $\nu_\tau$ masses have been performed by 
studying decays like $\pi\to \mu \bar{\nu}_\mu$.
The resulting bounds, $m_{\nu_{\mu,\tau}}\circa{<} \MeV$ are very loose.
Notice that $\beta$-decay experiments probe anti-neutrinos.
If one does not trust CPT and allows neutrinos and anti-neutrinos to have different masses,
 the looser bound $m_{\nu_e}<200\eV$ applies to neutrinos.


Finally,~\cite{AtomicNu} explores the futuristic possibility of studying atomic decays
into $\nu\bar\nu\gamma$, which would be convenient since atomic energy differences
are comparable to neutrino masses, allowing to discriminate Majorana from Dirac masses.
Tentative ideas for blocking the dominant purely electromagnetic rate 
and for enhancing the neutrino rate are discussed~\cite{AtomicNu}.


\section{Neutrino-less double $\beta$ decay}\label{0nu2beta}
\index{Neutrino-less double $\beta$ decay}
A few nuclei can only decay through double-$\beta$ decay,
that at the nucleon level corresponds to two simultaneous
$n\to pe\bar\nu_e$ decays, see fig.\fig{FeynBeta}b.
This is e.g.\ the case of  $^{76}_{32}{\rm Ge}$, that cannot $\beta$-decay to $^{76}_{33}{\rm As}$
because it is heavier.
It can only jump to the lighter ${}^{76}_{34}{\rm Se}$:
$$^{76}{\rm Ge} \to  {}^{76}{\rm Se} \, ee \,\bar{\nu}_e \bar{\nu}_e\qquad
(Q=2038.6\,{\rm keV}).$$
Since it is a second order weak process, $^{76}_{32}{\rm Ge}$ has a very long life-time, $\tau\sim 10^{21}\,{\rm yr}$~\cite{0nu2betarev}.
The measurement  of such electron energy spectrum seems to provide the only direct
confirmation of the fact that $\bar\nu_e$ obey the Pauli exclusion principle~\cite{bosonicnu}, 
as predicted by the only sensible theory.

\smallskip

If neutrinos have Majorana masses, the alternative
neutrino-less double $\beta$ decay ($0\nu2\beta$) decay 
$^{76}{\rm Ge} \to  {}^{76}{\rm Se} \, ee $
is also possible~\cite{0nu2betarev}.
Fig.\fig{FeynBeta}c shows the Feynman diagram for $0\nu2\beta$ at nuclear level.
The clashing arrows reflect the insertion of a Majorana mass in the virtual neutrino line,
that violates (electronic) lepton number by two 
unities, $\Delta L_e=2$.

In 1930 Pauli postulated the neutrino in order to explain why 
$\beta$-decay gives a continuous electron spectrum rather than a line.
This same kinematical feature now allows to distinguish $0\nu2\beta$ from ordinary $2\nu 2\beta$ decay:
as illustrated in fig.\fig{beta}b,
$0\nu 2\beta$ gives two electrons with total energy equal to $Q$,
while $2\nu 2\beta$ decay gives two electrons with a
continuous spectrum that extends up to $Q$.
In real life one has to fight with limited energy resolution and other backgrounds.

In general, double-$\beta$ decay processes are:
\beq\label{eq:AZ+2}
(A,Z)\to  (A,Z+2)+2\ e^- +n\,\bar\nu_e,\qquad   n=0,2
\eeq
The competing $\beta$ decay is kinematically forbidden for some even-even nuclei
(${}^{76}\mbox{Ge}$, ${}^{130}\mbox{Te}$, ${}^{100}\mbox{Mo}$,...)
that have ground levels arranged 
such that the levels ($A,Z)$ and $(A,Z+2)$ are below $(A,Z+1)$.
Notice that the processes of eq.\eq{AZ+2} are fine from the point of view of atomic physics:
two more protons and two more electrons appear.
This is not true for the analogous processes with emission of positrons and/or absorption of atomic electrons,
that therefore are expected to have lower rates and do not seem appropriate for achieving a
sufficient sensitivity to $m_{ee}$.

\medskip

\subsection{Connection between $0\nu2\beta$ and neutrino masses}
$0\nu2\beta$ is induced by Majorana neutrino masses, but can also be induced by alternative
more speculative new physics that violates lepton number;  one can show that
observation of $0\nu2\beta$ 
would imply that Majorana neutrino masses exist at some level.
Disregarding these alternative possibilities,
we here study the $\Gamma_{0\nu2\beta}$ decay rate induced by Majorana neutrino masses. 
Assuming that neutrino masses are much smaller than $Q$,
the $0\nu2\beta$ decay amplitude is proportional to $m_{ee}$, 
the $\nu_{eL} \nu_{eL}$ element of the neutrino mass matrix.
There seems to be no realistic way of probing other entries
of the Majorana neutrino mass matrix.
Assuming three Majorana neutrinos, $m_{ee}$ can be written in terms
of the neutrino masses $m_i$, mixing angles $\theta_{ij}$ and Majorana
CP-violating phases $\alpha,\beta$ as
\begin{equation}\label{eq:mee}
m_{ee} = \sum_i V_{ei}^2\ m_i =
\cos^2 \theta_{13}(m_1 e^{2i\beta}
\cos^2\theta_{12} +  m_2 e^{2i\alpha} \sin^2\theta_{12}) + m_3 
\sin^2\theta_{13}.
\end{equation}
$\Gamma_{0\nu2\beta}$ can be computed in terms of $\nu$ masses as
\beq\label{eq:Gamma0nu2beta}
{\Gamma_{0\nu2\beta} =G\ \cdot } 
\left|\mathscr{M} \cdot   m_{ee} /m_e\right|^2,\qquad
T_{1/2}^{0\nu2\beta} = \ln 2/\Gamma_{0\nu2\beta}
\eeq
where $G$ is the known phase space factor and
$\mathscr{M}$ is the nuclear $0\nu2\beta$ matrix element (see table~\ref{tab:0nu2betaM}), 
plagued by a sizable theoretical uncertainty,
and we used the fact that neutrino masses $m_i$ are much smaller than $Q$.\footnote{
A more generic approximated expression,
that might be useful in models with extra heavy sterile neutrinos,
is obtained by replacing $m_{ee} = \sum _i V_{ei}^2 m_i$
with $\sum_i V_{ei}^2 m_i^2 /(1+m_i^2/p^2)$ where $p\sim 100\MeV$ is
the momentum of the exchanged virtual neutrino.}

\begin{table}[t]\small
$$\hspace{-4mm}\begin{array}{c|ccccccc}
\hbox{nucleus} &
^{76}{\rm Ge} &
^{82}{\rm Se}&
^{100}{\rm Mo} &
^{130}{\rm Te} &
^{136}{\rm Xe}&
^{150}{\rm Nd}\\ \hline
\hbox{$Q$ in keV} &  2039 &  2995 & 3034  &  2529&2476&3367 \\
\hbox{$T_{1/2}^{2\nu2\beta}$ in $10^{20}$ yr} & 15\pm1 & 0.92\pm0.07 & .071\pm.004 & 9\pm1 &>8 & .082\pm .009\\[2mm]
\hbox{$G$ in $10^{-14}/$yr} & 0.63 & 2.73 & 11.3 & 4.14 &  4.37 & 19.4\\
\mathscr{M} \hbox{[\v{S}FRVE 2007]} & 3.3\div 5.7 & 2.8\div 5.1 & 2.2 \div 4.6 & 2.3 \div 4.3& 1.2\div 2.8 &\\
\mathscr{M} \hbox{~~~~~~[CS 2009]} & 4.0\div 6.6 & 2.8\div 4.6 & 2.7 \div 4.8 & 3.0\div 5.4 &2.1 \div 3.7 & \\
\mathscr{M} \hbox{~[MPCN 2008]} & 2.3 & 2.2 & & 2.1 & 1.8 &\\
\mathscr{M} \hbox{~~~~~~~[BI 2009]} & 5.5 & 4.4 & 3.7 & 4.1 &&2.3 \\
\mathscr{M}_0 \hbox{~~~[SMK 1990]}&4.2 & 4.0 & 1.3 & 3.6 & 1.7 &0.6\\ \hline
\hbox{90\%CL bound on} & 190, 160 &  4.4 &  5.8 & 30 & 4.5 & 0.036\\
\hbox{$T_{1/2}^{0\nu2\beta}$ in $10^{23}$ yr} &\hbox{HM, IGEX} &\hbox{NEMO-3} & \hbox{NEMO-3}  & \hbox{\sc Cuoricino} & \hbox{DAMA}&\hbox{NEMO-3} \\[2mm]
\hbox{\color{blue}bound on $m_{ee}/h$} &\color{blue} 0.35, 0.38\eV &\color{blue}1.2 \eV& \color{blue}1.5 \eV&\color{blue}0.40\eV & \color{blue}2.2 \eV&\color{blue}31\eV\\
\hline
\hbox{planned experiment} & \hbox{\footnotesize GERDA,Majorana} & \hbox{\sc SuperNemo} & \hbox{MOON} &  \hbox{CUORE} & \hbox{EXO} & \hbox{SNO+}\\
\hbox{$T_{1/2}^{0\nu2\beta}$ goal in $10^{26}\,\yr$} &2\to 60 &1\div 2&17&2\div 6 &0.6\to  8 \\
\end{array}$$
\caption[Data table for $0\nu2\beta$]{\em Summary of the main candidate nuclei for $0\nu2\beta$, their properties, the nuclear matrix elements~\cite{staudt90}
the present experimental status, the future prospects.
The factor $h\sim 1$ reminds that $0\nu2\beta$ elements are uncertain
($h=1$ corresponds to the matrix elements $\mathscr{M}=\mathscr{M}_0$).
\label{tab:0nu2betaM}}
\end{table}

Different calculations find values of $\mathscr{M}$ different by factors of ${\cal O}$(few)~\cite{staudt90}.
This probably over-estimates the theoretical uncertainty, as not all nuclear models are equally accurate.
More recent computations tried to provide error estimates.
In part, uncertainties come from the axial nucleon coupling $g_A$, which inside a nucleus could be $g_A\approx 1$
differing from the value measured in vacuum $g_A\approx 1.25$.
The main issue is estimating and reducing the uncertainty due to nuclear physics
inherent in the standard approximation techniques, mostly based on the
Quasiparticle Random Phase Approximation or on the Shell Model,
or sometimes on different approaches (Interacting Boson Model, ...).
A step in this direction can be done by comparing data on processes generated by
operators similar to the one relevant for $0\nu2\beta$ decay:
$\beta$-decay, $\beta$-capture, $2\nu2\beta$.
However this is not enough to validate a nuclear model, because
only $0\nu2\beta$ (that arises when two different nucleons exchange
a virtual neutrino, that thereby has a virtual momentum comparable to
the inverse size of the nucleus, $\sim 100\MeV$) probes the structure of the nucleus;
on the contrary $2\nu2\beta$ decay is just two independent decays with emission of neutrinos with
small momentum $\sim {\rm keV}$, such that its rate does not involve nuclear form factors.

For definiteness we adopt the $0\nu2\beta$ nuclear
matrix elements $\mathscr{M}_0$ computed by SMK (Staudt et al.)~\cite{staudt90} and listed in table~\ref{tab:0nu2betaM}.
%
To use a different calculation with matrix element $\mathscr{M}$ (also listed in table~\ref{tab:0nu2betaM}), 
just rescale by the factor $h=\mathscr{M}_0/\mathscr{M}$,
which depends on the nucleus studied, obtaining 
\beq\label{eq:0nu2beta:h}
{\Gamma_{0\nu2\beta} =G\ \cdot } 
\left|\mathscr{M}_0  \cdot  m_{ee}/m_e h \right|^2.
\eeq
We always explicit the factors $h$ when quoting an experimental result on $0\nu 2\beta$.


\subsection{Why improving on $0\nu 2\beta$ is difficult}
The number of $0\nu2\beta$ events that can be observed is given by
\beq
N_{\rm sig}={\cal T}\cdot \Gamma_{0\nu 2\beta}\cdot f\cdot N\cdot \epsilon
\eeq
where 
${\cal T }$ is the observation live-time;
$\Gamma_{0\nu 2\beta}$ is the $0\nu2\beta$ rate;
$\epsilon\le1$ is the efficiency in the detection of electrons;
$N$  is the total number of nuclei;
$f$ is the isotopic fraction that contributes to $0\nu2\beta$
(e.g. the isotope $^{76}$Ge  
is a fraction $f\sim 7\%$  of natural germanium).

From this formula, one might  think that the sensitivity to $\Gamma_{0\nu2\beta}$
scales linearly with
the mass of the detector and with the time ${\cal T}$.
This is true only neglecting the background.
Assuming that the background in the signal window
$\Delta E$ scales in a similar way one gets\footnote{Other cases are possible:
e.g.\ a surface contamination might exist and would not scale with $N$.}
$$
N_{\rm bkg}={\cal T}\ \Delta E\ \frac{d\Gamma_b}{dE}\ N
$$
(here, ${d\Gamma_b}/{dE}$ is the background rate per atom per energy interval
and it is supposed to be a constant). 
To compare the performances of different detectors, we can introduce 
a figure of merit  $F$, which has to be as large as possible.
The simplest definition is 
the ratio between the number of events $N_s$ with the 
Poisson fluctuation of the background, $\sqrt{N_{\rm bkg}}$:
$$
F= \frac{N_{\rm sig}}{\sqrt{N_{\rm bkg}}}=
\Gamma_{0\nu 2\beta}\cdot f\cdot 
\epsilon \cdot \sqrt{\frac{ N\cdot  {\cal T}}{\Delta E \cdot
{d\Gamma_b}/{dE}}}
$$ 
The dependence on $m_{ee}$ and on the nuclear matrix element is quadratic, 
while the dependence on the parameters we can control experimentally is 
milder:
the isotopic fraction $f$ and the efficiency enter linearly;
the mass, ${\cal T}$ and background rate only as a square root. 
In order to increase the sensitivity in $m_{ee}$ by a factor 10
one needs to make the experiment $10^4$ bigger,
or to improve it.

\begin{figure}[t]
\centerline{\includegraphics[width=15cm]{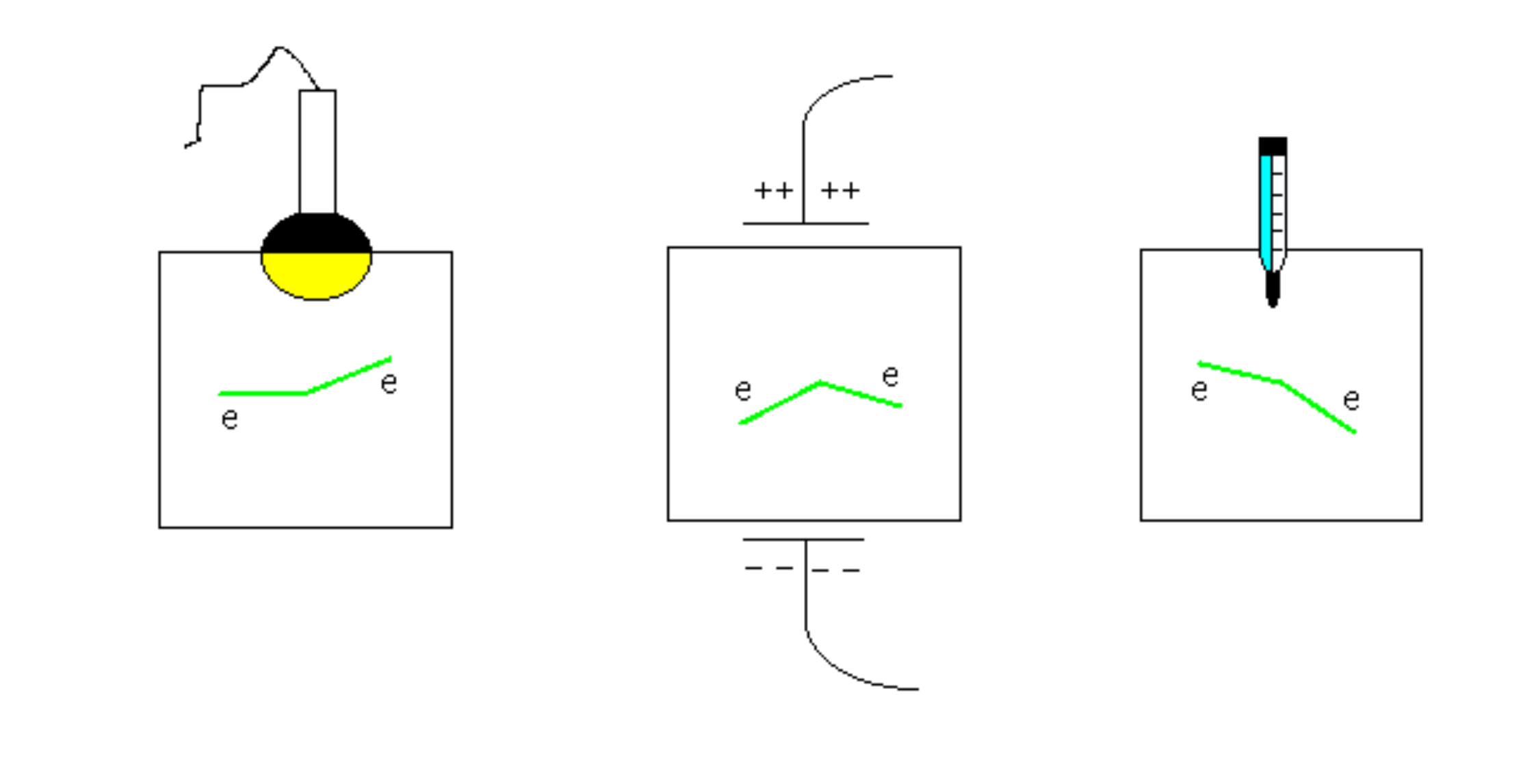}}
\vspace{-1cm}
\caption[$0\nu2\beta$ techniques]{\em Sketch of various possibilities to measure 
the energy of $2\beta$ decays:  (a) tracking-calorimetry, (b), charge-collection, (c) bolometric.}
\label{fig:termo}
\end{figure}

%

\begin{figure}[t]
$$\hspace{-8mm}
\includegraphics[width=5.5cm]{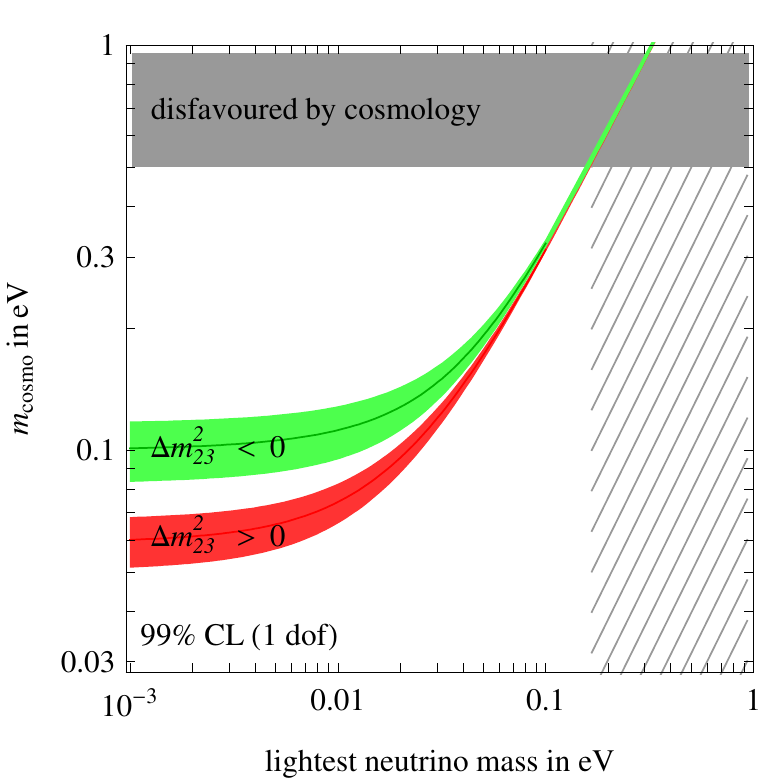}\quad
\includegraphics[width=5.5cm]{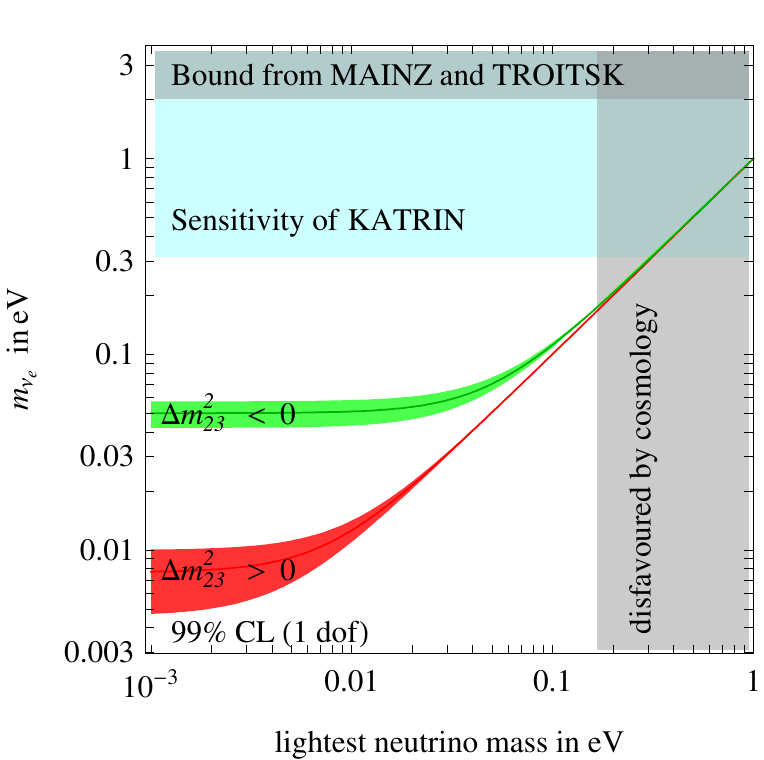}\quad
\includegraphics[width=5.5cm]{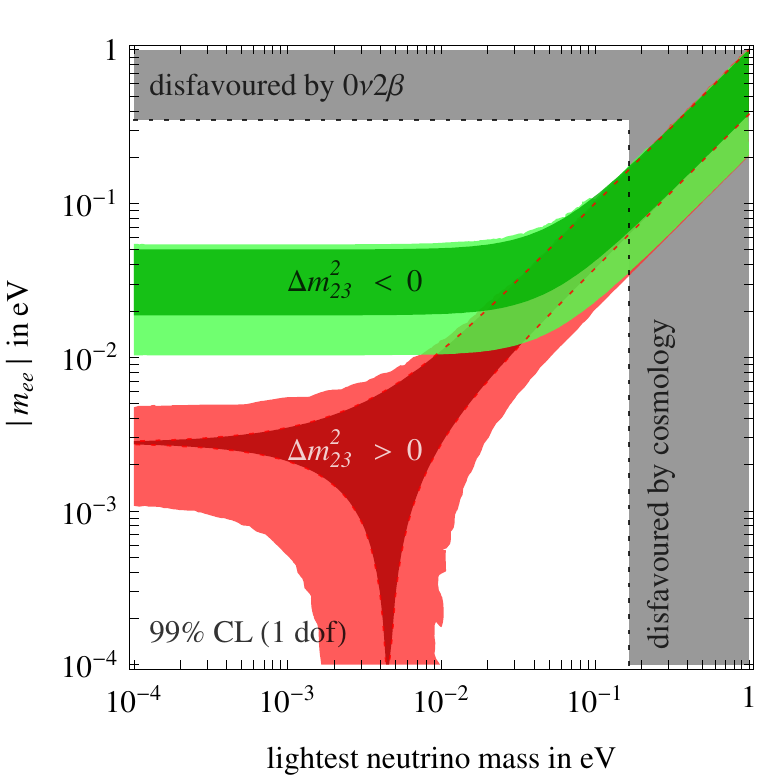}
$$
\caption[Expectations for non-oscillation experiments]{\label{fig:OscNonOsc}\em $99\%$ CL expected ranges as function of the lightest neutrino mass for the parameters:
$m_{\rm cosmo}=m_1+m_2+m_3$ probed by cosmology (fig.\fig{OscNonOsc}a), 
$m_{\nu_e}\equiv (m\cdot m^\dagger)^{1/2}_{ee}$ probed by  $\beta$-decay (fig.\fig{OscNonOsc}b),
$|m_{ee}|$ probed by $0\nu2\beta$ (fig.\fig{OscNonOsc}c).
$\Delta m^2_{23}>0$ corresponds to normal hierarchy
($m_{\rm lightest}=m_1$) and $\Delta m^2_{23}<0$ corresponds to inverted hierarchy
($m_{\rm lightest}=m_3$), see fig.\fig{spectra}.
The darker regions show how the ranges would shrink if
the present best-fit values of oscillation parameters were 
confirmed with negligible error.}
\end{figure}

As should be clear from the discussion  above, we absolutely 
need to measure the  energy well. The sketch
in fig.~\ref{fig:termo} illustrates some basic concepts to achieve this goal.
\begin{enumerate}
\item[(a)] The first concept,
implemented by the NEMO3 collaboration,
 aims at
tracking the single electrons well.
It would allow a significant suppression of the background, but
it did not yield (yet) very precise
energy measurements, such that the limiting factor is
discriminating the $0\nu2\beta$ line from the $2\nu2\beta$ continuum.
A positive signal would allow to check if the energy spectra of the
single electrons are the ones characteristic of a light massive neutrino exchange,
or of some other source of $0\nu2\beta$.

\item[(b)] The second concept, proposed 
in 1967 by Fiorini {\em et al.}~\cite{HM}, aims at collecting the ionization charge
produced by the electrons, with MeV-scale energy.
Experiments using  ${}^{76}$Ge  yield
the best existing limit (from HM and {IGEX}~\cite{HM}).
This technique will be implemented in future experiments, such as GERDA and maybe
Majorana.

\item[(c)] Fiorini et al.\ push the 
bolometer concept 
with tellurium: its isotope of interest has a large isotopic fraction in nature. 
This produced the next 
better result (from the terminated {\sc Cuoricino} experiment, to be enlarged to CUORE).
\end{enumerate}
Many other experiments and proposals are based on (various combinations
of) these concepts and other important considerations
(background control, isotopic enrichment, 
double tag, etc.).
\label{multi-site}
The so called ``{\em pulse shape discrimination}'' is a good example 
of how the background can be reduced in
${}^{76}$Ge detectors; in the terminology above, 
it might be classified as a rough ``electron tracking''.
In $0\nu2\beta$ events the energy is deposited by two electrons in a single point.
Background from $\gamma$ radiation 
deposits monochromatic energy in the crystal, 
producing a line in the energy spectrum, 
at energies that can be dangerously close to the $0\nu2\beta$ line.
However, $\gamma$ tend to manifest as multi-site events, making a few 
Compton scatterings, until their energy is so low that $\gamma$ get photoelectrically absorbed.
The electric pulse from charge collection of multi-site events
has on average a different time structure from single-site events:
the HM collaboration~\cite{evid} tried to exploit this difference
to suppress the background  by a
factor ${\cal O}(2)$ (IGEX also employs the same technique).

If a signal is seen, measuring the energy and/or angular distributions of the
events (as say in NEMO3) and/or related modes of decay such as
electron capture or double positron emission 
(say with a setup as in COBRA) would allow to test 
if $0\nu 2\beta $ is due to neutrino masses or to some other speculative 
source, such as new gauge interactions among right-handed fermions.

\subsection{Present $0\nu2\beta$ experiments}
As clear from table~\ref{tab:0nu2betaM} the main present experiments are HM, IGEX and {\sc Cuoricino}.
Their main features can be summarized as follows.
$$\begin{array}{l|ccc}
\hbox{Experiments~\cite{HM}}& \hbox{HM} & \hbox{IGEX} & \hbox{\sc Cuoricino}\\ \hline
\hbox{Nucleus} & {}^{76}{\rm Ge}& {}^{76}{\rm Ge}& {}^{130}{\rm Te}\\
\hbox{Exposure in $10^{25}$ nuclei$\cdot$yr} & 25&7&5.5\\
\hbox{Isotopic fraction $f$} & 0.86\hbox{ (enriched)}&0.86\hbox{ (enriched)}&0.34\hbox{ (natural)}   \\
\hbox{Efficiency $\epsilon$} &0.5 &0.7 & 0.84\\
\hbox{Energy resolution $\sigma_E$} & 1.6\keV &1.7\keV & \sim 3\keV \\ \hline
\hbox{Total events $n$} & 21 & 9.6 & \sim 70\\
\hbox{Expected background $b$} & 20.4\pm 1.6 & 17.2\pm 2 & s<10.7\\
\hbox{Predicted signal $s$}&76 |m_{ee}/h\eV|^2&23|m_{ee}/h\eV|^2& 21|m_{ee}/h \eV|^2
\end{array}$$
We reported the number of events and expected background
in a $\pm 3\sigma_E$ region around the $Q$ value of the $0\nu 2\beta$.
The Poisson likelihood  of having $s$ signal events is
${\cal L}(s)\propto e^{-s} (b+s)^n$ and $\chi^2 = - 2\ln{\cal L}$.
IGEX and {\sc Cuoricino} observe a number of events slightly below the expected background,
while the opposite happens for HM.
The constraints are reported in table~\ref{tab:0nu2betaM},
where $h=1$ if one assumes the $0\nu2\beta$ nuclear matrix elements of~\cite{staudt90}.
One can do a more precise analysis of the energy spectrum.
This will be discussed in section~\ref{HM} in connection with the HM hint.

\bigskip

Assuming that neutrino masses are of Majorana type
(Dirac neutrino masses would not induce the $L$-violating $0\nu2\beta$ decay)
one can partially infer $|m_{ee}|$ from oscillation data~\cite{0nu2betaOsc}.
Its explicit expression is given in eq.\eq{mee}.
Besides oscillation parameters, which have been partially measured,
$m_{ee}$ depends on the unknown Majorana phases $\alpha,\beta$
and on the absolute neutrino mass scale,
conveniently parameterized by the lightest neutrino mass.
Presently this is also unknown,  but $\beta$-decay and cosmology can measure it.
Fig.\fig{OscNonOsc}c shows the allowed range of $|m_{ee}|$.

In the simplest case of quasi-degenerate neutrinos with mass $m_{\nu}$
the three non-oscillation parameters $m_{\nu_e}$, $m_{\rm cosmo}$ and $m_{ee}$
are given by
\begin{equation}
\label{eq:neqh}
m_{\nu_e} = m_\nu,\qquad
m_{\rm cosmo}=3 m_{\nu},\qquad
0.24\,m_{\nu}<|m_{ee}|<m_{\nu} \hbox{ at $99\%$ C.L.}
\end{equation}
The lower bound  on $|m_{ee}|$ holds
thanks to the fact that solar data exclude a maximal solar mixing
and that CHOOZ requires a small $\theta_{13}$.
Therefore the upper bound on $|m_{ee}|$ implies the constraint 
$m_{\nu}/h<1.1\,\eV$ at $99\%$ C.L., with $h\sim1$ defined in eq.\eq{0nu2beta:h}.


%% file: review_hints.tex
\chapter{Unconfirmed anomalies}\label{anomalies}
In this section we review anomalous experimental results
that might be due to rare statistical fluctuations,
or to mistakes, or be the first hints of new discoveries.
Since this is not established physics
we unavoidably touch controversial issues.
Rather than presenting an acritical list of claims
we emphasize the controversial points that must be clarified.
Hopefully future work will lead to definite conclusions, 
maybe confirming one or more of these anomalies.

\section{Heidelberg-Moscow}\label{HM}
\begin{figure}[t]
$$\hspace{-4mm}\includegraphics[height=5cm]{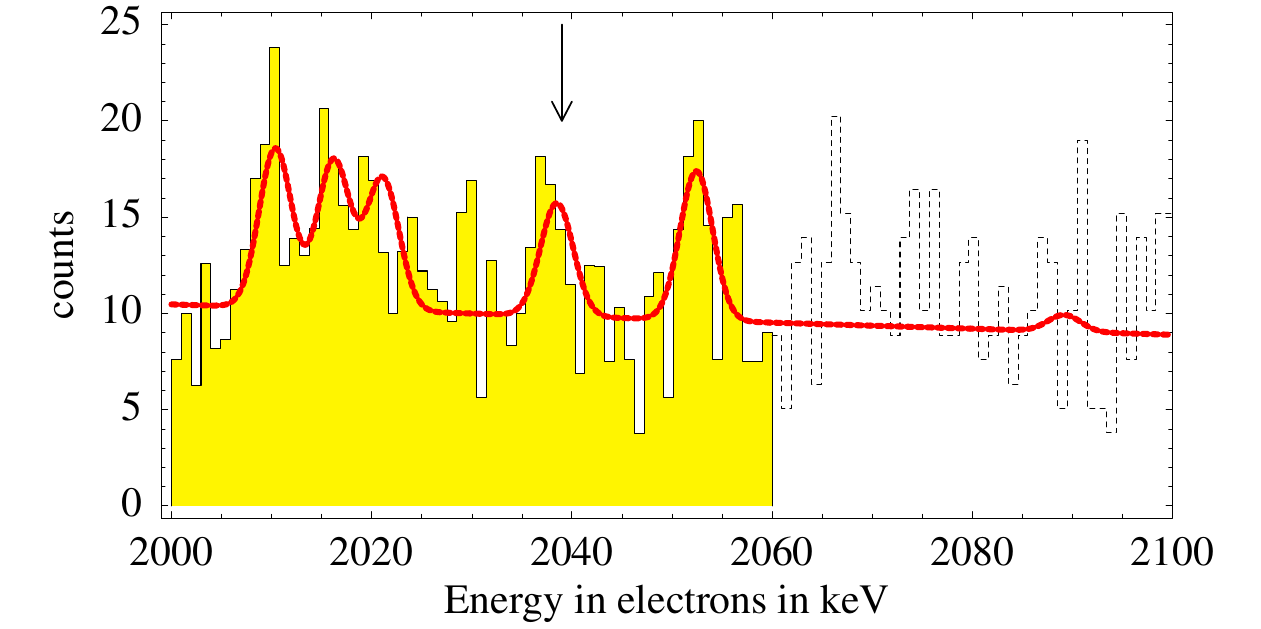}\includegraphics[height=5cm]{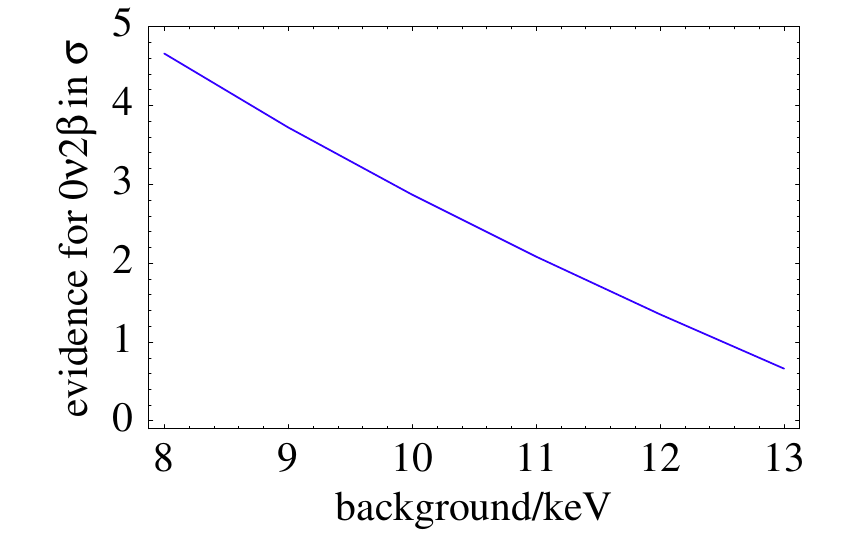}$$
\caption[HM data]{\label{fig:Klapdor2004}\em Fig.\fig{Klapdor2004}a: the latest HM data
($71.7\kg\cdot{\rm yr}$)
used to claim a $4.2\sigma$ evidence for $0\nu2\beta$.
Fig.\fig{Klapdor2004}b: the statistical significance of the $0\nu2\beta$ signal, as
function of the assumed flat component of the background.}
\end{figure}

\index{Heidelberg-Moscow}
We recall basic facts about $0\nu2\beta$ signals in $^{76}_{32}{\rm Ge}$ experiments, 
discussed in section~\ref{0nu2beta}.
$0\nu2\beta$ gives two electrons with total kinetic energy equal to the $Q$ value of the decay.
The HM collaboration reports the event rate as function of the total electron energy looking for
the following $0\nu2\beta$ signal:
\begin{quote}
a peak at $Q=2038.6\,\mbox{\rm keV}$ 
with known width,  $\sigma_E \approx 1.6\,\mbox{\rm keV}$ given by the
energy resolution, emerging over
the $\beta\beta$ and other backgrounds, which are not fully known.

\end{quote}
While the HM collaboration used their data  to set a bound on $|m_{ee}|$, 
some members of the HM collaboration reinterpreted the data 
as a $4.2\sigma$ evidence for $0\nu2\beta$,
and two members as a $6.2\sigma$ evidence~\cite{evid}.
Indeed a hint of a $0\nu2\beta$ peak (indicated by the arrow)
is visible in the most recent data plotted in fig.\fig{Klapdor2004}a.
In these latest results the peak is more visible than in  
latest published HM data~\cite{HM},
partly thanks to higher statistics (increased from 53.9 to 71.7 kg yr)
and partly thanks to an `improved analysis').

This claim is controversial, mainly because
one needs to fully understand the background before being
confident that a signal has been seen. 
In order to allow a better focus on this key issue,
we present fig.\fig{Klapdor2004}b, that should be uncontroversial.
It shows  the statistical significance of the $0\nu2\beta$ signal as function of
$b$, the unknown {\em true} level of
quasi-flat background around the  $Q$-value of $0\nu2\beta$.

The crucial point under debate is:
 how large is $b$?
 The HM collaboration earlier claimed~\cite{HM}
$b=(13.6\pm0.7){\rm events}/(71.7\,{\rm kg}\,{\rm yr}\cdot{\rm keV})$.
In such a case the statistical significance of 
the signal would be less than $1\sigma$, see fig.\fig{Klapdor2004}b.
This can be considered as the upper bound on $b$ computed assuming that all
events in a wider range around $Q$ come from a quasi-flat background.

A statistically significant hint for $0\nu2\beta$ is obtained if one can show that $b$ is lower.
The continuous line in fig.\fig{Klapdor2004}a shows a fit of HM data
using a tentative model of the background,
assumed to have a quasi-flat component (mainly due to `natural' and `cosmogenic' radioactivity)
plus some peaks due to faint $\gamma$ lines of $^{214}$Bi, which is
a radioactive impurity present in the apparatus
(from the $^{238}$U decay chain).
Their positions and intensities can be estimated from tables of nuclear decays;
however they are modified by ${\cal O}(1)$ factors by detector-related effects which
depend on the unknown localization of $^{214}$Bi.
The fit in fig.\fig{Klapdor2004}a is performed by 
allowing the intensity of each line to freely vary.
In this way part of the background is interpreted as $^{214}$Bi peaks,
thereby reducing the quasi-flat component.
We find that the statistical significance of the $0\nu2\beta$ signal is about $2.7\sigma$.

\smallskip

Some details in its implementation prevent this analysis from fully
reaching its goal, which is determining $b$ from regions with no peaks.
1) he latest data have been published only below 2060 keV.
(Above 2060 keV in fig.\fig{Klapdor2004}a we plotted  plotted HM data,
artificially rescaled to account for the larger statistics.
Including the old data above 2060 keV in
the fit would reduce the significance of the signal down to about $2.2\sigma$.
2) 
HM data contain hints of extra unidentified spurious peaks at specific energies (at 2030 keV
and above 2060 keV).
Fitting data assuming that these extra peaks can
be present at arbitrary energies with arbitrary intensities
reduces $b$ and enhances the statistical significance of the signal.

%

\bigskip

As discussed at page~\pageref{multi-site}, HM can partially suppress the multi-site $\gamma$ backgrounds 
via `pulse shape discrimination' such that the $0\nu2\beta$ peak can become more clearly visible.
In the most recent claim, the evidence increased to $6\sigma$ (up to systematic uncertainties)
but the number of authors decreased to 2~\cite{evid}.

According to other members of the HM collaboration
(paper by  A.M. Bakalyarov et al.\ in~\cite{evid}))
some HM detectors find extra unidentified peaks in runs where
pulses under discriminator threshold are included in the data-set:
omitting these runs the hint for a $0\nu2\beta$ peak disappears.
This claim is refuted by the members of the HM collaboration
which claim a signal.

\bigskip

In conclusion, we think that the latest HM data (below 2060 keV)
contain a $2.7 \sigma$ hint for $0\nu2\beta$.
Higher statistical significance can be obtained making
different assumptions about the background.
 It is crucial to understand how large is its quasi-flat
 component around $Q$.
 The natural interpretation of the signal would be
 in terms of quasi-degenerate neutrino masses of $(0.1\div0.9)\eV$.

Various future experiments plan to test the claim of~\cite{evid}.
In our view, a discussion of what should be considered as a convincing evidence for $0\nu2\beta$,
is useful, because any  experiment (past and future) needs to confront with this issue.
Observing $0\nu2\beta$ with different nuclei seems really advisable for two reasons:
to be fully sure that the signal is not faked by a spurious line,
and in view of theoretical uncertainties on $0\nu2\beta$ matrix elements.


\begin{table}[t]
{\small $$\begin{array}{ccccll}
\hbox{Experiment} &  \hbox{baseline} & \hbox{$\nu$ energy range}  & \hbox{channel} &\multicolumn{2}{c}{ \hbox{result}}\\ \hline
\hbox{\sc MiniBoone (FNAL)} &  541 \m & (300 - 3000)\,{\rm MeV}&\nu_\mu\to\nu_e  & P \circa{<}0.001 & \Delta m^2 > 0.03\eV^2\\
\hbox{LSND (Los Alamos)} & \hbox{30\m}& (10- 50)\MeV & \bar\nu_\mu\to\bar\nu_e & \multicolumn{2}{c}{\color{red}P = (2.6\pm0.8) 10^{-3}}\\
&&&\nu_\mu\to\nu_e& \multicolumn{2}{c}{P = (1.0\pm1.6) 10^{-3}}\\
 \hbox{{\sc Karmen} (Rutherford)} & 17.5 \m & (10- 50)\MeV & \bar\nu_\mu\to\bar\nu_e & P<0.00065 & \Delta m^2 <0.05\eV^2\\
 \hbox{\sc Nomad (CERN)} & 835\m & (1-40)\GeV &\nu_\mu\to\nu_\tau &P<0.00017&
\Delta m^2 >0.7\eV^2\\
&&&\nu_e\to\nu_\tau &P<0.0075&\Delta m^2 >5.9\eV^2\\
&&&\nu_\mu\to\nu_e&P<0.0006&\Delta m^2 >0.4\eV^2\\
\hbox{\sc Chorus (CERN)}& 823\m & (1-40)\GeV & \nu_\mu \to \nu_\tau  & P <0.00034 &\Delta m^2 > 0.6\eV^2\\
&&&\nu_e\to\nu_\tau & P<0.026 &
\Delta m^2 > 7.5\eV^2\\
\hbox{CDHS (CERN)} & 130\m/885\m & (2-100)\GeV &\nu_\mu\to\nu_\mu &P>0.95 &
\Delta m^2 > 0.25 \eV^2
\end{array}
$$}
\caption[Short base-line experiments]{\label{tab:shortbeams}\em
{\bf Results of the main short base-line  neutrino experiments}.
As discussed in section~\ref{LimitingRegimes},
bounds are given assuming vacuum oscillations of 2 neutrinos
and reporting the $90\%$ C.L.\ bound on the oscillation probability $P$ for large $\Delta m^2$, 
and the $90\%$ C.L.\ bound on $\Delta m^2$ for $P=1$ (appearance) or $P=0$ (disappearance).}
\end{table}

\section{LSND and MiniBoone}\label{LSND}\index{LSND}
We start with a brief summary of the present situation.
The LSND~\cite{LSND} experiment claimed a $\bar\nu_\mu\to \bar\nu_e$ signal.
The {\sc Karmen} experiment, similar to LSND, did not confirm the signal but could not exclude it.
Assuming that the LSND anomaly is due to oscillations (an hypothesis indirectly disfavored by other experiments), the 
{\sc MiniBoone} experiment~\cite{MiniBoone} searched for $\nu_\mu \to \nu_e$ at the same $L/E_\nu$ as LSND
without finding the signal suggested by LSND. However, {\sc MiniBoone} preliminary data
show a $\nu_e$ excess at lower energy.
By running in $\bar\nu$ mode,  {\sc MiniBoone} does not find any anomaly but does not have
the sensitivity to excluded the LSND anomaly.

\smallskip

We now describe more precisely the relevant short-baseline experiments, listed in table~\ref{tab:shortbeams}

\smallskip

In the LSND~\cite{LSND} and {\sc Karmen}~\cite{Karmen} experiments,
a proton beam is used to produce $\pi^+$, that decay  as
$$\pi^+\to\mu^+\nu_\mu,\qquad
\mu^+\to e^+\nu_e\bar\nu_\mu$$
generating  $\bar\nu_\mu,\nu_\mu$ and $\nu_e$ neutrinos.
The resulting neutrino beam
also contains a small $\bar\nu_e$ contamination,
about $\bar\nu_e/\bar\nu_\mu \circa{<}10^{-3}$.
In {\sc Karmen} both $\pi^+$ and $\mu^+$ decay at rest,
so that the SM prediction for the neutrino energy spectra can be easily computed:
$\nu_\mu$ are monochromatic with energy $29.8\MeV$,
while $\nu_e$ and $\bar\nu_\mu$ have a continuos spectrum up to
$52.8\MeV$.
In LSND most of the neutrinos are produced by $\pi^+$ and $\mu^+$ decays at rest.
Decays-in-flight of $\pi^+$ produce some
 flux of $\nu_\mu$ with higher energy, that has been used
for $\nu_\mu\to\nu_e$ searches.

The search for
$\bar\nu_\mu\to\bar\nu_e$ is performed
using the detection reaction $\bar\nu_e p\to n e^+$,
that has a large cross section (section~\ref{sigmahad}).
The detector tries to identify both the $e^+$ and the $n$
(via the $2.2\MeV$ $\gamma$ line obtained
when $n$ is captured by a proton).
These experiments are more sensitive to
oscillations than older experiments,
that used higher neutrino energy (see table~\ref{tab:shortbeams}).

\begin{figure*}[t]
$$\hspace{-5mm}
\includegraphics[width=83mm]{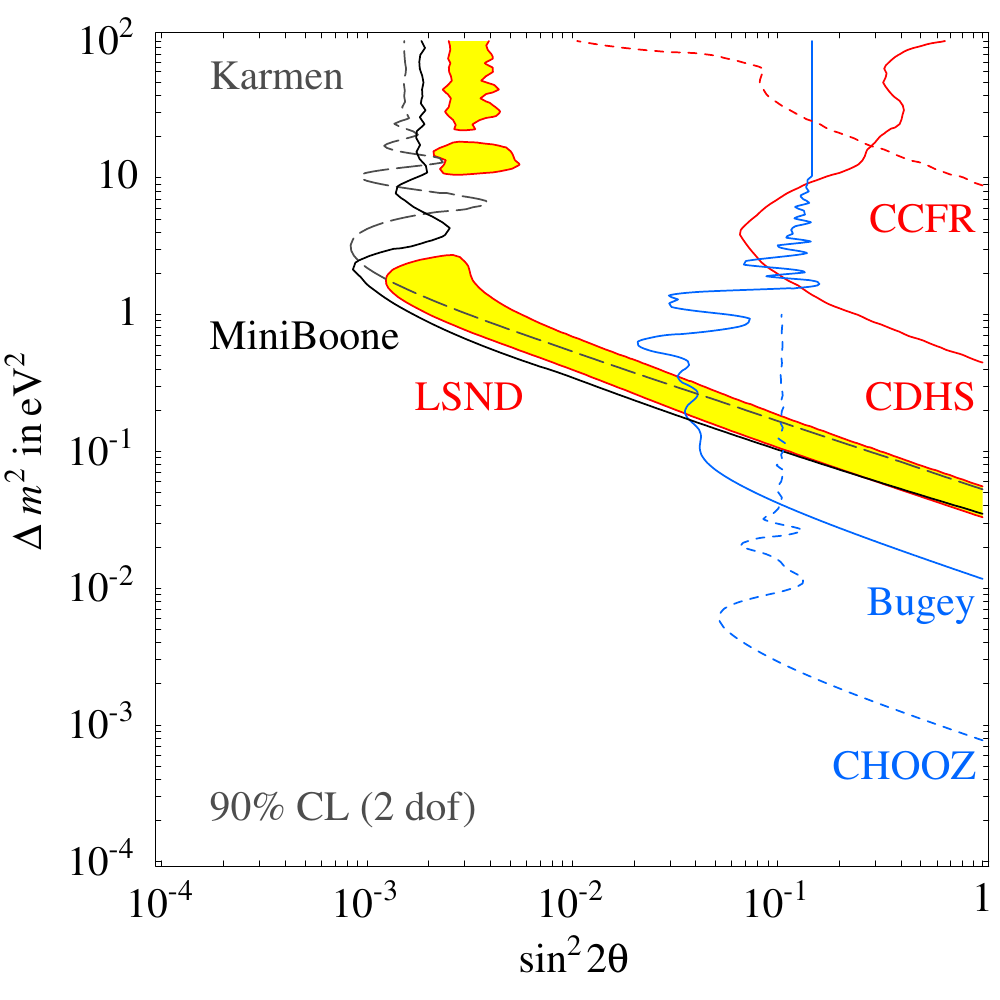} \hspace{1cm}
\includegraphics[width=83mm]{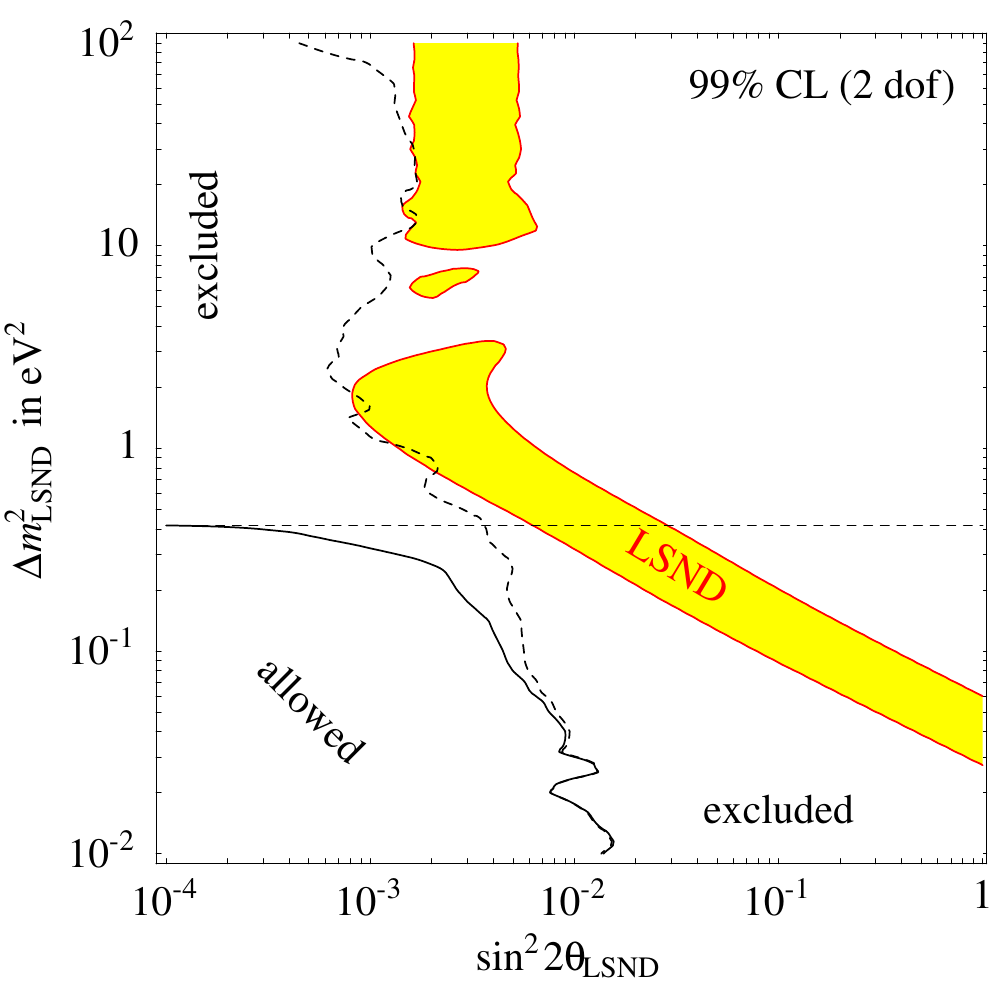}$$
\parbox{0.48\textwidth}{\caption[Oscillation data in the $\nu_e,\nu_\mu$ sector]{\em 
{\bf Data in the $\nu_e,\nu_\mu$ sector}.
The mixing angle $\theta$ on the horizontal axis  is  different for the different experiments.
Bound from {\sc MiniBoone},
{{\sc Karmen}} and LSND region (shaded) for $\nu_\mu\to\nu_e$.
{\color{blu} Bounds from {\sc Bugey} and  {\sc Chooz} ($\bar{\nu}_e$ disappearance)},
{\color{rossos} {\sc CDHS} ($\nubarnu_\mu$  disappearance),
{\sc CCFR} ($\nu_\mu$  disappearance).
All at $90\%$ CL (2 dof).}
\label{fig:exps}}}\hfill
\parbox{0.48\textwidth}{\caption[LSND fit]{\em {\bf 3+1 oscillations}.
The region of the $(\Delta m^2,\sin^2 2\theta_{e\mu})$ plane favored 
by LSND
is compared with the combined constraint from cosmology and $\nu$ experiments
(continuos lines).
Dropping the cosmological bound on $\nu$ masses (horizontal line),
the dashed line shows the constraints from $\nu$ experiments only.
All at $99\%$ CL (2 dof).
\label{fig:split}}}
\end{figure*}

LSND finds an evidence for $\bar\nu_\mu\to\bar\nu_e$,
that ranges between 3 to $7\sigma$
depending on how data are analyzed
(the final publication claims a $3.8\sigma$ excess in the total number of $\bar\nu_e$ events;
our analyses are made with the `official' $\chi^2$ provided by the LSND collaboration,
where $\chi^2_{\rm best~fit}-\chi^2_{\rm no~oscillation} \approx 5^2$).
This happens because LSND has a poor signal/background ratio:
choosing the selection cuts as in~\cite{LSNDKarmen}
the LSND sample contains
1000 background events and less than 100 signal events,
distinguished only on a statistical basis.
The statistical significance of the LSND signal depends
on how cuts are chosen, and relies on
the assumption that all sources of background have been correctly computed.
The main backgrounds are cosmic rays
and $\nu_e$ misidentification.
The final LSND results for the average oscillation probabilities are reported in table~\ref{tab:shortbeams}.
The LSND anomaly can be interpreted as due to oscillations.
Fitting LSND data alone in a two-flavour context gives the best-fit regions
shown in fig.\fig{split}a.

\smallskip

{\sc Karmen} finds $15$ events versus an expected background of $15.8$ events.
{\sc Karmen} has a few times less statistics than LSND and has a pulsed beam,
allowing to reduce the cosmic ray background ({\sc Karmen} also has a better shield)
and $\nu_e$ misidentification
(due to a nuclear decay with a life-time different than the one characteristic of $n$ capture).
Furthermore {\sc Karmen} has a baseline somewhat shorter than LSND.
At the end {\sc Karmen} excludes a significant part, but not all, of the
($\Delta m^2,\theta$) range suggested by LSND, as shown in fig.\fig{split}a.
The region that survives is the one at smaller $\Delta m^2$:
here the longer base-line makes LSND more sensitive than {\sc Karmen}.

\medskip

{\sc MiniBoone}~\cite{MiniBoone} was designed to test the oscillation interpretation of the LSND anomaly.
{\sc MiniBoone} employs
a pulsed neutrino beam  generated by $\pi\to \mu^+\nu_\mu$ decays in flight
 (muons being stopped) and
 initially searched for $\nu_\mu\to \nu_e$ at the same $L/E_\nu$ as LSND, but with both $L$ and $E_\nu$ increased by about one order of magnitude:
$L=541\m$ and $E_\nu$ peaked around $700\MeV$.
This allows to improve the signal/background ratio with respect to LSND.
Furthermore the choice $E_\nu \sim m_p$ allows to reconstruct the neutrino energy from the measured
energy and direction of the scattered particles.
At the {\sc MiniBoone} energy the uncertainty on the cross sections used for $\nu_e$ detection is about $10\%$.

{\sc MiniBoone} does not see $\nu_\mu\to \nu_e$ appearance at the $L/E_\nu$ suggested by LSND:
assuming vacuum oscillations of two neutrinos {\sc MiniBoone} contradicts the LSND anomaly at 98\% C.L.~\cite{MiniBoone}.

However, {\sc MiniBoone} finds a $\sim 3\sigma$ excess of $\nu_e$ at higher $L/E_\nu$,
in the lowest energy bins, $E_\nu < 475\MeV$, where the background is higher.
Vacuum oscillations cannot produce a signal with this spectral dependence.
The excess might be due to the $\gamma Z \omega$ coupling generated by the chiral anomaly one loop diagram,
where the $\omega$ couples to baryons,
the $Z$ to neutrinos, and the $\gamma$ fakes the electron produced by the usual $\nu_e N\to e N'$ scattering.

 {\sc MiniBoone} also performed $\bar\nu_\mu\to\bar\nu_e$  searches:
 like {\sc Karmen},  {\sc MiniBoone} does not see any excess, but is not sensitive enough to test the LSND anomaly.

\medskip

%

\subsection{Theoretical interpretations}
Many ideas have been proposed and excluded by newer experimental data.
A lot of activity had been devoted in trying to interpret the LSND anomaly as oscillations of one extra sterile neutrino
with a 3+1 mass spectrum, i.e.\ splitted from the 3 active neutrinos.
This is now excluded directly by  {\sc MiniBoone}, and indirectly by combining other
bounds about $\nu_\mu$ and $\nu_e$ oscillations, as illustrated in fig.\fig{exps}
and fig.\fig{split}.
One needs to assume that $\nu$ and $\bar\nu$ behave differently.
Keeping in mind that the case for the LSND $\nu$ anomaly and for the {\sc MiniBoone}
$\bar\nu$ anomalies is weak,  one might want to
explore which new physics can explain them~\cite{LSNDpheno}.

One possibility is maybe adding various sterile neutrinos, and fine-tuning CP-violating parameters.

A satisfactory global fit is possible by introducing one sterile neutrino and
allowing neutrinos and anti-neutrinos to have different masses and mixings.
One can fit the LSND anomaly in anti-neutrinos
without introducing any effect in neutrinos
and thereby avoiding constraints from {\sc MiniBoone} and other
experiments performed with neutrinos.

Otherwise, it is possible to add one extra sterile neutrino and
assume that the Lorentz symmetry is broken, such that it has a
speed-of-light different from the one of SM particles.
(This can be realized assuming that SM particles are confined on a 
a brane, and that  an extra sterile neutrino can freely travel in the extra dimensions).
Oscillations are effectively described
by the usual effective Hamiltonian given by eq.\eq{m+V} plus
an extra term $\propto E_\nu\, \diag(0,0,0,1)$ in the basis $\{\nu_e,\nu_\mu,\nu_\tau,\nu_{\rm s}\}$~\cite{LSNDpheno}.
The new term can give an MSW-like resonance at some energy $E_*$: by choosing
$E_* \sim 400\MeV$ one can fit the LSND and {\sc MiniBoone} anomalies,
at the expense of some conflict with solar or atmospheric experiments that probed
similar energies.

Equivalently, the extra Lorentz-breaking term could arise as a matter effect,
is the extra sterile neutrino interacts with some cosmological background
with a cosmological density  somewhat smaller than photons.
By adjusting the parameters such that the sterile neutrino feels a matter potential
$A_{\rm s}$ about $10^3$ times larger than the standard active matter potential
one can fit LSND and {\sc MiniBoone}.
Matter effects can differentiate neutrinos from anti-neutrinos in the usual way. 
The cosmological problem of the standard 3+1 scenario
(overproduction of sterile neutrinos via oscillations) is avoided
because $A_{\rm s}$ grows linearly with the temperature $T$:
at neutrino decoupling the active/sterile mixing angle in matter is suppressed enough that
the sterile abundance (estimated similarly to eq.\eq{sterileOscAb}) is small enough.

Finally, one can try to build models where the neutrino mass parameters are not
constant in space or time.

In summary, wild theoretical speculations are now needed to interpret the LSND and/or {\sc MiniBoone} anomalies.




\begin{figure}
\parbox{8cm}{\includegraphics[width=8cm]{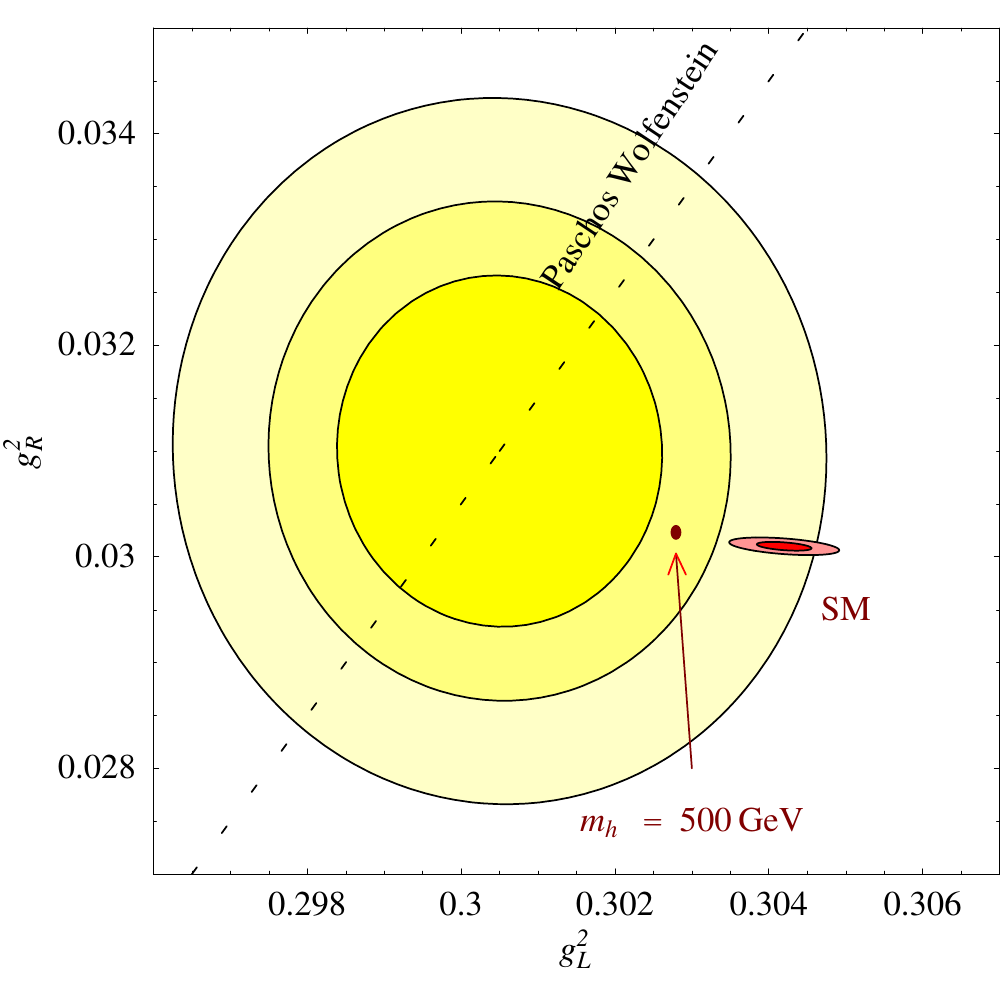}}\hfill
\parbox{6cm}{\caption[NuTeV data]{\label{fig:NuTeV}\em The SM prediction for ($g_L^2, g_R^2$) at $68,99\%$ CL and the
NuTeV determination, at $68,90,99\%$ CL.
The NuTeV central value moves
along the PW line using  different
sets of parton distribution functions that assume $s=\bar{s}$ and $u^p=d^n$.}}
\end{figure}

\section{NuTeV}\index{NuTeV}\label{NuTeV}
The NuTeV collaboration~\cite{NuTeV} reported a $\sim 3\sigma$ anomaly in
the NC/CC ratio of deep-inelastic muon-neutrino/nucleon scattering.
The effective $\nu_\mu$ coupling to left-handed quarks is found to be about $1\%$
lower than the best fit SM prediction.
LEP experiments found that charged lepton couplings agree with SM predictions within
{\em per-mille} accuracy.
See~\cite{NuTeVexps} for other related experiments.

\smallskip

The NuTeV experiment sent both a $\nu_\mu$ and a $\bar\nu_\mu$ beam
(obtained from the FermiLab Tevatron) on an iron target. 
Neutrinos had energy $E_\nu\sim 100\GeV$ and
the average transferred momentum was $\langle q^2\rangle \approx -20\GeV^2$. 
Events were detected by a calorimeter.
The muon produced in CC events gives a long track,
while the hadrons in NC events give a short track.
In this way the NuTeV collaboration could statistically distinguish NC from CC events.
The neutrino energy spectrum was computed by a MonteCarlo simulation.
The ratios of neutral--current (NC) to
charged--current (CC)
deep-inelastic neutrino--nucleon
scattering total cross--sections, $R_\nu$ and $R_{\bar \nu}$,
 are free from the uncertainties on the neutrino fluxes 
and contain the most interesting pieces of information.
We recall the tree-level SM prediction for these quantities.
The $\nu_\mu$-quark effective Lagrangian predicted by the SM at tree level is
given by eq.\eq{nuq}
in terms of the $Z$ couplings
$g_{Aq}$ ($q=\{u,d,s,\ldots\}$, $A = \{L,R\}$) listed in table~\ref{tab:gAi}  at page~\pageref{tab:gAi}.
Including only first generation quarks, for an
isoscalar target, and to leading order, $R_\nu$ and $R_{\bar \nu}$ are given by
\begin{eqnsystem}{sys:Rnu}
R_\nu &\equiv& \frac{\sigma(\nu {\cal N}\to \nu X)}{\sigma(\nu {\cal N}\to \mu X)} =
\frac{(3 g_L^2 + g_R^2)q +  (3 g_R^2 + g_L^2)\bar q}{3 q +\bar q} = g_L^2 + r g_R^2\\
R_{\bar{\nu}} &\equiv& \frac{\sigma(\bar\nu {\cal N}\to \bar\nu X)}{\sigma(\bar\nu {\cal N}\to \bar\mu X)} =
 \frac{(3 g_R^2 + g_L^2)q +(3g_L^2 + g_R^2)\bar q}{q +3\bar q} = g_L^2 + \frac{1}{r} g_R^2,
\end{eqnsystem}
where $q$ and $\bar q$ denote  the fraction of the nucleon
momentum carried by quarks and antiquarks, respectively.
For an isoscalar target, $q=(u+d)/2$, and
we have defined
\begin{equation}
r \equiv  \frac{\sigma(\bar{\nu}{\cal N}\to \bar\mu
X)}{\sigma({\nu}{\cal N}\to \mu X)}
=\frac{3 \bar{q} +q}{3q+\bar{q}}
\end{equation}
and
\begin{equation}
g_L^2 \equiv g_{Lu}^2 + g_{Ld}^2 = \frac{1}{2}-\sin^2\theta_{\rm W}+\frac{5}{9}\sin^4\theta_{\rm W},\qquad
g_R^2\equiv  g_{Ru}^2 + g_{Rd}^2 = \frac{5}{9}\sin^4\theta_{\rm W}.
\end{equation}
The observables $R_\nu^{\rm exp}$ and $R_{\bar{\nu}}^{\rm exp}$
measured at NuTeV differ from the ideal observables in eq.~(\ref{sys:Rnu}). 
Total cross--sections can only be determined up
to  experimental cuts and uncertainties, such as those related to
the spectrum of the neutrino beam, the contamination of the $\nu_\mu$
beam by electron neutrinos,  and
the efficiency of NC/CC discrimination.
Once all these effects are  taken into account,
the NuTeV data can be presented as a
measurement of 
\begin{equation}\label{eq:NuTeVgLgR}
g_L^2 = 0.3005\pm 0.0014\qquad\hbox{and}\qquad
g_R^2=0.0310\pm0.0011,
\end{equation}
where errors include both statistical and systematic uncertainties.
According to\eq{NuTeVgLgR}
$g_L$ is $\sim 3\sigma$ below its SM prediction.

\medskip

The difference of the effective couplings $g^2_L-g^2_R$
(`Paschos--Wolfenstein ratio'~\cite{PW}) is
subject to smaller theoretical and systematic uncertainties than
the individual couplings:
\begin{equation}\label{eq:PW}
R_{\rm PW} \equiv\frac{R_\nu - r R_{\bar{\nu}}}{1-r} =
\frac{\sigma(\nu {\cal N}\to \nu X)-\sigma(\bar\nu {\cal N}\to
\bar\nu X)}{\sigma(\nu {\cal N}\to \ell X) - \sigma(\bar{\nu}{\cal N}\to \bar{\ell}X)}=
 g_L^2- g_R^2 = \frac{1}{2}-\sin^2 \theta_{\rm W}.
\end{equation}
NuTeV is consistent with previous experiments, such as CCFR~\cite{CCFR}.
CCFR had a beam containing both $\nu_\mu$ and $\bar\nu_\mu$, while NuTeV has two separate 
$\nu_\mu$ and $\bar\nu_\mu$ beams.
This is the main improvement because allows
to get rid of the unprecisely known partonic structure of the nucleon using $R_{\rm PW}$,
which, under the above assumptions, is independent of $q$ and $\bar{q}$.
The NuTeV value of $R_{\rm PW}$ is $\sim 3\sigma$ below its SM prediction.

\medskip

Doing  precision physics with iron is a delicate task.
One needs to carefully study 
if the apparent anomaly could be due
to under-estimated errors, or
to some neglected effect.
Progress concerning the following points would be welcome
\begin{itemize}
\item {\bf NLO QCD corrections} have been neglected in the NuTeV analysis,
and cancel out in the ideal PW observable;
This cancellation remains strong enough in the true PW-like observable~\cite{NuTeVnewold}.

\item 
{\bf QED} and electroweak loops give a few $\%$ correction to $R_{\rm PW}$
and cannot explain the anomaly if  they have been included taking into account experimental cuts.
Without knowing them, the impact of recent precise computations cannot be determined.

\item {\bf Nuclear effects} 
could affect $R_\nu$ and $R_{\bar \nu}$,
but only at low momentum transfer and apparently not in
a way that allows to reconcile NuTeV data with the SM~\cite{NuTeVnewold}.
These effects are partially automatically included in the NuTeV analysis,
based on  their own parton distributions
obtained fitting only iron data.

\item {\bf Parton distributions}:
$q(x)$ are extracted from
global fits, usually performed under two simplifying assumptions:
$s(x) = \bar{s}(x)$ and $u^p(x) = d^n(x)$.
These approximations could be not accurate enough, at the level of precision reached by NuTeV.
In presence of a momentum asymmetry $q^-  = \int_0^1 x [q(x) -\bar{q}(x)] dx$
the ideal PW observable shifts as
$$R_{\rm PW} = \frac{1}{2}-\sin^2\theta_{\rm W} +
 \hbox{(EW corrections)} + (1.3 + \hbox{QCD corrections}) (u^- - d^-  -  s^-)$$
According to na\"{\i}ve estimates,
$u^- -d^- \sim \max[ (m_u-m_d)/\Lambda_{\rm QCD},\alpha_{\rm em}]$ and $s^-\sim s,\bar{s}$,
these effects could account for the NuTeV anomaly.

\item Isospin is broken by quark masses and by electromagnetism.
Electromagnetism gives an effect with the correct sign:
$u^--d^-<0$ because $u$ quarks have bigger charge than $d$ quarks and
therefore radiate more photons.
Some detailed computations performed replacing QCD with more
tractable phenomenological models
suggest that, due to cancellations, 
isospin-violating effects are somewhat too small.
It is not clear if a QCD computation would lead to the same conclusion.
An {\bf isospin-violating} interpretation of the NuTeV anomaly is compatible with other available data.

\item   Since a nucleon contains 3 quarks (rather than three antiquarks)
one expects that $s$ and $\bar{s}$ carry comparable (but not equal)
fractions of the total nucleon momentum.
Non perturbative fluctuations like $p \leftrightarrow K\Lambda$
are expected to give $s$ harder than $\bar{s}$
 since $K$ is lighter than $\Lambda$.
Indeed $s^- >0$ could explain the NuTeV anomaly.
Some model computations suggest that $s^-$ is too small,
but again it is not clear how reliable they are.
The $s^-$ issue can also be addressed relying on 
inclusive DIS data and on charm-production scattering data:
after some controversy recent global fits find $-10^{-3}<s^- < 4\cdot10^{-3}$
giving a hint for a
{\bf strange momentum asymmetry} of the desired sign and magnitude.
An accurate analysis of dimuon data performed by the NuTeV collaboration
finds $s^-=+(2.0\pm1.4)~10^{-3}$.

\end{itemize}
In conclusion, testing and possibly excluding such SM `systematic effects' which can produce the NuTeV anomaly
seems to be a difficult job.
Therefore, it is useful to speculate about possible new physics interpretations which might have cleaner signatures~\cite{NuTeVnewold}.
Unfortunately, no particularly compelling new physics with distinctive signatures has been found.
The problem is that $\bar\ell Z{\ell}$ and $\bar\ell W \nu$ couplings
(where $\ell$ denotes charged leptons)
agree with the SM and have been
measured about 10 times more accurately than $\bar\nu Z\nu$ couplings.
Proposals which overcome this problem look exotic,
while more plausible possibilities work only if one
deals with constraints in a `generous' way or
introduces and fine-tunes enough free parameters.

For example, 
mixing the $Z$ boson with an extra $Z'$ boson modifies NC neutrino couplings,
but also NC couplings of leptons ($\ell_L$ and $\nu$ are unified in the same ${\rm SU}(2)_L$ doublet)
conflicting with precision data.
New physics that only affects the gauge boson propagators cannot fit the NuTeV anomaly due to the same constraints.
Neutrino oscillations are not compatible with other experiments.
A reduction of neutrino couplings due to a $\sim 1\%$ mixing with sterile singlets 
does not work, because CC neutrino couplings have been too precisely tested
by $\mu$ decay together with precision data.
Combinations of the above effects with enough unknown parameters can work\footnote{For example,
mixing with a sterile singlet together with a heavy higgs  fits all data if one also assumes
that some extra unspecified  new physics affects $M_W$ and therefore ignores this measurement~\cite{NuTeVnewold}.
A failure of the SM fit of electroweak data
would support the case for new physics,
but at the moment we do not see any convincing problem.}.
The NuTeV anomaly can be fitted by adding to the SM Lagrangian the specific $\SU(2)_L$ invariant
effective operator 
$-(0.024\pm0.009)2\sqrt{2}G_{\rm F}[\bar{L}_2 \gamma_\mu L_2][\bar{Q}_1 \gamma^\mu Q_1]$ (1,2 are generation indexes).
The new physics which could generate it can either be heavy with sizable couplings (so that future colliders should see it)
or light with small couplings
(e.g.\ a $Z'$ with few GeV mass and negligible mixing with the $Z$).

On the experimental side~\cite{NuTeVexps}  it is possible to
measure alternative processes 
where the possible new physics behind the NuTeV anomaly
could manifest.
Facilities built for long-baseline neutrino experiments
will allow to repeat the NuTeV experiment, possibly at larger
neutrino energies.
Future reactor experiments could measure $\bar\nu_e e$ couplings
as accurately as NuTeV measured 
 $\nu_\mu N$ couplings.
 The E158 experiment is measuring the weak angle $\theta_{\rm W}$
  in $ee$ collisions, at energies even lower than NuTeV.

\medskip

In conclusion,  there are various possible
SM and new physics interpretations of the NuTeV anomaly.
More work is needed to identify the true one.

%% file: review_cosmo.tex
\chapter{Neutrinos in cosmology}\label{cosmology}
In absence of new physics, neutrinos are light, weakly interacting and
essentially stable so that relic big bang neutrinos (`CMB neutrinos') must still exist today
and be, after photons, the most abundant particle species in the universe.
Cosmology allows to count CMB neutrinos and to study some of their properties because
neutrinos significantly influenced observed relics of the past evolution of the universe.
We will discuss stages where this happened,
presenting the predictions of standard cosmology 

\begin{itemize}

\item {\em Big Bang Nucleosynthesis} (section~\ref{nucleos}).
The relative number of protons and neutrons
(and of few other light nuclei) has been fixed
by weak interactions involving neutrinos,
that decoupled below temperatures $T\sim \MeV$ (about one minute after the big-bang).
Controversial measurements of their primordial abundances are consistent
with SM predictions (3 neutrinos in thermal equilibrium at $T\sim \MeV$).

\item {\em Cosmic Microwave Background} (section~\ref{CMB}).
Photons decoupled at $T\sim 0.3\eV$ (about $0.3$ Myr later)
when electrons and protons formed neutral hydrogen.
Measurements of anisotropies in photon CMB 
are now testing many aspects of  cosmology,
including neutrinos.
Present data indicate that CMB neutrinos were present at $T\sim 0.3\eV$.

\item {\em Large Scale Structures} (section~\ref{CMB}).
Gravity tends to cluster non relativistic particles
(normal matter and cold dark matter), 
while relativistic non-interacting particles suppress this process. 
Massive neutrinos became non relativistic when $E_\nu\sim T \circa{<} m_\nu\sim 300\,{\rm K}$,
leaving a small imprint in the measurable amount of clustering of galaxies.
Present data give bounds on $m_\nu$ competitive with laboratory experiments.


\item {\em Thermal Leptogenesis} (section~\ref{leptogenesi}). 
The tiny excess of matter over anti-matter~\cite{WMAP}
\begin{equation}
\label{eq:nB}
\eta\equiv \frac{n_B - n_{\bar B}}{n_\gamma} = \frac{n_B}{n_\gamma} =(6.15\pm0.25)10^{-10}
\end{equation}
might have been generated in decays of right-handed neutrinos with mass maybe around $10^{10}\GeV$.
We do not know how to test this speculative mechanism.

\end{itemize}
The hottest relics observed so fare were produced only at $T\circa{<} \MeV$ 
--- an energy region
explored with cosmic rays and colliders more than 50 years ago.
However cosmology is sensitive to kinds of new physics
which cannot be probed by such experiments.
So far cosmological data are consistent with the SM, but
significant progress is expected in the near future.

%

\section{Big-bang nucleosynthesis}\label{nucleos}\index{Big-bang nucleosynthesis}
We discuss the reasonable agreement with data of  the standard BBN scenario, 
 and the consequent bounds on non-standard neutrino properties.  
Detailed discussions can be found in the relevant literature~\cite{bbn:history,BBNdata, bbnosc}.

\medskip

Big-bang nucleosynthesis
is the theory of abundances of light nuclei.
We start presenting the predictions of standard cosmology.
The early universe at temperatures $m_e \ll T\ll m_p$ contained
$\gamma,e,\bar{e},\nu_{e,\mu,\tau},\bar \nu_{e,\mu,\tau}$ plus a minor component
of protons and neutrons.
Reacting species were in quasi-homogeneous phase 
(as suggested by simplest inflationary scenarios and supported by CMB data), 
with the 3 light neutrinos in thermal equilibrium. 
Abundances have been determined by different processes,
all happened around $T\sim \MeV$, because $\MeV$ is 
\begin{enumerate}
\item  the typical binding energy of light nuclei (e.g. $m_n-m_p = 1.293\MeV$).
\item the temperature at which electrons become non-relativistic.
This event indirectly affected BBN.

\item the neutrino decoupling temperature.
In fact, the neutrino interaction rate 
(i.e.\ the number of collisions experienced by a neutrino per unit time)
was
$\Gamma_\nu \sim \sigma(\nu e) \cdot n_\nu \sim G_{\rm F}^2 T^2\cdot T^3$.
Comparing it with the expansion rate 
$H \sim T^2/M_{\rm Pl}$ 
gives
$\Gamma_\nu/H \sim (T/T^{\rm dec}_\nu)^3$:
neutrinos decoupled below $T_\nu^{\rm dec}\sim g^{1/6} (G_{\rm F}^2 M_{\rm Pl})^{-1/3}
\sim \MeV$.\footnote{The precise value of $H$ depends on the number of neutrinos
$N_\nu$ as follows:
$H = (8\pi \rho/3)^{1/2}/M_{\rm Pl}=1.66 g^{1/2}  T^2/M_{\rm Pl}$ 
where $\rho = g \pi^2 T^4/30$ is the energy density 
and $g = 2 + \frac{7}{8}(4 + 2N_\nu)$ is the number of SM spin degrees of freedom at $T\gg m_e$.}
\end{enumerate}
Weak scatterings involving $\nu_e$
 ($\nu_e n\leftrightarrow ep$ and
 $\bar{\nu}_e p\leftrightarrow \bar{e} n$) 
 kept the number ratio between protons and neutrons
 in thermal equilibrium,
 $n_n(T)/n_p(T) = e^{-(m_n-m_p)/T}$.
 Neutrinos decoupled about one second after the big-bang,
freezing $n_n/n_p\sim e^{-(m_n-m_p)/T_\nu^{\rm dec}}\sim 1/6$.
Therefore $e\bar{e}\to \gamma$ annihilations at $T\sim m_e$  heated
photons but not neutrinos.

\begin{figure}[t]
$$\raisebox{1cm}{\hbox{\includegraphics[height=7cm]{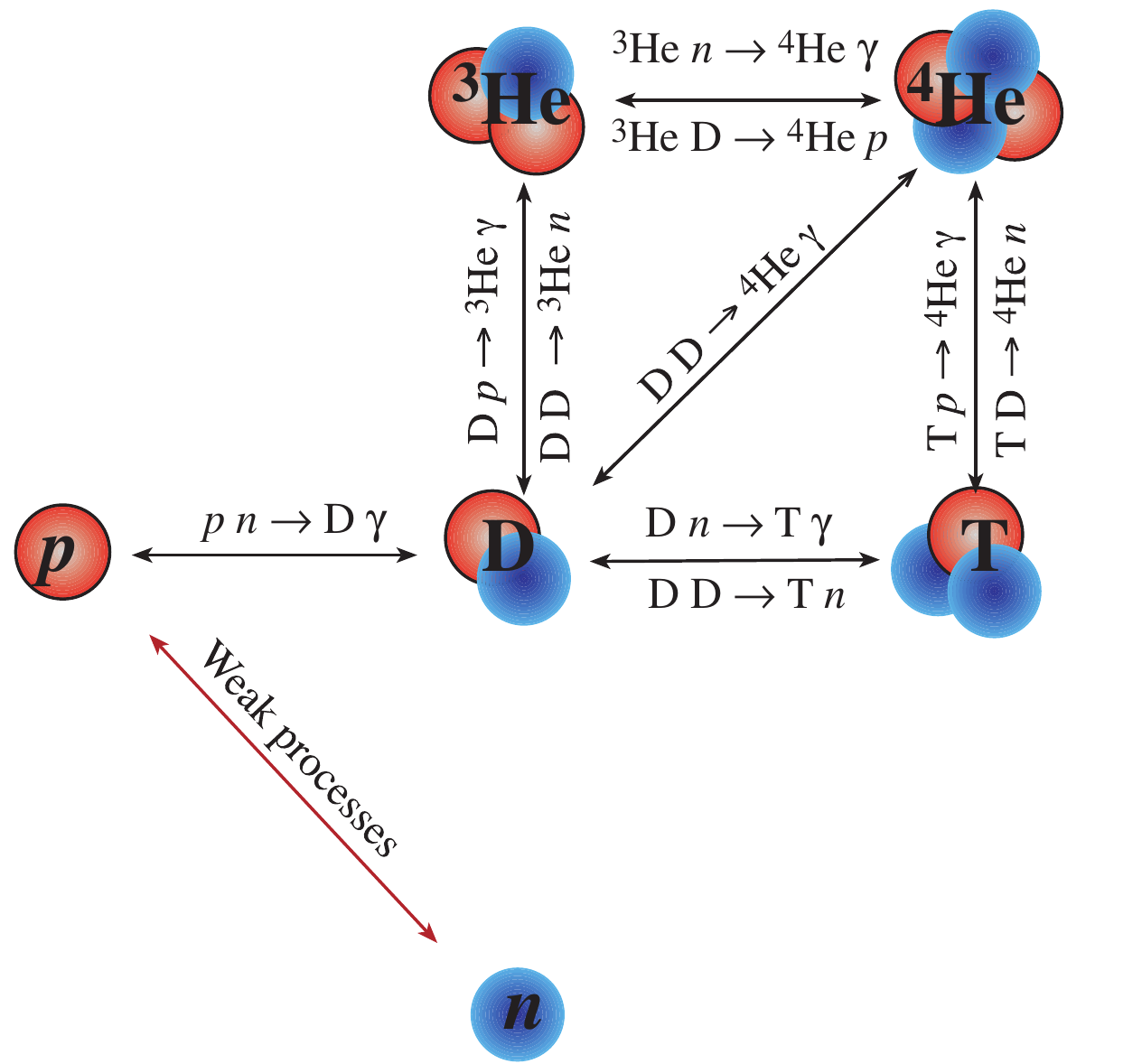}}}\hspace{2cm}
\includegraphics[width=8cm,height=8cm]{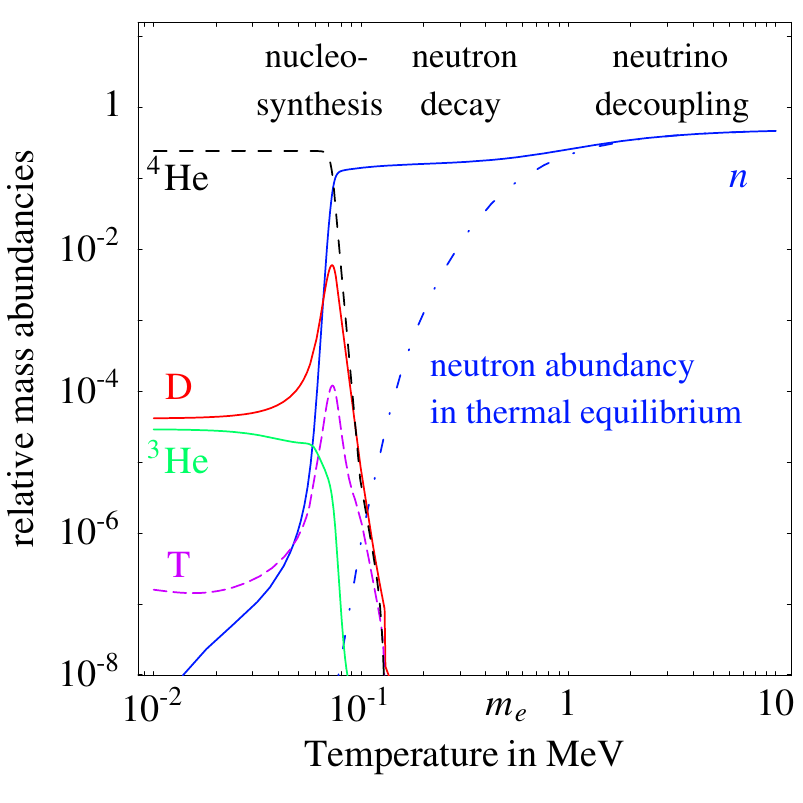}$$
\caption[BBN basics]{\label{fig:BBN}\em Left: main reactions relevant for BBN.
Right: evolution of the main nuclear components. }
\end{figure}

Neutrons are still free and can decay:
$n_n/n_p$ decreases down to about 1/7
when, at $T\approx 0.07\MeV$ (see fig.\fig{BBN}b)
the universe is sufficiently cold that neutrons get bound in nuclei.
Almost all neutrons form $^4{\rm He}$ because it is the light nucleus
with the largest binding energy ($B=28.3\MeV$),
and production of heavier nuclei 
(like $^7$Li,  there are no stable isotopes with $5,6,7$ or $8$ nucleons)
 is strongly suppressed by Coulombian repulsion.
Therefore the prediction for the $^4$He mass fraction is\footnote{An important detail.
Let us consider e.g.\ deuterium, the lightest nucleus produced by $ pn\leftrightarrow D\gamma$ reactions.
Since $\eta = n_B/n_\gamma \ll 1$ deuterium is formed not at $T\sim B$
($B=2.22\MeV$ is the small binding energy of deuterium)
but at $T\sim B/\ln(n_\gamma/n_B)\sim 0.1\MeV$.
Time in seconds and temperature in MeV are related by $t\sim 1/T^2$.
$Y_p$ depends slightly on $n_B/n_\gamma$ because its smallness delays nucleosynthesis, 
giving more time to neutrons to decay.}
$Y_p \approx {2 n_n}/(n_n+n_p) \approx 1/4$.
Around the central values of $n_B$ and $N_\nu$ one has
\beq\label{eq:He}\color{blus}
Y_p = 0.248+0.0096\ln\frac{n_B/n_\gamma}{6~10^{-10}} + 0.013 (N_\nu^{^4\rm He}-3).\eeq
Numerical factors follow from our semi-quantitative arguments.
We reported the precise  values obtained from public codes that describe the nuclear network,
giving  predictions accurate at the per-mille level.
The apex `$^4$He' on $N_\nu$ is present because we want to consider
generic kinds of new physics that modify $Y_p$
(e.g.\ a sterile neutrino that thermalizes during BBN):
we parameterize their correction to $Y_p$ 
in terms of an observable `effective numbers of neutrinos', 
 $N_\nu^{^4{\rm He}}$,
univocally defined by the inversion of eq.\eq{He}.
We will define in an analogous way a few other cosmological observables.

\smallskip

All other nuclei have small abundances, that depend
on powers of $\eta\equiv n_B/n_\gamma$ and on $N_\nu$
essentially trough the combination $\eta/\sqrt{g_*}$,
where $g_* = 4+(7/4)(11/3)^{4/3} N_\nu$ is the number of
relativistic dof after $e\bar e$ annihilations.
Of particular importance is the deuterium abundancy, predicted to be
\beq\label{eq:Deuterio}
 \frac{Y_{\rm D}}{Y_{\rm H}} \approx (2.75\pm 0.13)~10^{-5}~\frac{1+0.11~(N_\nu^{\rm D}-3)}{(\eta/6.15~10^{-10})^{1.6}}\eeq
with an uncertainty mainly induced from nuclear cross-sections.
The observed abundances of $^4$He,  deuterium, $^7$Li
can all be reproduced for $\eta\sim\hbox{few}\cdot 10^{-10}$.
CMB and LSS data presently give 
an independent measurement
of $\eta$, reported in eq.\eq{nB},
which is consistent with (and more precise than) BBN,
if we assume that the present standard cosmological model
(flat $\Lambda$CDM, \ldots) is correct.

We are here interested in testing the expansion rate,
parameterized in eq.s\eq{He},\eq{Deuterio} by $N_\nu-3$,
the number of extra relativistic species in thermal equilibrium at $T\sim \MeV$.
The (conservative) observational values
for the $^4$He and D abundancy imply~\cite{BBNdata}
\beq\begin{array}{rclcrcl}\label{eq:BBndata}
 Y_p &=& 0.25\pm 0.01     &\quad\Rightarrow\quad&
  \color{blu} N_\nu^{^4\rm He} &\color{blu} =&\color{blu} 3\pm0.7    \\
Y_{\rm D}/Y_{\rm H}& =&  (2.8 \pm 0.5)\,10^{-5}     &\Rightarrow& 
\color{blu}    N_\nu^{\rm D}&\color{blu} \approx& \color{blu} 3\pm 2
\end{array}\eeq
In order to discriminate $N_\nu=3$ from $N_\nu=4$ it is necessary to either
$i)$ somewhat improve the determination of $Y_p$, or to
$ii)$ improve on $Y_{\rm D}$, somewhat  improve on
$\eta$ and on the theoretical uncertainty on $Y_{\rm D}$.

Today the error budget is dominated by the observational uncertainties.
Measurements of the primordial abundances  $Y_d$ and $Y_{\rm D}$ are
difficult and controversial.
One of the problems is that, after BBN,
 nuclear abundances have been modified
  by stars and other astrophysical processes.

\begin{figure}
\begin{center}
\includegraphics[width=0.8\textwidth,height=0.5\textwidth]{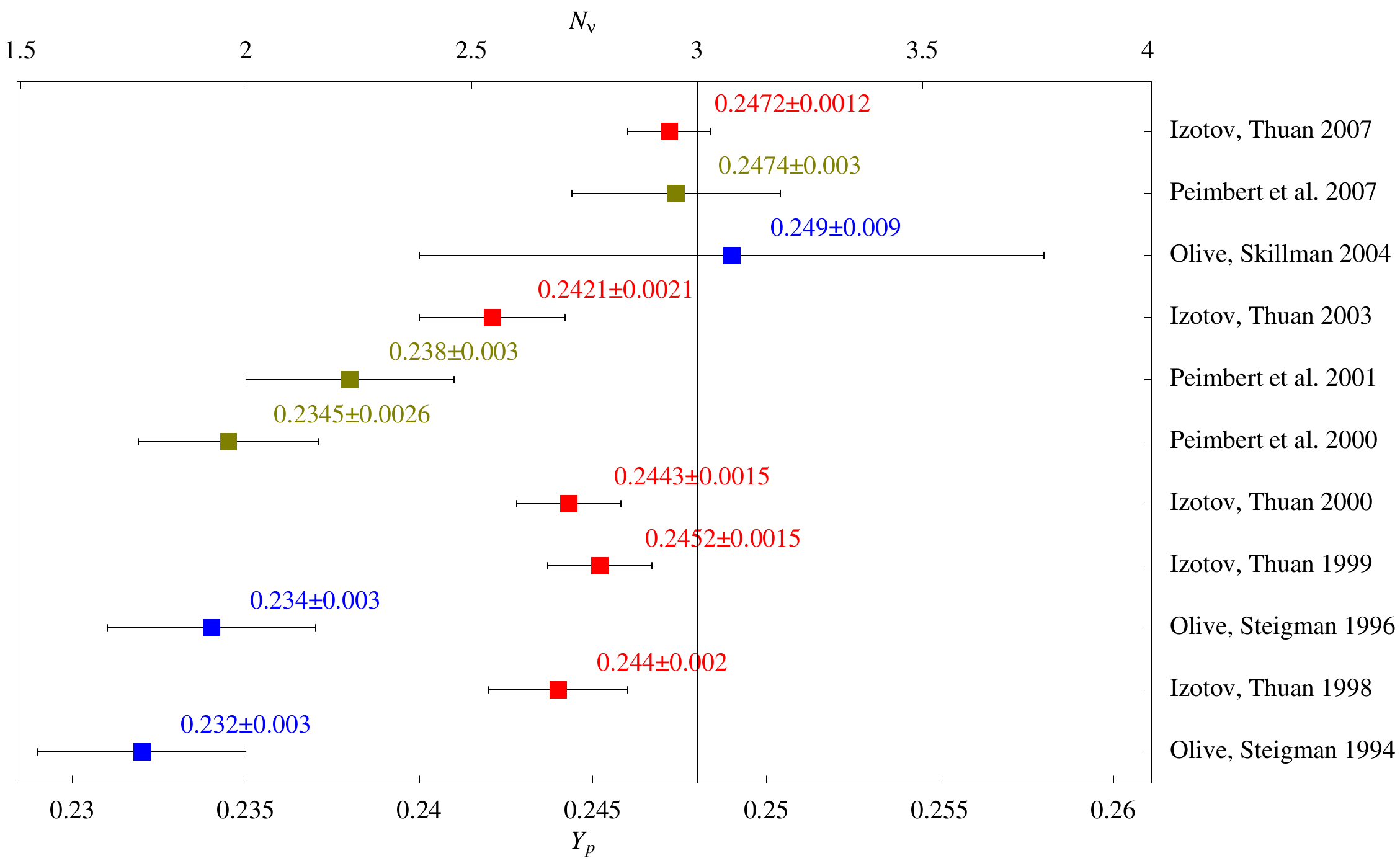}
\caption[Measurements of $Y_p$]{\label{fig:BBNdata}\em Recent measurements of the $^4{\rm He}$ mass fraction $Y_p$.
The lower axis shows the corresponding number of neutrinos at BBN.}
\end{center}
\end{figure}

\paragraph{Helium-4} Since stars produce $^4$He together with heavier elements (`metals')
 one reconstructs the primordial  $^4$He abundancy by measuring
it in differently contamined environments and extrapolating to zero metallicity.
 Of particular importance are the metal-poor HII regions:
gas clouds ionized by young stars.
Measuring de-excitation light spectra
one deduces the relative abundancy of
ionized and doubly ionized $^4$He with respect to ionized hydrogen.
In order to reconstruct the $^4$He/H relative abundancy one needs
stellar population models able of controlling
systematic corrections due to stellar absorption,
to different ionization regions for H and $^4$He,
to Balmer lines excited by collisional processes
and to temperature fluctuations.
We think that our eq.\eq{BBndata} correctly summarizes the present situation.
Adding more digits would suggest that $Y_p$ has an established
central value and uncertainty.
Unfortunately this is not the case: it is difficult to quantify the above
systematical uncertainties.
Some authors report smaller errors or smaller central values, such that
$N_\nu^{^4\rm He}=4$ looks already excluded:
fig.\fig{BBNdata} shows a compilation of recent measurements.
According to some authors, future improvements are possible.

\paragraph{Deuterium}
Alternatively, future measurements of the deuterium abundancy
might give the most stringent test of the number of neutrinos during BBN.
Deuterium has different and apparently less problematic systematic uncertainties:
it is not produced but only destroyed by stars,
and can be partly hidden in HD molecules.
It is measured from quasar absorption line systems  at high redshift,
which are presumably not contaminated by stars.
Several groups quote $\sim 10\%$ uncertainties 
but obtain $\sim 50\%$ discrepancies among their central values.
 Finding and studying well-suited high-redshift systems
 might result in precise determinations of the D abundancy.
   
%

\paragraph{Other nuclei}
Tritium decays.
The $^7$Li abundancy seems to be 3 times lower
than the value $Y_{^7{\rm Li}}\approx 4~10^{-10}$
predicted by BBN assuming eq.\eq{nB}.
This might be due to systematic uncertainties.
Other primordial abundance ($^3$He,\ldots) are poorly determined.

\medskip

The message is that  present data 
seem to agree in a satisfactory way with standard BBN predictions,
but are not yet accurate enough to discriminate
$N_\nu=3$ (SM) from $N_\nu=4$ (SM plus one extra thermalized fermion) or from
$N_\nu = 3+4/7$ (SM plus one extra thermalized scalar).
This could be done by future measurements of the $^4$He or deuterium abundances:
we subscribe the conclusion of Sarkar's review in \cite{reviews} 
``... measurements [of abundances] ought to costitute
a {\em key} programme for cosmology, with the same 
priority as, say, the measurements $H_0$ or $\Lambda$''.

\medskip

It is important to emphasize that one should not say
`$N_\nu=5$ is excluded' but rather
`$N_\nu=5$ is not compatible with standard cosmology'.
Standard cosmology has not been tested, and many more or less motivated non-standard 
BBN scenarios have been proposed.
There could be further reprocessing of
nuclear elements, e.g.\ by a very early 
population of stars; 
neutrinos  could have a non thermal distribution (e.g.\ in scenarios
with very low reheating temperature) or 
carry leptonic asymmetries; 
the distribution of light 
nuclear species (or even, the baryon number distribution itself) 
might be non-homogeneous; 
late decaying particles can modify 
the cosmological scenario where BBN takes place; 
large magnetic fields might exist 
during nucleosynthesis; fundamental constants 
may vary; neutrinos might have non standard properties;
other light species might populate the Universe; etc. 

\smallskip

In view of this situation it is important to test if $N_\nu=3$
with as many observables as possible.
We now discuss what we can learn from CMB
anisotropies.


\section[CMB and LSS]{Cosmic microwave background and large-scale structures}\label{CMB}\index{CMB neutrinos}
Photons decoupled from matter when the universe cooled enough that
electrons and protons formed neutral hydrogen.
The last scattering occurred when the universe had temperature
$T\approx 0.3\eV$, age $\approx2\cdot 10^{5}\yr$ and 
was about $1+z\sim 1000$ times smaller than today.
Namely, the observable horizon had a size of about a $\hbox{Mpc} \equiv 3.26~10^{6}\,\hbox{ly}$,
comparable to the typical distance between  galaxies today.
Measuring the fossil CMB radiation teaches a lot of things about the early universe.
In particular, the pattern of observed small anisotropies in the CMB temperature, 
$\delta T/T\sim 10^{-5}$, depends
on primordial fluctuations (possibly produced during a period of inflation),
on photon/baryon `acoustic' oscillations occurred around CMB decoupling,
and on the later evolution of the universe.
Although one has to study all these effects in order to  disentangle them,
we here concentrate on the r\^ole played by neutrinos.

\smallskip

As discussed in section~\ref{nucleos},
neutrinos decoupled at  a temperature $T\sim \hbox{few} \cdot m_e$.
Therefore electron/positron annihilations that occurred at $T\circa{<} m_e$
transferred their energy into photons but 
negligibly to neutrinos.
As a consequence the temperature of CMB neutrinos is predicted to be
{\color{blu}$T_\nu  = (4/11)^{1/3} T $} 
where $T $ is the CMB photon temperature.\footnote{We present the computation.
 After neutrino decoupling, 
 the neutrino temperature decreases as $T_\nu\propto 1/R$.
Since electron/positron annihilations proceed in thermal equilibrium with photons,
their total entropy $(s_e + s_\gamma)R^3$ remains constant.
Consequently $(T_\nu/T)^3 = (2+4s_e(T)/s_\gamma(T))/(2+4\cdot 7/8)$.
At temperatures above $m_e$ $T_\nu = T$ and the total energy density is
$\rho = g\pi^2 T^4/30$ with  $g = 43/4$.
At temperatures below $m_e$ we can neglect $s_e$  so that
$T_\nu= T(4/11)^{1/3}$ and $g=2+\frac{7}{8}2\cdot 3\cdot (\frac{4}{11})^{4/3}= 3.36$.
More precise computation take into account that at $T\sim m_e$
neutrinos are not completely decoupled,
especially the most energetic ones. As a consequence
$e\bar{e}$ annihilations also generate
a few additional $\nu_e$
(partially converted 
into $\nu_{\mu,\tau}$ by solar and atmospheric
oscillations) distorting their energy spectrum.}
Today $T=2.73\K$ so that $\color{blu} T_\nu = 1.96\K=0.17 \meV$, 
which corresponds to a present neutrino number density 
{\color{blu}$n_{\nu_i} =n_{\bar\nu_i} = 3 n_\gamma/22 = 56/\cm^3$}.
Since $T_\nu$ is smaller than the neutrino mass
scale suggested by solar and atmospheric oscillations,
CMB neutrinos are today mostly non-relativistic. 
Their present contribution to the total energy density is
\beq\label{eq:Omeganu}
\Omega_\nu \equiv \frac{\rho_\nu}{\rho_{\rm critical}} = 
\frac{\sum_i m_{\nu_i} n_{\nu_i}}{3H^2/8\pi G_{\rm N}} =\frac{10^{-3}}{h^2} \frac{\sum_i m_{\nu_i}}{0.1\eV}\eeq
where, as usual, the present Hubble constant is written as
$H=100h\, \hbox{km/s\,Mpc}$ with $h\approx 0.7$.
Neutrino masses have a minor effect (not yet observed),
so that it is convenient, in first approximation, to consider neutrinos as massless.
The old observation that neutrinos cannot be Dark Matter (because Dark Matter is `cold' rather than `hot') already
implies $\Omega_\nu < \Omega_{\rm m}$ i.e.\ $m_\nu < 0.15 \eV$.
We will discuss how stronger bound on $\Omega_\nu$ and consequently on $m_\nu$ are obtained by computing the
small effects of neutrino masses and comparing them with present observations.

\medskip

In this approximation the total energy density in relativistic particles (`radiation')
around recombination, $T\sim 0.3\eV$, is predicted to be
\beq\label{eq:NnuCMB} 
\rho_{\rm rad} = \rho_\gamma + \rho_\nu = \bigg[1 + \frac{7}{8}\bigg(\frac{4}{11}\bigg)^{4/3} N^{\rm CMB}_\nu \bigg]\rho_\gamma\eeq
with $N^{\rm CMB}_\nu = 3$
(small corrections from the approximation above give $N^{\rm CMB}_\nu = 3.04$).
In presence of new physics the SM formula eq.\eq{NnuCMB} is no longer true,
but still used as the definition
of the effective parameter $N^{\rm CMB}_\nu $.

\begin{figure}
$$\includegraphics[width=14cm]{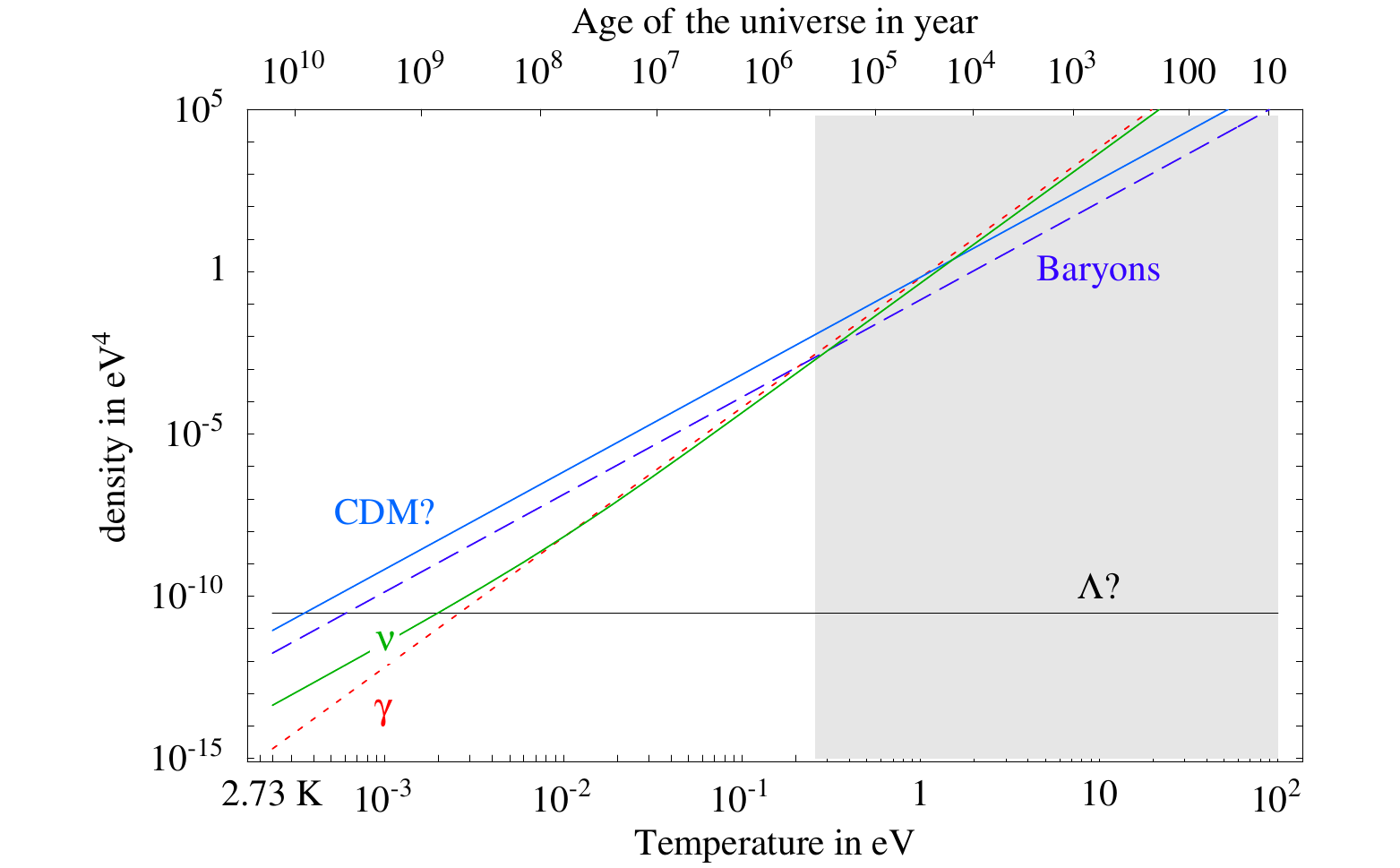}$$
\caption[Components of the early universe]{\label{fig:rhocosmo}\em Evolution of the average energy densities of
photons, neutrinos, baryons, Cold Dark Matter and cosmological constant $\Lambda$.
We assumed that today
$h\approx 0.7$, $\Omega_{\rm tot} = 1$,
$\Omega_\Lambda \approx 0.7$, $\Omega_{\rm CDM} \approx 0.3$.
Neutrinos are relativistic at $T\gg m_\nu$ ($\rho_\nu \approx T^4$),
and non-relativistic at lower temperatures ($\rho_\nu \approx m_\nu T^3$).
Shading covers the epoch before $\gamma$ decoupling.
}
\end{figure}

\begin{figure}[t]
$$\includegraphics[width=0.45\textwidth]{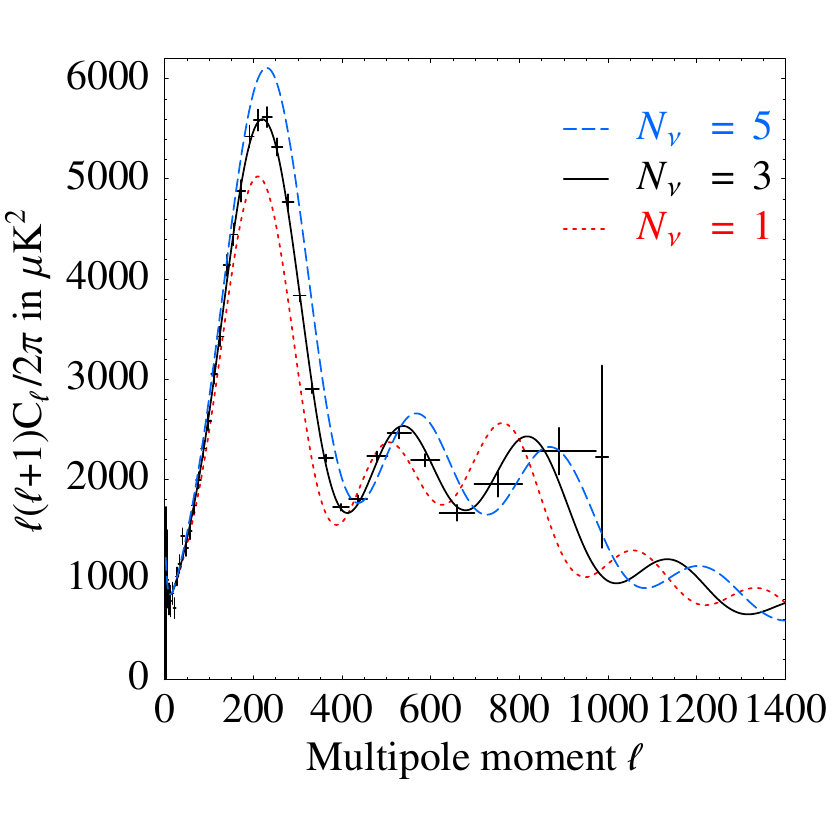}\qquad
\includegraphics[width=0.45\textwidth]{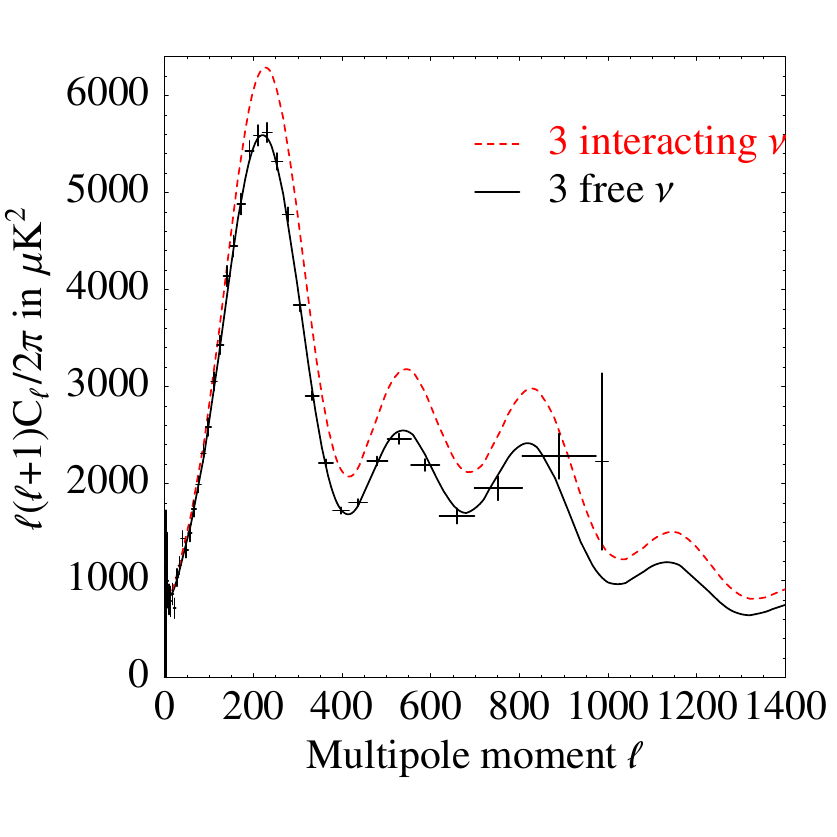}$$
\caption[Neutrinos and CMB]{\label{fig:CMBnu}\em How CMB anisotropies
(parameterized in terms of the conventional $C_\ell$)
depend, at fixed values of other cosmological parameters,
 on a) the energy density in relativistic free-streaming species
(parameterized in terms of the conventional `number of neutrinos' $N_\nu$);
b) the fraction of free vs interacting neutrinos.
The crosses are the WMAP data.
}
\end{figure}

We recall that the total energy density 
$\rho = \rho_{\Lambda} + \rho_{\rm CDM} + \rho_\gamma + \rho_\nu + \rho_{\rm baryons}+\cdots$
started being dominated by non-relativistic matter rather than by relativistic radiation at
$T\sim \eV$, slightly before recombination, see fig.\fig{rhocosmo}.
Measurements of CMB anisotropies allow to reconstruct $N_\nu$ in two different ways:
a) from the total energy density in relativistic particles, $\rho_{\rm rad}$,
that significantly contributes to the measurable expansion rate around recombination, since
recombination happened slightly later than  the transition from a radiation to a matter-dominated;
b) the energy density in freely moving relativistic particles (like neutrinos, and unlike photons)
can be reconstructed, as they  smooth out inhomogeneities, as discussed in the next section.
These effects are shown in fig.\fig{CMBnu} where we plot how CMB anisotropies change
by varying, at fixed best-fit values of all other
cosmological parameters, i) $N_\nu$; b) the fraction $p$ of freely-moving neutrinos.
The standard scenario predicts $N_\nu=3$ and $p=1$.
Global fits take into account that a non-standard neutrino cosmology can be
masked by variations of other cosmological parameters: one usually considers
$\Lambda$CDM flat models. In particular, variations of $N_\nu$ can be
to a great extent compensated by variations in the Hubble constant
and in the DM density, 
such that present data do not allow a precise determination of $N_\nu$.
Data are reaching the necessary sensitivity;
at the moment different analyses find somewhat contradictory results~\cite{WMAP, nuScalar,cosmomnu}:
a reasonable summary is:
\beq N_\nu^{\rm CMB} \approx  5\pm 1,\qquad
p \approx 1\pm 0.3.\eeq

We remark that these cosmological data cannot measure the relative weight of each neutrino flavour,
and cannot discriminate neutrinos from other speculative freely-moving relativistic particles.

\subsection{Neutrino free-stream}\label{FreeStream}\index{Cosmology!power spectrum}
We here describe how it is possible to test if the energy $\rho_\nu$
is carried by freely-moving relativistic particles.
This allows to discriminate between massless vs massive  and between free vs interacting 
neutrinos.

Gravity tends to increase the initial fluctuations in the CDM density:
this clustering process finally forms  galaxies and the other structures we observe today.
Relativistic particles with mean free path larger 
than the horizon  (such as SM neutrinos) freely move (`free stream')
and suppress this clustering process.
In presence of new physics, neutrinos could instead interact among themselves or with new particles:
in such a case neutrinos would tend
to cluster in structures with size comparable to their Jeans length.

The presence or absence of neutrino free-streaming shifts the positions of the
acoustic peaks of the CMB photon radiation.
In this way it is possible to measure one more observable, $N_\nu^{\rm FS}$,
which tests possible extra interactions among neutrinos.
Present data mildly favor the standard scenario~\cite{nuScalar}.

\medskip

Neutrino free-stream is also prevented by neutrino masses:
at $T\circa{<}m_\nu$ neutrinos become non relativistic 
and slow down so that in a Hubble time travel only in a fraction $\sim v/c$ of the horizon.
Therefore massive neutrinos only reduce clustering at such small scales.
Neutrino masses determine two different things:
1) the temperature at which neutrinos cease to be non-relativistic,
which controls the length on which neutrinos travel reducing clustering;
2) the fraction of energy carried by neutrinos, see eq.\eq{Omeganu},
which controls how much neutrinos can smooth inhomogeneities.


Assuming standard cosmology LSS and CMB data imply and upper bound on $\Omega_\nu$, 
the energy density in neutrinos~\cite{cosmomnu}:
\beq \label{eq:cosmoboundonnumasses}
\Omega_\nu h^2\circa{<}0.6 \cdot 10^{-2}\qquad\hbox{i.e.}
\qquad \sum m_{\nu_i} \circa{<} 0.6\eV\qquad
\hbox{ at $99.9\%$ C.L.}\eeq
which is the presently dominant upper bound on neutrino masses. 
CMB data alone give a safer bound on neutrino masses 
which is almost one order of magnitude weaker;
in the global fit CMB data are mainly needed to determine other cosmological parameters.
We have chosen a large C.L.\  because presently statistical fluctuations
(or systematic shifts?) make the constraint slightly stronger than the sensitivity.
The precise final bound depends on how one deals with systematic uncertainties in LSS data:
more aggressive strategies give stronger but less reliable bounds.
In presence of extra sterile neutrinos the constraint
depends on their abundances and can get slightly relaxed.

\bigskip

For the reader not satisfied by our short qualitative discussion 
we here re-discuss the above issues in a more quantitative way.
A standard Newtonian computation shows that small
fluctuations in the DM density
 $\delta_{\rm DM}(x) = \delta\rho_{\rm DM}(x)/\rho_{\rm DM}\ll 1$ evolve according to~\cite{Kolb}
\beq\label{eq:DMevo}
\ddot\delta_{\rm DM} + 2H\dot\delta_{\rm DM} = 4\pi G_{\rm N}\,\delta \rho\label{eq:deltaDM} \eeq
where the source term $\delta\rho$
 is the fluctuation around the average  {\em total} density $\rho$.
 For our purposes $\rho=\rho_{\rm DM}+\rho_\nu$ and,
 as long as $T\gg m_\nu$, the relativistic motion of neutrinos basically
 prevents neutrino clustering: $\delta\rho_\nu =0$.
 Therefore the evolution equation becomes
\beq \label{eq:delta2}\ddot\delta_{\rm DM} + 2H\dot\delta_{\rm DM} = 4\pi G_{\rm N}\rho(1-f_\nu)\delta_{\rm DM}\eeq
 where $f_\nu\equiv \rho_\nu/\rho_{\rm DM} $.
 Our universe has critical density $\rho=3H^2/8\pi G_{\rm N}$ and
 during matter domination $H\simeq 2/3t$: for $f_\nu=0$
 one finds a growing solution $\delta_{\rm DM}\propto t^{2/3}\propto a \propto T^{-1}$, 
 where $a$ is the expanding scale-factor of the universe
 (normalized to be $a=1$ today).
 From matter domination to today primordial 
 fluctuations increased by a factor $\eV/T_{\rm now}\sim 5000$
 producing the observed LSS.
 
A $f_\nu>0$ suppresses the growth of DM fluctuations:  
assuming e.g.\ a constant $f_\nu$ the solution is
\beq \delta_{\rm DM}(t) \propto a(t)^p\qquad\hbox{with}\qquad
p=\frac{\sqrt{1+24(1-f_\nu)}-1}{4}.  
\eeq
Fluctuations do not grow if $f_\nu=1$ i.e.\  in a universe dominated by relativistic particles.
This is why we only consider the matter-dominated era.
We can also neglect the fact that during the latest stage of evolution
 the universe gets dominated by some form of vacuum energy
(maybe a cosmological constant): it reduces the
late-time growth of fluctuations by $\sim 25\%$ only.

 \medskip
 
\begin{figure}
$$\includegraphics[width=12cm]{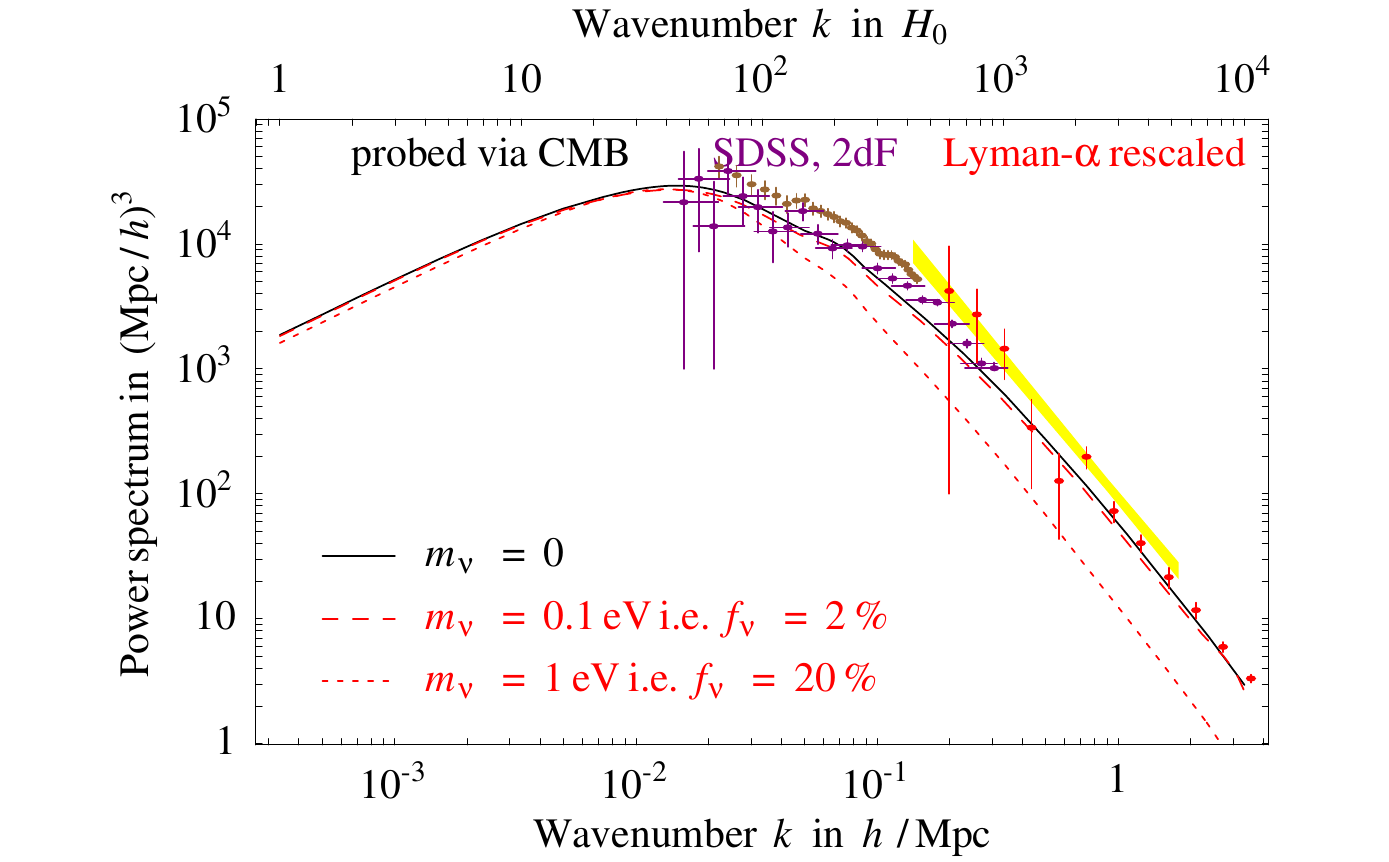}$$
\caption[Matter power spectrum]{\em The matter power spectrum $P(k)$ predicted by the best-fit
{\rm $\Lambda$CDM} cosmological model (continuous curve) and how neutrino masses affect it
(dashed curves).  Measurements at different scales 
have been performed with different techniques, that slightly overlap.
The data points do not show the overall uncertainty that plagues
galaxy surveys (SDSS, 2dF) at intermediate scales and especially Lyman-$\alpha$ data
at smaller scales i.e.\ at larger $k$.
\label{fig:Powermnu}}
\end{figure}

 Neutrinos started to be non relativistic after $a>a_{\rm NR}\approx T_\nu^0/m_\nu = 
 1.6~10^{-4}\eV/m_\nu$; in the phase with non relativistic neutrinos  one has a
 constant
   $f_\nu= \Omega_\nu/\Omega_{\rm DM}=\Omega_{\rm DM}^{-1}\cdot \sum m_\nu/94\eV$. 
A small $f_\nu$ can have a sizable impact because it slows the growth of fluctuations 
for a long time, since $a=a_{\rm NR}$ to today, $a=1$:
neutrino masses reduce the power spectrum $P$ (proportional to $\delta^2$) by 
\beq \label{eq:e-8f}
\frac{P(m_\nu)}{P(m_\nu=0)}= a_{\rm NR}^{2(1-p)}\approx
e^{-8f_\nu}.\eeq
The last form is a simple numerical interpolation.
This widely reported formula is the maximal effect, realized only on small scales
(right side of fig.\fig{Powermnu}).
To fully understand the effect of neutrino masses on different scales, we need
two further steps inside cosmology.
\begin{itemize}
\item Fluctuations $\delta(t,x)$ are decomposed in Fourier components $\delta(t,k)$
(where the wavenumber $k$ is conveniently chosen to be comoving, 
i.e.\ it corresponds to a wavelength 
$\lambda(t)=2\pi a(t)/ka(t)$ that expands together with the universe)
and the scale-dependent power spectrum is $P(k)\propto \delta(k)^2$.
We will not need the proportionality factor, defined such that the adimensional combination
$\sqrt{k^3 P(k)/2\pi^2}$ is the relative inhomogeneity $\delta$ on scale $k$.

\item Evolution of neutrino inhomogeneities $\delta_\nu$ is described by an equation analogous 
to\eq{deltaDM} with an extra pressure term:
\beq \ddot\delta_\nu + 2H\dot\delta_\nu = 4\pi G_{\rm N}~ \delta\rho - 
(\frac{kc_{\rm s}}{a})^2 \delta_\nu\eeq
where $c_{\rm s}$ is the average neutrino velocity
(a precise computation must study the evolution of the neutrino angular and energy distribution).
At $a\circa{>}a_{\rm NR}$ neutrinos become non-relativistic and 
$c_{\rm s} \approx T_\nu/m_\nu = T_\nu^0/m_\nu a$.
We previously assumed that neutrinos do not cluster: this is true
when $k$ is so large than the pressure term dominates over the gravitational terms
i.e.\  for  $k>k_{\rm Jeans}(a) \approx H_0 a^{1/2} m_\nu/T_\nu^0$.
\end{itemize}
Eq.\eq{e-8f} can now be generalized by repeating the previous arguments in the
different clustering regimes. The result is
\beq \label{eq:RP(k)}
\frac{P(m_\nu,k)}{P(m_\nu=0,k)}\approx \left\{\begin{array}{ll}
1 & k\circa{<}k_{\rm NR}\\
(k_{\rm NR}/k)^{4(1-p)} & k_{\rm NR}\circa{<}k\circa{<}k_0\\
(k_{\rm NR}/k_0)^{4(1-p)} & k\circa{>}k_0
\end{array}\right.\eeq
and depends on two scales:
$$k_{\rm NR} = k_{\rm Jeans}(a=a_{\rm NR}) \approx 60H_0 \sqrt{m_\nu/\eV},\qquad
k_0\equiv k_{\rm Jeans}(a=1) \approx 5000H_0\, (m_\nu/\eV).$$
At $k\circa{>}k_0$ one recovers eq.\eq{e-8f}.





Fig.\fig{Powermnu} shows the result of a full numerical computation.
Neutrino masses have a detectable effect  if $m_\nu\circa{>}0.1 \eV$;
optimistically future cosmological measurements of $P(k)$ might be sensitive
to the atmospheric mass splitting.

Neutrino masses have a minor impact on CMB anisotropies,
basically because photons last scattered at $T\sim 0.3\eV$, 
before that neutrino masses started to be an important factor.



\subsection{Caveats}
We explained the basic physics, but we could not here review 
the theoretical and the experimental techniques necessary to derive these results.
 However these results should not be accepted acritically:
 we now emphasize their main limitations.

 On the theoretical side, global fits are usually performed
 assuming `minimal cosmology':
$\Lambda$CDM in presence of primordial adiabatic fluctuations with constant spectral index.
Sometimes  critical density and/or scale-invariant spectrum is also assumed.
 This model has only been partially tested and confirmed.
 Fig.\fig{rhocosmo} shows the evolution of the main contributions to the average density:
a question mark is added on `$\Lambda$' and `CDM' because these components 
have not yet been directly tested.
A kink in the spectral index could simulate or mask the small effect of neutrino masses.
Minimal cosmology assumes no kink, but non-minimal models can easily produce it
(e.g.\ adding a jump in the inflaton potential or in other more elegant ways).
In conclusion, {\em results of  precision analyses of CMB and LSS data 
do not follow only from data but also
  rely on theoretical assumptions}.

 Furthermore, the evolution of  fluctuations can be analytically computed until they are
 a small corrections to an homogeneous background, such as around photon decoupling.
 Today this is true only on large enough scales,  $k\circa{<} 0.3/{\rm Mpc}\approx 1000 H_0$,
 while on small scales there are large deviations from the smooth approximation
 (e.g.\ galaxies are $\sim10^5$ more dense than the average universe).
Numerical simulations or approximations are needed to study the non-linear evolution of sizable 
inhomogeneities.
Due to this reason one sometimes prefers not to include the
Lyman-$\alpha$ data in fig.\fig{Powermnu}, that probe the power spectrum
on small scales dangerously close to the non-linear regime where the effects of neutrino masses are more sizable.

 \medskip
 
 On the experimental side there are some problematic systematics.
 Presently we can only measure the density of luminous matter,
and predict fluctuations in the total density.
Most analyses are performed assuming that
relative inhomogeneities in the density of galaxies are proportional (not necessarily equal) to
relative inhomogeneities in the total energy density,
with a scale-independent  proportionality factor $b$ (`bias').
Some analyses get stronger bounds by assuming models for the bias, 
that can be partially validated using the data.
Future measurements of gravitational lensing of CMB light
 and/or of light generated
 by far galaxies should allow to directly measure the total density  with greater accuracy.
 In this way, it might be possible to see the cosmological effects of neutrino masses,
 and measure them with an error a few times smaller than the atmospheric mass scale.
 This could allow to discriminate between normal and inverted neutrino mass hierarchy.

In the near future we expect improved measurements of the observables described above.
In longer terms (maybe in the next 20 years), new techniques might increase the sensitivity
of cosmology down to $\sum m_\nu \sim |\Delta m^2_{\rm atm}|^{1/2}\sim 0.05\eV$
{\em and} avoid the previous caveats: 
measuring gravitational lensing felt by CMB and by light emitted by first galaxies
should allow to reconstruct the time evolution of the power spectrum of all matter components
(unless this information will turn out to be masked by foregrounds).
This measurement would allow, for example, to discriminate the dynamical effect of neutrino masses
from the previously mentioned spectral kink.
Future maps of the 21cm hydrogen line might similarly allow to infer the history
of matter clustering up to $z\circa{<}50$, with maybe a sensitivity to neutrino masses down to $0.01\eV$~\cite{21cm}.
The next main experiment is {\sc Planck}, to be launched in 2008: it 
could measure $\sum m_\nu$ with $\pm0.15\eV$ uncertainty~\cite{cosmomnu}.

\subsection{Direct detection of neutrino CMB}\index{CMB neutrinos}
It would be interesting to directly test the background of relic CMB neutrinos.
Although neutrinos decoupled earlier than photons, being massive neutrino traveled less, so
that their surface of last scattering is closer to us than that of photons.
Extra light particles might distort it, giving
neutrino decay, anomalous neutrino interactions, extra oscillation channels...
Since relic neutrinos are today mostly non relativistic,
their effects significantly depend on whether neutrinos have
 Majorana or Dirac masses.
Furthermore massive neutrinos slower than the escape velocity cluster in galaxies.
In our galaxy $v_{\rm escape}\approx 500\km/\sec$,
which can be
comparable to the average neutrino velocity $\langle v_\nu\rangle\simeq (8 T_\nu/\pi m_i)^{1/2}
=(6200\km/\sec) (\eV/m_\nu)^{1/2}$.
If neutrinos have Dirac masses, gravitational interactions populate their non-interacting right-handed helicity,
because gravitational forces change the direction but not the spin of neutrinos.
Since neutrinos are fermions, local overdensities $n_{\nu_i}^{\rm local}$
are limited by the Pauli exclusion principle
(at most one neutrino in the phase-space volume $\Delta x \,\Delta p \sim \hbar$).
In fact, assuming that all levels get occupied up to $p\sim m_\nu v_{\rm escape}$\footnote{For comparison, the Fermi-Dirac distribution roughly corresponds
to having levels occupied up to $p\sim T_\nu$.}
one has
\beq n_{\nu_i}^{\rm local}-n_{\nu_i} \circa{<} \frac{(m_{\nu_i} v_{\rm escape})^3}{12\pi^2}=
\frac{5000}{\cm^3} \bigg(\frac{m_{\nu_i}}{\eV} \frac{v_{\rm escape}}{500\km}\bigg)
\eeq
where we inserted order one factors~\cite{detectingCMBnu}.
Precise simulations find an over-density of $50\%$ for $m_{\nu_i}=0.15\eV$.
Since the earth moves with velocity $v\sim 10^{-3}c$, this implies a flux of about
$10^{10\hbox{--}11}/{\rm cm}^2{\rm s}$.

\medskip

The detection of CMB neutrinos is a very challenging goal,
because their  interaction rates
are much below present experimental capabilities.
Some possibilities might become realistic in the future~\cite{detectingCMBnu}:
\begin{itemize}

\item {\em Neutrino capture on $\beta$-decaying nuclei}:
\beq
 \nu_e ~ A(Z) \to A(Z-1)~e^-,\qquad
 \bar\nu_e ~ A(Z)~ \to A(Z+1)~e^+.
  \eeq
One chooses nuclei such that the process is kinematically allowed
even without the initial-state CMB $\nu_e$:
the event rate is therefore not suppressed by the negligibly small $\nubarnu_e$ energy;
what matters is that $\nubarnu_e$ carries the lepton number that allows the reaction.
A rate of  about 10 events/year can be obtained with 100 g of H$^{3}$.
The price to be paid is that the initial nucleus radioactively decays emitting $e^\pm$,
giving a huge background.
The signal is a monochromatic $e^\pm$ with energy $\Delta E = 2m_\nu$ 
above the maximal energy of the continuous $\beta$-decay spectrum
(the factor 2 in $\Delta E$ gets reduced when taking into account neutrino mixings).
In practice this means that one must achieve an energy resolution
better than $m_\nu$. Since $m_\nu \circa{<}\eV$ this is a very demanding requirement:
the Katrin $\beta$-decay experiment (section~\ref{betadecay}) aims at an energy resolution of 1 eV.

\item {\em Coherent scattering}.
The neutrino-nucleus cross section is so small,
$\sigma\sim G_{\rm F}^2 \max(E_\nu,m_\nu)^2/\pi$,
that CMB neutrinos give something like
one event per year in a detector with mass $10^{16}\kg$.

A better signal is obtained exploiting the fact that 
CMB neutrinos have a macroscopic de Broglie wavelength,
$\lambda =1/p = (0.02\cm)\hbox{meV}/p$.
Therefore one can envisage elastic cross sections enhanced by the
square of the number $N$ of nucleons in a volume of size $R\circa{<}\lambda/2\pi$.
Coherence on even larger scales can be achieved using foam-like or laminated materials,
or simply many grains of size $R$.
Since each collision is accompanied by a momentum transfer $\Delta p = 2p$,
a flux $\Phi_\nu \sim n_\nu^{\rm local} v$ of non-relativistic neutrinos
produces an acceleration
\beq\label{eq:aCoherent}  a \sim \Phi_\nu \sigma_{\nu N} \Delta  p \frac{N^2}{M}\sim
10^{-28}\frac{\cm^2}{\sec} \frac{n_\nu^{\rm local}}{10^3/\cm^3} \frac{10^{-3}}{v}\frac{\rho}{\hbox{g}/\cm^3}
\bigg(\frac{R}{1/(m_\nu v)}\bigg)^3
 \eeq
 where $ v = |\langle \vec v_{\nu} - \vec v_\oplus \rangle|$ is the mean relative velocity between
 CMB neutrinos and the target,
 $\rho$ is density of the target and $M$ is its mass.
 The minimal acceleration that can be detected at present is $10^{-13} \cm/\s^2$,
 orders of magnitude above the expected signal.
 Solar neutrinos and possibly Dark Matter produce a much larger background.

The estimate\eq{aCoherent} is correct for Dirac neutrinos, that can have vectorial couplings.
Majorana neutrinos only have axial couplings, that in the non-relativistic limit
give effects suppressed by $v^2\sim 10^{-6}$ for an unpolarized target, 
or by $v\sim 10^{-3}$ for a polarized target.

\item {\em The Stodolski effect}.  
The energy of an electron receives an extra contribution (analogous to the MSW effect)
$\delta E\sim G_{\rm F} \vec{s}\cdot \vec{v}
( n_\nu-n_{\bar{\nu}})$
that depends on the
direction of its spin $\vec{s}$ with respect to the neutrino wind $\vec v$.
Notice that $\delta E$ is linear in $G_{\rm F}$ but
suppressed if the neutrino asymmetry 
$\eta_\nu = ( n_\nu-n_{\bar{\nu}})/n_\gamma$ is small.
We do not show the dependence on the Dirac or Majorana nature of neutrinos.
The effect manifests as a torque $\tau\sim N_e \,\delta E$
acting on magnetized macroscopic
object with $N_e$ polarized electrons.
Therefore an object with size $R$, 
one polarized electron per atom, atomic number $A$,
mass $M\sim N_e m_e A$
feels a linear acceleration
\beq a\sim \frac{\tau}{MR}\sim10^{-28}\frac{\cm}{\sec^2}\frac{100}{A}
\frac{\cm}{R}\frac{v_\nu}{10^{-3}}\eta_\nu
\eeq
about 15 orders of magnitude below the present sensitivity.

\item {\em Scattering of ultra-high-energy cosmic rays} (CR).
Cosmic rays with energy $E=M_Z^2/2m_{\nu_i}=4~10^{21}\GeV (\eV/m_{\nu_i})$
scatter on relic neutrinos with cross section enhanced by the $Z$-peak
resonance, $\sigma \simeq 2\pi \sqrt{2} G_{\rm F}$.
Their mean free path $1/(n_\nu \sigma)$ is about two orders of magnitude larger that the Hubble distance. 
Therefore one can search for 
a few $\%$ absorption dip in the spectrum of UHE CR,
and for the consequent scattering products (protons, photons,...).
This process could produce cosmic rays above the  GZK cut-off~\cite{GZK}.

Scattering with relic neutrinos of particle beams produced in  
foreseeable future accelerators does not give interesting rates.

\end{itemize}

%

\subsection{Neutrinos and the vacuum energy}\index{Cosmology!dark energy}
Cosmological data suggest that some contribution $\rho_{\rm DE}$
to the total energy density $\rho$ of the universe
remains roughly constant and is becoming today the dominant contribution, see fig.\fig{rhocosmo}.
It is known as `Dark Energy' (DE) because we do not know what it is.
The DE might be a cosmological constant,
$\rho_{\rm DE} = \rho_\Lambda = (2.3~\meV)^4$.
This interpretation involves only one free parameter and is so far consistent with data.
However it is puzzling that $\rho_\Lambda$ is so much smaller than
particle-physics energy scales that, according to theoretical prejudices, should contribute to it:
e.g.\ $\rho_\Lambda\sim 10^{-35}\Lambda_{\rm QCD}^4
\sim 10^{-50}v^4\sim 10^{-120}M_{\rm Pl}^4$.
All theoretical attempts of understanding the smallness of $\rho_\Lambda$ failed.
This issue might be affected by antrophic selection: a $\rho_\Lambda$ 100 times bigger
would have prevented the formation of our galaxy.

Alternatively, one can imagine that due to some unknown reason
the cosmological constant is zero, 
that DE is due to something else possibly related to particle physics.
We here discuss neutrino masses, which are a good candidate
because happen to be comparable to $\rho_\Lambda^{1/4}$.

`Mass varying neutrinos'~\cite{DEnu} are the most direct possibility:
in the non relativistic limit $\rho_\nu \sim m_\nu T^3$,
so that a constant DE would be obtained if  $m_\nu \propto 1/T^3$.
Attempts of building models with this property assume that
neutrino masses $m_\nu$ depend on the vev of a light scalar field $\phi$ that evolves such
that the total energy density $\rho_\nu(\phi) + V(\phi)$ remains roughly constant.
There is a generic problem: a scalar $\phi$ that evolves tracking
the minimum of $\rho_\nu(\phi) + V(\phi)$ also generates a long-range force
among neutrinos that causes their clustering, 
leaving a vanishing cosmological constant in the space outside of the neutrino clusters.
Furthermore quantum corrections to the potential $V(\phi)$ 
are much larger than the needed $V(\phi)$.
Without caring of such theoretical problems,
one can even speculate that the LSND anomaly is due
to the fact that  neutrino masses in rock are larger than
neutrino masses in air.

Alternatively, one can try to see what can be achieved in
 a self-consistent theoretical framework~\cite{DEnu}.
 Pseudo-Goldstone scalars are naturally light:
a scalar field with vev $f$ that spontaneously breaks a global symmetry,
also broken explicitly at a low energy $M$,
gets a potential $V(\phi)\sim M^4 \cos (\phi/f)+{\rm cte}$.
If $M^4\sim \rho_{\rm DE}$ and $f\sim M_{\rm Pl}$
the scalar is so light, $m_\phi\sim \rho_{\rm DE}^{1/2}/M_{\rm Pl}\sim 10^{-33}\eV$,
that only today $H^{-1}\sim m_\phi$ and $\phi$ starts falling towards the minimum of its potential.
Unfortunately it seems that $\phi$ cannot couple to SM particles, because
such couplings would destroy its lightness and/or would lead to unobserved
long range forces.
Interestingly, it is possible to couple $\phi$ to {\em right-handed} neutrinos such that
the identification $M\sim m_\nu$ has oscillation signatures.
Models where Majorana  masses of left-handed neutrinos depend on $\phi$ fail,
because Majorana $LL$  masses come either from
a non renormalizable interaction
or from exchange of heavy new particles (see section~\ref{Neutrino}),
resulting in either case in $M\gg m_\nu$.
Successful models are obtained by assuming that only
the Majorana $RR$ mass $M$ of light right-handed neutrino(s) depends on $\phi$,
because this is obtained from a renormalizable Yukawa interaction.
A detailed analysis shows that
neutrino masses can be obtained by adding $LR$ Dirac masses but no $LL$ Majorana masses,
resulting in a particular sub-case of mixed Majorana/Dirac.
(The opposite possibility is also trivially allowed, but does not seem phenomenologically interesting).

 Recently, possible space or time variations of fundamental constants 
attracted some interest. 
We remark that if the vacuum energy is small thanks to a fine-tuning done once and for all,
detectable variations of e.g.\  $\alpha_{\rm em}$ or $m_e/m_p$ are not compatible with its
observed smallness.
The above discussion shows that neutrino masses are the only fundamental constants that can 
vary without generating a too large vacuum energy.

\section{Baryogenesis through leptogenesis}\label{leptogenesi}\index{Leptogenesis}
The present baryon density of eq.\eq{nB}\footnote{We discuss how this number
is measured.
The photon density directly follows from the measurement
of the temperature of the CMB. Measuring baryons is more difficult,
because only a fraction of baryon formed stars and other luminous objects.
As discussed in section~\ref{nucleos}, 
BBN predictions depend on $n_B/n_\gamma$.
In particular the  presence of many more photons than baryons
delays big-bang nucleosynthesis (e.g.\ by enhancing the reaction
$pn\leftrightarrow \hbox{D}\gamma$ 
in the $\leftarrow$ direction)
and therefore directly follows from the primordial helium-4 abundancy.
The accurate determination in eq.\eq{nB} of $n_B/n_\gamma$ follows
from global fits of WMAP data about CMB anisotropies~\cite{WMAP}.}
can be obtained from an hot big-bang as the result of a
small excess of baryons over anti-baryons.
We would like to understand why,
when at $T\circa{<}m_p$ matter  almost completely annihilated with anti-matter,
we survived thanks to the `almost'
$$n_B - n_{\bar B}\propto 1000000001-1000000000 = 1.$$
This might be the initial condition at the beginning of the big-bang,
but it would a surprisingly small excess.
In inflationary models it is regarded as a surprisingly large excess,
since inflation erases initial conditions.

Assuming that the hot-big-bang started with zero baryon asymmetry
at some temperature $T\gg m_p$, can the baryon asymmetry
can be generated dynamically in the subsequent evolution?
Once that one realizes that this is an interesting issue (this was done by Sakharov), 
the answer is almost obvious:   yes,  provided that at some stage~\cite{Baryogenesis}
\begin{itemize}
\item[1.] baryon number $B$ is violated; 
\item[2.]  C and CP are violated
(otherwise baryons and antibaryons are generated in the same rate);
\item[3.]  the universe was not in thermal equilibrium
 (since we believe that CPT is conserved, particles and
antiparticles have the same mass, and therefore in thermal equilibrium have the same abundance).
\end{itemize}
A large amount of theoretical and experimental work showed that
{\em within the SM these conditions are not fulfilled}.
At first sight one might guess that the only problem is 1.;
in reality 2.\ and 3.\ are problematic.

\smallskip 

\begin{itemize}
\item[1.] Within the SM $B$ is violated in a non trivial way~\cite{Baryogenesis},
thanks to quantum anomalies combined with non-perturbative $\SU(2)_L$ processes known as sphalerons\footnote{We just state the result, because we are not able of providing an intuitive explanation of the basic physics,
which needs a deep understanding of advanced quantum field theory.
The $\eta'$ mass is the only observed consequence of this sector, and it is due
to the QCD analogous of electroweak sphalerons.  Therefore there should
be no doubt that sphalerons really exist, and
this is almost all what one needs to know to understand leptogenesis quantitatively.}:
the anomalous $B$ and $L$ symmetry are violated, while
$B-L$ is a conserved anomaly-free symmetry.
At temperatures $T \ll 100\GeV$ sphaleron rates are negligible because
suppressed by a tunneling factor 
$ e^{-2\pi /\alpha_2}$: tritium decays at an unobservably slow rate.\footnote{Sphalerons
do not induce proton decay, because there are 3 generations
and all of them must be involved in sphaleron processes.}
At $T\circa{>}100\GeV$ thermal tunneling gives fast $B+L$-violating
sphaleron processes:
the space-time density of sphaleron interactions, $\gamma\sim \alpha_2^5 T^4$, 
 is faster than the expansion rate of the universe up to temperatures of about $T\sim 10^{12}\GeV$~\cite{Baryogenesis}.

 \item[3.]
SM baryogenesis is not possible due to the lack of out-of equilibrium conditions.
The electroweak phase transition was regarded as a potential out-of equilibrium stage, but 
experiments now demand a higgs mass $m_h \circa{>}115\GeV$~\cite{LEP},
and SM computations of the Higgs thermal potential show that, for $m_h \circa{>} 70\GeV$, 
the higgs vev shifts smoothly from $\langle h\rangle = 0$ to
$\langle h\rangle = v$ as long as the universe cools down below $T\sim m_h$.

\item[2.]
In any case,
the amount of CP violation provided by the CKM phase would have been too small for generating the
observed baryon asymmetry. 
 \end{itemize}

Many extensions of the SM could generate the observed $n_B$.
`Baryogenesis at the electroweak phase transition'
needs new particles coupled to the higgs 
in order to obtain a out-of-equilibrium phase transition
and to provide extra sources of CP violation.
This already disfavored possibility will be tested at future accelerators.
`Baryogenesis from decays of GUT particles'
seems to conflict with non-observation of magnetic monopoles, at least in simplest models.
Furthermore minimal GUT model do not violate $B-L$,
so that sphaleron processes would later wash out the possibly generated baryon asymmetry.

\medskip

The existence of sphalerons changes the rules of the game and
suggests {\em baryogenesis through leptogenesis}:
lepton number might be violated by some non SM physics,
giving rise to a lepton asymmetry, which is converted
into the observed baryon asymmetry by sphalerons.

This scenario can be realized in many different ways~\cite{Baryogenesis}.
Majorana neutrino masses violate lepton number and presumably CP,
but do not provide enough out-of-equilibrium processes.
The minimal successful implementation~\cite{leptogenesis} needs just the minimal
amount of new physics which can give the observed small neutrino masses
via the see-saw mechanism:
heavy right-handed neutrinos $N$ with masses $M$.
{\em `Baryogenesis via thermal  leptogenesis'}~\cite{leptogenesis} proceeds at $T\sim M$, when
out-of-equilibrium  (condition 3) CP-violating (condition 2) decays of heavy right-handed neutrinos
generate a lepton asymmetry, converted in baryon asymmetry by SM sphalerons (condition 1).


\begin{figure}
$$\includegraphics[width=14cm]{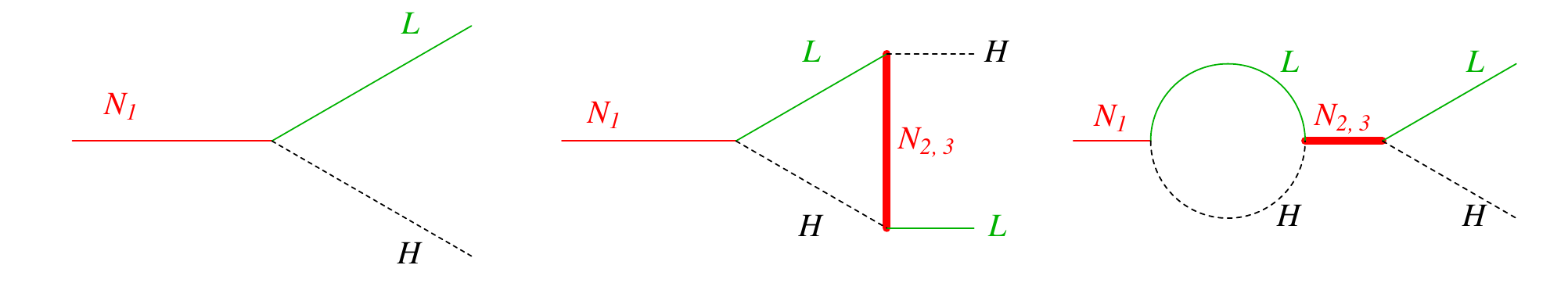}$$
\caption[CP violation in $N_1$ decays]{\em CP-violating $N_1$ decay.\label{fig:Ndecay}}
\end{figure}

\subsection{Thermal leptogenesis}
We now discuss the basic physics, obtaining estimates for the main results;
precise results are outlined in a series of footnotes.
The SM is extended by adding the heavy right-handed neutrinos suggested by see-saw models.
To get the essential points, we consider a simplified model with one lepton doublet $L$
(`one-flavor approximation')
and two right-handed neutrinos,
that we name `$N_1$' and `$N_{2,3}$', with 
$N_1$ lighter than $N_{2,3}$.
The relevant Lagrangian is
\beq\label{eq:seesaw1f}
\Lag = \Lag_{\rm SM}+ \bar N_{1,2,3} i \ds\, N_{1,2,3} +
\lambda_1 N_1 HL + \lambda_{2,3} N_{2,3} HL + \frac{M_1}{2} N_1^2 + \frac{M_{2,3}}{2} N^{2}_{2,3}+\hbox{h.c.}\eeq
By redefining the phases of the $N_1$, $N_{2,3}$, $L$ fields one can set $M_1$, $M_{2,3}$, $\lambda_1$ real
leaving an ineliminabile CP-violating phase in $\lambda_{2,3}=|\lambda_{2,3}|e^{-i\delta}$.


\begin{figure}[t]
$$\includegraphics[width=\textwidth]{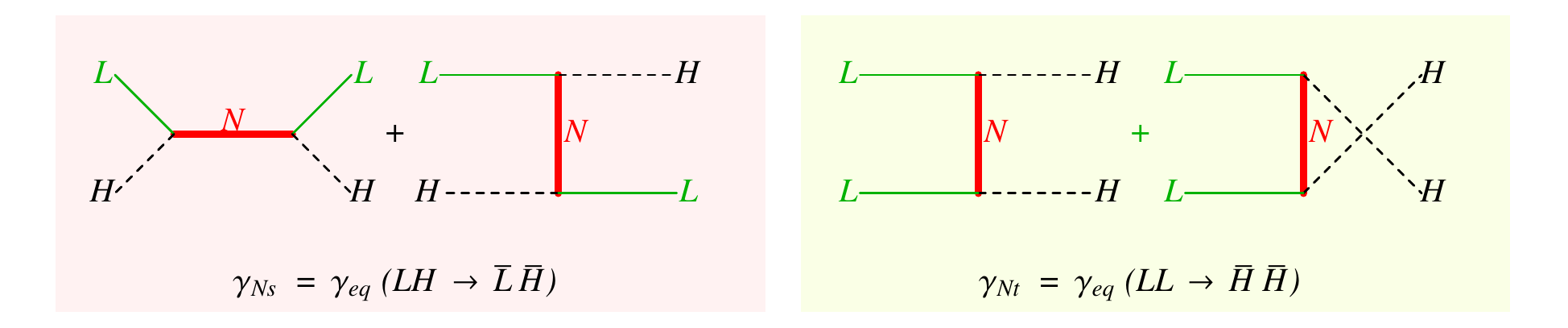}$$
\caption[$\Delta L=2$ scatterings]{\em Wash-out  $\Delta L = 2$ scatterings.
\label{fig:Nscatt2}}
\end{figure}

\smallskip

At tree-level, the decay width of $N_1$ is  $ \Gamma_1 = {\lambda^2_1 M_1}/{8\pi} $.
The interference between tree and loop diagrams shown in fig.\fig{Ndecay}
renders $N_1$ decays CP-asymmetric:
$$  \varepsilon_1 \equiv \frac{\Gamma(N_1\to L H)-\Gamma(N_1\to \bar{L} H^*)}{\Gamma(N_1\to L H)+\Gamma(N_1\to \bar{L} H^*)}\sim
\frac{1}{4\pi}\frac{M_1}{M_{2,3}}
\Im\lambda^{ 2}_{2,3}$$
In fact 
$$ \Gamma(N_1\to L H) \propto |\lambda_1 + A \lambda^*_1 \lambda^{2}_{2,3}|^2,\qquad
\Gamma(N_1\to \bar{L} H^*) \propto |\lambda^*_1 + A \lambda_1 \lambda^{2* }_{2,3}|^2$$
where $A$ is the complex CP-conserving loop factor.
In the limit $M_{2,3}\gg M_1$ the sum of the two one loop diagrams
reduces to an insertion of the $(LH)^2$ neutrino mass operator mediated by $N_{2,3}$:
therefore $A$ is suppressed by one power of $M_{2,3}$.
The intermediate states in the loop diagrams in fig.\fig{Ndecay} can be on shell;
therefore the Cutkosky rule~\cite{Cutkosky} guarantees that $A$ has an imaginary part.
Inserting the numerical factor valid for $M_{2,3}\gg M_1$
we can rewrite the CP-asymmetry as
\begin{equation}
\label{eq:epsilon1}{\color{blu}
\varepsilon_1 \simeq  \frac{3}{16\pi} \frac{\tilde m_{2,3} M_1}{v^2} \sin\delta} = 10^{-6}\frac{\tilde m_{2,3}}{0.05\eV}\frac{M_1}{10^{10}\GeV} \sin\delta
\end{equation}
where {\color{blu} $\tilde{m}_{2,3}\equiv |\lambda^{2}_{2,3}| v^2/M_{2,3}$} is
the contribution to the light neutrino mass mediated by $N_{2,3}$.\footnote{In
a generic see-saw model defined by the Lagrangian in eq.\eq{Lseesaw2},
the CP asymmetry in the decay of the lightest right-handed neutrino $N_1$ 
with mass $M_1< M_2<M_3$ can be written
as the sum of a $V$ertex contribution 
and of a $S$elf-energy contribution 
\begin{equation}\label{eq:eps}
\varepsilon_1=-\sum_{j=2,3}
  \frac{M_1 }{M_j }\frac{\Gamma_j }{M_j }\bigg(S_j + \frac{V_j}{2}\bigg)
\frac{ \hbox{Im}\,[ Y_{1j}^2 ]}{|Y |_{11} |Y |_{jj}}\,,
\end{equation}
where
\begin{equation}\label{eq:IGamma}
\frac{\Gamma_j}{M_j} = \frac{|Y |_{jj}}{8\pi}\equiv \frac{\tilde{m}_j M_j}{8\pi v^2}
\,,\qquad
S_j = \frac{M^2_j  \Delta M^2_{1j}}{(\Delta M^2_{1j})^2+M_1 ^2
   \Gamma_j ^2} \,, \qquad
V_j = 2 \frac{M^2_j }{M^2_1}
\bigg[ (1+\frac{M^2_j }{M^2_1})\log(1+\frac{M^2_1}{M^2_j })
- 1 \bigg]\,,
\end{equation}
with $\Delta M^2_{ij}\equiv M^2_j-M^2_i$ and $Y \equiv \lambda_N \cdot \lambda_N^\dagger$.
In the hierarchical limit $M_{2,3}/M_1\to\infty$ one has $S_{2,3}=1$
and $V_{2,3}=1$.
If $N_1$ and $N_2$ are almost degenerate
the CP-asymmetry is enhanced by a new effect, CP violation in $N_1N_2$ mixing,
which can give $|\varepsilon_1|\sim 1$.
This phenomenon is accounted by the factor $S_j$, which has a form
well known from the analogous $K^0\bar{K}^0$ system.
Neglecting this potential effect, the CP-asymmetry can be rewritten in a way
which emphasizes the flavour structure of the Yukawa couplings:
\be
\label{eq:epsN}
\varepsilon_1 = 
-\frac{1}{8 \pi}
\frac{ {\rm Im}\, {(Y\cdot f(M_N/M_1)\cdot Y^*)_{11}}}{Y_{11}}=
\frac{1}{8 \pi}\sum_{i >1}\frac{  {\rm Im} \,Y^2_{1i}}{Y_{11}}
f\left(\frac{M_i}{M_1}\right)\ee
In the SM
$$f(r) = r (1+r^2) \ln (1+r^{-2}) - r + {r}/(r^2-1) \simeq 3/2r \quad\hbox{for $r \gg 1$}
$$
and in supersymmetric extensions of the SM
$f(r) =
r \ln (1+r^{-2})+2r/(r^2-1) \simeq 3/r $.}
Other sources of CP violation (such as the CKM matrix)
which do not violate lepton number do not contribute to the above
CP asymmetry due to unitarity reasons.

The baryon asymmetry generated by $N_1$ decays can be written as
\begin{equation}
\label{eq:nB1}\color{blu}
\frac{n_B}{n_\gamma} = -1.14 \frac{ \varepsilon_1 \eta}{g_{\rm SM}}
\end{equation} 
where $ g_{\rm SM} = 118$ is the number of spin-degrees of freedom of SM particles
(present in the denominator of eq.\eq{nB1}
because only $N_1$ among the many other particles in the thermal bath
generates the asymmetry)
and $\color{blu}\eta $ is an efficiency factor that depends on 
how much out-of-equilibrium $N_1$-decays are.
We now discuss this issue.\footnote{A precise description can be achieved by solving the 
relevant set of Boltzmann equations, taking into account CP-violating $N_1$ decays and scatterings
as well as their inverse processes, as described in the literature~\cite{leptogenesis}.
We here give exact numerical results, presenting a semi-quantitative explanation of the main features.}
If $N_1$ decays and annihilations are slow enough,
the $N_1$ abundancy does not decrease according to the Boltzmann equilibrium statistics 
$\sim e^{-M_1/T}$ demanded by thermal equilibrium,
so that out-of-equilibrium $N_1$ decays can generate a lepton asymmetry.
Slow enough decay means $\Gamma_1 < H(T)$ at $T\sim M_1$, i.e.\
 $N_1$ lifetime longer than the inverse expansion rate.
Numerically
$$ \gamma \equiv {\Gamma_1\over H}\sim {\tilde{m}_1\over \tilde{m}^*}\qquad\hbox{where}\qquad
\tilde{m}^*\equiv \frac{256\sqrt{g_{\rm SM}} v^2}{3M_{\rm Pl}}=2.3~10^{-3}\eV$$
depends on cosmology.

\begin{figure}[t]
$$\includegraphics[width=8cm]{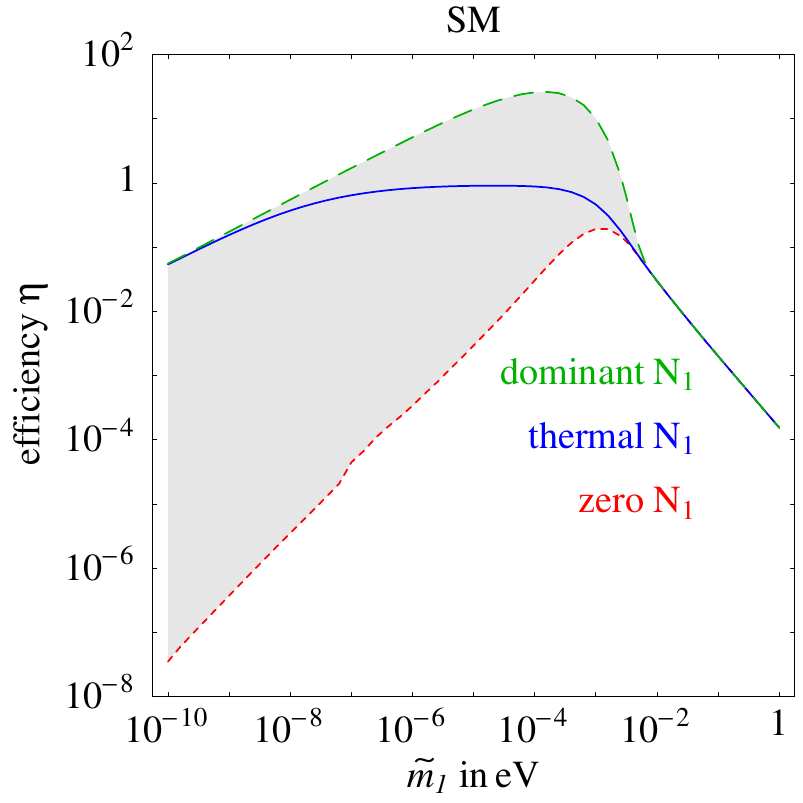}\hspace{1cm}
\includegraphics[width=8cm]{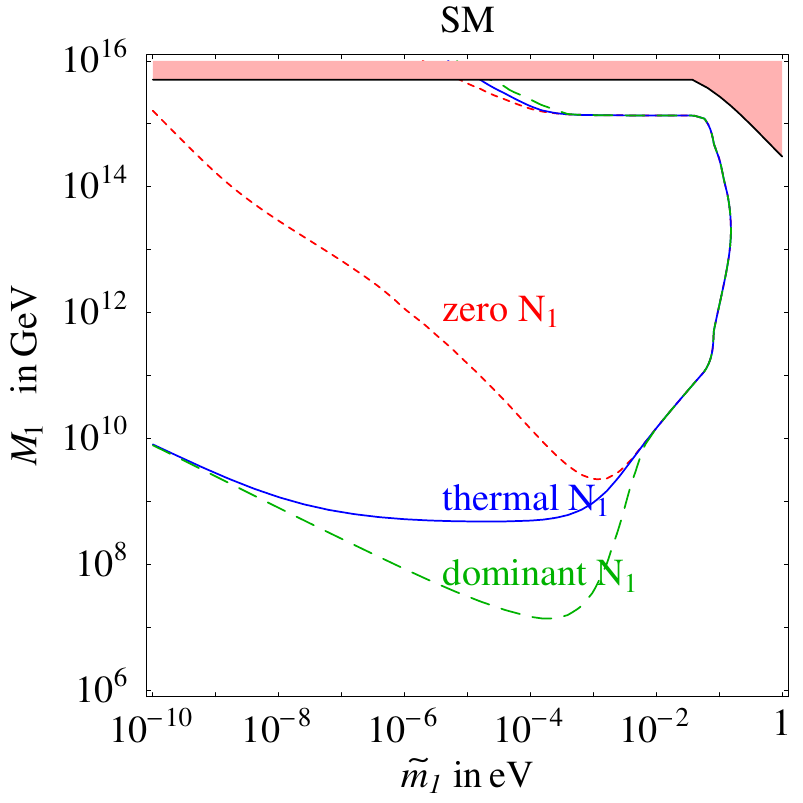}$$
\caption[Efficiency of leptogenesis]{\em {\bf Thermal leptogenesis}.
Fig.\fig{leptogEta}a: efficiency $\eta$ as function of 
$\tilde{m}_1$ for $M_1=10^{10}\GeV$
and for different assumptions about the initial $N_1$ abundancy:
for $\tilde{m}_1$ larger than a few {\rm meV} the efficiency is univocally predicted to be
$\eta(\tilde{m}_1) \approx 0.42 ({\rm meV}/\tilde{m}_1)^{1.15}$.
Fig.\fig{leptogEta}b: the regions in the $(\tilde{m}_1,M_1)$ plane
inside the curves can lead to successful leptogenesis.
\label{fig:leptogEta}}
\end{figure}

All the dependence on the mass and Yukawa couplings of $N_1$ is incorporated in 
$\color{blu}\tilde{m}_1\equiv \lambda^2_1 v^2/M_1$,
the contribution to the light neutrino mass mediated by $N_1$.
Unfortunately $\tilde{m}_1$ and $\tilde{m}_{2,3}$ are only related to the observed atmospheric and solar mass
splittings in a model-dependent way.
Unless neutrinos are almost degenerate
(and unless there are cancellations)
$\tilde{m}_1$ and $\tilde{m}_{2,3}$ are smaller than $m_{\rm atm}\approx 0.05\eV$.
\medskip

If $\gamma\ll 1$ (i.e.\ $N_1$ decays strongly out-of-equilibrium) then $\eta=1$.
If instead $\gamma \gg 1$ the lepton asymmetry is suppressed by $\eta\sim1/\gamma$.
The reason is that $\Delta L=1$ wash-out interactions 
(mainly $N_1$ decays and inverse-decays)
have $N_1$ as an external state
and therefore at low temperature $T< M_1$ their thermally-averaged rates are suppressed by a Boltzmann factor:
$\gamma_{\Delta L=1}(T < M_1)\approx \gamma  e^{-M_1/T} $. 
The $N_1$ quanta that decay when $\gamma_{\Delta L = 1}<1$, i.e.\ at $T< M_1/\ln \gamma$,
give rise to unwashed leptonic asymmetry.
At this stage the $N_1$ abundancy is suppressed by the Boltzmann factor $e^{-M_1/T} = 1/\gamma$.
In conclusion, the suppression factor is approximately given by
$$\eta \sim \min(1,\tilde{m}^*/\tilde{m}_1)\qquad
\hbox{(if $N_1$ initially have thermal abundancy)}.$$
Fig.\fig{Nscatt2} shows additional $\Delta L=2$ washing interactions mediated by $N_{1,2,3}$.
Dropping the contribution mediated by on-shell $N_1$
(already included as successive inverse decays and decays,
$H^*\bar{L}\leftrightarrow N_1 \leftrightarrow  HL$)
these scatterings
are generated by the same dimension 5 operators that give rise to Majorana neutrino masses.
Their thermally-averaged interaction rates are relevant only at
$M_1\circa{>}10^{14}\GeV$, when neutrinos have ${\cal O}(1)$ Yukawa couplings.
In such a case these interactions wash out the baryon asymmetry exponentially,
because are not suppressed at $T\circa{<} M_1$ by
the small $N_1$ abundancy.

\medskip

So far we assumed that right-handed neutrinos have thermal abundancy.
If instead $N_1$ start with zero abundancy and are generated
only by the scattering processes previously discussed,
at $T\sim M_1$ they reach thermal abundancy only if $\gamma \gg 1$.
Their abundancy is otherwise suppressed by a $\gamma$ factor.
Therefore the efficiency factor is approximatively given by
$$\eta \sim \min(\tilde{m}_1/\tilde{m}^*,\tilde{m}^*/\tilde{m}_1)\qquad
\hbox{(if $N_1$ initially have zero  abundancy)}.$$
If instead $N_1$ dominate the energy density of the universe, 
the suppression factor $1/g_{\rm SM}$ in eq.\eq{nB1} no longer applies, and
the efficiency factor can reach $\eta\sim g_{\rm SM}$.

These estimates agree with the results of detailed numerical
computations, shown in fig.\fig{leptogEta}a, performed
in one-flavor approximation and assuming that the $N_{2,3}$ contributions
to the baryon asymmetry can be fully neglected.\footnote{We here outline these computations.
The evolution of the total density of right-handed neutrinos,
$Y_{N_1} \equiv n_{N_1}/s$ and of the lepton asymmetry, $Y_{\cal L} \equiv (n_L - n_{\bar L})/s$
(measured in units of the total entropy density $s$)
as function of $z\equiv M_{N_1}/T$ is described by the Boltzmann equations
\beq\label{eq:BoltzmannLeptog}
zHs \frac{dY_{N_1}}{dz} = -\gamma_D \bigg(\frac{Y_{N_1}}{Y_{N_1}^{\rm eq}}-1\bigg),
\qquad
zHs \frac{dY_{\cal L}}{dz} =\gamma_D \bigg[\varepsilon_{1} 
\bigg(\frac{Y_{N_1}}{Y_{N_1}^{\rm eq}}-1\bigg) - 
\frac{Y_{\cal L}}{2Y_{L}^{\rm eq}}\bigg]
\eeq
where we kept only the dominant processes, 
$HL\leftrightarrow N_1\leftrightarrow H^*\bar{L}$,
with thermally averaged spacetime density $\gamma_D$.
The suffix `eq' denotes the values in thermal equilibrium.
The first three factors in the equation for $Y_{\cal L}$ correspond to three Sakharov conditions:
violation of lepton number ($\gamma_D$), of CP 
($\varepsilon_{1}$), 
departure from thermal equilibrium ($Y_{N_1}\neq Y_{N_1}^{\rm eq}$).
An important subtlety: decays and inverse-decays
$N_1\leftrightarrow HL,H^*\bar{L}$ alone generate a lepton asymmetry even in thermal equilibrium:
if $N_1$ decays preferentially produce leptons,
then CPT-invariance implies that inverse decay preferentially destroy anti-leptons.
Boltzmann equations that respect the Sacharov conditions are obtained 
after including the non-resonant part of $\Delta L=2$ scatterings
$HL\leftrightarrow H^*\bar{L}$.

State-of-the-art computations include various, but not all, subdominant effects.}
If $\tilde{m}\gg \tilde{m}^*$ (in particular if $\tilde{m}=m_{\rm atm}$ or
$m_{\rm sun}$) 
the efficiency does not depend on initial conditions
because $N_1$ is close to thermal equilibrium
and, in one-flavor approximation, washes-out a possible pre-existing asymmetry
(generated e.g.\ by the heavier $N_{2,3}$, more later).


\subsection{Leptogenesis constraints on neutrino masses}

Assuming that the observed baryon asymmetry is due to thermal leptogenesis
and that right-handed neutrinos are hierarchical,
one can derive interesting constraints on the masses of left-handed and right-handed neutrinos.
We remark that these results, discussed in the rest of this section,
{\em hold only under these untested assumptions} and therefore
are not true bounds.
We start reporting the clean presentation and precise results
obtained thanks to standard simplifying assumptions,
and later discuss their limitations.

\medskip

Assuming $M_1\ll M_{2,3}$ and neglecting terms suppressed by powers
of $M_1/M_{2,3}$,  
the two one-loop diagrams in fig.\fig{Ndecay} reduce to insertion of the neutrino mass operator
$(LH)^2$. Therefore the total CP-asymmetry $ \varepsilon_1$ in $N_1$ decays
(which sources leptogenesis in one-flavor approximation)
is directly related to neutrino masses,
and e.g.\ $ \varepsilon_1$ vanishes if neutrinos are degenerate. In fact, 
using the parametrization in eq.\eq{seesawParamCI}
$$ \varepsilon_1 = 
\frac{3}{16\pi}\frac{M_1}{v^2}
\frac{\hbox{Im} (\lambda_N \cdot m_\nu^*\cdot \lambda_N^T)_{11}}{(\lambda_N\lambda_N^\dagger)_{11}}=
 \frac{3}{16\pi}\frac{M_1}{v^2}
 \frac{\sum_i m_{\nu_i}^2 \hbox{Im}\, R_{1i}^2}{\sum_i m_{\nu_i} |R_{1i}|^2}$$
gives an upper bound~\cite{leptogenesisBounds}:
\beq\label{eq:DI} \color{blus} | \varepsilon_1| \le
 \frac{3}{16\pi}\frac{M_1}{v^2}(m_{\nu_3} - m_{\nu_1}).\eeq
Here $m_{\nu_3}$ ($m_{\nu_1}$) denotes the mass of the heaviest (lightest) neutrino.
In the limit $m_{\nu_1} \ll m_{\nu_3}$ the bound is trivial and holds when the CP asymmetry
is controlled by the neutrino mass operator, independently of
the specific source  that generates it (this implicitly means that
the contribution to neutrino masses non mediated by $N_1$ is mediated
by something much heavier than $N_1$).
The factor $m_{\nu_3} - m_{\nu_1}$ is non-trivial and
specific to the see-saw model with 3 right-handed neutrinos.
It can be intuitively understood as follows: 3 right-handed neutrinos can produce
the limiting case of equal neutrino masses 
only in the following way:
each $N_i$  gives the same mass to one neutrino mass eigenstate.
Since they are orthogonal in flavour space,
the CP-asymmetry of eq.\eq{DI} vanishes due to flavour orthogonality:
this is the origin of the $m_{\nu_3} - m_{\nu_1}$ suppression factor.

\medskip

\paragraph{Constraint on right-handed neutrino masses.}
The bound\eq{DI} becomes more stringent if right-handed neutrinos are light.
Combined with a precise SM computation of thermal leptogenesis 
(performed neglecting flavour issues)
and with the measured values of the baryon asymmetry and of the neutrino masses it implies that
\beq\label{eq:mNbound}
M_1 >  \left\{\begin{array}{rl}
2.4\times 10^{9} \GeV    &   \hbox{if $N_1$ has zero} \\
 4.9\times 10^{8}\GeV     &\hbox{if $N_1$ has thermal} \\  
1.7\times 10^{7}\GeV     &\hbox{if $N_1$ has dominant}
\end{array}\right.
\hbox{initial abundancy at $T\gg M_1 $.}
\eeq
We now discuss how this bound changes when simplifying approximations are dropped.

Flavor can affect the dynamics of leptogenesis, 
typically relaxing these bounds by a ${\cal O}(2)$ factor,
if $N_1$ couples to a lepton doublet which is not close to a flavor eigenstate.\footnote{A full presentation of the flavored dynamics of leptogenesis  is presently impractical because 
flavor adds many unknown parameters: e.g.\
we do not know which combination of $L_e$, $L_\mu$, $L_\tau$ is the lepton doublet
$L$ coupled to $N_1$ in the see-saw Lagrangian, eq.\eq{seesaw1f}.
The Boltzmann equation for $Y_{{\cal L}}$ must be generalized to an evolution
equation for the $3\times 3$ matrix density $\rho$ of lepton asymmetries in each flavor. 
However, it simplifies to qualitatively different behaviors in different ranges of $M_1$:
\begin{itemize}
\item If $M_1\circa{>}10^{12}\GeV$ the SM lepton Yukawa couplings induce scattering
rates much slower than the expansion rate $H$ at $T\sim M_1$:
quantum coherence among different flavors survives undamped and
the main new effect is that lepton asymmetries generated by $N_2,N_3$ decays 
can be washed-out by processes involving $N_1$ 
only along the combination of flavors to which $N_1$ couples.

\item 
If $M_1\circa{<}10^{10}\GeV$, the $\lambda_\mu$ and $\lambda_\tau$ Yukawa couplings induce scattering rates faster than  $H$ at $T\sim M_1$
  and damp quantum coherence in $\rho$: the matrix equation for $\rho$ reduces to 3
  Boltzmann equations for the asymmetries in the $\ell=\{e,\mu,\tau\}$ flavors.
  Neglecting a mild flavor mixing (induced by sphaleronic scatterings)
  these equations have the following form: 
  the first equation in\eq{BoltzmannLeptog} for $Y_{N_1}$ remains unchanged, and
  the second equation splits into three equations for $Y_{{\cal L}_{e}},Y_{{\cal L}_{\mu}},Y_{{\cal L}_{\tau}}$ with the CP-violating term and the wash-out term restricted to each flavor,
  as intuitively expected.
  Therefore, in each flavor $\ell$ one has a different 
 CP asymmetry $\varepsilon_{1\ell}$ and 
  efficiency $\eta_\ell(\tilde{m}_1,\tilde{m}_{1\ell})$,
 that now depends on two parameters: the flavor independent 
 $\tilde{m}_1$ that tells the $N_1$ decay rate, 
 and flavor-dependent $\tilde{m}_{1e,\mu,\tau}\equiv |\lambda_{1\ell}^2| v^2/M_1$ that parameterize wash-out on flavor $\ell$.
It can be approximated as
$\eta_\ell(\tilde{m}_1,\tilde{m}_{1\ell})\approx \eta(\tilde{m}_{1\ell})$, 
 where $\eta$ is the one-flavor efficiency, plotted in fig\fig{leptogEta}a. Therefore eq.\eq{nB1} gets replaced by
 \beq n_B/n_\gamma \approx  -1.14 \sum_{\ell} \varepsilon_{1\ell} \cdot \eta(\tilde{m}_{1\ell})/g_{\rm SM}.
\eeq  
 For large mixing angles one typically has $\tilde{m}_{1\ell}\sim \tilde{m}_1/{\cal O}(2)$,
enhancing the efficiency and weakening the bound\eq{mNbound} as mentioned in the text. 
\item Something intermediate happens if $10^{10}\GeV\circa{<}M_1\circa{<}10^{12}\GeV$:
quantum coherence stays undamped only among $\mu$ and $e$.

\end{itemize}
In supersymmetric models the critical values of $M$ become larger,
because the charged lepton Yukawa couplings
are enhanced by a $\simeq \tan\beta$ factors.}

More importantly, the CP asymmetry can be 
much larger than in eq.\eq{DI}
if right-handed neutrinos are either not much hierarchical
($M_1/M_{2,3}\circa{>} 0.01$ is enough)
or quasi-degenerate.
In both cases $M_1$ can be lighter than in eq.\eq{mNbound};
one can even have leptogenesis at the TeV scale.
Unfortunately not even this case can be tested at accelerators
because  right-handed neutrinos are too weakly coupled,
and the amount of baryon asymmetry depends on their tiny mass difference.

One can of course avoid the above bound by abandoning the minimal scenario, 
e.g.\ adding a coupling $N_1 LH'$
where $H'$ is an extra scalar doublet that does not get a vev:
it contributes to leptogenesis in the usual way, without
giving neutrino masses at tree level.

\paragraph{Constraint on left-handed neutrino masses.}
The bound\eq{DI} becomes more stringent if neutrinos are quasi-degenerate,
and implies that neutrinos must be lighter than about $0.2\eV$.
Since thermal leptogenesis depends on $\tilde{m}\equiv (\lambda_N\lambda_N^\dagger)_{11} v^2/M_1=
\sum_i m_{\nu_i} |R_{1i}|^2$,
this bound can be improved by computing the upper bound on $ \varepsilon_1$ at given $\tilde{m}$~\cite{leptogenesisBounds}
and maximizing $n_B$ with respect to $\tilde{m}$.
The final constraint is $\color{blu}m_{\nu_3} < 0.15\eV$ combining experimental
uncertainties at $3\sigma$.

One can invent very unnatural flavor matrices such that flavor distorts the
usual dynamics of leptogenesis relaxing this constraint.
More importantly, this constraint holds under the 
unnatural assumption 
that hierarchical right-handed neutrinos
give quasi-degenerate left-handed neutrinos:
good taste suggests that quasi-degenerate
neutrinos are more naturally  produced by quasi-degenerate
right-handed neutrinos.
In such a case the constraint on neutrino masses gets relaxed because
$ \varepsilon_1$ can be resonantly enhanced, and because $ \varepsilon_1$
no longer need to vanish when neutrinos are degenerate.
The result depends on possible reasons that naturally 
give rise to quasi-degenerate left-handed neutrinos.
With conservative assumptions, a $\sim 10\%$ quasi-degeneracy among $N_{1,2,3}$
allows neutrinos heavier than an eV compatibly with leptogenesis.
A constraint $m_{\nu_3}\circa{<}0.5\eV$ arises instead if some flavour symmetry (e.g.\ SO(3)) keeps 
right-handed neutrinos 
as degenerate as left-handed neutrinos.

\subsection{Leptogenesis in alternative models of neutrino masses}
Adding  {\em supersymmetry} does not significantly affect leptogenesis~\cite{leptogenesis2}:
\begin{itemize}
\item If $M_1\gg  m_{\rm SUSY}$ one can ignore SUSY breaking effects;
right-handed neutrinos and sneutrinos
have equal masses, equal decay rates and equal CP-asymmetries:
both $\Gamma_{1}$ and $\varepsilon_1$ become 2 times larger,
because there are new decay channels.
Eq.\eq{nB1} remains almost unchanged, because adding spartners roughly doubles
both the number of particles that produce the baryon asymmetry and
the number of particles that share it.
As a consequence of more CP-asymmetry compensated by more wash-out,
the constraints on right-handed and left-handed neutrino masses
discussed in the non-supersymmetric case remain essentially unchanged.
\item  If instead $M_1$ is not much larger than the scale of SUSY-breaking 
(presumed to be below 1 TeV),
complex soft terms give new contributions to the CP-asymmetry.
Being related to supersymmetry breaking rather than to neutrino masses,
the bound on the CP-asymmetry of eq.\eq{DI} no longer holds
and $M_1$ lighter than in eq.\eq{mNbound} is allowed.
This scenario is named `{\em soft leptogenesis}':
At larger $M_1$ soft terms can still be relevant,
but only in a fine-tuned range of parameters
with an anomalously small $B$-term associated to the $N_1$ mass,
by generating a mixing between the CP-odd and the CP-even component
of the lightest right-handed sneutrino.
\end{itemize}

\medskip

Alternatively, leptogenesis is typically produced by decays of the lightest 
particle that violates lepton number: it might be not a right-handed neutrino.
As discussed in section~\ref{Neutrino}, two other particles can mediate Majorana
neutrino masses at tree level:
fermion SU(2)$_L$ triplets and scalar  SU(2)$_L$ triplets.
Unlike in the singlet case, triplets have gauge interactions that
keep their abundancy  close to thermal equilibrium
(potentially conflicting with the 3rd Sakharov condition)
such that their efficiency is univocally predicted: 
quantitative analyses have shown that it can large enough~\cite{leptogenesis2}.
\begin{itemize}
\item {\em Leptogenesis from decays of a fermion triplet}.
Three fermion triplets $N_{1,2,3}$ is the only possibility which, with as few parameters as the singlet
model, can lead to successful leptogenesis.
If $N_1$ decays are slower than $N_1 N_1$
annihilations, one can approximate leptogenesis in three steps:
1) At $T\gg M_1$ gauge interactions thermalize the triplet abundancy.
2)  At $T\sim M_1$ gauge scatterings partially annihilate triplets.
A standard freeze-out estimate (see e.g.~\cite{Kolb}) shows that
a fraction $\eta \sim M_1/10^{12}\GeV$ survives to gauge annihilations.
3) At $T\ll M_1$ triplets freely decay generating a lepton asymmetry.
If instead decays are sufficiently fast that compete with annihilations,
the efficiency of thermal leptogenesis is comparable to the one
of the right-handed neutrino case,
even at small values of $M_1\sim \TeV$.
In conclusion, the efficiency is approximatively given by
\beq
\eta(\hbox{fermion triplet})\approx \min\left[1,\frac{H}{\Gamma},\frac{M}{10^{12}\GeV}\max(1,\frac{\Gamma}{H})\right].
\eeq

\item {\em Leptogenesis from decays of a scalar triplet}.
One scalar triplet with mass $M_T$
is the extension of the SM that can mediate generic neutrino masses 
having the minimal number of beyond-the-SM parameters;
however it does not lead to a large enough CP asymmetry.
Successful leptogenesis from scalar triplet decays
is possible in presence of other sources of CP violation,
that, in the simplest scenario,  might be contributions to neutrino masses mediated by heavier particles.\footnote{In supersymmetric models complex soft terms offer another plausible scenario.}
Writing the neutrino mass matrix $\mb{m}_\nu$ as the 
sum of the triplet contribution $\mb{m}_T$,
plus an extra contribution $\mb{m}_H$ mediated by other much $H$eavier particles:
$\mb{m}_\nu = \mT + \mb{m}_H$, 
the lepton asymmetry produced per decay is:
\beq\label{eq:DIT} 
\varepsilon_L  \equiv   \frac{\Gamma({\bar{T}}\rightarrow LL)-
\Gamma(T\rightarrow \bar{L}\bar{L})}{\Gamma_T} =
 \frac{1}{4\pi}\frac{M_T}{v^2}\sqrt{B_L B_H}
 \frac{{\rm Im}\,{\rm Tr}\, \mT^\dagger \mb{m}_H}{\tilde{m}_T}
 \eeq
 where $\tilde{m}_T^2\equiv  {\rm Tr}\,\mT^\dagger\mT$. 
 One can derive an upper bound on $|\varepsilon_L$ noticing that the
  last factor is less than  $\sqrt{\sum_i  m_{\nu_i}^2}$.
 In terms of the parameters of eq.\eq{Ltriplet}, the decay widths are 
 \beq\Gamma(T\to LL) =\frac{M_T}{16\pi} 
{\rm Tr}\, \mb{\lambda}_L \mb{\lambda}_L^\dagger=B_L \Gamma_T ,\qquad
\Gamma(T\to \bar H\bar H) =\frac{M_T}{16\pi} 
\lambda_H \lambda_H^\dagger=B_H \Gamma_T ,
\eeq
The efficiency can be estimated similarly to the fermion triplet case:
\beq
 \eta(\hbox{scalar triplet})\approx \min\!\left[1,\frac{H}{\min(\Gamma_L,\!\Gamma_H)},\frac{M_T}{10^{12}\GeV}\max(1,\frac{\Gamma_L\!+\!\Gamma_H}{H})\right].
 \eeq
 Notice that despite the presence of gauge interactions, that tend to maintain the triplet abundancy
very close to thermal equilibrium, one can have even maximal efficiency, $\eta\sim 1$ for any $M_T$
if 
i) one of the two decay rates ($T\to \bar L\bar L$ or $T\to HH$)
is faster than the annihilation rate;
ii)  the other one is slower than the expansion rate.
Thanks to i) annihilations are ineffective: triplets decay before annihilating.
Thanks to ii) fast decays do not produce
a strong washout of the lepton asymmetry (and consequently a small efficiency $\eta$),
because  lepton number is violated only by the contemporaneous presence
of the two $T\to \bar L\bar L$ and $T\to HH$ processes.
\end{itemize}
It is worth mentioning that 

Finally, if neutrinos have {\em Dirac masses}, 
one can invent models where  the neutrino Yukawa coupling
are mediated by tree level exchange of some extra particle,
whose $L$-conserving decays produce leptogenesis, by generating
equal and opposite lepton asymmetries in right-handed neutrino and
in left-handed leptons~\cite{leptogenesis2}.

\medskip

\subsection{Testing leptogenesis?}
At the moment nobody knows how to test if leptogenesis produced the baryon asymmetry.
Direct tests seem impossible; indirect tests seem unlikely, because there are
too many unknown high energy parameters
and only one observable.\footnote{In principle we could also measure
the three lepton flavour asymmetries.
But in practice we can just barely see that CMB neutrinos exist
(indirectly, via their gravitational effect on the CMB) and
their lepton asymmetry is just known to be $(n_L-n_{\bar{L}})/n_\gamma=0.07\pm0.05$~\cite{WMAP},
with an uncertainty 8 orders of magnitude larger than the baryon asymmetry.}
One can hope of obtaining more concrete results
within restrictive flavour models (see section~\ref{nu:t13}).
For example, the minimal see-saw model with 2 right-handed neutrinos described by eq.\eq{predsA}~\cite{minimalseesaw}
has only one CP-violating phase and
apparently allows to relate CP-violation in neutrino oscillations ($\phi$)
to CP-violation in leptogenesis.
Unfortunately not even the sign of $\phi$ can be predicted since
CP-violation in leptogenesis
(including its sign) depends on the unknown ratio between the two right-handed neutrino masses.
As discussed in section~\ref{LFV}, 
we can hope that quantum corrections imprint high-energy neutrino Yukawa couplings 
into measurable  slepton masses.
In this case the minimal see-saw model would be testable.
However, in general, these extra observables are not enough for a full reconstruction and test of the high-energy model of neutrino masses. 

Maybe future experiments will discover supersymmetry, LFV in charged leptons,
and will confirm that neutrino masses violate lepton number and CP,
and we will be able to convincingly argue that this can be considered as circumstantial
evidence for see-saw and thermal leptogenesis.
Archeology is not an exact science.

%% file: review_astro.tex
\chapter{Neutrinos in astrophysics}\label{supernova}
This section discusses neutrinos from astrophysical sources,
and the effects due to oscillations.
The most interesting possibilities,
putting aside those  discussed previously, 
are (1)~the $\sim 10$~MeV supernova neutrinos, 
(2)~the $1\div100$~TeV neutrinos from galactic 
sources, like young supernova remnants, 
and (3)~the ultra-high energy neutrinos from 
AGN, GRB or other unknown sources. 
These correspond to different techniques of detection,
namely (1)~underground detectors as those used to study 
solar neutrinos or proton decay;
(2)~dedicated underwater or under-ice observatories,
possibly as big as a cubic km in size;
(3)~observatories of estensive air showers. 
The field of neutrino astronomy has been 
pioneered by the 20 events from SN1987A \cite{sn1987a:data}, 
and these are the only observations of this type we have so far.
These observations were preceded by a 
few inspired theoretical works and 
stimulated a considerable number of papers;
here we quote just a review and a book~\cite{randb}.
They certainly did renew the tight link 
among astronomy, astrophysics, nuclear and particle physics.

It can be said in general that there 
are no solid expectations for the fluxes of 
the neutrinos we consider in this section.
Moreover, we should be ready for surprises,
as always when opening a new window on the cosmos.
Thus, without denying the important r\^ole of 
the various effects due to neutrino masses and mixing,
the primary goal of present (experimental and theoretical) 
investigations are just the astrophysical sources of neutrinos.
Keeping these considerations in mind, we begin by 
describing the possible goals of  `neutrino astronomy' and 
giving an overview of the expectations in section~\ref{subs:pns}.
Supernova neutrinos are discussed in section~\ref{subs:ccs}.
Neutrinos with higher energy (from tens of GeV to $10^{11}\GeV$) are discussed in section~\ref{AstroNu}.
In section~\ref{subs:enm} we outline the r\^ole of neutrino 
masses and mixing.

\section{Neutrino sources\label{subs:pns}}
In this section we recall the  connection of neutrino and 
photon astronomy and give a list of possible 
neutrino sources.

\subsection{Multi-messenger astronomy}
We can study astrophysical sources by  detecting the different kinds of particles that they
emit, at different energies.  Let us compare neutrinos to other messenger particles.
Since neutrinos are neutral and stable 
they are not deflected by magnetic fields.
Thus they point to the astrophysical site 
of production, just as the photons do.\footnote{Another interesting 
probe are very high energy neutrons coming from 
the galactic center, that reach us thanks to 
relativistic time dilatation:
$c\cdot \tau =8.5\mbox{ kpc}\cdot E/10^{18}\mbox{eV}$. 
At even higher energies, protons 
are no longer significantly deflected by galactic magnetic fields
and keep memory of their sources 
as can be seen from the Larmor radius 
$R_L=3\mbox{ kpc}\cdot E/10^{19}\mbox{eV}\cdot B/3\mu{\mbox{G}}$
in a typical galactic magnetic field (the extragalactic fields
are unknown, but expected to be on nG scale).}
Actually, this is the most important 
signal  to identify high energy neutrinos. 

The method used for low-energy supernova 
neutrinos is the self-trigger  
of a large number of events (let us recall that 
some low-energy neutrino reactions,  like $\nu e\to\nu e$, 
are mildly directional, giving us a chance to identify the 
direction of the supernova).

Neutrinos can be used to probe astrophysical
sources and to obtain information complementary 
to the one from light. E.g., around the sources of 
galactic or extragalactic cosmic rays (CR)
$\pi^\pm,\pi^0 $ are formed; their decay  yields 
$\gamma$ and neutrinos in comparable amounts.
Neutrinos are much more difficult to observe
than photons, especially at low energy,  
but for the same reason they are not 
easily reprocessed at sources 
or absorbed during the propagation. We 
recall in particular that
(a)~the $\gamma$ radiation from core collapse of a supernova is not 
{\em directly} related to the neutrino radiation, it is produced 
in later phases and it is consequently much smaller; 
(b)~it is difficult to identify 
the $\gamma$'s from $\pi^0$ decays, and to distinguish them 
from those of electromagnetic origin; 
(c)~finally, photons with energies above 100~TeV
originated outside our galaxy do not reach us, due to $\gamma\gamma\to e^+e^-$ scatterings
on background IR photons.

\begin{figure}[t]
\begin{center}
\includegraphics[width=0.8\textwidth]{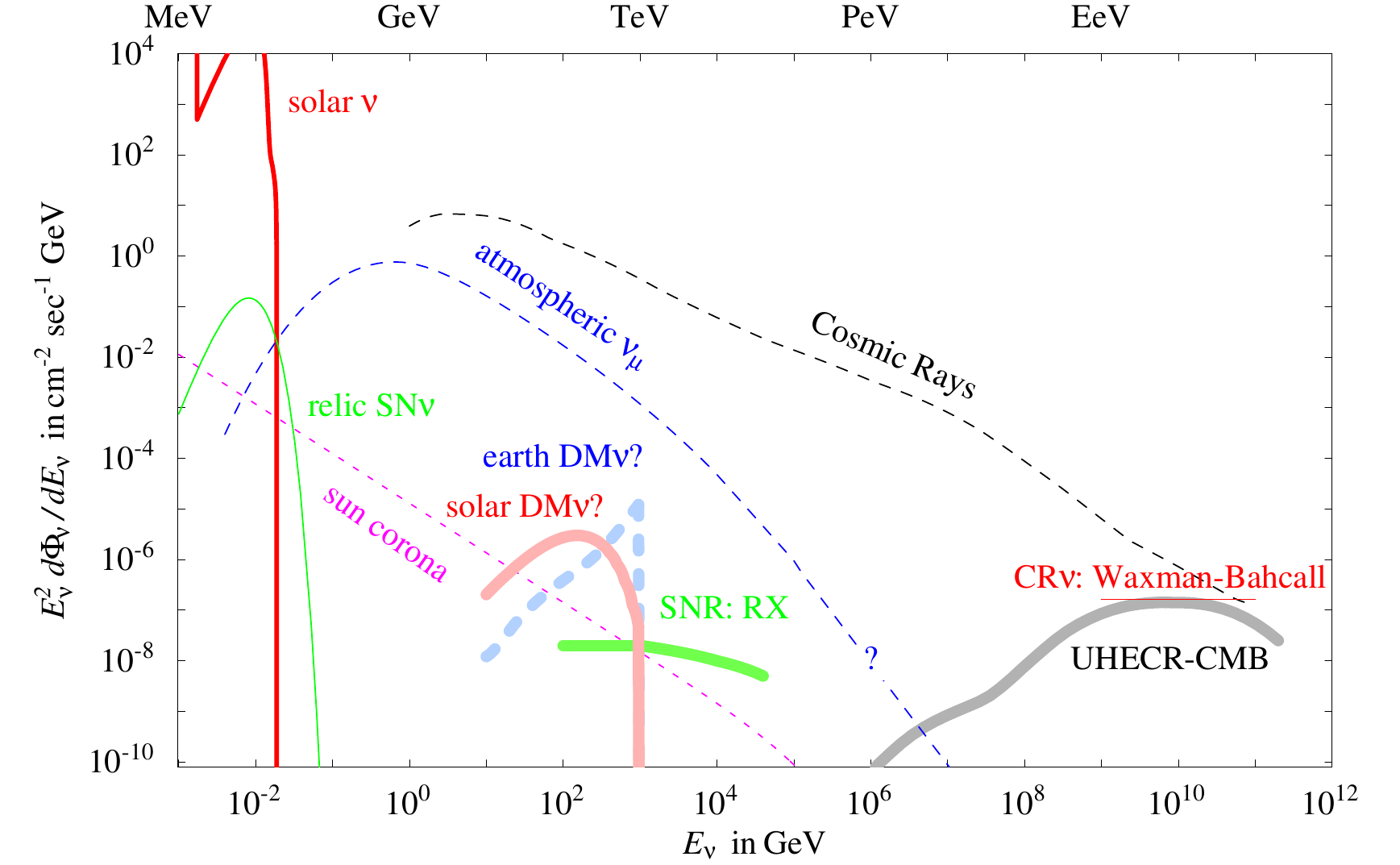}
\caption[Neutrinos from astrophysical sources]{\em Plausible optimistic sketch of
neutrino fluxes from astrophysical sources.
Present experimental constraints (not shown) are somewhat above unseen sources:
from left to right one has
neutrinos from relic SN, from DM annihilations in the sun or earth, from SN remnants,
from CR.
\label{fig:AstroNu}}
\end{center}
\end{figure}

\subsection{A list of the main possibilities\label{sec:pns2}}

Fig.\fig{AstroNu} presents an overview of the main possibilities, here listed.

\paragraph{Solar neutrinos} \hspace{-2ex} (already observed and 
here discussed in section~\ref{sun}) are $\nu_e$ at production, and 
have energy $E_\nu =(0.1\div 20) \MeV$. 
There is little doubt that this is `$\nu$-astronomy'.
Low energy $\nu$ experiments Gallex/GNO and SAGE
proofed that the $pp$-chain is the main energy source
and discovered $\nu_e\to\nu_{\mu,\tau}$ oscillations,
thanks also to the important r\^ole of some theorists, like
J.N.~Bahcall, and G.T.~Zatsepin and V.A.~Kuzmin, who advocated 
solar neutrino astronomy. The physics 
at the center of the sun (density, 
temperature) has been probed, with results consistent
with the more precise helioseismologycal probes.
The agreement of presenta data 
with expectations is excellent, but 
future precise studies of pp-neutrinos could teach us new
surprising lessons on astrophysics or in particle physics.
This will be done in existing experiments as Borexino, KamLAND,
SAGE, and future projects aiming at (real time) 
detection of low energy neutrinos.

\paragraph{Atmospheric neutrinos} \hspace{-2ex} (already observed 
and here discussed in section~\ref{atm}) are electron and 
muon (anti)neutrinos and have energy $E_\nu \sim(0.05\div 1000) \GeV$.
The study of atmospheric neutrinos 
originates as a branch  of cosmic rays studies. 
The investigation of CR secondaries, like $\gamma$, muons and $\nu$, 
permits us to understand CR spectra and their interactions with 
the earth atmosphere (which is not that different from possible sites 
of  production of CR). 
The main existing experiment
for atmospheric neutrinos is  Super-Kamiokande,
which will be hopefully followed by new larger experiment.
It is a bit puzzling 
that the most recent calculations of neutrino fluxes~\cite{flukanew} suggest that some 
features of the observations (like the total $\nubarnu_e$ flux) are not perfectly 
accounted for, even including oscillations.
New measurements of CR fluxes (possibly with new techniques)
and of the relevant hadronic cross sections (HARP experiment) 
are in this connection rather important.
Atmospheric muons and neutrinos are the most important 
background for neutrino astronomy.

\label{geonu}


\paragraph{Geoneutrinos} \hspace{-2ex} 
(first observations, target of future 
investigations) have energy $E_\nu \sim \MeV$. 
They originate from radioactive decays of 
uranium, thorium and potassium, that 
are expected to accumulate in the continental crust:
their total abundance in the mantle is expected to be only comparable
to the total abundance in the continental crust.
According to earth models~\cite{PREM,Geonu} radioactivity
produces a large fraction of heat released by the Earth,
and initially melted iron, that started falling to the center
getting much more heated by gravitational energy. 
The information carried by neutrinos 
is not accessible to geological investigations
and allows to test the above part of the Earth model.
The $\bar{\nu}_e$ from uranium and thorium decay 
can be seen with IBD; the $\bar{\nu}_e$ (and a few ${\nu}_e$)
from ${}^{40}$K cannot, due to their low energy 
(although they are presumed to be the main part 
of neutrino radiation). 
The accumulation of several kton$\times$years of exposure will
permit a significant test of the current 
theoretical predictions.
About 1/3 of geoneutrinos are expected to arrive from distances between 10 and 100 km,
  about 1/4 from 100 to 1000 km,
  and the remaining fraction from more than 1000 km.
Geoneutrinos can be investigated at KamLAND~\cite{Geonu}, that obtained 
a hint of $9\pm 6$ events, and Borexino, that observed $9.9^{+4.1}_{-3.4}$ events~\cite{Borexino}
which has a negligible background and is located far from nuclear reactors.
While geo-$\bar\nu_e$ have geological interest,
 it seems not possible to use  them for oscillation studies because
 their total flux can be predicted only up to a
  maybe $20\%$ uncertainty.

\paragraph{Neutrinos from Dark Matter annihilation} \hspace{-2ex} 
(searched) might exists with energy below $10\GeV$ to $1\TeV$~\cite{DMnu}.
DM might be a thermal relic of WIMP with mass 
$m_{\rm DM}\sim (T_{\rm now}\cdot M_{\rm Pl})^{1/2}\sim 100\GeV$.
DM particles accumulate around  the center of the earth and of the sun, 
and important overdensities  can also arise in central regions
of the galactic halo  or around the galactic black hole.
Their annihilation products involve neutrinos
with energy below $m_{\rm DM}$, that can be dominantly
observed as 
up-going muons induced by $\nubarnu_\mu$.
Baksan, Baikal, SK, MACRO, AMANDA and IceCUBE have produced upper limits~\cite{IceCUBE}.
These search will be continued with the future detectors 
mentioned in the next items, and will be 
ultimately limited by the atmospheric neutrino background.

\paragraph{Neutrinos from galactic sources} \hspace{-2ex}
(very actively searched) 
with energies around $1\div100$~TeV~\cite{ginzBer}. 
This is an interesting window for observation 
of cosmic sources, since the  atmospheric neutrino background
(from $\nu_\mu$ and $\bar{\nu}_\mu$) 
is overwhelming at lower energies, while above several 
100~TeV the Earth becomes opaque to neutrinos. 
The observation of point or diffuse sources 
is a very important goal: e.g.\ $\nu$ and $\gamma$ astronomy 
at these high energies can shed light on the origin of CR 
around the knee and below, that is known to
be of galactic origin. 
The existing limits come from MACRO, Super-Kamiokande, KGF, SOUDAN, 
Baikal, AMANDA~\cite{DiffuseNuBounds}.

\paragraph{Ultra-High-Energy neutrinos} 
\hspace{-2ex} (hopes for close future) have energies up to $10^{11}\GeV$,
that is nowadays the ultimate frontier in energy~\cite{ExtraGalactic}. 
One guaranteed source are existing UHE CR that occasionally 
interact with CMB $\gamma$ and produce $\pi^\pm$ that decay into $\nu$:
the flux of resulting `GZK neutrinos' is however significantly uncertain.
 Furthermore, it is expected that astrophysical accelerators of CR
 (AGN and GRB are some plausible candidates) also directly produce $\nu$.
Other more speculative sources of neutrinos discussed in the literature include
decay of superheavy quasi-stable particles, topological defects, etc.
UHE protons and nuclei are detected looking at the air showers that they produce 
after hitting the earth atmosphere.
Neutrinos have smaller cross sections, 
and therefore in these experiments would dominantly manifest as
quasi-horizontal air showers, 
searched for
by AGASA,  Hi-Res, EAS-TOP, and other experiments~\cite{HorizontalAirShowersExps}.
Significant progress is expected from AUGER~\cite{AUGER}  and ANITA~\cite{ANITA}. 
There are also attempts of detection 
using acoustic detectors or by monitoring the atmosphere from a satellite~\cite{JEMEUSO}.
More possibilities for observations arise by the production of $\tau$ 
leptons~\cite{fargion}. 
The {\sc IceCUBE} experiment~\cite{IceCUBE} will search for UHE neutrinos that cross the earth.

\paragraph{Supernova neutrinos} \hspace{-2ex} (observed in one occasion) 
have energy $E_\nu\circa{<}$100~MeV, include neutrinos of all types and 
will be discussed in more details in the next section.
This type of search promises big dividends
in astro/physics currency: core collapse SN 
are a possible source of infrared, visible, $X$, 
and $\gamma$ radiation, and possibly of gravitational waves;
they are of key importance for origin of galactic CR,
galactic reprocessing of elements; etc. 
Many operating neutrino detectors like SK, 
SNO, LVD, KamLAND, Baksan, MiniBOONE, 
AMANDA/IceCUBE could be blessed by the 
next galactic supernova and other detectors like 
Borexino or ICARUS could contribute to 
galactic supernov\ae{} monitoring in the future.

\paragraph{Relic supernova neutrinos} \hspace{-2ex} 
are neutrinos emitted by past core collapse supernov\ae.
Present experiments reached a sensitivity comparable to the expected rate~\cite{SNrelic}.

\section{Core collapse supernov\ae\label{subs:ccs}}\index{Supernov\ae}
In this section, we discuss neutrinos from supernov\ae{}.
More precisely, we will 
always be concerned with {\em core collapse} supernovae,
and we focus mostly on galactic events.
In section~\ref{cc1} we present the possible type of observable events,
the expected SN rate  and other general astronomical facts. 
In section~\ref{cc2} we describe the theoretical scenario for the gravitational
collapse and explosion, concentrating mostly on the `delayed scenario'.
In section~\ref{cc3} we summarize the expectations for the neutrino fluxes.

\subsection{General considerations\label{cc1}}
\paragraph{Galactic, extragalactic, and relic supernovae} Let us begin 
by discussing the various types of supernovae that can lead to an observable
signal in existing and future detectors.
\begin{enumerate}[(1)]
\item The hope of existing neutrino telescopes is the explosion of a 
galactic supernova, for the simple fact that the 
$1/D^2$ scaling of the flux is severe. For reference,
in water or scintillator detectors
one expects roughly
$300$ $\nu$-events per kton for a distance $D=10$ kpc.
\footnote{We recall that $\hbox{pc}=  3.26 \,\hbox{light-year}$ and that
our galaxy has a size of some 15 kpc, and 
we are located at  8.5 kpc from its center.
One  could expect that the chances of getting a supernova 
where matter is more abundant are higher
(the galactic center), but
one can also object that younger matter, conducive to SN formation,
lies elsewhere (in the spiral arms). 
However, we are unaware of the existence of
a `catalog of explosive stars of our galaxy', or of calculations 
of weighted matter distributions of our galaxy.} 

\item  Extragalactic neutrino astronomy begun several years ago with
SN1987A. In principle, one could profit of the 
wealth of galaxies around us (say, those in the `local group') to 
get events at human-scale pace. 
In practice, this is difficult, because 
core collapse SN takes place only in
spiral or irregular galaxies, and not in 
elliptical ones.\footnote{Their stellar population is 
older, and star forming regions 
are absent or very rare; in a sense, the stars 
of 10-40 solar masses are a problem of youth.} 
The only other large spiral galaxy of the local group 
is Andromeda (M31) but its mass is presumably only
half of our galaxy and its distance is about 700 kpc.
In the previous example, one would get 
60 events in a future Mton {W}\v{C} detector,
assuming $100\%$ efficiency.
Perhaps, the best chance would be another `big' SN 
from the Large Magellanic cloud (an irregular galaxy) 
but the odds for such an event are not high.

\item Another interesting possibility is the 
search for relic neutrinos emitted by past supernov\ae.
The practical method is to select an energy window around 
$20\div 40$ MeV,  where atmospheric and other neutrino 
backgrounds are small, and searching for an accumulation of neutrino 
events with more-or-less known energy distribution. 
The best limit has been obtained by SK~\cite{SNrelic}, 
and the sensitivity is approaching the signal 
suggested by astrophysical models. In principle,
one can suppress the main background (muons produced 
below the \v{C}erenkov threshold) by identifying the neutron from inverse beta 
decay. This could be possible 
by loading the water with an appropriate nucleus~\cite{gadzook}, 
that should absorb the neutron and yield visible $\gamma$ 
eventually.\footnote{Neutron identification by
$p+ n\to {\rm D}+\gamma(2.2\,{\rm MeV})$ was proved in scintillators 
(furthermore, no \v{C}erenkov threshold 
impedes); however no existing 
or planned scintillator has a mass larger than 1 kton.}
\end{enumerate}

\paragraph{Astronomical and other observations}
Supernov\ae{} (SN) are divided in two classes according to their light spectra:
the first class comprises SN~Ia 
that have very regular light curves (among the best 
`standard candles' for cosmology); 
the second comprises the rest (SN~II, Ib and Ic, 
depending on the presence of H and He lines in the spectra)
with a wide variety of light curves (e.g., the 
observed optical luminosities vary by 
at least three orders of magnitudes). 
We are interested in the second class of objects, that are
thought to originate from the gravitational collapse 
of a very massive star, $M>(6\div 10)\  M_\odot$, that possibly leads to
the formation of neutron stars  and black holes.

About 2000 supernov\ae{} have been observed \cite{turatto},
and their frequency depending on the type of galaxy has been 
studied on statistical basis. This makes it possible to predict 
the rate of occurrence  of core-collapse supernov\ae{} in our 
Galaxy:
\begin{equation}\label{eq:SNrate}
R_{\rm SN}=1/(30\div 70\,\mbox{yr}) .
\end{equation}
The main uncertainty is that the galactic type of 
the Milky Way (Sb or Sbc) is not firmly known. 
Numbers in the range 1~SN in 10 to 100 years have been also 
deduced on other basis, e.g.\ from the pulsar formation rate
or from theoretical estimations of stellar lifetime
{\em plus} stellar statistics,  but they do not seem to be very 
reliable at present.
A lower limit on occurrence of SN in the Milky Way can be 
put from the fact that no $\nu$-telescope
observe any event yet. 
The oldest telescope with galactic reach 
is Baksan \cite{baksan}, working since 30 June 1980 with 90\%
live-time. If we assume that since 1980
there was no main holes in time coverage, we arrive at 
a total live-time of about $T=25$~yr in which no galactic SN
has been seen. Poisson statistics implies
\begin{equation}\exp(-T R_{\rm SN})> 1-\hbox{C.L.}\qquad \hbox{i.e.}\qquad
R_{\rm SN}<\frac{1}{11\mbox{yr}} \mbox{ at }90\%\,\hbox{C.L.}
\end{equation}
Often, one recalls the possibility that SN events take place 
in optically obscured regions of our galaxy;
however, one should also recall that, beside $\nu$s, there 
are other manners to investigate the occurrence of such 
a phenomenon, e.g.\ from the released infrared radiation.

The large radiation in neutrinos
 in a stellar collapse is expected to be accompanied 
by other observable phenomena, 
such as gravitational waves (if the collapse deviates enough from
spherical symmetry)
and of course light (after that the shock 
wave emerges from the stellar mantle), $\gamma$ radiation 
(after that the radioactive elements emerge from the core),
{etc}. Gravitational radiation, 
is an important goal of future observations, and could indicate
the beginning of the gravitational collapse.

\subsection{Gravitational collapse and the explosion\label{cc2}}
We here present the expectations on the gravitational collapse and
discuss the unclear points of the general picture we have of this 
phenomenon. For what regards astrophysics, the most puzzling aspect is 
certainly that the current simulations miss to reproduce the explosion~\cite{SNexplosions}.

Before summarizing precise results we explain the basic physics.
In order to have a discussion as simple as possible
we drop factors of order unity.
We explicitly include  $\hbar$, $c$ and order one
factors  when this makes the physics more transparent.
We find it convenient to use 
standard symbols for a star with 
mass $M$ and radius $R$, volume $V\sim R^3$,
composed by $N = M/m_n\sim 10^{57} \,M/M_{\odot}$ 
nucleons (with mass $m_n$) and $\sim N$ electrons
(with mass $m_e$), 
with density $\rho\sim M/R^3$ and number 
density $n=N/V $. A single particle occupies an 
average volume $v\sim R^3/N\sim 1/n$ and has energy $u$.  
For the latest stage of the life of a massive star, 
it is important to recall that localized fermions are subject to Fermi motion:
due to the Pauli indetermination relation $\Delta x\,\Delta p=2 \pi \hbar$, 
$N$ fermions at zero temperature fill all states till the Fermi
momentum
\begin{equation}
p_F=\hbar (3\pi^2 N/V)^{1/3} \sim \hbar\;  n^{1/3}, 
\mbox{ where }n=\frac{N}{V}.
\label{fermim}
\end{equation}
This motion originates a pressure (known as `degeneracy' or `quantum pressure')
\begin{equation}
P\sim \frac{1}{v}\cdot u(p_F)=n\cdot ( \sqrt{ (m_e c^2)^2+(c p_F)^2}-m_e c^2)=
\left\{
\begin{array}{ll}
n^{5/3}\ \hbar^2/m_e & \mbox{ non-relativistic}\\
n^{4/3}\ \hbar c & \mbox{ relativistic}
\end{array}
\right.
\end{equation}
that increases with the increasing density $\rho=m_n\cdot n$.

\paragraph{Stellar evolution and structure of the presupernova}
Stars form because a large enough cloud of particles is unstable under gravity.
The cloud contracts and the gravitational potential energy 
gets converted into kinetic energy, $N\cdot u = GM^2/R$,
heating the gas. When the gas becomes hot enough, 
the nuclear reaction\eq{stella} begins to burn hydrogen into helium.
The nuclear energy stops the contraction: one gets a star that shines in
a quasi-equilibrium state between gravity and nuclear forces.
When all the hydrogen in the core has been burned, 
the star contracts and, if enough massive,  becomes enough 
hot to burn helium into carbon. 
After few of such steps (that are increasingly 
rapid and violent) very massive stars 
form an inert iron core, that cannot burn any more because 
$^{56}{\rm Fe}$ is the most stable of all nuclei
(this is the true ground state of QCD!).
The core is surrounded by shells of unburned lighter nuclei.
Occasionally this configuration is 
referred as `onion layers' structure. However one should 
realize that the innermost layers can be quite `rough' 
(due for instance to the presence of explosive 
nuclear reactions that occur in the latest stages of the 
life of the star) and the 
outermost ones can be expelled by 
intense stellar winds (actually, it is thought that
SN~Ib and Ic are such a type of objects).

\medskip

\index{Density!supernova}
Leaving for a moment the star and thinking about neutrinos,
in order to compute matter effects one needs to know
the profile density of the precursor of the supernova (`pre-supernova' star). 
This requires a good modelization of the
electronic density $\rho_e(x)=\rho(x) Y_e(x)$ of the 
pre-supernova, especially at densities around $(10\div 100)\,{\rm g}/{\rm cm}^3$
(MSW  `solar' resonance) 
and $(500\div2000)\,{\rm g}/{\rm cm}^3$
(MSW  `atmospheric' resonance). 
For orientation, a pre-supernova mantle  density 
$\rho \sim 4\cdot 10^4 (r_0/r)^3 \,{\rm g}/{\rm cm}^3$
with $r_0=10^4$~km and $Y_e\sim 1/2$ can be used.
If extra `sterile' neutrinos exist, one needs to know
the two functions
$\rho(x)$ and  $Y_e(x)$ separately, and recall that 
in the deleptonized core, $Y_e$ can be rather small;
this can give rise to additional MSW effects.

\paragraph{Gravitational instability of large mass stars}
The star remains stable if the force due to pressure
($F \sim P\cdot R^2$) compensates the attractive force 
due to gravity ($F \sim GM^2/R^2$) i.e.\ if $P = P_{\rm gravity} \approx  GM^2/R^4$.
Assuming an equation of state $P(\rho)$ of the form
$p=K \rho^\gamma$  it is possible to reach a stable configuration 
at a certain radius when $\gamma > 4/3$.
We now show that: 1) for relatively small stars 
the quantum pressure due to non-relativistic electrons 
has $\gamma=5/3$ and therefore supports the stars; 
2) if the mass of the star 
is $M>(6\div 10)M_\odot$, 
the iron core can become hot enough that
electrons attain relativistic velocities,  
which reduces $\gamma=4/3$ and
originates a gravitational instability.
\begin{itemize}
\item[1)] For non-relativistic electrons 
$u= p_F^2/2m_e \sim n^{2/3}\hbar^2/m_e$.
Therefore the quantum pressure of electrons is
\beq P \sim 
K\rho^{\gamma}\qquad\hbox{where}\qquad K \sim\frac{\hbar^2}{m_e m_n^{5/3}}\quad\hbox{and}\quad
\gamma=\frac{5}{3}.
\eeq
The above estimation also shows that the quantum pressure 
of the other heavier fermions (that 
instead dominate the mass density) is negligible.
In conclusion, if the electrons are non-relativistic, 
quantum pressure compensates
gravitational compression, and give rise to a stable 
configuration, a `white dwarf' (equating $P$ to $P_{\rm gravity}$ one 
realizes a characteristic feature of degenerate stars:
the radius decreases with the mass as $R\propto  M^{-1/3}$).

\item[2)] However, the Fermi momentum increases with the mass of the star,
until electrons become relativistic, $u \circa{>} m_ec^2$.
The energy of relativistic electrons is
$u \sim c p_F\sim n^{1/3}\hbar $ so that the quantum pressure becomes
\beq P \sim 
K\rho^{\gamma} 
\qquad\hbox{where}\qquad K = \frac{\hbar c}{m_n^{4/3}}\quad\hbox{and}\quad \gamma=\frac{4}{3}.\eeq
Thus, the quantum pressure $P$ scales with $1/R^4$ 
just as the gravitational 
pressure; but while the first increases as $M^{4/3}$, the second 
increases faster, as $M^2$. This implies that there is 
a limiting mass, the Chandrasekhar mass
$M \circa{>} M_{\rm Ch}\sim (\hbar c/G)^{3/2}/m_n^2$
($M_{\rm Ch} = 1.4 M_{\odot}$ including  order one factors)
above which the quasi-free electrons are unable to 
produce a stable configuration, and the star collapses
under its weight. 
\end{itemize}
Let us consider the situation when the  
inert iron core reaches a mass $M\sim M_{\rm Ch}$, 
with radius $\sim 8 \cdot 10^{3} $ km. 
The collapse begins because at a certain point an increase
of the temperature leads to a {\em decrease} of the pressure: 
the  loss of kinetic energy 
is due to the onset of iron photodissociation reaction
and of electron neutrino  production, $ep\to n\nu_e$.
The latter reaction gives rise to an initial burst 
of electron neutrinos, and this is occasionally referred 
as the {\em `infall phase'}. 
During this stage $\nu_e$ carry away some small  fraction of
the total energy (simulations suggest a fraction of 
$\%$ at most) and a part of the stored 
leptonic (electron) number. After some 100 ms, the inner part of the 
core (`inner core') contains mostly neutrons 
plus a soup of particle/antiparticle with 
an increasing temperature $T\gg m_e$.
The particles lighter than $T$ are photons, electrons, 
perhaps muons, and $\nu_{e,\mu,\tau}$. With increasing density, 
even neutrinos are momentarily trapped in the 
collapsing star.

\paragraph{Rebounce (and explosion?)}
When the inner part of the core (about 0.6 $M_\odot$ according 
to simulations) reaches nuclear density, the collapse gets halted by 
the quantum pressure of neutrons (and remaining protons)
that are non relativistic and therefore have $\gamma = 5/3$.
An enormous amount of gravitational 
binding energy gets converted into kinetic energy.
The radius of the core is about 
$R\sim \hbox{Fermi}\ N^{1/3} \sim (10\div 20) \mbox{ km}$
(1 Fermi = $10^{-13}$~cm).
The rebounce of the inner core generates 
an outward-going shock wave, but at this point  
existing simulations meet a serious difficulty.  
In fact, the simulated shock wave loses energy
due to iron dissociation on the way through the `outer core', 
it slows down, and eventually stalls.
Said in simpler terms, in practically all simulations
the rebounce is too weak to generate the 
explosion we see.
A popular working hypothesis is that
the shock wave is rejuvenated by the outflowing neutrinos,
which comprise the bulk of the kinetic energy. 
Indeed, the neutrinos trapped in the inner core will diffuse out;
they hit and push the stalling matter from below in a way probably 
crucial for finally getting an explosion. 
This is called the `delayed explosion' scenario. 
In some computer simulations, the delayed explosion  
happens in a fraction of a second,
during the so called {\em `accretion phase'}. A successful explosion
requires that at least about 10\% of the energy 
emitted in neutrinos is transferred to the shock 
wave. The `delayed explosion' is 
regarded by many as the most promising  
scenario to obtain a successful understanding of the explosion,
but it is fair to recall that {\em ab initio}  simulations 
are  still unable to obtain explosions.
In summary, we have not a `standard SN model' yet.
This could be due to a very complex dynamics; it 
could indicate that some ingredient is missing 
(such as an essential r\^ole of asphericity, of rotation, 
of magnetic fields, etc),
or that there is nothing like a `standard explosion';
or perhaps it could be a hint that several core collapse do not lead to
an optical burst (and explosion); or even it might 
be a hint for new physics.

\paragraph{Cooling}
The first second after the core collapse
is probably of key importance to understand supernova explosion. 
However, the main part of the
emission of neutrinos is supposed to take place later,
in the {\em `cooling phase'} (or `thermal',
or `Kelvin-Helmotz' phase). In this phase, 
the proto-neutron star cools and contracts in
a quasi equilibrium state.\footnote{In other words, 
we do not know yet the details of explosion but 
we can argue that we do not need to know them 
to describe the bulk neutrino emission.}  
We begin the description of this phase 
by evaluating the total energy radiated in neutrinos
(based on macroscopic arguments) and conclude 
estimating the luminosity, temperature, and time of the collapse
based on simple minded microscopic
arguments. 
A detailed discussion of expected neutrino
fluxes is left to section~\ref{cc3}.

\medskip

The core is so dense and hot that neutrinos are efficiently 
produced and partly trapped.
Other SM particles are much more strongly trapped, while
gravitons are so weakly coupled that are neither trapped nor 
significantly produced.
As a result neutrinos are 
able to carry away the energy.
The stellar collapse is supposed to possibly lead to the 
formation of a neutron star, occasionally seen as a pulsar 
with mass $M_{{\rm NS}} = (1\div 2)\, M_{\odot} $, 
and radius $R_{{\rm NS}} \approx 15\ {\rm km}\ (M_{\odot}/M_{\rm NS})^{1/3}$
(the scaling is due to the degenerate character of the equation of state,
as for white dwarfs).
Including a factor of order one from the 
expected distribution of the nuclear matter, we 
come to the following estimation of the released binding energy:
$$
{\cal E}_B\approx \frac{ 3 M_{\rm NS}^2}{7 R_{\rm NS}}=
(1\div 5)\cdot 10^{53}\mbox{ erg}.
$$
One expects that  the overwhelming part 
of ${\cal E}_B$ is carried away by neutrinos.\footnote{Let us 
compare ${\cal E}_B$ with other relevant energies. 
This is much bigger than the observed kinetic energy of the ejecta,
$E_{{\rm kin}} \sim 10^{51} $ erg
(a typical velocity of the shock wave is $v\sim 5000\km/\s$).
It is also much bigger than the energy needed 
to dissociate the outer iron core 
$\sim 2.2\MeV\cdot 0.6 M_{\odot}/m_n =2\cdot 10^{51}$ erg
(the mass of ${}^{56}$Fe is 123 MeV smaller 
than $13 m_\alpha +4 m_n$).
The energy seen in photons is very small, 
$E_{\rm  lum} \approx 10^{49}\,\hbox{erg}  \approx 10^{-4}\, {\cal E}_B$
(sufficient to outshine the host galaxy though!).
The energy emitted in  gravitational waves depends on the 
detailed dynamics of the collapse. A na\"{\i}ve 
guess is $G_N (M_{\rm core} v^2/2)^2/R\sim  
(v/c)^4\cdot {\cal E}_B\ll {\cal E}_B$.} 
Neutrinos are produced with energy
comparable to the temperature of the environment,
$u = {\cal E}_B/{N} \sim 100\MeV$:
neutrinos thermalize by random walking in the interior,
because their  mean free path is
$\ell \sim 1/n\sigma\sim 1/(G_{\rm F}^2 T^5) \sim 10\cm \,(100 \MeV/T)^5$
is smaller than the core.
The diffusion time across a distance $R_{\rm NS}$ is $\tau \sim R_{\rm NS}^2/\ell  c \sim 10\,{\rm s}$:
this sets the time-scale for the cooling of the core.

Neutrinos around the outer and cooler part of the supernova,
named `neutrinosphere', escape cooling the neutron star.
Therefore in thermal approximation only 
the temperature and density profile of the outer part of the SN
are needed to compute neutrino fluxes.
Simple estimations indicate that neutrinos exit with temperature
$T \sim 10\MeV$.
The instantaneous luminosity is 
${\cal L} \sim R^2 T^4\sim 10^{52}$ erg/sec, thus the six types of 
neutrinos need about 10 seconds to carry out all energy.
In the next section we present results of precise studies 
of neutrino interactions.

\medskip

Let us finally comment on the possibility of producing  a black hole. 
If also neutrons become ultra-relativistic, $T\gg m_n$,
their equation of state is the standard one of massless quarks, $P=\rho/3$
(since 
we are not able of solving non perturbative QCD, we do 
not precisely know what
happens when $T\sim m_n$, but we know that at $T\gg m_n$ the relevant 
degrees of freedom are quarks.
The exponent $\gamma=1$  in $P\propto \rho^\gamma$
can also be derived 
adapting the estimation done for electrons,
taking into account that now $\rho = u n$, rather than $\rho = m_n n$),
such that gravity wins over pressure and
the ultra-relativistic neutrons collapse into a black hole.
This might abruptly terminate neutrino emission.

\begin{figure}[t]
\begin{center}
\includegraphics[width=16cm,height=5cm]{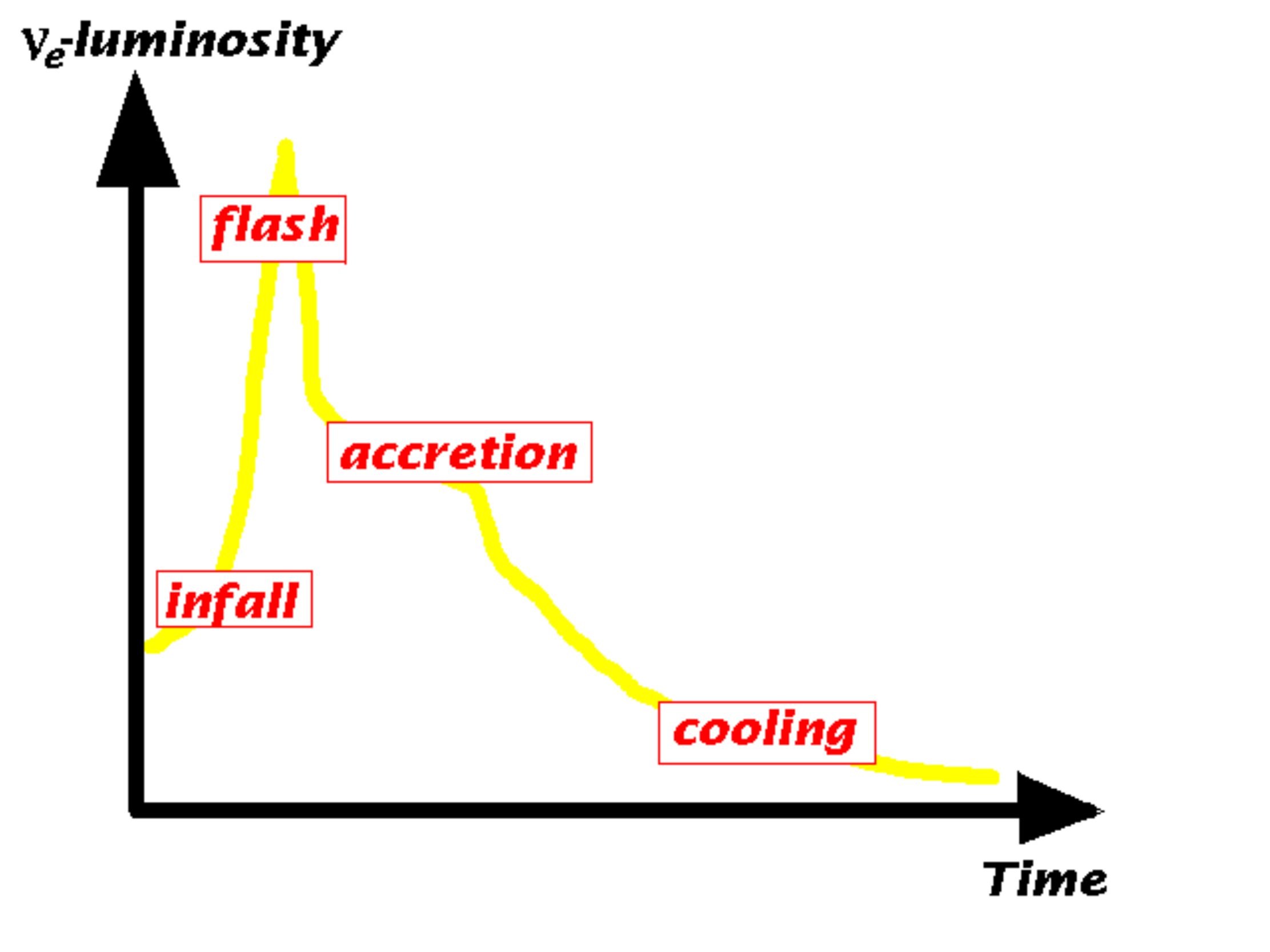}
\caption[Luminosity of $\nu_e$ from supernova]{\em Sketchy plot of the behavior of the $\nu_e$ luminosity, for 
the delayed scenario.
\label{fig:nuelumin}}
\end{center}
\end{figure}

\subsection{Neutrinos from core collapse supernovae\label{cc3}}

\paragraph{General properties of emitted neutrinos}
In the delayed scenario the collapse has four main phases.
Correspondingly, we distinguish between two early neutrino emissions,
named ``infall'' (or deleptonization, or 
early neutronization) and ``flash'' and 
the two late phases of ``accretion'' and ``cooling'',
as summarized in table~\ref{tab:del}.
The most uncertain phase is certainly the  one of ``accretion'',
that, together with ``cooling'', 
accounts for most of the energetics.
Perhaps, a fair estimation of uncertainties is around 100\%. 
In support of this (apparently too conservative) statement, 
we recall that we have no {\em ab initio}
calculations of these fluxes, and alternative 
scenarios have been proposed.
Furthermore, the calculations 
that tried to estimate the effect of rotation~\cite{SNexplosions}
found very different fluxes, and in particular a severe suppression
of muon and tau neutrinos.

\smallskip

We now  describe detailed expectations for SN neutrinos.
First we discuss the general 
characteristics and present a phenomenological survey,  
and in next paragraph we discuss how their 
luminosity, energy spectrum, and  possible 
non-thermal effects can be parameterized.
We can distinguish are 3 types of neutrinos:
$$\nu_e, \qquad \bar\nu_e\qquad\hbox{and}\qquad
\nu_x \equiv \nu_\mu , \bar\nu_\mu ,\nu_\tau,\bar\nu_\tau$$
Indeed $\nu_\mu$, $\bar{\nu}_\mu$, $\nu_\tau$ and $\bar{\nu}_\tau$ should have a very similar distribution
because $\nu_\mu$ and $\nu_\tau$ are produced by NC in the same manner
(muons are expected to be present only in the innermost core), and
$\nu_{\mu,\tau}$ and $\bar{\nu}_{\mu,\tau}$ have similar properties.
Different numerical calculations find values for the
`average' neutrino 
energies $\langle E_{\nu}\rangle$
and total emitted energy ${\cal E}_{\nu}$
in the following ranges:
\beq
\begin{array}{ll}
\langle E_{\nu_e} \rangle=10\div12 \mbox{ MeV },\qquad &
{\cal E}_{\nu_e}=10\div 30\ \%\ {\cal E}_B, \\
\langle E_{\bar{\nu}_e} \rangle=12\div 18 \mbox{ MeV }, &
{\cal E}_{\bar{\nu}_e} =10\div 30\ \%\ {\cal E}_B,\\ 
\langle E_{\nu_x} \rangle=15\div 28 \mbox{ MeV }, &
{\cal E}_{\nu_x} =20\div 10\ \%\ {\cal E}_B.
\end{array}\label{rrranges}
\eeq
The general reason of the hierarchy in the average energies 
is that neutrinos that interact
more decouple in more external regions of the star.
The approximate equality between the energy radiated in the 
various neutrino types found in several numerical calculations
has been  named `equipartition', but in
our understanding, there is no profound 
reason behind this result. It is important to recall 
that till recently it was commonly assumed a 
rather strict equipartition and a large hierarchy of 
average energies between the various neutrinos,  
while recent calculations suggest 
that the hierarchy of average energies is modest, and the equipartition
is satisfied only up to factor of 2.

\begin{table}[t]
\begin{center}
\begin{tabular}{|c|c|c|c|c|}
\hline
Phase & Neutrinos & Description & Duration & \% of ${\cal E}_B$ \\
\hline\hline
infall & $\nu_e$ & \parbox{7cm}{Collapse. $ep\to \nu_e n$.\\
$\nu$-trapping increases.} &
$\sim 100$ msec & $<1\%$\\[3mm]
\hline
flash & $\nu_e$ & \parbox{7cm}{$t\equiv 0$.
Bounce. Flash  when \\
$\nu$-sphere is reached.} & few msec & $\sim 1$\% \\[3mm]
\hline
accretion & all &
\parbox{7cm}{Stall. $e^+e^-\to \nu_x \bar{\nu}_x$.
Explosion \\
resumed (by neutrino heating?).} &
$\circa{<}$ sec
& $(10\div 20)$\% \\[3mm]
\hline
cooling & all &\parbox{7cm}{Proto neutron-star cools and contracts.}  &
$(10\div 100)$ sec &  $(80\div 90)$\% \\
\hline
\end{tabular}
\end{center}
\caption[Neutrinos from a supernova]{\em Brief description of the phases for neutrino emission 
in the delayed scenario for the explosion of core collapse SN.
$\nu_x$ indicates non-electron $\nu$ a $\bar\nu$.
\label{tab:del}}
\end{table}

This is important for oscillations signals:   oscillation play an important r\^ole only if
 $\nubarnu_e$ fluxes are different from $\nubarnu_x$  fluxes:
 if instead the energy distribution is the same and equipartition
is exact, there is no oscillation effect. 

\paragraph{Parameterized fluxes}
Let us begin recalling some definitions:
$$\begin{array}{rl}
\hbox{flux}\!\!&= F = dN/{dt\, dS\, dE} ,\\
\hbox{fluence}\!\! &=\Phi = dN/dS\, dE= \int F\,dt \\
\hbox{luminosity} \!\!&= {\cal L} = {d{\cal E}}/{dt} = \int E \,F \,dS\, dE.
\end{array}$$
Assuming that the emission is isotropic 
(which should be true up to $\sim 10\%$ corrections, or 
presumably less during cooling phase), we can 
describe the flux at earth as:
\begin{equation}
F=\frac{{\cal L}}{4\pi D^2}\times \frac{n(E/T,\xi)}{T^2}
\end{equation}
where $T$ is the temperature, $\xi$ one (or more) parameter(s) that describes 
the deviations from exact thermal distribution, and $D$ the distance from the SN.
The most common choices for the energy distributions are:
\begin{equation}
n(x,\xi)\propto 
\left\{
\begin{array}{ll}
x^2/(1+e^{x-\eta}) & \xi=\eta\neq 0, \\
x^2 e^{-(x/x_0)^2}/(1+e^x ) & \xi=x_0\neq \infty, \\
x^\alpha\cdot e^{-x} & \xi=\alpha\neq 2 .
\end{array}
\right.
\end{equation}
namely, 
modified Fermi-Dirac or Maxwell-Boltzmann distributions.
The normalization is  $\int x\cdot n(x,\xi) dx=1$, that 
implies that the instantaneous 
luminosity is ${\cal L}$ as it should be. In these terms,  
we can formulate the (non-trivial)
goal of reconstructing experimentally (or to calculate) 
three functions of time for each type of neutrino: ${\cal L},T,\xi$.
The total energy radiated ${\cal E}$ is often used
in place of the luminosity ${\cal L}$.

From now on  for definiteness we assume the 
Fermi-Dirac distribution.
In a simplified description, the full distributions can be 
characterized by a few parameters.  The most important ones are
\begin{itemize}
\item[$-$]  ${\cal E}_B$, the total energy radiated (binding energy);
\item[$-$] $T_{\bar e}$, the temperature of antineutrinos, which can be easily measured;
\item[$-$] $\kappa\equiv T_x/T_{\bar e}$;
\item[$-$] $f=f_e=f_{\bar e}$, the energy fraction in
electron (anti)neutrinos.
One has $f_x=(1-2 f)/4$, with
$f=1/6$ in the special case of `equipartition'.
\item[$-$] an effective `pinching' 
parameter $\eta$ equal for all types of neutrinos
(that is not expected to be  
accurate, but could be adequate). 
\end{itemize}
Usually $T_e$ is a less important 
parameter to describe the neutrino signal,
and can be estimated from the above parameters
by assigning a `reasonable' condition on the emitted lepton number
$N_e-N_{\bar e}$. In formulae,
\begin{equation}
T_e=T_{\bar e}/
[1+  (N_e-N_{\bar e}) 
(T_{\bar e} F_3(\eta)/F_2(\eta))\, /\, (f {\cal E}_B)].
\end{equation}
where $F_n(\eta)\equiv \int_0^\infty dx~x^n/(1+e^{x-\eta})$.

\begin{figure}[t]
$$\includegraphics[width=5.5cm,angle=270]{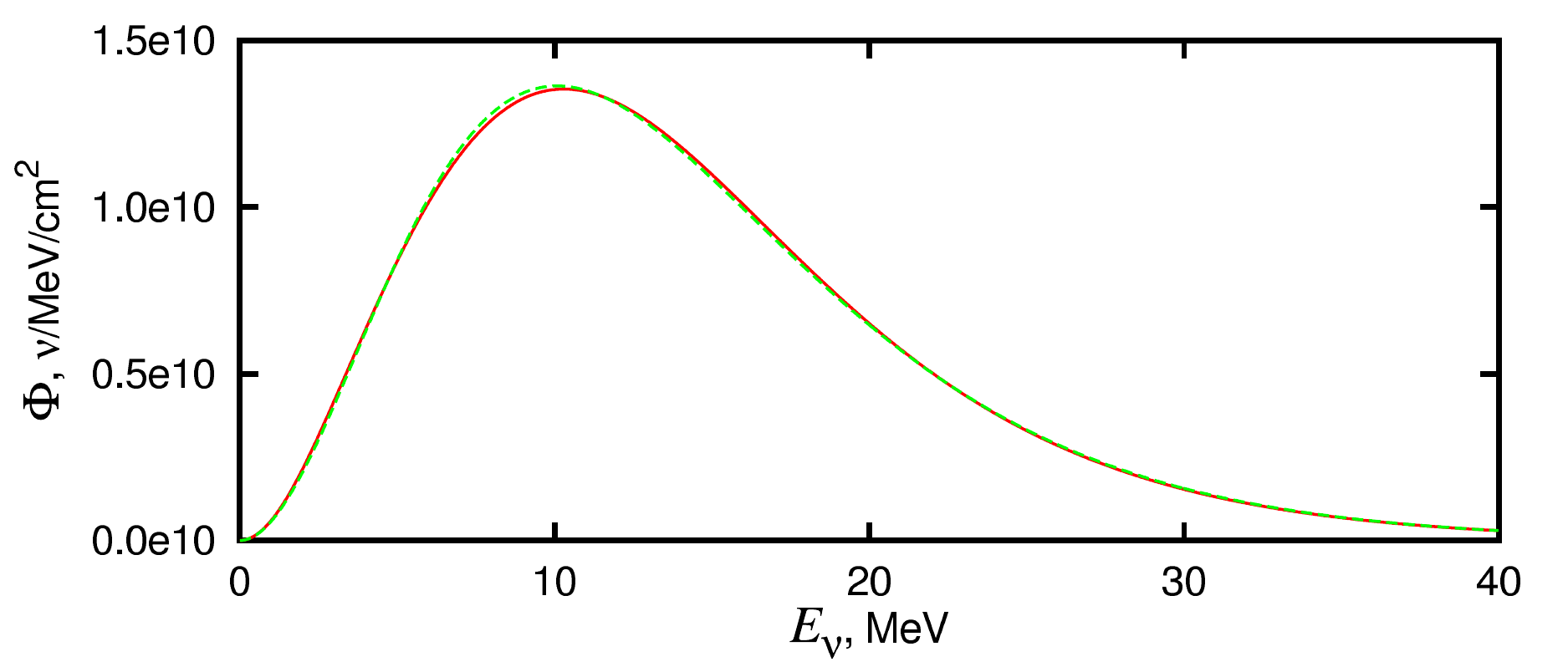}$$
\caption[Fluence of supernova neutrinos]{\em Comparison of the 
fluence obtained integrating the neutrino flux in time (red line) 
with an effective fluence described by appropriately chosen 
average (time independent) parameters (green line);
see text for an accurate description.
The agreement between the two curves is at the \% level.
\label{fig:fluenze}}
\end{figure}

\medskip

At a further level of refinement, we 
describe time dependent 
features distinguishing between `cooling' 
and `accretion' neutrinos.
The time dependence of the temperatures and luminosities
can be approximated as
\begin{equation}
T_i(t)=T_i(0)/(1+t/\tau_c)^{\beta_c},\qquad {\cal L}_i \propto T_i^4
\label{tscale}
\end{equation}
with $\tau_c\sim (10\div100)$ sec and the exponent $\beta_c$ 
have to be extracted from the data or computed.

On top of that, we can model an hypothetical accretion phase with duration $\tau_a$,
adding, for $t<\tau_a$,
another rather luminous phase, presumably with 
a marked non-thermal behavior ($\eta\neq 0$)
and with its own effective temperature.
Since the efficiency of energy transfer 
to matter is not large, $\nubarnu_e$ should carry 
a sizable  fraction of energy;
$\nu_x$ are of little use to revive the shock,
but presumably, only few of them are produced in this phase.

A simple model for the cooling 
employs the concept of `neutrino-spheres'
with radii $R_i$. Black-body emission gives:
\begin{equation}
{\cal L}_i=\frac{F_3(\eta_i)}{2 \pi} \cdot
R^2_i(t)\cdot T_i^4(t)\sim 5.2\times 10^{51}\ 
\frac{\mbox{erg}}{\mbox{sec}}
\left( \frac{R_i}{15\,\mbox{km}} \right)^2
\left( \frac{T_i}{4.5\,\mbox{MeV}} \right)^4
\end{equation}
for $i=e,\bar{e},\mu...$ ($\hbar=c=k=1$ here).
The numerical value of the luminosity, 
obtained assuming $\eta=0$, leads 
to an energy $\sim 5\times 10^{52}$~erg radiated 
in 10 seconds, when we use reasonable values of the radius 
and temperature.   
Similarly, one can model the accretion
phase by suggesting that the non-thermal 
neutrino production is from $e^\pm$ interactions 
with the accreting matter \cite{ll}.
This implies that the fluxes are proportional 
to the cross sections: thus, their scaling with energy should 
be more similar to $E_\nu^4$ than to $E_\nu^2$,
namely the deviation from a thermal spectrum 
should be larger during accretion.

In order to show the use of these formulae, 
we conclude with a numerical example.
Let us consider a time dependent flux, 
with temperature and radius 
of `neutrino-sphere' 
starting from the values $T(0)=5$~MeV and $R(0)=10$~km
and diminishing with time 
as in eq.~(\ref{tscale}) with $\tau_c=10$~s and $\beta_c=3/4$;
non-thermal effects are described by assigning
$\eta=2$. We can readily calculate 
the energy radiated, about $6\times 10^{52}$~erg 
and the average neutrino energy, around $14.2$~MeV,
which are both reasonable values. 
The fluence that we obtain 
integrating the flux in time can be  
parameterized by a 
distribution with `average' parameters 
$T\sim 4.45$~MeV and $\eta\sim 0.24$
(that give the right values of 
$\langle E_\nu\rangle$ and $\langle E^2_\nu\rangle$).
The good agreement between the two descriptions is illustrated
in fig.~\ref{fig:fluenze}.



\subsection{SN1987A}\label{cc4}
Galactic neutrino astronomy is still to begin, whereas,
curiously enough, extragalactic neutrino astronomy has already begun. 
At 7:36 (UT) of 23 february 1987, a number of experiments 
detected a neutrino signal from an atypically energetic stellar collapse 
occurred 170000 years ago in the Large Magellanic Cloud.
These experiments are: KamiokandeII (12 events), IMB (8 events), possibly 
Baksan (6 events)
and perhaps LSD~\cite{sn1987a:data}.

SN1987A neutrinos were not particularly useful to learn 
on oscillations, 
due to poor statistics and to astrophysical uncertainties.
The most reasonable hypothesis is that the $\sim 20$ neutrinos observed
are due to $\bar{\nu}_e $~\cite{randb}. 
In fact, $\sigma(\bar{\nu}_e p\to ne^+)$ is two orders of magnitude larger than
the cross sections of the other $\nu$ and $\bar{\nu}$, 
and it gives $e^+$ with almost isotropic angular distribution
(while scattering on electrons gives a forward-peaked $e$).
There is a reasonable agreement with expectations.
This is true in particular for the duration of  the emission, 
for the total amount of energy  released in $\bar{\nu}_e$, 
and for their mean energy.\footnote{The 
light yield, the observations of $\gamma$ radiation
from radioactive species and the astronomical properties 
of the precursor do not contradict the general theoretical 
picture, even though all of these show 
rather peculiar features.}
Poor statistics does now allow to discriminate well the temperature 
from the total energy:
a reduction in temperature can be compensated by an increase in flux.

Attempts of extracting as much information as possible by performing 
event-by-event fits are also limited by the fact that,
at a closer sight, a number of puzzling features appear. 
We here recall those  pertinent to neutrinos: 
the agreement between IMB and KII data is less than perfect; 
the average energy is on the low side of theoretical expectations; 
the temporal sequence of KII events is  rather non-uniform and 
the first event recorded at KII seem to point
back to the source (which would suggest a
unexpected identification of a $\nu_e$);
the LSD detector at Mont Blanc recorded other 5 neutrinos
but some five hours before the main signal.
The latter point is probably the most troublesome.
Together with a hint for gravitational
radiation obtained at Geograv, it could be interpreted 
as a manifestation of collapse in 2 stages, see~\cite{2stage}.

An analysis of KII, IMB and Baksan data yielded
some support  not only for a `standard' collapse, but also 
of a luminous accretion-like phase in the first half-a-second
of the neutrino emission~\cite{ll}.

The low average energy of KII and Baksan events lead many people to 
remark that the inclusion of solar oscillations with large mixing angle
transforms some hotter $\bar{\nu}_{\mu,\tau}$ into $\bar\nu_e$
thereby increasing their expected average energy 
and worsening the agreement with data~\cite{NuSN}.
This argument weakens if the hierarchy 
of temperatures is modest~\cite{raf} 
or if the originally produced flux of $\bar{\nu}_{\mu,\tau}$ is small.

\begin{figure}[t]
\begin{center}
\includegraphics[width=9cm,angle=270]{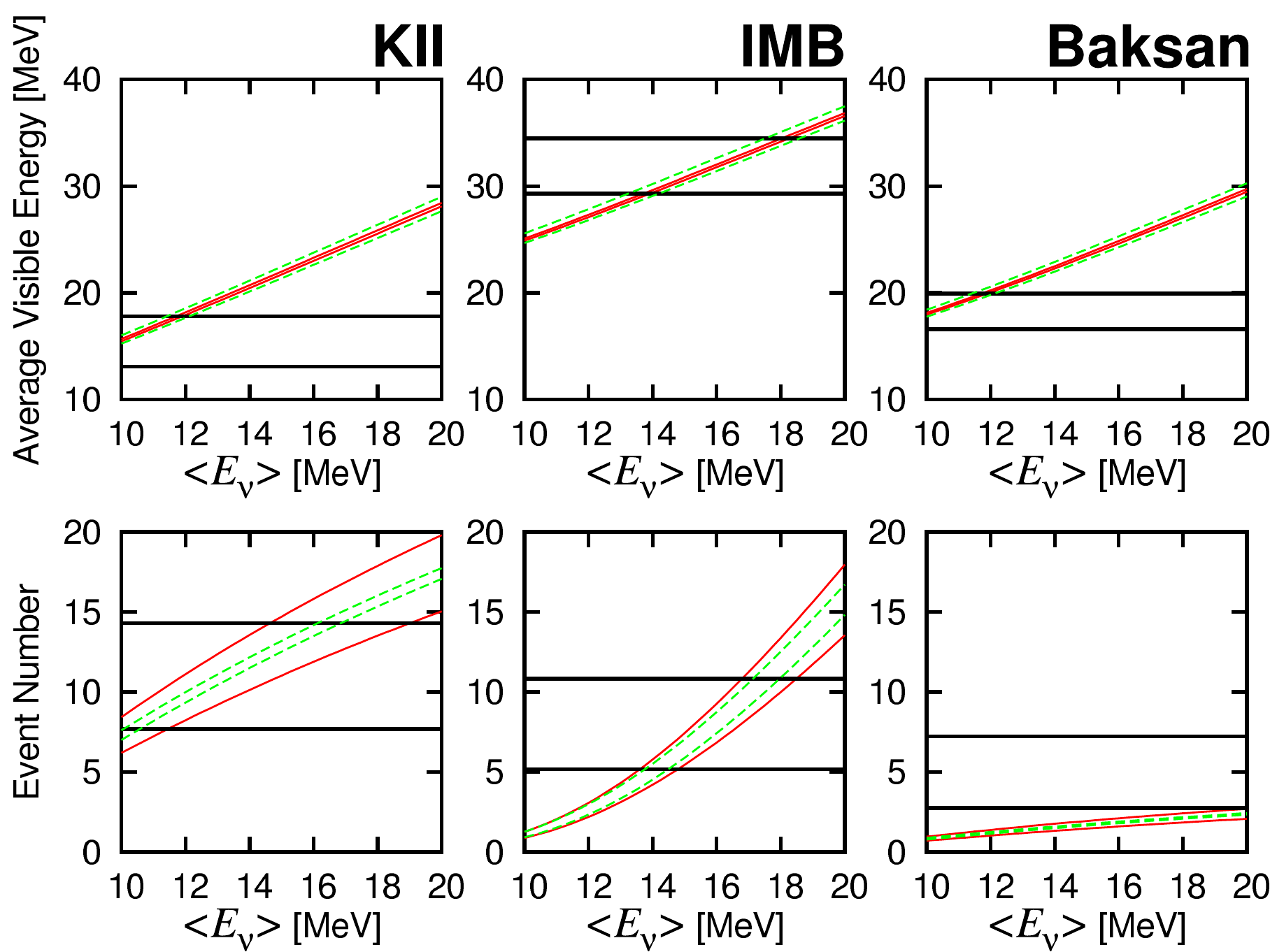}
\end{center}
\vspace{-4mm}
\caption[SN1987A data]{\em 
Comparison of observations 
(horizontal strips) and expectations calculated in the IBD hypothesis~\cite{ianns}.
The 3 upper panels show the average visible energy, the lower panels show 
the number of events, assuming  ${\cal E}_{\bar e}=4\cdot  10^{52}\,{\rm erg}$.
In each panel we show 4 expectations:
the continuous red lines 
correspond to a variation of the 
energy ${\cal E}_x$ radiated in  $\bar{\nu}_{\mu,\tau}$ parameter 
in the range $(2\div 6)\cdot 10^{52}\,{\rm  erg}$. The  dashed green  lines
correspond to a variation of the average energy $\langle E_x\rangle$
in the range 
$(1\div1.2)\cdot \langle E_{\bar e}\rangle$. 
\label{fig:kib}}
\end{figure}

However,  the most common view is to neglect these discrepancies 
and puzzles in consideration of the small number of neutrino events 
at our disposal, interpret all events as $\bar{\nu}_e$,
and draw no firm conclusion on the impact of oscillations on SN1987A events.
In figure~\ref{fig:kib}, we show that in this assumption there is 
a reasonable agreement between expectations and the observations,
for certain values of the astrophysical parameters that describe neutrino emission 
that fall in the expected range given in eq.~\ref{rrranges}.

\subsection{Future observations of supernova neutrinos\label{cc5}}
The galactic supernova rate is at best comparable to the 
inverse lifetime of a physicist, see eq.\eq{SNrate}.
A duty time of a minute per century makes difficult to 
build dedicated detectors,
but allows to detect SN neutrinos using other neutrino experiments.
Presently, at least one adequate neutrino experiment is kept 
running at any moment.
With several detectors located in different parts of the world, 
some of them could receive the next SN explosion during night,
allowing to probe also earth matter effects.

By measuring neutrinos emitted at the next supernova explosions,
we will probably learn more on oscillations and test non 
standard neutrino properties,
especially if we will have an accurate theory of  supernova explosions.
Surely, the next supernova neutrino burst will lead 
great advances in the astrophysics of the core collapse.

What we would like to know from next  galactic supernova?
It is not difficult to compile a list of wishes:
we would like to have detailed information on 
{\em time} and  {\em energy} distributions,
and study the signal in all possible {\em flavors}.
In practice, we can measure 
the spectrum of $\bar\nu_e$ very well;
the spectrum of electrons scattered by $\nu_e e$;
the total neutrino rate by neutral current.\footnote{All 
reactions of practical use  
do not allow to measure the
energy of individual  neutral 
current events, but only their total number and 
time distribution.
A possible exception is $\nu p\to \nu p$~\cite{johnb},
that could be studied by scintillators 
with very low threshold like Borexino~\cite{Borexino}.}
The possible goals are a detailed picture of 
the various phases of the collapse and of the explosion;
in particular the study of the first second (accretion?) 
seems to be of tremendous interest for astrophysics. 
Furthermore, it would be important to test  the 
flavor  distributions of the emitted neutrinos (e.g.\ the 
amount of $\nu_e$ and $\bar{\nu}_e$ during accretion).

\begin{figure}[t]
$$\includegraphics[width=8cm]{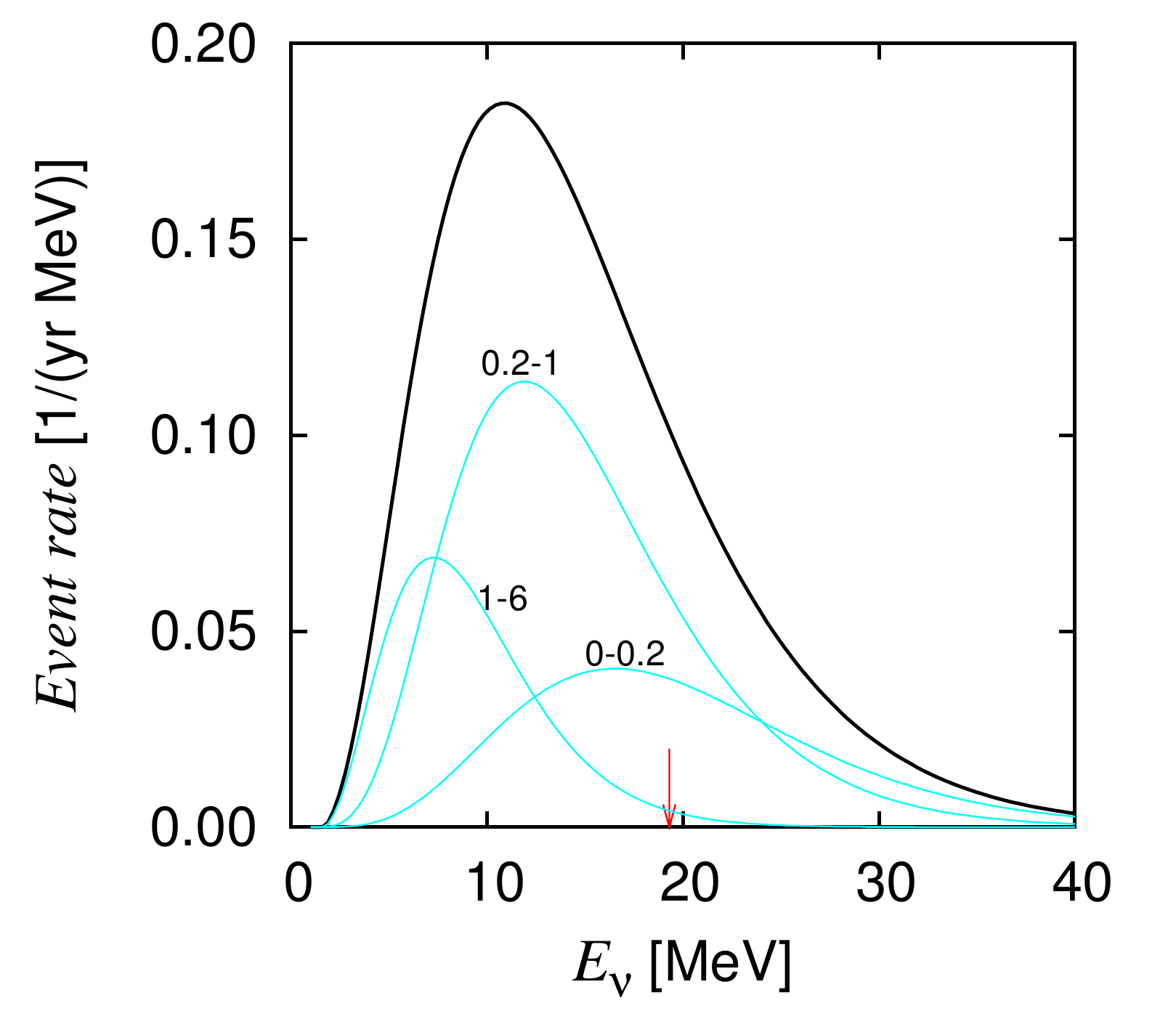}$$
\vskip-3mm
\caption[Distribution of SN relic neutrinos]{\label{fig:relitti}\em 
Differential rate for IBD interaction of SN relic neutrinos, for the
parameters assumed in the text. The three lower curves 
correspond to the contributions coming from the 
indicated regions of redshift. The arrow indicates the threshold in the SK detector.}
\end{figure}

\subsection{Relic supernova neutrinos}\label{SNrelic}
Observing $\nubarnu$ emitted by past core-collapse supernov\ae{} would allow
to test star formation models and, possibly, other aspects of cosmology such as dark energy~\cite{SNrelic}.
The curve denoted as `relic SN$\nu$' in fig.\fig{AstroNu} shows a typical expectation.

Again, thanks to the large IBD cross section, $\bar\nu_e$ give the most promising signal.
The number of IBD events generated in a detector with $N_p$ protons in a time observation
$T$ above a threshold $E_{\rm th}$ is
\beq N =N_p T   \int_{E_{\rm th}}^{E_{\rm max}}   dE_\nu \, \sigma(E_\nu)\cdot 
\frac{c}{H_0} \int_{0}^{z_{\rm max}} dz 
\frac{R(z)\ }{\sqrt{\Omega_\Lambda+\Omega_{\rm DM}(1+z)^3}} \frac{dN_{\bar\nu_e}((1+z) E_\nu)}{dE_\nu}
\eeq
where the upper limits (maximum energy $E_{\rm max} \sim 40\div60$ MeV 
and redshift $z_{\rm max} \sim 5\div 6$) are not crucial parameters, and 
the cosmological parameters are $H_0\approx70$ km/sec/Mpc, 
$\Omega_{\Lambda}\approx 0.7$, $\Omega_{\rm DM}\approx 0.3$. 
The crucial quantity is the rate of core collapse 
supernov\ae{} $R$ as function of the redshift $z$. We assume that it can be approximated as
\beq
R(z)\approx R(0)\times\left\{ 
\begin{array}{cr}
(1+z)^2 & \mbox{if $z<1$}\\
4 & \mbox{if $z\ge 1$}
\end{array}\right.\eeq
Its present value is $R(0)=f_{\rm SN}\cdot R_*(0)$,
where $R_*(0)$ is the present cosmic 
rate of star formation,
and $f_{\rm SN}$ is the fraction of stars that become core collapse supernov\ae.
For reference we take $ R_*(0)=0.02\cdot M_\odot/({\rm Mpc}^3\cdot {\rm  yr})$ 
and $f_{\rm SN}=1 \%/M_\odot$.
The uncertainty in  $R(0)$ is the dominant one; other uncertainties are less critical: 
the distribution in $z$ is tested by observations at low redshifts; 
the closest supernov\ae{} produce the largest part of the signal
and thus the point where the slope is modified ($z=1$) is not very relevant
(unless $R$ grows with $z$ more rapidly than what we assumed).

The event rate is roughly proportional to the energy released and to
the average energy of SN neutrinos. 
We assume that  each burst carries $4\cdot 10^{52}$ erg, 
with energy spectrum 
$dN/dE \propto  E^3 \exp(-4 E/14\MeV)$.
Although there is no SN explosion theory, these assumptions 
agree with SN1987A observations.
With these values, one expects that relic SN $\bar\nu_e$ give
0.7 events per year in a SK-like detector
(fiducial mass of 22.5 kton, water target, unit efficiency above $E_{\rm th}=19.3\MeV$), 
half of which come from $z<0.2$.
The total number of
events is 3, half of which would come from $z<0.5$, see figure \ref{fig:relitti}.

SK produced a  strong limit that already 
excludes the most optimistic  models 
for star formation:
a rate 3 times larger than the one quoted here starts being incompatible with SK data.
Since this means that a moderate improvement over the present SK sensitivity
could allow to observe SN relic $\bar\nu_e$,
let us discuss how it can be achieved.

Backgrounds (rather than statistics) start to be the main limitation.
The search strategy is based on the different energy spectra of the signal and of the backgrounds.
Atmospheric $\bar\nu_e$ provide an essentially irreducible background above $60\MeV$.
At lower energy, the main background is produced by $\mu^\pm$ that enter the detector
without producing \v{C}erenkov light because too slow.
This background is not  present in a scintillator detector, such as KamLAND, LVD and Borexino,
that however have a significantly smaller mass than SK.
In big W\v{C} detectors, such as SK, it can be avoided by detecting not only the $e^+$ produced
by IBD events $\bar\nu_e p \to e^+ n$, but also the neutron (namely, doubly tagging the signal).
This can be achieved by adding, for instance, Gadolinium to the SK detector~\cite{scireac}.

\section{Effects of neutrino oscillations\label{subs:enm}} 
At first sight,  astrophysical sites offer 
ideal conditions to study how neutrinos propagate.
We have enormous distances 
(e.g., the distance of the galactic center
is about $2.5\times 10^{20}$~m), 
a wide variety of energies (just think to the 
extension in energy of the cosmic ray spectrum), magnetic fields of 
huge intensity (e.g., $10^{12}$ G on the surface of 
a neutron star) or of wide extension 
(e.g., intergalactic  fields), etc.
This allows to probe oscillations, matter effects,
and possibly magnetic moment and finite lifetime.
However,  these propagation effects have to be disentangled from
the the properties of the neutrino source
(for instance  supernov\ae{} are discussed in section~\ref{sec:pns2}),
that are plagued by sizeable astrophysical uncertainties.

We here mostly focus on the scenario
suggested by solar and atmospheric data:
oscillations of 3 neutrinos.
These `active'  oscillations do not affect the total rate of NC events 
(as well known after  SNO solar neutrino results~\cite{SNO}): 
the total fluxes 
$F_{e}^0+F_\mu^0+F_\tau^0$ and 
$F_{\bar{e}}^0+F_{\bar{\mu}}^0+F_{\bar{\tau}}^0$ 
remain unaffected.
In this section we study
neutrinos with energy $(1\div 100)\MeV$ produced in high density environments,
such as in core collapse supernov\ae, in neutron
star mergers, or in accretion processes on a black hole. 
For concreteness we specifically consider the case of supernova neutrinos,
but our considerations have more general validity.

\subsection{Oscillation of supernova neutrinos}\index{Oscillation!in supernov\ae}

Since the $\mu$ and $\tau$ fluxes at production are supposed to be identical,
two functions are needed to describe how oscillations affect CC rates:  
$P_{ee}(E_\nu)$ and $P_{\bar{e}\bar{e}}(E_\nu)$, 
the electron neutrino/antineutrino survival 
probabilities.
This can be seen by rewriting the general expression
$F_e=P_{ee} F_e^0+P_{\mu e} F_\mu^0+P_{\tau e} F_\tau^0$ as
$F_e=P_{ee} F_e^0+(P_{\mu e} +P_{\tau e}) F_{\mu,\tau}^0$ and 
recalling that $1=P_{ee}+P_{\mu e} +P_{\tau e}$.
In order to calculate $P_{ee}$, one has to solve the
evolution equation of eq.\eq{m+V}.
Inserting numbers, the effective  hamiltonian is
\begin{equation}\label{HeffSN}
H =2.534\cdot V \frac{\mathrm{diag}(m_i^2)}{E_\nu} V^\dagger \pm
3.868\cdot 10^{-7}\ \rho Y_e\cdot \mathrm{diag}(1,0,0)
\label{heff}
\end{equation}
where the sign $+$ holds for neutrinos and $-$ for $\bar\nu$.
Neutrino masses $m_i$ are in eV, the neutrino energy $E_\nu$ in  MeV,
the density $\rho$ in g/cm$^3$ and the effective hamiltonian is in 1/m.
The supernova density $\rho$ and the 
electron fraction $Y_e=N_p/(N_p+N_n)$ must be taken
from some pre-supernova model, including 
the modifications due to the shock wave when needed.
Neutrinos are produces in the region where the matter term dominates:
one might think that the flavor-diagonal matter effects block oscillations,
that only occur when neutrinos reach the outer region
where the matter density is low enough to have MSW resonances.

\begin{figure}[t]
$$\includegraphics[width=\textwidth]{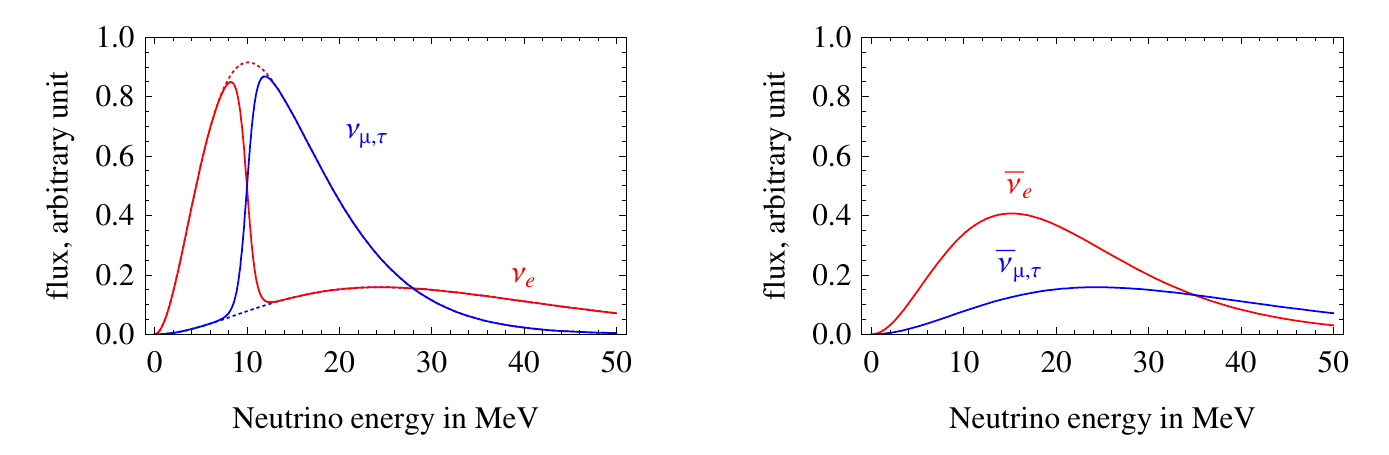}$$
\vspace{-10mm}
\caption[Spectral split of SN neutrinos]{\label{fig:SpectralSplit}\em 
Spectral split of supernova neutrinos, present if neutrinos have inverted hierarchy.
Later, neutrinos are affected by the usual MSW resonances, vacuum and Earth oscillations.}
\end{figure}

\paragraph{Collective oscillations}
On the contrary, it was realized in~\cite{CollectiveOsc} that oscillations occur also in the inner region, where
the neutrino density is so high that neutrinos themselves act as a non trivial background to their propagation.
The resulting set of non-linear differential equations exhibits an unexpected collective behavior
when, like in a supernova, a large neutrino/antineutrino is present.
The density profile is uncertain and different for different supernov\ae{}; 
we here make the plausible assumption that collective effects occur only before MSW resonances, 
thereby effectively modifying the initial neutrino energy spectra later distorted by the `traditional' oscillations.
The bottom line is the following: if neutrinos have inverted mass hierarchy and if $\theta_{13}>0$
(even if it is very small), neutrinos experience a `{\em spectral split}' at energies
$E_\nu > E_{\rm cr}\approx 10\MeV$:  the flux of $\nu_e$ gets interchanged with the flux of $\nu_{\mu,\tau}$,
see fig.\fig{SpectralSplit}.
Roughly noting happens to anti-neutrinos.
Despite the low energy and the inconvenient flavor, detectors might be able to observe this feature in future SN explosions.

This and other features were first noticed in extensive numerical simulations.
The neutrino matter effect is proportional to $\sqrt{2}G_{\rm F} (N_\nu+ N_{\bar\nu})(1-\cos\vartheta)$,
where $\vartheta$ is the neutrino-neutrino crossing angle, 
such that one needs to introduce bins in direction, energy and flavor,
obtaining a huge system of coupled `stiff' non-linear differential equations.
The spectral split feature is still present if they are approximated with a simpler system, 
neutrino crossings are assumed to occur at a single averaged angle.
It is convenient to introduce a ``Bloch vector'' notation, i.e.\ the $2\times 2$ neutrino and anti-neutrino
density matrices at energy $E$
are parameterized as $\rho_E = N_\nu (1 + \vec{P}_E\cdot\vec{\sigma})/2$ and
$\bar\rho_E = N_{\bar\nu} (1 + \vec{\bar P}_E\cdot\vec{\sigma})/2$.
Then evolution equations have the form
\beq \dot{\vec{P}}_E = (\omega \vec B +A \hat{z} + \mu \vec D)\times \vec{P}_E,\qquad
 \dot{\vec{\bar{P}}}_E = (-\omega \vec B + A \hat{z} + \mu \vec D)\times \vec{\bar P}_E\eeq
 where
 $A = \sqrt{2} G_{\rm F} N_e$,
 $\mu\equiv \sqrt{2}G_{\rm F} (N_\nu+ N_{\bar\nu})$,
 $\omega = \Delta m^2_{\rm atm}/2E_\nu$,
 $\vec{B}\equiv  \hat{x}  \sin2\theta_{13}\mp \hat{z} \cos2\theta_{13}\approx \mp\hat{z}$
 (the upper (lower) sign applies to normal (inverted) hierarchy),
 and
 \beq\vec{D} \equiv \frac{1}{N_\nu + N_{\bar \nu}} \int dE_\nu \left(\frac{dN_\nu}{dE_\nu} \vec P_E+ \frac{dN_{\bar\nu}}{dE_\nu}\vec{\bar{P}}_E\right)\eeq
couples different energies.
The combination $\vec D\cdot\vec B\approx \mp D_z$ is conserved, and it essentially is the electron lepton number, $ N_{\nu_e}- N_{\bar\nu_e}$.
Na\"{\i}vely the usual matter term, proportional to $A$, should block oscillations;
instead it turns out to be essentially irrelevant.
These equations can be shown to be formally analogous to the ones that describe a gyroscopic pendulum in the gravitational field,
and this analogy helps in understanding why neutrinos exhibit non-trivial collective behaviors,
such as `bipolar oscillations', `nutations', etc.
For inverted hierarchy, the system is like a pendulum that starts from the upper unstable position and that, 
for any $\theta_{13}>0$, eventually finds a way of falling down to the stable position.
This is only limited by the conservation of $D_z$, giving rise to the spectral split above the critical energy $E_{\rm cr}\approx 10\MeV$ fixed by
\beq\int_{E_{\rm cr}}^\infty dE_\nu \left(\frac{dN_{\nu_e}}{dE_\nu} - \frac{dN_{\nu_{\mu,\tau}}}{dE_\nu}\right)=
\int_0^\infty dE_\nu \left(\frac{dN_{\bar\nu_e}}{dE_\nu} - \frac{dN_{\bar\nu_{\mu,\tau}}}{dE_\nu}\right).\eeq
Fig.\fig{SpectralSplit} shows the resulting spectra, that can be seen as initial conditions for the later usual oscillations.

\paragraph{MSW resonances in the supernova}
Neutrinos exiting from a supernova can encounter two MSW resonances:
the `solar resonance' (described by the 
oscillations parameters $\Delta m^2_{12}$ and $\theta_{12}$)
is very much likely to be adiabatic, like in the solar case;
the `atmospheric resonance' 
(described by  $\Delta m^2_{13}$ and $\theta_{13}$)
is adiabatic if $\theta_{13}$ is large enough.
More quantitatively, assuming a typical pre-supernova profile 
$\rho\sim 2\ 10^4\ (10^4 \mbox{km}/ r)^3\,{\rm g/cm}^3$
and $Y\sim 1/2$ and using the adiabaticity condition described in section~\ref{MSWresonance},
the atmospheric resonance 
is adiabatic ($P_C\simeq 0$) if
$\theta_{13}\gg  \sqrt{E_\nu/\Delta m^2_{\rm atm}\,{\rm km}}\sim 1^\circ$
and is non adiabatic $(P_C \simeq 1)$ if $\theta_{13}\ll 0.1^\circ$.
Taking into account that neutrinos experience the solar resonance,
and that neutrinos (anti-neutrinos) experience the atmospheric resonance
if neutrinos have normal (inverted) hierarchy, we get:
\begin{equation}\label{eq:Psupernova}
\begin{array}{l|cc}
&\hbox{normal hierarchy} & \hbox{inverted hierarchy}\\ \hline 
P_{\bar{e}\bar{e}} &V_{e1}^2 & (1-P_C) V_{e3}^2 + P_C V_{e1}^2 \\
P_{{e}{e}} & (1-P_C) V_{e3}^2 + P_C V_{e2}^2  &V_{e2}^2
\end{array}
\end{equation}
This leads to the following limiting possibilities:
\begin{itemize}
\item[A)] If $\theta_{13}$ is small enough that $P_C\simeq 1$ one has, for normal and inverted hierarchy:
$$F_{\bar e} = F_{\bar e}^0 \cos^2\theta_{12}  +F_{\bar \mu,\bar \tau}^0  \sin^2\theta_{12}\qquad
F_{e} =F^0_{e}   \sin^2\theta_{12} + F^0_{\mu,\tau} \cos^2\theta_{12}.$$

\item[B)] If $\theta_{13}$ is large enough that $P_C\simeq 0$ and neutrinos have normal hierarchy one has:
$$F_{\bar e} \hbox{ as in A)},\qquad  F_{ e} = F^0_{\mu,\tau}.$$

\item[C)] If $\theta_{13}$ is large enough that $P_C\simeq 0$ and neutrinos have inverted hierarchy one has:
$$F_{e}  \hbox{ as in A)},\qquad   F_{\bar e} = F^0_{\bar\mu,\bar\tau}.$$
\end{itemize}
The case of inverted hierarchy gives rise to a more sizable variation with $\theta_{13}$, 
and this variation takes place in anti-neutrinos,
that can be better observed thanks to IBD.

\paragraph{Shock wave}
An interesting possibility is that  $P_C$ could be time dependent, 
because the shock wave can cross the layer where the MSW resonance takes place,
increasing the density gradient:
depending on $\theta_{13}$ and to the type of neutrino mass hierarchy
this effects possibly leads to observable signals in  future 
large neutrino detectors.

\paragraph{Earth matter effect}\index{Oscillation!in the earth}
If (anti)neutrinos cross the Earth before hitting the detector, 
new oscillations occur and previous expressions are slightly modified. 
To understand the features of these oscillations we consider 
a constant density, say, the Earth mantle. 
Earth matter effects
can be accounted by performing the following replacements in eq.\eq{Psupernova}
\beq V_{e2}^2\to   V_{e2}^2 + \delta P,\qquad\hbox{and}\qquad
V_{e1}^2\to   V_{e1}^2 - \delta P\eeq
such that Earth matter effects are sensitive to $\theta_{13}$ and to the type of neutrino mass hierarchy.
By explicitly solving the evolution equation one finds 
\begin{equation}
\delta P=
\varepsilon 
\cdot \frac{\sin^2(  \Delta m^2_{12} L/4 E_\nu\ \sqrt{(1+\varepsilon)^2-4 \varepsilon\cos^2\theta_{12}} )}
{(1+\varepsilon)^2-4 \varepsilon\cos^2\theta_{12}}\ 
\sin^2 2\theta_{12}
\end{equation}
This $\delta P$ holds for neutrinos; for anti-neutrinos one must replace $\theta_{12}\to \pi/2-\theta_{12}$.
The quantity $\varepsilon$ is
\begin{equation}
\varepsilon=
\frac{\sqrt{2} G_{\rm F} N_{e}}{\Delta m^2_{\rm sun}/2 E_\nu}
\approx 5.5\ \% \frac{E_\nu}{20\MeV}\frac{\rho}{3\,{\rm g/cm}^3}
\end{equation}
having assumed $Y_e=1/2$ and the measured $\Delta m^2_{\rm sun}$.
This effect is small for solar neutrinos, that have $E_\nu < 20\MeV$.
Supernova neutrinos are expected to reach larger energies,
and earth matter effect can become detectably large.
At first order in $\varepsilon$ we can approximate 
the fraction with $\sin^2($vacuum  oscillation phase).
The typical oscillation length is (see eq.\eq{lambdaosc})
$\lambda\approx 600\km$, somewhat smaller than the radius of the earth.
In conclusion,  earth matter effects are best seen at higher $E_\nu$ and
if the supernova is seen just below the horizon.

\begin{table}[t]
\begin{center}
\begin{tabular}{c|cc}
& Normal mass hierarchy & Inverted mass hierarchy\\ \hline
Small $\theta_{13}\circa{<} 10^{-5}$ & Earth + burst & Earth + burst + spectral split\\
Large $\theta_{13}\circa{>} 10^{-3}$ & Earth & Shock + burst + spectral split
\end{tabular}
\end{center}
\vspace{-5mm}
\caption[Signals of oscillations of SN neutrinos]{\em Possible oscillation signatures in
supernova neutrinos, depending on the value of $\theta_{13}$ and on the type of mass hierarchy:
Earth-matter effects; shock wave effects in the
  $\bar\nu_e$ spectra; the $\nu_e$ burst and the spectral split of $\nu_e \leftrightarrow \nu_{\mu,\tau}$.
     \label{tab:Tomas}}
\end{table}

\paragraph{Summary of oscillation effects}
Table \ref{tab:Tomas} summarizes the discussion:
depending on the value of $\theta_{13}$ and on the type of mass hierarchy
we list the possible oscillation signatures: the Earth matter effect for $\bar{\nu}_e$ (detectable via IBD); 
the presence of modifications due to the shock wave 
for $\bar{\nu}_e$ (detectable via IBD); the absence of a neutronization 
$\nu_e$ burst (detectable via a reaction with a suitable nucleus or via ES);
the presence of the $\nu_e \leftrightarrow \nu_{\mu,\tau}$ split in energy spectra.
In order to make quantitative statements
one would need a definite theory of neutrino emission: 
the first two signals disappear if all antineutrinos are emitted with similar distributions;
even the neutronization burst signal could be modified if rotation has an important
role in the collapse.

\paragraph{Other possible effects}
Let us recall for completeness some  possible extra effects.
In presence of a polarized (magnetized) underlying medium, 
the MSW term in eq.\eq{m+V} is different (it does 
not reduce to the average weak charge). 
Suitable neutrino magnetic moment can permit transitions 
from neutrino to antineutrino states. 
If extra sterile neutrinos with keV-scale masses exist,
they can encounter MSW resonances in the inner region where neutrinos 
are trapped.
Lighter sterile neutrinos give rise to extra MSW resonances in the outer region.

\subsection{Oscillations from cosmic sources}
High-energy neutrinos are expected to be produced in diffuse media, such that 
matter corrections to oscillations are negligible;
if these neutrinos cross the Earth, matter effects suppress oscillations and
absorption can be relevant.
The path-length is so large that vacuum oscillations are in the averaged regime,
where oscillation probabilities are  described by eq.\eq{Paveraged}:
inserting the observed  mixing angles one finds
\begin{equation}
P_{\ell\ell'}=\sum_{i=1}^3 |V_{\ell i} V_{\ell' i} |^2
\approx
\left( 
\begin{array}{ccc}
0.6 & 0.2 & 0.2 \\
0.2 & 0.4 & 0.4 \\
0.2 & 0.4 & 0.4 
\end{array}
\right).
\end{equation}
Since $\theta_{23}\approx \pi/4$ the muon and tau fluxes 
become almost equal at the detector.


The observation of neutrinos from cosmological distances, 
if possible, could allow us to search for new effects possibly produced 
by neutrino decay (see section~\ref{Decay}), or by slow neutrino oscillations into
extra sterile neutrinos (see section~\ref{Sterile}).

However, in the typical situation of the $\pi$ decay chain,
the initial flavor ratio is $e:\mu:\tau =  1:2:0$ and atmospheric oscillations
transform it into  $1:1:1$,
which is unaffected by new effects.
Therefore it is hoped that other natural sources of neutrinos with a more favorable flavor composition exist.
For example, $n\to p e\bar\nu_e$ decays might give a sizable contribution
to some source of cosmic ray neutrinos~\cite{n->nu}.

\subsection{Oscillations and interactions for neutrinos from DM annihilations}
Annihilations (with cross section $\sigma v$)
or decays (with decay rate $\Gamma$) of DM particles with mass $m_{\rm DM}$
accumulated around the Galactic Center, the
core of the sun
or of the earth can produce detectable fluxes of neutrinos~\cite{DMnu} with energy below $m_{\rm DM}$, here called DM$\nu$.
Assuming that DM arises as thermal relict of a weakly interacting massive particle
points to $\sigma v \approx 3~10^{-26}{\rm cm}^3/{\rm sec}$, which is typical
of weak scale particles, $m_{\rm DM}\sim 10\GeV \div 10\TeV$.
Detectably large neutrinos fluxes can be obtained.

The flux of $\nu_f$  from the Galactic Center is given
in terms of the line.of.sight integral of the DM density $\rho(\vec x)$:
\beq 
\label{gammaflux}
\frac{d \Phi_f}{dE~d\Omega} = \frac{1}{4\pi}
\int_{\rm l.o.s.} ds~P_{fi}
\left[\frac{\sigma v}{2}\left( \frac{\rho}{m_{\rm DM}}\right)^2  \left.\frac{dN_i}{dE}\right|_{\rm ann} + \Gamma \frac{\rho}{m_{\rm DM}}
 \left.\frac{dN_i}{dE}\right|_{\rm dec}  \right] .
\eeq
where the oscillation probability can be approximated with its averaged vacuum oscillations limit, eq.\eq{Paveraged},
and $dN/dE$ is the neutrino energy spectrum produced per annihilation or per decay.
Present data imply $\sigma v\circa{<}10^{-(21\div 22)}\cm^3/{\rm sec}$
and $1/\Gamma \circa{>}10^{24}\,{\rm sec} (m_{\rm DM}/\TeV)$, assuming
a typical $dN/dE$ such that an order one fraction of the DM mass goes into neutrinos.
Observations of $\gamma, e^+,\bar p$ cosmic rays
provide somewhat stronger constraints, unless DM is very heavy $m_{\rm DM}\circa{>}10\TeV$
or DM only gives neutrinos.

\medskip

DM$\nu$ from the sun and from the earth suffer an additional uncertainty by about one order
of magnitude, because they depend on the DM density inside these bodies.
Solar fluxes are expected to be larger than terrestrial fluxes, but not by a significant factor:
the sun is bigger and better captures DM, but the earth core is closer.
Furthermore, the DM capture rate in the earth is enhanced if $m_{\rm DM}$ is comparable
to the mass of some heavy element in the earth,
while this is not possible in the sun that is dominantly composed of H and He.
In the sun the capture rate is typically in equilibrium with the annihilation rate,
while this is not the case for the earth: the present annihilation rate depends
on past history.

Captured  DM particles must have negligible velocity, below the escape velocity
(at most $11\km/\sec$ in the earth, and at most $620\km/\sec$ in the sun).
The density distribution of DM particles within the sun or the earth
is approximatively given by~\cite{DMnu}
\beq\label{eq:RDM} \label{eq:n(r)} n(r) = n_0 \exp(-r^2/R_{\rm DM}^2) \qquad 
R_{\rm DM} =\frac{R}{\sqrt{\beta m_{\rm DM}}}
\eeq
where $r$ is the radial coordinate, 
$\beta = 2\pi G_{\rm F} \rho_0 R^2/3T_0$,
$\rho_0$ and $T_0$ are the central density and temperature of the body  (sun or earth)
and  $R$ is its radius. Numerically
$\beta = 1.76/\GeV$ for the earth
and $\beta=98.3/\GeV$ for the sun.

The energy and flavour spectrum of DM$\nu$ depends on the possible
DM annihilation channels.
The main possibilities (including only known particles) are
$\nu\bar\nu$, $\tau\bar\tau$, $b\bar b$, $W^+W^-$, $ZZ$, $t\bar{t}$.
The last three channels are kinematically allowed only if $m_{\rm DM}>M_W,M_Z,m_t$ respectively.
Decays into too long lived or too strongly interacting particles (such as $e,\mu,\pi$),
that thereby get stopped by interactions with matter before decaying,
do not produce energetic neutrinos.
Interactions with matter are not important for $W,Z,t$,
that directly decay into `prompt' neutrinos as well as into $b, \tau$, etc,
whose decays produce more neutrinos.
In order to compute the flux of neutrinos produced by decays of $b, \tau$
one must take into account that they loose some energy before decaying
due to interactions with surrounding matter.

If the DM particle is a Majorana fermion, like the neutralino in SUSY models,
its annihilation rate in a fermion/anti-fermion pair with mass $m_f$
is suppressed by an helicity factor
$m_f^2/m_{\rm DM}^2$.
This means that $b\bar{b}$ and $\tau\bar{\tau}$ channels are relevant only for $m_{\rm DM}<M_W$,
and that  annihilation into $\nu\bar\nu$ is negligible.
This is not the case if the DM particle has instead  spin 1: 
annihilations into two neutrinos with energy $m_{\rm DM}$ can have significant branching ratio.

\begin{figure}[t]
$$\includegraphics[width=\textwidth]{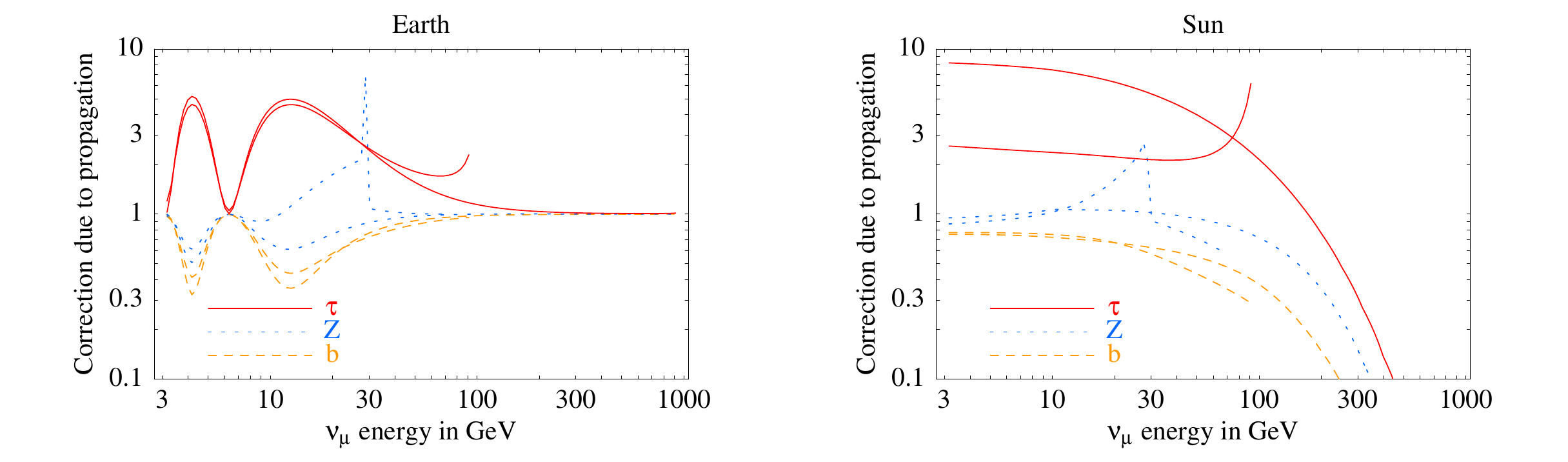}$$
\vspace{-10mm}
\caption[Neutrinos from DM annihilations]{\label{fig:OscNoOsc}\em 
Modifications of neutrino fluxes from DM annihilations due to propagation. 
The figures show the ratio of $\nu_\mu$ fluxes `with'  over `without' the effects of neutrino propagation (oscillations, absorptions, regeneration). The lines refer to neutrinos from DM annihilations into $\tau\bar\tau$ (continuous line), $ZZ$ (dotted) and $b\bar{b}$ (dashed), for $m_{\rm DM}= \{100,1000\}\GeV$ (distinguishable by the corresponding maximum neutrino energy).}
\end{figure}

Different DM annihilation channels result in DM$\nu$ with a characteristic energy spectrum and
typical energy $E_\nu \sim m_{\rm DM}/\hbox{(2 or 3)}$.
One expects $F_{e} = F_{\mu} = F_{\bar e}=F_{\bar\mu}\neq
F_{\tau}=F_{\bar\tau}$.
The energy and flavour spectra can be significantly affected by oscillations:
e.g.\ $\tau$ decays dominantly produce $\nu_\tau$, which are converted into
$\nu_\mu$ by atmospheric oscillations.
Absorption is relevant for neutrinos produced around the center of the sun
with energy above $100\GeV$ (and above $\sim 10^5\GeV$ in the earth).
NC processes reduce the energy but not the number of $\nubarnu_{e,\mu,\tau}$.
CC processes have cross sections a few times higher than NC processes, and 
do not directly produce a flux of secondary neutrinos.
This happens indirectly only in the case of $\nubarnu_\tau$,
that generate $\tau^\pm$ with left polarization, that decay back mainly into $\nubarnu_\tau$
with lower energy.
Section~\ref{OscAbs} presented the formalism appropriate for computing the combined effect of oscillations and absorption, that amounts to a ${\cal O}(0.1\div 10)$ correction,
as illustrated in fig.\fig{OscNoOsc}. The main effects are
\begin{itemize}
\item[$\oplus$] DM$\nu$ from the Earth are affected by atmospheric 
$\nubarnu_\mu\leftrightarrow\nubarnu_\tau$
oscillations at energies $E_\nu\circa{<}100\GeV$, and by absorption at $E_\nu\circa{>}10\TeV$.

\item[$\odot$] DM$\nu$ from the Sun of any flavor and any energy
are affected by averaged `solar' and `atmospheric' oscillations.
Furthermore, absorption suppresses neutrinos with
$E_\nu\circa{>}100\GeV$, that are partially converted
(by NC and by $\nubarnu_\tau$ regeneration) into
lower energy neutrinos.
Neutrinos with energy $E_\nu\gg 100\GeV$ approach a well-defined
limit spectrum,
roughly equal to $d\Phi_\nu/dE\propto e^{-E/100\GeV}$.
\end{itemize}
So far no DM$\nu$ have been seen, and atmospheric
neutrinos are the main background.
A less important similar background is  produced in the corona (i.e.\ the `atmosphere') of the sun,
that reach us after experiencing averaged oscillations, see fig.\fig{AstroNu}.
Therefore one needs big experiments with good angular resolution,
such that one can point to the known source of DM$\nu$.
Experiments can achieve a resolution of few degrees and be ultimately
limited by the typical scattering angle  $\sim (m_N/E_\nu)^{1/2}$ of $\nu N\to \ell N'$ processes.
In the case of DM$\nu$ from the earth, the size of the production region results in a comparable
angular spread.  
Therefore experiments give stronger constraints on the DM$\nu$ fluxes for larger $m_{\rm DM}$.
Future experiments are discussed in section~\ref{IceCUBE}.

\section[Galactic and extra-galactic neutrinos]{Galactic and extra-galactic neutrinos}\label{NuAstro}\label{AstroNu}

\subsection{Large volume neutrino detectors}\label{IceCUBE}
\index{IceCUBE}\index{Km3NeT}
We here discuss the near future experiments that aim at starting neutrino astronomy
by discovering neutrinos with energies from $10^2$ up to $10^{11}\GeV$.

Present bounds and astrophysical expectations discussed in the next sections
suggest that these experiments need a ${\rm km}^3$-scale volume.
Detectors with surface and volume so much bigger than SK can be built by 
inserting strings of  photo-multipliers
in the sea water or ice, used as target material.
To perform a \v{C}erenkov experiment a transparent enough medium
is needed, and one has to cope with 
luminescent fishes in water and bubbles of air in ice.
In view of the CR background, neutrino telescopes are
mainly sensitive to neutrinos that cross the earth and come from below.
Neutrinos with energies above $100\TeV$ get attenuated by the earth,
such that the most promising signal becomes quasi-horizontal events
and depends on local geography.

The first such experiments will be {\sc IceCUBE}~\cite{ICECUBE,IceCUBE},
a km$^3$ under-ice detector under construction at the south pole.
Since its location is not convenient for observing neutrinos from
the galactic center,\footnote{Most conventional telescopes can only look the northern sky because
built in the northern hemisphere. It is ironic that  {\sc IceCUBE} has the same
limitation because detects neutrinos at the south pole.
It might be possible to reject $\sim 99\%$ of the background of down-going atmospheric $\nubarnu_\mu$ by 
vetoing a correlation with $\mu^\pm$.

We recall that the declination $\delta$ is measured with respect to the earth rotation plane,
and that the galactic center is located at $\delta =-29^\circ$.
The sun is located at $7.9\pm0.4\,{\rm kpc}$ from the galactic center
(and 34 pc above the central plane of the Milky way) rotating 
with an orbital speed of 217 km/s.}
 the {\sc Antares}, NEMO, NESTOR 
collaborations are planning to build a km$^3$-scale under-water
detector named {\sc Km3NeT} in the northern hemisphere, somewhere in the
Mediterranean see.

The main signal in these experiments is the  \v{C}erenkov light generated
by up-going muons produced by $\nubarnu_\mu q \to \mu^\pm X$
CC interactions of $\nubarnu_\mu$.
The neutrino direction can be reconstructed up to a minimal uncertainty 
$\delta \theta\sim \sqrt{m_N/E_\nu}\sim 1.5^\circ/\sqrt{E_\nu\,{\rm in~TeV}}$
(due to the kinematics of neutrino/nucleon scatterings,
and to the subsequent $\mu$ interactions with matter).
At large enough energies the experimental uncertainty becomes dominant:
about $1^\circ$ in IceCUBE and about $0.3^\circ$ in the {\sc Antares} project
(for comparison, the angular size of the sun and moon is $0.5^\circ$).
A small $\delta \theta$ is crucial for discriminating point sources from the
atmospheric background. The energy spectrum and the time dependence
(in the case of impulsive sources) provide additional handles.
The atmospheric background at IceCUBE is about 
$0.4\nu/({\rm yr}{\rm km}^2 \cdot{\rm degree}^2)$ at energies larger than a TeV,
and 5 times smaller above 10 TeV.

NC interactions of $\nubarnu_{e,\mu,\tau}$ give  a hadronic
shower which can be too small to reach the photo-multipliers.
CC interactions of $\nubarnu_e$ give a bigger electromagnetic shower.
CC interactions of $\nubarnu_\tau$ give a hadronic shower and a $\tau$,
which decays immediately if $E_\tau\circa{<}10^6\GeV$,
decays within the detector if $E_\tau\circa{<}10^{7\div 7.5}\GeV$
and leaves a track similar to a $\mu$ track if it has higher energy.

These experiments will reach a km$^3$ volume at the expense of a poor granularity:
a bigger experiment has a bigger distance among  photo-multipliers so that
particles scattered by neutrinos start to be seen above an energy threshold of few tenths of GeV,
and energy cannot be measured accurately.
For example, one expects a $\circa{>} 30\%$ uncertainty around TeV energies.

There are ideas about how to build a $10\,{\rm km}$ scale experiment
that detects acoustic or radio waves (rather than \v{C}erenkov light)
produced in interactions of neutrinos with $E_\nu \gg10^{15}\eV$~\cite{RadioAcoustic}.

\bigskip

Finally, we emphasize that if these experiments will collect
UHE neutrino events, by measuring their zenith angular distribution
will allow to reconstruct the neutrino cross
section with nucleons at UHE energies
even if the total neutrino flux is unknown~\cite{UHEsigma}.
Many authors studied how exotic new physics (such as extra dimensions and
TeV-scale quantum gravity)
could affect this cross section.
To explain how this measurement is possible, let us give some useful numbers.
The density of the atmosphere decreases roughly as $e^{-h/8.4\,{\rm km}}$,
and  $99\%$ of the atmosphere mass lies in its lower 30 km.
The depth of the atmosphere is $L_{\downarrow}= 0.01033 \kmwe$
when crossed vertically, and $L_{\to}=0.36\kmwe$ when crossed horizontally;
the depth of the earth crossed along a diameter 
is $L_\oplus = 1.1~10^5 \kmwe$.
(For the sun one has $L_\odot=3~10^{7}\kmwe$).
These quantities set the absolute scale for cross section measurements:
the interaction length of a particle with cross section $\sigma$ on nucleons is
$L=1.7~10^4\,{\rm kmwe}\, ({\rm nb}/\sigma)$.
Protons have $\sigma \approx 0.15\,{\rm b}$ and thereby interact in the upper atmosphere:
this is the observation that tells that CR are dominantly hadrons.
According to the SM, UHE neutrinos have a much smaller 
$\sigma\approx 10 \,{\rm nb} \cdot (E_\nu/10^{18}\eV)^{0.4}$,
such that their interaction length is a few orders of magnitude bigger than the
atmosphere depth, and smaller than the Earth depth.
For UHE neutrinos one therefore expects
$L_\downarrow,L_\to \ll L \ll L_\oplus$: 
the number of atmospheric events is
proportional to $\sigma$ (and is maximal around the horizontal),
while the number of up-going events is roughly inversely proportional to $\sigma$
(because a too large $\sigma$ attenuates the neutrino flux).

\subsection{Galactic cosmic-ray neutrinos}\label{cc6}
In this and in the next section we discuss expectations for neutrinos fluxes related
to cosmic rays.
Although there are plausible CR theories, sources of cosmic rays have not been established ---
indeed this is something we hope to learn by observing the associated high energy neutrinos.
The fact that the slope of the energy spectrum of CR,
$dN/dE\propto E^n$ with $n\sim 3$,
becomes slightly harder around $E\approx 10^{10}\GeV$ (the `ankle')
suggests that there are two different populations of cosmic rays:
those below the ankle  are believed to be of galactic origin,
while CR above the ankle  are believed to have extragalactic origin,
because our galaxy apparently does not have powerful enough galactic sources
and/or does not have a magnetic field intense enough to trap CR at highest energies,
nor to make them isotropic.\footnote{CR above the ankle show no clustering around the galactic center.
CR much below the ankle, believed to come from the galactic center, are isotropized by
the galactic magnetic field $B$. This effect start disappearing around the ankle,
depending on how intense $B$ is and on the $q/m$ of CR, which might be
protons or heavy nuclei.
CR just below the ankle apparently show an indication
of clustering around the galactic center.}
The above imprecise statements reflect our present limited knowledge.
It is widely believed that CR arise from protons or nuclei accelerated
by expanding clouds of gas and magnetic fields
via some version of a mechanism called `diffusive shock acceleration',
which naturally produces a power law CR spectrum.

In this section we focus on galactic neutrinos~\cite{ginzBer}:
the most promising sources are 
young supernova remnants (SNR).
Alternative sources are small objects with very large magnetic fields,
such as   young supernov\ae{}, 
pulsar wind nebul\ae, microquasars.

We recall the main theoretical reason why SNR 
are considered as possible sources of neutrinos. 
Ginzburg and Syrovatskii observed that the Milky Way radiates CR at a rate 
${\cal L}_{\rm CR}=V_{\rm CR}\rho_{\rm CR}/\tau_{\rm CR}\sim 10^{41}$~erg/s
(where the volume, the lifetime and the average density of CR are 
all rather uncertain):
this energy loss is compensated if any 30 year
a new SNR (of any type)
converts 10\% of its kinetic 
energy (of the order of 1 foe = $10^{51}$ erg) 
into CR. The expanding clouds of gas and magnetic fields
remnant of SN explosions can last for thousands of years and
there is little doubt that a maximum energy $E_{\rm max}\sim 10^{5}\GeV$ can be achieved, 
and there are active discussions if non  linear amplifications
of the magnetic fields allow to reach the `knee' energy, around $10^{6\div 7}\GeV$.
Reaching the `ankle' is considered more problematic.

Let us come to the present observations and expectations. 
The observation by HESS~\cite{HESS}
of the southern sky with $\gamma$ rays at the TeV and above 
revealed two bright SNR, RX J1713.7-3946 and RX J0852.0-4622 (alas Vela Junior),
and the observed spectra below 10 TeV look to be 
power laws with hard spectral indices:
$\Phi_\gamma = dN_\gamma/dE \propto E^{-\Gamma}$ with $\Gamma\sim 2$,
the value suggested by the Fermi acceleration mechanism.
Since these sources are expected to be transparent to $\gamma$ radiation,
a {\em tentative} identification of these $\gamma$ as `{\rm hadronic}' 
implies that we should expect neutrinos.\footnote{Before working out the
implications of this $\gamma\leftrightarrow \nu$ connection, we discuss its main caveats.
1) $\gamma$ might instead be generated by alternative $e \gamma$ processes that generate no $\nu$:
inverse Compton or synchrotron emission.
2) If sources are not transparent to $\gamma$ we would instead under-estimate the flux of neutrinos.
While galactic $\gamma$ with PeV energies reach us unattenuated, 
this does not hold for higher energy extra-galactic $\gamma$.
This explains why we need two separate sections with two separate arguments:
the $\gamma\leftrightarrow \nu$ connection in this section for galactic CR$\nu$,
and another argument in the next section for extra-galactic CR$\nu$.}
To convert the measured $\gamma$ flux into an expected neutrino flux, one first 
get the $\pi$ flux from the simple $\pi^0\to \gamma\gamma$ kinematics:
$$ \Phi_{\pi^+}(E) = \Phi_{\pi^-}(E) = \Phi_{\pi^0}(E) = - \frac{E}{2} \frac{d\Phi_\gamma}{dE}.$$
One expects equal numbers of $\pi^+,\pi^-,\pi^0$ because many $\pi$ are produced in each collision.
Next, the neutrino flux produced by $\pi^\pm \to \mu^\pm \nubarnu_\mu$ decays is
\beq 
\Phi_{\nu_\mu} (E)= \int_{E/(1-r)}^\infty \frac{dE'~\Phi_{\pi^+}(E')}{(1-r)E'} = \frac{\Phi_\gamma(E/(1-r))}{2(1-r)}
\eeq
where $r=(m_\mu/m_\pi)^2$. Finally, one has to take into account $\mu^\pm$ decays and oscillations.
Extrapolating the $\gamma$ spectra till several hundred TeV suggests signals of about  
$10\div 15$ induced-muon events per km$^2$ per year. However, this should 
be reduced by about $50\%$ if the neutrino spectrum drops below the
assumed power law around 10 TeV (which would be consistent with the 
present expectations on $E_{\rm max}$, since the expected neutrino energy is 
1/20 the energy of primary CR) and by a similar factor 
to account for finite detection efficiency. 
Thus, we are lead to 
expect that we need a rather effective rejection of the background 
and a large exposure in order to see these sources.
The total number of signal events is comparable to the 
total number of background events with the same direction
produced by atmospheric neutrinos:
CR$\nu$ can be observable if they have a hard energy spectrum and
dominate over the atmospheric neutrino background at large energies.
The $\gamma$ rays from the RX J1713.7-3946 SNR have been observed~\cite{HESS} till 100 TeV:
above $\sim 10\TeV$ their energy spectrum softens, deviating from a power law.
This suggests that the protons well above 100 TeV left
the SNR already and detection of CR$\nu$ will be challenging. 
The other intense source of $\gamma$ rays is the younger Vela Jr.,
that has been observed~\cite{HESS} till $\sim 20 \TeV$, and the $\gamma$ 
energy spectrum not show clearly the presence of a cutoff. 
This fact enforces the hope that CR$\nu$ can be observable.


\begin{table}
\begin{center}
\begin{tabular}{c|ccccc}
Name & TeV $\gamma$ observation & declination & distance & size & age \\ \hline
Vela Junior & up to 10 TeV (HESS~\cite{HESS}) & $-46^\circ 22'$ & 0.2 kpc & $2^\circ$ & 680 yr\\  
RX J1713.7-3946  & up to 100 TeV (HESS~\cite{HESS}) & $-39^\circ 46'$ & 1 kpc & $1^\circ$ & 1600 yr\\  
SN 1006 & bound close to expected signal  & $-41^\circ 53'$ & 2 kpc & $36'$ & 1000 yr\\  
Cas A & HEGRA (maybe) & $+58^\circ 08'$ & 3 kpc & $6'$ & 320 yr\\  
\end{tabular}
\caption[Candidate CR$\nu$ sources]{\em Most promising supernova remnants candidate source of galactic CR$\nu$. 
Notice that their distance and age are not reliably determined.
The angular size is at most comparable with the angular resolution.
}
\end{center}\end{table}

These considerations are not meant to be conclusive, but rather to illustrate the possible 
relevance of theoretical expectations and (in the case of SNR) 
of $\gamma$ ray observations. 
There could be other intense sources of neutrinos but not
of photons, generically denoted as ``hidden sources''
(e.g., a microquasar  or a black hole inside a star
which screens the electromagnetic radiation).


\subsection{Extragalactic cosmic-ray neutrinos}
As discussed in the previous section, extra-galactic sources should be the origin 
of the cosmic rays observed at the highest energies,
from $E\sim E_{\rm GZK}\sim 10^{10.7}\GeV$ down to the `ankle'  at $E\sim 10^{9.9}\GeV$ 
(or less likely down to $10^9\GeV$, or, unlikely, down to the `knee' at  $E\sim  10^{7}\GeV$;
future neutrino data might clarify this issue).
The measured CR energy density of these extragalactic CR
 is comparable to the total energy density emitted 
so far by some known candidate extra-galactic sources, especially 
$\gamma$ ray bursts and active galactic nuclei (AGN), supporting the view that these
are the sources of extragalactic CR.\footnote{AUGER~\cite{AUGER} observed angular correlations
between CR of highest energies with AGN up to a distance of 100 Mpc.

A region with size $R$ and magnetic field $B$ can contain ultra-relativistic particles of charge $q$ up to energies $E = qRB$: to reach large energies one therefore needs a large $BR$, 
and this is what makes circular colliders, such as LHC, expensive.
Taking into account that natural accelerators are not expected to be $100\%$ efficient
(e.g.\ energy losses exclude normal neutron stars from the list of candidates)
it is not clear if some astrophysical source can accelerate CR up to the maximal observed energies, or if 
some new physics contribution is needed,  such as decays of quasi-stable ultra heavy relic particles.}
The Fermi mechanism for acceleration of CR suggests that the CR
spectral index at production is close to 2,
i.e.\ that their energy is logarithmically distributed among different energy scales.
Consequently the total energy in CR depends only 
logarithmically on the uncertain extremes of their energy range,
and equating it to the measured energy density of extragalactic CR 
allows to reliably  normalize the CR flux at production:
\beq \frac{dU}{dV} = \frac{4\pi}{c} \int dE~ E \frac{dN_{\rm CR}^{\rm prod}}{dE}\approx 3~10^{-19}\frac{{\rm erg}}{{\rm cm}^3}. \eeq
One gets
\beq
E^2 \frac{dN_{\rm CR}^{\rm prod}}{dE}\approx \frac{10^{10}\GeV}{{\rm km}^2\cdot{\rm year}\cdot{\rm sr}}
\approx 3~10^{-8}\frac{\GeV}{{\rm cm}^2\cdot{\rm sec}\cdot{\rm sr}}
\eeq
and likely extends below the ankle,
where we can only observe the larger flux of galactic CR
trapped by the galactic magnetic field.
Propagation trough turbulent magnetic fields presumably
converts the CR spectral index towards the measured one, close to 3.

We can now connect $dN_{\rm CR}^{\rm prod}/dE$ to neutrinos.
Protons $p$ accelerated in the region with intense magnetic fields interact with the ambient radiation $\gamma$:
\beq \label{eq:pgamma->Delta}
p~\gamma \to \Delta^+ \to \left\{\begin{array}{l} \pi^0~ p\\ \pi^+ n\end{array}\right.  .\eeq
Roughly the same energy goes into nucleons, $\pi^0$, $\pi^+$, that end up in CR, $\gamma$, $\nu$ respectively.
The clean $\gamma \leftrightarrow \nu$ connection discussed in the previous section does
not hold because, before reaching us from extra-galactic distances,
the $\gamma$ energy degrades down to GeV energies, below
the threshold for $\gamma_{\rm UHE}\gamma_E \to e^+ e^-$ where $\gamma_E$ is 
the electric field of heavy nuclei.
While the secondary protons may remain trapped in the
acceleration region, equal numbers of neutrons, neutral and charged pions escape/decay.

Therefore, one expects that 
transparent sources emit roughly the same energy in CR and in $\nu$:
${d N_\gamma}/{dE}\sim{dN^{\rm prod}_{\rm CR}}/{dE}$.
This argument was made more precise by Waxman and Bahcall~\cite{ExtraGalactic}, that after including some
order one factors (related to oscillations, to kinematics, to redshift) obtained the expected
$\nu_\mu+\bar\nu_\mu$ flux:
 \beq E_\nu^2 \frac{d N_\nu}{dE_\nu}\approx 1.3~10^{-8}\frac{\GeV}{{\rm cm}^2\cdot{\rm sec}\cdot{\rm sr}}.\eeq
 This corresponds to the line indicated as `CR$\nu$' in fig.\fig{AstroNu}, and 
would give tens of events per km$^2$ per yr in the IceCUBE detector.
The same flux is expected in $e$ and $\tau$ flavors.
As in the previous section, this would be an underestimate if sources were opaque (i.e.\ sources transparent only to neutrinos).
The factor ${\rm sr}=1$ indicates that the flux is per unit of solid angle.

\bigskip

Fig.\fig{AstroNu} shows another, possibly subdominant, flux of UHE neutrinos,
denoted as `UHE$\,$CR-CMB', because they are generated by collision
of UHE protons with CMB radiation.
Indeed the the cross section for proton scattering on CMB photons
(that have known energy $E_\gamma \sim 3\K$ and density, see section~\ref{CMB})
is large above the kinematical threshold of the $p\gamma \to \Delta$ reaction in eq.\eq{pgamma->Delta},
i.e.\ at $E_p > E_{\rm GZK}\equiv (m_\Delta^2 - m_p^2)/4 E_\gamma\approx 5~10^{10}\GeV$.
The $\Delta$, with mass $m_\Delta \approx 1.3 m_p$, is the lightest
hadron that allows such a resonant scattering.
The $\Delta^+$ 
mostly decays into $p\pi^0$ and $ n \pi^+$: the two particles
in the final state roughly have only half of the energy of the initial proton.
Charged pions decay into neutrinos generating the  `UHE$\,$CR-CMB' flux.
It is significantly uncertain, and its possible relevance is not guaranteed.


\index{AGASA}\index{HiRes}\index{$Z$ burst}\index{AUGER}
The above discussion also implies that 
protons with energy above $E_{\rm GZK}$ (`Greisen-Zatsepin-Kuzmin cut-off')
should be absorbed before reaching us,
unless produced by unseen nearby sources.
Analogous arguments hold for other hadronic particles
that might compose UHE cosmic rays
(their composition  is not known).
This seems confirmed by the {\sc Auger} observations~\cite{AUGER}.

%% file: review_flavour.tex
\chapter{Understanding neutrino masses}\label{flavour}
Most of the fundamental parameters known 
so far are fermion masses, or in other words 
{\em flavor parameters}. 
We measure quark, lepton and neutrino 
masses and mixings with the hope  of 
understanding (sooner or later) why they are what 
they are. This is the flavour problem.

\medskip
Let us describe the content of this section.
We begin in section \ref{FlavourDiscussion}
with a discussion of the present 
obstructions and difficulties towards a theory of flavor.
In section~\ref{nu:centric} we describe some
patterns that could emerge from
the observed $\nu$ masses and mixings and how
they can be easily explained with flavor models, without getting predictions.
More ambitious approaches based on grand unification 
are discussed in section~\ref{nu:gut} and their successes, 
strictures and failures are outlined.  
Finally, we review in section~\ref{FlavourTest} how these models can be tested,
focussing on models that make tentative predictions for $\theta_{12}$ and $\theta_{13}$.

\section{The problem of the flavour problem}\label{FlavourDiscussion}
Here, we try to diagnosis the difficulties that are blocking our
theoretical and experimental attempts of understanding flavour.
On the theoretical side, understanding the values of the fundamental parameters in QFT is particularly difficult because
they receive ultraviolet divergent quantum correction.
Without knowing which high-energy theory provides the physical cut-off  (possibly related to quantum gravity),
the practicable way to get some control over 
the parameters is recognizing possible symmetries that relate
different parameters.
Examples of  very concrete progress achieved thanks to this approach are well known:
\begin{itemize}
\item Around 1970, thanks to {\bf gauge invariance},
theorists have been very useful in understanding experiments and in establishing the SM
along the following road:
$$\hbox{photon}\to\hbox{gauge invariance}\to \hbox{gluons}, Z,W.$$
The most generic gauge invariant renormalizable Lagrangian that
can be written with the SM fields contains 18 apparently fundamental 
parameters;
13 of them describe flavour.\footnote{The cosmological constant and the
QCD $\theta$ parameter are two extra parameters, not included in the 
above counting. The first parameter is apparently crucial to 
understand the expansion of the Universe. 
The QCD $\theta$ parameter
gives rise to unobserved CP-violating effects:
present bounds on the neutron EDM imply $\theta\circa{<}{10}^{-9}$,
while one would na\"{\i}vely expect $\theta\sim 1$.
Although this issue might have some connection with the flavour problem,
we will not consider it here.}
Further success along this line could be obtained by 
assuming that the SM is the low-energy limit of a 
`unification' theory, with gauge group SU(5)  or SO(10)
broken at large energy $M_{\rm GUT}\sim 10^{16}\GeV$.

\item {\bf Supersymmetry} could become another predictive symmetry. 
Although all searches for supersymmetric signals gave so far null results,
weak-scale supersymmetry remains one possible interpretation to the
Higgs mass hierarchy problem.\footnote{Weak-scale supersymmetry
is also motivated by gauge unification.
Non super-symmetric SU(5) makes one wrong
prediction for the SM gauge couplings.
A successful prediction is obtained  assuming  MSSM supersymmetric particles
at the weak-scale.
The probability that this agreement happens
accidentally is $\sim\%$ 
according to `reasonable' estimates of theoretical uncertainties.}
(It should be added for fairness 
that supersymmetry introduces new flavor problems).


\end{itemize}
However, lepton and quark masses and mixings show no clear pattern
that indicates the possible symmetry behind them,
and in the SM are simply described 
by a list of 13 mysterious numbers.

\medskip

It is hard to obtain predictions in the flavour sector,
because flavour extensions of the SM often 
involve many more unknown parameters than the SM.
Neglecting neutrino masses, the lepton kinetic terms and Yukawa couplings
are described by three $3\times 3$ matrices which contain 36 real parameters;
but only 3 of them are measurable at low energy: $m_e$, $m_\mu$ and $m_\tau$.
Similarly SM quarks are described by 63 real parameter;
but only 10 of them are measurable at low energy (6 masses, 3 mixings, 1 CP-violating phase).
Since $3\ll 36$ and $10\ll 63$, only very restrictive symmetries or assumptions give testable predictions.\footnote{A related technical issue is: given a prediction for the flavor matrices, 
how can one extract the physics from it, getting rid of the unobservable extra parameters?
This could be done analytically with 2 generations, but with $3\times 3$ matrices one needs
to employ either approximations (many observed mass ratios and mixing angles are small),
or try to get some information inventing specific combinations of the mass matrices,
designed such that they are invariant under flavor reparametrizations and
related to some physical quantity, e.g.\ CP violation.}

By  postulating properly broken flavour symmetries (possibly in a context
where quarks and leptons are unified)
one can `explain'  the hierarchy
 $m_e\ll m_\mu\ll m_\tau$ and the similar one in quarks,
in terms of few small symmetry-breaking parameter and dozens
of unknown order-one parameters. 
See~\cite{attempts} for the first attempts.
Unfortunately it is possible 
to achieve the same in many different ways, and 
these efforts resulted in thousands of papers
(one every few days since many years)
with no observable consequence, up to rare exceptions.

\medskip

This situation is reminiscent of the outcome of another possible approach:
although quantum gravity likely gives effects too small to be observed,
it has been theoretically investigated hoping that it leads to a unique `theory of everything',
or at least to a predictive theory.
String/$M$ theory seemed a promising attempt, that apparently leads
to one simple theory in 11 dimensions.
Unfortunately, SM physics is neither simple nor 11 dimensional, and
there are so many different ways of getting rid of the extra dimensions
(e.g.\ there are 5 string models in 10 dimensions)
that after reaching 4 dimensions one has $10^{{\cal O}(500)}$ string models and
predictivity is lost.
Indeed in these string models the complicated 
physics that we see at low energy arises mostly thanks to
a complicated higher dimensional geography.

\medskip

The pattern of SM fundamental parameters suggests one deeper 
reason behind the apparently uselessly vast `landscape' 
of flavour models and of string models.
The SM allows the existence of many stable nuclei,
but this richness arises  because (relevant combinations of)
$\alpha_{\rm em}$, $\Lambda_{\rm QCD}$ and of
 the electron, up- and down-quark masses happen to have special values.
Different values would not lead to the complex chemistry that allows our existence.
This and analogous considerations concerning the cosmological constant lead to 
speculate about {\em anthropic selection}~\cite{anthropic}:
some  apparently  fundamental parameters could (in some way) take many different values, 
and we happen to observe one atypical set of values that allows for the existence of observers.
Like geography, flavour would not be a fundamental issue and the key point
that we should understand would be: in which way
fundamental parameters can take different set of values?
The simplest way consists in writing models where the potential has many different local minima,
by e.g.\ employing many heavy scalar fields (that in string models parameterize the unknown
higher-dimensional geography).
Other ways (e.g.\ employing light scalars with
position-dependent vevs) might lead to observable signals.

Maybe these speculations will lead to results\footnote{See~\cite{anthropic} for attempts.
E.g.\ neutrinos 
much heavier than 1eV reduce the growth of primordial fluctuations
(see eq.\eq{RP(k)}) reducing the number of galaxies.
If this is the only reason why neutrinos are light, neutrnos should not be much lighter than 1 eV.}
that will be summarized in future reviews.
We here review standard approaches, 
focussing on simple models of 
neutrino masses and mixings. Since neutrinos masses
(as well as quark and lepton masses of second and third generation)
do not seem to be too much anthropically relevant, 
one can hope that they reflect some property of the high energy theory.

\section{In the search for a pattern}\label{nu:centric}
Before writing flavour models that can explain the pattern of fermion masses, 
one must qualitatively interpret how this pattern looks like.
Charged leptons and quarks clearly exhibit large mass hierarchies and small mixing angles.
Neutrinos show a qualitatively different pattern, but the situation is not yet clear.
The possible symmetries inspired by neutrino observations include:  $L_e - L_\mu-L_\tau$~\cite{bhsw},
SO(3) (assuming quasi-degenerate neutrinos),
$\mu\leftrightarrow\tau$ permutations\footnote{A neutrino mass matrix $m$ symmetric under
$\mu\leftrightarrow\tau$ commutes with the permutation matrix
\beq\label{eq:Pmutau} P = \pmatrix{1&0&0\cr 0&0 &1\cr 0&1&0}.\eeq 
Therefore, the eigenstates of $m$ coincide with the eigenstates of $P$, 
implying maximal $\theta_{23}$ and $\theta_{13}=0$.
In charged leptons, the $\mu\leftrightarrow\tau$ symmetry is of course badly broken by $m_\tau \gg m_\mu$.}, 
S$_3$ $e\leftrightarrow \mu\leftrightarrow\tau$ permutations,
`quark-lepton complementarity'~\cite{sunCabibbo}, A$_4$ and S$_4$~\cite{A4}, to be discussed later.
The common problem of these approaches is that symmetry breaking terms are responsible of the finer structure:
unless they are  theoretically predicted one gets back to the generic mass matrix i.e.\ no predictions.

In this section we present two patterns possibly emerging from present data,
and their possible interpretation.

%
%
%
%
%
%
%
%

\begin{figure}
$$\includegraphics{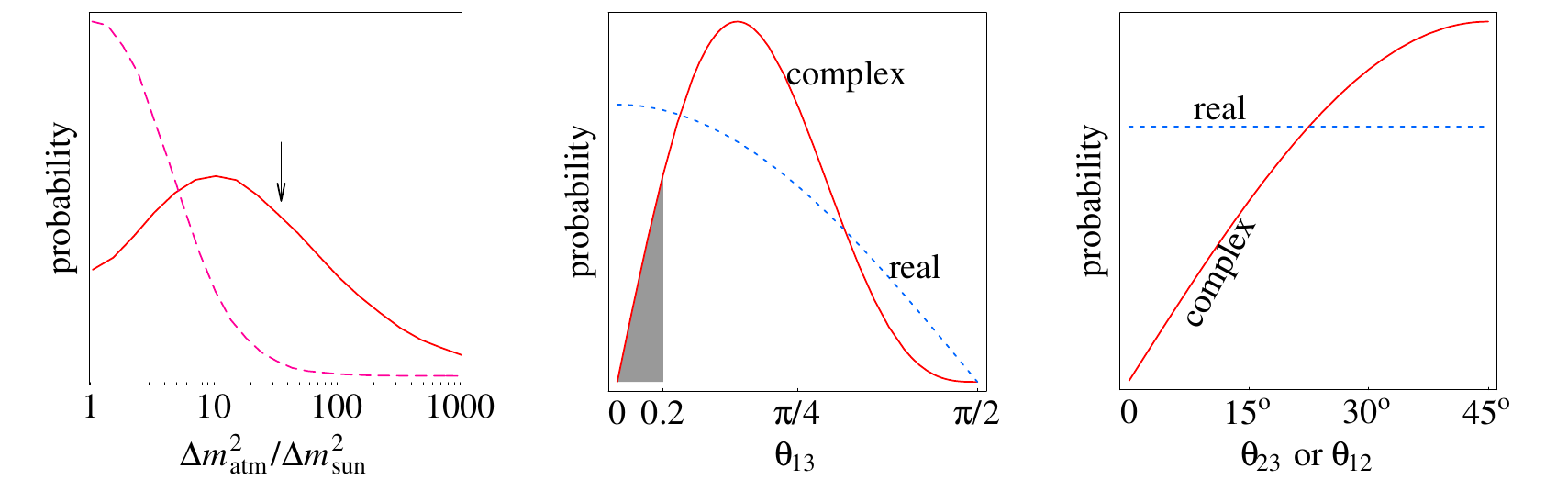}$$
\caption[Expectations from anarchy]{\label{fig:anarchy}\em {\bf Oscillation parameters expected from an anarchical 
 neutrino mass matrix}.
Fig.\fig{anarchy}a: see-saw (solid line), non see-saw (dashed) with complex couplings.
Fig.\fig{anarchy}b,c: complex couplings (solid), real couplings (dashed).}
\end{figure}

\subsection{Anarachy?}\label{Flavour:nu}
The obvious starting question is: is there any structure in 
the neutrino mass matrix suggested by experiments?
A neutrino mass matrix 
with generic ${\cal O}(0.03)\eV$ entries
would give all mixing angles of ${\cal O}(1)$ and all
$\Delta m^2 \sim 10^{-3}\eV^2$.
The experimental signals typical of this minimal case were
studied in~\cite{generic}, and have been not observed.
As a result,
experiments now require relatively small values of 
\beq \label{eq:small?}
\theta_{13}<\theta_{13\rm max} \approx 0.15\qquad\hbox{and of}\qquad
R=\Delta m^2_{\rm sun}/\Delta m^2_{\rm atm} =  (0.030\pm 0.005).\eeq
How likely these two small numbers
are accidentally produced by a 
structure-less (or `anarchical') mass matrix?

\smallskip

Some authors tried to estimate such probability in two different ways~\cite{anarchy}:
 i) by generating many random matrices and counting how frequently  they satisfy
the bounds of eq.\eq{small?};
ii) by computing the analogous frequency as function of some parameter $\epsilon$
(such that $\epsilon \ll 1$ generates the small $R$ and $\theta_{13}$ measured by experiments)
and testing if a small $\epsilon$ works better than $\epsilon=1$.
In Bayesian language (see appendix~\ref{Statistics}) these are two MonteCarlo procedures that compute two
probabilities; in both cases ${\cal O}(1)$ factors depend on the assumed arbitrary prior.
They are different because answer to two different questions: 
i) has the meaning of a gof test and 
ii) tests at which CL  $\epsilon =1$ is in the best-fit range of $\epsilon$.
As usual (see appendix~\ref{Statistics}) ii) is more reliable ---
less arbitrary factors affect a relative probability
than an absolute probability.
This kind of studies suggests that the observed pattern of neutrino masses and mixings
contains some kind of structure with $\sim 90\%$ probability:
a structure-less mass matrix is disfavored but not excluded.
More precisely, assuming that the modulus of each matrix element has flat probability 
in some interval, in fig.\fig{anarchy} we show the probability distribution of $R$,
of $\theta_{13}$ and of $\theta_{12,23}$.
We can understand analytically why
the probability scales in a non-trivial way with $R$ and with $\theta_{13}$.
\begin{itemize}
\item {\em Scaling with $R$}.
As expected, anarchical
neutrino mass matrices give a small $R$ with small probability $p\propto R^{1/2}$  (solid line of  fig.\fig{anarchy}a).
A larger probability $p\propto R^{1/4}$ (dashed line) is obtained in see-saw models,
i.e.\ by assuming that
neutrino masses are mediated by virtual exchange of
heavy right-handed neutrinos with anarchical Yukawa couplings and masses.

\item {\em Scaling with $\theta_{13}$}.
It is reasonable to assume that the probability distribution of the mixing angles 
is given by the Haar measure of the U(3) flavour group
(or SO(3) if one assumes that CP is conserved)\footnote{Since this result might seem obscure,
it is useful to remind that analogous
non trivial trigonometric factors arise in a well known case, where their
geometrical origin is more transparent.
Analyses of solar and atmospheric neutrino data
use the well known fact that a isotropic background is flat
in $d\Omega = d\cos\vartheta_{\rm zenith}\,d\varphi$,
where $\vartheta_{\rm zenith}$ and $\varphi$ are the usual polar coordinates.
The Haar measure analogously describes the `solid angle' on a (complex) hyper-sphere.}~\cite{anarchy}:
\beq
dp\propto \left\{
\begin{array}{ll}
d\cos^4\theta_{13}\,d \sin^2\theta_{12}\,d\sin^2 \theta_{23}\,d\phi      &   \hbox{complex couplings,}\\
d\sin\theta_{13}\,d\theta_{12}d\theta_{23}     &    \hbox{real couplings.}
\end{array}\right. 
\eeq
Assuming complex couplings, small values of $\theta_{13}$ have a non trivially small probability:
$p(\theta_{13}<\theta_{13\rm max}) = 1-\cos^4 \theta_{13\rm max}\simeq 2\theta_{13\rm max}^2$.
The quadratic dependence on $\theta_{13\rm max} $ is related to 
the fact that the CP-phase $\phi$ becomes non physical at $\theta_{13}=0$.
The probability that a random $\theta_{13}$ is accidentally compatible with the
 CHOOZ upper bound is $\int e^{-\chi^2(\theta_{13})/2} d\cos^4\theta_{13}=5\%$.
 Assuming real couplings gives a weaker  linear dependence of $p$
on $\theta_{13\rm max} $.
These results agree with the numerical scan of fig.\fig{anarchy}b.
\end{itemize}
Anarchy suggests that $\theta_{13}$ should be not much below its present experimental upper bound.
If this is the case, we will remain in an ambiguous situation:
measured masses and mixings disfavor anarchy but
do not clearly indicate the presence of some structure.

\subsection[Mass hierarchy with large $\theta$?]{Mass hierarchy between largely mixed states?}\label{Flavour:nu2}
We have seen that neutrino masses and mixings show two potentially small numbers ($\theta_{13}$ and 
$R= \Delta m^2_{\rm sun}/\Delta m^2_{\rm atm}$)
that one might want to explain by inventing appropriate flavour models.
While a small $\theta_{13}$ can be explained in many different ways,
a small $R$ (together with $\theta_{\rm atm}\sim 1$) is a very characteristic feature.
If not due to an accident, it is a strong indication because,
while it is easy to write matrices that give hierarchical masses with small mixing,
or large mixing without mass hierarchies,
only few Majorana neutrino mass matrices give both.
Indeed a large atmospheric mixing angle
between the {\em most splitted\/} neutrino states ($\Delta m^2_{\rm
atm}\gg \Delta m^2_{\rm sun}$) is naturally produced
by two plausible peculiar structures for the Majorana neutrino mass matrix
$\mb{m}_\nu$.
We write them in the limit 
$\Delta m^2_{\rm sun} = 0$ and $\theta_{13}=0$ and
working in the mass eigenstate basis of charged leptons.
\begin{itemize}
\item[(h)]  If $\Delta m^2_{\rm atm}>0$
(i.e.\ neutrinos have a {\bf hierarchical} spectrum)
one needs
the rank one matrix~\cite{dominant}
\beq\label{eq:textureh} \mb{m}_\nu \hbox{(h)}\propto  
\pmatrix{
0&0&0 \cr 0&s^2&sc\cr 0&sc&c^2}.\eeq
The see-saw mechanism can naturally generate such a rank 1 mass matrix:
if only one right-handed neutrino exists,
it gives mass to only one left-handed neutrino, leaving the other two massless.
Therefore the matrix (h)  is obtained when one right-handed neutrino,
coupled mostly to $L_\mu$ and $L_\tau$,
\begin{equation}\label{eq:seesawmotivato}
\lambda N (sL_\mu + c L_\tau)H_{{u}} + \frac{M}{2} N^2,\qquad
\end{equation}
gives the dominant contribution to the light neutrino masses.

\item[(i)]  If $\Delta m^2_{\rm atm}<0$
(i.e.\ neutrinos have  an {\bf inverted} spectrum)
one needs to justify why two of them are almost degenerate.
A rank two pseudo-Dirac matrix does the job~\cite{bhsw}:
\beq\label{eq:texturei}
\mb{m}_\nu\hbox{(i)} \propto\pmatrix{
0&s& c \cr s&0&0\cr c&0&0}.\eeq
The eigenvalues of this matrix are proportional to $0,1,-1$:
massive degenerate neutrinos have opposite CP-parity.
Other choices of the relative phase between the two
massive state do not suggest to a `natural' form for $\mb{m}_\nu$.

The matrix (i) can be justified by imposing a $L_e-L_\mu-L_\tau$ U(1) flavour symmetry,
that in the unbroken limit automatically implies $\Delta m^2_{\rm sun}=0$,
$\theta_{13}=0$ and $\theta_{12}=\pi/4$.
This last prediction is incompatible with
data, that demand a large but not maximal $\theta_{12},\theta_{12}+\theta_{13}<\pi/4$.
Although can find specific sources of breaking of  $L_e-L_\mu-L_\tau$ that  give a large correction
only to $\theta_{12}-\pi/4$~\cite{bhsw},
models for inverted hierarchy (that typically use $L_e-L_\mu-L_\tau$ as ingredient)
seem now less appealing.
One can assume that $\mb{m}_\nu\hbox{(i)}$ holds in a basis where the charged lepton mass matrix 
has non diagonal elements, that provide extra sources for mixing angles and reduce predictivity.

In the see-saw context $\mb{m}_\nu\hbox{(i)}$ can be generated 
by exchanging a pseudo-Dirac couple of right-handed neutrinos, $N$ and $N'$.
By appropriately extending the $L_e-L_\mu-L_\tau$ symmetry to $N$ and $N'$
one can justify the following see-saw model:
\beq
\lambda N (sL_\mu + cL_\tau)H_{U} +
\lambda' N' L_eH_{U} + M  NN'.\eeq

\item[(d)] Neutrinos 
could also have a quasi-{\bf degenerate} spectrum, but 
we know no simple way of justifying why this should be the case.
Various constructions have been discussed, mostly based on an SO(3) flavor symmetry.
\end{itemize}



%

%


\subsection{U(1) flavor models can reproduce all patterns}\label{nu:flav} 


The SM Lagrangian has a U(3)$^5$ flavour symmetry 
(rotations of the $E,L,U,D,Q$ families)
explicitly broken by the Yukawa couplings.
This structure implies a peculiar strong suppression of CP-violating effects
in $K$ mixing and decay,  a very strong suppression of $e,n$ electric dipoles, 
and no flavour-violating processes like $\mu\to e \gamma$ in leptons.
Present data agree with SM predictions, disfavoring different flavour structures
(e.g.\ things would be different if two Higgs doublets would give comparable contributions
to fermion masses).

One can try to understand flavour by assuming that
some subgroup of U(3)$^5$ is an exact flavour symmetry,
 spontaneously broken by the vev of some scalar $\phi$ named `flavon'.
The flavour symmetry could be global, or gauged, or some theoretically more fashionable variant.
All them work in almost the same way.

The minimal choice
 of the flavour symmetry is an U(1)$_{\rm F}$ subgroup of U(3)$^5$.
One proceeds by assigning arbitrary ${\rm U}(1)_F$ charges $q_F$ to the various fields $F$.
Without loss of generality, we can assume that the flavon $\phi$ has  U(1)$_{\rm F}$ charge $+1$
(so that $\phi^*$ has charge $-1$).
We assume that $\phi$ is neutral under the SM gauge group $G_{\rm SM}$, 
since we want to break $G_{\rm SM}$ only by an ordinary Higgs field.
\footnote{One could instead employ some discrete subgroup of U(1)$_{\rm F}$,
or some larger non-abelian group, such as an U(2).
Other approaches do not employ a flavour group:
for example, assuming that the SM fermions are localized in different places in hypothetical extra dimensions,
small fermion masses could be due to a small overlap between fermion wave-functions~\cite{GSW}.}

In the non-renormalizable effective theory valid below some cut-off $\Lambda$
describing the SM fields plus the flavon $\phi$,
an operator ${\cal O}$ containing SM fields with total U(1)$_{\rm F}$ charge $q_{\cal O}$
(for example ${\cal O} = UQH$ has $q_{UQH} = q_U + q_Q + q_H$)
typically appears with coefficient of order $|\phi/\Lambda|^{q_{\cal O}}$.
All small numbers are expected to arise from the smallness of $\lambda \equiv \phi/\Lambda$.
One can write explicit renormalizable models containing many new particles with mass $\sim \Lambda$
that mediate all the couplings~\cite{seesaw}.
Since many models exist, each with many free parameters,
it is often convenient to focus of how
the flavour symmetry restricts the low-energy effective theory.
Of course, one can invent specific examples of a full theory
that gives atypical outcomes (e.g.\ some operator could be absent).

Restricting to the sub-case of integer charges, the Yukawa couplings are expected to be of order
\begin{equation}\label{eq:SMflavon}
 \lambda_U^{ij} \sim \lambda^{|q_{U_i} + q_{Q_j} + q_H|},\qquad
\lambda_D^{ij} \sim \lambda^{|q_{D_i} + q_{Q_j} - q_H|},\qquad
\lambda_E^{ij} \sim \lambda^{|q_{E_i} + q_{L_j} - q_H|}.
\end{equation}
The rules of the game become a bit different in  {\bf supersymmetric} extensions of the SM:
 the flavon $\phi$ has a fermionic partner $\tilde\phi$:
consequently it is no longer possible to generate a flavon of charge $-1$ by complex conjugation.
The precise statement is: the superpotential $\mathscr{W} $
is a holomorphic function of the superfields.
When building a supersymmetric extension of the SM, due to this fact one introduces
two Higgs doublets with opposite hypercharge: $H_U$ coupled to up-quarks,
and $H_D$, coupled to down-quarks and leptons.
The MSSM superpotential then contains all needed couplings
$$\mathscr{W} =\lambda_E ~  ELH_D + \lambda_D DQH_D  +   \lambda_U \,  UQH_U + \mu H_U H_D.
  $$ 
One can similarly introduce two flavons with opposite flavour charge,
$\varphi_+$ and $\varphi_-$. 
In general they have different vev.
Writing and minimizing explicit supersymmetric potentials,
one can see that often one of the two flavons gets negligible vev.
This means that\eq{SMflavon} has to modified allowing for $\lambda_+\neq \lambda_-$:
the sign of $q_{\cal O}$ becomes important.

\begin{table}
$$
\begin{array}{ccccc}
\hline 
\hbox{U(1) charges} &\hbox{anarchy} & \hbox{small $\theta_{13}$} & \hbox{normal hierarchy} & \hbox{inverted hierarchy}\\ \hline
q_{Q_i}=q_{U_i}=q_{E_i}       & 3,2,0  & 2,1,0  & 3,2,0  & 3,2,0 \\
q_{L_i}=q_{D_i} &  0,0,0& 1,0,0 & 2,0,0&  1, -1,-1\\
q_{N_i}         & 0,0,0 & 2,1,0 & 1,-1,0 & -1,1,0 \\
q_{H_{\rm u}},q_{H_{\rm d}} & 0,0 & 0,0 & 0,0 & 0,0\\ \hline
R=\Delta m^2_{\rm sun}/\Delta m^2_{\rm atm} & {\cal O}(1)  & {\cal O}(1)  & {\cal O}(\lambda^4)  & {\cal O}(\lambda^2) \\
\theta_{13}& {\cal O}(1)  & {\cal O}(\lambda)  & {\cal O}(\lambda^2)  & {\cal O}(1) \\
\theta_{12}& {\cal O}(1)  & {\cal O}(\lambda)  & {\cal O}(1)  & {\cal O}(1) \\
\theta_{23} & {\cal O}(1)  & {\cal O}(1)  & {\cal O}(1)  & {\cal O}(1) \\
\end{array}$$
  \caption[Sample of U(1) see-saw models]{\label{tab:models}\em Sample of {\rm SU(5)}-unified see-saw models with 
 realistic {\rm U(1)} flavour symmetries.  
We here aim at simplicity, rather than at fully reproducing quark and lepton masses and mixings up to ${\cal O}(1)$ 
uncertainties.
Although in the second model $\theta_{12}$ is typically small, the same accident that gives $R\ll1$ 
can also give $\theta_{12}\sim 1$.}
\end{table}

\bigskip

One can easily invent assignments of flavour charges under an U(1)$_{\rm F}$ flavour group that
qualitatively reproduce all masses and mixings
of charged leptons and quarks. 
Table~\ref{tab:models} exemplifies four U(1) flavour models that roughly reproduce 
four neutrino patterns possibly suggested by present data: i) anarchy; ii) small $\theta_{13}$;
iii) small $\Delta m^2_{\rm sun}\ll \Delta m^2_{\rm atm}$;
iv) inverted hierarchy.
These models are all compatible with SU(5) unification.

It is interesting to discuss  how anarchy or quasi-anarchy in neutrinos is easily compatible
with the hierarchical structure in charged leptons and quarks within SU(5) unification.
Concerning e.g.\ anarchy one can just assume that the flavor U(1) symmetry does not distinguish
the $\bar 5$ multiplets that contain neutrinos (such that they have equal U(1) charges, e.g.\ 
$q_{L_i} = q_{D_i}=q_{\bar{5}_i} = \{0,0,0\}$)
and assign appropriate charges to the $10_i$ multiplets, such that the relevant mass matrices have the form:
$$ m_\nu\propto {\cal O}\pmatrix{1&1&1\cr1&1&1\cr1&1&1},\qquad
m_E\sim m_D^T \propto{\cal O}\pmatrix{\lambda^3 & \lambda^3 & \lambda^3\cr\lambda^2&\lambda^2&\lambda^2\cr 1&1&1},\qquad
m_U\propto {\cal O}\pmatrix{\lambda^6 & \lambda^5 & \lambda^3\cr\lambda^5&\lambda^4&\lambda^2\cr \lambda^3&\lambda^2&1}$$
which reproduces reasonably well the order of magnitude of
all observed masses and mixings, thanks to the experimental fact that 
up quarks have a larger hierarchy than down quarks and leptons.

The drawback of these models is that they can reproduce the
qualitative aspects of all possible patterns up to  ${\cal O}(1)$ factors, 
but they do not make predictions, nor prefer one specific pattern.

\section[Unified flavour models and $\nu$ masses]{Unified flavour models and neutrino masses}\label{nu:gut}
Gauge-unified extensions of the SM are considered more appealing and
potentially more predictive also in the flavour sector.
One can say that the indication that neutrinos are massive
is one of the successes of SO(10) \cite{so10} and other GUTs.
In certain models, the scale of right handed 
neutrinos is connected to the unification scale, 
and the texture of neutrino masses  to the texture of charged fermion masses. 
Right-handed neutrinos (or scalar triplets) permit to implement 
leptogenesis (section~\ref{leptogenesi}) and, in supersymmetric GUT, 
leptonic flavor transition such as $\mu\to e\gamma$ can be detectably large (section~\ref{LFV}).

We concentrate the discussion on a few selected points:
in section \ref{sm:ana} we outline 
features the SM that permit to appreciate GUT better. 
SU(5) models are discussed in section~\ref{sm:su5}.
In section \ref{sm:so10} we concentrate on SO(10): we describe the situation of a simple  
model and offer a brief overview of other models.

\subsection{SM and GUTs: common aspects and differences}\label{sm:ana}
It is natural to consider the SM as the prototype gauge theory from where
we can start to give a meaning to fermion masses. The
features of the SM that are relevant to the present discussion are:
\begin{enumerate}
\item In the SM there is only one massive parameter, the scale of $\SU(2)_L\otimes{\rm 
U}(1)_Y$ symmetry breaking; the masses of all particles (gauge bosons, 
higgs particle, quarks and leptons) are proportional to this unique scale. 
\end{enumerate}
This first property is lost when the SM is extended adding
right-handed neutrinos (or the other non-chiral particles, that can mediate
tree-level neutrino masses).
While the same situation holds in SU(5)-unified extensions of the SM,
in other cases the high scale behind neutrino masses
is related to a gauge scale.
For example, the mass of right-handed neutrinos breaks the
${\rm U}(1)_{B-L}$ and/or $\SU(2)_R$ symmetries
contained in Left-Right, Pati-Salam or SO(10) models:
$$ \SU(3)_c\otimes\SU(2)_L\otimes\SU(2)_R\otimes{\rm U}(1)_{B-L}
\subset \SU(4)_c \otimes \SU(2)_L\otimes\SU(2)_R\subset {\rm SO}(10).$$
In these models all quark, lepton and neutrino masses can be controlled
by gauge-symmetry breaking scales.
\begin{itemize}
\item[2.] The SM contains are 3 replic\ae{} of the 5 types of fermions 
$Q,U,D,L,E$ that form a family. The kinetic terms of the fermions have 
a global symmetry  U(3)$^5$, 
called the {\em flavor group}.
\end{itemize}
The structure of the flavour group simplifies in unified models. 
For instance, the existence of right handed neutrino in a family
becomes a necessity in SO(10) where all fermions of a family
(including $\nu_R$) fit in the 16-dimensional spinorial representation. 
The flavor group of SO(10) is just U(3).

\begin{itemize}
\item[3.] In the SM fermion masses break explicitly the U(3)$^5$ flavour group, but
some U(1) subgroups remain unbroken: lepton flavour $L_e$, $L_\mu$, $L_\tau$
and $B-L$.
\end{itemize}
Both neutrino masses and gauge unification break lepton flavour.

%

\index{SU(5)}
\subsection{SU(5) and fermion masses}\label{sm:su5}
 In {SU(5)} models, 
the SM fermions are unified in
$\bar 5_i = D_i\oplus L_i$ and 
$10_i = Q_i\oplus U_i\oplus E_i$ SU(5) multiplets,
while right-handed neutrinos remain gauge singlets.
Therefore  $L$ and $D$ have the same U(1)$_{\rm F}$ charge,
at least in simpler models;
in general, U(1)$_{\rm F}$ charges can be shifted by SU(5)-breaking effects.
Extra assumptions are needed to get results.
E.g.\ assuming that Higgs doublets lie in a $5$ representation,
in the limit of unbroken {\rm SU(5)} one gets 
{\color{blu} $\lambda_D = \lambda_E^T$},
which roughly resembles the observed pattern.
Among the consequent predictions for the Yukawa couplings renormalized at the GUT scale
$$\lambda_\tau = \lambda_b,\qquad \lambda_\mu = \lambda_s,\qquad\lambda_e = \lambda_d$$
the first one can be acceptable in supersymmetric SU(5), 
while the lighter fermion masses rather satisfy
$\lambda_\mu \approx 3 \lambda_s$ and $\lambda_e \approx \lambda_d/3$.
These two 3 factors can be nicely justified in non minimal SU(5) models~\cite{GY}
built in such a way that (i) $\lambda_D^{11} = \lambda_E^{11} = 0$;
(ii) SU(5)-breaking effects generate $\lambda_D^{22}=-N_C\lambda^{22}_E$
(where $N_C = 3$ is the number of colors).
The 12 element of $\lambda_D$ gives a contribution 
$\sqrt{m_d/m_s}$ to the Cabibbo angle, which happens
to be close to the measured value.
We can therefore expect that the corresponding 12 element in $\lambda_E$ is diagonalized by a 12 rotation with angle
\beq\label{eq:GYnu}\Delta\theta_{12} = \sqrt{m_e / m_\mu}.\eeq
Combined with the (unknown) matrix that diagonalizes the neutrino mass matrix,
this often results into a contribution to the physical $\theta_{13}$ mixing angle.

\medskip

Let us now discuss the compatibility of neutrino masses and mixings with SU(5) unification.
It is difficult put not impossible
to invent flavour models compatible with SU(5), 
if one thinks that a flavour model should explain the smallness of $\theta_{13},R=\Delta m^2_{\rm sun}/\Delta m^2_{\rm atm}$.
The reason is that $R,\theta_{13}$ are much larger than analogous
mass ratios and mixing angles in quarks and leptons.
Table~\ref{tab:models} shows a few examples.
If instead one thinks that the smallness of $\theta_{13}$ and $R$ does
not require a dedicated interpretation and
accepts a `anarchical' neutrino mass matrix, this is immediately married with SU(5):
one just assumes that the flavour symmetry acts equally on $\bar{5}_{1,2,3}$
and assigns appropriate flavour charges to $10_{1,2,3}$. 
The first column of table~\ref{tab:models} shows one example.
A qualitative pattern emerges: 
no mass hierarchy among neutrinos and a strong mass hierarchy among up-quarks
implies (in agreement with data) an intermediate mass hierarchy among down-quarks and 
among charged leptons and relatively small quark mixing angles.

In conclusion, SU(5) looks easily compatible with neutrino data but does not suggest precise predictions.
Roughly the opposite happens in the SO(10) case.


\index{SO(10)}
\subsection{SO(10) and fermion masses}\label{sm:so10}
Right-handed neutrinos become massive after that SO(10) 
(and in particular its subgroups U(1)$_{B-L}$ and $\SU(2)_R$)
break down to the SM gauge group.
The right-handed neutrino mass operator has the form
$16~16~\langle \overline{126}\rangle$,
where the $\overline{126}$ could be either an elementary Higgs or
a composite field (e.g.\ the product of two $\overline{16}$).

The simplest SO(10) models, with the Higgs entirely contained
in the minimal 10 representation,
predict $\lambda_N=\lambda_E = \lambda_D = \lambda_U$
and symmetric Yukawa matrices,
in qualitative contrast with the observed data:
up-quarks exhibit a stronger mass hierarchy than leptons and down-quarks,
neutrinos have large mixing angles while quarks have small mixings.

This unwanted prediction can be avoided by assuming that
Yukawas arise from effective higher dimensional operators containing SO(10)-breaking vevs~\cite{otherso10},
but going in this direction does not make full use of SO(10).
A SU(3) flavour model based on rank-1 matrices 
has been proposed and studied in detail 
in the last paper of \cite{otherso10}.
These models tend to give relatively small values of $\theta_{13}$.


An alternative road, that exploits the SO(10) symmetry, consists in assigning 
the Higgs doublet to non-minimal SO(10) representations~\cite{so10susy}:
10 and ${\overline{126}}$.
One often prefers to build a supersymmetric model, such that also a  ${126}$ is needed.\footnote{The $\overline{126}$ has other
advantages.  Its vev directly controls right-handed neutrino masses, and
preserves the matter parity, often introduced in supersymmetric models
to avoid unseen effects. \label{foo:nro}}
Restricting to renormalizable couplings only the superpotential contains two Yukawa matrices
$$
\mathscr{W} = \lambda_{ij}^{10} 16_i 16_j 10 + \lambda_{ij}^{126} 16_i 16_j \overline{126}
$$
that generate the quark and lepton Yukawa matrices as
\begin{equation}\begin{array}{rclrcl}
\lambda_U &=& \lambda^{10} \cos\alpha_U + \lambda^{126} \sin\alpha_U \qquad&
\lambda_D &=& \lambda^{10} \cos\alpha_D + \lambda^{126}\sin\alpha_D \\
\lambda_E &=& \lambda^{10} \cos\alpha_D -3 \lambda^{126}\sin\alpha_D &
\lambda_N &=& \lambda^{10} \cos\alpha_U -3 \lambda^{126}\sin\alpha_U 
\end{array}
\end{equation}
where $\alpha_D,\alpha_U$  are complex angles that parametrize how the light Higgs is contained
in the {10}, ${126}$ representations.
Right-handed neutrino masses are proportional to $\lambda^{126}_{ij}$ and 
to $\langle\overline{126}\rangle$, such that all flavour matrixes 
are predicted in terms of two flavour matrices and a few other parameters.


The couplings to the ${10}$ lead to the adequate expectation of  
$b/\tau$ unification, the couplings to the  $\overline{126}$ permit to give large Majorana
neutrino masses to the right handed neutrinos and to correct
for the strict $\lambda_D=\lambda^T_E$ equality for all 3 families.
This model offers the hope to reproduce the observed 
fermion masses and mixing. All existing fits based on
such a model agree to predict that $\theta_{13}$ should be rather large.
However, at this stage, the model is not complete,
since the  ${\overline{126}}$ Higgs can spontaneously break
SO(10) only down to SU(5).
Adding just another higgs field (the ${{210}}$) allows to break SO(10) to
the SM gauge group  and
to produce the desired mixed composition of the MSSM higgs 
fields (i.e. $\alpha_{U,D}\neq 0$).\footnote{This concludes the definition of the model, 
that has been termed `minimal' due to the small number of higgs fields
and of free supersymmetric parameters (26).
The less appealing points of this model are the fact that 
non-renormalizable terms are ignored;
the scale where non-perturbativity  is lost is rather close to the grand unification scale;
simplicity (and simple string models) prefer smaller SO(10) representations;
no special  mechanism explains the lightness of the higgs doublet (apart from a suitable fine-tuning of parameters);
some fine-tunings among $\lambda^{10}$ and $\lambda^{126}$ are needed to reproduced fermion masses.
None of these objections seems to point to a real contradiction, and in
our view they are to a certain extent counterbalanced by the fact that
this model does not need any flavour symmetry,
is predictive for  proton decay, leptogenesis and lepton flavor violation as well.} 
The existing fits of fermion masses 
have been performed allowing 
complete freedom in the choice of higgs composition and of the 
neutrino mass scale. 
However, it seems that this freedom does not 
exist in the full model with the ${{210}}$.
More specifically, it seems possible to have the neutrinos with masses 
larger than $v^2/M_{\rm GUT}$ only in certain points 
of the parameter space where: 
1) charged fermion masses cannot be reproduced;
2) supersymmetric unification of gauge couplings fails.
Even if  deeper investigations will confirm that this is real conclusion,
this model remains an example of how SO(10) allows to obtain predictions.



\section{Testing flavour models}\label{FlavourTest}
The true problem is how to test speculations about flavour symmetries.
There are two possibilities:\begin{itemize}
\item[1)] {\bf Direct tests}: the model predicts some new physical process.
However flavour data agree with SM predictions, pushing the flavour scale $\Lambda$
to so large energies that direct tests seem not feasible.
In fact, U(1)$_{\rm F}$ flavour symmetries allow operators like
$$\frac{c_1}{\Lambda^2}  |D_1 Q_1|^2+\frac{c_2}{\Lambda^2} |D_2 Q_2|^2$$
that therefore are expected to appear with coefficients $c_i\sim 1$.
Rotating the $Q,D$ fields 
by an angle $\theta_{\rm C}\sim 0.2$
one reaches the mass eigenstate basis, where
the above operators contains a term
$\sim {\theta_{\rm C}^2 (c_1 - c_2)} |s\, d|^2/\Lambda^2$
that contributes to mixing and CP-violation in the $K^0\bar{K}^0$ system,
such that $\Lambda\circa{>} 10^3 \TeV$.
Appropriate unbroken non-abelian symmetries like U(2) could force $c_1 = c_2$, avoiding
this unseen new-physics effect. However: 
\begin{itemize}
\item[a)]  flavour symmetries must be broken; 
spontaneous breaking generates light Goldstone bosons
(that become longitudinal polarizations of massive gauge bosons,
in case the flavour symmetry is gauged)
that in the non-abelian case mediate unseen flavour-changing processes.

\item[b)]  no flavour symmetry compatible with the electron mass term $m_e~ELH^*$
can forbid the operator $c\cdot  m_e(E \gamma_{\mu\nu} F_{\mu\nu} L)H^*/\Lambda^2$
that would induce an unobserved electric dipole for the electron.
This implies $\Lambda\circa{>}10^2\TeV$, and an analogous
constraints comes from the neutron electric dipole.
(Of course the situation can be better in specific models, e.g.\ if
CP is broken in some appropriate way such that $c$ is real).

\end{itemize}
In many models the flavour scale $\Lambda$ is identified with the unification or Planck scales.

\item[2)] {\bf  Indirect tests}: the model predicts a relation between the SM parameters.
The predictive power of a model can be na\"{\i}vely estimated as
$$\hbox{number of predictions} \approx \hbox{number of SM parameters}-\hbox{number of parameters} $$
and is usually negative.
This estimate is sometimes too pessimistic:
some parameters of the model could be irrelevant because small (e.g.\ $13$ entries of Yukawa matrices),
or could not affect all observables (e.g.\ lepton and quark mass matrices could be two separate sectors).

\end{itemize}

\medskip

\medskip

\subsection{Predictions for $\theta_{12}$}
The solar neutrino mixing angle is measured to be large but not maximal.
Some  non-trivial information might be contained in  $\theta_{12}$,
and several speculations have been invoked to explain its  value.
We here review precise quantitative predictions,
that can be tested by future more precise measurements of $\theta_{12}$.
One should keep in mind that these predictions contain a significant amount of
postdiction:
initially, most models predicted a maximal $\theta_{12}=\pi/4$, and
after that experiments showed that $\theta_{12}$ is less than maximal
it was possible to construct models that explain it.
\begin{enumerate}
\item $\theta_{12}=35.3^\circ$.
{\em Tri-bi-maximal} neutrino mixing is the most popular proposal.
It is based on the idea that $\theta_{13}=0$, $\theta_{23}=\pi/4$, $\tan^2\theta_{12}=1/2$
are compatible with present data and correspond to the following clean structure of neutrino mass eigenstates:
\beq \label{eq:32}\nu_3 = \frac{\nu_\tau - \nu_\mu}{\sqrt{2}},\qquad
\nu_2 = \frac{\nu_e + \nu_\mu+\nu_\tau}{\sqrt{3}}\eeq
Models that try to explain tri-bi-maximal mixing in terms of {broken} flavour symmetries
have been realized using the A$_4$ and S$_4$ discrete groups~\cite{A4}.
Indeed\eq{32} is invariant under $\nu_i\to P_{ij}\cdot\nu_j$
(where $P$ is the $\mu\leftrightarrow \tau$ permutation matrix of eq.\eq{mutau})
as well as under $\nu_i \to S_{ij} \cdot \nu_j$
where $S$ is a $3\times 3$ matrix with diagonal elements $-1/3$ and off-diagonal elements $2/3$.
The diagonal charged lepton mass matrix $m_E^\dagger m_E$ is invariant under $L_i\to T_{ij}L_j$
with $T = \diag(1,e^{2\pi i/3},e^{4\pi i/3})$.
The transformations $S,T$ obey $S^2 = T^3 = (ST)^3=1$ which defines the discrete
tetrahedral A$_4$ group~\cite{A4} (even permutations of 4 elements): 
it has representations with dimensions 2 and 3,
and, more importantly, three inequivalent singlet representations.
A$_4$ can be broken in a way that naturally  implies `tri-bi-maximal mixing' in some leading-order approximations.
The transformations $S,T,A$ define the group S$_4$ (permutations of 4 elements) which also
found use in model building.
Deviations that lead to $\theta_{13}\neq 0$ are not predicted, unless one
assumes that some particular source dominates.

\item $\theta_{12}= (31.9\pm0.1)^\circ$.
{\em Quark/lepton complementarity}~\cite{sunCabibbo} is the observation that
$\theta_{12} = \pi/4 - \theta_{\rm C}$ provides an acceptable
value of $\theta_{12}$. It is not clear why the Cabibbo angle $\theta_{\rm C}$ should have 
something to do with neutrino physics.
Even in the 23 sector the sum of the (small) quark mixing plus the atmospheric mixing can be
maximal. 

\item $\theta_{12}=31.7^\circ$.
This {\em golden prediction}~\cite{Golden}
$\cot\theta_{12}=\varphi$, where $\varphi=(1+\sqrt{5})/2$ is the golden ratio,
follows from the simple texture 
\beq
m_\nu = \left(
\begin{array}{ c  c c}
0 &m&0 \\
m &m&0\\
0&0&m_{\rm atm}
\end{array}
\right), \;\;\;\;\;\;
\lambda_E = \pmatrix{\lambda_e & 0 &0 \cr
0& \lambda_\mu/\sqrt{2} & 
\lambda_\mu/\sqrt{2}\cr
0& -\lambda_\tau/\sqrt{2} & 
\lambda_\tau/\sqrt{2}}.
\eeq
A similar prediction might hold in the quark sector, $\cot\theta_C =\varphi^3$,
giving rise to quark/lepton complementarity, $\theta_{12}+\theta_C = \pi/4$.


\item $\theta_{12}=36^\circ=\pi/5$, that can be obtained from the dihedral group $D_{10}$~\cite{Rodejohann}.

\item $\theta_{12}=30.4^\circ$, that follows from the tetra-maximal ansatz~\cite{4maximal}:
\beq\label{eq:4maximal}
V=R_{23}(\pi/4,\pi/2)\cdot R_{13}(\pi/4)\cdot R_{12}(\pi/4)\cdot R_{13}(-\pi/4)\eeq
where $R_{ij}$ is defined below eq.\eq{Vunitary}  and
 $R_{23}(\theta,\phi)\equiv \diag(1,1,e^{i\phi})\cdot R_{23}(\theta)\cdot \diag(1,1,e^{-i\phi})$.
\end{enumerate}


\subsection{Predictions for $\theta_{13}$}\label{nu:t13}
There are two qualitative reasons to expect that $\theta_{13}$ is detectably large.
First, adding to the hierarchical texture\eq{textureh} small perturbations 
that generate the solar $\nu_e/\nu_{\mu,\tau}$ mixing and mass
typically generates also a contribution to $\theta_{13}$ of order
\beq \theta_{13}\sim \theta_{\rm sun} \sqrt{R}\sim 0.1
\qquad\hbox{where}\qquad R\equiv\frac{\Delta m^2_{\rm sun}}{ \Delta m^2_{\rm atm}}.
\eeq
Of course the most generic perturbation gives
the most generic hierarchical mass matrix, such that
there are no generic predictions for $\theta_{13}$.
However a $\theta_{13}$ much smaller than 0.1
needs a cancellation,
which should be considered as unlikely unless
it follows from some underlying symmetry reason.

Second, SU(5) unification suggests that $\lambda_E$ is non diagonal,
and possibly leads to the Cabibbo-like $e/\mu$ angle in eq.\eq{GYnu}.
Combined with the roughly maximal $\nu_\mu/\nu_\tau$ atmospheric mixing,
it induces a contribution to $\theta_{13}$ of order~\cite{GY}
$$\theta_{13}\approx \sqrt{\frac{m_e}{2m_\mu}} \approx  0.05.$$

Since these qualitative  predictions are not constraining enough,
in order to make real progress we need testable quantitative {\em predictions}.
With the exception of neutrinos, all SM flavour parameters have already been measured, 
and we do not expect a significant reduction of the experimental errors in the near future.
Future models will make {\em postdictions}, that unlike predictions are never badly 
wrong.  Although this is not common practice we here ignore postdictions.\footnote{We recall
that many models that now `naturally reproduce all observed data',
previously  reproduced the `Small Mixing Angle' solution to the solar neutrino problem
when it seemed favored.
Similarly, all competitors of the SM were able of postdicting the photon;
what made the difference is that the SM correctly predicted the $Z$ boson.}

We now review concrete predictions for the not yet measured quantities:
$\theta_{13}$, $|m_{ee}|$,
deviations of $\theta_{23}$ from $\pi/4$, and the CP-violating phase $\phi$.
Since it seems likely that only $\theta_{13}$ will be precisely measured
in the close future,
we focus on models that make {\em quantitative predictions} for $\theta_{13}$.
\begin{itemize}
\item 
We ignore predictions up to unknown ${\cal O}(1)$ factors.
We ignore predictions already disfavored by data.
We ignore models that predict
some combination of $\theta_{13}$ and of CP-violating phases,
since it is not of immediate experimental interest.
\end{itemize}
Predictive models need bold assumptions.
Most models are based on `texture zeros' i.e.\ they assume that only a few
elements of the lepton mass matrices are non vanishing.
These attempts could lead us to recognize that some elements might really be
negligibly small; this would constitute a valuable step towards a theory of flavour.
However, estimates of the future experimental accuracy and of the number of possible textures
show that too many different predictions are possible:  
whatever will be the true value of $\theta_{13}$, few textures will probably be able of predicting it.
Therefore, in order to get some useful result that experiments can disproof we need to be more selective.
\begin{itemize}
\item We focus on predictive models which look more attractive,
either because `theoretically motivated' or because explain the
observed smallness of $\theta_{13}$ and of
$R=\Delta m^2_{\rm sun}/ \Delta m^2_{\rm atm}$ in a `natural' way,
without a fine-tuning of the parameters.
\end{itemize}
This last criterium could be misleading and is certainly subjective,
but we do not know any better attempt.
Here is a list of attempts that satisfy to these requirements.
In almost all cases
analogous predictions with $\theta_{23}\to \pi/2- \theta_{23}$ are obtained by
building analogous models with $\mu\leftrightarrow\tau$ replaced.
For the moment the atmospheric mixing is consistent with maximal
and the two kind of possibilities cannot be distinguished.\footnote{
The predictions quoted in this section 
have been computed applying standard propagation of errors
to the simplified data in table~\ref{tab:tab1}
(in order to avoid singular Jacobians we use with $\theta_{23} = \pi/4 \pm 0.06$
in place of $\sin^2 2\theta_{23}$).
Within a factor of 2 this fast approximation agrees with a careful marginalization
of the joint probability density.}
We follow the notations of section~\ref{Neutrino}: 
$\lambda_N$ ($\lambda_E$) is the matrix of neutrino (charged lepton)
Yukawa couplings of see-saw models, 
$m_\nu$ ($M_N$) the symmetric mass matrix of left-handed (right-handed) neutrinos.

\begin{figure}
$$\includegraphics{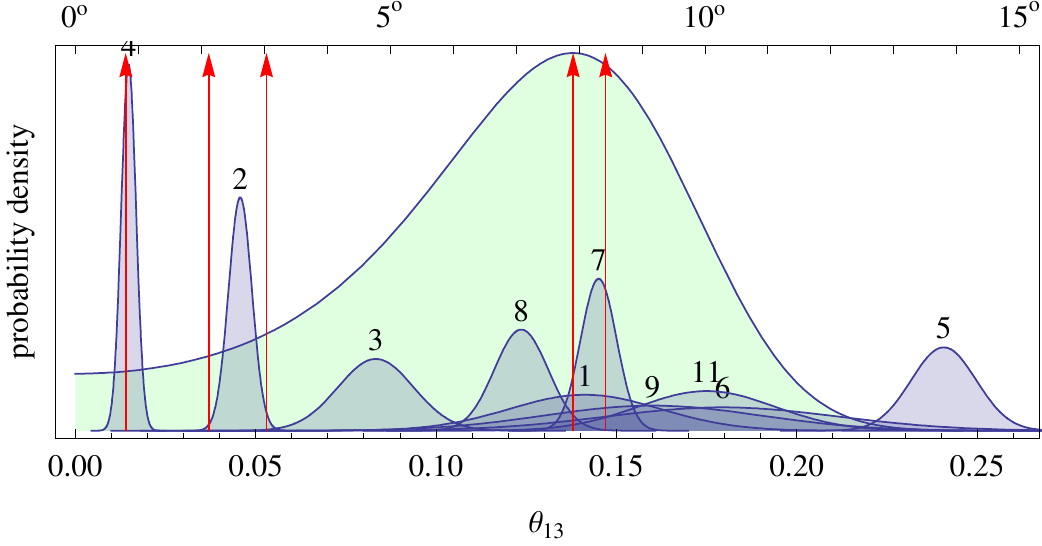}$$
\caption[Expectations from anarchy]{\label{fig:anarchy}\em {\bf $\theta_{13}$ prediction contest}.
Each Gaussian bell is one prediction for $\theta_{13}$, as listed in section~\ref{nu:t13} with the number written over the bell.
Widths are determined according to uncertainties; predictions with negligible uncertainty are plotted as arrows.
The winner model will be the one with the highest overlap with the experimental probability of $\theta_{13}$,
plotted as a green curve. Models that make no predictions are ignored.}
\end{figure}

\begin{enumerate}
\item 
Assuming 
$$
\lambda_E = \pmatrix{*&0&0\cr 0&*&0\cr 0&0&*} \qquad\hbox{and}\qquad
m_\nu = \pmatrix{0 & 0 & * \cr 0&*&*\cr *&*&*}$$
(where `0' denotes a vanishing entry and
`*' a generic non-vanishing entry) gives~\cite{predsA}.
\begin{equation}\label{eq:predsA}
\theta_{13}\simeq \frac{\tan2\theta_{12}}{2}\tan\theta_{23}(R\cos2\theta_{12})^{1/2}=0.135\pm0.02.\end{equation}
Such pattern does not need an underlying see-saw model,
but can be realized from the following see-saw texture:
$$ \lambda_N = \pmatrix{ * &0&0\cr 0&*&0\cr 0&0&*},\qquad
M_N = \pmatrix{*&*&*\cr *&*&0\cr *&0&0}.$$

\item Assuming
$$ m_\nu = \pmatrix{0&*&0\cr *&0&*\cr 0&*&*},\qquad
\lambda_E = \pmatrix{*&0&0\cr 0&*&*\cr 0&0&*}$$
gives
\begin{equation}\label{eq:predsB}
\theta_{13}\simeq \frac{\tan2\theta_{12}}{2}(R\cos2\theta_{12})^{3/4}=0.044\pm0.003.
\end{equation}
This can be realized as a see-saw texture:
$$ \lambda_N = \pmatrix{ * &0&*\cr 0&*&0\cr 0&0&*},\qquad
M_N = \pmatrix{0&* &0\cr *&0&0\cr 0&0&*}.$$

\item There is one `most minimal' see-saw texture not reducible to a texture for $m_\nu$.
Assuming two right-handed neutrinos with
$$ \lambda_N = \pmatrix{ * &*&0\cr 0&*&*},\qquad
M_N = \pmatrix{* &0\cr 0 & *},\qquad
\lambda_E = \pmatrix{*&0&0\cr 0&*&0\cr 0&0&*} $$
gives~\cite{minimalseesaw}: 
\begin{equation}\label{eq:predsX}
\theta_{13}\simeq
\frac{\sqrt{R}}{2} \sin2\theta_{12}\,\tan\theta_{23}=0.083\pm0.010
,\qquad m_{ee} = m_{\rm sun}\sin^2 \theta_{12}.
\end{equation}
\item In the previous case, one can instead assume $M_N=0$ and get Dirac neutrino masses with
$\theta_{13}\simeq (R/2) \sin 2\theta_{12}\tan\theta_{23}\simeq 0.015\pm 0.002$.


\item A different prediction is obtained by assuming an alternative form for 
the right-handed neutrino mass matrix~\cite{minimalnonFT}:
\beq M_N = \pmatrix{* &*\cr * & 0}\qquad\hbox{gives}\qquad
\theta_{13} \simeq R^{1/4} \sin\theta_{12}=0.224\pm0.013.\eeq
This prediction can be alternatively obtained from a $M_N$ with
vanishing diagonal entries, combined with a
non diagonal $\lambda_E$.

\item One can easily invent simple neutrino mass matrices
diagonalized by
$R_{23}(\theta_{23})R_{12}(\pi/4)$, see e.g.\ eq.\eq{texturei}.
Since experiments excluded maximal solar mixing,
this cannot be the only contribution to the neutrino mixing matrix.
Diagonalization of the charged lepton mass matrix generically gives another contribution.
Assuming that it only gives a 12 rotation, with angle $\Delta\theta_{12}$ results into
$V=R_{12}(\Delta \theta_{12}) \cdot R_{23}(\theta_{23})R_{12}(\pi/4)$.
When rewritten in the standard parametrization\eq{Vunitary}, such $V$ corresponds
to having $\theta_{13}\neq \pi/4$ and
$\theta_{13}\neq0$ related by~\cite{13nuell}
\beq\label{eq:pred13nuell}
\sin \theta_{13}=\tan\theta_{23}\cdot \tan(\theta_{12}-\pi/4) = 0.20\pm 0.04.\eeq

\item A texture that assumes some texture zeroes and some
strict equalities among non vanishing entries
predicts~\cite{HARRISON}
\beq\sin\theta_{13}=\sqrt{2R/3} = 0.145\pm 0.006
.\eeq

\item Another texture that assumes some texture zeroes and strict equalities among non vanishing entries~\cite{0606142} predicts
\beq \theta_{13}=\sqrt{R} \tan\theta_{12}= 0.120 \pm 0.008.\eeq

\item Assuming that the product of the neutrino mixing matrix times the quark mixing
matrix has a zero in the 13 entry implies $\theta_{13}=0.16^{+0.01}_{-0.04}$~\cite{sunCabibbo}.

\item Tetra-maximal mixing~\cite{4maximal} predicts $\theta_{13}=0.147$.

\item  A `$T'$' model~\cite{FramptonT'} predicts $\theta_{13}=0.175\pm0.019$.
\end{enumerate}
We now list predictions for $\theta_{13}$ coming from models or textures  based on SO(10). 
  These models typically can reproduce all quarks and lepton masses
  and mixings in terms of a restricted set parameters, that often need to be fine-tuned.
The predictions are  $\theta_{13}\simeq 0.037$~\cite{BO},
$\theta_{13} \simeq 0.014$~\cite{AlBarr},
$\theta_{13}\simeq 0.138$~\cite{Moha05},
$\theta_{13}\simeq m_e m_\mu/\sqrt{2}m_\tau^2= 1.1~10^{-5}$~\cite{Matsuda},
$\theta_{13}\simeq \theta_{\rm C}/3\sqrt{2}=0.053$~\cite{Kingtheta13}.

%% file: review_behind.tex
\chapter{Behind neutrino masses?}\label{extra}
We here discuss possible speculative new phenomena suggested or possibly
related to neutrino masses. Observing extra phenomena might allow us to identify 
which new physics generates the observed neutrino masses.
Effective-Lagrangian arguments suggested that the main  low-energy
manifestation of heavy new physics that violates lepton number
is the Majorana neutrino mass operator:
neutrinos become massive and other neutrino properties are negligibly affected.
Experiments are now confirming this view, by indicating that the new physics responsible of the
solar (section~\ref{sun}) and atmospheric (section~\ref{atm}) anomalies is neutrino oscillations.
New particles lighter than about a TeV
can evade the above expectation giving rise to new phenomena.
\begin{itemize}
\item In section~\ref{MagneticMoment} we discuss the phenomenology of the main subleading
effective operator: {\em neutrino magnetic moments}.
\end{itemize}
Some specific kinds of new particles can be so light that  direct searches are possible.
Studying the behavior of the existing light particles
(photons, gravitons and neutrinos) allows to probe the possible existence of
new light particles.
We here review what can be done with neutrinos.
They are sensitive to:
\begin{itemize}
\item New {\em light neutral fermions} can naturally interact only with neutrinos,
behaving as  `sterile neutrinos', and giving extra oscillation effects as discussed in section~\ref{Sterile}.

\item 
New {\em light neutral scalars}  can interact with neutrinos,
provided that new light neutral fermions also exist.
As discussed in section~\ref{Scalars} the main signals are 
a modified neutrino cosmology, neutrino decay, new matter effects.


\end{itemize}
Furthermore, we discuss two specific possibilities of new physics motivated by solutions to the hierarchy problem:
\begin{itemize}
\item In {\em supersymmetric see-saw models} quantum corrections induced by neutrino Yukawa couplings
affect slepton masses and might result in detectably large rates for  $\mu\to e\gamma$ or other lepton-flavour-violating processes (section~\ref{LFV}).
In section~\ref{RGE} we discuss quantum corrections to neutrino masses.

\item In section~\ref{extradims} we discuss how (after forbidding in some way Majorana masses)
{\em neutrinos in extra dimensions} can get small Dirac masses with potentially unusual phenomenology.
\end{itemize}

\section{Neutrino electro-magnetic dipoles}\label{MagneticMoment}\index{Magnetic dipole}
\index{Electric dipole}
As discussed in section~\ref{Neutrino},
 neutrinos might have Majorana and/or Dirac masses.
An analogous distinction arises for their electro-magnetic moments.
In general, neutral fermions $\psi_i$ can interact with photons via electro-magnetic moments $\mu_{ij}$,
described by the following dimension 5 Lagrangian operator
\beq\label{eq:mu}\sum_{ij} \mu_{ij}\,[ \psi_i \gamma_{\mu\nu}\psi_j] F^{\mu\nu} + \hbox{h.c.}\eeq
where $\psi_i$ are Weyl fermions and $\mu_{ij} = -\mu_{ji}$ has dimensions mass$^{-1}$.
For $i=j$ one has $\mu_{ii}=0$.

\begin{figure}
\begin{center}
\includegraphics[width=\textwidth]{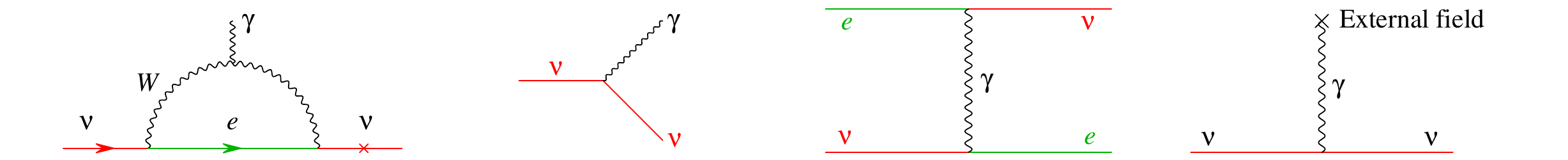}
\caption[BSM neutrino effects]{\label{fig:MagneticMoment}\em 
(a) Neutrino masses induce a magnetic moment.
(b) Neutrino decay.
(c) Neutrino/electron scattering.
(d) Interaction with an external magnetic field.}
\end{center}
\end{figure}

\begin{itemize}
\item Neutrinos might have only left-handed polarizations.
In such a case neutrinos can only have $\Delta L=2$ Majorana masses (section~\ref{Majorana}).
Analogously, neutrinos can only have $\Delta L=2$ `Majorana-like' 
flavour-violating
electro-magnetic dipoles
$\mu_{e\mu}$, $\mu_{e\tau}$, $\mu_{\mu\tau}$, as 
can be seen inserting $\psi_i = \{\nu_e,\nu_\mu,\nu_\tau\}$
in eq.\eq{mu}.

\item Neutrinos might have both left and right-handed components.
In such a case neutrinos can have 
$\Delta L=2$ Majorana masses of $LL$ type,
$\Delta L=2$ Majorana masses of $RR$ type,
and $\Delta L=0$ Dirac masses of $LR$ type.
Only Dirac masses are allowed if lepton number is imposed (section~\ref{Dirac}).

Analogously, there can be $LL$, $RR$ and $LR$ 
electro-magnetic dipoles. Indeed the $6\times 6$ asymmetric matrix $\mu_{ij}$
can be
decomposed into three $3\times 3$ sub-matrices as
\beq\mu_{ij} = \pmatrix{\mu^{LL}_{\ell\ell'} & \mu^{LR}_{\ell\ell'}\cr - \mu^{LR}_{\ell\ell'} & \mu^{RR}_{\ell\ell'}}\qquad
\hbox{in the basis}\qquad
\psi_i = \{\nu_\ell^L,\nu^R_\ell\},\qquad
\ell=\{e,\mu,\tau\}.\eeq
If lepton number is imposed  only the $LR$ dipoles are allowed.  
Their flavour-diagonal components $\mu_{ii}^{LR}$ do not need to vanish
and can be decomposed into
the CP-conserving magnetic dipole $ \hbox{Re}\,\mu_{ii}^{LR}$
and the CP-violating electric dipole $\hbox{Im}\,\mu_{ii}^{LR}$.
\end{itemize}
In both cases neutrino masses $m_\nu$ together with electro-weak interactions
produce tiny electro-magnetic dipoles (fig.\fig{MagneticMoment}a)~\cite{MagneticMoment}:
\beq \mu_{\nu} \sim \frac{3eG_{\rm F}}{8\sqrt{2}\pi^2} m_\nu =
{3\cdot 10^{-20}}\mu_{\rm B} \frac{m_\nu}{0.1\eV}\eeq
where $\mu_{\rm B} =e\hbar/2m_e c$ is the Bohr magneton.
The expected dipoles are about 9 orders of magnitude below the present bounds discussed below,
but new physics can give much larger dipoles.

Neutrino electro-magnetic dipoles can produce a variety of processes:
\begin{enumerate}
\item Neutrino decay  (fig.\fig{MagneticMoment}b), $\Gamma_\nu(\nu\to\nu'\gamma)\sim m_\nu^3 \mu_\nu^2$.

\item Scattering on electrons (fig.\fig{MagneticMoment}c), with 
differential cross section~\cite{MagneticMoment} 
\beq
\frac{d\sigma}{dT}(\nubarnu e \to\nubarnu e)= 
\frac{\pi\alpha_{\rm em}^2|\mu_\nu^2|}{\mu_B^2 m_e^2}\bigg(\frac{1}{T}-\frac{1}{E_\nu}\bigg)
.\eeq
The enhancement at small electron recoil energy $T$ is limited by the detection threshold
and ultimately by the presence of atomic energy levels.
Experiments have been performed with atmospheric, reactor and solar neutrinos.
Dipoles that involve $\nu_e$ are more strongly constrained:
$\color{blu}\mu_\nu <0.32~10^{-10}\mu_{\rm B}$ at 90\% C.L.~\cite{MagneticMoment}.
The next round of experiments might improve the sensitivity down to $10^{-12}\mu_{\rm B}$.

\item If neutrinos have a right-handed component, it can get populated due to
precession of the neutrino spin 
in an external magnetic field (fig.\fig{MagneticMoment}d).
The consequent cooling of supernov\ae{} and of  red-giant stars
allows to set the bound $\color{blu}\mu_\nu\circa{<} 10^{-12}\mu_{\rm B}$,
subject to sizable astrophysical uncertainties~\cite{MagneticMoment}.

\item Similarly, the combined effect of neutrino masses and of
the interaction with the magnetic fields of the sun
can convert solar $\nu_e$ into anti-neutrinos.
Global fits of solar neutrino data, supplemented  with models of the solar magnetic field,
suggest $\color{blu}\mu_\nu \circa{<} 10^{-10\div 12}\mu_{\rm B}$.
This bound does not apply if $\nu \to \bar\nu$ transitions are suppressed because 
the solar magnetic field is too weak (in the inner radiative zone)
and too turbulent (in the outer convective zone)~\cite{MagneticMoment}.

\item Magnetic moments generate contributions to neutrino masses:
by assuming that this effect is not too unnaturally large than observed neutrino masses,
one gets $\mu_\nu \circa{<} 10^{-12\div 15}\mu_B$: the stronger constraint applies if neutrinos
have Dirac masses.

\end{enumerate}


\section{Neutrinos and light fermions}\label{Sterile}\index{Neutrino!sterile?}
The SM contains fermions ($e,\nu, u,d,\ldots$)
variously charged under electric, weak, strong interactions:
fermions $\nu_R$ neutral under all SM gauge interactions
might exist without giving any observable effect in collider experiments.
The relevant terms in the SU(2)$_L$-invariant  effective Lagrangian that describes active
neutrinos $\nu$ together with extra light `{\em sterile neutrinos}' $\nu_R$ are
\beq\label{eq:Lsterile} \frac{m_{LL}}{2} \frac{(LH)^2}{v^2}+ \frac{m_{RR}}{2}\nu_R^2 + {m_{LR}}\nu_R L\frac{H}{v}  + \hbox{h.c.}\eeq
$H$ is the higgs doublet with vacuum expectation value $(0,v)$.
The first dimension-5 operator gives Majorana $\nu$ masses $m_{LL}$ and is naturally small
if lepton number is broken at a high-energy scale.
The second term gives Majorana $\nu_R$ masses $m_{RR}$,
and the third term Dirac $\nu_L\nu_R$ masses $m_{LR}$:
one needs to understand why $m_{LR}$ and $m_{RR}$ are small~\cite{Sterile}.

\subsection{Motivations}
A new light particle would probably be a discovery of fundamental importance,
because it lightness is likely related to some fundamental principle,
as it is the case for the known light particles,  the photon, the neutrinos and the graviton.
Attempts of guessing physics beyond the SM
from first principles motivate a number of fermions which might
have $\TeV^2/M_{\rm Pl}$ masses and
behave as sterile neutrinos.
A few candidates are
axino, branino, dilatino, familino, Goldstino, Majorino, modulino,
radino.
These ambitious approaches so far do not give
useful predictions on the flavour parameters in the effective Lagrangian of eq.\eq{Lsterile}.

More specific models can be more predictive.
Unification of matter fermions into SO(10) 16 or $E_6$ 27 representations 
predicts extra singlets, which however
generically receive GUT-scale masses.
It is easy to invent ad-hoc discrete or continuous symmetries that keep a fermion light.
One can use only ingredients already present in the SM.
For example, the extra fermions can be forced to be light assuming that they are chiral
under some extra gauge symmetry (that could possibly become non perturbative at some
QCD-like scale, and give composite sterile neutrinos).
Alternatively, the extra fermions may be light for
the same reason why neutrinos are light in the SM.
Following this point of view up to its extreme, one can add to the SM a set of `mirror particles',
obtaining 3 sterile neutrinos.

\bigskip

A few theoretically favored patterns emerge from rather 
general naturalness  considerations.
We consider the most generic mass matrix with $LL$, $RR$ and $LR$ mass terms.
If $m_{LL}$ dominates one obtains light sterile neutrinos with mass $m_{\rm s}\ll m_{\rm a}$ and
active/sterile mixings
$\theta_{\rm s}^2 \sim m_{\rm s}/m_{\rm a}$.
If $m_{RR}$ dominates,  sterile neutrinos are heavy with $\theta_{\rm s}^2 \sim m_{\rm a}/m_{\rm s}$.
If $m_{LR}$ dominates one obtains quasi-Dirac neutrinos that split into couples.

\bigskip

From a phenomenological point of view, sterile neutrinos have been the standard 
`emergence exit' that allowed to fit
many puzzling results in particle physics, astronomy, cosmology.
We list some open questions;
most (maybe all) of them will likely be sooner or later understood without invoking new physics.
Sterile neutrinos could be the physics behind
the LSND anomaly (section~\ref{LSND}),
the NuTeV anomaly  (section~\ref{NuTeV}),
behave as dark matter (see in~\cite{SterilePheno}),
their decays could generate the diffuse ionization of our galaxy,
can revive the shock wave that finally manifests as SN fireworks (section~\ref{subs:ccs}),
can help r-process nucleosynthesis by reducing $Y_e$
in presence of magnetic fields can generate the observed pulsar motion,
can generate anomalous time modulations in solar neutrino rates~\cite{Pee(t)}.

\begin{figure}[p]
$$\hspace{-7mm}\includegraphics[width=18cm]{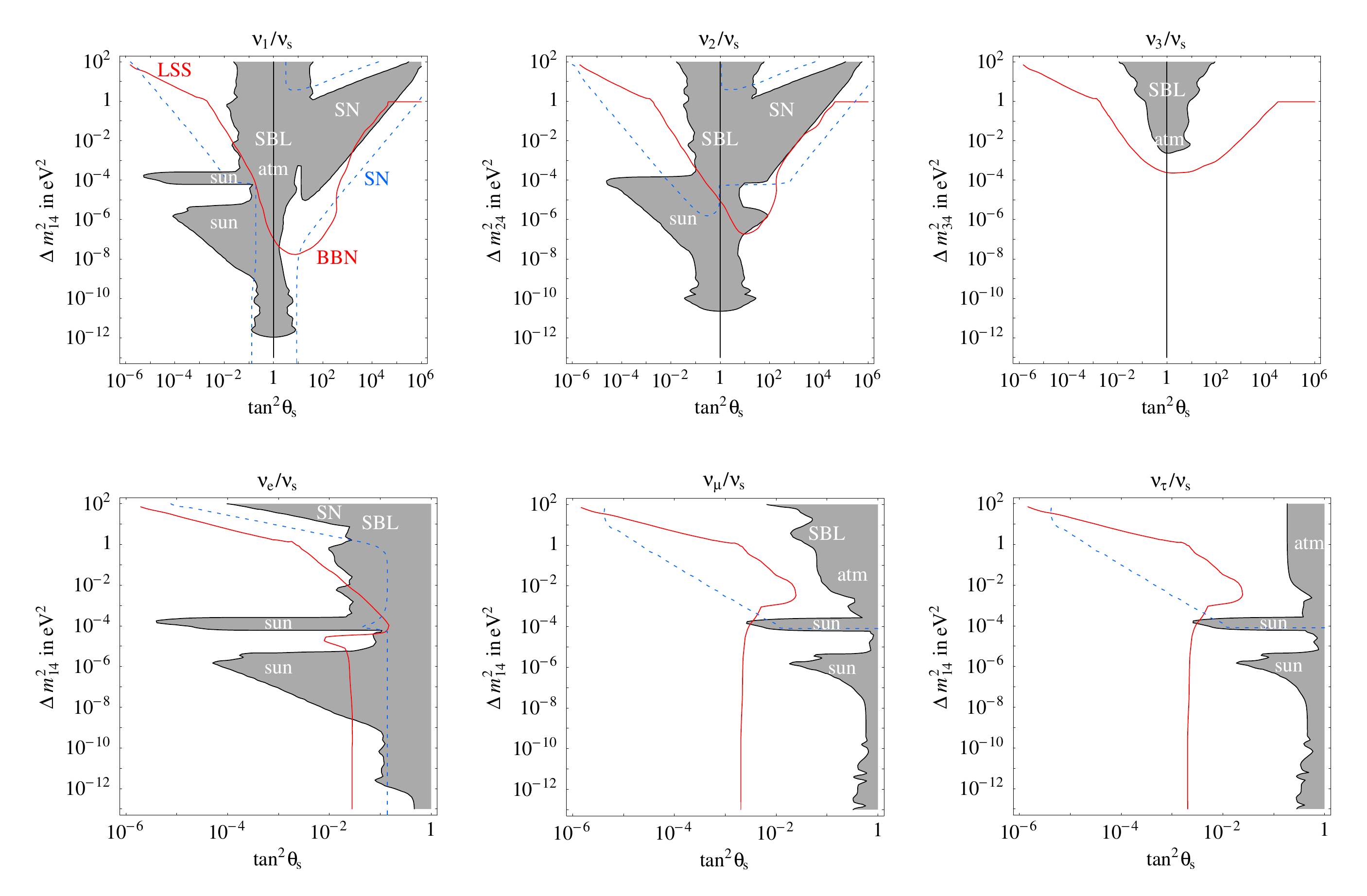}$$
\caption[Summary of sterile neutrino effects]{\label{fig:all}\em {\bf Summary of sterile neutrino effects}.
The shaded region is excluded at $99\%$ C.L.\ (2 dof)
by solar or atmospheric or reactor or short base-line experiments.
We shaded as excluded also regions where sterile neutrinos
suppress  the SN1987A $\bar\nu_e$ rate by more than $70\%$.
This rate is suppressed by more than $20\%$ inside the {\color{blue} dashed blue line},
that can be explored at the next SN explosion if it will be possible to
understand the  collapse well enough.
Within standard cosmology,
the region above the {\color{rossos} red continuous line} is disfavored (maybe already excluded) by BBN and LSS.}
\end{figure}

\begin{figure}
$$\includegraphics[width=7cm]{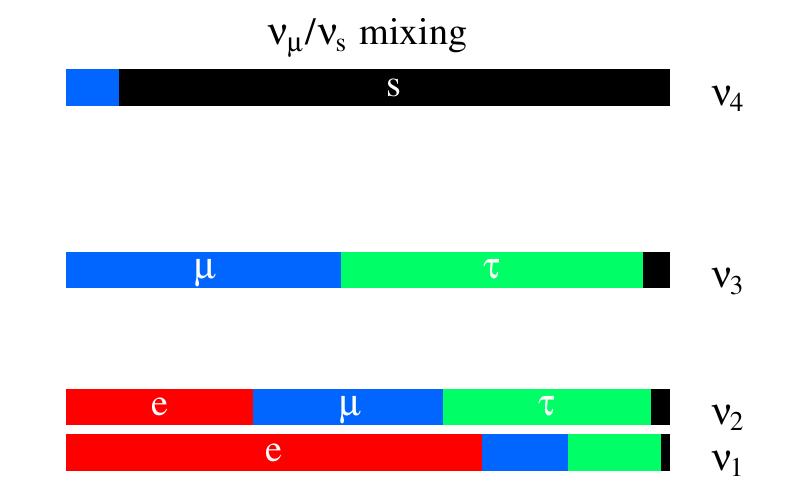}\qquad
\includegraphics[width=7cm]{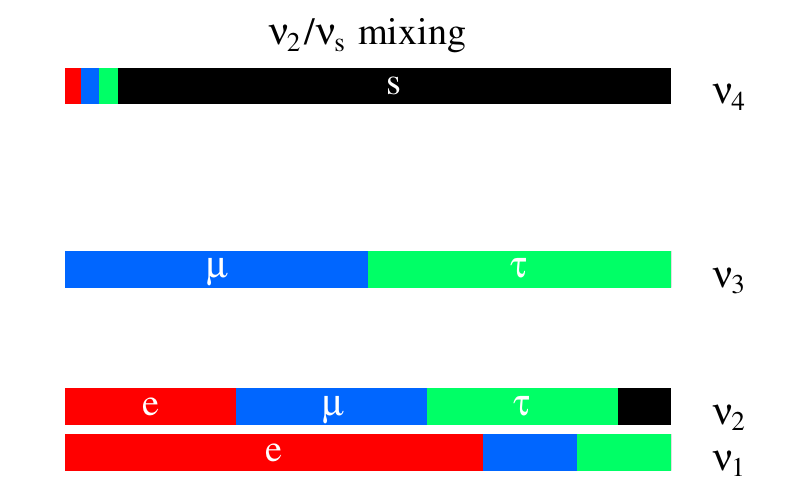}$$
\caption[Basic four neutrino spectra]{\label{fig:spettrias}\em {\bf 
Basic kinds of four neutrino mass spectra}. 
Left: sterile mixing with a flavour eigenstate ($\nu_\mu$ in the picture).
Right: sterile mixing with a mass eigenstate ($\nu_2$ in the picture).
}
\end{figure}

\subsection{Active/sterile mixings}
We now review the most promising ways to probe the existence of 
eV-scale sterile neutrinos.
Most probes are based on a careful study of
natural sources of neutrinos (the universe, the sun, supernov\ae, cosmic rays,...)
which have their own peculiar capabilities and limitations.
The sensitivity of some of these probes is enhanced by MSW resonances~\cite{MSW}.
In cosmology, active $\nu$ and $\bar\nu$ encounter a 
MSW resonance with sterile neutrinos lighter than active ones.
Roughly the same happens to supernova $\bar\nu_e$.
On the contrary, solar $\nu_e$ encounter a MSW resonance with sterile neutrinos heavier than active neutrinos. 
The present constraints are compared in figures\fig{all}, where we
assumed normal hierarchy of active neutrinos, 
a negligibly small $\theta_{13}$,
and added one extra sterile neutrino with arbitrary mass $m_4$.
We focus on the following six mixing patterns which
cover the qualitatively different possibilities:
$$\nu_{\rm s}/\nu_e,\quad
\nu_{\rm s}/ \nu_\mu,\quad
\nu_{\rm s}/\nu_\tau,\quad
\nu_{\rm s}/\nu_1,\quad
\nu_{\rm s}/\nu_2\quad\hbox{and}\quad \nu_{\rm s}/\nu_3$$
where $\nu_{1,2,3}$ are the mass eigenstates in absence of sterile mixing.
We remark one qualitative difference among the two classes of mixings:
\begin{itemize}

\item  {\em Mixing with a flavour eigenstate}  (depicted in fig.\fig{spettrias}a):
$\nu_{\rm s}/\nu_\ell$ ($\ell =e$ or $\mu$ or $\tau$).
The sterile neutrino oscillates into a well defined flavour
at 3 different $\Delta m^2 = m_4^2 - m_{1,2,3}^2$
(which cannot all be smaller than the observed splittings $\Delta m^2_{\rm sun,atm}$).

\item {\em Mixing with a mass eigenstate}  (depicted in fig.\fig{spettrias}b):
$\nu_{\rm s}/\nu_i$ ($i=1$ or 2 or 3).
The sterile neutrino oscillates into a neutrino of mixed flavour
at a single $\Delta m^2 = m_4^2 - m_i^2$, which can be arbitrarily small.
\end{itemize}
We now summarize the main constraints and the most promising signals.

\medskip

\begin{figure}
$$\includegraphics[width=10cm]{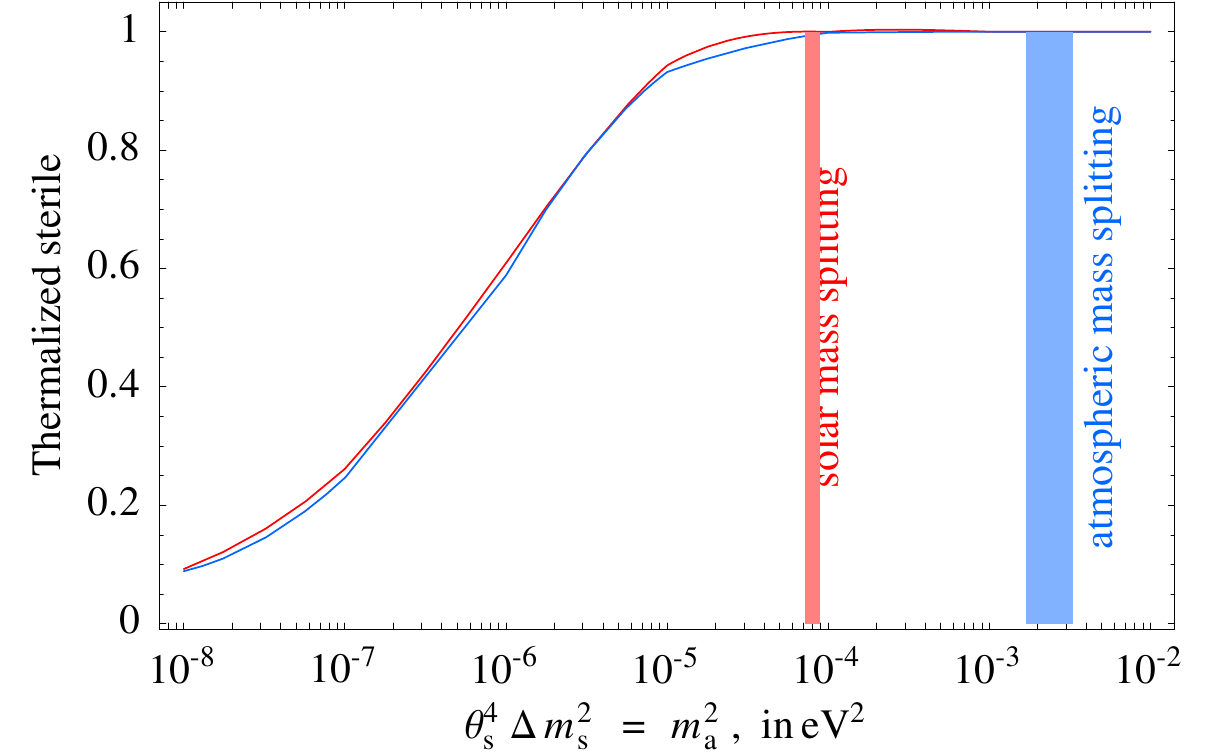}$$
\caption[Sterile neutrino abundancy]{\label{fig:sterileOscAbfig}\em The number density of a heavy sterile neutrino
thermalized through oscillations as function of $m_{\rm a}$, its contribution to active neutrino masses.}
\end{figure}

\subsection{Cosmology}
Compatibility with {\bf standard BBN} constrains 
oscillations into sterile neutrinos dominantly occurred at temperatures $T\circa{>} 0.1\MeV$.
The main observable is the {\em helium-4} primordial abundancy,
that we parameterize in terms of an effective number of neutrinos $N_\nu^{^4{\rm He}}$,
see eq.\eq{He}.
As discussed in section~\ref{nucleos},
present data do not exclude the presence of one extra thermalized neutrino, 
$N_\nu^{^4{\rm He}}=4$.
The helium-4 abundancy is sensitive to two different sterile effects: 
i.) increase of the total neutrino density; ii.) depletion of electron-neutrinos.
Section~\ref{OscUniverse} presented the formalism appropriate for precisely computing these effects,
that we now discuss in a semi-quantitative way.
\begin{enumerate}[i.)]
\item  Sterile neutrinos thermalize via averaged oscillations
(regime C of section~\ref{LimitingRegimes}).
The sterile interaction rate at temperature $T\ll M_Z$
is given by $\Gamma_{\rm s} \sim (\theta^m_{\rm s})^2 \Gamma_\nu$
where $\Gamma_\nu$ is the interaction rate of an active neutrino,
and $\theta_{\rm s}^m(T)\ll 1$ is the active/sterile mixing angle in matter at temperature $T$:
\beq
 \frac{\Gamma_{\rm s}}{H} \approx( \theta_{\rm s}^m)^2 (\frac{T}{4\,{\rm MeV}})^3\qquad
\theta_{\rm s}^m \approx \frac{\theta_{\rm s}}{1+T^6 G_{\rm F}^2/\alpha \Delta m^2}.\eeq
$\Gamma_{\rm s}/H$ is maximal at  $T_* \approx 10\,{\rm MeV} ({\Delta m^2}/{\eV^2})^{1/6}$
and the amount of thermalized sterile is
\beq \label{eq:sterileOscAb}
\Delta N_\nu \approx \left.\min(1,\frac{\Gamma_{\rm s}}{H}\right|_{\rm max})\approx \min(1,10^3 \theta_{\rm s}^2(\frac{\Delta m^2}{\eV^2})^{1/2}).
\eeq
The effect has a minor dependence on the flavour of the active neutrino involved in active/sterile mixing
and becomes negligible at $\Delta m^2\circa{<} 10^{-5}\eV^2$.
Our estimate does not take into account the possibility of resonant enhancement of active/sterile mixing,
possible only if $\Delta m^2<0$ i.e.\  if the `mostly sterile' neutrino is lighter than some `mostly active' neutrino.
The contribution to active neutrino masses mediated by a much heavier sterile neutrino is $m_{\rm a} = \theta_{\rm s}^2 m_{\rm s}$,
so that eq.\eq{sterileOscAb} is directly controlled by $m_{\rm a}$: fig.\fig{sterileOscAbfig} shows the precise result.

\item The second effect makes BBN sensitive to active/sterile  oscillations down to $\Delta m^2\sim 10^{-8}\eV^2$.
If $T_*$ is smaller than the neutrino decoupling temperature,
sterile neutrinos are produced by depleting active neutrinos
with the total neutrino density remaining constant.
Since there are many more neutrinos than nucleons, 
decoupled neutrinos (i.e.\ interaction rate of neutrinos with nucleons
smaller than the expansion rate) still play an active r\^ole
(i.e.\ interaction rate of nucleons with neutrinos larger than the expansion rate):
the $p/n$ ratio is determined by
$\nu_e n\leftrightarrow ep$ and $\bar\nu_e p\leftrightarrow \bar{e}n$ scatterings.
A reduced $\nubarnu_e$ number density affects BBN
increasing the effective $N_\nu^{^4{\rm He}}$ parameter.
\end{enumerate}
However, if BBN were non-standard, a modified density of electron neutrinos
could compensate the sterile corrections to $N_\nu^{^4{\rm He}}$:
for example  helium-4 constraints on sterile oscillations can be evaded by
allowing a  large neutrino asymmetry.

For all these reasons it is important to measure a second BBN effect.
Measurements of the {\em deuterium} primordial abundancy are less affected by systematic uncertainties:
in the future it might be possible to improve its measurement obtaining an
uncertainty on the effective parameter $N_\nu^{\rm D}$ (precisely defined in eq.\eq{Deuterio})
significantly below 1, 
possibly making deuterium the most significant BBN probe.
However, the deuterium abundancy is less sensitive than the helium-4 abundancy
to $\nu_e$ depletion and therefore
to  values of $\Delta m^2$ below $10^{-5}\eV^2$.

\bigskip
Future studies of CMB acoustic oscillations
should allow to precisely measure the  total neutrino density $N_\nu^{\rm CMB}$
at recombination ($T\sim \eV$) with $\pm0.2$ ({\sc Planck}) or
 maybe $\pm0.05$ ({\sc CMBpol}) error~\cite{WMAP}.
Sterile neutrinos affect  $N_\nu^{\rm CMB}$ only if  $\Delta m^2\circa{>}10^{-5}\eV^2$;
in such a case
$N_\nu^{\rm CMB}\approx N_\nu^{^4{\rm He}}\approx N_\nu^{\rm D}$.

As discussed in section~\ref{FreeStream}, 
{Large Scale Structure} data are sensitive to neutrino masses.
From a cosmological point of view, the main parameter that controls this kind of
effects is the energy density in (massive) neutrinos, $\Omega_\nu$.
Sterile neutrinos with masses up to about 1 keV cosmologically behave as
warm dark matter: present LSS data demand $\Omega_\nu\circa{<}0.01$~\cite{WMAP}
(This parameter alone does not fully encode all relevant physics:
e.g.\ a smaller non-thermal population of heavier sterile neutrinos 
behaves differently than a larger population of lighter sterile neutrinos).
This LSS constraint  complements the BBN constraint on sterile oscillations
as illustrated in fig.\fig{BBN}.


\subsection{Solar neutrinos}
Solar $\nu_e$ experiments have explored sterile oscillations not testable by cosmology,
thanks to two different effects.
(1) MSW resonances make solar $\nu_e$ sensitive to small active/sterile mixing
and $\Delta m^2\circa{>}10^{-8}\eV^2$.
(2) With large mixing, solar $\nu_e$ are sensitive down to $\Delta m^2\sim 10^{-12}\eV^2$.
Future experiments will explore new aspects of the solar neutrino anomaly,
allowing to measure in a redundant way the active oscillation parameters
or to discover a new anomaly.
We emphasize one qualitative point.
Due to  LMA oscillations, neutrinos exit from the sun as almost pure $\nu_2$
at energies $E_\nu  \circa{>}  \hbox{few MeV}$ (section~\ref{sun}).
Neutrinos with these energies have been precisely studied by SNO and SK,
but are not affected by sterile oscillations if they mostly involve $\nu_1$.
This can happen either because $\nu_{\rm s}$ mixes with $\nu_1$ or because
it is quasi-degenerate to it.
Therefore  there is a whole class of sterile effects which 
manifest only at $E_\nu \circa{<}  \hbox{\rm few MeV}$ ---
an energy range explored so far only by  Gallium experiments~\cite{Ga}.
Future precise measurement of solar $\nu_e$ at sub-MeV energies
will allow to significantly
extend searches for active/sterile effects.
Part of these extended region will be 
tested by Borexino~\cite{Borexino}, where a sterile neutrino can manifest 
as day/night variations, or as seasonal variations,
or even by reducing the total rate.

\subsection{Supernov\ae}
Supernova neutrinos 
will be good probes of sterile oscillations because
have a different pattern of MSW resonances
and a longer base-line than solar $\nu_e$.
Consequently supernova $\bar\nu_e$ are more sensitive than solar $\nu_e$
in two main cases: 
(a) small $\Delta m^2\circa{>}10^{-18}\eV^2$ with large $\theta_{\rm s}$;
(b) $\nu_{\rm s}$ lighter than $\nu_1$ with small mixing.
Oscillations into one sterile neutrino can reduce the $\bar\nu_e$ rate by up to $80\%$
and, in a more restricted range of oscillation parameters,
vary the average $\bar\nu_e$ energy by $30\%$.
SN1987A data agreed with expectations.
Future SN experiments can perform quantitative tests, but
it is not clear how to deal with theoretical uncertainties.
Neutrinos emitted by past supernov\ae, or by other extragalactic sources,
do not seem to allow more sensitive searches for sterile neutrinos~\cite{SNrelic}.

Sterile neutrinos could explain two possible anomalies:
\begin{itemize}
\item {\bf Pulsar kicks}~\cite{PulsarKicks}. 
 Pulsars are neutron stars with intense magnetic fields $B\sim 10^{12}$ Gauss
and rotation periods from a few msec to a few seconds (presumably,
about 10 msec when generated from core collapse supernov\ae).

 Observations show that a population of pulsars (typically found away from the galactic plane) 
 move with peculiar velocities $v\sim 1000\,{\rm km/s}$ ($mv^2/2 \sim 10^{49}\,{\rm erg}$)
 about one order of magnitude
 larger than what suggested by models of asymmetric SN explosions.
 Since SN explosions emit almost all energy in neutrinos, 
 this anomaly might be explained by  a few $\%$ asymmetry in neutrino production.
 Normal neutrinos, even if produced asymmetrically, diffuse within the core, resulting a symmetric emission.
 In presence of sterile neutrinos 
 with keV-scale masses and $\theta_{\rm s}\circa{<}10^{-7}$,
 the following mechanism  can produce the asymmetry: 
 in presence of very strong and axially oriented magnetic fields, 
 possibly present in neutron stars, the MSW potential includes a contribution (at one loop, from neutrino scattering on the polarized medium) which depends on the relative angle between the neutrino momentum and the magnetic field making neutrino emission asymmetric.

\item {\bf $r$-process nucleosynthesis}~\cite{r-process}. 
Nuclei can capture ambient neutrons forming heavier nuclei that are however unstable.
Within supernov\ae{} ejecta the neutron flux is so high that capture is more $r$apid than decay,
possibly leading to the formation of the observed heavy elements.
However, detailed studies find that this process is prevented 
by $\nu_e$, which convert neutrons into protons
($\nu_e n \to p e$) leading to the formation of $^4_2$He nuclei, rather than of neutron-rich nuclei.
One possibility of overcoming this problem is that $\nu_e$ are depleted by 
$\nu_e \to \nu_{\rm s}$ resonant oscillations in the mantle of the star.
This can happen, without dramatically reducing the observed $\bar\nu_e$ SN flux, 
for active/sterile oscillation parameters in the range
 $\Delta m^2 \sim (1 \div 10^2) \eV^2$ and 
$\sin^2 2 \theta_{\rm s} \sim 10^{-3} \div \textrm{few} \cdot 10^{-1}$.

\end{itemize}

\subsection{Atmospheric neutrinos}
{Atmospheric} experiments (SK, MACRO, K2K) indirectly exclude
active/sterile oscillations with $\Delta m^2\circa{>}10^{-3\div 4}\eV^2$
and $\tan^2\theta_{\rm s}\circa{>}0.1\div0.2$.
Up to minor differences, this applies to all flavours.
{Terrestrial experiments} that mainly probed disappearance of
$\bar\nu_e$ and $\nu_\mu$ ({\sc Chooz}, CDHS,\ldots)
exclude active/sterile  mixings with these flavours for $\tan^2\theta_{\rm s}\circa{>}0.03$
and $\Delta m^2\circa{>}10^{-3}\eV^2$.
Therefore future short-baseline experiments can search for sterile effects with smaller $\theta_{\rm s}$.
Possible signals are $\bar\nu_e$ disappearance in reactor experiments,
a deficit of NC events or $\nu_\tau$ appearance in beam experiments.
Within standard cosmology these effects can be probed by CMB and BBN,
which already disfavor them.

%



\subsection{A minimal scenario}
The sterile neutrinos might be the 3 right-handed neutrinos employed by the see-saw mechanism~\cite{seesaw}. This corresponds to eq.\eq{Lsterile} with $m_{LL}=0$, obtaining a renormalizable extension of the SM. In the usual scenario right-handed neutrinos are assumed to be very heavy (up to about $10^{14}\GeV$) and can
mediate Majorana neutrino masses (see section~\ref{ExtraSinglets})
and produce baryogenesis via leptogenesis (see section~\ref{leptogenesi})
but cannot be directly observed.

One can hope that right-handed neutrinos are instead light enough to give detectable effects,
and possibly explain some hints. 
This restricted scenario predicts no $0\nu2\beta$ (because $m_{LL}=0$: lepton number is violated only by right-handed neutrino masses) if right-handed
neutrinos are lighter than the $Q$-value of $0\nu2\beta$ ($Q\sim\hbox{few MeV}$).

There is a  small window, around a few keV, where a sterile neutrino could be all observed DM~\cite{SterilePheno}.
Particles lighter than a keV behave as warm dark matter, while  sterile neutrinos heavier than a few keV decay too fast (into $\nu\bar\nu\nu$ at tree level thanks to mixing with active neutrinos,
and into $\nu\gamma$ at loop level:
the visible $\gamma$ has monochromatic energy $E_\gamma\approx m_{\rm DM}/2$, 
giving the main experimental signal of sterile neutrinos as DM).
The small window is already excluded unless sterile neutrinos happen to have kinetic energies
somewhat smaller than active neutrinos.
Notice that eq.\eq{sterileOscAb} implies that the two sterile neutrinos that mediate the `atmospheric' and `solar' mass splitting
 are not good DM candidates (because their abundancy would be too large, $N_\nu\approx 1$): 
 one has to rely on third sterile neutrino with essentially unknown couplings~\cite{SterilePheno}.
 Furthermore, this scenario allows successful baryogenesis  if the two `known' 
 sterile neutrinos are quasi-degenerate around the GeV scale.


\section{Neutrinos and light scalars}\label{Decay}\label{Scalars}
\index{Neutrino!decay?}\index{Light scalars}
Electromagnetic gauge invariance and Lorenz invariance allow
Yukawa couplings $g_{ij}$ between neutrinos and a hypothetical
light scalar $X$,
$g_{ij}\,\nu_i\nu_j X/2+\hbox{h.c}$.
The electroweak invariant version of this operator has dimension 6:
$X(LH)^2$, such that electroweak gauge invariance suppresses $g_{ij}$ down to
negligibly small values, unless $g_{ij}$ are mediated by a particle
lighter than the weak scale. 
Apparently, the only possibility that can lead to sizable $g_{ij}$ consists
in having also light sterile neutrino(s) $\nu_R$.
Indeed sterile neutrinos can have a Yukawa coupling $g_{RR}$ to $X$  not 
suppressed by electroweak gauge invariance, and the interactions
\beq \frac{\nu_R^2}{2} (m_{RR}+g_{RR}  X ) + m_{LR} \frac{LH}{v} \nu_R + m_{LL}\frac{(LH)^2}{2v^2} +
\hbox{h.c.}\eeq
generate
the Yukawa coupling $g = g_{RR} m_{LR}^2/2 m_{RR}^2$ between neutrinos and $X$
after integrating out $\nu_R = - m_{LR}\nu/(m_{RR} + g_{RR}X)$.
Introducing more than one sterile neutrino, one can similarly generate a Yukawa
coupling between active neutrinos, sterile neutrinos and $X$.
Although a sizable $g$ needs light sterile neutrinos, we here study the phenomenology
of the Yukawa couplings $g$ in isolation from sterile neutrinos.

\medskip

Another theoretical issue is: why $X$ is light?
One plausible sub-class of models provides a neat answer and  implies a special phenomenology.
The scalar $X$ might be light because it is a (pseudo)-Goldstone boson 
if its vev spontaneously breaks a global lepton number symmetry;
if neutrino masses $m_{ij} = g_{ij}\langle X \rangle$
arise only from the coupling $g_{ij} \nu_i\nu_j X/2$,
$X$ couples to neutrino mass eigenstates such that there is no neutrino decay in vacuum.
Neutrino decay in matter is possible, because
matter effects lead to couplings between different mass eigenstates.

Different areas of physics allows to test $g_{ij}$ couplings.
We start from the simplest and less sensitive probes, moving to the
more sensitive and subtle probes.

 \subsection{Rare decays}
Emission of $X$ scalars would affect 
decays of $\pi$ and $K$ into $e\bar\nu_e$ or $\mu\bar\nu_\mu$,
resulting in the constraint $|g_{ij}|\circa{<}10^{-2}$
for all $ij\neq\tau\tau$~\cite{NuDecay}.
These decays should likely be studied in the full theory, 
rather than using the effective $g_{ij}$ couplings.
$0\nu2\beta$ gives a stronger constraint on $|g_{ee}|\circa{<}10^{-4}$.

\subsection{Neutrino decay}
A minimal source of neutrino decay is provided by
neutrino masses, together with one loop SM effects.
However, estimates show that the resulting life-time is so long that neutrinos
are practically stable even on astrophysical and cosmological time-scales:
\beq\Gamma(\nu_i\to \gamma\nu_j)\sim \frac{e^2 g_2^4}{(4\pi)^5}\frac{m_i^5 m_\tau^2}{M_W^6}\sim
\frac{1}{10^{40}\hbox{yr}}\bigg(\frac{m_i}{0.1\eV}\bigg)^5.\eeq
In the limit $m_i\gg m_j$ neutrinos with Majorana masses decay two times faster than Dirac neutrinos~\cite{NuDecay}.

Therefore a possible observation of neutrino decay would imply
physics beyond the SM, and more specifically a coupling to some new 
very light or massless particle $X$. 
Assuming that $X$ has spin 0,
the Lagrangian coupling $g_{ij}\,\nu_i\nu_j X/2+\hbox{h.c.}$ 
(where $g_{ij}$ is a symmetric flavour matrix, here written in the basis of neutrino
mass eigenstates $\nu_i$)
gives
\beq\Gamma(\nu_i \to X \nu_j) = \Gamma(\nu_i\to X \bar\nu_j) = \frac{g^2_{ij}}{32\pi} m_i = 
\frac{g^2_{ij}}{0.40\,{\rm mm}}\frac{m_i}{0.05\eV}
\eeq
for $m_i\gg m_j$. Taking into account Lorentz dilatation,
the life-time of an ultrarelativistic neutrino is 
$$\tau = \frac{E_\nu}{m_i}\Gamma = \frac{8000\km}{g^2_{ij}} \frac{E_\nu}{\GeV}\frac{2.5\cdot 10^{-3}\eV^2}{m_i^2}$$
i.e.\ relativity tells that sensitive probes of neutrino decay must have a large $L/E_\nu$,
like in the case of oscillations.
Neutrinos have been observed at the following values of $L/E_\nu$:
$$\begin{array}{rrcl}
\hbox{atmospheric:} &10^{4}\,\hbox{km}/300\MeV &\sim& 10^{-10}\,\hbox{s}/\eV,\\
\hbox{solar:} &500\,\hbox{sec}/5\MeV &\sim& 10^{-4}\,\hbox{s}/\eV,\\
\hbox{supernova:} &10\,\hbox{kpc}/10\MeV &\sim& 10^{5}\,\hbox{s}/\eV.
\end{array}$$
Present supernova data are not conclusive enough to derive constraints.
Atmospheric data only imply $|g_{ij}|\circa{<}4\pi$, which is
anyhow required by perturbativity.
Solar data suggest $|g_{12}|\circa{<}10^{-3}$.

\subsection{CMB and interacting neutrinos}
As discussed in section~\ref{FreeStream} the presence of freely-moving relativistic neutrinos 
suppresses  CMB anisotropies. 
The situation changes if the interaction rate $\Gamma$ of neutrinos with scalars $X$ is
faster than  $H$, the expansion rate at recombination  ($T_{\rm rec}\sim 0.3\eV$).
While a detailed analysis is needed to study the case $\Gamma\sim H$,
if $\Gamma\gg H$ neutrinos and scalars form an interacting $\nu/X$ fluid
and damp CMB anisotropies less effectively~\cite{nuScalar}.

The evolution of its inhomogeneities can be precisely described by an equation somewhat analogous
to eq.\eq{DMevo}, that applies to DM inhomogeneities.
The main differences are that the $\nu/X$ fluid would be relativistic
($w=1/3$) and with a non-vanishing sound speed, $c_{\rm s}^2 = 1/3$.
The total energy density of the $\nu/X$ fluid is parameterized, in the standard cosmological
language, in terms of an  `effective number of neutrinos' $N_\nu^I$,
where $I$ stands for $I$nteracting. $N^I_\nu$ can be bigger than $3$ if the
scalar(s) $X$ carry some energy, and smaller than 3 if less neutrinos 
significantly interact with $X$; the remaining neutrinos behave normally.
These are the minimal parameters that characterize the cosmological behavior of the system.
Neutrino masses and the $X$ mass(es) can be additional parameters,
and lead to time-dependent $N^I_nu,w,c_{\rm s}^2$.

Present global fits of cosmological data 
indicate a $(3\div 4)\sigma$ preference for the standard scenario
 of freely moving neutrinos~\cite{nuScalar}, and future data should settle the issue.
This effect can give the most  sensitive probe to the $g_{ij}$ couplings.
Indeed the
 $\nu_j \nu_j \leftrightarrow XX$
 processes have a rate $\Gamma \sim g_{ij}^4 T$
 comparable to the expansion rate $H\sim T^2/M_{\rm Pl}$ for
  $g_{ij}\circa{<} (T_{\rm rec} M_{\rm Pl})^{1/4}\sim 10^{-7}$.
  If neutrino decay $\nu_i \to \nu_j X$ ($i\neq j$) is kinematically allowed because
  $X$ is light enough,
its rate $\Gamma\sim g_{ij}^2 m_i$ is comparable to the universe expansion rate
   at recombination for    $g_{ij}\circa{<} (T_{\rm rec} M_{\rm Pl})^{1/2}\sim 10^{-13}$,
   allowing to probe very small values of the off-diagonal couplings $g_{ij}$.
 \index{Cosmology!and neutrino decay}

 \medskip

\section{Quantum corrections to neutrino masses}\label{RGE}
Quantum corrections shift the coefficient $\kappa$ of
the dimension-5 effective operator
\begin{equation}
\label{eq:LLHH}
\frac{\kappa_{ij}}{4} (L_i H) (L_j H)
\end{equation}
that induces Majorana neutrino masses $m_{ij} = \kappa_{ij} v^2/2$
with $v=\langle H\rangle = 174\GeV$.
We now compute these quantum effects.

\begin{figure}
$$\includegraphics{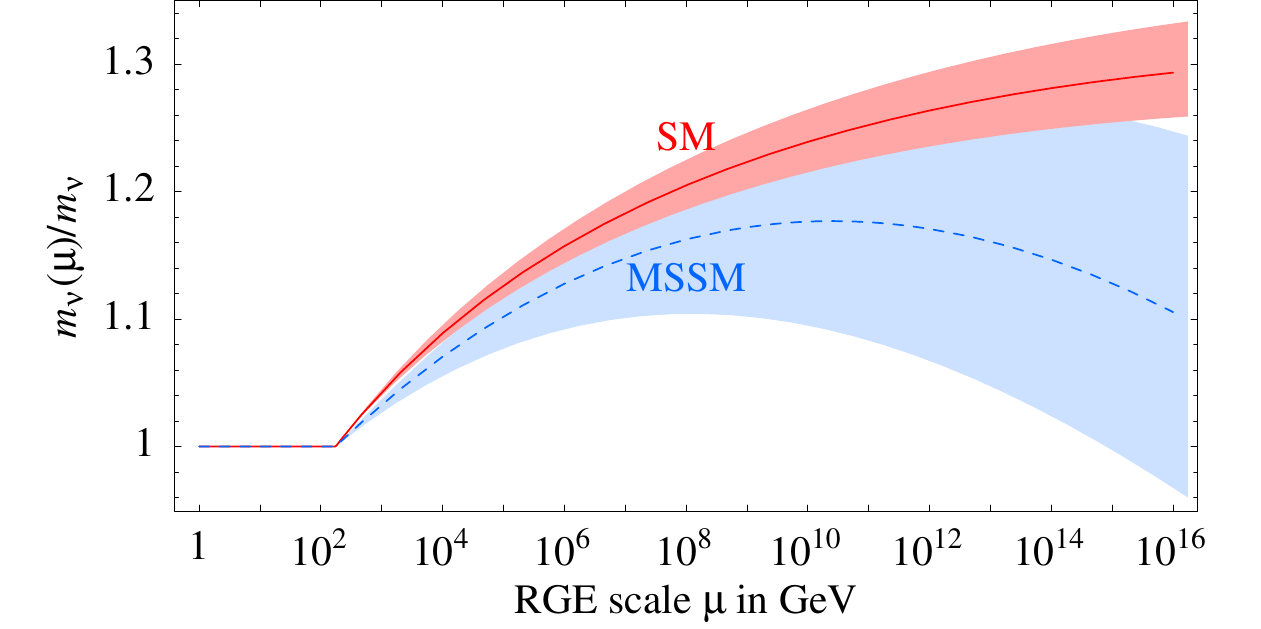}$$
\caption[RGE running of neutrino masses]{\label{fig:RGE}\em Universal running of $m_\nu$ in the SM and in the MSSM.
The bands indicate the present uncertainty.}
\end{figure}

\subsection{MSSM}
We start from the MSSM where the computation can be easily
performed since
superpotential interactions (like the operator in\eq{LLHH}) do not receive direct
quantum corrections~\cite{WB}.
The kinetic term of a generic chiral superfield $\Phi$ 
(with Yukawa interactions $\lambda$
and gauge interactions $g$) receive quantum corrections
that correct it into
$$\Lag_{\rm kin}= Z_\Phi^2 (\Phi^\dagger \Phi)\qquad\hbox{where}\qquad
Z_\Phi =  1 + (g_I^2 c_I^\Phi - \lambda^2 )t
+ {\cal O}(g^2 t)^2,\qquad
t\equiv  \frac{\ln \mu/\Lambda}{(4\pi)^2}.
$$
$\Lambda$ ($\mu)$ is a supersymmetric UV (IR) cutoff and
$c_I^\Phi$ are Casimir factors
($c_2^{L,H} = 3/2$ and $c_Y^{L,H} =1/2$ for Higgs and lepton doublets).
Rewriting the Lagrangian in terms of renormalized
fields with standard kinetic terms, $\Phi\to \Phi/Z_\Phi$,
the coupling $\kappa$ gets corrected into
\begin{equation}
\label{eq:unres}
\kappa_{ij}(\mu) = \kappa(\Lambda)/Z_{L_i}^2Z_{L_j}Z_H^2.
\end{equation}
So, RGE corrections due to gauge interactions only give an overall rescaling,
and RGE corrections due to the $\tau$ Yukawa coupling do not qualitatively change neutrino masses.
Suppose e.g.\ that only $\mu+\tau$ has a mass (such that the neutrino mass matrix has rank 1):
after adding RGE corrections due to the $\tau$ Yukawa coupling one still have rank 1 mass matrix,
for a combination of neutrinos slightly different from $\mu+\tau$.

These expressions are accurate only if $g^2 t\ll 1$.
When $\mu \sim M_Z$ and $\Lambda\sim M_{\rm Pl}$
this condition is violated by too large logarithms $\ln\Lambda/\mu$.
An accurate expression can be obtained by resumming
all quantum corrections of ${\cal O}(g^2 t)^n$.
This is accomplished using renormalization group techniques,
by transforming\eq{unres} into the differential RGE equation
\begin{equation}
\label{eq:RGEMSSM}{\color{blus}
\frac{d\kappa}{dt}= \kappa (-2g_Y^2 - 6 g_2^2+6\lambda_t^2) 
+\kappa \cdot (\lambda_E^\dagger \cdot \lambda_E)^T+
 (\lambda_E^\dagger  \cdot \lambda_E)\cdot \kappa }  +
 {\cal O}\bigg(\frac{g^4}{(4\pi)^2}\bigg)
\qquad\hbox{(MSSM)}.
\end{equation}

\subsection{SM}
In the SM, RGE equations maintain the same structure with different
numerical factors~\cite{SMRGE}:
\begin{equation}
\label{eq:RGESM}{\color{blus}
\frac{d\kappa}{dt}= \kappa (\lambda  - 3 g_2^2+6\lambda_t^2) 
-\frac{3}{2}\bigg[\kappa \cdot (\lambda_E^\dagger \cdot \lambda_E)^T+
 (\lambda_E^\dagger  \cdot \lambda_E)\cdot \kappa\bigg]}+
 {\cal O}\bigg(\frac{g^4}{(4\pi)^2}\bigg) \qquad\hbox{(SM)} 
\end{equation}
The lepton and higgs couplings $\lambda_E$ and $\lambda$ are defined in\eq{LSM}.
Neglecting the small $e$ and $\mu$ Yukawa couplings,
the solution to these RGE equations is
$$\kappa(\mu) = r \pmatrix{\kappa_{ee} & \kappa_{e\mu} & y \kappa_{e\tau}\cr
\kappa_{\mu e} & \kappa_{\mu\mu} & y \kappa_{e\tau}\cr
y \kappa_{\tau e} & y\kappa_{\tau\mu} & y^2 \kappa_{\tau\tau}},\qquad
$$
where
$$ r(\mu) = \left\{\begin{array}{ll}
\exp\bigg[\int (\lambda - 3g_2^2+ 6 \lambda_t^2)dt\bigg] & \hbox{SM}\\
\exp\bigg[\int (-2 g_Y^2 -6 g_2^2+6\lambda_t^2)dt\bigg] & \hbox{MSSM}
\end{array}
\right. \qquad
y(\mu) = \left\{\begin{array}{ll}
\exp\bigg[-\frac{3}{2}\int \lambda_\tau^2 dt\bigg] & \hbox{SM}\\
\exp\bigg[\int \lambda_\tau^2\, dt\bigg] & \hbox{MSSM}
\end{array}
\right. 
$$
The overall rescaling $r(\mu)$ is plotted in fig.\fig{RGE} where we have assumed $\alpha_3 (M_Z)= 0.118$,
a pole top mass of $175\GeV$, $m_h = 115\GeV$ in the SM and
moderately large $\tan\beta$ in the MSSM.

The $y$ term is flavour dependent
and renormalizes neutrino masses and 
mixing angles.\footnote{One can write RGE equations for neutrino masses and mixings,
however they have a structure much more complicated than
RGE equations for the neutrino mass matrix.
The `run-and-diagonalize' approach
(i.e.\ solve the above RGE equations and at the end diagonalize the neutrino mass matrix)
is usually much more convenient than the 
`diagonalize-and-run' approach, not discussed here.}
In the SM, this effect is  numerically small, $y \approx 1-10^{-6}\ln (\mu/M_Z)$,
and can be neglected unless one considers fine-tuned neutrino mass matrices, unstable under
these small RGE corrections.
For example, if one puts by hand large equal masses to all neutrinos,
small RGE corrections can generate a sizable mass splitting.
In the MSSM  $\tan\beta$ can be so large that $\lambda_\tau\sim 1$;
in this case RGE equations contain extra terms.

\smallskip

Present data allow one massless neutrino;
if the other two neutrinos have Majorana masses with the values
demanded by solar and atmospheric data,
its masslessness is not protected by a symmetry, 
and indeed a non-zero mass
is generated by quantum corrections at two loop and higher order.
In the SM one roughly gets
$m_1 \sim m_3  \lambda_\tau^4/(4\pi)^4 \sim 10^{-11}\eV$,
and in the MSSM a similar results applies, with a $\tan^4\beta$ enhancement
coming from $\lambda^4_\tau$~\cite{SMRGE}.

\subsection{Quantum correction to see-saw models}
See-saw contain extra renormalization effects at energies
above the masses of right-handed neutrinos, generated by the  neutrino Yukawa
couplings $\lambda_N$.
The technique described above would allow to easily compute
these effects in  supersymmetric see-saw models,
taking into account that the quantum corrections $Z_N$ 
(renormalizations of kinetic terms of heavy right-handed neutrinos)
do not affect light neutrinos.
Explicit expressions are given in the literature.
In practice $\lambda_N$ are unknown so that we do not learn much
by RGE-evolving them up and down.
For example, some neutrino Yukawa couplings could be large enough
that the solar mixing angle, renormalized at the unification scale, becomes maximal
or achieves any other value.


\medskip

In the next section we study a different class of quantum corrections
present in  supersymmetric see-saw models, that can have detectable effects.

\section[Lepton-flavour violation and SUSY]{Lepton-flavour violation and supersymmetry}\label{LFV}
An unpleasant feature of the see-saw mechanism is that
we do not see how it can be realistically tested, i.e.\ 
how it could became true physics rather than remaining a plausible speculation.

In general, if the observed solar and atmospheric anomalies are due to neutrino masses,
lepton-flavour-violating (LFV) decays such as $\mu\to e\gamma$ and $\tau\to \mu \gamma$
must be present at some level.
However, if the effective theory at energies
below the new physics that generates neutrino masses is the SM,
then lepton flavour is violated only by non-renormalizable interactions
and the resulting rates are of order 
$\BR(\mu \to e \gamma) \sim (m_\mu m_{e \mu}/M_W^2)^2 \sim 10^{-50}$
where $m_{e\mu}$ is the $e \mu$ element of the neutrino mass matrix.
This would be hopelessly below the present experimental bound $\BR(\mu \to e \gamma)\circa{<} 10^{-11}$ and  below any possible future improvement,
as summarized in table\tab{bounds}.
See-saw models give a concrete example of this general fact:
one can precisely compute the $\mu\to e \gamma$ decay amplitude,
confirming that it is unobservably small because proportional to 
$\lambda_N^2/M_N \propto m_\nu$.
Furthermore, the smallness of the observed neutrino masses, $m_\nu\sim \lambda^2_N v^2/M_N$, suggests that directly observing the right-handed neutrinos is also impossible,
either because $M_N$ is too heavy, or because $\lambda_N$ is too small.


\begin{table}[t]
$$\begin{array}{|rcllc|rcllc|} \hline
\multicolumn{4}{|c}{\hbox{\color{blus} present bound\color{black}}} & \hbox{\color{blus} future?\color{black}} &
\multicolumn{4}{c}{\hbox{\color{blus} present bound\color{black}}}  & \hbox{\color{blus} future?\color{black}}
\\ \hline 
d_N &<& 3.0~10^{-26}\ecm &\cite{dNexp}&     10^{-27}\ecm &
\hbox{BR}(\tau\to \mu \gamma) &<& 4.5~10^{-8}&\cite{expTauDec} & 10^{-9}
\\ 
d_e &<& 1.5~10^{-27}\ecm &\cite{deexp}&  10^{-29}\ecm &
\hbox{BR}(\mu\to e \gamma) &<& 1.2~10^{-11}&\cite{muegExp} &10^{-14}
\\ 
d_{^{199}\rm Hg} &<& 1.8~10^{-28}\ecm    &\cite{dHgexp}&  & 
\hbox{BR}(\mu\to e\bar{e}e) &<& 1.0~10^{-12}&\cite{mueExp} &10^{-16}
\\ 
d_\mu &<& 1.5~10^{-19}\ecm &\cite{dmuexp}&  10^{-25}\ecm &
\hbox{CR}(\mu\to e \hbox{ in Ti}) &<& 6.1~10^{-13}&\cite{expMuCapt} & 10^{-18}
\\
\hline
\end{array}$$\color{black}
\caption[Experimental bounds on CP and lepton-flavor violating processes]{\em Compilation of $90\%$ {\em CL} bounds on {\rm
CP}-violating and lepton-flavor violating processes.}
\label{tab:bounds}
\end{table}

If supersymmetric particles exists at the weak scale, things can be very different~\cite{nuLFV}.
In the context of the Minimal Supersymmetric Standard Model (MSSM)
radiative corrections induced by $\lambda_N$  affect supersymmetry-breaking
slepton mass terms, if they are already present in the Lagrangian at energies above $M_N$
(alternatively supersymmetry-breaking could instead be transmitted to MSSM particles at energies below $M_N$, 
where right-handed neutrinos no longer exists).
The crucial difference between the SM and the MSSM is that the SM remembers
of the existence of very heavy right-handed neutrinos only trough non-renormalizable operators
like $(LH)^2/\Lambda$ (that give rise to neutrino masses).
The MSSM contains more renormalizable terms, like slepton masses $\mb{m}^2_{\tilde{L}}\tilde{L}^* \tilde{L}$,
where right-handed neutrinos can leave their imprint.
For example, the correction to the $3\times 3$ mass matrix of
left-handed sleptons is
\begin{equation}\label{eq:slepton}
\mb{m}^2_{\tilde{L}} = m_0^2 \One  -\frac{3m_0^2}{(4\pi)^2} 
\mb{Y}_N +\cdots\qquad\hbox{where}\qquad 
\mb{Y}_N\equiv \mb{\lambda}_N^\dagger
\ln(\frac{\MGUT^2}{\mb{MM}^\dagger})
\mb{\lambda}_N
\end{equation}
having  neglected
$A$-terms and ${\cal O}(\lambda_N^4)$ effects and assumed
universal soft terms at $\MGUT$\footnote{In the MSSM lepton flavour is not
an accidental symmetry as in the SM, and some unknown mechanism must suppress lepton/slepton mixing
down to an acceptable level.
This problem motivates the assumption of universal soft-terms at $\MGUT$,
although nothing guarantees it.
Rather, in GUT models the unified top quark Yukawa coupling distorts universal soft terms,
giving rise to other LFV effects related to GUT physics rather than to neutrino physics.}.  In this approximation, the experimental
bounds from $\ell_i\to \ell_j \gamma$ decays are saturated for 
\begin{equation}\label{eq:Ynu}
[\mb{Y}_N]_{\tau \mu} ,[\mb{Y}_N]_{\tau e} \sim 10^{1\pm 1},\qquad
[\mb{Y}_N]_{\mu 
e} \sim 10^{-1\pm 1}.\end{equation}
The precise value depends on the so far unknown masses of supersymmetric particles.
Large neutrino couplings (e.g.\ $\lambda_N \sim \lambda_t$) could give $\mu\to e \gamma$ or
$\tau\to \mu \gamma$ just below their experimental bounds,
while smaller neutrino couplings (e.g.\ $\lambda_\nu \sim \lambda_\tau$) would give no significant effect.
However, we have no idea of which value $Y_N\sim \lambda_N^2$ should have
(neutrino masses only tell us the value of $\lambda_N^2/M_N$), so that
{\em see-saw models make no testable prediction}.
In fact, the $\mb{M}_N,\mb{\lambda}_N$ and $\mb{\lambda}_E$ matrices that describe
the supersymmetric see-saw
contain 15 real parameters and 6 CP-violating phases.  
At low energy, in
the mass eigenstate basis of the leptons, 3 real parameters describe
the lepton masses, and both the neutrino and the left-handed slepton
mass matrices are described by 6 real parameters and 3 CP-violating phases.
Since $(15+6) = (3+0) + (6+3)+(6+3)$ we see 
that the generic see-saw mechanism has too many free parameters to allow predictions:
any
pattern of lepton and neutrino masses is compatible with any pattern of
radiatively-generated flavor violations in left-handed slepton
masses.

\begin{figure}
$$\includegraphics{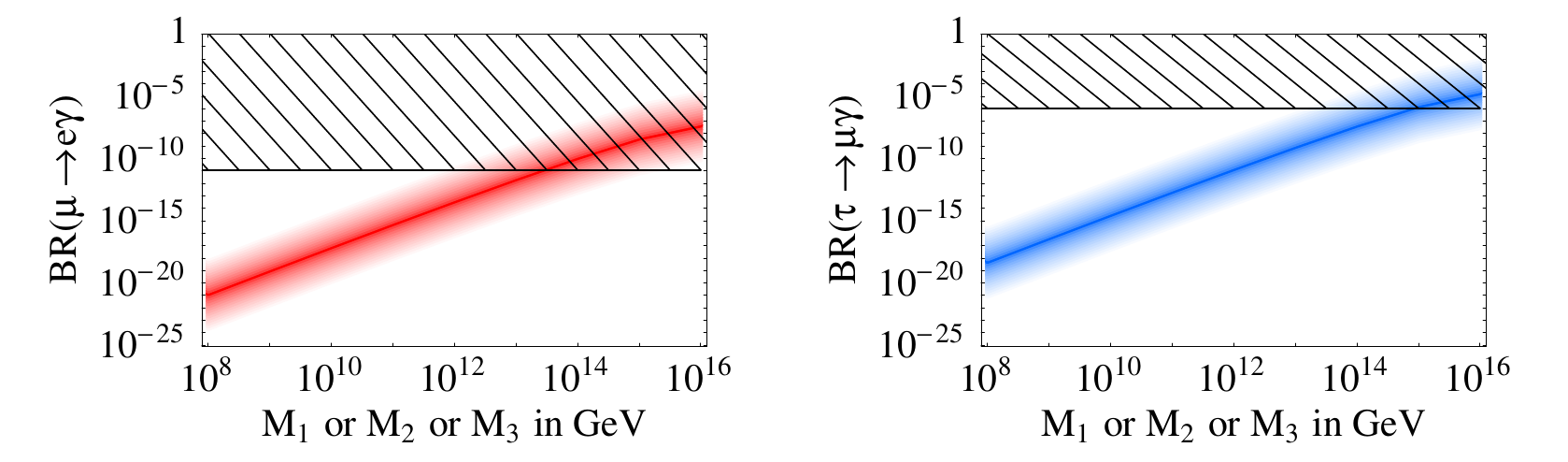}$$
\caption[SUSY see-saw predictions for LFV]{\label{fig:LFVnu}\em Predictions for $\mu\to e\gamma$ and $\tau\to\mu\gamma$ rates
induced by Yukawa neutrinos and supersymmetry under assumptions discussed in the text.
The masses of right-handed neutrinos are unknown;
leptogenesis suggests $M_1\sim 10^9\GeV$.}
\end{figure}

In order to get more concrete results, we
assume that the Yukawa couplings of the neutrinos
follow the same pattern observed in their masses:
large $\mu/\tau$ mixing in the `atmospheric' sector
and large $e/\mu/\tau$ mixing in the `solar' sector.
This is not demanded by data and plausible counterexamples exist~\cite{AFM}.
Having fixed the mixing angles with this arbitrary assumption, 
we can plot in fig.\fig{LFVnu} the predicted $\mu\to e\gamma$ and $\tau\to\mu\gamma$
rates as function of the remaining unknown parameters,
the right-handed neutrino masses $M_i$.
Leptogenesis suggests 
that the lightest $\nu_R$ has a mass around $10^9\GeV$ (see section~\ref{leptogenesi}).
Our present ignorance of sparticle masses
induces an additional subdominant uncertainty of about 5 orders of magnitude,
represented in fig.\fig{LFVnu} by the thickness of the lines.

At low energy, lepton-flavour violation is transmitted
from SUSY sparticles to SM particles mostly as an effective dipole operator.
This implies characteristic correlations between $\mu\to e \gamma$, $\mu\to e\bar{e}e$, $\mu\to e$ rates.

In conclusion, in supersymmetric see-saw models
large neutrino Yukawa {\em couplings} (e.g.\ $\lambda_N\sim \lambda_t$)
can generate sizable $\mu\to e\gamma$ and $\tau\to\mu\gamma$ rates.
These effects cannot be predicted in terms of measured neutrino {\em masses}.
This means that, if these effects will be measured and if one can get convinced that they are
generated only by neutrino Yukawa couplings,
we would learn some information about neutrino Yukawa couplings.
We cannot realistically hope that in this way it will be possible to experimentally reconstruct
all see-saw parameters; 
a realistic summary of prospects is:
$$
\begin{array}{cl}
m_1, m_2, m_3 & \hbox{Two $\Delta m^2$ already measured;  $m_1$ measurable only if large.}\\
\theta_{12},\theta_{23},\theta_{13} & \hbox{$\theta_{12}$ and $\theta_{23}$ already 
measured, $\theta_{13}$ measurable.}\\
\phi,\alpha,\beta & \hbox{$\phi$ measurable from oscillations; one combination of $\alpha,\beta$ affects $0\nu2\beta$.} \\[2mm]
[Y_N]_{ee,\mu\mu,\tau\tau} & \hbox{Two measurable from slepton mass splittings}.\\
|Y_N|_{e\mu,\mu\tau,e\tau} &\hbox{All measurable from LFV rates.}\\
{\rm arg}\, [Y_N]_{e\mu,\mu\tau,e\tau} & \hbox{One combination from electron electric dipole?}
\end{array}$$
Measuring CP-violating phases is particularly important if one hopes
to test whether leptogenesis generated the observed baryon asymmetry.
However various cancellations make the computation of electric dipoles
induced by neutrino Yukawas tricky, and its result small.
The reason behind these cancellations is that the Lagrangian has a
${\rm U}(3)_L\otimes{\rm U}(3)_E\otimes{\rm U}(3)_N$ flavour symmetry,
broken explicitly by the Yukawa couplings and right-handed neutrino masses.
This restricts the way in which Yukawa couplings can combine to give a $3\times 3$ flavour matrix of EDMs, that transforms in the $(3,3,1)$ representation.
Detailed computations find that a detectable $d_e$ can only
arise if $Y_N$ has large, order one, entries and if $\tan\beta$ is large.

In SUSY models LFV rates, and in particular the small electric dipoles, can receive other contributions not related to neutrino Yukawa couplings: this would complicate the situation.

SUSY models where 
neutrino masses are mediated by one scalar triplet (rather than by right handed neutrinos) allow
to predict the relative ratio of different LFV rates if the whole neutrino mass matrix is known.

\section{Neutrinos in extra dimensions}\label{extradims}\index{Neutrino!in extra dimensions?}
The electric force between an electron and a proton is about 30 orders of magnitude
stronger than their gravitational force.
Supersymmetry can explain this huge number by relating the Fermi scale
to the scale of supersymmetry breaking,
that can naturally be much smaller than the Planck mass.

Different interpretations exist.
If gravity propagates in $d$ flat extra dimensions of radius $R$,
while SM particles are confined on a 3+1 dimensional (mem)brane, the Newton force at 
$r \ll R$
is proportional to $1/r^{2+d}$.
Depending on $R$ and $d$, gravity can become strong at values of
$1/r\sim M_D$ below the Planck scale.
Choosing $M_D\sim \TeV$ --- the lowest value allowed by collider experiments ---
one could try to address the Higgs mass hierarchy problem in this context.
The experimental success of the SM disfavors this drastic possibility,  but
only experiments that will start in $\sim$2008 can exclude it.
In the meantime, many authors explore its various 
phenomenological consequences.
We review the ones concerning neutrinos.

\medskip

Extra dimensions suggest a non-standard scenario  for neutrinos.
If right-handed neutrinos propagate in the extra dimensions,
their Yukawa couplings to SM neutrinos
are suppressed by $M_D/M_{\rm Pl}$,
 like gravitational couplings.
If one also assumes that Majorana neutrino masses are negligible
(for example by imposing conservation of lepton number, that is no longer
a good accidental symmetry as in the SM),
this scenario gives small Dirac neutrino masses $m_\nu \sim v M_D/M_{\rm Pl}$,
somewhat smaller than what suggested by neutrino data.
In non minimal models (e.g.\ with right-handed neutrinos propagating in
$\delta<d$ extra dimensions, or with warped extra dimensions) the clean connection with gravity is lost and
one can get any desired $m_\nu$.
For concreteness, we consider $4+\delta$-dimensional massless fermions
$\Psi_i(x_\mu,y)$ which, inside their components,
contain the  ``right-handed neutrinos'' $\nu_{Ri}$
($i=1,2,3$ is the generation index). The fermions $\Psi_i$
interact in our brane located at $y=0$, through their components $\nu_{Ri}$,  
with the lepton doublet $L_i$ in a way that conserves total 
lepton number. \begin{equation}
\label{eq:5action}
\Lag = [\Lag_{\rm SM} - 
L_i \lambda_{ij} \nu_{Rj}H   +\hbox{h.c.}]\delta(y) + \bar{\Psi}_i i\Dsl\, \Psi_i
\end{equation}
where $i\Dsl$ is a $4+\delta$ dimensional Dirac operator,
$H$ is the SM Higgs doublet  and $\lambda$ is a matrix
of Yukawa couplings with dimensions ${\rm (mass)}^{-\delta/2}$. 
As manifest from~(\ref{eq:5action}), $\lambda$ can be made 
diagonal without loss of generality at the price of introducing the usual 
unitary neutrino mixing matrix $V$.
Using the KK decomposition
\begin{equation}
\Psi_i(x,\vec{y})= \frac{1}{\sqrt{V}}\sum_{\vec{n}} \Psi_{\vec{n}i}(x) \exp
\left(\frac{i \vec{n}\cdot\vec{y}}{R}\right) ,
\end{equation}
where $V \equiv (2\pi R)^\delta$ is the volume of the extra dimensions, and
performing the $d^\delta y$ integration, eq.~(\ref{eq:5action}) yields the 4-dimensional
neutrino Lagrangian
\begin{equation}
\Lag_\nu= \bar{L}_i i\ds \,L_i +\sum_{\vec{n}}\bar{\Psi}_{\vec{n}i} \left(i\ds\,-
\frac{\vec{n}\cdot \vec{\gamma}}{R}\right) \Psi_{\vec{n}i} - 
\left[ \frac{\lambda_{ij}}{\sqrt{V}}
L_i \nu_{R\vec{n}j}  H +\hbox{h.c.} \right] ,
\label{lagrquat}
\end{equation}
where $\vec{\gamma}$ are the extra-dimensional Dirac matrices.
After electroweak symmetry breaking, neutrinos 
obtain a {\color{blus} Dirac mass matrix
$m_\nu = \lambda v/\sqrt{V}$},
which are small if extra dimensions have a large volume.
More interestingly, the above Lagrangian also describes the KK excitations $\nu_{Rn}$ ($n=1,2,3,\ldots$) of the right-handed neutrinos,
which behave as a tower of sterile neutrinos with masses 
$m_n \sim n/R$ and mixing angles $\color{blu} \theta_n\sim m_\nu/m_n$
with the active ones.
 For simplicity we wrote explicit formul\ae{} in the case of the simplest extra
dimension, a $\delta$-dimensional torus with equal radii $R$,
but the phenomenon is quite general.

\subsection{Oscillation signals of extra-dimensional neutrinos}
If at least one extra dimension is large enough, some KK excitations can be light enough,
to give detectable effects in neutrino oscillation experiments: $m_n\circa{<}\eV$.
Since KK neutrinos are a particular form of sterile neutrinos, 
experimental data tell that they cannot be the source of the solar or atmospheric anomalies.
One could try to use KK's to fit the LSND anomaly by choosing $1/R\sim \eV$.
In such a case the lightest KK $\nu_{1R}$ gives the dominant effect
obtaining a scheme qualitatively similar to the `3+1' scheme (see section~\ref{LSND}).
However:
1)  the `3+1' scheme has some tension with
disappearance experiments and cosmology:
these problems become slightly more serious in the `3+KK' scheme.
2) unlike in the `3+1' model, the minimal `3+KK' model {\em predicts} 
the mixing angle between $\nu_{1R}$ and the SM neutrinos $\nu_\ell$ to be
$ \pi V_{\ell 3} m_{\rm atm} R/\sqrt{3}$
(assuming a hierarchical spectrum of active neutrinos;
the other spectra are more problematic), which is too small in view of the 
{\sc Chooz} bound on $V_{e3}$.
3) As discussed in the next section,
$1/R\sim \eV$ is not compatible with the observed SN1987A neutrino burst.

These problems can be alleviated by inventing less predictive 
non minimal models, obtained e.g.\ by adding 5-dimensional neutrino mass terms $\mu$.
At the end, one can introduce extra unknown parameters $\mu,\mu'$
that modify the predictions for the masses of the KK into
$m_n^2\sim \mu^2 + (n/R)^2$ or into $m_n\sim \mu + n/R$
and the predictions for the mixing angles into
$\theta_{n}\sim \mu'/m_n$,
such that they are no longer related to neutrino masses.

\medskip

In conclusion, oscillations into KK cannot be the source of observed
oscillations, and do not give new appealing interpretations to existing anomalous results.
If $1/R\gg\eV$ the KK neutrinos have too small mixing angles  to give  detectable effects in future neutrino oscillation experiments.
Still, there are two places where their effects could be seen.

\subsection{Supernova signals of extra-dimensional neutrinos}
The SN1987A observation puts severe constraints on extra dimensional right-handed neutrinos.
This is because KK neutrinos can carry away from the supernova too much energy in invisible channels,
thus weakening the observed neutrino burst in an unacceptable way.
The energy loss rate in invisible channels, $W_{\rm inv}$, must be less
than about $10^{19}\,{\rm erg}/{\rm gr}\cdot\sec$ for typical average conditions of a SN core with a
density $\rho\approx 3\cdot 10^{14}{\rm gr}/\cm^3$ and a temperature $T\approx 30\MeV$
(see section~\ref{subs:ccs}).
KK neutrinos can be produced
both by incoherent scatterings and by coherent oscillations.

\medskip

The rate of {\em incoherent production} of a single KK state is $\Gamma_{\rm inc}\sim (m_\nu/E_\nu)^2\Gamma_\nu$,
where $m_\nu$ is the largest neutrino mass,
$E_\nu\sim 3T\sim 100\MeV$ is a typical neutrino energy and 
$\Gamma_\nu\approx G_{\rm F}^2n_N E^2$
is the collision rate of a standard neutrino in the supernova core, in terms of the
nucleon density $n_N$.
Multiplying for the number of KK states, approximately $(RE)^\delta$, the corresponding energy loss rate can be estimated as
$W_{\rm inv}\approx (m/E)^2 (RE)^\delta \cdot W_\nu$, where $W_\nu\approx 10^{27}{\rm erg}/{\rm gr}\cdot\sec$ is the energy loss rate
produced by $\Gamma_\nu$ if the standard neutrinos were not trapped.
Requiring $W_{\rm inv}/W_\nu \circa{<}10^{-8}$ gives
\begin{equation}\label{eq:R-coherent}
R<0.2\,\hbox{mm} \cdot 10^{-11(1-1/\delta)} \left(\frac{m^2}{10^{-3}\eV^2}\right)^{-1/\delta}.
\end{equation}
Only in the case of a single large extra dimension, $\delta = 1$,
the KK can be light enough to directly affect neutrino physics.

\medskip

We now consider the effect of {\em coherent oscillations} of the active neutrino state
associated with the mass $m_\nu$ into its KK tower.
Depending on the sign of the matter potential $A\sim 10\eV$ in the supernova core, MSW resonant oscillations will take place
either in the neutrino or in the antineutrino channel, but not in both. 
KK states lighter than $m_n\approx n/R\approx (AE_\nu)^{1/2}\approx
10^{4}\eV$ are involved in resonant transitions.
The effect of these oscillations can be reliably estimated for small $\xi=m_\nu R$,
because due to the small vacuum mixing angle  between the $n$-th KK state and the standard
neutrino, $\theta_n\approx \xi/n$, 
the width of the $n$-th resonance is smaller than the separation between two contiguous resonances.
As a consequence, the survival probability for a standard neutrino (or antineutrino) produced in the core can be computed
as the product of the survival probabilities in each resonance crossed during a mean free path,
$P_{\nu\nu}\approx \prod_n P_n$.
The survival probability at each level crossing is
given by the Landau-Zener approximation (section~\ref{MSWresonance}),
$P_n\approx e^{-\pi \gamma_n/2}$, where
$$\gamma_n \approx \left.\frac{4\xi^2}{R^2E}\frac{V}{dV/dr}\right|_{\rm res}$$
is the adiabaticity parameter at the $n$-th resonance crossing.
Approximating $V/(dV/dr)$ with the radius of the core, $r_{\rm core}\approx 10\km$, $\gamma_n$ is of order
$m^2_\nu/(10^{-3}\eV^2)$ and is approximately $n$-independent.
Hence the individual $P_n$ are expected to deviate sensibly from unity.
The condition $P_{\nu\nu}\circa{<}10^{-8}$ can be satisfied only
if all KK are heavy enough that no resonance is crossed,  {\color{blu}$R\circa{<}\,$\AA}.

\medskip

If this condition is violated the SN gets strongly modified, 
possibly in an interesting and still allowed way.
Depending on the sign of $V$ (which can be different in different regions of the SN)
either $\nu$ or $\bar\nu$ (but not both) promptly get converted into KK
which escape from the SN.  This process varies  the chemical composition
of the SN, and therefore the matter potential $V$.
This fast loss of energy stops either when $V=0$ is reached
(if $\nu_e$ are involved)
or due to Pauli blocking
(when roughly  all $\nu_{\mu,\tau}$ or all $\bar\nu_{\mu,\tau}$ have escaped).
In this way about one half of the SN energy gets lost in KK.
The remaining half is emitted in neutrinos, with non standard flavour ratios
(typically more $\nu$ than $\bar\nu$).  
The constraint on the radius of the extra dimension
gets relaxed at least down to {\color{blu}$R\circa{<} 10\,$\AA}: at larger $R$
non-resonant oscillations produce too many KK.
Detailed predictions have been computed only in non-minimal models,
where $R$ can be significantly larger.

\begin{figure}
$$\includegraphics{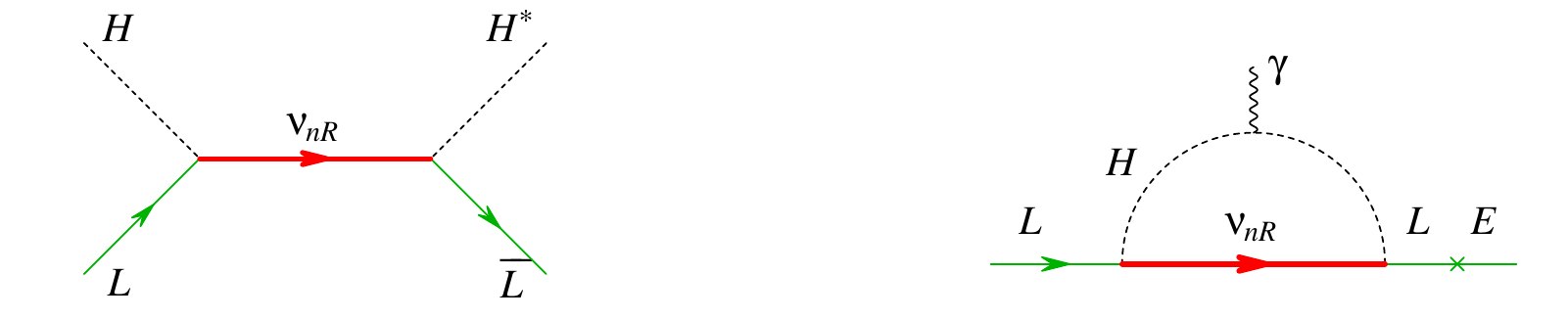}$$
\caption[Effects of neutrinos in extra dimensions]{\em  Tree-level and one-loop virtual effects of extra-dimensional right-handed neutrinos.
\label{fig:FeynExtrad}}
\end{figure}

\subsection{Virtual signals of extra-dimensional neutrinos}\label{dim6seesaw}
Virtual exchange of right-handed neutrinos at tree level
gives the dimension-six operator (see fig.\fig{FeynExtrad}a)
\begin{equation}\label{eq:TreeLevel}
\Lag_{\rm tree} =  \epsilon_{ij}~
2\sqrt{2}G_F (H^\dagger \bar{L}_i)i\ds (HL_j),
\end{equation}
where  $\epsilon_{ij}=\epsilon_{ji}^*$ are dimensionless couplings.
In the usual four-dimensional scenario, this operator arises with negligibly small coefficient,
suppressed by the large squared masses of right-handed neutrinos.
The  hierarchy problem motivates TeV-scale quantum gravity, suggesting
$\epsilon_{ij}\sim  (v/\TeV )^2\sim 10^{-{\rm few}}$.
In the minimal model of eq.\eq{5action} 
$\epsilon_{ij}\propto ( m_\nu m_\nu^\dagger)_{ij}$, with
an ultraviolet divergent proportionality factor, that can be only  estimated by introducing some arbitrary cut-off.
Exploiting the predicted flavour structure,
the six $\epsilon_{ij}$ parameters can be expressed in terms of a single unknown $\epsilon$ as
\begin{equation}\label{eq:minimal}\begin{array}{ll}
\displaystyle
\epsilon_{\tau\tau}\approx \epsilon_{\mu\mu}\approx \epsilon_{\mu\tau} \approx\frac{\epsilon}{2},\qquad&
\displaystyle
\epsilon_{e e} \approx \left( |V_{e3}|^2 + \frac{\Delta m^2_{\rm sun}}{3\Delta m^2_{\rm atm}} \right)\epsilon , \\[5mm]
\displaystyle
\epsilon_{e \mu} \approx \left(\frac{|V_{e3}|}{\sqrt{2}} + 
\frac{e^{-i\phi}\Delta m^2_{\rm sun}}{3\Delta m^2_{\rm atm}} \right)\epsilon ,\qquad&
\displaystyle
\epsilon_{e \tau} \approx \left(\frac{|V_{e3}|}{\sqrt{2}} - 
\frac{e^{-i\phi}\Delta m^2_{\rm sun}}{3\Delta m^2_{\rm atm}} \right) 
\epsilon,
\end{array}
\end{equation}
where we assumed normal hierarchy, maximal atmospheric mixing
and $\tan^2\theta_{12}=1/2$.

The  operators in eq.\eq{TreeLevel} induce
potentially large flavour transitions $P(\nu_i\to
\nu_j)\sim |\epsilon_{ij}^2|$  at ${\cal O}(L^0)$ (i.e.\ at very short
baselines $L\ll E_{\nu}/\Delta m^2$),
and CP-violating effects at ${\cal O}(L^1)$ 
(rather than at ${\cal O}(L^2)$ and ${\cal O}(L^3)$ as in ordinary oscillations). 
At one-loop level, other operators are 
generated, giving rise to rare muon and tau processes that violate 
lepton flavour (see fig.\fig{FeynExtrad}b). Again, the coefficients of these operators  are cut-off dependent 
and can only be estimated;
they are roughly suppressed by a four-dimensional loop factor $\sim \alpha/4\pi$
with respect to the tree level operator in\eq{TreeLevel}.
Their flavour structure is again dictated by the $\epsilon_{ij}$ parameters.
Effects in charged leptons are suppressed by a one loop factor with respect to
effects in neutrinos:
\begin{equation}\label{eq:5dvirtual}
P(\nu_i\to \nu_j; L\approx 0) \sim |\epsilon_{ij}^2|,\qquad
{\rm BR}(\ell_i\to \ell_j \gamma) \sim |e\epsilon_{ij}/4\pi|^2, \qquad
(i\neq j).
\end{equation}
Present bounds from neutrino experiments  and
from charged lepton processes are summarized in table~1.
Due to the loop factor,
detectable neutrino effects are compatible with lepton flavour violating bounds.\footnote{$\SU(2)_L$-invariant non renormalizable operators
parameterize the low-energy effects of
generic new physics
too heavy for being directly probed.
Various papers~\cite{nuNRO} studied how generic NRO could affect future neutrino experiments.
However, a typical NRO affects both neutrinos and charged leptons at tree level
(since they are unified into lepton doublets $L$).
In such a case, present bounds on charged leptons prevent detectable effects in future neutrino experiments,
except maybe in $\nu_\mu\leftrightarrow\nu_\tau$ transitions
The operator of eq.\eq{TreeLevel} is the simplest exception.
Therefore it is interesting to study more generally which
kind of new physics can induce it.
It arises if lepton doublets mix at the TeV scale with ``right-handed neutrinos''.
In four dimensions such models are not motivated by neutrino masses;
conservation of lepton number (or something equivalent)
must be imposed to avoid too large neutrino masses~(see in~\cite{NuSUSY}).
Models with right-handed neutrinos in a warped extra dimension~\cite{NuSUSY} also induce the operator in\eq{TreeLevel},
but the connection between neutrino masses and $\epsilon_{ij}$ is extremely model dependent.}
Values of $\epsilon_{ij}\circa{>}10^{-4}$ (including possible CP-violating phases)
will be probed by future neutrino experiments.
In the minimal model, $\mu,\tau,\pi$ decays and precision LEP data imply $\epsilon \circa{<} 0.004$.

\medskip

\begin{table}[t]
$$
\begin{array}{|c|lclc|lclc|}\hline
& 
\multicolumn{4}{|c|}{\hbox{\color{blus} present {\bf bounds} from experiments with\color{black}}} & 
\multicolumn{4}{|c|}{\hbox{\color{blus} future {\bf sensitivity} from experiments with\color{black}}} \\ \hline
&
\multicolumn{2}{|c}{\hbox{\color{blus} neutrinos\color{black}}}&
\multicolumn{2}{|c}{\hbox{\color{blus} charged leptons\color{black}}}&
\multicolumn{2}{|c}{\hbox{\color{blus} neutrinos\color{black}}}&
\multicolumn{2}{|c|}{\hbox{\color{blus} charged leptons\color{black}}}\\ \hline
\color{blus}|\epsilon_{e\mu}|\color{black} &  
<0.05 &  \hbox{({\sc {\sc Karmen}})} &
\circa{<} 10^{-4} &  (\mu\to e \gamma) &
 \sim 10^{-4} &  \hbox{($\nu$ factory)} &
 \sim 10^{-5\div 6} &  (\mu~{\rm decays}) \\
\color{blus}|\epsilon_{e\tau}|\color{black} &  
<0.09 &  \hbox{(NOMAD)} &
\circa{<} 10^{-1} &  (\tau\to e \gamma) &
 \sim 10^{-4} &  \hbox{($\nu$ factory)} &
 \sim 10^{-1\div 2} &  (\tau\to e\gamma) \\
\color{blus}|\epsilon_{\mu\tau}|\color{black} &  
<0.013 &  \hbox{(NOMAD)} &
\circa{<} 10^{-1} &  (\tau\to \mu \gamma) &
 \sim 10^{-4} &  \hbox{($\nu$ factory)} &
 \sim 10^{-1\div 2} &  (\tau\to\mu\gamma) \\
\color{blus}\epsilon_{ee}\color{black} &  
<0.025 &  \hbox{(reactors)} &
\circa{<} 10^{-3} &  (Z\hbox{ data}) &
 \sim 10^{-3} &  \hbox{($\nu$ factory)} &
 \sim 10^{-3\div 4} &  (Z\hbox{ data}) \\
\color{blus}\epsilon_{\mu\mu}\color{black} &  
~~~\hbox{---} &&
\circa{<} 10^{-3}&  (Z\hbox{ data}) &
 \sim10^{-3} &  \hbox{($\nu$ factory)} &
 \sim 10^{-3\div4} &  (Z\hbox{ data}) \\
\color{blus}\epsilon_{\tau\tau}\color{black} &  
~~~\hbox{---}  &&
\circa{<} 10^{-2} &  (Z\to\nu\bar\nu) &
~~~\hbox{---} &&
 \sim 10^{-3} &  (Z\to\nu\bar\nu) \\ \hline
\end{array}$$
\caption[Bounds and sensitivity to anomalous neutrino couplings]{\em Bounds  and future sensitivity reaches for the $\epsilon_{ij}$ coefficients defined by eq.\eq{TreeLevel}.}\end{table}

Notice that the same low-energy physics can be obtained without extra dimensions,
by mixing SM neutrinos with 
sterile neutrinos with TeV-scale pseudo-Dirac masses.
The particular form of their mass matrix 
guarantees that lepton number
is conserved, avoiding too large neutrino masses.
In such renormalizable 4-dimensional models the
estimates of eq.\eq{5dvirtual} can be replaced by precise results,
which depend on the Yukawa couplings.

\begin{figure}
\begin{center}
\includegraphics[width=0.8\textwidth]{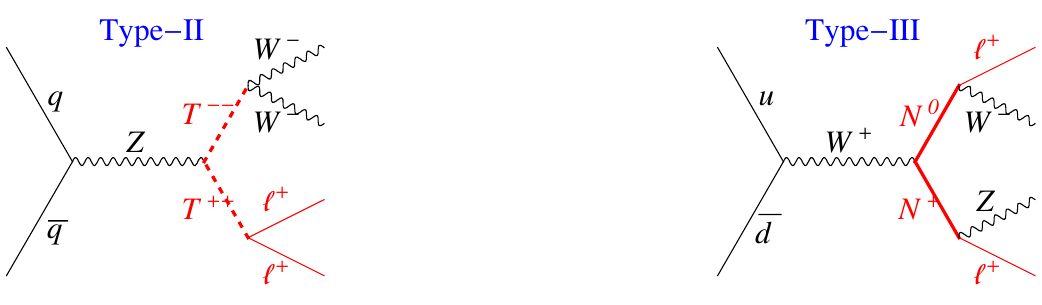}
\caption[see-saw at LHC]{\label{fig:LHCseesaw}\em 
Lepton-number-violating signals of type-II and type-III see-saw at the LHC $pp$ collider.}
\end{center}
\end{figure}

\section{Testing neutrino masses at LHC?}\label{LHC}
We have seen that supersymmetric or extra dimensional see-saw models have some promising testable consequence.
We here discuss the less promising signals of the minimal scenarios.

Type I, II and III see-saw cannot be distinguished as long as we can only observe
their dominant low energy manifestation: all of them generate the same unique $(LH)^2/2\Lambda_L$ dimension 5 operators,
that give Majorana neutrino masses.

\medskip

The $(LH)^2$ operator also induces couplings of two left-handed leptons with the higgs
or with heavy SM vectors, but all the resulting rates are uninterestingly small for the observed $\Lambda_L \sim 10^{14}\GeV$:
e.g.\ the cross-section for the lepton-number-violating scattering
$ee\to W^- W^-$ is $\sigma \sim 1/\Lambda_L^2$ above the kinematical threshold.

\medskip

Going to higher dimension 6 order, see-saw models generate different operators, that however likely give undetectably small effects.
Type-I see-saw generates the $(H^\dagger \bar{L})i\ds (HL)$ operator studied in section~\ref{dim6seesaw},
that manifests as flavor violating neutrino interactions~\cite{nuNRO}.
Type-III see-saw gives the $(H^\dagger \tau^a \bar{L})\Ds(H \tau^a L)$, that also give flavor violating interactions for charged leptons.
Type-II see-saw gives flavor violating $|L\tau^a L|^2$ four-fermion interactions, as well as operators involving the higgs doublet $H$.
At loop level, see-saw models induce the lepton-number-violating operator 
$\epsilon\cdot 2\sqrt{2}G_{\rm F} (\bar \mu \gamma_\mu P_L e)^2$:
the non-observation of muonium $M=e^-\mu^+$ oscillations into anti-muonium $\bar M = e^+ \mu^-$
only constrains  its coefficient to be $|\epsilon| < 3~10^{-5}$.

\medskip

Some of the particles that mediate neutrino masses might be light enough to be produced at colliders, such as LHC.
In such a case their contribution to quantum correction to the higgs mass squared $m_H^2$ would be naturally smaller than $m_H^2$.
However no detectable effect arises if they only have the small couplings needed to mediate the small neutrino masses.
The scalar or fermion $\SU(2)_L$ triplets of type-II and type-III see-saw models also have gauge interactions, fully predicted
by theory.
The right-handed neutrinos of type-I see-saw are neutral under SM gauge interactions, so that they are undetectable.
Still it is possible that extra vector bosons exist at the TeV scale, where the electroweak $\SU(2)_L\otimes{\rm U}(1)_Y$
gauge symmetry of the SM might be extended into $\SU(3)_L\otimes {\rm U}(1)$, or adding an extra ${\rm U}(1)_{B-L}$ vector,
possibly associated with  $\SU(2)_R$ vectors.
The resulting signals are model-dependent and qualitatively similar to the well-defined signals of type-II or III see-saw,
that we now discuss.

Gauge interactions lead to pair production of the various components
of the fermion $N=\{N^0,N^\pm\}$ or of the scalar $T=\{T^0,T^\pm,T^{\pm\pm}\}$ triplets.
The production cross-sections depend strongly on $M$:  for $M=1\TeV$ one has
 $\sigma(pp\to NN) \sim 1\,{\rm fb}$ and $\sigma(pp\to TT^*)\sim 0.1\,{\rm fb}$,
 smaller because scalars are not produced in $s$-wave.

The small interactions that lead to neutrino masses also lead to triplet decays.
The life-time of a fermion triplet with mass $M$
that gives a contribution $\tilde{m}_1$ to neutrino masses is
\beq \tau_{N_0} = \tau_{N_\pm} = \frac{8\pi v^2}{\tilde{m}_1 M^2}=1.5\cm
\frac{\rm meV}{\tilde{m}_1} \Big(\frac{100\GeV}{M}\Big)^2\label{eq:displacedvertex}\eeq
up to corrections suppressed by $M_Z^2/M$: the smallness of neutrino masses can lead to
decay vertices detectably displaced from the production point.
The main decay modes are $N^\pm \to \ell^\pm Z,\nubarnu_\ell W^\pm$ and $N^0\to W^+\ell^-,W^-\ell^+$,
all with comparable branching ratios.

In view of the type-II see-saw Lagrangian \eq{Ltriplet}, scalar triplets 
have two decay modes: $\Gamma(T^{\pm\pm}\to \ell^\pm\ell^\pm)\sim \lambda_T^2 M/4\pi$
and $\Gamma(T^{\pm\pm}\to W^\pm W^\pm)\sim \lambda_H^2 M/4\pi$.
If $\lambda_T\sim \lambda_H$ the triplet life-time is comparable to eq.\eq{displacedvertex},
otherwise it can be much shorter.
Lepton number is violated only when both couplings $\lambda_T$ and $\lambda_H$ are present,
so that a lepton-number-violating signal arises when both decays happen, as in the left diagram in fig.\fig{LHCseesaw}:
given that the each of the heavy SM vectors dominantly decays into two quarks, the main signal is
$pp\to \ell^\pm\ell^\pm 4j$.
The right diagram in fig.\fig{LHCseesaw} shows how type-III see-saw can produce the same final states.
Appropriate cuts can suppress the SM backgrounds to these processes:
apparent violation of lepton number arises when two neutrinos carry away lepton number and undetectably small
missing transverse energy.
In type-II see-saw the invariant mass of the two same-sign leptons is peaked at $M$.
Other decay modes give rise to less notable signatures.

Triplets can produce baryogengesis via thermal leptogenesis only if heavier than 1.6 TeV~\cite{leptogenesis2},
beyond the discovery reach of LHC.

Virtual corrections to precision data do not seem particularly significant:
neglecting the small coupligs that lead to neutrino masses,
the gauge interactions of the scalar or fermion triplets lead to minor corrections
to the $W$ and $Y$ parameters, as well as to the anomalous magnetic moment of the muon.

Finally, it is possible that, unlike what suggested by the smallness of neutrino masses,
some see-saw couplings are sizable even for $M\sim \TeV$.
Within type-II see-saw one has $m_\nu = \lambda_T \lambda_H v^2/M$
so that one of the two couplings can be large if the other is very small.
Within type-I and type-III see-saw, one could have in the Lagrangian a structure of the form
\beq  M N_1N_2 + \lambda_1 N_1LH + \lambda_2 N_2 LH\eeq
giving $m_\nu = \lambda_1 \lambda_2 v^2/M$ such that again one of the two couplings could be large.
Such large couplings could lead to detectable signatures, e.g.\ to dimension-6 operators not suppressed by 
$\Lambda_L=M/\lambda_1\lambda_2 \sim 10^{14}\GeV$.

\chapter{Applications}\label{Applications}
Techniques developed for high-energy physics resulted into practical applications,
but it seems hard to find practical applications for particles that decay in $10^{-20}\,{\rm sec}$.
From this point of view, neutrinos might be an exception: unlike
 other particles not present in everyday life  (such as muons, $Z$, $W$, etc...)
neutrinos are light and stable.
Furthermore, neutrinos can do things that ordinary particles present in everyday life ($\gamma, e, p ,n$)
cannot do:  neutrinos can cross the earth or the sun without being absorbed.
 For practical purposes neutrinos have too much of this property:
 neutrino interaction rate are so small that neutrino detectors are big and expensive.
Interaction rates grow with energy,
but neutrino sources are more easily found at lower energies. 
Thanks to developments of detection techniques, 
 neutrino physics already started to be relevant outside fundamental physics,
  and some more or less practical applications have been proposed~\cite{Applications}.
  Although short, this section exists.
 
 Detectors such as CHOOZ and KamLAND
 have monitored the activity of  nuclear reactors.
 A web of Mton-size $\bar\nu_e$ detectors could be used to discover and locate hidden nuclear reactors
 (as well as producing interesting physics): its cost is not out-of-scale if seen as a military project
 rather than as an experiment.
 A few detectors, placed around a small country, can control its activity.
 
One detector, placed close to one nuclear power plant can see enough events
(a typical number is $10^{5}$ per day in a 10 ton detector at a distance of 10 m)
that allows to daily measure its total power within few $\%$ precision and
the composition of its fission elements (from the energy spectrum;
however  presently the $^{239}$Pu percentage would be poorly measured).

 KamLAND observed $\bar\nu_e$ emitted by radioactive material inside the earth~\cite{Geonu},
 with results that agree with earth models: such MeV-energy
 neutrinos can be used to measure the density of the earth core.
A km-scale detector such as IceCUBE~\cite{IceCUBE} can show that the earth has a denser core
(possibly with $5\sigma$ confidence level) from the absorption of
atmospheric neutrinos with energies $\sim 10\TeV$.

 Solar experiments confirmed predictions of solar models,
 in particular the temperature around the center of the sun.
Future experiments can test the expected low-energy features
of the solar neutrino spectrum and search for possible
 time variations of the solar neutrino rate.
 SK and SNO can see where the sun is during the night.
 More importantly,
 earth matter effects are predicted to increase by a few $\%$ the rate
 of solar neutrinos detected during the night.
Future big experiments could see this effect; furthermore
matter inhomogeneities (e.g.\ oil instead of rock) located below the detector
give rise to extra matter effects: a huge neutrino detector
(installed e.g.\ in a submarine) has been discussed as a technique for searching oil.

Neutrino beams have been proposed as a tool to send informations towards
submarines in deep sea, that can today be reached only by Hz scale radio-waves,
with a correspondingly low information transfer rate.

More unrealistically, neutrino beams with energy of about 1000 TeV has
been discussed as a tool for destroying nuclear weapons.

Neutrinos also have unwanted effects.
A circular $\mu^-\mu^+$ collider would allow higher energies
than circular $e^-e^+$ colliders, because energy losses due to photon radiation
get suppressed by a $(m_e/m_\mu)^4$ factor.
However $\mu^\pm$ decay into $e^\pm$ and $\nubarnu$,
giving rise to technical problems that can possibly be solved,
but also to high-energy neutrino radiation which cannot be shielded and
which would be too dangerous.

Finally, it is often said that supersymmetry was discovered thanks to strings;
indeed the third re-discovery  happened in the context of a string model.
Similarly, the second re-discovery happened thanks to $\nu_e$,
in an (incorrect) attempt to interpret it as the massless Goldstone boson of spontaneously broken supersymmetry~\cite{nuSUSY}.




%% file: review_appendici.tex
\def\baselinestretch{0.91}

\chapter{Acronyms}\label{Acronyms}
For clarity, we list many TAA
(Technical Acronyms and Abbreviations)
employed in neutrino physics and in our review.

\begin{multicols}{2}
\begin{itemize}\small
\item[$0\nu2\beta$] Neutrino-less double beta (decay).
\item[AGN] Active Galactic Nucleus.
\item[BR] Branching Ratio.
\item[CMB] Cosmic Microwave Background.
\item[CC] Charged Current: scattering that transforms $\nu$ into charged leptons.
\item[CL] Confidence Level.
\item[CNGS] CERN to Gran Sasso long baseline experiment.
\item[CP] Charge Conjugation times Parity.
\item[CPT] CP times Time reversal.
\item[CR] Cosmic Rays.
\item[DIS] Deep Inelastic Scattering.
\item[DM] Dark Matter.
\item[DM$\nu$] Neutrinos from DM annihilations.
\item[dof] degree of freedom.
\item[EDM] Electric Dipole Moment.
\item[ES] Electron Scattering (of neutrinos).
\item[FWHM] Full Width at Half Maximum (of a bell).
\item[FS] Free Streaming.
\item[GNO] Gallium Neutrino Observatory: a solar neutrino experiment.
\item[GRB] Gamma Ray Burst.
\item[GUT] Grand Unified Theory.
\item[IBD] Inverse Beta Decay, $\bar\nu_e p\to e^+ n$.
\item[KamLAND]  Kamioka Liquid scintillator Anti-Neutrino Detector.
\item[K2K] KEK to Kamioka long-baseline experiment.
\item[LEP] the most recent $e\bar{e}$ collider at CERN.
\item[LFV] Lepton Flavour Violation.
\item[LHC] Large (or Last) Hadron Collider.
\item[LMA] Large Mixing Angle:  the solution to the solar neutrino anomaly.
\item[LOW] a solution of the solar neutrino anomaly excluded by data.
\item[LSND] A reactor experiment that claims $\bar{\nu}_\mu\to\bar{\nu}_e$ oscillations.
\item[LSZ]  Landau-St{\"u}ckelberg-Zener level crossing probability.
\item[MINOS] FermiLab to Homestake  long-baseline experiment.
\item[MSSM] Minimal Supersymmetric Standard Model.
\item[MSW] Matter corrections to neutrino oscillations.
\item[Mton] $10^{9}\,{\rm kg}$.
\item[NC] Neutral Current: scattering that does not transform $\nu$ into charged leptons.
\item[NRO] Non Renormalizable Operator.
\item[NS] Neutron Star.
\item[PDF] Parton Distribution Function.
\item[PDF] Portable Document Format.
\item[PDF] Probability Distribution Function.
\item[pot] protons on target.
\item[QFT] (Relativistic) Quantum Field Theory.
\item[QVO] Quasi-Vacuum Oscillations: a solution of the solar neutrino anomaly excluded by data.
\item[SK] SuperKamiokande: japanese experiment.
\item[SM] Standard Model of particles.
\item[SMA] Small Mixing Angle: a solution to the solar neutrino anomaly excluded by data.
\item[SN] Supernova.
\item[SNR] Supernova Remnant.
\item[SNO] Sudbury Neutrino Observatory: a solar experiment.
\item[SNR] SuperNova Remnant.
\item[SSM] Standard Solar Model.
\item[SUSY] SUperSYmmetry.
\item[T2K] Tokai to Kamioka: Japanese super-beam experiment.
\item[tof] time of flight.
\item[UHE] Ultra High Energy.
\item[UNO] Underground Nucelon/Neutrino Observatory: project for a Mton detector.
\item[vev] vacuum expectation value.
\item[VO] Vacuum Oscillations: a solution to the solar neutrino anomaly excluded by data.
\item[W\v{C}] Water \v{C}erenkov: experimental technique.
\item[WMAP] Wilkinson Microwave Anisotropy Probe.
\item[WKB] Wentzel-Kramers-Brillouin semiclassical approximation.
\end{itemize}
\end{multicols}

\def\baselinestretch{1}

\chapter{Statistics}\label{Statistics}\index{Statistics}

The physics that we would
like to learn is indirectly contained in
some set of data.
Since experiments need years and money,
it is important to extract as much as information as possible
with a correct and efficient use of statistics.
Usually, this issue is crucial when new discoveries start emerging from experiments,
and becomes less relevant after that data become so good that they speak by themselves.

According to a diffuse wrong impression,
statistics is some difficult mathematics proofed centuries ago,
so that a physicist interested in physics can apply it without understanding what he is doing.
In real analyses, where non-standard situations occur,
such attitude leads to violations of common sense at a numerically significant level.
While experimental results are usually presented in a way which looks and is correct,
reanalyses sometimes treat statistics in bizarre or inefficient ways.
Statistical inference indeed is a delicate subject.
Various subtle points lead to interesting philosophical discussions
with no fully satisfactory solution.
All details are carefully discussed in standard presentations,
at the price of obscuring the numerically important part.
We explain only the simple main tools,
skipping all subtle points that are numerically irrelevant in most practical circumstances.
In order to be concrete, we use a terminology specific
of reanalyses of neutrino data --- but the following discussions have a more general validity.
We address the two following topics:
\begin{enumerate}
  \item  In section~\ref{inference} we explain how,
given any theory,  data can be used to infer the values of its free parameters.

  \item Once 1.\ is done, we want to know
if the theory describes data in a satisfactory way, 
or if the data 
contain an indication for
some {generic} new physics.
This issue is discussed in section~\ref{gof}.
\end{enumerate}


\section{Statistical inference}\label{inference}
We proceed in 3 steps. 
Readers who consider step 2 unsatisfactory can read step 2$'$.

\paragraph{1.  Likelihood.}
We define the conditional probability $p(A|B)$ as the probability of $A$ assuming that $B$ is true.
The starting point of any analysis is
that we know $p(\hbox{data}|\hbox{theory})$: i.e.\
we can predict the probability distribution of observables
assuming any given theory.
This function is named `likelihood' and describes what the experiment is doing.
Experimental results should be reported in a way that
allows to compute it.
A typical example is
$p({\rm data}|{\rm theory})=
p(R|\Delta m^2,\theta) $
where $\Delta m^2,\theta$ are neutrino oscillation parameters
that we want to infer from $n$ measured neutrino rates $R$.
It is convenient to write $p$ in terms of a `chi-square' $\chi^2$ because
in many cases it is a simple quadratic function of $R$
\begin{equation}
\label{eq:chi2}
p({\rm data}|{\rm theory})=
p(R|\Delta m^2,\theta) = \frac{\exp[-\chi^2/2]}{(2 \pi)^{n/2} \sqrt{\det \sigma^2}}\qquad
\chi^2 = (R-T)\cdot \frac{1}{\sigma^2}\cdot (R-T)
\end{equation}
where 
$T(\Delta m^2,\theta)$ are the predicted rates
and $\sigma^2(T)$ is  the $n\times n$ error matrix.

We now discuss in which cases 
the likelihood is gaussian as function of $R$, so that we can use\eq{chi2}.
The  `central limit theorem' shows
that a random variable subject to many comparable statistical fluctuations
follows a gaussian distribution.
In some cases other distributions are appropriate 
(Possionian for low counting rates, etc.).
Often one does not know which bell to use.
This is the case of  `theoretical errors'
that arise e.g.\ when predictions can only be computed
with a limited accuracy.
It is pointless to debate if these errors should be described by
a gaussian or by a flat distribution
with the same standard deviation $\sigma$.
The real issue is honestly {\em estimating} $\sigma$:
doubling or halving the error is sometimes a matter of opinion.
In such cases we choose a Gaussian because 
working with other shapes would be harder:
approximating any bell with a Gaussian leads to important simple analytical results.
This approximation may give
an unrealistically low probability of many-$\sigma$ fluctuations;
however we do not really need to carefully distinguish a
$10\sigma$ from a $20\sigma$ evidence: both are well established.

\paragraph{2.  Bayesian inference.}
We have $p({\rm data}|{\rm theory})$ and we
want $p(\hbox{theory}|\hbox{data})$, i.e.\
we want to know the probability distribution that describes
our knowledge of $\Delta m^2,\theta$, assuming that the data are true.
Once we have $p(\Delta m^2,\theta|R)$ it is obvious how to use it:
for example the $90\%$ CL region
is the set of $\theta$ and $\Delta m^2$ values with higher relative probability
 that encloses $90\%$ of the total probability.

Up to a normalization factor $p(\hbox{theory}|\hbox{data})$ is given by the `Bayes theorem'~\cite{Statistics}
\begin{equation}
\label{eq:Bayes}
p(\hbox{theory}|\hbox{data}) \propto p(\hbox{data}|\hbox{theory})p(\hbox{theory})
\end{equation}
easily demonstrated using
 elementary properties of probability:
$$p(A\hbox{ and } B)=p(A|B)p(B)=p(B|A)p(A).$$
Eq.\eq{Bayes} dictates how the information on the theory
is updated from $p(\hbox{theory})$ to 
$p(\hbox{theory}|\hbox{data})$ when new experimental information
$p(\hbox{data}|\hbox{theory})$ becomes available.
In order to employ\eq{Bayes} for the first time, one needs to
assume some arbitrary {\em a priori} probability distribution $p(\hbox{theory})$,
called `prior'.
There is no way of choosing a non-informative prior.
For example a prior density probability flat  in $\tan^2\theta$ is different
from a prior flat in $\sin^2 2\theta$,
since the Jacobian relating these two variables is not constant
(and divergent when $\theta\to\pi/4$).
When data get so accurate
that a `small' $\theta$ range is selected,
the initial arbitrariness becomes numerically irrelevant,
since any reasonable prior is roughly constant in a small $\theta$ range.

\paragraph{3. Gaussian approximation.}
When data get accurate, the $\chi^2$ 
can be well approximated,
around its minimum, with a quadratic function of the theoretical parameters.
In gaussian approximation the $90\%$ CL region (2 dof) is the ellipse
$$\Delta \chi^2 (\Delta m^2 ,\theta)< 4.61\qquad\hbox{where}\qquad
\Delta \chi^2 (\Delta m^2 ,\theta)\equiv 
 \chi^2 (\Delta m^2 ,\theta)-\chi^2_{\rm min}.$$
In general, with $N$ theoretical parameters $\theta_i$ one can
analytically integrate their probability distribution 
over the $N-1$ `angular' combinations:
the probability density of  $z=\Delta\chi^2$ is
\begin{equation}
\label{eq:chiq}
p_N(z) = \frac{2^{-N/2}}{\Gamma(N/2)}z^{N/2-1}e^{-z/2} 
\end{equation}
To get the cut on $\Delta\chi^2$ corresponding to any desired CL,
one has to perform the last integral in $dz$ numerically,
or read it from `tables of the $\chi^2$ with $N$ dof'.
Here are some frequently used values:
\beq\label{eq:CLsamples} \begin{array}{c|ccccc}
&  68.3\% & 90\% & 95\% & 99\% & 99.73\%  \\ \hline
 \hbox{1 dof}     &   1 & 2.71 & 3.84 & 6.63 & 9\cr 
 \hbox{2 dof} & 2.30 & 4.61 & 5.99 & 9.21 & 11.8\cr
 \hbox{3 dof} & 3.53 & 6.25 & 7.81 & 11.3 & 14.2
\end{array}\eeq
In general the PDF of $\theta_i$ has
a non-Gaussian shape, maybe with few local minima
(e.g.\ different solutions to the solar neutrino anomaly).
Nevertheless a cut on the $\Delta\chi^2$ at the value
that would be appropriate in the
gaussian limit is usually an excellent approximation.
Deviations from the gaussian approximation depend on the choice of the prior.
Locally, one can always choose $p(\hbox{theory})$ to enforce this gaussian  approximation.

\medskip

\begin{figure}[t]
$$
\includegraphics[width=0.6\textwidth]{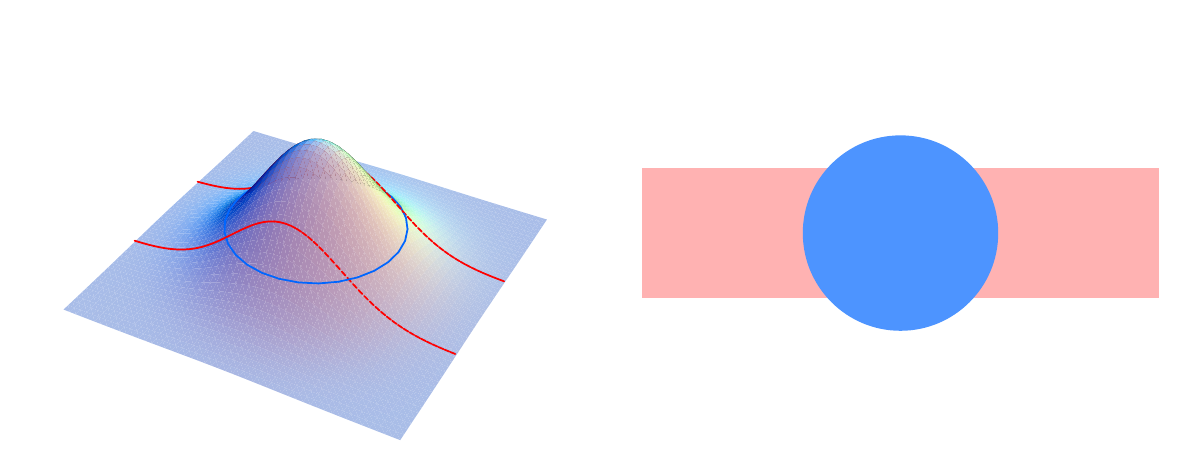}
$$
\caption[$\chi^2$ for different number of dof]{\em Two dimensional Gaussian:
both the circle and the rectangle enclose $68\%$ of the
total probability. The diameter is larger than the height of the rectangle.
Due to this trivial geometric effect, the $\Delta\chi^2$ depends on the number of dof,
as exemplified in eq.\eq{CLsamples}.\label{fig:CerchioStriscia}}
\end{figure}

We have discussed all essential points.
In various circumstances one likes to perform some extra standard manipulations,
that we now describe.
\begin{itemize}
\item Sometimes predictions depend on many unknown parameters,
and one is interested in only a few of them.
However a global fit gives a joint PDF for all parameters.
For example one wants to extract from solar data the 2
oscillation parameters $\theta$, $\Delta m^2$.
Solar fits involve other $k=N-2$ parameters $\lambda$ that describe the sun 
(e.g.\  the predicted neutrino rates depend on the flux of Boron neutrinos)
and the response
of the experimental apparata.
To focus on $\theta,\Delta m^2$ one considers
the probability marginalized with respect to the `nuisance' parameters $\lambda$:
\begin{equation}
\label{eq:marB}
 p(\theta,\Delta m^2|\hbox{data}) = \int d^{k}\lambda ~ p(\theta,\Delta m^2,\lambda|\hbox{data}) 
\end{equation}
In Gaussian approximation this reduces to
\begin{equation}
\label{eq:marG}
\chi^2(\theta,\Delta m^2) = \min_{\lambda} \chi^2(\theta,\Delta m^2,\lambda) 
\end{equation}
and best fit regions are found cutting $\Delta \chi^2(\theta,\Delta m^2)$ 
at the $\Delta \chi^2$ value appropriate for 2 dof ($\Delta m^2$ and $\theta$), not $N$ dof.
In fact, we now show how
minimization over Gaussian nuisance parameters
can be done analytically, 
and the result is equivalent to adding errors in quadrature.
It is convenient to parameterize the $\chi^2$ as follows.
The $k$ nuisance parameters $\lambda$  shift the predictions $T$ for the $n$
observables as $T= T_0 + M \cdot \lambda$
where $M$ is an $n\times k$ matrix.
Forming the differences $\Delta = R-T = \Delta_0 - M\cdot \lambda$ 
between the measured and predicted values,
one gets a $\chi^2$ of the form
\beq \chi^2 = \Delta^T  \cdot V_0  \cdot \Delta + \lambda^T  \cdot v  \cdot \lambda\eeq
where the symmetric matrices $V_0$ and $v$ describe uncertainties on $\Delta$ and on $\lambda$.
The $\chi^2$ is minimized for
$$\lambda = \lambda_{\rm min} = (v + M^T \cdot V_0 \cdot M)^{-1} \cdot  M^T \cdot V_0 \cdot\Delta_0$$
so that
$$\min_\lambda \chi^2= 
\Delta_0^T  \cdot V_0  \cdot \Delta_0 -\lambda_{\rm min}^T (v+M^T  \cdot V_0  \cdot M)\lambda_{\rm min}=
\Delta^T_0\cdot V \cdot \Delta_0$$
with
\beq \label{eq:quadrature}
V= V_0 - V_0  \cdot M  \cdot \frac{1}{v+M^T \cdot  V_0 \cdot  M}  \cdot M^T \cdot  V_0 = \bigg(\frac{1}{V_0} + M \cdot \frac{1}{v}  \cdot M^T\bigg)^{-1}.\eeq
The last expression for $V$ is well known:
Gaussian uncertainties can be combined by
summing in quadrature individual contributions,
and finally inverting the resulting
$n\times n$ covariance matrix
(where $n$ is the number of data points).
The first expression for $V$, or eq.\eq{marG},
shows how the same result can be obtained by inverting a $k\times k$ matrix
(where $k$ is the number of individual sources of uncertainty).
When $k$ or $n$ are large 
one approach (or a combination of the two)
might be significantly
simpler than the other one.

\item Sometimes  one is interested in all $N$ parameters, 
but a function of $N>2$ variables
cannot be published on a 2-dimensional sheet of paper.
Again, it is convenient to report marginalized probabilities.
For example one can extract from solar data
$p(\theta,\Delta m^2,\eta_{\rm s}|R)$ (where $\eta_{\rm s}$ is the sterile neutrino fraction)
and report $p(\theta,\Delta m^2|R)$ and $p(\eta_{\rm s}|R)$.
In gaussian approximation, the $68\%$ CL regions are
obtained by cutting the $\Delta \chi^2$ at the appropriate values:
$\Delta \chi^2(\theta,\Delta m^2)<2.30$ (2 dof)
$\Delta \chi^2(\eta_{\rm s}) < 1$ (1 dof).
Using $\Delta \chi^2 < 3.53$ (3 dof) is wrong,
as pictorially illustrated in fig.\fig{CerchioStriscia}.

\medskip

\item Sometimes one wants to know one specific function $f(\theta)$
of the parameters $\theta$ extracted from a fit.  Its PDF is
\begin{equation}
\label{eq:f}
\begin{array}{rll}
p(f) ~=& \displaystyle\int d^N\!\theta ~ p(\theta|R) ~\delta (f(\theta) - f) &\hbox{Bayesian result}\\
\chi^2(f) ~=&\displaystyle \min_{\theta: f(\theta)=f} \chi^2(\theta)      &    \hbox{Gaussian approximation (1 dof)}
\end{array}
\end{equation} 
Two comments are in order.
1) When some of the parameters $\theta$ have a `large' uncertainty
the Bayesian analysis
(assuming any reasonable prior) automatically disfavors values of $f$ that require a fine-tuning
of the unknown parameters.
On the contrary the gaussian approximation
does not introduce fine-tuning constraints.
2) When $f(\theta)$ is not a simple function, the minimization
demanded by gaussian approximation can be conveniently performed 
using a Lagrange multiplier. In practice one builds $\chi^2(\theta)+\lambda f(\theta)$
and computes the location of its minimum $\theta(\lambda)$ as function of $\lambda$.
By interpolating several points ($\chi^2(\theta(\lambda)),f(\lambda)$)
one gets the desired $\chi^2(f) $.

One concrete problem where these two comments are useful is
computing the range of the $0\nu2\beta$ observable $f=|m_{ee}|$ preferred by oscillation data.


\end{itemize}

\paragraph{2$'$. Neynman inference.}
Unlike steps 1 and 3, step 2 is not fully satisfactory because
introduces the arbitrary prior.
Various physicists prefer an alternative approach,
suggested by Neynman~\cite{Statistics}.\footnote{This is sometimes 
named {\em frequentist} approach, in order
to emphasize that it only needs a restrictive definition of probability
(based on the limiting frequency of repeated experiments)
in place of the more flexible definition
implicit in the Bayesian approach.
A frequentistic
definition of probability is adeguate for e.g.\ uncertainties due to quantum fluctuations,
but does not allow to include theoretical and systematic uncertainties
(e.g.\ higher order QCD corrections
do not fluctuate when the experiment is repeated).
We use the name `Neynman inference' because 
even with a Bayesian definition of probability
the Neynman construction 
can be applied (as an alternative to the Bayes theorem) to build confidence regions.}
It  is useful and practicable only in special cases.
In Gaussian approximation,
a proper Neynman construction
gives the same results of the much simpler Bayesian approach.
Therefore we just outline the basic points, without proofing them.

Knowing the likelihood  $p({\rm data}|{\rm
theory}) = p(R|\theta)$,
for each value of $\theta$ one can build {\em a} range of $R$
that contains $90\%$ of the probability.
If it contains the measured data point, that value of $\theta$ is accepted at $90\%$ CL,
because  $90\%$ of the families of ranges built in this way by repeating the same experiment
contain the unknown true value.
Even the Neynman inference contains an arbitrary choice:
one can build a $90\%$ range of $R$
in many different ways.
The simplest prescription is:
build the range including first values of $R$ with the highest
$p(R|\theta)$ (Crow-Gardner ordering~\cite{Statistics}).
This is inefficient and unstable: in Gaussian approximation
it reduces to a cut on the total $\chi^2$ with $n-N$ dof
(rather than on the $\Delta\chi^2$ with $N$ dof).

A better prescription is:
build the range including first values of $R$ with the highest
 \begin{equation}
\label{eq:FC}
p(R|\theta)/p(R|\theta_{\rm best}(R))
\end{equation} 
(named by physicists as Feldman-Cousins~\cite{Statistics} (FC) ordering,
although it was already described e.g.\ on the book by Kendall-Stuart)
where $\theta_{\rm best}(R)$ is the best-fit value of the theoretical parameters
as function of the possible experimental outcome $R$.
In\eq{FC} the likelihood $p$ is divided by the maximal value that $p$ can get
within the given theoretical model.
Assuming that the theory is right, when we get an 
experimental outcome that is unlikely for all possible values
of $\theta$, we know that a statistical fluctuation occurred and we can
correct for it.
In this way the FC ordering recognizes and
eliminates the statistical fluctuations that have nothing to do with the determination of the parameters.
In Gaussian approximation it reduces to the cut on the
$\Delta\chi^2$ discussed at point 3.
If irrelevant data are included in the data-set, they contribute to the total
$\chi^2$ but cancel out when computing the $\Delta\chi^2$.

In special situations the FC regions can appreciably different from their gaussian approximation.
For example, when fitting oscillation data at values of $\Delta m^2$ so large that the
oscillation probability practically depends only on the mixing angle $\theta$
(i.e.\ in region C of fig.\fig{tipico})
this technique recognizes that $\Delta m^2$ is no longer a relevant parameter 
and gives allowed regions that correspond to a cut on the $\Delta \chi^2$ with only 1 dof,
rather than with 2 dof.
When $n\circa{>} 10$ (i.e.\ there are more than $\sim 10 $ data $R$),
computing the FC regions exceeds the
numerical capabilities of present computers,
unless it is possible to reliably perform the necessary multi-dimensional integrations with  MonteCarlo techniques.
The FC construction looks intuitively obvious, when presented
as a MonteCarlo procedure.
The CL at which any given value of $\theta$ is accepted can be computed as follows:
\begin{itemize}
  \item[a)] Compute the value of $r_{\rm exp}=p(R|\theta)/p(R|\theta_{\rm best}(R))$
(the $\Delta\chi^2$, in gaussian approximation),
where $R$ is the experimental result.

  \item[b)] Assuming the desired value of $\theta$, generate a large enough sample
of simulated experimental results $R$.
For each $R$ compute its value of $r$.

\item[c)]  Compute how frequently one obtains $r>r_{\rm exp}$.
This is the CL.
\end{itemize}


\medskip

The extra manipulations described at point 3 can be performed
by inventing appropriate non-standard ordering prescription
or by adopting semi-Bayesian methods (e.g.\ marginalizing the likelihood).
In gaussian approximation the final FC results agree
with the much simpler Bayesian results previously discussed.

\section{Goodness-of-fit}\label{gof}
In the previous section we discussed how to use data to infer unknown theoretical parameters,
assuming that the theory is true.
We now discuss {\em how to test if the theory describes data in a good way}?
This is a more difficult question that does not has a fully satisfactory answer nor a
fully precise meaning.
But of course this is what we want to know.

The true answer is: {\em look carefully at data searching for possible problems}.
This seems `artigianal', but we do not know any `scientific'
statistical test that can be blindly applied
giving efficient answers in the general case.

 \smallskip

We first describe the frequently used Pearson $\chi^2$ test,
that provides one `objective' answer.
Let us consider $z=\min {\chi}^2$, the value of the total $\chi^2$ evaluated at the best-fit point.
If the theory and the data are correct, one expects
that eq.\eq{chiq} gives its probability distribution $p_N(z)$
where $N$ is the total number of degrees of freedom:
$$ N = \hbox{number of data points} - \hbox{number of unknown parameters}.$$
One usually reports the goodness-of-fit (gof), defined as
the probability of having a $z$ larger than the measured value.

For $N\gg1$ $p_N$ can be approximated by a Gaussian with
mean $\mu_N = N$ and standard deviation $\sigma_N = \sqrt{2N}$:
one expects $\min {\chi}^2=N\pm\sqrt{2N}$.
$\sigma_N$ grows with $N$ because
when $N\gg 1$ one should expect some rare statistical fluctuation in the data-set: e.g.\
it is not unlikely to find a $4\sigma$ anomaly in a set of 100 data.
If the anomaly is there because the theory makes a $4\sigma$ wrong prediction,
the $\chi^2$ test is not good enough to recognize that the theory is wrong.
A practical example: the solar neutrino data-set has $N\sim 100$ observables
and, according to the $\chi^2$ test,  some excluded solutions still have an acceptable gof.

A fit can be bad for different reasons.
Depending on the specific case, it can be easy to perform an alternative more efficient test.
In a real analysis one often encounters a mix of the following typical situations.
\begin{itemize}
\item[a)] A simple recipe allows to improve the efficiency of the Pearson $\chi^2$ test
while maintaining its general validity. 
{\em Perform the standard Pearson $\chi^2$ test including only the data that really test the theory}.
The improvement can be significant because data-sets are often chosen in the opposite way, including even irrelevant data.
Authors like to claim ``we perform a complete analysis'' and feel
not politically correct to drop results given by
less accurate experiments.
Indeed, fitting all data is the right way to  get the
best possible inferences on unknown parameters:
the tools discussed in the previous section automatically 
ignore statistical fluctuation unrelated to determination of parameters.
But putting all together is not the right way of checking if some gof problem is present.
For example, using many energy bins with width smaller than the energy resolution
does not affect best-fit regions, but affects the total $\chi^2$
deteriorating its efficiency.

\item[b)] {\em In some cases the fit is bad because different pieces of data
are mutually incompatible}, in any `reasonable' theory.
For example let us suppose that an experiment
receives a flux of neutrinos, known to be time-independent, finding
$100\pm\sqrt{100}$ events during the first year and $200\pm\sqrt{200}$ in the second year.
While a na\"{\i}ve Pearson $\chi^2$ test would tell that any theory is disfavored,
a `reasonable' result is obtained by fitting the total number of events,
as suggested in a).

It is interesting to notice that the rate per year if of course given by 
$\frac{1}{2}(300\pm\sqrt{300})$,
equal to the weighted average of the two annual rates.
This shows that, if the discrepancy between 100 and 200 is due to a statistical fluctuation,
it is wrong to enlarge the error on the average rate
(possibly following one of the proposed recipes).
Errors should be enlarged only if one thinks that there is something wrong in the experiment.

\item[c)] {\em In some cases the fit is bad because different pieces of data
are mutually incompatible within the assumed theory}.
In these cases one can perform a dedicated test.
For example, the SMA solution to the solar neutrino problem was first excluded
because it can not suppress the rates without also giving unseen spectral distortions.
Splitting the data into 2 sets (rates and spectra) and fitting them
separately, the SK collaboration obtained two disjoint best-fit regions,
thereby showing that SMA oscillations allowed by rates were not allowed by spectral data.

Since there are many ways of splitting data into different sets,
with this procedure one may obtain artificially low gof values by
choosing a particular splitting after having seen the data.
This is a well known practical problem of various statistical analyses.
Therefore the splitting should be justified by honest physical motivations.


\end{itemize}

%% file: review_ref.tex

\small
\paragraph{Acknowledgments}
We would like to thank our numerous collaborators; some of our  
recent papers arose as pieces of this review that needed original work.
Special thanks to
E.~Bellotti, R.~Bernabei, A.~Bettini, C.~Bucci,  
C.~Cattadori, F.~Cavanna, M.~Cirelli, O.~Cremonesi, S.~Dugad, M.~Frigerio,
W.~Fulgione, P.L.~Ghia, 
A.~Giuliani, D.~Grasso, W.~Hofmann, H.V.~Klapdor-Kleingrothaus, G.~Mills, O.~Palamara, G.~Riccobene 
for the help with experimental data and interpretations;  to F.~Aharonian,
J.F.~Beacom, C.~Distefano, G.~Fiorentini, N.~Fornengo, V.S.~Imshennik,
M.~Kachelrie\ss,
P.~Lipari, T.~Montaruli, D.K.~Nadyozhin, G.~Navarra, O.~Ryazhskaya,
K.~Sato, M.~Selvi for discussions and explanations regarding
astrophysics/astroparticle physics; to R.\ Barbieri,
G.~Battistoni, Z.~Berezhiani,
W.~Buchmuller, A.~Butkevich, M.~Chen, G.~Co, S.~Davidson, 
T.~Hambye, F.~Feruglio, K.~Kar, A.~Ianni, R.N.~Mohapatra,
B.~Mukhopadhyaya, M. Nishimura, E.~Paschos, N.~Paver,
S.~Petcov, M.~Picariello, M.~Raidal, A.~Romanino, R.~Rattazzi, S.~Rigolin,  
E.~Roulet, R.~Spaeti, R.~Thorne, P.~Vogel, T.~Yanagida for help with 
various theoretical aspects.
F.V.~is particularly grateful to V.S.~Berezinsky, 
S.~Bertolini, M.L.~Costantini, F.~Nesti, G.~Senjanovi\'c,
A.Yu.~Smirnov. 

Preliminary versions of this review have been available
since 2002 at  \url{www.cern.ch/astrumia/review.html},
presented at the
XI Seminario Nazionale di Fisica Teorica, Parma 2002,
lost in a disk crash in 2003, eaten by S.~Strumia in 2004, 
stolen in Madrid metro in 2005.

\footnotesize\newpage
\begin{multicols}{2}

\end{multicols}
